\documentclass[twoside,11pt,indexing]{book}                                   %

\usepackage{mon_de}
\usepackage{wrapfig}

\usepackage{makeidx}

\renewcommand{\thesection}{\arabic{chapter}.\arabic{section}} %
\renewcommand{\thetable}{\arabic{table}}
\renewcommand{\thefigure}{\arabic{figure}}

\usepackage{eso-pic}
\begin{document}
 
\AddToShipoutPicture*{{\includegraphics[scale=1.2,keepaspectratio]{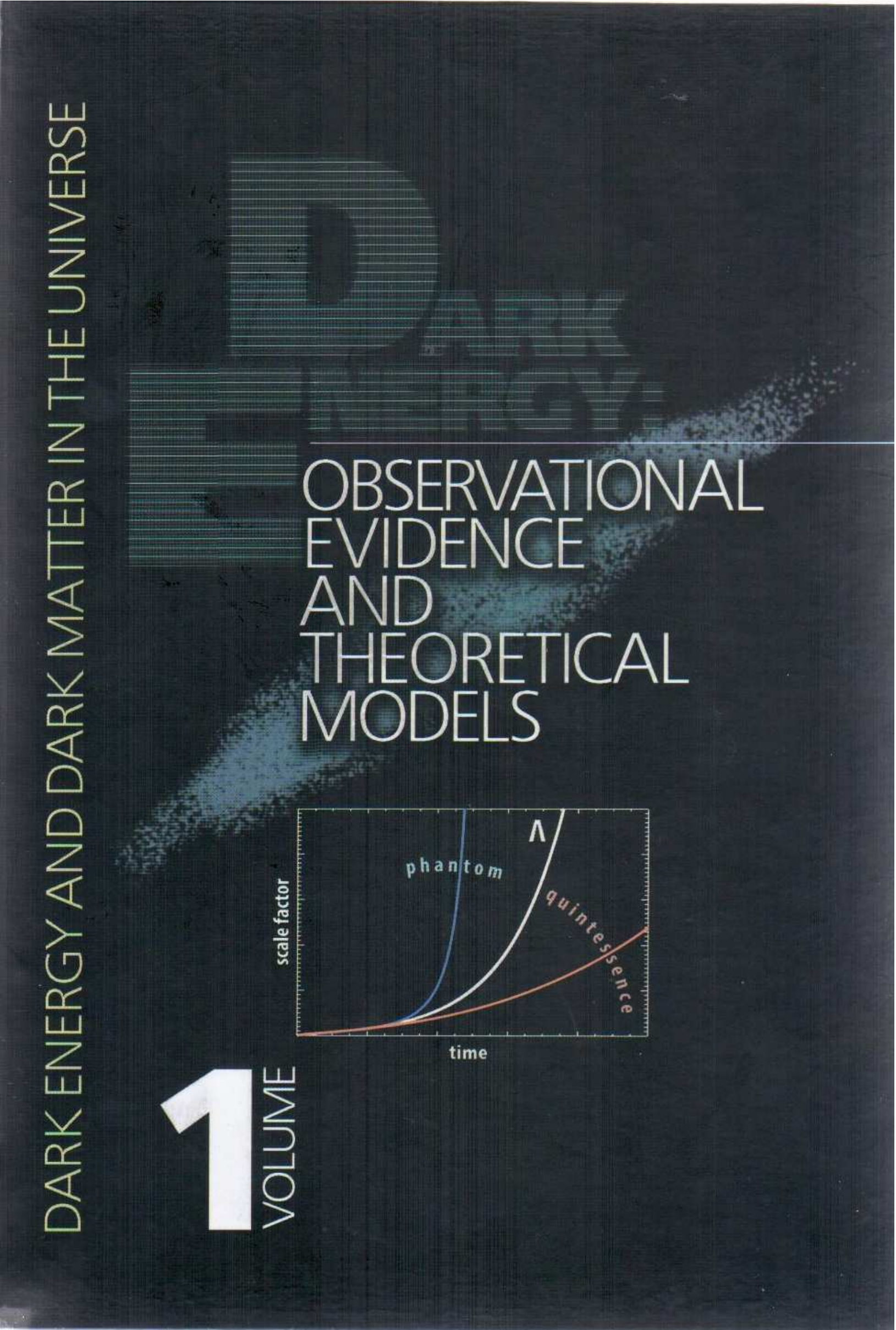}}} 

\setcounter{page}{1}%

\parindent=7mm

\newpage
\mbox{ }\thispagestyle{empty} 
...

\newpage
\mbox{ }\thispagestyle{empty} 

\begin{center}

\mbox{ }\vspace*{2cm}

{\LARGE\bfseries Dark energy and dark matter in the Universe}\\[3mm]
{\Large\bfseries in three volumes}
    \vspace*{1.2cm}

{\Large
  B.\,Novosyadlyj, V.\,Pelykh, Yu.\,Shtanov, A.\,Zhuk}
\vspace*{1cm}

   {\LARGE\bfseries DARK ENERGY:\\[5pt]
     OBSERVATIONAL EVIDENCE\\[1mm] AND THEORETICAL MODELS}\\[0.5cm]
  {\LARGE\bfseries Volume 1}\\[1cm]
   {\Large Editor V.\,Shulga\\
   \mbox{ }\vspace{-0.1cm}

  \rule{7.8cm}{0.5pt}\\[8pt]
  PROJECT\\
  <<UKRAINIAN SCIENTIFIC BOOK\\
  IN A FOREIGN LANGUAGE>>
  \rule{7.8cm}{0.5pt}\\

  \mbox{ }\vspace{1cm}

  KYIV $\cdot$ AKADEMPERIODYKA $\cdot$ 2013}

\end{center}

\newpage
\mbox{ }\thispagestyle{empty}

\vspace*{1.0cm}
\hspace*{3mm}\parbox{11.2cm}{\begin{raggedleft}
  {\sffamily
  {\normalsize NATIONAL ACADEMY OF SCIENCES OF UKRAINE}\\[1mm]
  {\small BOGOLYUBOV INSTITUTE FOR THEORETICAL PHYSICS\\[1mm]
  PIDSTRYHACH INSTITUTE FOR APPLIED PROBLEMS\\[-0.5mm] OF MECHANICS AND MATHEMATICS \\[1mm]
  RADIOASTRONOMICAL INSTITUTE\\[2mm]
  IVAN FRANKO NATIONAL UNIVERSITY OF LVIV\\[1mm]
  MECHNIKOV NATIONAL UNIVERSITY OF ODESSA

  }

\vspace{-0.2cm}
  \rule{11.1cm}{0.5pt}\linebreak
  \vspace{-0.4cm}

\selectlanguage{ukrainian}
{\normalsize ═└╓▓╬═└╦▄═└ └╩└─┼╠▓▀ ═└╙╩ ╙╩╨└п═╚}\\[1mm]
{\small ▓═╤╥╚╥╙╥ ╥┼╬╨┼╥╚╫═╬п ╘▓╟╚╩╚ │ь.\,╠.╠.\,┴╬├╬╦▐┴╬┬└\\[1mm]
▓═╤╥╚╥╙╥ ╧╨╚╩╦└─═╚╒ ╧╨╬┴╦┼╠ ╠┼╒└═▓╩╚\\[-0.5mm] ▓ ╠└╥┼╠└╥╚╩╚ │ь.\,▀.╤.\,╧▓─╤╥╨╚├└╫└\\[1mm]
╨└─▓╬└╤╥╨╬═╬╠▓╫═╚╔ ▓═╤╥╚╥╙╥\\[2mm]
╦▄┬▓┬╤▄╩╚╔ ═└╓▓╬═└╦▄═╚╔ ╙═▓┬┼╨╤╚╥┼╥ │ьхэ│\,▓┬└═└\,╘╨└═╩└\\[1mm]
╬─┼╤▄╩╚╔ ═└╓▓╬═└╦▄═╚╔ ╙═▓┬┼╨╤╚╥┼╥ │ьхэ│\,▓.▓.\,╠┼╫═╚╩╬┬└

}}

\end{raggedleft}}

\newpage
\mbox{ }\thispagestyle{empty} 

\vbox{\vspace*{-5.2cm}\hbox{\raisebox{0.6cm}{\parbox[b]{3cm}{\small
  \noindent UDK 524.8+539\\
  \noindent BBK 22.6+22.3\\
\hspace*{9mm}D20}}}}\vspace*{11mm}

{\small\noindent\textbf{Reviewers:}\\[1mm]
{\it S.I.\,VILCHYNSKIY}, Dr. Sc., Professor,  Head of Department of Quantum\\
Field Theory at Taras Shevchenko National
University of Kyiv\\[0.5mm]
{\it O.B.\,ZASLAVSKII}, Dr. Sc., Senior researcher, Leading researcher\\ at V.N.\,Karazin Kharkiv National University\\

\vspace*{7.0mm}

\noindent{\small\itshape{Approved for publication by the
    Scientific Council of the Bogolyubov Institute\\ for Theoretical
    Physics and Pidstryhach Institute for Applied Problems of
    Mechanics\\ and Mathematics of the National Academy of Sciences of
    Ukraine

    }\vspace*{3mm}

\noindent\bfseries\itshape{Publication was made possible  by a State
contract promoting\\ the production of scientific printed material}

}

\vspace*{33.0mm}

\noindent \hspace*{22.0mm}{\small
D20}~~\raisebox{0.36cm}{\parbox[t]{10.05cm}{\hspace*{7mm}{\small
{\bf Dark} energy and dark matter in the Universe: in  three
vo\-lu\-mes\,/\,Editor\,\,V.\,\,Shulga.\,\,---\,\,Vol.\,\,1.\,\,Dark\,\,energy:\,\,ob\-ser\-va\-tio\-nal
evidence and theoretical models\,/\,{Novosyadlyj\,\,B.,
Pelykh\,\,V., Shtanov\,\,Yu.,
 Zhuk\,\,A.~--- K.\,: Akademperiodyka, 2013. --- 380\,p.}  \\[1.5mm]
\hspace*{7mm}ISBN 978-966-360-239-4\\[1.0mm]
\hspace*{7mm}ISBN 978-966-360-240-0 (vol.\,1)\\[1.0mm]}
{\scriptsize\hspace*{7mm}The  book elucidates the current state of
the dark energy problem and pre\-sents the results of the authors,
who work in this area.  It describes the ob-\linebreak servational
evidence for the existence of dark energy, the methods and results
of constraining of its parameters, modeling of dark energy by scalar
fields, the space-times with extra spatial dimensions, especially
Kaluza---Klein models, the braneworld models with a single extra
dimension as well as the problems of positive definition of
gravitational energy in General Relativity, energy conditions and
consequences of their
violation in the presence of dark energy.\\
\hspace*{7mm}This monograph is intended for science professionals,
educators and graduate students, specializing in general relativity,
cosmology, field theory and particle physics. \vspace*{-0.5mm}

}

\begin{flushright}\vspace*{-4mm}
{\scriptsize\bf UDC 524.8+539\\[-1mm]
BBK 22.6+22.3}
\end{flushright}}}

\vspace*{4mm}\noindent \parbox[b]{8cm}{\bf ISBN 978-966-360-239-4\\
ISBN 978-966-360-240-0
(vol.\,1)}\hspace*{0.15cm}\parbox[b]{5.3cm}{\footnotesize\copyright\,\,Novosyadlyj B., Pelykh V.,\\
\phantom{\copyright\,\,}Shtanov Yu.,
 Zhuk  A., 2013\\
 \copyright\,\,Akademperiodyka, design, 2013}

}

\newpage


 {\parindent=0pt\hspace*{2.9cm}
\vbox{\vspace*{-1.6cm}\hbox{\large\raisebox{0pt}{{\par}}}
\vspace*{0.6cm}\hbox{$\blacksquare$}
 \vspace{-3.0mm}
\hbox{\rule{10cm}{1pt}}
\vspace{2mm}%
\raggedright{\bfseries\sffamily\Large  CONTENTS}\\%
\vspace*{-1mm} \hbox{\rule{10cm}{1pt}}\par \vspace{39mm}}}


\markboth{Contents$_{ }$}{Contents$_{ }$} 
\begin{picture}(10,10)
\put(121,-144.2){\bfseries\sffamily{5}}
\end{picture}

\thispagestyle{empty}

\vspace{-3.2cm}

\begin{picture}(30,30)
\put(7,-25.2){\sffamily\bf\Huge\sffamily 1} \put(0,-29.2)
{\small\sffamily
C\hspace{0.5mm}H\hspace{0.5mm}A\hspace{0.5mm}P\hspace{0.5mm}T\hspace{0.5mm}E\hspace{0.5mm}R}
\put(7,-123.2){\sffamily\bf\Huge\sffamily 2}%
\put(0,-127.2) {\small\sffamily
C\hspace{0.5mm}H\hspace{0.5mm}A\hspace{0.5mm}P\hspace{0.5mm}T\hspace{0.5mm}E\hspace{0.5mm}R}
\end{picture}
 {\small
\begin{tabbing}
AAAAAAAAAAAAAAAAAAAAAAAAxxx\=A \kill  %
\hspace{3cm}\parbox[b]{9.4cm}{{FOREWORD OF THE EDITOR}\dotfill}\>\`\,\,9  \\ [3mm]%
\hspace{3cm}\parbox[b]{9.4cm}{{ACKNOWLEDGMENTS}\dotfill}\>\`11  \\ [3mm]%
\hspace{3cm}\parbox[b]{9.4cm}{\normalsize\bf\sffamily Observational
evidence for dark energy}  \>\`{ } \\ [2mm]
\hspace{3cm}\parbox[b]{9.4cm}{{\bf 1.1.~}Introduction\dotfill}  \>\`13\\ [0.4mm]%
\hspace{3cm}\parbox[b]{9.4cm}{{\bf 1.2.~}Dynamics of expansion
of the homogeneous isotropic multicomponent Universe\dotfill}  \>\`16 \\ [0.4mm]%
\hspace{3cm}\parbox[b]{9.4cm}{{\bf 1.3.~}The luminosity distance~---
redshift
relation and SNe Ia evidence for dark energy\dotfill}  \>\`22 \\ [0.4mm]%
\hspace{3cm}\parbox[b]{9.4cm}{{\bf 1.4.~}The angular diameter distance~--- redshift relation and acoustic peak tests\dotfill}  \>\`30 \\ [0.4mm]%
\hspace{3cm}\parbox[b]{9.4cm}{{\bfseries 1.4.1.~}CMB acoustic peaks\dotfill}  \>\`30 \\ [0.4mm]%
\hspace{3cm}\parbox[b]{9.4cm}{{\bfseries 1.4.2.~}Baryon acoustic oscillations\dotfill}  \>\`38 \\ [0.4mm]%
\hspace{3cm}\parbox[b]{9.4cm}{{\bfseries 1.4.3.~}X-ray gas fraction in clusters\dotfill}  \>\`42 \\ [0.4mm]%
\hspace{3cm}\parbox[b]{9.4cm}{{\bf 1.5.~}Evidence for dark energy
from study of
large scale structure\dotfill}  \>\`45 \\ [0.4mm]%
\hspace{3cm}\parbox[b]{9.4cm}{{\bfseries 1.5.1.~}Linear power spectrum of matter density perturbations\dotfill}  \>\`45 \\ [0.4mm]%
\hspace{3cm}\parbox[b]{9.4cm}{{\bf 1.6.~}Angular power spectrum of
CMB temperature
  fluctua\-tions\dotfill}  \>\`55 \\ [0.4mm]%
\hspace{3cm}\parbox[b]{9.4cm}{{\bfseries 1.6.1.~}Integrated Sachs---Wolfe effect\dotfill}  \>\`55 \\ [0.4mm]%
\hspace{3cm}\parbox[b]{9.4cm}{{\bfseries 1.6.2.~}Weak gravitational lensing of CMB\dotfill}  \>\`60 \\ [0.4mm]%
\hspace{3cm}\parbox[b]{9.4cm}{{\bf 1.7.~}Age of the Universe\dotfill}  \>\`62 \\ [0.7mm]%
\hspace{3cm}\parbox[b]{9.4cm}{{\bf 1.8.~}Constraints on dark energy
parameters from
  combined data\dotfill}  \>\`64 \\ [0.7mm]%
\hspace{3cm}\parbox[b]{9.4cm}{{\bf 1.9.~}Summary\dotfill}  \>\`70 \\ [3mm]%

\hspace{3cm}\parbox[b]{9.4cm}{\normalsize\bf\sffamily Scalar field models of dark energy}  \>\`{ } \\ [2mm]%
\hspace{3cm}\parbox[b]{9.4cm}{{\bf 2.1.~}Introduction\dotfill}  \>\`72\\ [0.4mm]%
\hspace{3cm}\parbox[b]{9.4cm}{{\bf 2.2.~}Cosmological constant as
vacuum energy: ideas
and problems\dotfill}  \>\`73 \\ [0.4mm]%
\hspace{3cm}\parbox[b]{9.4cm}{{\bf 2.3.~}Scalar fields as dark
energy\dotfill}  \>\`74 \\ 

\hspace{3cm}\parbox[b]{9.2cm}{{\bf 2.4.~}Scalar
  perturbations of the scalar field and other components\dotfill}  \>\`77 \\ [0.55mm]%
\hspace{3cm}\parbox[b]{9.2cm}{{\bf 2.5.~}Specifying the scalar-field
models of dark
energy\dotfill}  \>\`82\\ [0.55mm]%
\hspace{3cm}\parbox[b]{9.2cm}{{\bfseries 2.5.1.~}Lagrangian\dotfill}  \>\`83 \\ [0.55mm]%
\hspace{3cm}\parbox[b]{9.2cm}{{\bfseries 2.5.2.~}Potential\dotfill}  \>\`83 \\ [0.55mm]%
\hspace{3cm}\parbox[b]{9.2cm}{{\bfseries 2.5.3.~}EoS parameter\dotfill}  \>\`86 \\ [0.55mm]%
\hspace{3cm}\parbox[b]{9.2cm}{{\bfseries 2.5.4.~}The effective sound speed\dotfill}  \>\`97 \\ [0.55mm]%
\hspace{3cm}\parbox[b]{9.2cm}{{\bf 2.6.~}Quintessential scalar
fields with barotropic EoS\dotfill}  \>\`100 \\ [0.55mm]%
\hspace{3cm}\parbox[b]{9.2cm}{{\bfseries 2.6.1.~}Classical scalar field\dotfill}  \>\`100 \\ [0.55mm]%
\hspace{3cm}\parbox[b]{9.2cm}{{\bfseries 2.6.2.~}Tachyonic scalar field\dotfill}  \>\`103 \\ [0.55mm]%
\hspace{3cm}\parbox[b]{9.2cm}{{\bfseries 2.6.3.~}Quintessential scalar fields in the phase plane\dotfill}  \>\`105 \\ [0.55mm]%
\hspace{3cm}\parbox[b]{9.2cm}{{\bfseries 2.6.4.~}Best-fit parameters of QSF\dotfill}  \>\`107 \\ [0.55mm]%
\hspace{3cm}\parbox[b]{9.2cm}{{\bf 2.7.~}Phantom scalar fields with barotropic EoS\dotfill}  \>\`113 \\ [0.55mm]%
\hspace{3cm}\parbox[b]{9.2cm}{{\bfseries 2.7.1.~}Gravitation instability of PSF and large scale struc\-ture formation\dotfill}  \>\`119 \\ [0.55mm]%
\hspace{3cm}\parbox[b]{9.2cm}{{\bfseries 2.7.2.~}Best-fit parameters of PSF\dotfill}  \>\`122 \\ [0.55mm]%
\hspace{3cm}\parbox[b]{9.2cm}{{\bf 2.8.~}Distinguishing of scalar
field models of
dark energy\dotfill}  \>\`123 \\[0.55mm] %
\hspace{3cm}\parbox[b]{9.2cm}{{\bf 2.9.~}Summary\dotfill}  \>\`130 \\ 
\vbox{\vspace*{-3mm}\begin{picture}(0,-30)
\put(7,-7.0){\sffamily\bf\Huge\sffamily 3} \put(0,-11.0)
{\small\sffamily
C\hspace{0.55mm}H\hspace{0.55mm}A\hspace{0.55mm}P\hspace{0.55mm}T\hspace{0.55mm}E\hspace{0.55mm}R}
\end{picture}}\\
\hspace{3cm}\parbox[b]{9.2cm}{\normalsize\bf\sffamily Kaluza---Klein models}  \>\`{} \\ [2mm]%
\hspace{3cm}\parbox[b]{9.2cm}{{\bf 3.1.~}Introduction\dotfill}  \>\`133 \\ [0.55mm]%
\hspace{3cm}\parbox[b]{9.2cm}{{\bf 3.2.~}Dimensional reduction,
stable compactification, gravita\-tional excitons, effective
cosmological
constant\dotfill}  \>\`136 \\ [0.55mm]%
\hspace{3cm}\parbox[b]{9.2cm}{{\bfseries 3.2.1.~}General setup\dotfill}  \>\`136 \\ [0.55mm]%
\hspace{3cm}\parbox[b]{9.2cm}{{\bfseries 3.2.2.~}Stable compactification with minimal scalar fields\dotfill}  \>\`140 \\ [0.55mm]%
\hspace{3cm}\parbox[b]{9.2cm}{{\bfseries 3.2.3.~}Perfect fluid: no-go theorem\dotfill}  \>\`142 \\ [0.55mm]%
\hspace{3cm}\parbox[b]{9.2cm}{{\bfseries 3.2.4.~}Conventional cosmology from multidimensional mo\-dels\dotfill}  \>\`145 \\ [0.55mm]%
\hspace{3cm}\parbox[b]{9.2cm}{{\bf 3.3.~}Abelian gauge fields in KK models, dimensional reduc\-tion\dotfill}  \>\`152 \\ [0.55mm]%
\hspace{3cm}\parbox[b]{9.2cm}{{\bf 3.4.~}Gravitational excitons and their cosmological and astro\-physical implications. Dark matter from extra  dimensions}  \>\`155 \\ [0.55mm]%
\hspace{3cm}\parbox[b]{9.2cm}{{\bfseries 3.4.1.~}Effective equation of motion for gravexcitons\dotfill}  \>\`156 \\ [0.55mm]%
\hspace{3cm}\parbox[b]{9.2cm}{{\bfseries 3.4.2.~}Light and ultra-light gravexcitons: $m_{\psi} \leq 10^{-2}$\,GeV}  \>\`158 \\ [0.55mm]%
\hspace{3cm}\parbox[b]{9.2cm}{{\bfseries 3.4.3.~}Heavy gravexcitons: $m_{\psi} \geq 10^{-2}$\,GeV\dotfill}  \>\`160 \\ [0.55mm]%
\hspace{3cm}\parbox[b]{9.2cm}{{\bfseries 3.4.4.~}Variation of the fine-structure constant\dotfill}  \>\`163 \\ [0.55mm]%
\hspace{3cm}\parbox[b]{9.2cm}{{\bfseries 3.4.5.~}Lorentz invariance violation\dotfill}  \>\`165 \\ [0.55mm]%
\hspace{3cm}\parbox[b]{9.2cm}{{\bf 3.5.~}Dark energy in cur\-va\-tu\-re-non-linear $f(R)$ multidimen\-sional cosmological models\dotfill}  \>\`170\\ [0.55mm]%
\hspace{3cm}\parbox[b]{9.2cm}{{\bfseries 3.5.1.~}Internal space stabilization for pure geometrical $f(R)$ models\dotfill}  \>\`171 \\ [0.55mm]%
\hspace{3cm}\parbox[b]{9.2cm}{{\bfseries 3.5.2.~}Dark energy in $f(R)$ models with form fields\dotfill}  \>\`177 \\ [0.55mm]%
\hspace{3cm}\parbox[b]{9.2cm}{{\bf 3.6.~}S$p$-branes. Dynamical dark energy from extra   dimen\-sions\dotfill}  \>\`183 \\ [0.4mm]%
\hspace{3cm}\parbox[b]{9.2cm}{{\bfseries 3.6.1.~}Dark energy in pure geometrical S$p$-brane model with hyperbolic internal space\dotfill}  \>\`187 \\ [0.4mm]%
\hspace{3cm}\parbox[b]{9.2cm}{{\bf 3.7.~}Problematic aspects of Kaluza---Klein models\dotfill}  \>\`192 \\ [0.4mm]%
\hspace{3cm}\parbox[b]{9.2cm}{{\bfseries 3.7.1.~}Equations of state in general case\dotfill}  \>\`195 \\ [0.4mm]%
\hspace{3cm}\parbox[b]{9.2cm}{{\bfseries 3.7.2.~}Latent solitons\dotfill}  \>\`197 \\ [0.4mm]%
\hspace{3cm}\parbox[b]{9.2cm}{{\bfseries 3.7.3.~}Experimental restrictions on the equations of state of a multidimensional perfect fluid\dotfill}  \>\`199 \\ [0.4mm]%
\hspace{3cm}\parbox[b]{9.2cm}{{\bf 3.8.~}Summary\dotfill}  \>\`203 \\ %
\vbox{\begin{picture}(0,-30) \put(7,-7.0){\sffamily\bf\Huge\sffamily
4} \put(0,-11.0) {\small\sffamily
C\hspace{0.55mm}H\hspace{0.55mm}A\hspace{0.55mm}P\hspace{0.55mm}T\hspace{0.55mm}E\hspace{0.55mm}R}
\end{picture}}\\
\hspace{3cm}\parbox[b]{9.2cm}{\normalsize\bf\sffamily Braneworld models}  \>\`{} \\ [2mm]%
\hspace{3cm}\parbox[b]{9.2cm}{{\bf 4.1.~}Introduction\dotfill}  \>\`206\\ [0.4mm]%
\hspace{3cm}\parbox[b]{9.2cm}{{\bf 4.2.~}General setup and notation\dotfill}  \>\`208 \\ [0.4mm]%
\hspace{3cm}\parbox[b]{9.2cm}{{\bf 4.3.~}Cosmological solutions\dotfill}  \>\`211 \\ [0.4mm]%
\hspace{3cm}\parbox[b]{9.2cm}{{\bf 4.4.~}Vacuum and static branes\dotfill}  \>\`214 \\ [0.4mm]%
\hspace{3cm}\parbox[b]{9.2cm}{{\bf 4.5.~}Properties of braneworld gravity\dotfill}  \>\`218 \\ [0.4mm]%
\hspace{3cm}\parbox[b]{9.2cm}{{\bf 4.6.~}Phantom property of braneworld dark energy\dotfill}  \>\`224 \\ [0.4mm]%
\hspace{3cm}\parbox[b]{9.2cm}{{\bf 4.7.~}Disappearing dark energy\dotfill}  \>\`232 \\ [0.4mm]%
\hspace{3cm}\parbox[b]{9.2cm}{{\bf 4.8.~}Cosmic mimicry\dotfill}  \>\`235 \\ [0.4mm]%
\hspace{3cm}\parbox[b]{9.2cm}{{\bf 4.9.~}Loitering\dotfill}  \>\`242 \\ [0.4mm]%
\hspace{3cm}\parbox[b]{9.2cm}{{\bfseries 4.9.1.~}Loitering Universe\dotfill}  \>\`242 \\ [0.4mm]%
\hspace{3cm}\parbox[b]{9.2cm}{{\bfseries 4.9.2.~}Loitering in braneworld models\dotfill}  \>\`243 \\ [0.4mm]%
\hspace{3cm}\parbox[b]{9.2cm}{{\bfseries 4.9.3.~}The parameter space in loitering models\dotfill}  \>\`250 \\ [0.4mm]%
\hspace{3cm}\parbox[b]{9.2cm}{{\bfseries 4.9.4.~}Inflation in braneworld models with loitering\dotfill}  \>\`252 \\ [0.4mm]%
\hspace{3cm}\parbox[b]{9.2cm}{{\bf 4.10.~}Quiescent singularities\dotfill}  \>\`254 \\ [0.4mm]%
\hspace{3cm}\parbox[b]{9.2cm}{{\bfseries 4.10.1.~}Homogeneous case\dotfill}  \>\`254 \\ [0.4mm]%
\hspace{3cm}\parbox[b]{9.2cm}{{\bfseries 4.10.2.~}Inhomogeneous case\dotfill}  \>\`258 \\ [0.4mm]%
\hspace{3cm}\parbox[b]{9.2cm}{{\bf 4.11.~}Asymmetric branes\dotfill}  \>\`262 \\ [0.4mm]%
\hspace{3cm}\parbox[b]{9.2cm}{{\bfseries 4.11.1.~}Induced cosmological constant on the brane\dotfill}  \>\`264 \\ [0.4mm]%
\hspace{3cm}\parbox[b]{9.2cm}{{\bfseries 4.11.2.~}Cosmic mimicry\dotfill}  \>\`265 \\ [0.4mm]%
\hspace{3cm}\parbox[b]{9.2cm}{{\bfseries 4.11.3.~}Phantom branes\dotfill}  \>\`267 \\ [0.4mm]%
\hspace{3cm}\parbox[b]{9.2cm}{{\bfseries 4.11.4.~}Disappearing dark energy\dotfill}  \>\`269 \\ [0.4mm]%
\hspace{3cm}\parbox[b]{9.2cm}{{\bfseries 4.11.5.~}Quiescent singularities\dotfill}  \>\`271 \\ [0.4mm]%
\hspace{3cm}\parbox[b]{9.2cm}{{\bfseries 4.11.6.~}Stability issues\dotfill}  \>\`272 \\ [0.4mm]%
\hspace{3cm}\parbox[b]{9.2cm}{{\bf 4.12.~}Gravitational instability on the brane\dotfill}  \>\`272 \\ [0.4mm]%
\hspace{3cm}\parbox[b]{9.2cm}{{\bfseries 4.12.1.~}Scalar cosmological perturbations on the brane\dotfill}  \>\`273 \\ [0.4mm]%
\hspace{3cm}\parbox[b]{9.2cm}{{\bfseries 4.12.2.~}Simplified boundary conditions for scalar perturba\-tions\dotfill}  \>\`275 \\ [0.4mm]%
\hspace{3cm}\parbox[b]{9.2cm}{{\bfseries 4.12.3.~}Scalar perturbations in the DGP model\dotfill}  \>\`279 \\ [0.4mm]%
\hspace{3cm}\parbox[b]{9.2cm}{{\bfseries 4.12.4.~}Scalar perturbations in the mimicry model\dotfill}  \>\`283 \\ [0.4mm]%
\hspace{3cm}\parbox[b]{9.2cm}{{\bf 4.13.~}Perturbations of the bulk\dotfill}  \>\`288 \\ [0.4mm]%
\hspace{3cm}\parbox[b]{9.2cm}{{\bfseries 4.13.1.~}General system of equations\dotfill}  \>\`288 \\ [0.4mm]%
\hspace{3cm}\parbox[b]{9.2cm}{{\bfseries 4.13.2.~}Perturbations on the flat background bulk geometry\dotfill}  \>\`292 \\ [0.4mm]%
\hspace{3cm}\parbox[b]{9.2cm}{{\bfseries 4.13.3.~}Quasi-static approximation\dotfill}  \>\`295 \\ [0.4mm]%
\hspace{3cm}\parbox[b]{9.2cm}{{\bf 4.14.~}Summary\dotfill}
\>\`296\\
\hspace{3cm}\parbox[b]{9.2cm}{\normalsize\bf\sffamily Energy in
general relativity\\
  in view of spinor and tensor methods}  \>\`{} \\ [-2mm]%
\vbox{\vspace*{-8mm}\begin{picture}(0,-30)
\put(7,3.7){\sffamily\bf\Huge\sffamily 5} \put(0,-0.5)
{\small\sffamily
C\hspace{0.55mm}H\hspace{0.55mm}A\hspace{0.55mm}P\hspace{0.55mm}T\hspace{0.55mm}E\hspace{0.55mm}R}
\end{picture}}\\
\hspace{3cm}\parbox[b]{9.2cm}{{\bf 5.1.~}Introduction\dotfill}  \>\`302\\ [0.4mm]%
\hspace{3cm}\parbox[b]{9.2cm}{{\bf 5.2.~}Connection between spinor and tensor methods in the positive energy problem\dotfill}  \>\`304 \\ [0.4mm]%
\hspace{3cm}\parbox[b]{9.2cm}{{\bfseries 5.2.1.~}Sen---Witten equation in Petrov type  $N$ space-time}  \>\`305 \\ [0.4mm]%
\hspace{3cm}\parbox[b]{9.2cm}{{\bfseries 5.2.2.~}Nodal surfaces of selfadjoint elliptic  second order equa\-tions\dotfill}  \>\`305 \\ [0.4mm]%
\hspace{3cm}\parbox[b]{9.2cm}{{\bfseries 5.2.3.~}The solutions of Sen---Witten equation have no zeros\dotfill}  \>\`306 \\ [0.4mm]%
\hspace{3cm}\parbox[b]{9.2cm}{{\bfseries 5.2.4.~}Sen---Witten equations and SOF\dotfill}  \>\`308 \\ [0.4mm]%
\hspace{3cm}\parbox[b]{9.2cm}{{\bfseries 5.2.5.~}Conclusion\dotfill}  \>\`310 \\ [0.4mm]%
\hspace{3cm}\parbox[b]{9.2cm}{{\bf 5.3.~}Nodal points of elliptic equations and system of equa\-tions\dotfill}  \>\`311 \\ [0.4mm]%
\hspace{3cm}\parbox[b]{9.2cm}{{\bfseries 5.3.1.~}Conditions for the absence of nodal points\dotfill}  \>\`313 \\ [0.4mm]%
\hspace{3cm}\parbox[b]{9.2cm}{{\bfseries 5.3.2.~}The conditions of nodal points absence for the solu\-tions of Sen---Witten equation\dotfill}  \>\`317 \\ [0.4mm]%
\hspace{3cm}\parbox[b]{9.2cm}{{\bfseries 5.3.3.~}Towards Sen---Witten equation, special orthonormal frame and preferred time variables\dotfill}  \>\`318 \\ [0.4mm]%
\hspace{3cm}\parbox[b]{9.2cm}{{\bf 5.4.~}Sen---Witten orthonormal three-frame and gravitational energy quasilocalization\dotfill}  \>\`321 \\ [0.4mm]%
\hspace{3cm}\parbox[b]{9.2cm}{{\bfseries 5.4.1.~}Direct link between the 4-covariant spinor 3-form and the Einstein Hamiltonian\dotfill}  \>\`323 \\ [0.4mm]%
\hspace{3cm}\parbox[b]{9.2cm}{{\bfseries 5.4.2.~}In which cases the conditions of Theorem 4 are ful\-filled?\dotfill}  \>\`329 \\ [0.4mm]%
\hspace{3cm}\parbox[b]{9.2cm}{{\bf 5.5.~}Summary\dotfill}  \>\`331 \\ [3mm]%

\hspace{3cm}\parbox[b]{9.2cm}{\normalsize\bf\sffamily APPENDIX A. Friedmann\vspace*{-0.7mm} equations\\ for the multicomponent scalar-field model}  \>\`{333} \\ [3mm]%

\hspace{3cm}\parbox[b]{9.2cm}{\normalsize\bf\sffamily APPENDIX B. Mathematical\vspace*{-0.7mm} details\\ of the braneworld model}  \>\`{336} \\ [2mm]%
\hspace{3cm}\parbox[b]{9.2cm}{{\bf B.1.~}Variational problem in the
presence of
the brane\dotfill}  \>\`336\\ [0.4mm]%
\hspace{3cm}\parbox[b]{9.2cm}{{\bf B.2.~}Graphical representation of
the brane
evolutiong\dotfill}  \>\`338 \\ [0.4mm]%
\hspace{3cm}\parbox[b]{9.2cm}{{\bf B.3.~}Background cosmological
solution in the
bulk\dotfill}  \>\`339 \\ [0.4mm]%
\hspace{3cm}\parbox[b]{9.2cm}{{\bf B.4.~}Scalar perturbation of the bulk metric\dotfill}  \>\`340 \\ [3mm]%

\hspace{3cm}\parbox[b]{9.2cm}{{BIBLIOGRAPHY}\dotfill}\>\`341  \\ [3mm]%

\hspace{3cm}\parbox[b]{9.2cm}{{INDEX}\dotfill}\>\`374  \\ [3mm]%

\hspace{3cm}\parbox[b]{9.2cm}{{ABOUT THE AUTHORS}\dotfill}\>\`378  \\ %

\end{tabbing}}

\newpage

 {\parindent=0pt\hspace*{2.9cm}
\vbox{\vspace*{-1.6cm}\hbox{\large\raisebox{0pt}{{\par}}}
\vspace*{0.6cm}\hbox{$\blacksquare$}
 \vspace{-3.0mm}
\hbox{\rule{10cm}{1pt}}
\vspace{2mm}%
\raggedright{\bfseries\sffamily\Large  FOREWORD OF THE EDITOR}\\%
\vspace*{-1mm} \hbox{\rule{10cm}{1pt}}\par \vspace{39mm}}}

\begin{wrapfigure}{l}{2.6cm}
\vspace*{-5.9cm}{\includegraphics[width=3.0cm]{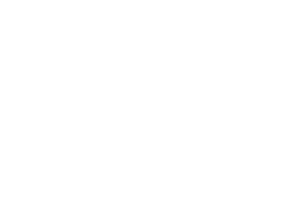}}\vskip11.5cm
\end{wrapfigure}

\markboth{Foreword of the Editor$_{ }$}{Foreword of the Editor$_{ }$} 
\begin{picture}(10,10)
\put(121,-144.2){\bfseries\sffamily{9}}
\end{picture}

\thispagestyle{empty}

\vspace{-0.9cm}

\noindent\small  Current science of the extraterrestrial world,
based on observations and physical theories, during the last decades
has collected many independent types of evidence to certify that
about 95\,\% of the energy-mass content of our Universe is dark and
invisible, and its physical nature is unknown. There are also many
arguments that it is composed of two ingredients: dark matter, which
facilitates clustering of the baryonic matter, and dark energy,
which is almost uniform and is responsible for the accelerated
expansion of the Universe. Scientific teams of physicists,
astrophysicists and cosmologists all over the world endeavor to
unveil these mysterious components, which today dominate in the
average density of the Universe and determine the physical
properties and evolution of our Universe. This task becomes
extremely important for the elaboration of particle physics beyond
the Standard Model.\baselineskip=12.5pt

In 2007, the programme of the National Academy of Sciences of
Ukraine ``Investigations of the structure and composition of the
Universe, hidden mass and dark energy (Cosmomicrophy\-sics)'' was
initiated, aimed at consolidating the efforts of different
scientific teams in several research institutes and universities
wor\-king in the field of astroparticle physics and theory of
gravity and investigating the dark components of the Universe. The
results of six-year activity in the framework of this programme will
be presented in the  three-volume edition ``Dark energy and dark
matter in the Universe''. The present book is the first of the
volume series and is devoted to the problem of dark energy. It
describes the current state of the problem of dark energy as well as
the contribution of the\, authors\, to\, this\, issue.
 Their investigation in this area was supported in part by
the Cosmomicrophy\-sics programme during the last six years, but the
book also con\-tains their previous results obtained in the
framework of the state projects in their respective institutions as
well as other projects and grants which are mentioned in the
acknowledgments. The first and second chapters of the book are
written by Bohdan Novosyadlyj, the third one by Alexander Zhuk, the
fourth one by Yuri Shtanov and the fifth one by Volodymyr Pelykh.

{This monography is dedicated to the memory of our colleagues and
friends Peter Fomin (1930---2011), Anatolij Minakov (1949---2012)
and Victor Vakulik (1953---2012), distinguished Ukrainian physicists
who had passed away when this book was being written. Peter Fomin
made significant contributions to the quantum field theory and
quantum cosmology; he is the author of the idea of quantum birth of
the Universe. Anatolij Minakov laid the foundations for the
application of the theory of gravitational lensing to the
interpretation of gravitational lens systems and estimation of dark
matter abundance. Victor Vakulik was an unsurpassed specialist in
computer simulation and a brilliant interpreter of astrophysical
\mbox{observational data.}}

 \vspace*{3mm}

\mbox{\hspace*{7.4cm}{\it \footnotesize V.\,SHULGA}}\\
\mbox{\hspace*{5.9cm}{\raisebox{-13mm}[0cm][0cm]{\includegraphics[width=4cm]{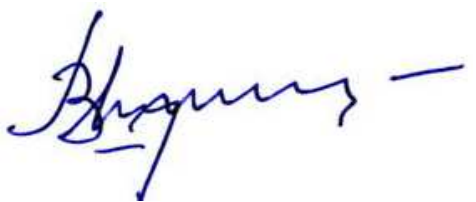}}}}

\newpage

 {\parindent=0pt\hspace*{2.9cm}
\vbox{\vspace*{-1.6cm}\hbox{\large\raisebox{0pt}{{\par}}}
\vspace*{0.6cm}\hbox{$\blacksquare$}
 \vspace{-3.0mm}
\hbox{\rule{10cm}{1pt}}
\vspace{2mm}%
\raggedright{\bfseries\sffamily\Large  ACKNOWLEDGMENTS}\\%
\vspace*{-1mm} \hbox{\rule{10cm}{1pt}}\par \vspace{39mm}}}

\begin{wrapfigure}{l}{2.6cm}
\vspace*{-5.9cm}{\includegraphics[width=3.0cm]{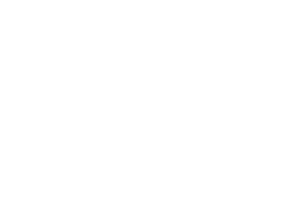}}\vskip13.2cm
\end{wrapfigure}

\markboth{Acknowledgments$_{ }$}{Acknowledgments$_{ }$} 
\begin{picture}(10,10)
\put(120,-144.2){\bfseries\sffamily{}}
\end{picture}

\thispagestyle{empty}

\vspace{-2.2cm}
 \noindent\small We are grateful to the National Academy of Sciences of
Ukraine for financial support of the research program
``Investigations of the structure and composition of the Universe,
hidden mass and dark energy (Cosmomicrophysics)'', within which this
monograph is written. We are thankful to the editor of this
monograph academician Valerij Shulga for the suggestion of its
writing, participation in the formation of its content and
organization of its publication. We also thank academician Yaroslav
Yatskiv for his generous support of this project. We also appreciate
helpful and highly professional assistance from the staff of the
academic publisher ``Akademperiodyka''.\baselineskip=12.5pt

We are deeply grateful to our co-authors for productive long-term
collaboration, the results of which form the basis of the monograph.
Among them: Ste\-pan Apu\-ne\-vych, Vik\-tor Baukh, Val\-dir
Be\-ze\-rra, Ul\-rich Ble\-yer, Ale\-xey Cho\-pov\-sky, Luis
Cris\-pino, Ruth Dur\-rer, Ma\-xim Ein\-gorn, Seyed Hos\-sein Fakhr,
Pe\-ter Fo\-min, Ste\-phan Gott\-l{\"o}ber, Uwe G{\"u}nther,
Vla\-dimir Iva\-shchuk, Tina\-tin Kahnia\-shvili, Uwe Kas\-per,
Svet\-lana Kris\-kiv, Yurij Ku\-li\-nich, Vla\-di\-mir Lu\-kash,
Orival de Me\-dei\-ros, Vi\-ta\-ly Mel\-ni\-kov, Paulo Mo\-niz,
Mar\-tin Rai\-ner, Car\-los Ro\-mero, Va\-run Sahni, Tamer\-lan
Sai\-dov, Olga Ser\-gijenko, Ar\-man Sha\-fieloo, Ale\-xey
Sta\-ro\-bin\-sky, Ale\-xey To\-po\-ren\-sky, Petr Tre\-tya\-kov,
Ri\-car\-do Val\-dar\-nini, Ale\-xan\-der \mbox{Viznyuk.}

Last but not least, we thank our families for their unconditional
and devoted support and understanding.

Our research was supported by the university and institution
projects of the Mini\-stry of Edu\-ca\-tion and Scien\-ce and
Na\-tio\-nal Aca\-demy of Scien\-ces of Uk\-rai\-ne as well as by
grants from the SCOPES pro\-gramme financed by the Swiss
Na\-tio\-nal Scie\-nce Foun\-da\-tion, the To\-mal\-la
Foun\-da\-tion, the DAAD pro\-gram\-me of Ger\-ma\-ny, the
Bra\-zi\-lian spon\-soring agency CAPES, the Abdus Salam ICTP, the
SISSA, the Mini\-stry of Science and Tech\-no\-logy of India, the
IUCAA (India) and the Theory Division of the CERN visitor
programs.\vspace*{-4.0mm}

\begin{flushright}
{\footnotesize{\it B.\,NOVOSYADLYJ,\\ V.\,PELYKH, Yu.\,SHTANOV, A.\,ZHUK}\\[0.5mm]
Lviv\,---\,Kyiv\,---\,Odessa}
\end{flushright}

\normalsize

\newpage
\vspace*{7mm}\hspace*{0.8cm}\parbox[b]{7.2cm}{\raggedleft{\it \Large To the memory\\
of
prominent Ukrainian scientists,\\ our colleagues and friends\\
Peter Fomin, Anatolij Minakov\\ and Victor Vakulik

}}  \thispagestyle{empty}

\newpage

\setcounter{chapter}{0}
\chapter{\label{s:0}  OBSERVATIONAL EVIDENCE\\[1mm] FOR DARK ENERGY}\markboth{CHAPTER 1.\,\,Observational evidence for dark energy}{CHAPTER 1.\,\,Observational evidence for dark energy}
\thispagestyle{empty}\vspace*{-12mm}

\begin{wrapfigure}{l}{2.6cm}
\vspace*{-5.9cm}{\includegraphics[width=3.0cm]{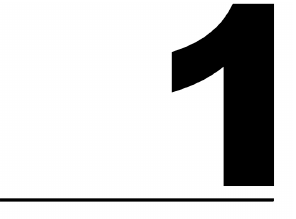}}\vskip17.2cm
\end{wrapfigure}
\vspace*{10mm}

 \setcounter{section}{1} \vspace*{-5mm}
\hspace*{3cm}\section*{\hspace*{-3cm}1.1.\,\,Introduction$_{}$}\label{ch1-Intr}\begin{picture}(10,10)
\put(125,-128){\bfseries\sffamily{\,13}}
\end{picture}

\begin{figure}
  \vskip1mm
  \centering
  \includegraphics[width=10.5cm]{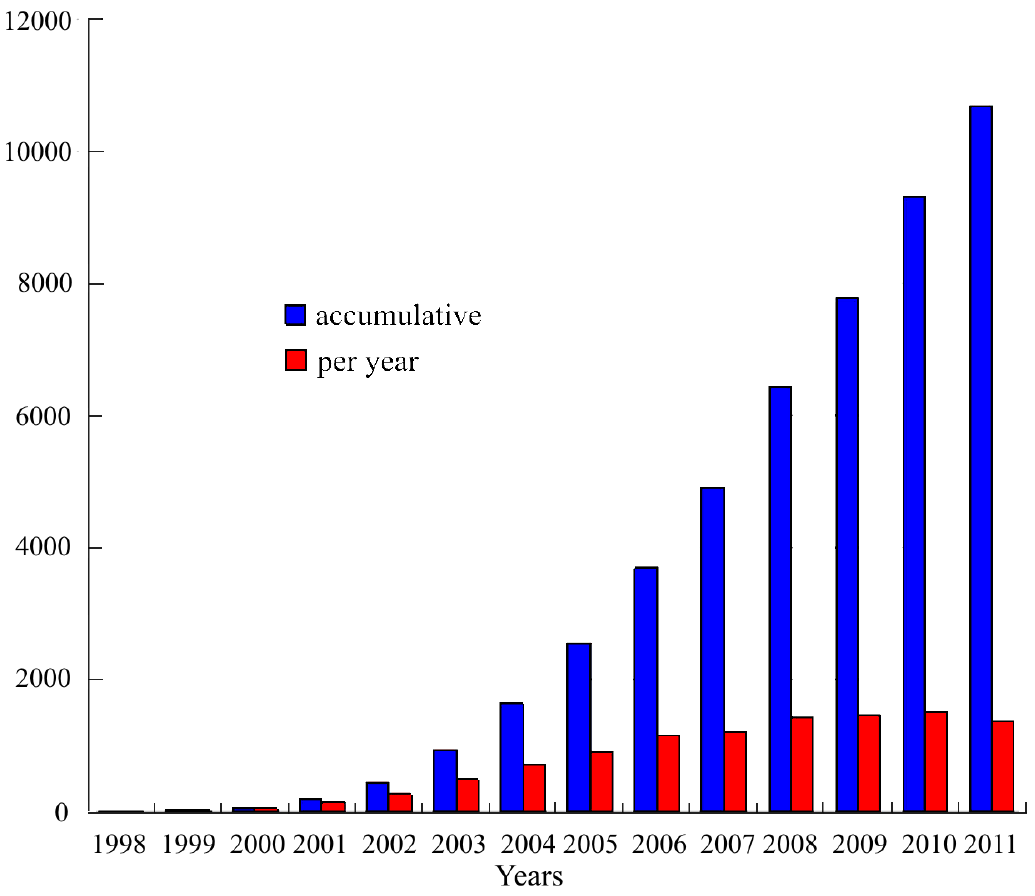}
\vskip-3mm
  \parbox[b]{10.5cm}{\caption{Number of papers (accumulative and per year) in SAO/NASA
    Astrophysics Data System with key words ``dark energy'' in title
    or abstract}}\vspace*{-1.5mm}
  \label{pn}
\end{figure}

\vspace*{-1.1cm} \noindent The term ``dark energy'' has appeared in
the titles and abst\-racts of scientific papers in 1998 after
announcements about discovery of the accelerated expansion
\index{accelerated expansion}  the Universe, made by two teams
practically simultaneously~--- Supernova Cosmology Project
\cite{Perlmutter1998,Perlmutter1999} \index{Supernova Cosmology
Project (SCP)} and High-Z Supernova Search \index{High-Z Supernova
Search} \cite{Riess1998,Schmidt1998}.  The first application of this
term in the con\-ven\-tio\-nal now meaning we have found in the
papers \mbox{[5---7]}, one year before M.\,Turner and M.\,White
called this essence ``smooth com\-po\-nent'' \cite{Turner1997} and
P.\,Steinhardt ``quin\-tessence'' \cite{Steinhardt1997}.
\index{quintessence} Up to now the number of papers devoted to this
subject has amounted to more then ten thousands and continues to
grow in about two thousands per year during last years
(Fig.~\ref{pn}), that signifies the actuality of dark energy problem
and its importance for fundamental phy\-sics, astrophysics and
cosmology.

Dark energy stands for the wide spectrum of new physical substances
capable to provide the accelerated expansion of the Universe. But
not only that. It must be noted, that cosmology was looking for such
essence for years. Indeed, measurements of peculiar velocities of
galaxies carried out in the second half of 80s and first half of 90s
showed, that total matter density (baryons and cold dark matter, CDM) %
\index{cold dark matter (CDM)}%
of our Universe is 0.4\,$\pm$\,0.1 of critical one \cite{Dekel1994}.
On
the other hand, all measurements of the Hubble constant %
\index{Hubble constant}%
showed, that it is larger than
$60\,\text{km}/\text{s}$\,$\cdot$\,$\text{Mpc}$ \cite{Mould1995}. It
meant, that the age of the maximally open Universe does not exceed
13 billion years, while the age of oldest stars in the globular
clusters was estimated to be 13.5\,$\pm$\,2 billion years
\cite{Jimenez1996}.  So, open models of the Universe have not
enough margin of safety, but models with cosmological constant have: %
\index{cosmological constant} %
even at the upper limit of Hubble constant estimation~---
$85\,\text{km}/\text{s}$\,$\cdot$\,$\text{Mpc}$~--- the age of the
Universe is
about 14 billion years. %
\index{age of Universe} %
The matter dominated model with low 3-space cur\-va\-tu\-re was
ruled out by these data at high confidential level.

Other indications of the presence of unusual component in our
Universe
came from the study of large scale structure formation. %
\index{large scale structure}%
Bahcall et al. \cite{Bahcall1997} have revealed the massive rich
clusters of galaxies at redshifts $z>0.5$, %
\index{galaxy clusters} %
such early appearance of them in the COBE normalized
\cite{COBE1992,Bunn1997} hierarchical \mbox{scenarios} of large
scale
structure formation is possible if growth factor %
\index{growth factor} %
of density \mbox{perturbations} %
\index{density perturbations} %
is lower than one in the standard CDM model (SCDM) %
\index{standard CDM model (SCDM)} %
with density dominated by cold dark matter and zero 3-space
cur\-va\-tu\-re. It is such in open CDM (OCDM) %
\index{open CDM (OCDM) model} %
and CDM with cosmological constant ($\Lambda$CDM) models. The last
ones  had higher margin of safety than former. The measurements of
power spectra of matter density perturbations \cite{Peacock1997},
X-ray cluster temperature function \cite{Henry1991} and galaxy bulk
flows %
\index{bulk flow} %
\cite{Dekel1994b} also preferred $\Lambda$CDM model %
\index{LambdaCDM model ($\Lambda$CDM)} %
\cite{Kahniashvili1996,Valdarnini1998}. But the science lacked for
the direct test for such model~--- determination of deceleration
parameter $q_0$, which is positive in the matter dominated models
(decelerated expansion) and negative in the $\Lambda$ dominated ones
(accelerated expansion). Its early estimations, listed in review
\cite{Sandage1988}, were very rough, $-1.3\le q_0\le 1.6$, and gave
no possibility to constraint the models. Improving this test became
the
key task for Hubble Space Telescope (HST) %
\index{Hubble Space Telescope (HST)} %
and largest ground-based telescopes installed in 90s. And this
problem was successfully solved using Type Ia supernovae (SNe Ia)
\index{Supernova Ia} as standard candles for magnitude-redshift
relation.

The 1998 became the prominent year for cosmology. The High-Z
Su\-per\-no\-va Search %
\index{High-Z Supernova Search} %
team on the base of spectral and photometric \mbox{investigations}
of 16 high-redshift and 34 nearby SNe Ia claimed \cite{Riess1998}
that
deceleration parameter $q_0<0$ %
\index{deceleration parameter} %
at confidential levels from 99.5\,\% (2.8$\sigma$) to 99.9\,\%
(3.9$\sigma$) for \mbox{$q_0=$}
\mbox{$=\Omega_m/2-\Omega_{\Lambda}$}. These authors found also that
more uncertain estimate for its value is
$q_0=-1\pm0.4$.  The Supernova Cosmology Project %
\index{Supernova Cosmology Project (SCP)} %
team during ten-year observations had discovered 42 SNe Ia at
redshifts between 0.18 and 0.83 which were photometrically and
spectroscopically investigated carefully. After complete analysis of
the data the team claimed \cite{Perlmutter1999} that cosmological
constant %
\index{cosmological constant} %
is non-zero and positive with the confidence of 99\,\%, including
the identified systematic uncertainties. They obtained the relation
which approximates the joint probability distribution of matter and
cosmological constant density parameters:
$0.8\Omega_m+0.6\Omega_{\Lambda}\approx-0.2\pm0.1$. It means that
deceleration parameter $q_0\approx -5\Omega_m/6-1/3\pm1/6$ and is
surely negative. So, the accelerated expansion of the Universe was
strongly preferred by magnitude-redshift data for SNe Ia. But for
assurance other proofs based on the other observations should be
realized. The large scale structure evidence had not enough
strin\-gency yet, since different scenarios then were under
discussions~--- its forming from adiabatic or iso-cur\-va\-tu\-re
primordial perturbations, topological defects, cosmological strings
or other seeds. In the last two years of XX century the precise
measurements of cosmic microwave
background (CMB) temperature anisotropy %
\index{cosmic microwave background (CMB)} %
realized in the BOOMERanG %
\cite{Boomerang} %
and MAXIMA \cite{Maxima} stratospheric experiments revealed the
acoustic peaks %
\index{MAXIMA}%
\index{acoustic peaks} %
in its angular power spectrum predicted by scenario of large scale
structure formation from adiabatic primordial perturbations. %
\index{primordial perturbation} %
So, the large scale structure data jointly with CMB anisotropy %
\index{CMB anisotropy} %
have become the powerful test for dark energy models. During the
first decade of XXI century the numerous data proving the existence
of dark energy have been obtained. For the discovery of the
accelerating expansion of the Universe through observations of
distant supernovae Saul Perlmutter, Adam G.\,Riess and Brian
P.\,Schmidt were honored with the Nobel Prize in Physics for 2011.

On the other hand theorists have proposed many candidates for dark
energy which well match the observational data
[24---44] So, the important problem of nowadays cosmology is
developing of key tests for distingui\-shing between different
models of dark energy and constraining the values of their
parameters. The state-of-art observational evidence for dark energy
and const\-raints on its parameters are discussed in this chapter.

\newpage

\section[\!Dynamics of expansion of the
  homogeneous isotropic multicomponent Universe]{\!Dynamics of expansion
of the
  homogeneous\\ \hspace*{-0.9cm}isotropic multicomponent Universe}
\label{ch1-sec1}

\hspace*{3cm}The base of cosmology as physical science is General
Relativity founded by Albert Einstein \cite{Einstein1916}. He was
also the first
who in 1917 introdu\-ced the new essence, cosmological constant %
\index{cosmological constant} %
\cite{Einstein1917}, which acts against gravitation
attraction\,\footnote{\,A. Einstein introduced it in order to
realize a
  static universe in the framework of General Relativity, but after
  discussion with A.\,Friedmann and discovery of expansion of the
  Universe by E.\,Hubble he abandoned this idea.} and now is
considered as the simplest candidate for dark energy.  The other
fundamental conceptual base of standard cosmology is the
cosmological
principle %
\index{cosmological principle} %
\index{large scale structure}%
which states that on the large scales the Universe is homogeneous
and isotropic. The geometric properties of such Universe define the
general form of space-time line element as %
\begin{equation}
  \label{ds}
  ds^2=g_{\mu\nu}dx^{\mu}dx^{\nu}=a^2(\eta)\left(-d\eta^2+dr^2+\chi^2(r)(d\vartheta^2+\sin^2\vartheta
  d\varphi^2)\right)\!,\!\!\!\!
\end{equation}
where $g_{\mu\nu}$ is a metric tensor, %
\index{metric tensor}%
$\eta$ is the conformal time %
\index{conformal time} %
 defined by relation $cdt=$ $=a(\eta)d\eta$,
$a(\eta)$ is the scale factor, $r,\,\vartheta,\,\varphi$ are
spherical coordinates in 3-space with a  cur\-va\-tu\-re $K$ and
\begin{equation}
\label{chir} \chi(r) =
\begin{cases}
  \displaystyle\frac{1}{\sqrt{K}} \sin\sqrt{K}r,  & K>0~~(\text{spherical 3-space}),\\
  r,      & K=0~~(\text{flat 3-space}),\\
  \displaystyle\frac{1}{\sqrt{|K|}}\sinh\sqrt{|K|}r, & K<0~~(\text{hyperspherical
  3-space}).
  \end{cases}\!\!\!\!\!
\end{equation}.

Here and below the Greek indices ($\nu,\, \mu,\, \mbox{...}$) run
from 0 to 3 and the Latin ones ($i,\,j,\,\mbox{...}$) run from 1 to
3. Henceforth we put $c=1$, so the time variable $t\equiv x_0$ as
well as conformal time $\eta$ have the dimension of a length. We
follow also usual convention that terms with the same upper and
lower indices are summed over.  The space-time with metric
(\ref{ds}) is called
Friedmann---Robertson---Walker (FRW) %
\index{FRW metric} %
one in memoriam of the first investigations of homogeneous and
isotropic solutions of Einstein's equations by A.\,Friedman in 1922
\cite{Friedmann1922} and distance-redshift relation in the expanding
Universe\,\footnote{\,Physical properties of the non-stationary
Universe
  with FRW metric were first analyzed by G.\,Le\-mai\-tre in 1927
  \cite{Lemaitre1927}} by H.\,Ro\-bert\-son in 1928
\cite{Robertson1928} and A.\,Walker in 1933 \cite{Walker1933}.

Today we know that our Universe is filled with matter-energy and
fields, which can be classified as relativistic component (cosmic
background radiation and relic neutrino), non-relativistic one
(baryon and dark matter) and dark energy of unknown nature which
accelerates its expansion. Each of them can be described by
energy-momentum
tensor %
\index{energy-momentum tensor} %
of perfect fluid
\begin{equation}
  \label{emt}
  T{_{\mu\nu}}{_{N}}=\left(\rho_{N}+p_{N}\right)u{_{\mu}}{_{N}}u{_{\nu}}{_{N}}+p_{N}g_{\mu\nu},
\end{equation}
that is consequence of symmetry of FRW space-time. Here $\rho_{N}$
is energy density and $p_{N}$ is pressure, which are defined as
time- and space-like eigenvalues of $T_{\mu\,N}^{\nu}$
correspondingly, $u^{\mu}_{N}=(-a,0,0,0)$ is four-velocity of the
fluid in como\-ving coordinates, and $N$ stands for relativistic
$(r)$, matter $(m)$ or
dark energy $(de)$ components. The space-time metric %
\index{space-time metric} %
(\ref{ds}) and energy-momentum tensors (\ref{emt}) of each component
are used for analysis of expansion dynamics, physical phenomena and
distant-redshift relations of homogeneous isotropic Universe, called
also cosmological background. In the theory of large scale structure
formation they have more general forms, that will be discussed
later.

Assuming that the interaction between all components is only
gravita\-tio\-nal, each of them should satisfy the differential %
\index{conservation law} %
energy-momentum conservation law separately:\vspace*{-3mm}
\begin{equation}
  \label{demcl}
  T^{\nu}_{\mu;\nu N}=0.\vspace*{-1mm}
\end{equation}

Hereafter ``;'' denotes the covariant derivative with respect to the
coordinate with given index in the space with metric $g_{ij}$. In
the homogeneous Universe the density $\rho_{N}$ and pressure $p_{N}$
of
perfect fluid are functions of time only, so the equation of state %
\index{equation of state (EoS)} %
(EoS) can be presented in the simple form\vspace*{-1mm}
\begin{equation}
  \label{eos}
  p_{N}(\eta)=w_{N}(\eta)\rho_{N}(\eta),\vspace*{-1mm}
\end{equation}
where $w_{N}$ is called EoS parameter, it can be constant or
time-variable, that depends on physical properties of the component
$(N)$.

In the space-time (\ref{ds}) the differential energy-momentum
conservation law (\ref{demcl}) gives\vspace*{-3mm}
\begin{equation}
  \label{rho'}
  \dot{\rho}_{N}=-3\frac{\dot a}{a} \rho_{N}(1+w_{N}),
\end{equation}
here and below a dot denotes the derivative with respect to the
conformal \mbox{time, $\dot{\;} \equiv d/d\eta$.}

For the non-relativistic matter $w_{m}=0$ and for the relativistic
one $w_{r}=$ $=1/3$. For these cases the equation (\ref{rho'}) can
be easily integrated to obtain the time dependences of density of
these components in the form\vspace*{-2mm}
\begin{equation}
  \label{rho_mr}
  \rho_{m}=\rho_{m}^{(0)}(a_0/a)^{3},\quad \rho_{r}=\rho_{r}^{(0)}(a_0/a)^{4},\vspace*{-1mm}
\end{equation} %
\index{EoS parameter}where index ``0'' in parentheses and without
them denotes the current values of corresponding variables. The EoS
parameter of dark energy is unknown and can be constant or
time-variable.  In the general case the integral of equation
(\ref{rho'}) for dark energy is as follows\vspace*{-2mm}
\begin{equation}
\label{rho_de}
\rho_{de}=\rho_{de}^{(0)}(a_0/a)^{3(1+\tilde{w}_{de})},
\end{equation}
where $\tilde{w}_{de}=w_{de}$ for the constant EoS parameter
$w_{de}$ and\vspace*{-3mm}
\begin{equation}
\tilde{w}_{de}=\frac{1}{\ln{(a/a_0)}}\int\limits_{a_0}^{a}{w_{de}d\ln{a}}
\label{tilde_w_de}\vspace*{-5mm}
\end{equation}
for the time-variable one.

In this chapter we consider mainly the dark energy with constant EoS
parameter $w_{de}$. The case $w_{de}=-1$ corresponds to the well
studied $\Lambda$CDM model with $\rho_{de}=\rho_{\Lambda}$~=~const.
When $w_{de}>-1$ the density of DE monotonically decreases with time
(often
called quintessence), %
\index{quintessence} %
in the opposite case ($w_{de}<-1$) it increases. Since in the last
case the density starts from zero at $a=0$ it is dubbed ``phantom''
\cite{Caldwell2002}. %
\index{phantom dark energy} %

The Einstein equations relate the Ricci tensor $R_{\mu\nu}$ %
\index{Enistein equations} %
\index{tensor Ricci} %
to the total energy-momentum tensor of all components as follows
\begin{equation}
  \label{eeqs}
  R_{\mu\nu}-\frac{1}{2}g_{\mu\nu}R=8\pi G\sum_{N}T{_{\mu\nu}}{_{N}},
\end{equation}
where $R\equiv g^{\alpha\beta}R_{\alpha\beta}$ is scalar
cur\-va\-tu\-re of
space-time. For the space-time with metric (\ref{ds}) they become %
\begin{gather}
  \left(\!\frac{\dot{a}}{a}\!\right)^{\!2}  +  K  =   \frac{8\pi G}{3} a^2 \sum_{N}\rho_{N},  \label{Freq1}\\
  \frac{\ddot{a}}{a} - \left(\!\frac{\dot{a}}{a}\!\right)^{\!2} = - \frac{4\pi
    G}{3}\,a^2 \sum_{N}\rho_{N}(1+3w_{N}), \label{Freq2}
\end{gather}
which can be integrated for $a(\eta)$ when right-hand sides (r.h.s.)
are defined.

The first term in left-hand side (l.h.s.) of (\ref{Freq1}) is the
rate of expansion and in l.h.s.\ of (\ref{Freq2}) is acceleration
($d^2a/dt^2$). So, the dynamics of cosmic expansion is determined by
specifying the properties of matter-energy components \mbox{of the
Universe.}

It is convenient to introduce the dimensionless parameters %
\index{density parameter} %
of matter density $\Omega_m$, radiation density $\Omega_{r}$, dark
energy density $\Omega_{de}$ and cur\-va\-tu\-re of 3-space
$\Omega_K$ as follows
\begin{equation}
\label{def1}
\begin{split}
  & \Omega_m \equiv \left(\!\frac{8\pi G\rho_m
      a^2}{3(\dot{a}/a)^2}\!\right)_{\!\eta_0}\!\!, \quad
  \Omega_{r} \equiv \left(\!\frac{8\pi G\rho_{r} a^2}{3 (\dot{a}/a)^2}\!\right)_{\!\eta_0}\!\!,
  \\[3mm]
  & \Omega_{de} \equiv \left(\!\frac{8\pi G\rho_{de}
      a^2}{3(\dot{a}/a)^2}\!\right)_{\!\eta_0}\!\!, \quad \Omega_K \equiv
  \left(\!\frac{-K}{(\dot{a}/a)^2}\!\right)_{\!\eta_0}\!\!.
\end{split}
\end{equation}

The density parameter of relativistic component $\Omega_{r}$ at
current epoch is the sum of photon and neutrino ones and equals
\begin{equation}
  \label{Om_r}
  \Omega_{r}= 4.17\!\cdot\!10^{-5}\left(\!\frac{1+\rho_{\nu}/\rho_{\gamma}}{1.6813}\!\right)\!\left(\!\frac{T_0}{2.726}\!\right)^{\!\!4}
  \approx 4.17\!\cdot\!10^{-5}h^{-2}.
\end{equation}

The equations (\ref{Freq1}), (\ref{Freq2}) in this notation
become\vspace*{-2mm}
\begin{align}
  H =  H_0\left(\!\Omega_r \frac{a_0^4}{a^4} +  \Omega_m\frac{a_0^3}{a^3}+\Omega_K\frac{a_0^2}{a^2} +\Omega_{de}(\frac{a_0}{a})^{3(1+w_{de})}\!\right)^{\!{1}/{2}}\!\!,\label{H} \\
  q = \frac{H_0^2}{H^2}\left(\!\Omega_r \frac{a_0^4}{a^4} +
    \frac{1}{2}\Omega_m\frac{a_0^3}{a^3}
    +\frac{1}{2}(1+3w_{de})\Omega_{de}\left(\!\frac{a_0}{a}\!\right)^{\!3(1+w_{de})}\!\right)\!,\!\!\! \label{q}
\end{align}\vspace*{-4mm}

\noindent where $H\equiv a^{-1}da/dt=a^{-2}da/d\eta=\dot{a}/a^2$ is
Hubble
parameter %
\index{Hubble parameter} %
at any time $\eta$ and $q\equiv-\left(a\ddot{a}/\dot{a}^2- 1\right)$
is deceleration parameter.

At the current epoch $\eta=\eta_0$ the Friedmann equations %
\index{Friedmann equations} %
(\ref{H}), (\ref{q}) are simplified to the form\vspace*{-3mm}
\begin{equation}
  \Omega_r+\Omega_m+\Omega_K+\Omega_{de}=1, \label{sumom}
\end{equation}\vspace*{-6mm}
\begin{equation}
  q_0=\Omega_r+\frac{1}{2}\Omega_m+\frac{1}{2}(1+3w_{de})\Omega_{de}. \label{q0}
\end{equation}

\begin{figure}
\vskip1mm \noindent\raisebox{0.1cm}{\parbox[b]{5.0cm} {\caption{The
lines of zero decele\-ration parameter\index{deceleration parameter}
    $w_{de}\!=\!-(1-\Omega_K)/$ $/(1-\Omega_m-\Omega_K)/3$ for flat (solid
    line), closed (dotted line) and open (dashed line) models. Below
    these lines there is range of values of $w_{de}$ and $\Omega_m$ for
    \mbox{which} the expansion is accelerated. The rectangle shows
    observationally constrained range for these
    parameters}}}\hspace{0.5cm}\includegraphics[width=7.5cm]{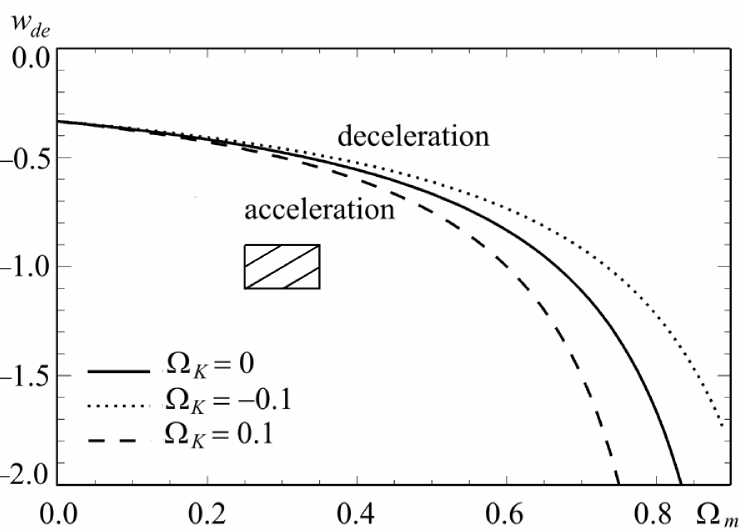}\vspace*{-7mm}
  \label{acc_zone}
\end{figure}

If the Universe is filled only with matter and radiation  then
deceleration parameter is always positive ($d^2a/dt^2<0$) for any
cur\-va\-tu\-re, as one can see from (\ref{q})---(\ref{q0}). And
when only third component with\vspace*{-1mm}
\begin{equation}
  w_{de}<-\frac{1}{3}\frac{1-\Omega_K}{1-\Omega_m-\Omega_K} \label{w_lim}\vspace*{-1mm}
\end{equation} %
\index{deceleration parameter}is present then deceleration parameter
$q_0$ at current epoch could be negative ($d^2a/dt^2>0$). In the
last expression we omit $\Omega_r$, since it is substantively lower
than $\Omega_m$. In Fig.~\ref{acc_zone} the ranges of values of
parameters for which expansion of the Universe is accelerated are
shown.

\index{cosmological constant} %
If the dark energy is cosmological constant $\Lambda$, for which
$w_{de}=-1$ and $\Omega_\Lambda \equiv \Lambda/3H_0^2$, then the
accelerated expansion of the Universe will take place under
condition $\Omega_m<2(1-\Omega_K)/3$. In the models with vanishing
cur\-va\-tu\-re (standard $\Lambda$CDM ones) matter density
parameter must be lower than $0.66$. The dependence of deceleration
parameter $q_0$ on $\Omega_m$ for different $\Omega_K$ and $w_{de}$
at current epoch
is shown in Fig.~1.3.

\begin{figure}
  \includegraphics[width=6.6cm]{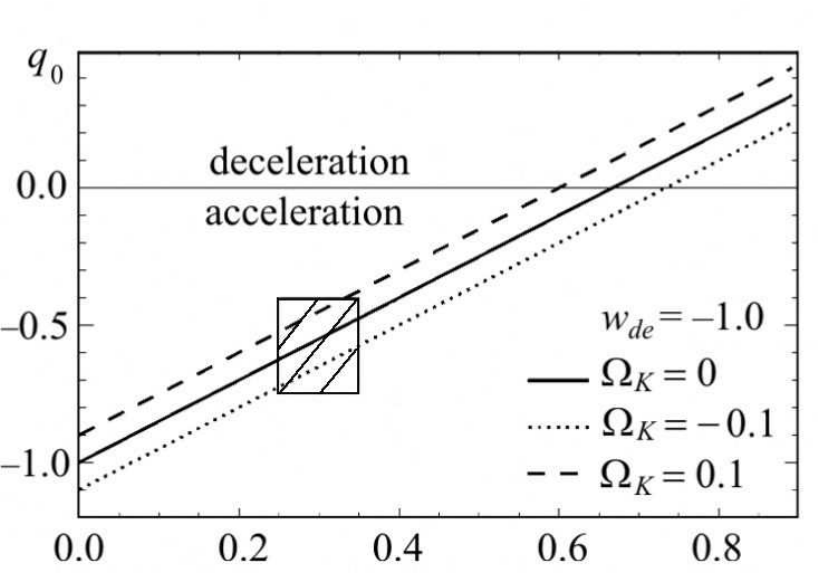} \hspace*{-3mm} \includegraphics[width=6.6cm]{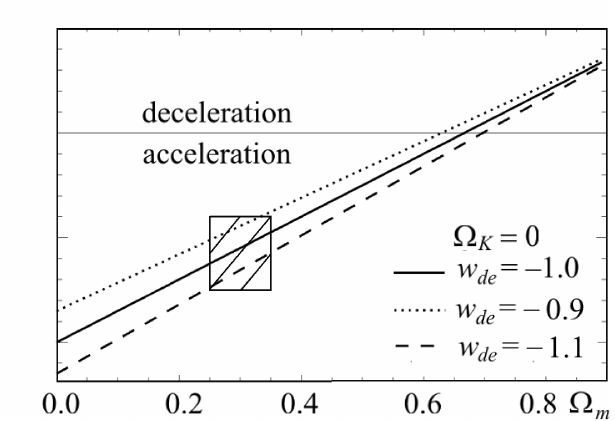}
  \vskip-1mm\caption{The dependence of current value of deceleration parameter
    on $\Omega_m$ for different $\Omega_K$ (left panel) and different
    $w_{de}$ (right one). The rectangles show observational
    constraints on $q_0$ and $\Omega_m$ from \cite{Perlmutter1999}}\vspace*{2mm}
  \label{q0_Kw}
\end{figure}

\begin{figure}[h!]
  \includegraphics[width=6.5cm,height=4.7cm]{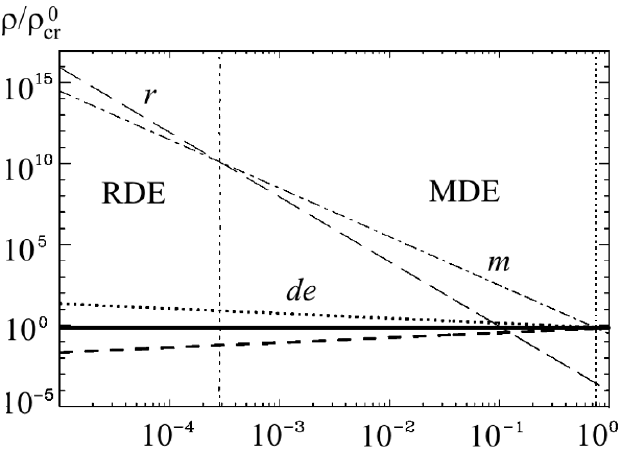}\hspace*{0.4cm}\includegraphics[width=6.1cm,height=4.6cm]{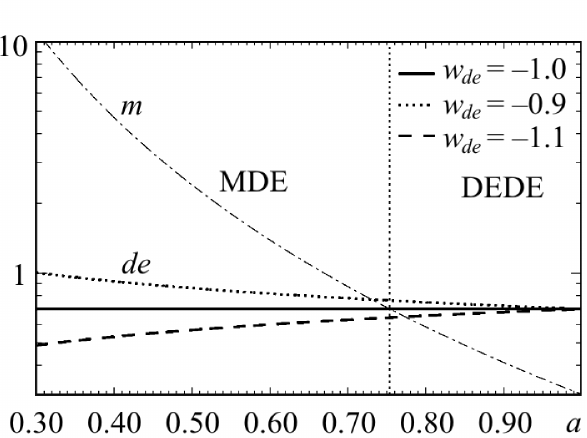}
  \vskip-1mm
\caption{The evolution of energy density of relativistic ($r$),
matter
  ($m$) and dark energy $w=$~const ($de$) components. RDE~--- radiation
  dominated epoch, MDE~--- matter dominated epoch and DEDE~--- dark energy
  dominated epoch. %
  \index{dark energy dominated (DED) epoch} %
  In the left panel the plot is in log-log scale from early epoch to
  current one, in the right panel the plot is in norm-log scale and
  illustrates the late epoch. All lines correspond to model with
  $\Omega_m=0.3$ and $\Omega_{de}=0.7$. The MDE-DEDE crossing line is
  shown for $\Lambda$-model ($w_{de}=-1$)}\vspace*{1mm}
\label{rhormde_w}
\end{figure}

\index{multicomponent Universe}The dynamics of expansion of the
multicomponent Universe is different in different epochs, that
follows from equations (\ref{H}), (\ref{q}). Really, for $w_{de}$,
which satisfies condition (\ref{w_lim}), the dynamical history can
be divided into three periods (Fig.~1.4):

 {\bf radiation dominated} (RD) epoch, %
\index{radiation dominated (RD) epoch} %
  when $\rho_r\gg\rho_m\gg\rho_{de}$ and $q=1$; it was at $a\ll
  a_{eq}^{(r-m)}$, where $a_{eq}^{(r-m)}=4.17\!\cdot\!10^{-5}\Omega_m^{-1}h^{-2}$ is scale factor at radiation-matter
  equality;

{\bf matter dominated} (MD) epoch, %
  \index{matter dominated (MD) epoch} %
  when $\rho_m\gg\rho_r$ and $\rho_m\gg\rho_{de}$; then $q=0.5$; it
  was at $a_{eq}^{(r-m)}\ll a\ll a_{eq}^{(m-de)}$, where
  $a_{eq}^{(m-de)}=\left(\Omega_m/\Omega_{de}\right)^{-1/3w_{de}}$ is
  scale factor at matter~--- dark energy equality;

{\bf dark energy dominated} (DED) epoch, %
  \index{dark energy dominated (DED) epoch} %
  when $\rho_{de}\gg\rho_m\gg\rho_r$ at $a>$ $>a_{eq}^{(m-de)}$; it is
  epoch of accelerated expansion %
  \index{accelerated expansion} %
  of the Universe. In the case of $w_{de}=$const it will last forever.

In the model with realistic parameters \mbox{$\Omega_m=0.3$},
\mbox{$\Omega_{de}=0.7$}, \mbox{$h=0.7$}, \mbox{$w_{de}=-1$} the
scale factors (redshifts) related to epoch change-overs  (shown in
Fig.\,1.\ref{rhormde_w}) are \mbox{$a_{eq}^{(r-m)}=2.9\cdot\!
10^{-4}$ ($z_{eq}^{(r-m)}=3500$)} and \mbox{$a_{eq}^{(m-de)}=0.75$}
($z_{eq}^{(m-de)}=0.32$).

The measurements of $q_0$ give the possibility to connect
$\Omega_m$, $\Omega_{de}$ and $w_{de}$ by relation (\ref{q0}), where
$\Omega_r$ can be omitted. In the flat Universe, for which
$\Omega_{de}=1-\Omega_m$, it can be reduced to the relation between
$w_{de}$ and $\Omega_{de}$:
\begin{equation}
\Omega_{de}=(2q_0-1)/3w_{de}. \label{om_de}
\end{equation}

\begin{figure}
  \vskip1mm
  \includegraphics[width=6.5cm]{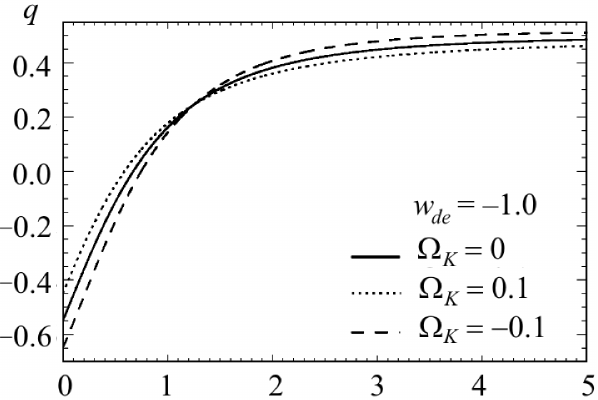}\includegraphics[width=6.5cm]{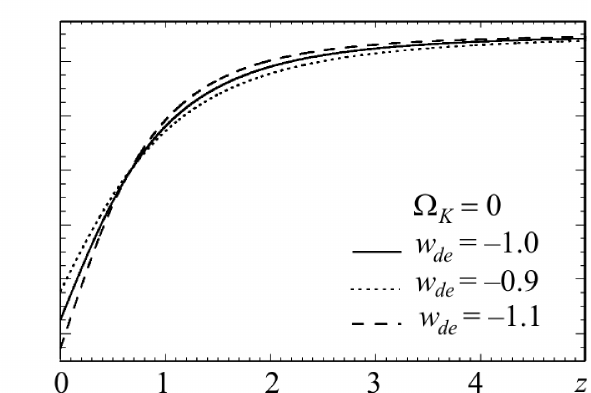}
  \vskip-2mm
    \caption{The dependence of deceleration parameter $q$ on $z$ for
    cosmological models with different $\Omega_K$ (left panel) and
    $w_{de}$ (right one). Here values of other parameters  are:
    $\Omega_m=0.3$, $\Omega_{de}=1-\Omega_K-\Omega_m$, $H_0=70$~km/s/Mpc}
  \label{q_Kw}
\end{figure}

And only for the flat $\Lambda$-model we can estimate
$\Omega_{\Lambda}$ from measurements of $q_0$:
$\Omega_{\Lambda}=(1-2q_0)/3$. For joint estimation of $w_{de}$ and
$\Omega_{de}$ the measurements of $q$ at other redshifts are
required. For example, if we have estimations of $q_0$ and redshift
$z_{q=0}$ where decelerated expansion was changed by accelerated one
($q=0$), then EoS parameter %
\index{EoS parameter} %
can be estimated from transcendent equation
\begin{equation}
  (1+z_{q=0})^{3w_{de}}=\frac{2q_0-1-3w_{de}}{(1+3w_{de})(2q_0-1)},
\label{zq0}
\end{equation}
and $\Omega_{de}$ from equation (\ref{om_de}). In the general case
for estimation of $\Omega_m$, $\Omega_{de}$ and $w_{de}$ the form of
$q(z)$ dependence in the range $0<z<5$, shown in Fig.~1.\ref{q_Kw},
must be measured. It can be done by realization of tests
``luminosity distance~--- redshift'' for sources with known
luminosities or ``angular
diameter distance~--- redshift'' for sources with known diameters. %
\index{angular diameter distance} %

\newpage

\section[\!The luminosity distance~--- redshift
relation and SNe Ia evidence]{\!The luminosity distance~--- redshift
relation\\ \hspace*{-0.9cm}and SNe Ia evidence for dark
energy\label{ch1-sec2}}

\hspace*{3cm}The luminosity distance %
\index{luminosity distance} %
is defined by
\begin{equation}
 d^2_L \equiv \frac{L}{4\pi F},
\end{equation}
where $L$ is absolute luminosity of a source and $F$ is an observed
flux. The source is at comoving distance $r$ %
\index{comoving distance} %
and has emitted photons at $\eta$, which observer from the Earth
detects in the point $r=0$ at current moment of time $\eta_0$.

\begin{figure}
  \vskip1mm
  \includegraphics[width=6.7cm]{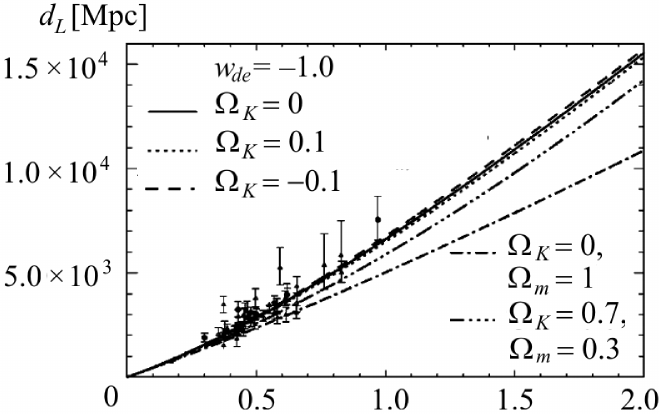}\hspace*{0.3cm}\vspace*{0.5mm}\includegraphics[width=6.0cm]{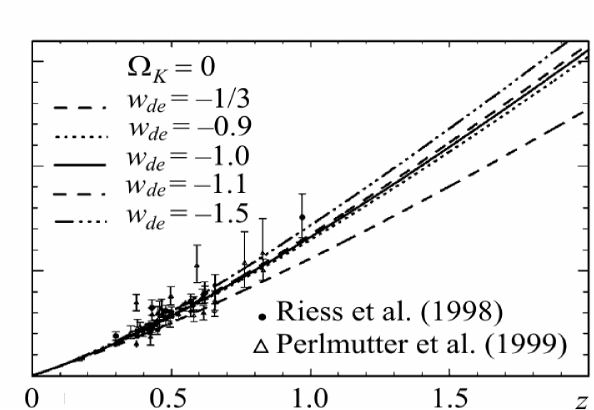}
  \vskip-2mm
   \caption{Left panel: The dependence of luminosity distance $d_{L}$
    on redshift for cosmological models with different values of
    $\Omega_K,\,\Omega_{\Lambda},\,\Omega_m$: (0, 0, 1), (0, 0.7,
    0.3),  (--0.1, 0.8, 0.3), \mbox{(0.1, 0.6, 0.3)}, (0.7, 0, 0.3).
    Right panel: The same for cosmological models with
    $\Omega_K=0$, $\Omega_{de}=0.7$, $\Omega_m=0.3$ and different values
    of $w_{de}$. In  both panels the luminosity distances to
    high-$z$ supernovae Ia derived from the data presented in
    \cite{Riess1998,Perlmutter1999} are shown by signs.  (It is
    assumed that $H_0=70$~km/s/Mpc for all calculations
    here)}\vspace*{-3mm}
  \label{dlum_ch1}
\end{figure}

Taking into account the redshifting of each photon detected by
observer from the Earth and lowering the photons arrival rate, both
by the factor $a(\eta_0)/a(\eta)=1+z$, in the FRW space-time
(\ref{ds}) the luminosity distance $d_L$ of a source and its
redshift are bound by relation
\begin{equation}
\label{dl}
  d_{L}=(1+z)\chi\left(\int\limits_0^z\frac{dz'}{H(z')}\!\right)\!\!,
\end{equation}
where function $\chi$ is defined by expression (\ref{chir}) and
$H(z')$ by equation (\ref{H}). The computed $d_L(z)$-dependences for
models with different values of $\Omega_m$, $\Omega_K$,
$\Omega_{de}$ and $w_{de}$ are shown in Fig.~1.\ref{dlum_ch1}. There
also shown for comparison the luminosity distances to high-$z$ SNe
Ia derived from their moduli distances and corrected magnitudes
presented in Table 5 of \cite{Riess1998} and Table 1 of
\cite{Perlmutter1999} correspondingly\,\footnote{\,These papers
contain the most accurate data
  in 1998---1999 supporting the existence of dark energy and become the
  base for awarding of Nobel Prise in physics to S.\,Perlmutter,
  A.\,Riess and B.\,Shmidt in 2011.}. They certainly indicate that
models with dark energy are strongly preferred by the SNe Ia
observations.

One can see, that models with values of parameters $\Omega_K$ and
$w_{de}$ from the ranges [--0.1, 0.1] and [--0.9, --1.1]
correspondingly are only slightly distinguished by $d_L(z)$ for
$0\le z\le5$ since it is integral of $H^{-1}(z)$ over $z$.
Fig.~1.\ref{q_Kw} illustrates that
the deceleration parameter, %
\index{deceleration parameter} %
which is defined by ratio of the first derivative of $H(z)$ with
respect to redshift $z$ and $H(z)$,
\begin{equation}
\label{q_Hz} q(z)=\frac{z+1}{H(z)}\frac{dH(z)}{dz}-1
\end{equation}
is more sensitive to value of EoS
parameter. %
\index{EoS parameter} %
One expects, that higher derivatives of $H(z)$ with respect to
redshift $z$ are also more sensitive to value of EoS parameter. It
is convenient to introduce the dimensionless parameters analogical
to the deceleration parameter $q(t)\equiv
-\displaystyle\frac{1}{a(t)H^2(t)}\frac{d^2a(t)}{dt^2}$:
\begin{equation*}
  j(t)= \frac{1}{a(t)H^3(t)}\frac{d^3a(t)}{dt^3}, \quad s(t)=\frac{1}{a(t)H^4(t)}\frac{d^4a(t)}{dt^4},
\end{equation*}
dubbed {\it jerk} and {\it snap} (see \cite{Visser2004,Weinberg2008} %
\index{jerk} %
\index{snap} %
and citing therein), which are represented by second and third
derivatives of $H(z)$ with respect to $z$ as follows\vspace*{-2mm}
\begin{gather}
  j(z)= q^2(z)+\frac{(z+1)^2}{H(z)}\frac{d^2H(z)}{dz^2}, \label{j_z}\\
  s(z)=
  -q^3(z)-[4q(z)+3]\frac{(z+1)^2}{H(z)}\frac{d^2H(z)}{dz^2}-\frac{(z+1)^3}{H(z)}\frac{d^3H(z)}{dz^3}.\!\!\!\!\!\!\! \label{s_z}
\end{gather}\vspace*{-5mm}

\begin{figure*}
  \vskip1mm
  \includegraphics[width=13cm]{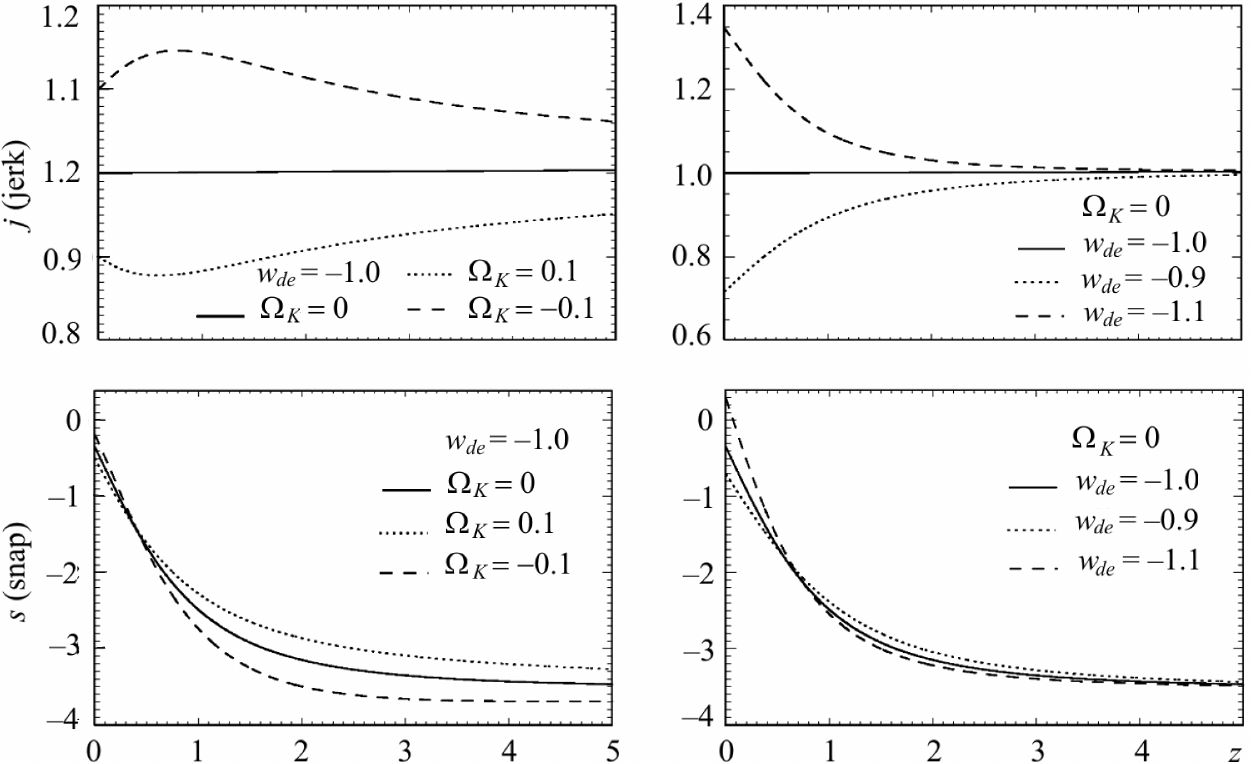}
  \vskip-2mm
  \caption{The dependence of jerk parameter $j$ (top panels) and snap
    parameter $s$ (bottom panels) on redshift $z$ for cosmological
    models with different $\Omega_K$ (left panel) and $w_{de}$ (right
    one).  The values of other parameters are the same as in
    Fig.~1.\ref{q_Kw}}
  \label{jerk-snap}
\end{figure*}

They are shown in Fig.~1.\ref{jerk-snap}. Indeed, they can be used
for accurate determinations of $w_{de}$ if high-precision
measurements of $d_L$ will be possible, since
$H(z)=\displaystyle\left[\frac{d}{dz}\left(\!\frac{d_L}{z+1}\!\right)\!\right]^{\!-1}$
for the flat Universe for example. Unfortunately, the contemporary
observational data give the possibility to establish $q$ at current
epoch only, which is the coefficient of second order term in the
power series of $d_L$ on $z$ in the vicinity of $z=0$\vspace*{-2mm}
\[
d_L(z) =  \frac{c}{H_0}\bigg\{\! z + \frac{1}{2}\left[1-q_0\right]
{z^2}-\frac{1}{6}\left[1-q_0-3q_0^2+j_0+\Omega_K\right]
 z^3+\hspace*{0.5cm}
\]\vspace*{-7mm}
\begin{equation}
  \hspace*{0.2cm}+ \frac{1}{24}\left[2-2q_0-15q_0^2-15q_0^3+5j_0+10q_0j_0+s_0 +
    \Omega_K\right] z^4 \!\!+ O(z^5) \!\bigg\}\!.\!\!\!\!
\end{equation}\vspace*{-3mm}

This expansion of $d_L$ in $z$ up to fourth order
term\,\footnote{\,It
  have been deduced firstly by M.\,Visser \cite{Visser2004}.  The term
  of third order in $z$ was previously calculated by T.\,Chiba and
  T.\,Nakamura \cite{Chiba1998}, the first two terms are Weinberg's
  version of Hubble law \cite{Weinberg1972}.} approximates the exact
expression (\ref{dl}) with error $\le$$ 1\,\%$ up to $z\approx 0.5$
for all models in Fig.~1.\ref{dlum_ch1} excluding OMD one with
$\Omega_K=0.7$, for which this level of approximation accuracy is
overcome at $z\sim 0.2$. The including of third order term in the
value of $d_L$ at $z\sim0.1$ gives 3---7\,\% of value of second one,
the including of fourth one is at the level of few tenthes of
percent, so, their estimation on the base of current data on SNe Ia
luminosity distances is impossible. Enough precision of observations
for obtaining the information on the variation of $q(z)$ and $j(z)$
will be reached at the end of current decade, as it is expected.

The observed flux and absolute luminosity of sources in astrophysics
traditionally are presented by visible and absolute magnitudes $m$
and $M$, which are connected to the luminosity distance $d_L$ by
well known
relation %
\index{luminosity distance} %
\begin{equation}
\label{dist_mod} (m-M)=5\log{d_L}+25.
\end{equation}
Their difference in the l.h.s. is called distance modulus and is
marked by $\mu$.

The sources with known luminosity $L$ or absolute magnitude $M$ are
called standard candles\index{standard candles}, they give
possibility to measure the distances, that is crucial for
astrophysics and cosmology. The best standard candles in cosmology
are Type Ia supernovae (SNe Ia), \index{Supernova Ia} which are
thermonuclear explosions of near-Chandrasekhar mass carbon-oxygen
white dwarfs residing in binary systems. The word ``the best'' means
``the better than any other one'', nothing more. They are not ideal
standard candles, since their raw peak brightnesses vary by factor
two, this limits their cosmological applications. The key
astrophysical development in the cosmological use of SNe Ia was the
realization that their luminosities could be further standardized
using empirical relationships: SN Ia light-curve width~--- SN Ia
luminosity \cite{Phillips1993, Kasen2007} and SN Ia color~--- SN Ia
luminosity \cite{Tripp1998}. Light curves can be described by
stretch parameter $s$, which stretches or tightens a template of
light curve
to match an observed one. %
\index{light-curve width}

The correction of observed peak magnitudes in the most cosmological
applications is as follows
\begin{equation}
  \label{m_corr}
 m_B^{(corr)}=m_B^{(obs)}+\alpha(s-1)-\beta C^{(obs)},
\end{equation}
where $\alpha$ and $\beta$ are parameters, which must be determined
for sample of SNe Ia. As a rule, the magnitudes are measured in the
standard B and V bands and the color is their difference,
$C^{(obs)}=m_B^{(obs)}-m_V^{(obs)}$. For using the measured
$m_B^{(obs)}$ and $m_V^{(obs)}$ of distant SNe Ia in the
cosmological context they must be corrected also for redshifting of
their spectral energy distributions as well as extinction along the
line of sight. A
few methods have been developed for SNe Ia light-curve fitting, %
\index{light-curve fitting}%
which were called the stretch one [59---61], CMAGIC \cite{Wang2003},
BATM \cite{Tonry2003}, $\Delta m_{15}$ \cite{Hamuy1996,Prieto2006},
Spectral Adaptive Light curve Template (SALT) %
\index{Spectral Adaptive Light curve Template (SALT)} %
\cite{Guy2005} and its improvement SALT2 \cite{Guy2007}, Multicolor
Light Curve Shape (MLCS) %
\index{Multicolor Light Curve Shape (MLCS)} %
\cite{Riess1996} and its improvement MLCS2k2 \cite{Jha2007}, SiFTO
\cite{Conley2008} etc. They give quite similar results for
reasonably large sample of supernovae. The typical precision
achieved in SN Ia distance estimations is $\sim$5---7\,\% which make
them crucial \mbox{objects
for cosmology.}%
\index{Multicolor Light Curve Shape (MLCS)}%

\begin{figure}
  \vskip1mm
  \includegraphics[width=13cm]{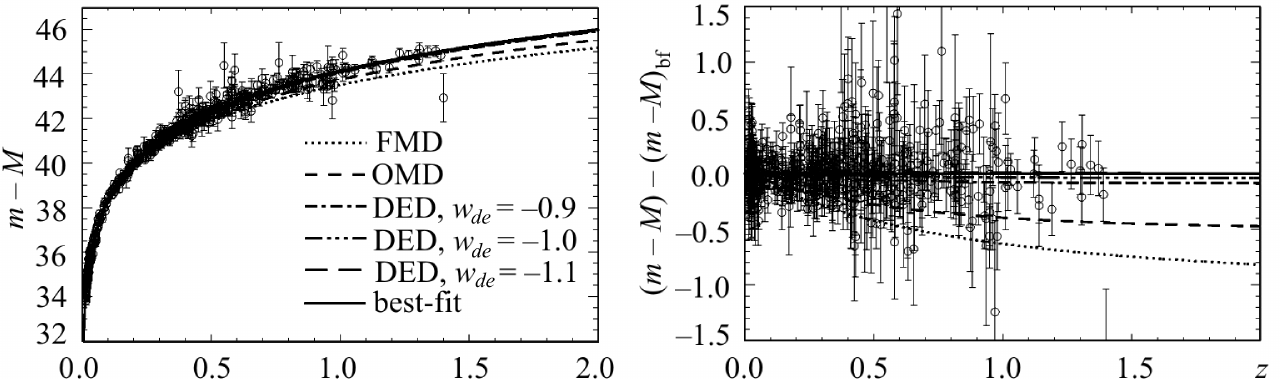}
  \vskip-2mm
    \caption{Left panel: the distance moduli $(m-M)(z)$ for 557 SNe Ia
    from Union2 compilation \cite{SNUnion2} (signs) and for five
    models (lines): flat matter dominated (FMD, $\Omega_m=1.0$), %
    \index{flat matter dominated (FMD) model} %
    open matter dominated (OMD, $\Omega_m=0.3$) and dark energy
    dominated (DED, $\Omega_m=0.3$, $\Omega_{de}=0.7$) with
    $w_{de}=-0.9$, $w_{de}=-1.0$ and $w_{de}=-1.1$. Right panel: the
    residuals of the distance modulus relative to the best-fit model
    for SNe Ia data ($\Omega_{m}=0.2875$, $w_{de}=-1.0486$,
    $h=0.7013$) as well as for other models}
  \label{dl_union2}
\end{figure}

Since 1998 a lot of observational projects for search and rigorous
investigations of Type Ia supernovae were realized using the most
advanced telescopes of the world. Up to now about thousand SNe Ia at
\mbox{$0.1\le z\le1.6$} were discovered and photometrically and
spectroscopically investigated carefully. They have been collected
by high-redshift surveys including Hubble Space Telescope (HST)
\index{Hubble Space Telescope (HST)} \cite{HSTa,HSTb}, SuperNova
Legacy Survey (SNLS) \index{SuperNova Legacy Survey (SNLS)}
\cite{SNLS}, Equation of State: SupErNovae trace Cosmic Expansion
(ESSENCE) \index{Equation of State: SupErNovae trace Cosmic
Expansion
  (ESSENCE)}
\cite{ESSENCEa,ESSENCEb}, Super Nova Sloan Digital Sky Survey (SN
SDSS) \index{SuperNova Sloan Digital Sky Survey (SN SDSS)}
\cite{Sako2008,Kessler2009} ones. The first unified sample of SNe Ia
``Union'' (HST\,+\,SNLS\,+\,ESSENCE) \cite{SNUnion} counted 307
selected supernovae, the second one~--- ``Union2''
\cite{SNUnion2}~--- counts 557 supernovae eligible for cosmological
applications. Their luminosity distance moduli, calculated
\mbox{using} SALT2 method for light-curve fitting,
\index{light-curve fitting} versus redshifts are shown in the left
panel of Fig.~1.\ref{dl_union2}. The best-fit curve corresponds to
the model with \index{Union sample} parameters\,\footnote{\,They
have been determined for the data
  \cite{SNUnion2} using Levenberg---Marquardt method \cite{NumRec} and
  relations (\ref{dl})---(\ref{dist_mod}).} $\Omega_{de}=0.2875$,
$w_{m}=-1.0486$, $h=0.7013$, the total $\chi^2$ for it, calculated
as
\begin{equation}
  \label{chi2}
  \chi^2=\sum_{i=1}^{N_{tot}}\frac{(\mu_i-\mu_{bf})^2}{\Delta\mu_i^2},
\end{equation}
is equal to 542.6 and the standard deviation, called also RMS of the
Hubble residuals,
$\sigma=\sqrt{\sum_{i=1}^{N_{tot}}(\mu_i-\mu_{bf})^2/N_{tot}}$, is
0.28. Here $\mu_i$ is luminosity distance modulus of $i$th supernova
from Union2 sample, $\Delta\mu_i$ is statistical error of its
determination, $\mu_{bf}$ is the theoretical distance modulus at the
redshift of $i$th supernova for best-fit model.

The dependences of luminous distance modulus on redshifts in
different
cosmological models~--- flat matter dominated (FMD, $\Omega_m=1.0$), %
\index{open matter dominated (OMD) model|(} %
open matter dominated (OMD, $\Omega_m=0.3$) and three dark energy
dominated (DED, $\Omega_m=$ $=0.3$, $\Omega_{de}=0.7$) %
\index{dark energy dominated (DED) model} %
with $w_{de}=-0.9$, $w_{de}=-1.0$ and $w_{de}=-1.1$ ones are shown
in Fig.~1.\ref{dl_union2} by lines for comparison. The $\chi^2$'s
for them are 2047, 1182, 569, 547, 544 correspondingly ($\mu_{mod}$
instead of $\mu_{bf}$ in formula (\ref{chi2})). In the right panel
of Fig.~1.\ref{dl_union2} the residuals of the distance modulus
relative to the best-fit curve for SNe Ia data as well as other
models are shown. One can see,

\begin{wrapfigure}{r}{6.5cm}
\vspace*{-3mm} \noindent\includegraphics[width=6.5cm]{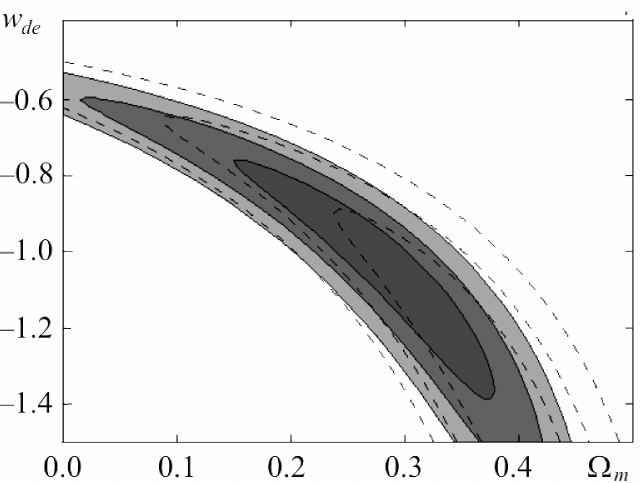}
\raisebox{0.2cm}{\parbox[b]{6.5cm}{\caption{68.3\,\%, 95.4\,\% and
99.7\,\% \mbox{confidence} regions of the
    $\Omega_m-w_{de}$ plane from SNe Ia alone from Union (dashed
    contours) and Union2 (shaded contours) compilations (from~\cite{SNUnion2})\label{union_constraints}}
  }}\vspace*{-7mm}
\end{wrapfigure}

\noindent that SNe Ia luminosity distance~--- redshift test strongly
prefers the dark energy dominated cosmological models. The 68.3\,\%,
95.4\,\% and 99.7\,\% confidence regions of the
\mbox{$\Omega_m-w_{de}$}
plane from SNe Ia alone from Union (dashed contours) and Union2 %
\index{Union sample} %
(shaded contours) compilations are shown in
Fig.~1.\ref{union_constraints}. It supports conclusion that
cosmological models without dark energy are ruled out at high
confidence level. But
allowable range for values of EoS parameter is too wide yet %
\index{EoS parameter} %
($-1.4<w_{de}<-0.75 $ at 1$\sigma$ C.L.) for distinguishing the type
of dark energy,
  so, the extension of SNe Ia sample and reducing of
identified systematic errors remain crucial problem of SNe Ia
observations and astrophysics (see excellent review of M.\,Sullivan
in \cite{Sullivan2010}).

\setcounter{figure}{9}
\begin{figure}
  \vskip1mm
  \includegraphics[width=13cm]{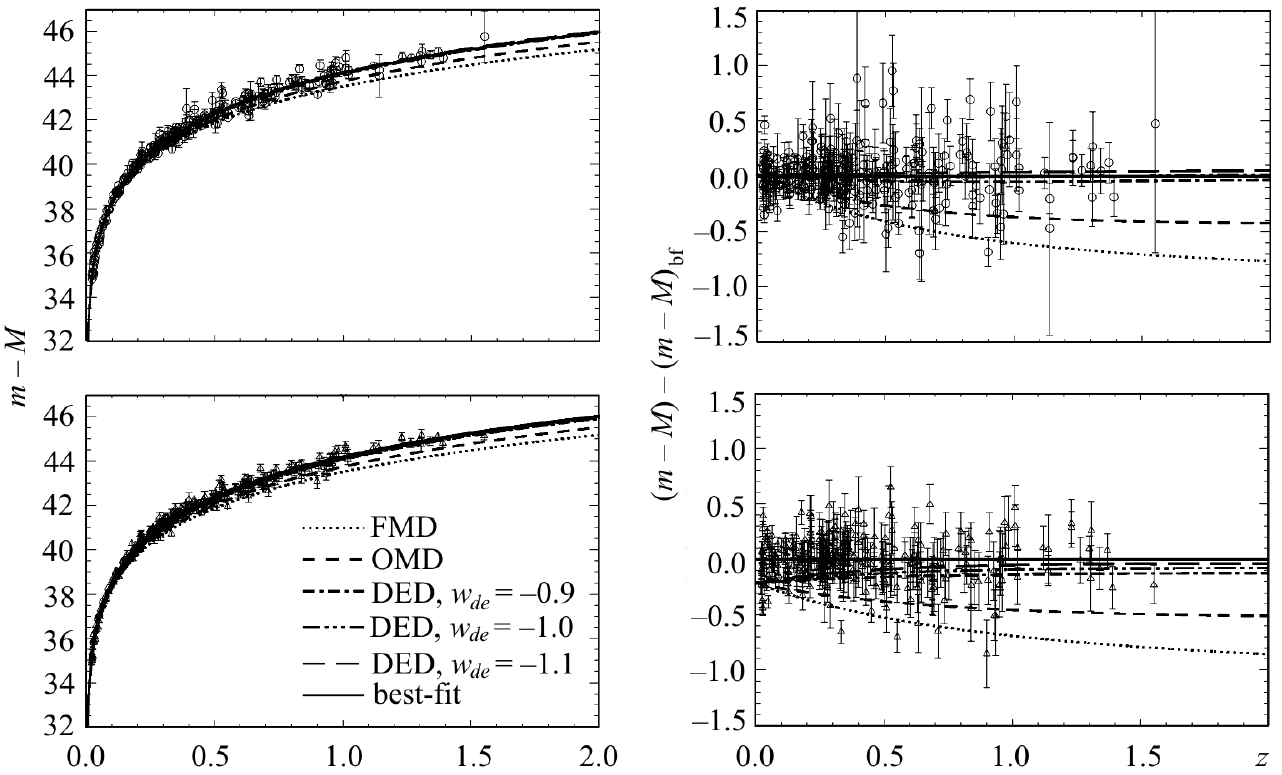}
  \vskip-2mm
   \caption{Left panel: the distance moduli $(m-M)(z)$ for 288 SNe Ia
    from SDSS compilation \cite{Kessler2009} (signs) and for five
    models (lines): flat matter dominated (FMD, $\Omega_m=1.0$), open
    matter dominated (OMD, $\Omega_m=0.3$) and dark energy dominated
    (DED, $\Omega_m=0.3$, $\Omega_{de}=0.7$) with $w_{de}=-0.9$,
    $w_{de}=-1.0$ and $w_{de}=-1.1$. Right panel: the residuals of the
    distance moduli relative to the best-fit models for SNe Ia data
    as well as for other models. In the top panels the luminosity
    distance moduli are calculated using SALT2 light-curve fitting %
    \index{light-curve fitting}%
    method, in bottom ones~--- using MLCS2k2 method}
  \label{dl_sdss}
\end{figure}

The homogeneous sample of 103 SNe Ia at the redshifts range
$0.04<z<$ $<0.42$ was selected from SDSS-II SuperNova Survey and
combined
with SNe Ia from HST, %
\index{Hubble Space Telescope (HST)} %
SNLS and ESSENCE surveys in order to reduce the statistical and
systematical errors in the estimation of their luminosity distance
moduli \cite{Kessler2009}. %
\index{Equation of State: SupErNovae trace Cosmic Expansion
  (ESSENCE)} %
In the left panel of Fig.~1.\ref{dl_sdss} the luminosity distance
moduli versus redshifts are presented for 288 SNe Ia from this
compilation, called SDSS SN one. Two methods for SNe Ia light-curve
fitting have
been used there~--- SALT2 (open circles) and MLCS2k2 (open triangles) in %
\index{Multicolor Light Curve Shape (MLCS)|(}%
order to compare them. The best-fit curve to SDSS SN SALT2
corresponds to the model with parameters $\Omega_{m}=$ $=0.323$,
$w_{de}=-1.1292$, $h=0.7046$, the total $\chi^2$ for it equals 559.5
and the standard deviation is 0.23, lower than for Union2 SALT2
supernova distance moduli. For the same FMD, OMD and DED
($w_{de}=-0.9$, $w_{de}=-1.0$,
$w_{de}=$ $=-1.1$) models, %
\index{flat matter dominated (FMD) model|)} %
shown by lines in Fig.~1.\ref{dl_sdss}, the $\chi^2$'s are
correspondingly 1972, 1626, 560 and 567. In the case of SDSS SN
MLCS2k2 supernova distance moduli the best-fit model has parameters:
$\Omega_{m}=0.3205$, $w_{de}=-0.7854$, $h=0.6344$. The total
$\chi^2$ for it equals 778.8 and the standard deviation is 0.21,
lowest in comparison with Union2 and SDSS SN SALT2 compilations. The
$\chi^2$'s for FMD, OMD and 3 DED models equal 4170, 2880, 1626,
1500, 1397 correspondingly. So, Union2 and SDSS SN SALT2
compilations are similar, while SDSS SN MLCS2k2 compilation prefer
essentially higher \mbox{$w_{de}$ and lower $H_0$.}

Combining the SDSS SN compilation with other cosmological
measure\-ments (BAO from SDSS LRG survey and CMB temperature
anisotropy from WMAP experiment) Kessler et al. (2009)
\cite{Kessler2009} have found that for spatially flat cosmological
model $w_{de} =-0.96\pm0.06(\text{stat})\pm0.12(\mathrm{syst})$,
$\Omega_m=0.265\,\pm$
$\pm\,0.016(\mathrm{stat})\pm0.025(\mathrm{syst})$ using the SALT2
and $w_{de} =-0.76\pm0.07(\mathrm{stat})\,\pm$
$\pm\,0.11(\mathrm{syst})$, $\Omega_m=0.307\pm
0.019(\mathrm{stat})\pm 0.023 (\mathrm{syst})$ using MLCS2k2 fitter.
So, the 1$\sigma$ confidence contours for $w_{de}$ and $\Omega_m$
for the same SN sample using SALT2 and MLCS2k2 fitters overlap only
partially. This means, that one of them or both have yet
unidentified systematic errors.  In the paper \cite{Kessler2009} the
differences between 2 methods of light curve fitting, SALT2 and
MLCS2k2, are thoroughly analyzed but convincing arguments for one or
the another are not given.

\begin{figure}
  \vskip1mm
  \includegraphics[width=13cm]{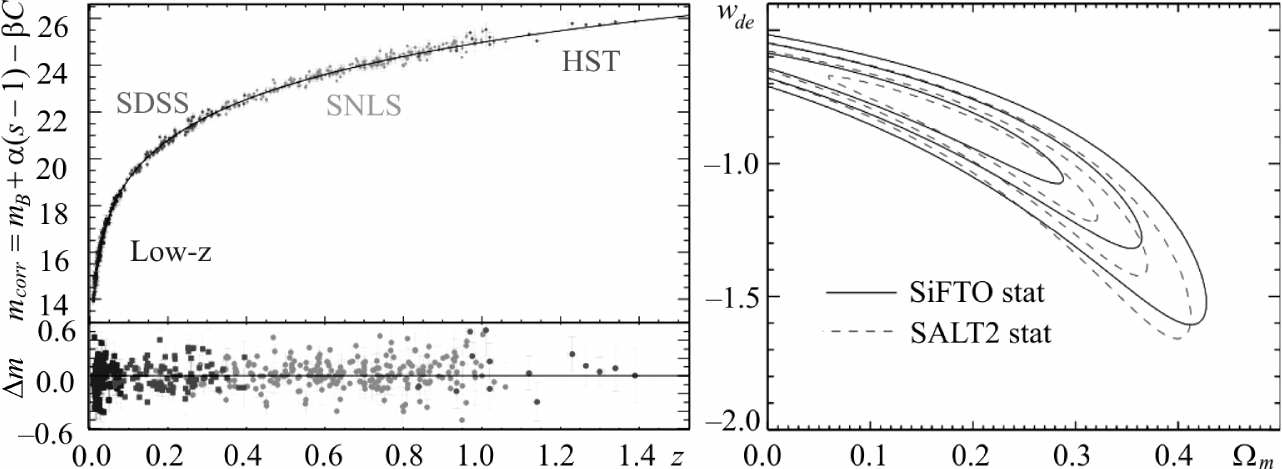}
  \vskip-2mm
   \caption{Left panel: the distance moduli $(m-M)(z)$ for 472 selected
    SNe Ia and residuals from best-fit curve $\Delta m=m_{corr}-m_{bf}$ (bottom). Right panel:
    68.3\,\%, 95.4\,\% and 99.7\,\% confidence regions of the
    $\Omega_m-w_{de}$ plane from SNe Ia alone (SNLS3 compilation with
    SALT2 and SiFTO fitters) assuming a flat universe and constant
    dark energy equation of state. (From \cite{Conley2011})}
  \label{snls3_constraints}
\end{figure}

The papers \cite{Guy2010,Conley2011,snls3} present the modern
high-quality sample of SNe Ia, based on 3-year SuperNova Legacy
Survey
(SNLS3) %
\index{SuperNova Legacy Survey (SNLS)} %
data including other mentioned above supernova samples. It contains
472 selected SNe Ia for cosmological applications with distance
moduli determined by updated versions of SALT2 and SiFTO light-curve
fitting %
\index{light-curve fitting}%
\index{light-curve fitter}%
methods. They use the same phenomenological correction formula
(\ref{m_corr}) but differ substantially in their detailed
parametrization of observables and in the procedures considered for
training and light-curve fitting (see for details \S\,4.3.1 and
\S\,4.3.2 in \cite{Guy2010}). The RMSs of the Hubble residuals are
0.17 for SNLS3 SALT2 supernovae and 0.15 for SNLS3 SiFTO ones,
essentially lower then for previous SNe Ia samples of similar
completeness. Comparison of SN-only statistical constraints on
$\Omega_m$, $w_{de}$ for SALT2 and SiFTO fitters assuming a flat
universe and constant EoS parameter is shown in the left panel of
Fig.~1.\ref{snls3_constraints}. In the right panel of
Fig.~1.\ref{snls3_constraints} 68.3\,\%, 95.4\,\% and 99.7\,\%
confidence regions of the $\Omega_m-w_{de}$ plane from SNe Ia alone
(SNLS3 compilation with SALT2 and SiFTO fitters) assuming a flat
universe and constant dark energy equation of state are shown. All
contours are prolate and convoluted, this indicates some degeneracy
of likelihood
function. %
\index{likelihood function}%
The median line can be approximated roughly by
\begin{equation}
  w_{de}^{(SN Ia)}\approx-0.651-0.122\Omega_m-5.076\Omega_m^2. \label{w_SNIa}
\end{equation}
The best-fit values of $w_{de}$ and $\Omega_m$ and their 1$\sigma$
confidential ranges are $w_{de} =$\linebreak
$=-0.95^{+0.17}_{-0.19}$, $\Omega_m=0.214^{+0.072}_{-0.097}$ for
SALT2 light-curve fitter and $w_{de} =$\linebreak
$=-0.85^{+0.14}_{-0.20}$, $\Omega_m=0.173^{+0.095}_{-0.098}$ for
SiFTO one \cite{Conley2011}.

One can conclude, that luminosity distance~--- redshift relation for
SNe Ia obtained by using SALT2 and SiFTO light-curve fitting methods
prefers $w_{de}$ in the 1$\sigma$ confidential range (--1.15,
--0.7), while using MLCS2k2 method in the range (--0.94, --0.6).
Therefore, the $\Lambda$-models are not excluded by SNe Ia alone
with SALT2 and SiFTO light-curve fitters, while SNe Ia with MLCS2k2
one exclude them at 1$\sigma$ confidential level.
\index{Multicolor Light Curve Shape (MLCS)|)} %

The discussions on advantages of different light-curve fitting
methods \index{light-curve fitting}  continue in the literature. In
spite of that, one can certainly conclude that data on SNe Ia
luminosity distances prefer the dark energy dominated cosmological
models. The low accuracy of current determination of $w_{de}$, which
is important for establishing of the nature of dark energy, does not
understate the main conclusion issued from SNe Ia luminosity
distance~--- redshift relation: it requires the cosmic acceleration
at $>$$99.999\,\%$ confidential level, including all systematic
effects \cite{Conley2011}.

Other observational evidence for existence of dark energy, based on
the luminosity distance~--- redshift relation, comes from the
investigations of gamma-ray bursts (GRBs). %
\index{gamma-ray bursts (GRBs)} %
Using peak energy~--- peak luminosity relation for 63 optically
identified GRBs Tsutsui et al. (2009) extend the Hubble diagram up
to $z=5.6$ \cite{Tsutsui2009} and show that dark energy dominated
models are preferable for these data too. Although constraints from
GRBs themselves are not so strong as from SNe Ia they are important
argument for existence of dark energy since the other class of
objects at higher redshifts supports that.

\newpage

\section[\!The angular
  diameter distance~--- redshift relation and acoustic peak tests]{\!\!The angular
  diameter distance~---\\ \hspace*{-0.9cm}redshift relation and acoustic peak tests}
\label{ch1-sec3}

\hspace*{3cm}The objects of known dimensions can be used as ``standard rulers'' %
\index{standard rulers} %
for measuring of angular diameter distances which are defined as
follows %
\index{angular diameter distance} %
\begin{equation}
  d_{A}\equiv\frac{D}{\Theta},
\end{equation}
where $D$ is known orthogonal to the line of sight actual size of
the object and $\Theta$ is its measured angular diameter. To relate
$d_A$
with redshift $z$ of the object astrophysicists use %
\index{FRW metric} %
FRW metric (\ref{ds}) for space-like interval at the sphere with
comoving radius $\chi(r)$ at the moment $\eta$ of emitting of
photons, which observer from the Earth detects at the current moment
of time $\eta_0$ and measures the angle $\Theta$ that subtends an
object of actual size $D=a(\eta)\chi(r)\Theta$
($d\eta=dr=$\linebreak $=d\phi=0$). Therefore, the angular diameter
distance~--- redshifts relation becomes as follows
\begin{equation}
  \label{dA}
  d_{A}=\frac{1}{(1+z)}\chi\left(\int\limits_0^z\frac{dz'}{H(z')}\!\right)\!,
\end{equation}
where function $\chi$ is defined by expression (\ref{chir}) and
$H(z')$ by equation (\ref{H}). Comparing this relation with
(\ref{dl}) one can see, that angular diameter distance and
luminosity one are related by simple ratio: $d_A=d_L/(1+z)^2$.

\index{baryon acoustic oscillations (BAO)} %
We know now that all astrophysical objects~--- galaxies, rich
clusters of galaxies and so on~--- are not ``standard rulers'' of
enough accuracy to be used for testing of cosmological models. Only
acoustic peaks in
the power spectra of CMB temperature fluctuations %
\index{CMB temperature fluctuations|(} %
and the baryon acoustic oscillations (BAO) in two-point correlation
function of matter density space distribution are. Since power
spectra and correlation function are statistical measures of
fluctuations and the tests based on their features are statistical
in their nature
too. %
\index{correlation function}%

\subsection{\!CMB acoustic peaks}
\label{ch1-subsec31}

\hspace*{3cm}\index{cosmic microwave background (CMB)|(}Mapping of
the CMB temperature sky with subdegree angular resolution and
$\Delta T/T\sim10^{-5}$ sensitivity has revealed the
acoustic peaks, %
\index{acoustic peaks} %
predicted by the adiabatic scenario of large scale structure %
\index{large scale structure}%
formation. The first such maps of small parts of the sky, obtained
in
the balloon experiments BOOMERanG \cite{Boomerang} %
\index{BOOMERanG} %
and MAXIMA \cite{Maxima}, %
\index{MAXIMA}%
have indicated that positions and amplitudes of acoustic peaks in
the
angular power spectrum %
\index{power spectrum}%
of CMB temperature fluctuations prefer the flat Universe with low
matter density, supporting the discovery of dark energy through
observations of distant supernova. The all-sky precise
measure\-ments of CMB anisotropy in the cosmic experiment WMAP
\cite{WMAP1a,WMAP7b,WMAP7c} %
\index{WMAP} %
opened up a new opportunity to determine the cosmological parameters
with high precision and provide another independent test for the
existence of dark energy.

The theory of CMB anisotropy was initiated by R.\,Sachs \& A.\,Wolfe %
\index{CMB anisotropy} %
\cite{Sachs1967} and J.\,Silk \cite{Silk1968}. The CMB radiation
comes from last scattering surface\linebreak
 at which electrons are
trapped by
hydrogen to form atoms, so-called recombination or decoupling epoch. %
\index{cosmological recombination} %
The thermal photons were tightly coupled to baryons by Thomson and
Compton scattering before the decoupling epoch at \mbox{$z_{dec}\sim
1000$}, but they could freely move to us after that. So, an adequate
calculation of the recombination process is crucial for modeling the
power spectrum of CMB temperature fluctuations and polarization. %
\index{power spectrum}%
The first analyses of recombination kinetics were carried out by
Zeldovich et al. (1968) \cite{Zeldovich1968} and Peebles (1968)
\cite{Peebles1968} in 1967. In subsequent papers [100---105] the
main\linebreak processes have been studied using the 3-level
approximation of hydrogen and helium atoms.

The most complete analysis of cosmological recombination processes
with taking into account the multi-level structure of hydrogen and
helium atoms ($\simeq$300 levels) and non-equilibrium
ionization-recombination kinetics has been performed by S.\,Seager,
D.\,Sasselov \& D.\,Scott \cite{Seager2000}.  Also all known plasma
thermal processes were taken into account therein. These authors
have
provided cosmological community with software RECFAST %
\index{RECFAST} %
\cite{Seager1999} which ensures the accuracy of calculation of
number density of electrons $\sim $1\,\%. However, the researches
aimed on improving the calculation of recombination and decoup\-ling
of the radiation from baryon plasma are still going on (see recent
papers by [108---111] and citing therein).  Development of
perturbations of number densities of ions and electrons stipulated
by scalar mode of cosmological fluctuations and physical processes
during recombination epoch have been analyzed by one of the authors
of this book \cite{Novosyadlyj2006}.  Last  improvements of
cosmological recombination calculations result into
new codes CosmoRec \cite{Chluba2010} %
\index{CosmoRec} %
and HyRec \cite{Ali-Haimoud2010}, %
\index{HyRec} %
which provide the subpercent accuracy of calculation of number
density of ionization fractions during and after recombination.

\begin{figure}
  \vskip1mm
  \centering
  \includegraphics[width=10.5cm]{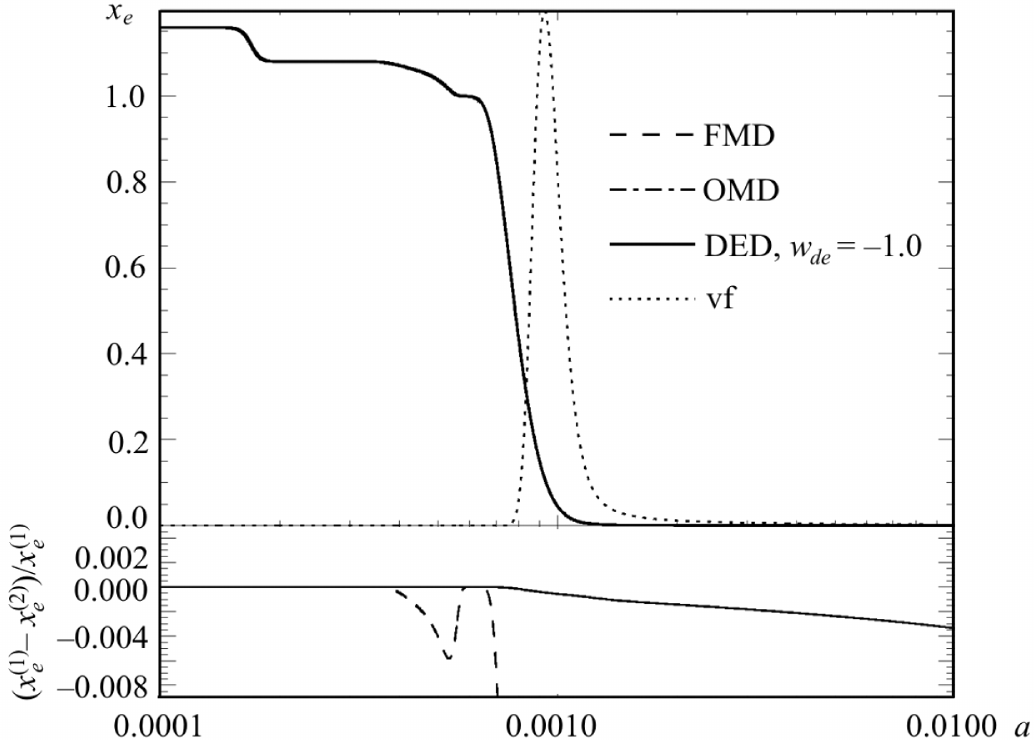}
  \vskip-2mm
  {\caption{Top panel: the dependences of relative number densities of
    free electrons on redshift in the flat matter dominated (FMD) %
    \index{flat matter dominated (FMD) model|(} %
    cosmological model (dashed line; $\Omega_{cdm}=0.95$,
    $\Omega_{b}=0.05$, $h=0.70$), open one (OMD) (dash-dotted
    line; $\Omega_{cdm}=0.25$, $\Omega_{b}=0.05$, $h=0.70$)
    and flat dark energy dominated (DED) one (solid line;
    $\Omega_{\Lambda}=0.70$, $w_{de}=-1$, $\Omega_{cdm}=0.25$,
    $\Omega_{b}=0.05$, $h=0.70$). Dotted line presents
    visibility function $d\tau /dz e^{-\tau}$ ($\times 270$). Bottom
    panel: the relative differences of $x_e$ are shown by solid line
    for (DED-OMD)/DED and dashed line for (DED-FMD)/DED\label{xe}}}

\end{figure}

The dependences of relative number density of free electrons
\mbox{$x_e\equiv$}\linebreak $\equiv n_e/(n_{H}+n_{He})$ on redshift
in the flat matter dominated (FMD) cosmological model
($\Omega_{cdm}=0.95$, $\Omega_{b}=0.05$, $h=0.70$), open one (OMD)
($\Omega_{cdm}=0.25$,
$\Omega_{b}=0.05$, $h=0.70$) %
\index{open matter dominated (OMD) model|)} %
and flat dark energy dominated (DED) one ($\Omega_{\Lambda}=0.70$,
$w_{de}=-1$, $\Omega_{cdm}=0.25$, $\Omega_{b}=0.05$, $h=0.70$)
computed by RECFAST are shown in Fig.~1.\ref{xe}. The difference
between $x_e(z)$ for DED and OMD does not exceed one percent and
between DED and FMD three percents during decoup\-ling epoch and
later, so the lines $x_e(z)$ are superimposed. The dotted line shows
visibility function $d\tau /dz e^{-\tau}$ ($\times 270$), peak of
which corresponds to the decoup-\linebreak ling moment $z_{dec}$.

Scalar mode of cosmological perturbations, which provide the large
scale structure formation, is also the main source for the CMB
temperature anisotro\-pies. %
\index{cosmic microwave background (CMB)} %
They generate CMB temperature fluctuations which can be written in
gauge-invariant form as a sum of four terms~--- the ordinary %
\index{Sachs---Wolfe effect} %
Sachs---Wolfe effect, the integrated Sachs---Wolfe term, the Doppler
term and the acoustic term \cite{Sachs1967}:\vspace*{-4mm}
\[
 \left(\!\frac{\Delta T}{T_0}\!\right)(\textbf{n}) =
 (\Phi-\Psi)(\eta_{dec},r_{dec},\textbf{n})-\int\limits_{\eta_{dec}}^{\eta_0}(\dot\Phi(\eta,r,\textbf{n})-\dot{\Psi}(\eta,r,\textbf{n}))d\eta
\,+ \]\vspace*{-3mm}
\begin{equation}
 \label{ani} +\,
 V_i(\eta_{dec},r_{dec})n^i+\frac{1}{4}D_{\gamma}(\eta_{dec},r_{dec},\textbf{n}).
\end{equation}

\index{Bardeen potentials} %
Here $\Phi$ and $\Psi$ are the Bardeen metric potentials
\cite{Bardeen1980}, $V_i$ is the baryon velocity and $D_{\gamma}$ is
a gauge-invariant variable for the radiation density fluctuations. A
dot denotes the partial derivative w.r.t.  conformal time $\eta$. %
\index{conformal time} %
For perfect fluids and for dust we have $\Psi=-\Phi$.  In Newtonian
limit the Bardeen potentials just reduce to the ordinary Newtonian
potential. For adiabatic perturbations %
\index{density perturbations} %
$\displaystyle\frac{1}{4}D_{\gamma} = \frac{1}{3}\delta_b -
\frac{5}{3}\Phi$, where $\delta_b$ is the magnitude of baryon matter
density perturbations, it corresponds to $\epsilon_m$ in Bardeen's
notation. The variable $\eta$ is conformal time, $r$ is comoving
radial coordinate and $ \textbf{n}$ is direction on the sky point
(or $\theta$, $\phi$) in metric
(\ref{ds}).  In the linear perturbation theory %
\index{perturbation theory}%
it is convenient to perform the Fourier transformation of all
spatially-dependent variables and use the equations for
corresponding Fourier amplitudes. It is convenient to present the
${\textbf
  n}$-dependence of ${\Delta T/T_0}(k, \textbf{n})$ in spherical
harmonic series
\begin{gather*}
  \frac{\Delta T}{T_0}(k,\textbf{n}) = \sum_{\ell,m}a_{\ell
    m}(k,\eta_0)Y_{\ell m}(\textbf{n}),~~ \langle a_{\ell m}
  a_{\ell^{\prime}m^{\prime}}^*\rangle= \delta_{\ell m}
  \delta_{\ell^{\prime}m^{\prime}}C_\ell, \\
  C_\ell = \frac{2}{\pi}\int\limits_0^{\infty}dk k^2\left(\!\frac{\Delta
      T}{T_0}\!\right)_{\!\ell}^{\!2}\!.
\end{gather*}

So, $C_\ell$ is angular power spectrum of temperature fluctuations %
\index{power spectrum}%
($\ell=\pi/\theta$).  For its computation in any cosmological model
the coupled system of Einstein---Boltzmann %
\index{Einstein---Boltzmann equations} %
equations for evolution of metric, density and velocity
perturbations must be solved. The complete system of such equations
for
multicomponent Universe %
\index{multicomponent Universe} %
(radiation, neutrinos, baryons and dark matter) as well as method of
their integration firstly were described by Ma \& Bertschinger
(1995) \cite{Ma1995}. These authors also have provided the
cosmological community with software COSMICS, which makes accurate
calculations of evolution density and velocity perturbations of all
components as well as perturbation of met\-rics in synchronous and
conformal-Newtonian gauges. %
\index{conformal Newtonian gauge} %
\index{synchronous gauge} %
This code was used by number of authors to calculate the transfer
function of density perturbations and power spectra of CMB
temperature fluctuations and polarization, in particular, by
\cite{cmbfast96,cmbfast99} for development of publicly available
software CMBFAST. %
\index{CMBFAST} %
Other improved publicly available cosmological codes which use
similar
approach are CMBEasy \cite{Doran2005}, %
\index{CMBEasy} %
CAMB \cite{camb,camb_source} %
\index{CAMB}%
and CLASS [122---124]%
\index{CLASS}. %
The CAMB code is included in CosmoMC software %
\cite{cosmomc,cosmomc_source}%
\index{CosmoMC}%
\index{Markov chain Monte Carlo (MCMC)} %
doing the fast Markov chain Monte Carlo exploration of cosmological
parameter space for an input set of data. It is widely used in
modern cosmology. We omit here the detailed discussion of theory of
CMB anisotropy because of its bulkiness, completeness of its
coverage in the cited above papers as well as availability of
numerous
books (\!\!\cite{Naselsky2003,Durrer2008} for example) and review papers %
\index{cosmological parameters} %
\index{CMB anisotropy} (\!\!\cite{Durrer2001,Novosyadlyj2007} for
example).

\begin{figure}
 \vskip1mm
  \centering
  \includegraphics[width=10.5cm]{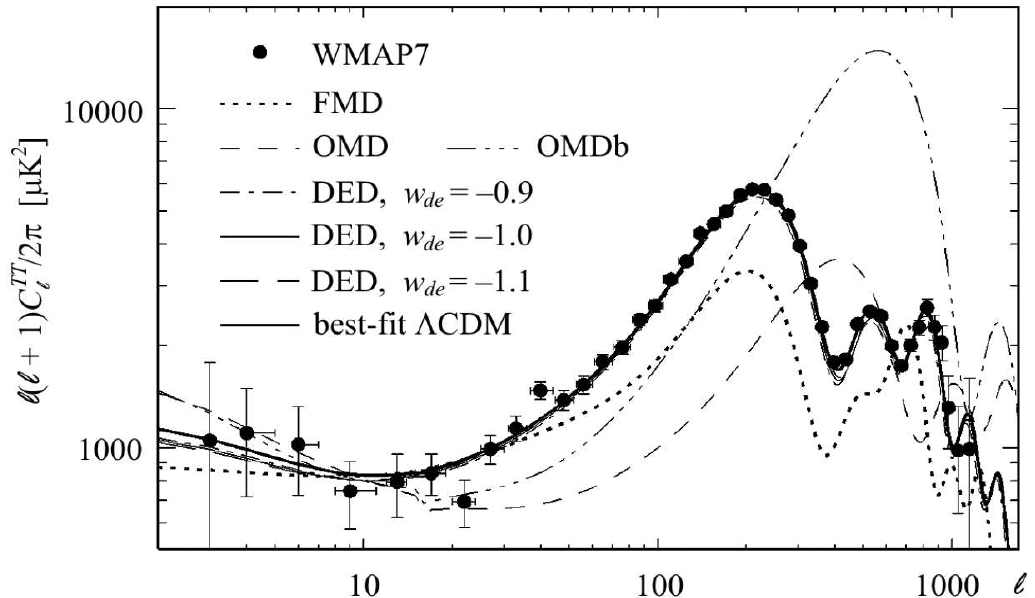}
  \vskip-2mm
    {\caption{The binned power spectra of CMB temperature fluctuations
    measured in experiment WMAP \cite{WMAP7a,WMAP7b,WMAP7c} (dots) and
    predicted by models (lines): flat cold dark matter dominated (FMD,
    $\Omega_b=0.05$, $\Omega_{cdm}=0.95$), open cold dark matter
    dominated (OMD, $\Omega_b=0.05$, $\Omega_{cdm}=0.25$), open baryon
    matter dominated model (OMDb, $\Omega_b=0.3$) and flat dark energy
    dominated (DED, $\Omega_b=0.05$, $\Omega_{cdm}=0.25$,
    $\Omega_{de}=0.7$) with $w_{de}=-0.9$, $w_{de}=-1.0$ and
    $w_{de}=-1.1$. The solid line is best-fit $\Lambda$CDM model %
    with parameters $\Omega_b=0.046$, $\Omega_{cdm}=0.234$,
    $\Omega_{\Lambda}=0.72$, $H_0=70.0$ km/s\,$\cdot$\,Mpc, $n_s=0.97$,
    $A_s=2.2\cdot10^{-9}$ and $z_{rei}=10.5$\label{wmap7_models}}}
\index{LambdaCDM model ($\Lambda$CDM)}%

\end{figure}

In Fig.~1.\ref{wmap7_models} the binned angular power spectrum of
CMB temperature fluc\-tua\-tions obtained on the base of 7-year WMAP
observations [93---95] as well as model ones calculated using CAMB
code \cite{camb,camb_source} are presented. The
solid line shows the CMB power spectrum %
\index{CMB power spectrum} %
in the flat $\Lambda$CDM model %
with best-fit parameters (minimal set) which we have determined
using the CosmoMC software \cite{Novosyadlyj2010}: baryon matter
density in units of critical one $\Omega_b=0.046$, cold dark matter
density
$\Omega_{cdm}=0.234$, cosmological constant density parameter %
\index{cosmological constant} %
\index{cosmological parameters} %
$\Omega_{\Lambda}=0.72$, Hubble constant $H_0=70.0$ km/s\,$\cdot$\,Mpc, %
\index{Hubble constant} %
spectral index of primordial power spectrum of scalar mode $n_s=0.97$, %
\index{power spectrum}%
amplitude of primordial spectrum $A_s=2.2\cdot10^{-9}$ and redshift
or reionization $z_{rei}=10.5$. The best-fit $\Lambda$CDM model
power spectrum passes all points of WMAP7 data (thick solid line)
($\chi^2_{min}=44.3$ for $N_{DoF}=39$). The other model spectra have
been calculated for reasonable values of cosmological parameters and
normalized to amplitude of 10th spherical harmonic multipole by
method proposed in \cite{Bunn1997}. One can see that all matter
dominated models (FMD, OMD, OMDb) strongly contradict the WMAP
observational spectrum ($\chi^2=116430,\,13879,\,66178$
correspondingly), that is visible by the naked eye. Only dark energy
dominated models (DED) can match the positions and amplitudes of
acoustic peaks in the angular power spectrum of CMB temperature
fluctuations (their lines are
superimposed with best-fit $\Lambda$CDM model one). %
\index{LambdaCDM model ($\Lambda$CDM)}%

To understand the numerical results, presented in
Fig.~1.\ref{wmap7_models}, %
\index{WMAP} %
let us use some analytic formulas and approximations of simulations
which have been developed in the papers [132---137].

A useful fitting formula for $z_{dec}$ is given by
\cite{Hu1996}:
\begin{equation}
  z_{dec}=\frac{1}{a_{dec}}-1=1048[1+0.00124\omega_b^{-0.738}][1+g_1\omega_m^{g_2}],
  \label{zdec}
\end{equation}

\noindent
where
\begin{equation*}
  g_1=0.0783\omega_b^{-0.238}[1+39.5\omega_b^{0.763}]^{-1},\quad
  g_2=0.56[1+21.1\omega_b^{1.81}]^{-1},
\end{equation*}
$\omega_b\equiv\Omega_bh^2$ and $\omega_m\equiv\Omega_mh^2$. In the
cosmological model with $\Omega_{cdm}=0.25$, $\Omega_{b}=0.05$,
$h=0.70$ the redshift of decoupling is $z_{dec}=1089$ (peak of
visibility function in Fig.~1.\ref{xe}).

The locations of the acoustic peaks %
\index{acoustic peaks} %
\index{CMB power spectrum} %
in the CMB power spectrum depend on the value of sound horizon at
decoupling epoch
\begin{equation}
  r_s(z_{dec})\equiv \frac{1}{1+z_{dec}}\int\limits_{z_{dec}}^{\infty}\frac{c_sdz}{H(z)}
  \label{rsi}
\end{equation}
and the angular diameter distance to the last scattering surface,
$d_A(z_{dec})$. Comparing with numerical calculations it was shown
(see [134---136] and references therein) that the spherical harmonic
which corresponds to the $m$-th acoustic peak is well approximated
by the relation
\begin{equation}
  \ell_{p_m}= (m-\phi_m)\pi\frac{d_A(z_{dec})}{r_s(z_{dec})},
  \label{lpm}
\end{equation}
where $\phi_m$ takes into account the shift of $m$-th peak from its
location in the idealized model which is caused by driving effects
from the decay of the gravitational potential.  Doran and Lilley
(2002) give the accurate analytic approximation in the form
\begin{equation}
  \phi_m={\bar\phi} - \delta\phi_m,
\end{equation}
where ${\bar\phi}$ is overall phase shift of the spectrum (or the
first peak) and $\delta\phi_m$ is a relative shift of each peak and
dip caused by the Doppler shift of the oscillating fluid.  For the
overall phase shift of the spectrum they found %
\begin{gather} {\bar\phi}=(1.466-0.466n_s)a_1r_*^{a_2},
  \intertext{ where }
  r_*\equiv
  \rho_{rad}(z_{dec})/\rho_{m}(z_{dec})=\frac{0.0416}{\omega_m}
  \left(\!\frac{1+\rho_{\nu}/\rho_{\gamma}}{1.6813}\!\right)\!\left(\!\frac{T_0}{2.726}\!\right)^{\!\!4}\!\left(\!\frac{z_{dec}}{1000}\!\right)\!\!\!\!\!
\end{gather}
is the ratio of radiation density to matter one at decoupling and
$$
a_1=0.286+0.626\omega_b,\quad
a_2=0.1786-6.308\omega_b+174.9\omega_b^2-1168\omega_b^3
$$
are fitting coefficients. Here and below the numbers in the
expressions are obtained for a present CMB temperature of
$T_0=2.726$~K and the ratio of densities of massless neutrinos and
photons $\rho_{\nu}/\rho_{\gamma}=0.6813$ for three massless
neutrino species ($f_{\nu}\equiv
\rho_{\nu}/(\rho_{\gamma}+\rho_{\nu})=0.405$).  All values can be
easily scaled to other values of $T_0$ and $f_{\nu}$.

The relative shift of the 1st acoustic peak is zero,
$\delta\phi_1=0$. For the 2nd one it is\vspace*{-2mm}
\begin{equation}
 \delta\phi_2=c_0-c_1r_*-c_2/r_*^{c_3}+0.05(n_s-1),
\end{equation}
with\vspace*{-3mm}
\begin{equation}
\begin{array}{c}
  \displaystyle c_0 = -0.1+0.213e^{-52\omega_b},~~ c_1 = 0.015+0.063e^{-3500\omega_b^2},
  \\[2mm]
\displaystyle   c_2 = 6\cdot10^{-6}+0.137(\omega_b-0.07)^2,~~ c_3 =
0.8+70\omega_b,
\end{array}
\end{equation}
and for the 3rd peak\vspace*{-3mm}
\begin{equation}
  \delta\phi_3=10-d_1r_*^{d_2}+0.08(n_s-1),
\end{equation}
with $d_1=9.97+3.3\omega_b$,
$d_2=0.0016+0.196\omega_b+2.25$\,$\cdot$\,$10^{-5}\omega_b^{-1}.$

The sound speed in the pre-recombination plasma is %
\index{sound speed} %
\[
c_s=c/\sqrt{3(1+R)}
\]\vspace*{-5mm}

\noindent with\vspace*{-2mm}
\begin{equation}
    R\equiv 3\rho_b/4\rho_{\gamma}=30315(T_0/2.726)^{-4}\omega_b a
\end{equation}
and scale factor is well approximated by\,\footnote{\,It is obtained
by
  integration of (\ref{H}) with $\Omega_K=\Omega_{de}=0$, which is
  allowable since at this epoch
  $\Omega_Ka^{-2}$, $\Omega_{de}a^{-3(1+w_{de})} \ll
  \Omega_ma^{-3}$, $\Omega_ra^{-4}$ in the realistic models.}
\begin{equation}
 a(\eta)=a_{eq}\left[\frac{\eta}{\eta_1}+\left(\!\frac{\eta}{2\eta_1}\!\right)^{\!\!2}\right]\!\!,\label{a_eta}
\end{equation}\vspace*{-5mm}

\noindent with\vspace*{-2mm}
\begin{equation}
  a_{eq}=\frac{4.17\cdot10^{-5}}{\omega_m}\left(\!\frac{1+\frac{\rho_{\nu}}{\rho_{\gamma}}}{1.6813}\!\right)\!
  \left(\!\frac{T_0}{2.726}\!\right)^{\!\!4}\!,\quad \eta_1\equiv
  \frac{\eta_{eq}}{2(\sqrt{2}-1)}=\frac{c}{H_0}\frac{\sqrt{\Omega_r}}{\Omega_m}\!\!\!\!\!
\end{equation}\vspace*{-4mm}

\noindent the integral for sound horizon (\ref{rsi}) can be reduced
to the analytic formula\vspace*{-1mm}
\begin{equation}
  r_s(z_{dec})=\frac{19.9}{\sqrt{\omega_b\omega_m}}\left(\!\frac{T_0}{2.726}\!\right)^{\!\!2}\ln\frac{\sqrt{1+R_{dec}}+\sqrt{R_{dec}+R_{eq}}}{1+\sqrt{R_{eq}}}\;
  \text{Mpc}.\!\!\!\!
\label{rs}
\end{equation}\vspace*{-3mm}

\index{acoustic peaks|(} %
The deviation of the acoustic extrema locations calculated using
formulas (\ref{lpm})---(\ref{rs}) from the values obtained by CAMB
code is $<$1\,\% for the first peak, $<$6\,\% for the second one and
$<$3\,\% for the third one (for DED models and somewhat worse for
other ones) in the sufficiently wide range of parameters.

\begin{figure}
\vskip1mm
    \raisebox{0.0cm}{\parbox[b]{6.5cm}{\caption{Dependences of locations of the 1st, 2nd and 3rd acoustic
    peaks on $\Omega_{de}$ for models with \mbox{$\Omega_m=0.3$},
    \mbox{$\Omega_b=0.05$}, \mbox{$n_s=-0.97$}, \mbox{$H_0=70$km/s\,$\cdot$\,Mpc} (solid
    lines). The 1$\sigma$ ran\-ges for them obtained from the WMAP7
    angular power spectrum of CMB temperature fluctuations
    \cite{WMAP7a} are shown by horizontal dotted lines\label{cmb_peaks}}}}\hspace{0.4cm}\includegraphics[width=6.1cm]{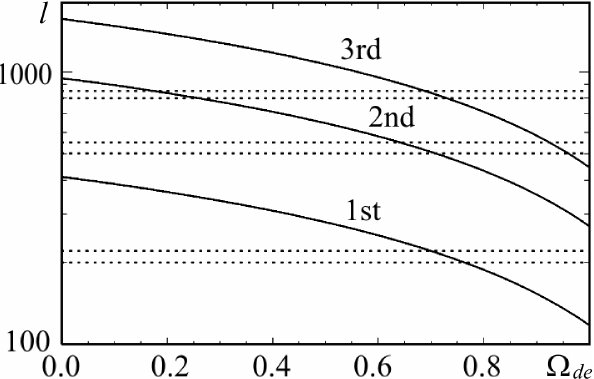}
\vspace*{-3mm}
\end{figure}

The dependences of locations of the first, second and third acoustic
peaks on $\Omega_{de}$ for models with $\Omega_m=0.3$,
$\Omega_b=0.05$, $n_s=-0.97$, $H_0=70\;$km/s\,$\cdot$\,Mpc are shown
in Fig.~1.\ref{cmb_peaks}.  The 1$\sigma$ ranges for them obtained
from the WMAP7 angular power spectrum of CMB temperature
fluctuations
\cite{WMAP7a,WMAP7c} are shown there too. %
\index{power spectrum|(}%
One can see that only model with $\Omega_{de}\approx0.7$ predicts
the peak locations which match the observational data. In
Table~1.\ref{tabl1} the locations of acoustic peaks are presented
for the FMD,
OMD and DED models. %
\index{flat matter dominated (FMD) model|)} %
The FMD and OMD models are ruled out by WMAP7 data while DED ones
well agree with them. The weak dependence of peak locations on
$w_{de}$ is recognized too. The DED models with $w_{de}\approx-1$
are preferable.
\index{acoustic peaks|)} %

As a rule for estimation of the cosmological parameters %
\index{cosmological parameters} %
\index{CMB anisotropy} %
including dark energy ones all data on CMB anisotropy are used, not
only data on peak positions. The form of the power spectrum of CMB
temperature fluctuations and its amplitude depend on practically all
cosmological parameters, the minimal set of which in the models with
dark energy contain eight ones: density parameter of baryons
$\Omega_{b}$, density parameter of cold dark matter $\Omega_{cdm}$,
density parameter of dark energy $\Omega_{de}$, EoS parameter of
dark
energy $w_{de}$, %
\index{EoS parameter} %
Hubble constant $H_0$, spectral index of initial matter density
power spectrum $n_s$ (scalar mode), amplitude of initial matter
density power spectrum $A_s$ and reionization optical depth
$\tau_{rei}$. So, determination of dark energy parameters using CMB
anisotropy data has sense jointly with other ones. It can be done by
maximization of the
likelihood function %
\index{likelihood function}\vspace*{-2mm}
\index{WMAP} %
\begin{equation*}
L(\mathbf{x};\theta_k)=\exp\left(\!-\frac{1}{2}(x_i-x_i^{th})C_{ij}(x_j-x_j^{th})\!\right)\!,
\end{equation*}\vspace*{-4mm}

\begin{table}[b]
\vspace*{-5mm} \noindent\parbox[b]{13cm}{ \caption{\bf Locations of
acoustic peaks: observations versus
models\hspace*{2.5cm}\label{tabl1}}}\vspace*{1mm} \tabcolsep10.3 pt

\noindent{\footnotesize
\begin{tabular}{|c|c|c|c|c|c|c|}
    \hline
\rule{0pt}{5mm}\raisebox{-1.7mm}[0cm][0cm]{\scriptsize Peaks} &
\rule{0pt}{5mm}\raisebox{-1.7mm}[0cm][0cm]{\scriptsize WMAP7} &
\rule{0pt}{5mm}\raisebox{-1.7mm}[0cm][0cm]{\scriptsize FMD}
&\rule{0pt}{5mm}\raisebox{-1.7mm}[0cm][0cm]{\scriptsize OMD}
&DED&DED&DED \\
    &       &     &   &{\scriptsize $w_{de}=-0.9$}&{\scriptsize $w_{de}=-1.0$}&{\scriptsize $w_{de}=-1.1$}
    \\[2mm]
       \hline
       \rule{0pt}{5mm}1st&210\,$\pm$\,10&201&\,\,\,412&216& 218& 220 \\
       2nd&526\,$\pm$\,24&426&\,\,\,947&499& 504& 508\\
        3rd& 825\,$\pm$\,25&764&1563&821& 829& 836\\[2mm]
    \hline
  \end{tabular}
  }
\end{table}

\noindent
where $\mathbf{x}$ is measured CMB anisotropy data, %
\index{CMB anisotropy} %
$\mathbf{x}^{th}$ is predicted in the model with parameters
$\theta_k$, $C_{ij}$ is covariance matrix.  Assuming a flat Universe
the number of free parameters is reduced to seven, since
$\Omega_{de}=1-\Omega_m$.

\begin{figure}
\vskip1mm \centering
\includegraphics[width=10.5cm]{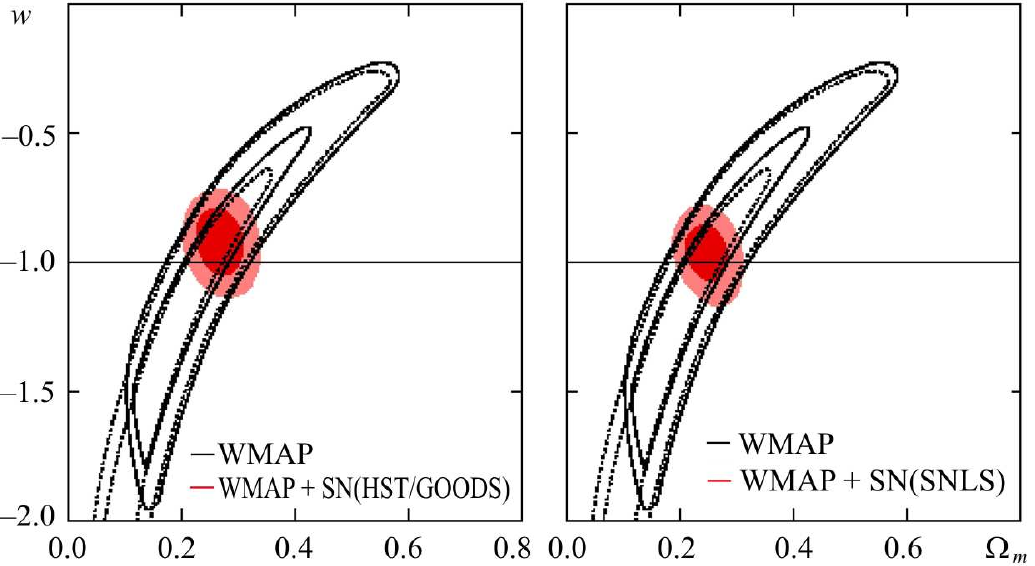}
  \vskip-2mm
    \caption{The 2D marginalized contours (68\,\% and 95\,\% CL) in the
    $\Omega_m-w$ plane for WMAP data only (lines) and WMAP data
    combined with SN Ia data (shaded regions).  The dashed lines show
    the contours for WMAP data only without prior on $H_{0}$ and solid
    ones with the assumed prior of $H_{0} < 100$~km/s/Mpc. (From
    \cite{WMAP3b})}\vspace*{-3mm}
  \label{w_omm_wmap3}
\end{figure}

Integrating $L(\mathbf{x};\theta_k)$ over $H_0$, $n_s$, $A_s$,
$\tau_{rei}$ and $\Omega_{b}$ and $\Omega_{cdm}$ for fixed their sum
$\Omega_m=\Omega_b+\Omega_{cdm}$ one can obtain 2-dimensional
marginalized likelihood function
$\mathcal{L}(\mathbf{x};\Omega_m,w_{de})$. The 2-dimensional
marginalized contours (68\,\% and 95\,\% CL) in the
$\Omega_m-w_{de}$ plane for WMAP data \cite{WMAP3b} are presented in
Fig.~1.\ref{w_omm_wmap3}. Their prolate and convoluted form
indicates some degeneracy in the $\Omega_m-w_{de}$ plane: the line
of maximal likelihood density distribution is\vspace*{-1mm}
\begin{equation}
  w_{de}^{(CMB)}\approx-2.61+9.45\Omega_m-10.45\Omega_m^2.\label{w_CMB}
\end{equation}\vspace*{-5mm}

\noindent It crosses with the similar approximation
$w_{de}^{(SNIa)}$\,(Eq.\,(\ref{w_SNIa})) in the point
$\Omega_m=0.23$, $w_{de}=-0.96$.

The best-fit values of $w_{de}$ and $\Omega_{de}$ and their
1$\sigma$ confidential ranges determined from the WMAP5 data only
are $w_{de} =-1.06^{+0.41}_{-0.42}$,
$\Omega_{de}=0.73^{+0.10}_{-0.11}$. Limits are significantly
improved when WMAP data are combined with SNe Ia data, that is
illustrated by shaded contours in Fig.~1.\ref{w_omm_wmap3}. In the
case of WMAP\-+\-SNLS data the best-fit values of dark energy
parameters are: $w_{de}=-0.97\pm0.07$ and $\Omega_{de}=0.73\pm0.03$.

Therefore, the acoustic peak locations in the angular power spectrum
of CMB temperature fluctuations are independent evidence for
existence
of dark energy or cosmological constant. %
\index{cosmological constant} %
\index{CMB temperature fluctuations|)}\vspace*{-2mm}

\subsection{\!Baryon acoustic oscillations}

\hspace*{3cm}Other realization of ``angular diameter distance~--- \mbox{redshift}'' test %
\index{angular diameter distance} %
is implemented by extraction of the baryon acoustic oscillations
(BAO) from the space distribution of galaxies which is described by
two-point correlation function %
\index{correlation function|(} %
or power spectrum of luminous matter density
\mbox{perturbations.}\looseness=-1

\index{baryon acoustic oscillations (BAO)} %
The idea about modulation of the spectrum of density perturbations
by prerecombination acoustic oscillations in the baryonic Universe
was announced first by A.\,Sakharov in 1965 \cite{Sakharov1965}.
That is why they are called sometimes in the literature Sakharov
oscillations. P.J.E.\,Peebles was first who has analyzed their
manifestation in the autocorrelation function of the mass
distribution
$\xi(r)$, %
has computed the spike in $\xi(r)$ at large $r$ caused by the
oscillation of the initial power spectrum of density perturbations
\cite{Peebles1981} %
\index{density perturbations} %
and has predicted the possibility of their detection on the base of
the galaxy sky surveys. The physics of BAO phenomena, the analysis
of numerical modeling as well as useful analytic approximations are
presented in the papers \cite{Hu1996,Eisenstein1998} and numerous
early and recent reviews and textbooks.

\begin{figure}
 \vskip1mm
  \includegraphics[width=13cm]{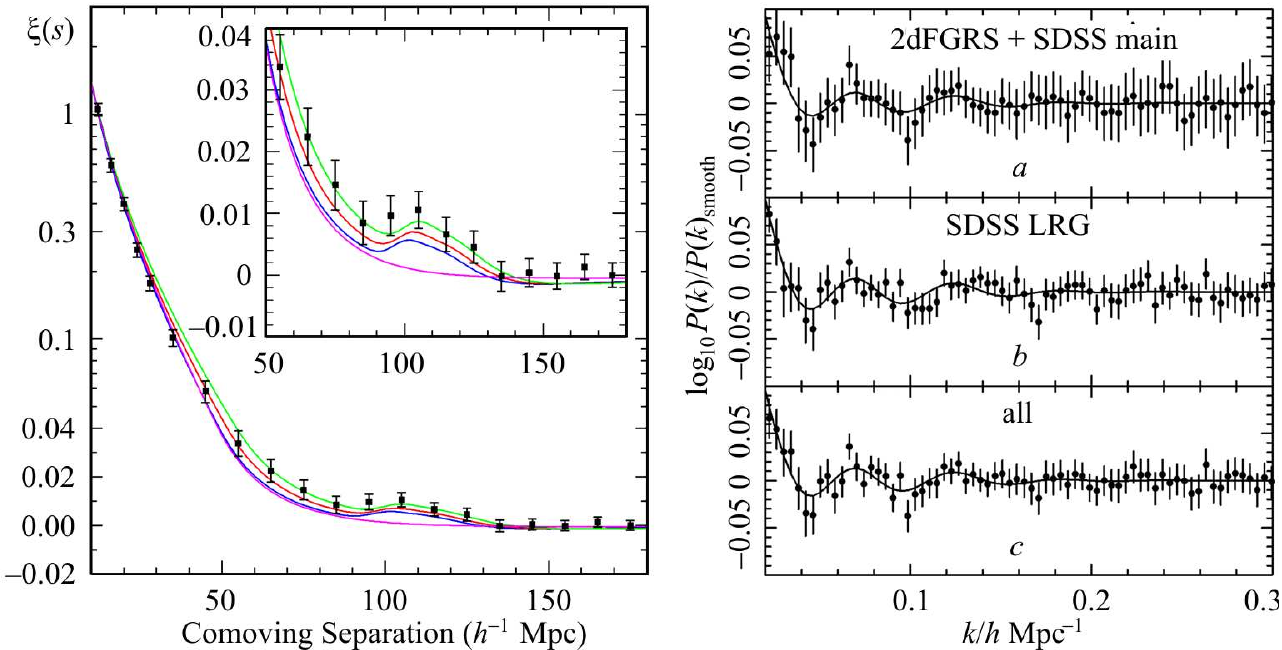}
  \vskip-2mm
    \caption{Left panel: The space two-point correlation function of the
    SDSS LRG sample.  The models are from top to bottom
    $\Omega_mh^2=0.12$, 0.13, 0.14, all with $\Omega_bh^2=0.024$ and
    $n=0.98$ and with a mild non-linear prescription folded in. For a
    pure CDM model the peak vanishes (lowest line at separation <$130h^{-1}$\,km/s\,$\cdot$\,Mpc). (From \cite{Eisenstein2005}). Right
    panel: BAO in power spectra from the combined SDSS and 2dFGRS main
    galaxies ({\it a}), from the SDSS DR5 LRG sample ({\it b}) and the combination
    of these two samples ({\it c}).  (From
    \cite{Percival2007})}\vspace*{-2mm}
  \label{bao}
\end{figure}

The first certain detection of the BAO signal was made by SDSS
\index{Sloan Digital Sky Survey (SDSS)} colla\-boration
\cite{Eisenstein2005} in 2005 using the two-point correlation
function of luminous red galaxies (left panel of Fig.~1.\ref{bao}).
Later, in 2007, they were detected in power spectra obtained from
the combined SDSS and 2dFGRS main galaxies samples, from the SDSS
DR5 LRG sample and the combination of these two samples
\cite{Percival2007} (right panel of Fig.~1.\ref{bao}).

The statistically significant bump at
$100h^{-1}\;$km/s\,$\cdot$\,Mpc scale in the redshift-space
correlation function or oscillations in the power spectrum of matter
density perturbations \index{density perturbations}
(Fig.~1.\ref{bao}) is the outcome of acoustic peak shown in
Fig.~1.\ref{wmap7_models}. It is manifestation of baryon-photon
plasma
oscillations before and during cosmological recombination epoch. %
\index{cosmological recombination} The main part of CMB radiation
comes to Earth from the last scattering
surface at $z_{dec}$ defined by maximum of visibility function %
\index{visibility function} %
\index{cosmic microwave background (CMB)|)} %
(Fig.~1.\ref{xe}), which in the DED-models with best-fit parameters
is $\approx$1090 (optical depth from current epoch to $z_{dec}$
caused by
Thomson scattering is $\tau_{dec}\approx 0.7$). %
\index{best-fit parameters} %

The fraction of free electrons at this moment was yet large enough
($x_e\approx$ $\approx 0.12$) and since the ratio of radiation
density to baryon matter one then was
$\rho_{rad}(z_{dec})/\rho_{b}(z_{dec})\approx 1.7$ the radiation
density perturbations drag the baryon matter density ones via
Compton and Coulomb interactions. It continues until the rate of
Compton scattering between photons and electrons becomes too low to
drag baryons. The drag epoch $z_{drag}$, defined as the time at
which the baryons are released from the Compton drag of the photons
in terms of a weighted integral over the Thomson scattering rate, is
well approximated by following expression obtained in
\cite{Eisenstein1998}:
\begin{equation}
  z_{drag} = 1291 \frac{\omega_m^{0.251}}{1 + 0.659\,\omega_m^{\,0.828} }
  [1 + b_1 \omega_b^{b_2}], \label{zdrag}
\end{equation}
where $b_1 = 0.313\,\omega_m^{-0.419} [1 + 0.607 \omega_m^{0.674} ]$
and $b_2 = 0.238\,\omega_m^{0.223}$. In the models with
$\Omega_b=0.05$, $\Omega_m=0.3$ and $h=0.7$ it equals 1026. %
\index{sound horizon} %
The sound horizon at $z_{drag}$, when baryons were released from the
Compton drag of photons, plays a crucial role in determination of
the location of baryon acoustic oscillations. It can be estimated
using expression (\ref{rs}), where $R_{dec}$ must be substituted by
$R_{drag}$. In the same DED model $r_s(z_{drag})=148$ Mpc in
comoving coordinates. It is well approximated by expression
\begin{equation}
  r_s(z_{drag})\approx\frac{44.5\ln(9.83/\Omega_mh^2)}{\sqrt{1+10(\Omega_bh^2)^{3/4}}},  \label{rs_appr}
\end{equation}
presented in \cite{Eisenstein1998}, where the approximation for the
first peak location in $k$-space is presented too:
\begin{equation}
  k_{BAO}\approx \frac{5\pi}{2r_s}(1+0.21\Omega_mh^2).  \label{k_BAO}
\end{equation}

In the FMD model it equals 0.083, in the OMD and DED ones with
presented above parameters it is 0.055. The linear scales
$\lambda_{BAO}=2\pi/k_{BAO}$, which correspond to these wave numbers
are 76 and 114 Mpc respectively. But we observe the angular and
redshift distributions of galaxies and deduce the linear scales from
the angular distance~--- redshift relation. So, it is possible to
measure the following ratios
\begin{align}
  \label{theta_s}
  \theta_s(z)=\frac{r_s(z_{drag})}{(1+z)d_A(z)}, \quad \Delta
  z_s(z)=\frac{r_s(z_{drag})H(z)}{c},
\end{align}
where $\theta_s(z)$ is measured angular size of physical length
$r_s(z_{drag})/(1+z)$, which lies orthogonal to the line of sight at
redshift $z$, and $\Delta z_s(z)$ is $z$-extension of the same
length when it lies along the light of sight.

\begin{figure}
\vskip1mm
    \raisebox{0.0cm}{\parbox[b]{4.5cm}{\caption{The BAO distance mea\-sure $D_V(z)$ for FMD, OMD and DED
    \index{flat matter dominated (FMD) model|(}
    models (lines) and observational \mbox{constraints} (sym\-bols) extracted
    from SDSS and 2dF galaxy redshift surveys
    \cite{Eisenstein2005,Percival2007}\label{fig_dv}}}}\hspace*{0.4cm}\includegraphics[width=8cm]{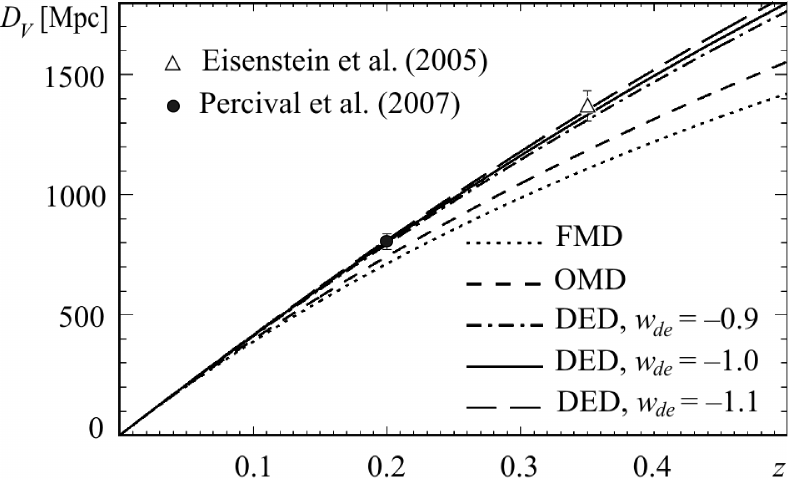}
\end{figure}

Using both measures (\ref{theta_s}) it is possible to obtain a
combined distance scale ratio which is related to spherically
averaged
correlation function or power spectrum: %
\index{power spectrum} %
\index{correlation function|)} %
\begin{equation*}
  \left[\theta_s^2(z)\Delta z_s(z)\right]^{1/3}\equiv \frac{r_s(z_{drag})}{[(1+z)^2d_A^2(z)c/H(z)]^{1/3}}.
\end{equation*}

\noindent In the paper \cite{Eisenstein2005} the following measure
was introduced
\begin{equation}
  D_V(z) = \left[(1+z)^2d_A^2(z)\frac{cz}{H(z)}\right]^{1/3}\!\!, \label{Dv}
\end{equation}
it is the dilation scale as the cube root of the product of the
radial dilation times the square of the transverse dilation. Its
$z$-dependence for FMD, OMD and DED models is presented in
Fig.~1.\ref{fig_dv}. The observational constraints for it extracted
from the SDSS Luminous Red Galaxies (SDSS LRG) survey
\cite{Eisenstein2005} and SDSS galaxy samples (Data Release 5)
combined with 2dF Galaxy
Redshift Survey (2dF GRS) data \cite{Percival2007} %
\index{Two-degree Field Galaxy Redshift survey (2dFGRS)} %
are presented there too.  One can see that they prefer DED models.

\begin{figure}
 \vskip1mm
  \includegraphics[width=8cm]{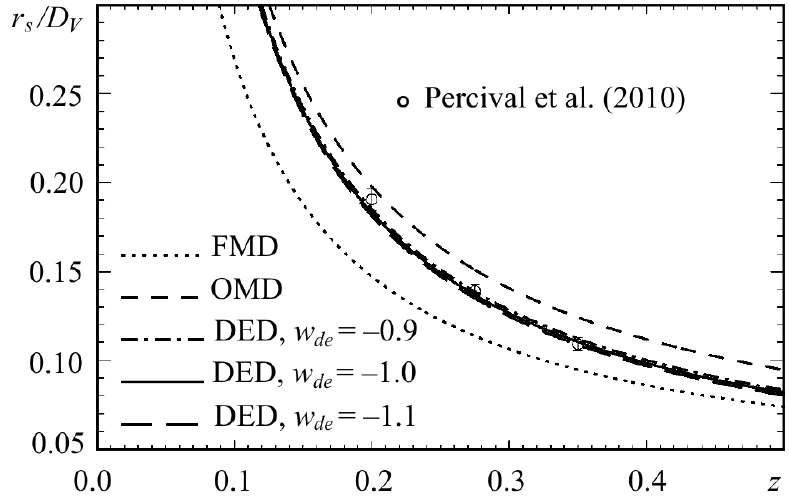}\hspace*{0.5cm}\raisebox{0.0cm}{\parbox[b]{4.4cm}{\caption{The BAO relative
  distance measure $r_s(z_{drag})/$ $D_V(z)$
    for FMD, OMD and DED models (li\-nes) and obser\-vational
    \mbox{constraints}
   \index{flat matter dominated (FMD) model|)}
    ex\-trac\-ted from SDSS DR7 and 2dF galaxy redshift sur\-veys
    \cite{Percival2010} (symbols)\label{fig_rbao}}}}
 \end{figure}

In the paper \cite{Percival2010} the data release 7 (DR7) of SDSS %
\index{Sloan Digital Sky Survey (SDSS)} %
galaxy survey combined with 2dF GRS data was used for measuring the
BAO signal in a series of redshift slices. The relative distance
measure $d_z=r_s(z_{drag})/D_V(z)$ was determined for $z=0.2,\,0.35$
and showed that DED models best matches these data
(Fig.~1.\ref{fig_rbao}).

\begin{wrapfigure}{r}{6.1cm}
\hspace*{1mm}\noindent\includegraphics[width=6cm]{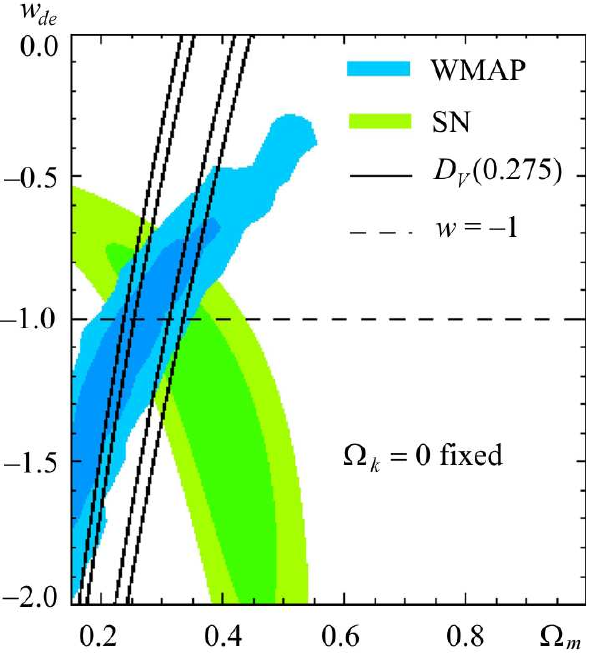}\vspace*{-2mm}\\
\hspace*{1mm}\raisebox{0.2cm}{\parbox[b]{6.0cm}{\caption{The 2D
marginalized contours (68\,\% and 95\,\% CL) in the
    $\Omega_m-w_{de}$ plane from BAO data (lines), WMAP and Union SN
    Ia. BAO constraints in $\Omega_m-w_{de}$ space deduced from the
    $r_s/D_V(0.275)$ relation (From \cite{Percival2010})\label{w_omm_bao}}
  }}\vspace*{-3mm}
\end{wrapfigure}

From the likelihood analysis Percival et al. (2010) found that the
BAO data alone weakly constraint the dark energy parameters in the
plane $\Omega_m-w_{de}$, that is shown in Fig.~1.\ref{w_omm_bao} by
lines. The strong degeneracy of $\Omega_m-w_{de}$ likelihood
distribution function give possibility to constraint the matter
density parameter $\Omega_m$ for fixed EoS parameter $w_{de}$
according to approximate
expression %
\index{EoS parameter}
\[
\Omega_m\approx0.282+0.0935(1+w_{de})\,+
\]\vspace*{-5mm}
\begin{equation}
  +\,0.015(1+w_{de})^2\pm[0.058+0.012(1+w_{de})]. \label{w_BAO}
\end{equation}
It crosses with similar approximation $w_{de}^{(SNIa)}$
(Eq.~\ref{w_SNIa}) in the point $\Omega_m\approx$ $\approx0.275$,
$w_{de}\approx-1.06$. The limits are significantly improved when BAO
data are combined with SN Ia or CMB data. Really, BAO data combined
with SN Ia ones give the best-fit values and 1$\sigma$ CL as follows
$\Omega_m=0.29\pm0.02$, $w_{de}=-0.97\pm0.11$, and combined with
WMAP5 data on CMB anisotropy they give $\Omega_m=0.283\pm0.026$,
$w_{de}=-0.97\pm0.17$.

Therefore, BAO data extracted from different galaxy surveys alone
prefer cosmological models with dark energy and combined with CMB
anisotropy and SN Ia distance moduli data significantly improve the
determination of dark energy parameters. %
\index{cosmic microwave background (CMB)|(} %
\index{CMB anisotropy|(} %

\subsection{\!X-ray gas fraction in clusters}

\hspace*{3cm}Other probe of the accelerated expansion of the
Universe based on the ``angular diameter distance~--- redshift''
relation is the measurement of the X-ray gas mass fraction,
\index{X-ray gas mass function} $f_{gas}$, in clusters of
\mbox{galaxies}\index{galaxy clusters} situated at different
redshifts. This $f_{gas}$~technique for determination of
cosmo\-logical parameters was proposed independently by S.\,Sasaki
\cite{Sasaki1996} and U.\,Pen \cite{Pen1997} in 1996. It was
improved and tested in papers [146---151] on the base of data of
Chandra observatory.

The first certain detection of cosmic acceleration using the
$f_{gas}$~technique was made by Allen et al. in 2004
\cite{Allen2004}
using Chandra %
\index{Chandra} %
observations of 26 hot (\mbox{$kT\gtrsim{5}$~keV}), X-ray luminous
(\mbox{$L_{bol}\gtrsim 10^{45}h_{70}^{-2}$~erg/s}), dynamically
relaxed clusters spanning the redshift range 0.07---0.9.  It led to
a $\sim $$3\sigma$ detection of the acceleration of expansion of the
Universe and the tight constraint on the mean

\noindent mass~density \mbox{$\Omega_{m}=0.25\pm0.04$} in excellent
agree\-ment with independent findings from SN Ia distance moduli,
CMB ani\-so\-tro\-py and galaxy \mbox{redshift} sur\-veys studies.
This constraint originates from the dependence of the $f_{gas}$
measurements, which are derived from the observed X-ray gas
temperature and density profiles, on the assumed distances to the
clusters, \mbox{$f_{gas}\propto d_A(z)^{1.5}$}. To understand the
origin of this de\-pen\-dence, consider a spherical region of
observed angular radius $\theta$ within which the mean gas mass
fraction is measured. The physical size, $R$, is related to the
angle $\theta$ as $R =
\theta d_{A}$.  The X-ray luminosity %
\index{X-ray luminosity} %
emitted from within this region, $L_{X}$, is related to the detected
flux, $F_{X}$, as $L_{X} = 4\pi d_{L}^2 F_{X}$, where $d_{L}$ is the
luminosity distance and $d_{A}=d_{L}/(1+z)^2$ is the angular
diameter
distance. %
\index{angular diameter distance|(} %
Since the X-ray emission is primarily due to collisional processes
(bremsstrahlung and line emission) and is optically thin, we may
also write $L_{X} \propto n^2 V$, where $n$ is the mean number
density of colliding gas particles and $V$ is the volume of the
emitting region, with $V=4\pi(\theta d_{ A})^3/3$.

Considering the cosmological distance %
\index{cosmological distance} %
dependences, we see that \mbox{$n \propto $} $\propto d_{ L}/d_{
A}^{1.5}$, and that the observed gas mass within the measurement
radius $M_{gas} \propto nV \propto d_{L}d_{ A}^{1.5}$. The total
mass, $M_{tot}$, determined from the X-ray data under the assumption
of hydrostatic equilibrium, $M_{tot} \propto d_{A}$.  Thus, the
X-ray gas mass fraction measured within angle $\theta$ is
$f_{gas}=M_{gas}/M_{tot} \propto d_{L}d_{A}^{0.5}$. The expectation
from non-radiative hydrodynamical simulations is that for the
largest ($kT\gtrsim 5$~keV), dynamically relaxed clusters and for
measurement radii beyond the innermost core ($r \gtrsim r_{2500}$,
where $r_{2500}$ is defined by condition $\rho_{gas}(r\le
r_{2500};z)\ge 2500\rho_{cr}(z)$), $f_{gas}$ should be approximately
constant with redshift. However, possible systematic variation of
$f_{gas}$ with redshift can be accounted for in a straightforward
manner, so long as the allowed range of such variation is
constrained by numerical simulations or other comple-\linebreak
mentary data.

The rigorous phenomenological expression for $f_{gas}^{ph}$ used for
testing of cos\-mo\-lo\-gy by observations is as
follows\vspace*{-3mm}
\begin{equation}
\label{fgas} f_{gas}^{ph}(z;\mathbf{P},\mathbf{Q}) = \frac{ K A
\gamma
  b_0(1+\alpha_b z)} {1+s_0(1+\alpha_s z) } \left(\!
  \frac{\Omega_{b}}{\Omega_{m}} \!\right)\!\left[
  \frac{d_{A}^{\Lambda}(z)}{d_{A}(z)} \right]^{1.5}\!,
\end{equation}
where
$\mathbf{P}=(\Omega_b,\,\Omega_m,\,\Omega_{de},\,w_{de},\,H_0)$ is
set of parameters in the cosmological model of interest, $d_{A}(z)$
is the angular diameter distance (\ref{dA}) computed for it,
$d_{A}^{\Lambda}(z)$ is the angular diameter distance %
\index{angular diameter distance|)} %
computed in the reference spatially-flat $\Lambda \mathrm{CDM}$ model %
\index{LambdaCDM model ($\Lambda$CDM)} %
(with $\Omega_{m}=0.3$, $\Omega_{de}=0.7$, $w_{de}=-1$, $h=0.7$),
and $\mathbf{Q}=$
$=(s_0,\,b_0,\,\alpha_s,\,\alpha_b,\,K,\,A,\,\gamma$) is set of
parameters related to modeling the cluster gas mass fraction. The
factor $K$ is a `calibration' constant that parameterizes residual
uncertainty in the accuracy of the instrument calibration and X-ray
modeling; the factor $A$ accounts for the change in angle subtended
by innermost core of cluster $r_{2500}$ as the underlying cosmology
is varied; the parameter $\gamma$ models non-thermal pressure
support in the clusters; the factor $b_0(1+\alpha_{b}z)$ is the
ratio by which the baryon fraction measured at the central part of
X-ray clusters is depleted with respect to the universal mean at
redshift $z$; the parameter $s_0(1+\alpha_\mathrm{s}z)$ models the
baryonic mass fraction in stars at redshift $z$. They are discussed
in depth in \cite{Allen2008}.  Thus, in the general case the
approach contains 12 parameters ($\mathbf{P}+\mathbf{Q}$) for
determination by matching $f_{gas}^{ph}$ to $f_{gas}^{obs}$. The
number of free parameters can be reduced to 8 if the flat Universe
is assumed and 4 cluster model parameters $K,\,A,\,\gamma,\,b_0$ are
presented by one combined $\tilde{K}$, which is their product
\cite{Samushia2008}.

Allen et al. (2008) collected 42 hot, X-ray luminous, dynamically
relaxed galaxy clusters spanning the redshift range $0.05<z<1.1$ and
measured for them $f_{gas}^{obs}$ using Chandra data (see for
details \cite{Allen2008}). Since the angular diameter distance
$d_{A}(z)$, and so $f_{gas}^{ph}$, depends on the assumed dark
energy model, one can compare predicted values of the gas mass
fraction with measurements for clusters at redshift $z_i$ by
constructing a \rule{0pt}{12pt}$\chi^2=\sum\limits_i(f_{\rm
  gas}^{ph}(z_i)-f_{gas}^{obs}(z_i))^2/\sigma_i^2$\vspace*{-1mm} function
($\sigma_i$ are the one standard deviation measurement errors and
the summation is over the 42 clusters) and  constrain parameters of
given dark energy models.

\begin{figure}
\vskip1mm
  \raisebox{0.0cm}{\parbox[b]{6cm}{\caption{The 68\,\% and 95\,\% confidence constraints in the
    $\Omega_m$-$w_{de}$ plane obtained from the analysis of the
    Chandra $f_{gas}$ data  using priors on baryonic
    content and Hubble constant
    (\mbox{$\Omega_bh^2=0.0214\pm0.0020$}
    \cite{Kirkman2003}, %
    \index{Hubble constant}%
    $h=0.72\pm0.08$ \cite{Freedman2001}). Also the independent results
    obtained from CMB data  using a weak uniform prior
    on $h$ ($0.2<h<2.0$) and SNIa data  are shown. The
    inner contours show the constraint obtained from all three
    datasets without any external priors (From~\cite{Allen2008})\label{w_omm_xray}}}}\hspace*{0.5cm}\includegraphics[width=6.5cm]{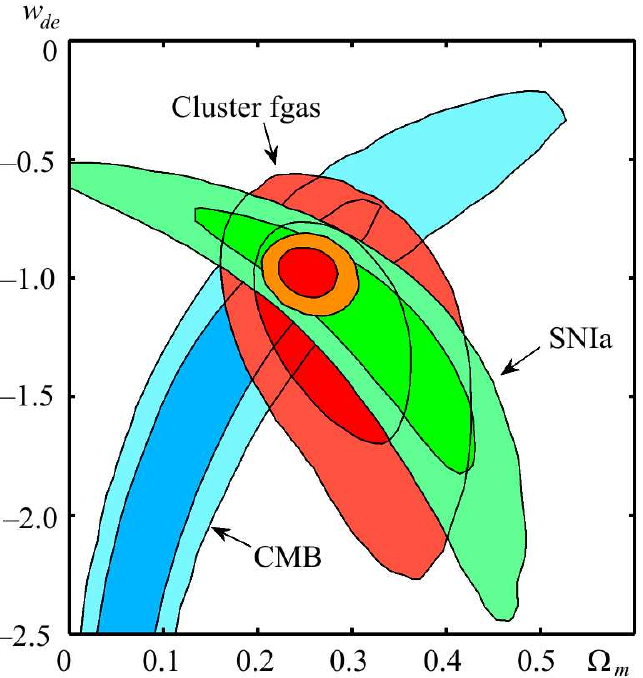}
 \end{figure}

\index{cosmological parameters} %
For determination of cosmological parameters by $f_{gas}$ technique
the authors of \cite{Allen2008,Samushia2008} have used a Markov
chain Monte Carlo method and priors on baryonic content and Hubble
constant. %
\index{Hubble constant}%
\index{Markov chain Monte Carlo (MCMC)}%
Analyzing the data for all 42 clusters, employing priors
\mbox{$\Omega_bh^2=0.0214\,\pm\,0.0020$} \cite{Kirkman2003} and
\mbox{$h=0.72\,\pm\,0.08$} \cite{Freedman2001} Allen et al. (2008)
have detected the effects of dark energy at $\sim $99.99\,\%\,C.L.
with $\Omega_m=0.28\pm0.06$ and $w_{de}=-1.14\pm0.31$ for a flat
cosmology with a constant dark energy equation of state. Practically
the same values of the dark energy parameters were obtained
independently by Samushia \& Ratra (2008). In
Fig.~1.\ref{w_omm_xray} the two-dimensional $\Omega_m$-$w_{de}$
marginalized over rest parameters contours from cluster X-ray gas
mass fraction alone as well as combined with CMB anisotropy and SN
Ia moduli distances data are presented. The const\-raints obtained
from all three data sets are as follows: $\Omega_m=0.253\pm0.021$
and $w_{de}=-0.98\pm0.07$.

\index{cosmic microwave background (CMB)|)}So, the measurements of
the ratio of baryonic-to-total mass, $f_{gas}$, in the largest,
dynamically relaxed galaxy clusters clearly detect the effects of
dark energy on the expansion of the Universe and constrain the
parameters of dark energy model via its effects on the
distance-redshift relation. The $f_{gas}$ data alone allow the range
of dark energy parameters which overlaps with ones constrained by
other data (see Fig.~1.\ref{w_omm_xray}). The accuracy is somewhat
less or comparable to that obtained from CMB anisotropy or SNe Ia
distance moduli method, but importance of this technique consists in
the fact that quite similar cosmological results are obtained from
the distance-redshift information for quite different class of
source
populations. %
\index{CMB anisotropy|)} %

\section[\!Evidence for dark energy from study of
large
  scale structure]{\!Evidence for dark energy\\\hspace*{-0.9cm}from study of large
  scale structure\label{ch1-sec4}}

\subsection{\!Linear power spectrum of matter\\\hspace*{-1.2cm}density perturbations\label{ch1-subsec41}}

\hspace*{3cm}The large scale structure of the Universe is formed by
the
gravitational growth of primordial perturbations %
\index{large scale structure}%
\index{primordial perturbation} %
generated presumably in the inflation epoch.  As it follows from the
pioneering work of E.\,Lifshits \cite{Lifshits1946} the temporal
behavior of amplitudes of the Fourier modes
\begin{equation*}
  \delta(\mathbf{k},\eta)\equiv\frac{1}{(2\pi)^3}\int e^{-i\mathbf{k}\mathbf{r}}\delta(\mathbf{r},\eta)d^3r,
\end{equation*}
of density perturbations in the real space
\begin{equation*}
  \delta(\mathbf{r},\eta)\equiv(\rho(\mathbf{r},\eta)-\bar{\rho}(\eta))/\bar{\rho}(\eta))
\end{equation*}
is result of competition of gravity and pressure gradient, thus,
depends on nature of density-dominating component and relation of
scale of perturbation to horizon scale.  In the case of spatial
isotropy the most interesting is dependence of Fourier modes of
density perturbations on the module of wave vector
$k\equiv|\mathbf{k}|$.  The amplitudes of superhorizon perturbations
($k\eta\ll1$) in the synchronous gauge %
\index{synchronous gauge} %
increase as $\delta(k,\eta)\propto \eta^2$ in RD epoch as well as in
MD one. But $\eta\propto a\propto t^{1/2}$ in RD epoch, and
$\eta\propto a^{1/2}\propto t^{1/3}$ in MD epoch. It means, that
starting from the radiation matter equality
\begin{equation}
  \eta_{eq}  =2(\sqrt{2}-1)\frac{c}{H_0}\frac{\sqrt{\Omega_r}}{\Omega_m}=\frac{16.05}{\Omega_mh^2},\label{eta_eq}
\end{equation}\vspace*{-3mm}
\begin{equation}  a_{eq}
  =1.619\cdot10^{-7}\eta_{eq}^2\Omega_mh^2=\frac{4.17\cdot10^{-5}}{\Omega_mh^2}, \label{a_eq}
\end{equation}
when $\bar{\rho}_r(\eta_{eq})=\bar{\rho}_m(\eta_{eq})$, the rate of
increasing of density perturbations is lower. The wave number that
corresponds to the horizon scale at the radiation matter equality is
as follows
\begin{equation}
  k_{eq}\equiv \eta_{eq}^{-1}=0.0623\Omega_mh^2.
  \label{keq}
\end{equation}

\begin{figure}
  \vskip1mm
  \includegraphics[width=13cm]{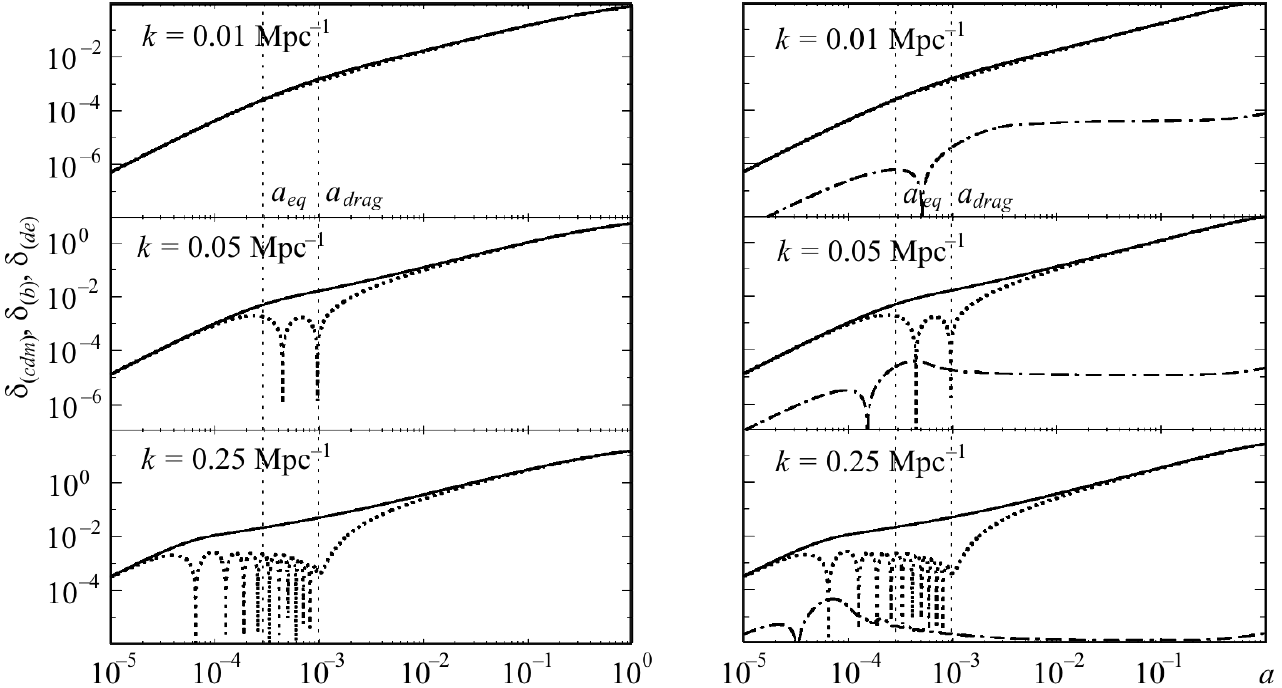}
  \vskip-2mm
    \caption{Evolution of Fourier amplitudes ($k=0.01,\,0.05,\,0.25$~Mpc$^{-1}$) of density perturbations of cold dark matter (solid
    line), baryonic matter (dotted line) and dark energy (dash-dotted
    line) computed by CAMB. In the left panel OMD model, in the right
    DED with $w_{de}=-0.9$ (parameters are the same as in
    Fig.~1.\ref{wmap7_models}). The vertical thin dotted lines show
    radiation matter equality moment, $a_{eq}$, and the drag one,
    $a_{drag}$, when the baryons are released from the Compton drag of
    the photons}
  \label{ddeb}
\end{figure}

The density perturbations of baryon-photon plasma with scale smaller
than horizon scale ($k>k_{eq}$) oscillate with approximately
constant amplitudes, $\delta_b\propto\cos{(k\eta/\sqrt{3})}$ till
$a_{drag}$, when the baryons are released from the Compton drag of
the photons due to the recombination (dotted line in
Fig.~1.\ref{ddeb}).

\index{density perturbations} %
The Fourier amplitudes of density perturbations of cold dark matter
in\-crea\-sed all time, since it is collisionless. But perturbations
with wave numbers $k\gg k_{eq}$ have entered the horizon long before
the radiation matter equality, at
$\eta_h=k^{-1}\ll\eta_{eq}=k_{eq}^{-1}$ and oscillations of
baryon-photon plasma effectively suppressed the increasing of dark
matter perturbations, so that at $\eta_h\ll \eta\ll$ $ \ll\eta_{eq}$
they grow logarithmically \cite{Hu1995}:
\begin{equation}
  \delta_{cdm}\propto \ln{\left(\!I_2\frac{a}{a_h}\!\right)}\!, \label{delta_ln}
\end{equation}
where $a_h\equiv a(\eta_h)$ can be estimated using (\ref{a_eta})
\begin{equation}
  a_h=2(\sqrt{2}-1)\frac{k_{eq}}{k}a_{eq}\left(\!\!1+\frac{(\sqrt{2}-1)}{2}\left(\!\frac{k_{eq}}{k}\!\right)^{\!\!2}\right)\approx2(\sqrt{2}-1)\frac{k_{eq}}{k}a_{eq}.\label{a_h}
\end{equation}
and the value of constant $I_2= 0.594 ( 1 - 0.631 f_\nu + 0.284
f_\nu^2 )$ depends on neu-\linebreak trino fraction in the
relativistic component $f_{\nu}\equiv \rho_\nu/\rho_r$
($f_{\nu}=0.405$ and $I_2=$ $=0.47$ for standard model).

In Fig.~1.\ref{ddeb} the evolution of density perturbations of cold
dark matter and baryonic matter with wave numbers
$k=0.01,\,0.05,\,0.25$\,Mpc$^{-1}$ is shown from before horizon
crossing up to current epoch for OMD (left panel) and DED (right
panel) models. For DED models it is shown also the evolution of dark
energy density perturbations. In the models with $w_{de}={\rm
const}<-1/3$ their amplitudes are essentially lower than
corresponding amplitudes for matter components and they do not leave
appreciable fingerprints in the form of matter power spectrum. This
figure illustrates also that dark matter is responsible for the
formation of galaxies and
large scale structure of the Universe. %
\index{large scale structure}%

Therefore, Fourier amplitudes of density perturbations of cold dark
matter with $k\le k_{eq}$ before radiation matter equality $a\le
a_{eq}$ increased all time $\propto a^2$ independently on $k$, while
perturbations with $k> k_{eq}$ changed their growth rate to
(\ref{delta_ln}) at $a\ge a_h$ and their amplitudes at $a_{eq}$
depend on $k$ additionally to the $k$-dependence of the primordial
power spectrum. The pure crossing horizon effect can be well
demonstrated by using transfer function defined as
\begin{equation*}
  T_{cdm}(k,a_{eq})=\frac{\delta_{cdm}(k,a_{eq})}{\delta_{cdm}(k_{1},a_{eq})}\frac{\delta_{cdm}(k_{1},a_{1})}{\delta_{cdm}(k,a_{1})},
\end{equation*}
where $k_1\ll k_h$, $a_1\ll a_h$. Roughly assuming
$\delta_{cdm}(k\le k_{eq},a\le a_{eq})=\delta_{cdm}(k\ge $ $\ge
k_{eq},a\le a_{h})=B(k)a^2$ and $\delta_{cdm}(k\ge k_{eq},a_h\le
a\le a_{eq})=B(k)(a_h^2+$ $+\ln(a/a_h))$, where $B(k)$ is primordial
amplitude, we obtain
\begin{equation*}
T_{cdm}(k,a_{eq})\!=\!1\,\text{ for }k\ll k_{eq}\text{ and
}T_{cdm}(k,a_{eq})\!=\!\left(\!\!\frac{k_{eq}}{k}\!\!\right)^{\!\!2}\!\!\left(\!\!1\!+\!\ln{\frac{k}{k_{eq}}}\!\right)\text{
for } k\gg k_{eq}.
\end{equation*}

If cold dark matter is dominating matter component then the
$k$-depen\-dence of transfer function slightly changes after
$\eta_{eq}$, in the opposite case the other effects must be taken
into account (Silk damping \cite{Silk1968} and photon Compton drag
\cite{Hu1996} for baryonic component, collisionless damping for hot
or warm dark matter components, etc.) which additionally suppress
the initial matter power spectrum\hspace*{0.1mm}\footnote{\,The
imprint of the baryon
  acoustic oscillations at Compton drag epoch on the matter power
  spectrum was discussed shortly in the  section 1.4.2.} at $k\gg
k_{eq}$.  But horizon crossing effect is \mbox{dominating} one at
scales related to the observed large scale structure of the Universe
and \index{large scale structure}accurately described by the linear
theory of cosmological perturbations in
multicomponent~universe.\index{multicomponent Universe}

If primordial power spectrum of scalar mode perturbations, generated
in the early Universe, is power law, $B^2(k)=Ak^{n_s}$, then the
initial (after recombination) one is %
\index{power spectrum} %
\begin{equation*}
  \mathcal{P}_m(k,a)\equiv \langle\delta_m(k,a)\delta_m^*(k,a)\rangle=Ak^{n_s}T_m^2(k,a),
\end{equation*}
where $A$ is normalization constant and $T_m(k,a_{drag})$ is the
transfer function of matter density perturbations
$\delta_m(k,a)=\Omega_b\delta_b(k,a)/\Omega_m+\Omega_{cdm}\delta_{cdm}(k,a)/\Omega_m$
($\Omega_m=\Omega_b+\Omega_{cdm}$), which takes into account all
processes  affecting the form of the spectrum up to end of the drag
epoch.  The initial power spectrum $\mathcal{P}_m(k,a)$ has peak at
some $k_{max}$ from $[0,\infty]$ for any $n_s$ from $0<n_s<4$, or at
$k_{max}\approx k_{eq}$ for $n_s\approx 1$. The exact value of peak
position depends on matter density $\Omega_m$, dimensionless Hubble
parameter $h$ and spectral index $n_s$ of primordial power spectrum.
Thus, the determination of peak position in the initial power
spectrum of matter density perturbations $\mathcal{P}_m(k,a)$ from
observations \mbox{fixes} the matter content for given values of $h$
and $n_s$. If 3-cur\-va\-tu\-re is known or constrained by other
observations then these data constrain also the dark energy content.
And not only the peak position of $\mathcal{P}_m(k,a)$ give
possibility to do that, but also its inclination and amplitude at
any $k>k_{eq}$ do so. The essence is that amplitude of the matter
density perturbations depends also on dynamics of expansion of the
Universe.

Indeed, the equation of evolution of matter density perturbations in
synch\-ro\-nous gauge at DM~--- DED epochs (see
\cite{Novosyadlyj2007} and
citing therein) is as follows %
\index{dark energy dominated (DED) epoch} %
\begin{equation}
  \ddot \delta_m + \frac{\dot a}{a}\dot \delta_m+\left(\!\frac{\ddot a}{a}-2\frac{\dot a^2}{a^2}-K\!\right)\delta_m=0, \label{delta_m_evol}
\end{equation}
where $a$ is solution of equation (\ref{H}). It shows that the rate
of amplitude growth depends on matter content, parameters of dark
energy and cur\-va\-tu\-re, but does not depend on scale of
perturbations. So, the form of initial power spectrum is practically
the same during the linear stage of evolution of matter density
perturbations at DM~--- DED epochs.

\begin{figure}
\vskip1mm \raisebox{0.0cm}{\parbox[b]{4.5cm}{\caption{Evolution of
matter density perturbations at MD~--- DED
    epochs %
    \index{dark energy dominated (DED) epoch} %
    \index{flat matter dominated (FMD) model|(} %
    in the FMD, OMD and DED models computed by CAMB for the same
    parameters as in Fig.~1.\ref{wmap7_models}. Amplitudes are
    arbitrarily normalized to $10^{-2}$ at $z=100$ ($a=0.01$)~--- the
    same for any scale\label{fig_dm}}}}\hspace*{0.5cm}\includegraphics[width=8cm]{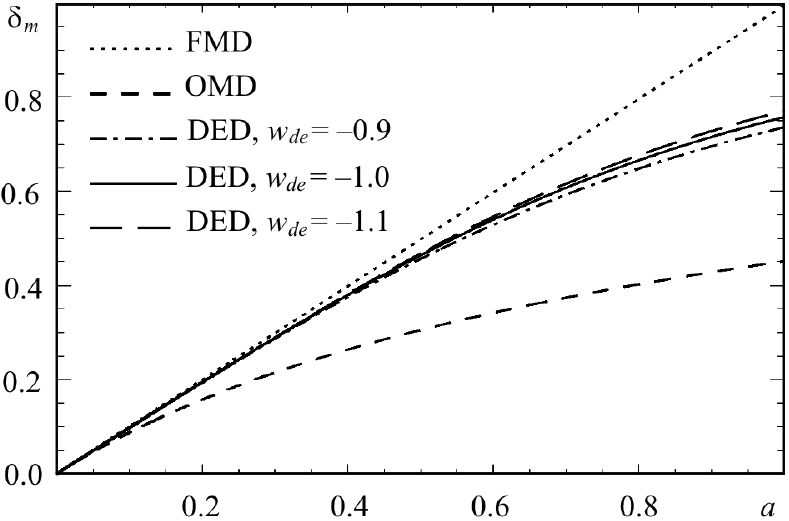}
\end{figure}

In Fig.~1.\ref{fig_dm} it is shown the evolution of matter density
perturbations from $z=100$ ($a=0.01$) up to the current epoch
($z=0$, $a=1$) for the FMD, OMD and DED models computed by CAMB with
the same parameters as in Fig.~1.\ref{wmap7_models}. The growth of
linear matter density perturbations at this epoch is approximated
with about one percent accuracy by simple expression
\cite{Linder2005}
\begin{equation}
  \delta_m(a)\propto \exp{\int\limits_0^a\Omega^{\gamma}_m(a')}d\ln{a'},
  \label{dm_appr}
\end{equation}
where
$\Omega_m(a)=\Omega_ma^{-3}/(\Omega_ma^{-3}+\Omega_{de}a^{-3(1+w_0)}+\Omega_Ka^{-2})$
and $\gamma$ is growth index which we suppose
equals\,\footnote{\,Linder in the paper \cite{Linder2005} proposed
  other best-fit values for growth index:
  $\gamma=0.545\,+$ $+\,0.05(1+w)~(z=1)$, but for $w=$~const models the
  approximation function (\ref{dm_appr}) with $\gamma=0.6$ fits the
  lines in Fig.~1.\ref{fig_dm} with percent accuracy for DED models as
  well as for OMD ones, while that $\sim$2---3\,\%.}  0.6. In the
literature some authors use the value $D(a)=\delta_m(a)/a$, dubbed
growth factor, %
\index{growth factor|(} %
which shows how the growth rate of matter density perturbations in
the models with dark energy or cur\-va\-tu\-re is retarded in
comparison with
one in the Einstein~--- de Sitter model, %
\index{Einstein-de Sitter model} %
in which $\delta_m(a)\propto a$.  Fig.~1.\ref{fig_dm} illustrates
the possibility of distinguishing of these models by studying of
matter clustering at different redshifts.

\begin{figure}
\vskip1mm \centering
  \includegraphics[width=10cm]{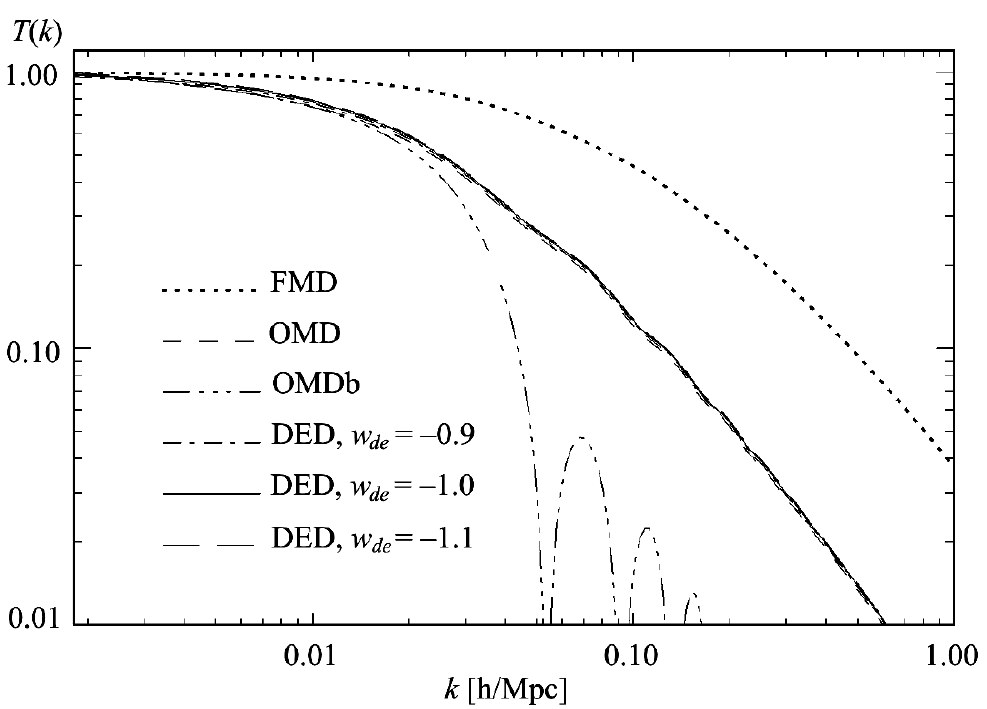}
  \vskip-2mm
  \parbox[b]{10cm}{\caption{Transfer functions of matter density perturbations at
    cur\-rent epoch in the FMD, OMD and DED models with the same
    para\-meters as in Fig.~1.\ref{wmap7_models} calculated using the CAMB~\cite{camb}\label{tkm}}}

\end{figure}

After the cosmological recombination the baryons are released from the %
\index{cosmological recombination} %
Compton drag of the photons and are free to fall inside the dark
matter potential wells and starting from $a\approx0.01$
($z\approx100$) their spectrum catches the dark matter one (see
Fig.~1.\ref{ddeb}). The form of transfer functions for both
components
\begin{equation}
  T_{(i)}(k,a)=\frac{\delta_{(i)}(k,a)} {\delta_{(i)}(k_{1},a)} \frac{\delta_{(i)}(k_{1},a_{1})} {\delta_{(i)}(k,a_{1})}, \label{Tk_i}
\end{equation}
where $a\gg0.01$, $k_1\ll k_h$, $a_1\ll a_h$ and $(i)$ notes here
$cdm$ or $b$, after that is~invariable\,\footnote{\,If the dark
matter is
  warm or some fraction is hot (massive active neutrinos)
  then the amplitude of density perturbations at scales lower
  free-streaming one decay \cite{Bond1983} (see also
  \cite{Novosyadlyj1999,Boyarsky2009} and citing therein.}.

In Fig.\,\,1.\ref{tkm} the transfer functions %
\index{transfer function} %
of matter density perturbations, \mbox{defined~as}
\begin{equation}
  T_{m}(k)=\frac{\Omega_{cdm}}{\Omega_m}T_{cdm}(k,1)+\frac{\Omega_{b}}{\Omega_m}T_{b}(k,1), \label{Tmk}
\end{equation}
at current epoch in the FMD, OMD and DED models with the same
parameters as in Fig.~1.\ref{wmap7_models} are shown. For comparison
it is shown also the transfer function for the open (low matter
density) baryonic dominated model with $\Omega_m=$ $=\Omega_b=0.3$
in which the baryon oscillations in $k$-space are frozen acoustic
oscillations which they were at the drag epoch. The lines, which
mark transfer functions in OMD and DED models with the same values
of $\Omega_{cdm}$ and $\Omega_b$,  overlap. Contrary, the transfer
functions in the models with different $\Omega_m$ (FMD and OMD or
DED models), or models with the same $\Omega_m$ but different
$\Omega_{cdm}$ and $\Omega_b$ (OMD and OMDb) are quite different.
So, transfer function of matter density perturbations is not
sensitive\,\footnote{\,The cases of
  affect of other dark energy models on the matter transfer functions
  are discussed in the next section.} to presence of cosmological
constant or dark energy with constant EoS parameter %
\index{EoS parameter} %
with value from vicinity of --1. It is explained by the essence of
transfer function, since it does not take into account the growth
factor which depends on expansion dynamics of the Universe
(Fig.~1.\ref{fig_dm}).

\begin{figure}
\vskip1mm
  \includegraphics[width=13cm]{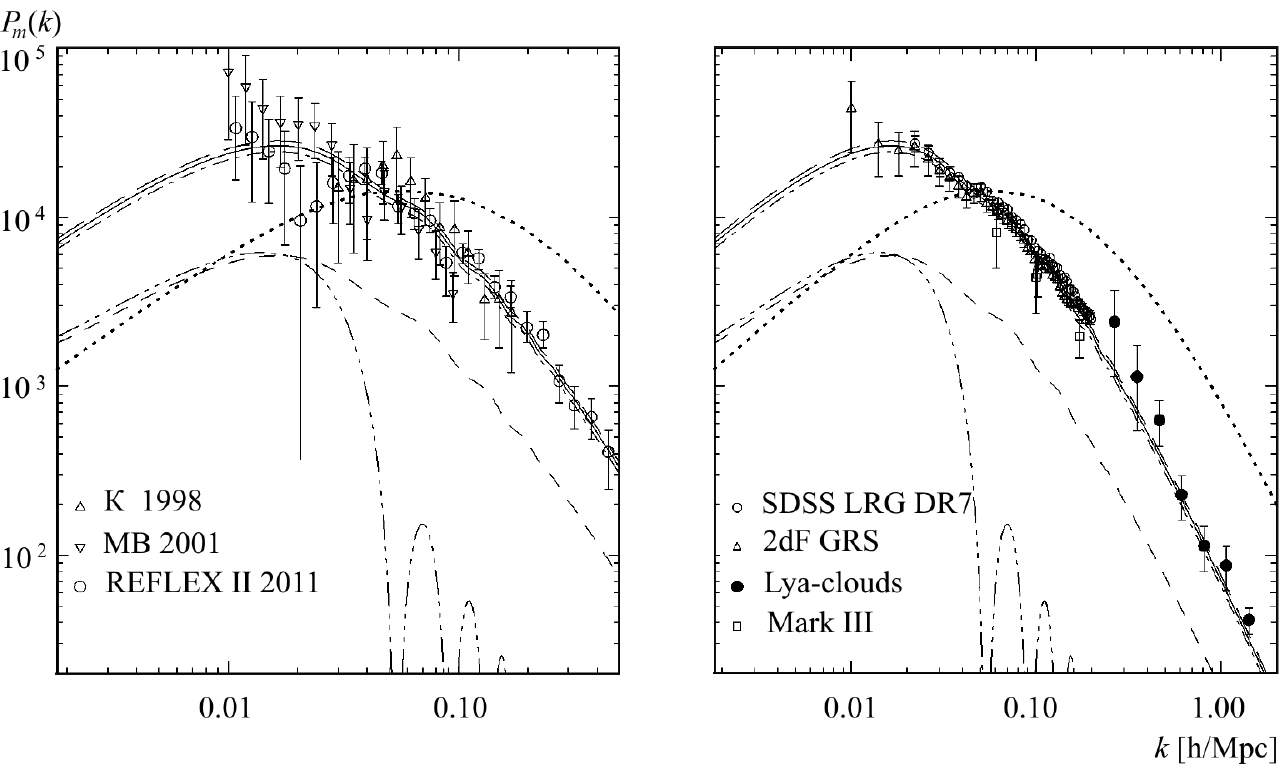}
  \vskip-2mm
  \caption{The linear power spectra in FMD, OMD, OMDb and DED models,
    which have been normalized at decoupling epoch to the amplitude of
    the angular power spectrum of CMB temperature fluctuations
    \cite{Bunn1997} obtained in the COBE experiment \cite{COBE1992},
    versus measured ones from the Abell/ACO (R 1998
    \cite{Retzlaff1998}, MB 2001 \cite{Miller2001}) and X-ray galaxy
    cluster catalogue REFLEX II \cite{REFLEX2} (left panel), galaxy
    ones (SDSS LRG DR7 \cite{Reid2010}, 2dF GRS \cite{Cole2005}),
    peculiar velocity field from Mark III catalogue
    \cite{Kolatt1997,Freudling1999} (in units of
    $\Omega_m^{1.2}\mathrm{h^{-3}}\mathrm{Mpc^3}$) and
    Ly$\alpha$-clouds along the line of sight to the distant quasars
    \cite{Croft1999} (right panel)}
  \label{pkm}
\end{figure}

\index{large scale structure}The characteristic of large scale
structure of the Universe, which can~be obtained directly from
observations, is power spectrum of matter density per\-turbations
\begin{equation}
  P_m(k)=Ak^{n_s}T_m^2(k)\left[\frac{\delta_m(1)}{\delta_m(a_i)}\right]^2\!, \label{Pmk_0}
\end{equation} %
\index{cosmic microwave background (CMB)|(}where $A$ is
normalization constant deduced, for example, from CMB
anisotropy data and the ratio in the brackets is growth factor of %
\index{CMB anisotropy} %
\index{growth factor|)} %
matter density perturbations, which can be calculated numerically or
using analytical approximation (\ref{dm_appr}). The power spectra
for the same models are presented in Fig.~1.\ref{pkm}. They are
normalized at decoupling epoch to the amplitude at horizon scale
deduced from CMB temperature fluctuations \cite{Bunn1997} (the same
normalization as in Fig.~1.\ref{wmap7_models}). %
\index{CMB temperature fluctuations} %

The normalized model power spectra are compared with observational
ones derived from the Abell/ACO %
\index{Abell-ACO} %
\cite{Retzlaff1998,Miller2001} and X-ray galaxy cluster catalogue
REFLEX II \cite{REFLEX2}, galaxy ones SDSS LRG DR7 \cite{Reid2010}
and 2dF GRS \cite{Cole2005}, peculiar velocity field from Mark III
catalogue \cite{Kolatt1997,Freudling1999} (in units
$\Omega_m^{1.2}\mathrm{h^{-3}}\mathrm{Mpc^3}$) and Ly$\alpha$-clouds %
\index{Ly$\alpha$-clouds}%
along the line of sight to the distant quasars \cite{Croft1999}.
They are related via bias factor $b_i$,
\begin{equation*}
  P_i(k)=b_i^2 P_m(k),
\end{equation*}
which depends on classes, scales and luminosities of objects
\cite{Bardeen1986} (index ``i'' marks the class of objects with
corresponding catalogue). In the left panel the cluster power
spectra are reduced by square bias factor 2.5, 4 and 3.5 for
Abell/ACO rich clusters with richness $R\ge0$ \cite{Retzlaff1998},
with richness $R\ge1$ \cite{Miller2001} and REFLEX II X-ray clusters
with $L_X>0.015\cdot\!10^{44}$ erg/s correspon\-ding\-ly. The values
of bias factors are well explained by the peaks statistics of random
Gaussian fields \cite{Bardeen1986}. The figures 14 and 15 in
\cite{REFLEX2} illustrate the dependence of bias factor on
luminosity for X-ray clusters. The spectra in the right panel are
unbiased. The amplitude of matter power spectrum derived from the
Mark III catalogue of peculiar velocities of galaxies
\cite{Kolatt1997,Freudling1999} presented in the units of
$\Omega_m^{1.2}\mathrm{h^{-3}}\mathrm{Mpc^3}$. So, in the OMD model
it is biased by the factor $\sim$14.8 ($b_{Mark\, III}\approx3.8$),
in the DED models it is biased by the factor $\sim$3.2 ($b_{Mark\,
  III}\approx1.6$). The value of bias factor for OMD model is too
large for galaxies from the point of view of the statistics of
random Gaussian fields \cite{Bardeen1986}. The amplitude of the
matter power spectrum measured from the Ly$\alpha$ forest at $z=2.5$
is recalculated to $z=0$ using (\ref{dm_appr}) in the DED model
(presented in Fig.~1.\ref{pkm}) and is practically unbiased
($b_{Ly\alpha}\sim1$). Recalculation to $z=0$ in the OMD model gives
biased clustering of Ly$\alpha$-clouds %
\index{Ly$\alpha$-clouds}%
with $b_{Ly\alpha}\sim2.3$ and, contrary, in the FMD model it gives
untibiased clustering with %
\index{flat matter dominated (FMD) model|)} %
$b_{Ly\alpha}\sim0.$4---0.6. For both models the bias parameters for
Ly$\alpha$-clouds are hard for explanation in the framework of
current
theory of large scale structure formation. %
\index{large scale structure}%
The most accurate measurements of matter power spectrum realized on
the base of galaxy redshift surveys SDSS and 2dF. They agree
perfectly with form and amplitude of the matter power spectrum in
DED models normalized to CMB data. The matter power spectrum of OMD
model  matches the SDSS LRG7 and 2dF LRG spectra with biasing
parameter $\sim $1.6. The both FMD and OMDb models are ruled out by
these data at high confidential level. The model spectra at small
scale, $k\ge0.1$, are corrected for non-linear evolution of density
perturbations in the
late epoch using HALOFIT %
\index{HALOFIT} %
approximation \cite{Smith2003}.

One can conclude, that the form and amplitude of the linear power
spectrum of DED model normalized at early epoch to CMB power
spectrum
obtained %
\index{CMB power spectrum} %
in the COBE or WMAP experiments\,\footnote{\,They have close
amplitudes at
  low spherical harmonics.}, evolved according to the theory of linear
perturbations through $\sim$13 billion years of MD~--- DED epochs, %
\index{dark energy dominated (DED) epoch} %
match well the observational power spectra derived from catalogue of
different types of objects at current epoch. Therefore, the data on
the power spectra of space inhomogeneities of different types of
objects, which are elements of the large scale structure %
\index{large scale structure} %
of the Universe, prefer the models with dark energy dominating by
density now.

The linear power spectrum of matter density perturbations is
important measurable characteristics of the large scale structure of
the Universe but not exclusive. Its moments, defined as
\begin{equation}
  \sigma^{2}_{j}(R_{ss})\equiv \frac{1}{2\pi^{2}}\int\limits_{0}^{\infty}P(k)W^2(kR_{ss})k^{2+2j}dk \quad (j=-1,0,1,2, \mbox{...}).
  \label{sigm_j}
\end{equation}
are measurable too. Here $W(R_{ss})$ is Fourier transformation of
window function of smoothing (averaging) scale $R_{ss}$. The
Gaussian and Heaviside (top-hat) smoothing are used most often. So,
the window functions can be as follows
\begin{equation}
  W_G(kR_{ss})=\exp{\left(\!-\frac{k^{2}R^2_{ss}}{2}\!\right)}\!, \nonumber
\end{equation}
for the Gaussian smoothing
$f_G(R_{ss})=\exp{(-(r-r{'})^{2}/2R_{ss}^{2})}$ in the real space,
or
\begin{equation}
  W_H(kR_{ss})=3\frac{\sin(kR_{ss})-(kR_{ss})\cos(kR_{ss})}{(kR_{ss})^{3}}\nonumber
\end{equation}
for top-hat one $f_H(R_{ss})=\Theta (1-{|r-r{'}|}/{R_{ss}})$, where
\begin{gather*}
  \Theta(x)=\left. \begin{cases}
      1, &  x \ge 0\nonumber \\
      0, &  x < 0\nonumber \\
    \end{cases}
  \right\}\!,
\end{gather*}
is Heaviside step function. The (0)-moment, $\sigma_0$, is r.m.s. of
matter density perturbation field smoothed by sphere with $R_{ss}$:
\begin{equation}
  \sigma_0(R_{ss}=\langle\delta^2(r,R_{ss}\rangle^{{1}/{2}},
 \label{deltarth}
 \end{equation}\vspace*{-5mm}

\noindent where\vspace*{-3mm}
\begin{equation}
 \delta(r,R_{ss}) \equiv \frac{3}{R^{3}_{ss}} \int\limits_{0}^{\infty}dr^{'}
 r^{'2}\delta(r^{'}) \Theta \left(\!1-\frac{|r-r{'}|}{R_{ss}}
 \!\right)\!,\nonumber
 \end{equation}

The (--1)-moment, $\sigma_{-1}$, is connected with r.m.s. of
peculiar velocity of galaxies in the sphere with radius $R_{ss}$,
dubbed
\textit{bulk flow}: %
\index{bulk flow} %
\begin{equation}
  \sigma_V(R_{ss})\equiv\langle V^2(R_{ss})\rangle^{{1}/{2}}=H_0f(1)\sigma_{-1},
 \label{deltavrth}
\end{equation}
where $f(a)$ is the ratio of growth function %
\index{growth function} %
for amplitude of velocity perturbations $V(a)$ and analogical growth
function for amplitude of density ones $\delta_m(a)$. Since, the
density perturbations and velocity ones are connected by Euler
equation %
\index{Euler equation} %
$-ikV=\dot\delta_m(a)$, so, their ratio is
$f(a)=d\ln{\delta_m(a)}/d\ln{a}$ and taking into account
(\ref{dm_appr}) $f(1)=\Omega_m^{\gamma}$. The ratio of other two
moments, $\sigma_1$ and $\sigma_2$, gives the characteristic scale
of
peaks in the Gaussian fluctuation field %
\index{Gaussian fluctuation} %
\begin{equation}
  R_{*}\equiv \sqrt{3} \cdot \frac{\sigma_{1}}{\sigma_{2}}.
\end{equation}

\begin{figure}
\vskip1mm
  \includegraphics[width=13cm]{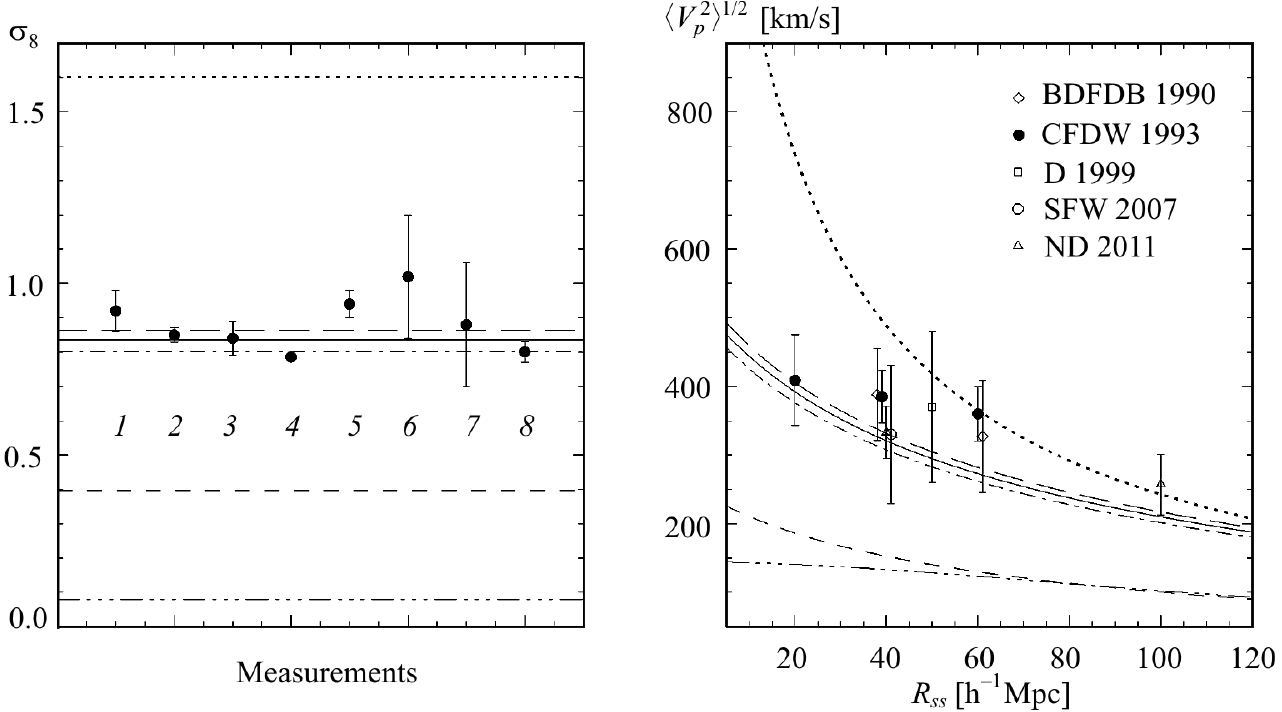}
  \vskip-2mm
\caption{Left panel: the r.m.s. of matter density perturbations
  $\sigma_8$ computed for the model spectra presented in
  Fig. 1.\ref{pkm} (horizontal lines). By the dotes its values from
  different measurements are shown: {\it 1}~--- SDSS and 2dF galaxies
  catalogues \cite{Tegmark2004,Cole2005,Eisenstein2005}, {\it 2}~---
  Ly$\alpha$ forest \cite{Tytler2004}, {\it 3}~--- Cosmic Shear
  \cite{Benjamin2007}, {\it 4}~--- cluster of galaxies catalogue
  \cite{Vikhlinin2008}, {\it 5}~--- Sunyaev---Zeldovich effect
  \cite{Reichardt2008}, {\it 6}~--- cosmic flows \cite{Feldman2010}, {\it 7}~--- CFI++
  Tully---Fisher catalogue \cite{Nusser2011}, {\it 8}~--- CMB angular power
  spectrum of temperature fluctuations \cite{WMAP7a,WMAP7c}. Right
  panel: bulk flows for different models and scales. By the signs its
  values from different measurements are shown: BDFDB 1990
  \cite{Bertschinger1990}, CFDW 1993 \cite{Courteau1993}, D 1999
  \cite{Dekel1999}, SFW 2007 \cite{Sarkar2007}, ND 2011
  \cite{Nusser2011} }
  \label{sigma}
\end{figure}

Most measurements of r.m.s. of matter density perturbations are
related to scale 8h$^{-1}$Mpc containing the mass of order of rich
clusters of galaxies. The recent ones have been carried out on the
base of different catalogs and types of objects (galaxies, clusters,
cosmic shear, Ly$\alpha$-clouds, Sunyaev---Zeldovich effect, cosmic
flows, CFI++ Tully---Fisher measurements, CMB anisotropy data)
\index{CMB anisotropy|(}\index{Ly$\alpha$-clouds} and are presented
in Fig.~1.\ref{sigma} (left panel). Some recent measurements of the
bulk flows at different scales are presented in the right panel of
Fig.~1.\ref{sigma}. One can see, that all measurements of these
moments prefer models with \mbox{dark energy.}

All techniques of obtaining the linear power spectrum of matter
density perturbations from the space distributions of any class of
objects have sequences of uncertainties, distortions and
contaminations which essentially lower its accuracy and scale
extension. It is because the tracers of large scale structure of the %
\index{large scale structure}%
Universe are luminous objects or dense absorption clouds of baryonic
matter at different distances, stages of nonlinear evolution,
chemical compositions, internal structure, dynamics, kinematics and
so on. The discreteness of their space distributions, volume and
luminosities limitations of their samples complicate their
cosmological interpretation too. Most from these problems are absent
when linear power spectrum is extracted from CMB anisotropy formed
in the early Universe.

\section[\!Angular power spectrum of CMB temperature
  fluctuations]{\!\!Angular power spectrum of CMB\\\hspace*{-0.95cm}temperature
  fluctuations\label{ch1-sec6}}

\subsection{\!\label{ch1-subsec51}Integrated Sachs---Wolfe effect}

\hspace*{3cm}\index{Integrated Sachs---Wolfe effect|(}In the
subsection {\ref{ch1-subsec31}} we have discussed briefly the
physical effects related to the formation of angular power spectrum %
\index{CMB power spectrum} %
\index{CMB temperature fluctuations|(} %
of CMB temperature fluctuations caused by scalar mode of
cosmological perturbations, main source of CMB anisotropy. We
emphasized there the importance of positions of acoustic peaks, %
\index{acoustic peaks} %
measurements of which are realization of ``angular diameter
distance~---
redshift'' test indicating the presence of dark energy. %
\index{angular diameter distance} %
The positions of troughs can be used for that as well as
\cite{Doran2002,Durrer2003}. The ratios of amplitudes of acoustic
peaks and deeps of troughs are sensitive to physical densities of
baryons and dark matter \cite{Doran2002,Durrer2003}, that is
illustrated by Fig.~1.\ref{wmap7_models}.

The amplitude and inclination of the CMB power spectrum at low %
\index{CMB power spectrum} %
spherical harmonics ($\ell\le20$) is sensitive to the presence of
dark energy or space cur\-va\-tu\-re via the late integrated
Sachs---Wolfe (ISW) effect\,\footnote{\,Producing the additional
temperature fluctuations of
  CMB by decaying of gravitational potential of large scale
  perturbation at the linear stage of its evolution. At the non-linear
  stage (formation of galaxy clusters, for example) it is called
  Rees---Sciama effect \cite{Rees1968}.} \cite{Sachs1967}, %
\index{Sachs---Wolfe effect} %
which is described by second term in the r.h.s. of expression
(\ref{ani}). The spherical $\ell$-harmonic of Fourier mode of
$\displaystyle\left(\!\frac{\Delta T}{T_0}\!\right)_k$, caused by
the late ISW effect, is  following:
\begin{equation}
  \left(\!\frac{\Delta T}{T_0}\!\right)_{k\ell}^{\!(ISW)} = -(2\ell+1)\int\limits^{a_{dec}}_{1}(\Psi'_k-\Phi'_k)j_{\ell}(kr)da,\label{dT_ISW}
\end{equation}
where $r(a)=\int^{a}_{1}(\tilde{a}^2H(\tilde{a}))^{-1}d\tilde{a}$ is
the comoving distance along the line of sight, $\Psi_k$ and $\Phi_k$
are amplitudes of Fourier modes of %
\index{Bardeen potentials} %
Bardeen metric potentials.

\index{dark energy dominated (DED) epoch|(} %
At the MD and DED epochs $\Psi_k=-\Phi_k$ and the amplitude of
$k$-mode of gravitational potential changes as
\begin{equation}
  (3K-k^2)\Psi_k(a)=\frac{3}{2}H^2_0\sum_i\Omega_i(a)\delta_i(k;a)/a,
\end{equation}

\begin{figure}[!tb]
  \vskip1mm
  \includegraphics[width=13cm]{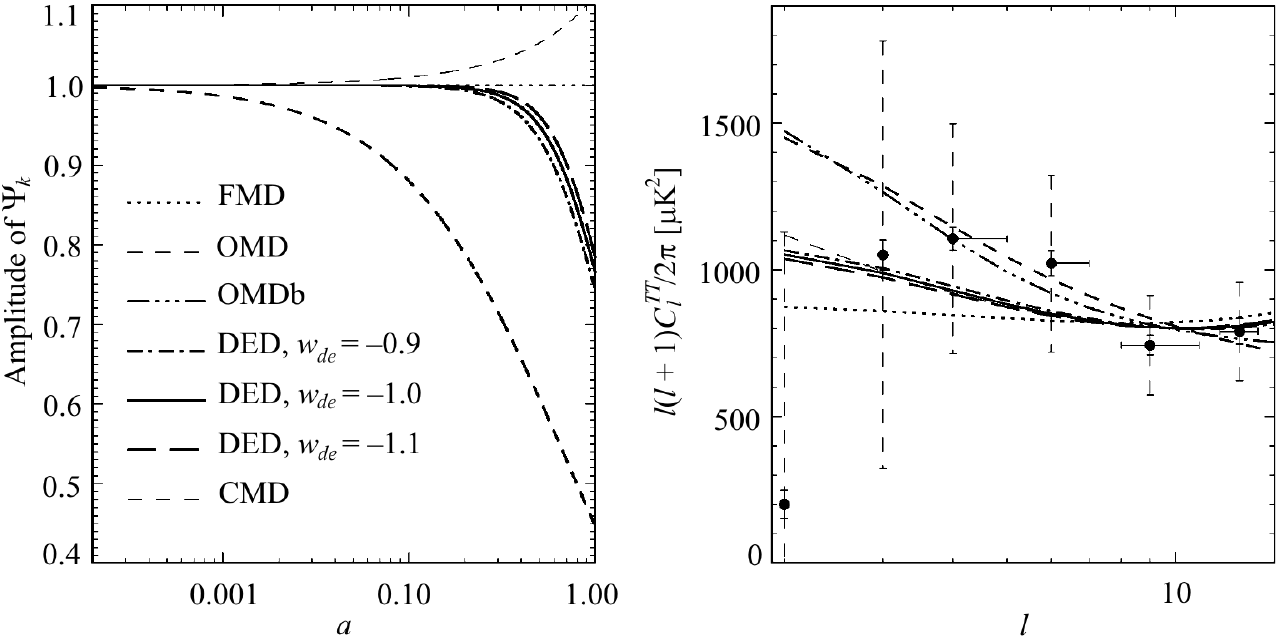}
  \vskip-2mm
  \caption{Left panel: Evolution of the amplitude of metric
    perturbation at MD~--- DED epochs in the FMD, OMD and DED models. %
    \index{dark energy dominated (DED) epoch|)}%
    \index{flat matter dominated (FMD) model|)}%
    The growing gravitational potential in the closed matter dominated
    model with $\Omega_m=1.2$ and $\Omega_{de}=0$ (CMD, thin short
    dash line) is shown for comparison. The lines of OMD and OMDb are
    overlapped. Right panel: The model angular power spectrum of CMB
    temperature fluctuations at low spherical harmonics where the late
    ISW effect gives contribution comparable with ordinary SF effect
    in the OMD and DED models (lines). The binned power spectrum from
    WMAP7 measurements \cite{WMAP7a,WMAP7c} is shown by dots. The
    1$\sigma$ errors shown by dashed lines are computed from diagonal
    terms of the Fisher matrix and include the measurement errors and
    cosmic variance, the errors shown by solid lines are measurements
    ones only multiplied by 10 for visibility}\vspace*{-2mm}
  \label{Psi_a}
\end{figure}

\noindent
that follows from Einstein equations for scalar perturbations %
\index{scalar perturbations}%
\index{Einstein equations} %
(see, for example, Eq.~62 in \cite{Novosyadlyj2007}). In the left
panel of Fig.~1.\ref{Psi_a} the evolution of amplitudes of
$\Psi_k(a)$
in FMD, OMD and DED models at MD~--- DED epochs is shown. %
\index{flat matter dominated (FMD) model|(} %
There the approximation (\ref{delta_m_evol}) for $\delta_m(a)$ was
used. The late ISW effect gives contribution to $\Delta T/T$ along
the line of sight to the last scattering surface where $\Psi'_k\ne
0$. In the FMD model it is absent, since $\Psi_k$ is constant. Its
value at decoupling equals $\Delta T/T$ caused by ordinary
Sachs---Wolfe (OSW) effect (the first term in the r.h.s. of
(\ref{ani})). In the OMD and DED models maximal rate of decaying of
$\Psi_k$ is at the current epoch, but taking into account the
properties of Bessel functions of $\ell>0$ order the main
contribution to $\Delta T/T$ is accumulated at $0.01\lesssim
z\lesssim 1$ with peak of integral over $k$ in (\ref{dT_ISW}) at
\mbox{$z\sim0.$3---0.4}, that was shown in our paper
\cite{Apunevych2000}. The later property causes the maximal effect
of ISW at small $k$ (large scales) and lowest $\ell$. The growing
gravitational potential in the closed matter dominated model with
$\Omega_m=1.2$ and $\Omega_{de}=0$ is shown there for complicity. It
must be noted here, that $\Psi'_k\ne0$ also in the vicinity of the
last scattering surface, this is caused by transition of dynamics of
expansion from radiation density dominated Universe to matter one
and oscillating of baryon-photon plasma at decoupling epoch. So, it
gives notable contribution into $\Delta T/T$ in the models with low
$\Omega_m$ and at scales close to the horizon scale at decoupling
epoch, that corresponds to spherical harmonics $\sim$100. It is
dubbed the early ISW effect and is related to the primary anisotropy
of CMB.

In the right panel of Fig.~1.\ref{Psi_a}  the model angular power
spectra of CMB temperature fluctuations  is shown for low spherical
harmonics where the late ISW effect gives maximal contribution to
$\Delta T/T$ in the OMD and DED models. %
In these models the late ISW contributes to the quadrupole component
of $\Delta T/T$ about $20\,\%$ in the DED models and about $60\,\%$
in the OMD with $\Omega_m\approx0.3$, So, the negative space
cur\-va\-tu\-re causes stronger ISW effect than dark energy for
models with similar $\Omega_m$. And contrary, the space
cur\-va\-tu\-re $|\Omega_K|\sim0.2$ gives the ISW contribution
comparable to ISW one in the DED models with $\Omega_{de}\sim0.7$
(in the figure it is shown for closed model). It  causes the
degeneracy in the space of parameters
$\Omega_m-\Omega_K-\Omega_{de}-w_{de}$ when CMB data alone are used
for determination of cosmological parameters. That is why the prior
on zero cur\-va\-tu\-re is applied as rule.  The binned power
spectrum from WMAP7 \index{WMAP} measurements \cite{WMAP7a,WMAP7c}
is shown by dots. The 1$\sigma$ errors shown by dashed lines are
computed from diagonal terms of the Fisher matrix and\linebreak
include the measurement errors and cosmic variance, the errors shown
by\linebreak vertical solid lines are measurement ones  multiplied
by 10 for visibility. Therefore, CMB anisotropy data in the range of
ISW effect are not enough accurate due to unremovable cosmic
variance for constraining the dark energy models. But since the
contribution of late ISW into $\Delta T/T$ forms
in the range of observable large scale structure of the Universe the %
\index{large scale structure}%
cross-correlation between them can be used to detect it and to
constraint the dark energy parame-\linebreak ters [183---186].

The essence of such approach consist in the fact that CMB anisotropy %
\index{CMB anisotropy|)} %
sky map, presented by $\displaystyle\left(\!\frac{\Delta
T}{T_0}\!\right)(\mathbf{n})$ in the form (\ref{ani}), contains the
ISW contribution from matter density perturbations,
$\delta_m(k;a,\textbf{n})$ at $z\lesssim 1$, which strongly
correlate with galaxies space inhomogeneities
$\delta_g(k;a,\textbf{n})=b_g(k;a)\delta_m(k;a,\textbf{n})$, deduced
from the galaxy sky surveys. For the comoving scales $\gtrsim
100$Mpc the bias factor $b_g$ varies weakly with scale and redshift
\cite{Percival2007} and the general assumption about its time and
scale independence is well-grounded. Other components of
(\ref{ani}), which are related with perturbations at decoupling
epoch, do not correlate with $\delta_g(k;a,\mathbf{n})$ at current
epoch and do not
contribute to cross-correlation function (CCF), %
\index{cross-correlation function (CCF)|(} %
defined as
\begin{equation}
  C^{Tg}(\vartheta) \equiv  \left\langle\! \frac{\Delta T}{T_0} (\mathbf{n}_1) \delta_g (\mathbf{n}_2) \! \!\right\rangle \label{C_Tgo}
\end{equation}
with the average carried over all the pairs at the same angular
distance\linebreak \mbox{$ \vartheta = | \mathbf {n}_1 - \mathbf
{n}_2 |$.}

It is possible to express this value in the harmonic space with the
use of the Legendre polynomials $\mathcal{P}_l$:\vspace*{-3mm}
\begin{equation}
  C^{Tg}(\vartheta) = \sum_{\ell=2}^{\infty} \frac {2\ell + 1} {4 \pi} C_{\ell}^{Tg}
  \mathcal{P}_{\ell} [\cos (\vartheta)], \label{C_Tgt}
\end{equation}\vspace*{-3mm}

\noindent where cross-correlation power spectra are given
by\vspace*{-1mm}
\begin{equation}
  C_{\ell}^{Tg}=\frac{2}{\pi}\int\limits_0^{\infty}dk k^2 P_m(k) I_{\ell}^{{ISW}}(k) I_{\ell}^{g}(k). \label{Cl_Tg}
\end{equation}\vspace*{-3mm}

The two integrands in the last expression are
respectively\vspace*{-1mm}
\begin{align}
  I_{\ell}^{ISW}(k) & =  2\int\limits_1^{a_{min}} e^{-\tau(a)} \left(\!\frac{\delta_m(a)}{a}\!\right)^{\!\prime}j_{\ell}[kr(a)] da, \label{Il_ISW}
  \\
  I_{\ell}^{g}(k) & = \int\limits_1^{a_{min}} b_g \mathcal{N}' (a)
  \delta_m (a) j_{\ell}[kr(a)] da, \label{Il_g}
\end{align}
where $ \Psi_k$ and $\delta_m(k,z) $ are the Fourier components of
the
gravitational potential and matter perturbations, %
\index{matter perturbations}%
$ j_{\ell} (x) $ are the spherical Bessel functions, $\mathcal{N}'$
is a selection function of the galaxy survey, $\delta_m(a)$ is
growth
function %
\index{growth function} %
of matter density perturbations at linear stage, $r(a)$ is the
comoving distance along the line of sight, $\tau(a)$ is the optical
depth along the line of sight caused by Thomson scattering by free
electrons and $a_{min}$ corresponds to $z_{max}$ of the survey.

Applying the (\ref{C_Tgo}) to the CMB and galaxies sky maps one can
obtain the observational CCF which contains the information about
parameters of our Universe. They can be constrained by comparison of
CCF function computed for the model with given parameters using
(\ref{C_Tgt})---(\ref{Il_g}).

\begin{figure}
 \vskip1mm
  \includegraphics[width=13cm]{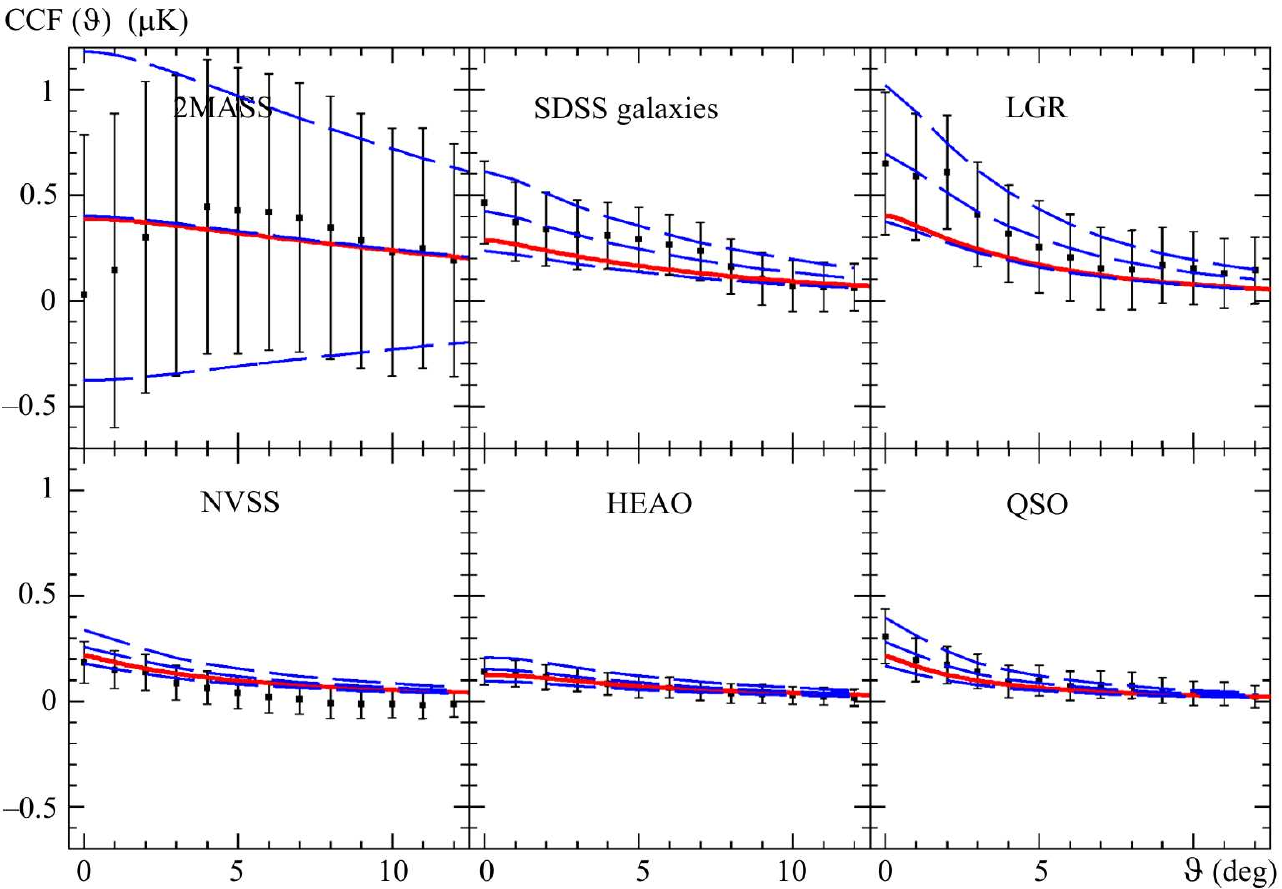}
  \vskip-2mm\caption{Measurements of the CCFs between the WMAP CMB maps and all
    available galaxy catalogue that trace the matter distribution with
    light from the whole range of the electromagnetic spectrum (black
    points), compared with the theory from WMAP best fit cosmology and
    the galactic bias from the literature (solid lines).  Their
    $1\sigma$ deviations are shown by dashed lines.  The errors are
    calculated from Monte Carlo simulations of temperature and
    density fluctuations (From \cite{Giannantonio2008})}\vspace*{-2mm}
  \label{ccf_isw}
\end{figure}

\begin{figure}
 \vskip1mm
  \includegraphics[width=13cm]{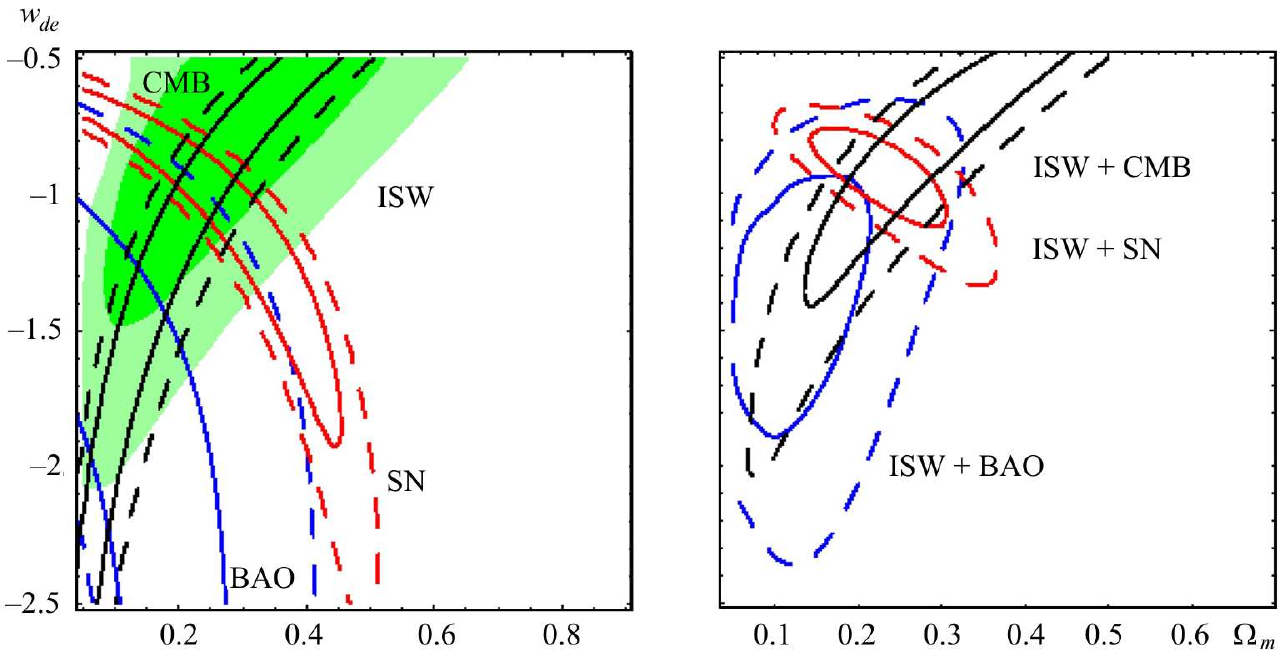}
  \vskip-2mm
    \caption{$\Omega_m-w_{de}$ constraints from measurements of the ISW
    effect (shadow range in the left panel). Constraints from other
    observations, including CMB, SNe Ia and BAO
    are shown for comparison. In the right panel the combined
    likelihoods using the ISW + each one of these other constraints
    are presented. In the both panels the 1 and 2 $ \sigma $ contours
    are shown by solid and dashed lines respectively (From
    \cite{Giannantonio2008})}\vspace*{-2mm}
  \label{w_omm_isw}
\end{figure}

Such approach has been used by several groups [187---194] to detect
the ISW effect using WMAP all sky maps of the CMB temperature
fluctuations and several maps of galaxy space distributions. The
cross-correlations between $\displaystyle\frac{\Delta T}{T_0}
(\mathbf{n})$ map and individual $\delta_m (\mathbf{n})$ maps with
different galaxy surveys that trace the matter distribution with
light from the whole range of the electromagnetic spectrum have been
detected at 2---3$\sigma$ significance. It indicates the rapid
slowing down in the growth of amplitude of density perturbations and
means the existence of dark energy in the flat Universe. The
combination of data from different surveys and redshifts essentially
reduces uncertainties and introduces important new constraints
\cite{Corasaniti2005,Gaztanaga2006}. The best surveys available for
this purpose include the following: the optical Sloan Digital Sky
Survey (SDSS), %
\index{Sloan Digital Sky Survey (SDSS)} %
the infrared 2 Micron All-Sky Survey (2MASS), %
\index{Two-Micron All-Sky Survey (2MASS)} %
the X-ray catalogue from the High Energy Astrophysical Observatory
(HEAO) %
\index{High Energy Astrophysical Observatory (HEAO)} %
and radio galaxy catalogue from the NRAO VLA Sky Survey (NVSS). %
\index{NRAO VLA Sky Survey (NVSS)} %
The high quality of the SDSS data allows us to extract some further
subsamples from it, consisting of Luminous Red Galaxies (LRG) and
quasars (QSO) in addition to the main galaxy sample. The results of
measurements of the CCFs between WMAP CMB maps and each from these %
\index{cross-correlation function (CCF)|)} %
catalogues obtained in \cite{Giannantonio2008} are presented in
Fig.~1.\ref{ccf_isw}. The analysis of auto- and cross-correlations
between all catalogues and including the full covariance matrix
between all data gave the possibility to authors of
\cite{Giannantonio2008} to increase the significance of the total
combined measurement of ISW effect up to 4.5$\sigma$ and to
constraint the $\Omega_m-w_{de}$ parameter space as it is shown in
Fig.~1.\ref{w_omm_isw}. One can see, that measurement of ISW effect
is a good probe for dark energy and independent evidence for its
presence
in our Universe, like SNe Ia, CMB acoustic peaks and BAOs. %
\index{acoustic peaks} %
Its combination with other detections of dark energy narrows
essentially confidence ranges of DE parameters (right panel of
Fig.~1.\ref{w_omm_isw}).

The analysis of the constraining power of future measurements of the
ISW effect on models of $w_{de}(z)$ was carried out in
\cite{Pogosian2005}. There it was demonstrated that the
cross-correlation of Planck CMB data and Large Synoptic Survey
Telescope galaxy catalogues will provide important independent
constraints on $w_{de}(z)$ at high $z$.\vspace*{-2mm}

\subsection{\!\label{ch1-subsec52}Weak gravitational lensing of CMB}

\hspace*{3cm}The measurements of gravitational lensing %
\index{weak gravitational lensing} %
of CMB tem\-pe\-ra\-tu\-re fluctuations by the foreground large scale %
\index{large scale structure}%
structure are powerful source of information about the geometry,
expansion history and dark components of the Universe (see excellent
reviews [199---201] and citing therein). The \mbox{first} detection
of len-sing signal at 3.4$\sigma$ significance was realized by Smith
et al. (2008) \cite{Smith2007} from cross-correlation of WMAP data
\cite{WMAP3a,WMAP3b} with $\approx$2 million radio sources founded
in the NRAO VLA Sky Survey (NVSS) \cite{Condon1998}. In the next
year Hirata et al. (2008) \cite{Hirata2008} for the same data set
announced the detection of len\-sing signal at 2.1$\sigma$
significance. They supported the result by cross-correlation of
WMAP3 data with SDSS LRG and quasar samples at
1.8$\sigma$ significance. Combining all three large scale structure %
\index{large scale structure}%
samples they stated the detection at 2.5$\sigma$ level. The obtained
there cross-correlation amplitude agrees well with one expected for
DED model with WMAP cosmological parameters.

\begin{figure}
  \vskip1mm \raisebox{0.0cm}{\parbox[b]{5.5cm}{\caption{The CMB lensing power spectra detected by Atacama Cos\-mo\-lo\-gy
    Telescope \cite{Das2011} (dots) and computed for two models
    (lines), which have practically indistinguishable $TT$ power
    spectra well matching the WMAP7 one ($\Lambda$CDM with
    $\Omega_{\Lambda}=0.73$, $\Omega_m=0.27$ and closed MD model with
    $\Omega_{\Lambda}=0$, $\Omega_m=1.29$) (From \cite{Sherwin2011})\label{grav_lens}}}}\hspace*{0.5cm}\includegraphics[width=7cm]{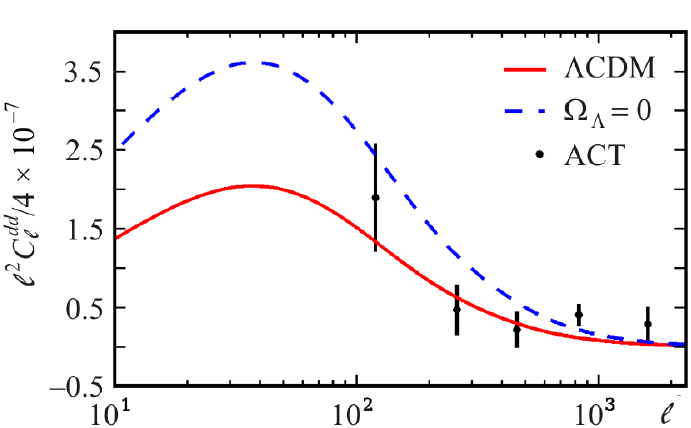}
\end{figure}

Recently, the lensing signal has been detected in CMB alone by
Atacama
Cosmology Telescope (ACT) %
\index{Atacama Cosmology Telescope (ACT)} %
\cite{Das2011} at 4$\sigma$ significance. The gravitational lensing
remaps the CMB temperature fluctuations on the sky as follows %
\index{CMB temperature fluctuations|)} \vspace*{-2mm}
\begin{equation*}
  T(\mathbf{\tilde{n}})=\tilde{T}(\mathbf{\tilde{n}}+\mathbf{\alpha(\tilde{n})}),
\end{equation*}
where $\mathbf{\alpha(\tilde{n})}$ is deflection field and tilde
denotes the unlensed quantities. It imprints a distinctive
non-Gaussian signature on the pattern of the microwave sky, which
can be measured by 4-point correlation function or power spectrum of
the convergence field $\kappa = \frac{1}{2} \nabla \mathbf{\alpha}$
in the form %
\index{correlation function}%
\[
    \label{eq:estimator}
    (2 \pi)^2 \delta(\mathbf{L}-\mathbf{L}') \hat C^{\kappa \kappa}_L
    = |N^{\kappa }(\mathbf{L})|^2
    \int\frac{d^2\mathbf{\ell}}{(2\pi)^2}
    \int \frac{d^2\mathbf{\ell}'}{(2\pi)^2}|g(\mathbf{\ell},\mathbf{L})|^2 \,\times
    \]
\[
    \times \left[\vphantom{\int} T^*(\mathbf{\ell})
      T^*(\mathbf{L}-\mathbf{\ell}) T(\mathbf{\ell}')
      T(\mathbf{L}'-\mathbf{\ell}') - \langle T^*(\mathbf{\ell})
      T^*(\mathbf{L}-\mathbf{\ell}) T(\mathbf{\ell}')
      T(\mathbf{L}'-\mathbf{\ell}') \rangle_\mathrm{Gauss}
      \right]\!,
 \]
where $\mathbf{\ell}, \mathbf{\ell}', \mathbf{L}, \mathbf{L}'$ are
coordinates in Fourier space in the flat-sky approximation, $g$
defines the filters that can be used to optimize signal-to-noise,
$N$ is a normalization, and the second term is the Gaussian part of
the 4-point correlation function. Subtraction from the full 4-point
function its Gaussian part gives the non-Gaussian lensing signal.

\index{Integrated Sachs---Wolfe effect|)}\index{power spectrum|)}The
power spectrum of the convergence field extracted from ACT
tem\-pe\-ra\-tu\-re maps in \cite{Das2011} is
presented\,\footnote{\,For details
  of pipeline of obtaining the lensing power spectrum from ACT data we
  refer to the original paper \cite{Das2011}.} in
Fig.~1.\ref{grav_lens}. There are also shown the predicted power
spectra of the convergence field for $\Lambda$CDM
($\Omega_{\Lambda}=0.73$, $\Omega_m=$ $=0.27$) and closed MD
($\Omega_{\Lambda}=0$, $\Omega_m=1.29$) models, which have
practically indistinguishable $TT$ power spectra \cite{Sherwin2011}.
They differ only at largest angular scales ($\ell<10$) due to the
ISW effect, but there cosmic variance is too large (see right panel
of Fig.~1.\ref{Psi_a}). Strong degeneracy in $\Omega_m-w_{de}$
parameter space constrained by CMB $TT$ power spectrum alone is
illustrated by prolate and convoluted form of confidence contours in
Fig.~1.\ref{w_omm_isw}. The addition of lensing data should break
this degeneracy.

\begin{figure}
\noindent\includegraphics[width=6.5cm]{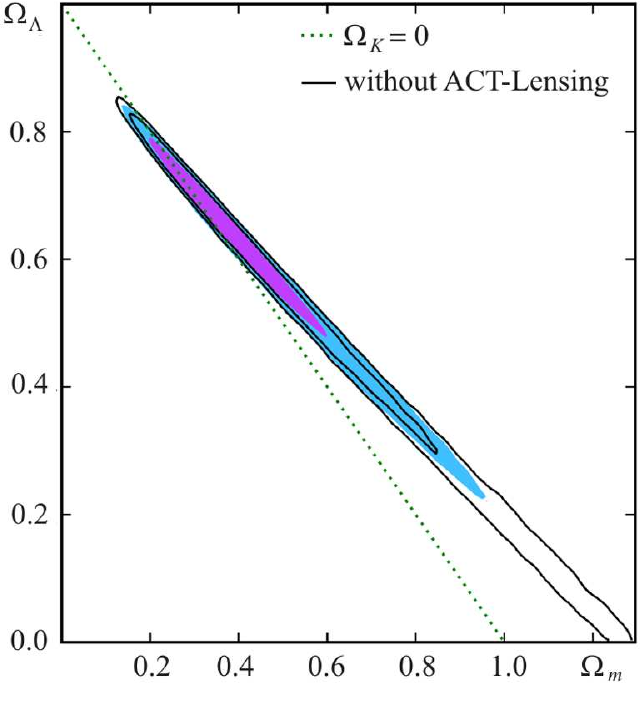}\hspace*{0.4cm}
\raisebox{0.0cm}{\parbox[b]{6.0cm}{\caption{\footnotesize
Two-dimensional marginalized posterior probability for
    $\Omega_m$ and $\Omega_\Lambda$ (68\,\% and 95\,\% C.L.). Shaded
    contours are for WMAP + ACT lensing, black lines are for WMAP
    only (From \cite{Sherwin2011})\label{w_omm_gr_lens}}

\vskip5mm \noindent  that ACT lensing data prefer mo\-del with dark
energy and serve as one more independent argument for its existence
and dominance \mbox{in density.}

\hspace*{0.7cm} In the paper \cite{Hollenstein2009} it was shown
that combining of
expected Planck %
\index{Planck mission} %
CMB data with the weak lensing survey of Euclid will give powerful
constraints on early dark energy and will be able to break
degeneracies in the~ parameter~ set~ inherent~ to~
the}}\vspace*{-7mm}
\end{figure}

Recently, these data have been used by Sherwin et al. (2011)
\cite{Sherwin2011} for con\-straining the
$\Omega_m-\Omega_{\Lambda}$ parameter space on the base of WMAP7
$TT$ and ACT lensing power spectra. The results are~ presented~ in~
Fig.\,1.\ref{w_omm_gr_lens}.~ One~ can~ see

\noindent  various observational channels. So, the current and
planned experiments for measurements of gravitational lensing
promise to become the crucial data in~unveiling the mystery of dark
sector of our Universe.\vspace*{3mm}

\section{\!Age of the Universe\label{ch1-sec7}}

\hspace*{3cm}\index{age of Universe|(}We have mentioned at the
beginning of this chapter the problem of agreement of the expansion
age of the Universe with age of the oldest stars of our galaxy.  The
estimations of the age of the oldest stars in the globular clusters,
obtained in the 90s of last century, are in the range
$13.5\pm2$~Gyrs \cite{Jimenez1996, Chaboyer1998a, Chaboyer1998b,
Carretta1999, Krauss2003}. The similar results were obtained using
other methods: white dwarf cooling sequence in globular clusters
($14.5\pm1.5$~Gyrs \cite{Renzini1996}, $12.7\pm0.7$~Gyrs
\cite{Hansen2002}) and content of U-238 in the old stars of halo of
the Milky Way ($14.0\pm2.4$~Gyrs \cite{Hill2002}). The lower limit
of such estimations is 11~Gyrs at 2$\sigma$ C.L. \cite{Krauss2003}.
The expansion age of the Universe should be larger. So, what we
\mbox{have here?}

The age of the Universe in the general cosmological model with given
parameters is as follows:
\begin{equation}
  t_0=\int\limits_0^1\frac{da}{aH(a)},
  \label{t_0}
\end{equation}

\noindent where $H(a)$ is given by Eq.~(\ref{H}). Since
$\Omega_r\sim10^{-5}$ and the main contribution to the integral
comes from the range $a>0.01$, the radiation term in (\ref{H}) can
be neglected. In the Einstein---de Sitter model
($\Omega_r=\Omega_K=\Omega_{de}=0$, $\Omega_m=1$) the age of the
Universe is defined by current value of
Hubble constant %
\index{Hubble constant}
\begin{equation}
  t_0=\frac{2}{3}\frac{1}{H_0}, \label{t_0_EdS}
\end{equation}
and equals $9.3\pm0.6$ Gyrs for $H_0=70\pm4$ km/s$\,\cdot\,$Mpc,
that is essentially lower than the age of the oldest stars of our
galaxy and, therefore, this model is ruled out by these
measurements.

In the open models without dark energy the equation (\ref{t_0}) is
integrated to give
\begin{equation}
  t_0=\frac{H_0^{-1}}{1-\Omega_m}\left[1+\frac{\Omega_m}{2\sqrt{1-\Omega_m}}\ln{\left(\!\frac{1-\sqrt{1-\Omega_m}}{1+\sqrt{1-\Omega_m}}\right)}\!\right]\!. \label{t_0_omd}
\end{equation}\vspace*{1mm}

\noindent In the limit $\Omega_m\rightarrow1$ we recover the value
(\ref{t_0_EdS}). In the limit $\Omega_m\rightarrow0$ we obtain also
finite value: $t_0\rightarrow H_0^{-1}$. For
$H_0=70\pm4$~km/s\,$\cdot$\,Mpc and $\Omega_m=$ $=0.3\pm0.1$ from
(\ref{t_0_omd}) we obtain $t_0=11.3\pm1.2$ Gyrs, that is at lower
limits of measurements of age of the oldest stars. These models
prefer uncomfortably low $\Omega_m$, that contradict its
measurements from peculiar velocity of galaxies, for example, and
large negative cur\-va\-tu\-re, that contradict data on positions of
CMB acoustic peaks.
\index{acoustic peaks|(} %

\begin{figure}[b]
 \vskip-2mm
 \raisebox{0.0cm}{\parbox[b]{5.5cm}{\caption{The dependences of age of the Universe on matter density in
    the models without dark energy (OMD, \mbox{$\Omega_{de}=0$}) and in the
    flat models (\mbox{$\Omega_K=0$}) with dark energy (DED). The dotted
    strip shows the age of the Universe estimated from the age of
    oldest white dwarfs, oldest stars of globular clusters and from
    the content of U-238 in the oldest stars of galaxy halo. The
    rectangle shows marginalized 1$\sigma$ constraints from combined
    data (Table 1.\ref{tab_params})\label{age_omm}}}}\hspace*{0.5cm}\includegraphics[width=7cm]{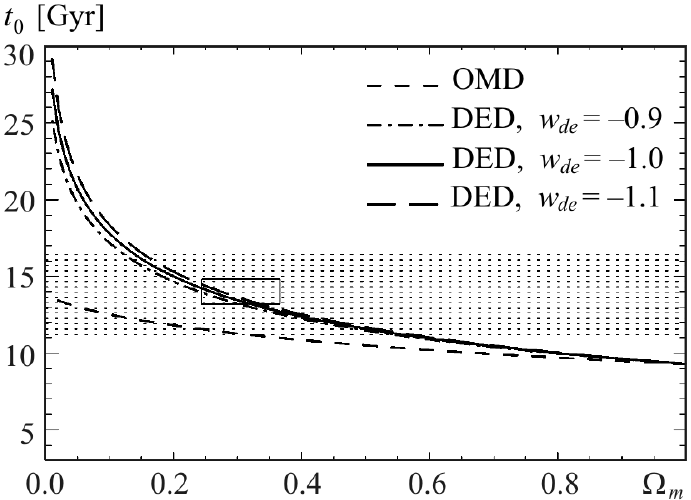}
\end{figure}

The equation (\ref{t_0}) has analytic solution also for flat
$\Lambda$CDM model: %
\index{LambdaCDM model ($\Lambda$CDM)} %
\begin{equation}
  t_0=\frac{H_0^{-1}}{3\sqrt{1-\Omega_m}}\ln{\left(\!\frac{1-\sqrt{1+\Omega_m}}{1-\sqrt{1-\Omega_m}}\right)}\!. \label{t_0_lcdm}
\end{equation}\vspace*{0.5mm}

\noindent In the limit $\Omega_m\rightarrow1$ we obtain the value
(\ref{t_0_EdS}) of Einstein---de Sitter model, but in the opposite
limit $\Omega_m\rightarrow0$ the cosmic age goes to infinity,
$t_0\rightarrow\infty$, which makes this model more suitable for
agreement with independent measurements of cosmic age (see
Fig.~1.\ref{age_omm}). Really, for the same range of parameters
(\mbox{$H_0=70\pm4$\,km/s\,$\cdot$\,Mpc} and $\Omega_m=0.3\pm0.1$)
the age of
expansion of the Universe is now in the interval 12---16~Gyrs. %
\index{age of Universe|)} %

In the models with non-$\Lambda$ dark energy, even in the simplest
among them~--- $w$CDM ($w={\rm const}\ne-1$), the integral in
(\ref{t_0}) can be calculated only numerically.  The results of
calculations for DED and OMD models are shown in
Fig.~1.\ref{age_omm}. Also the measurements of age of the oldest
stars are shown there. One can see, that the DED models can explain,
in principle, all independent measurements of
cosmic age.  One can see, that EoS parameter also affects the age of %
\index{EoS parameter} %
expansion of the Universe which increases with decreasing of
$w_{de}$. It was used by Krauss \& Chaboyer (2003) \cite{Krauss2003}
for constraining of dark energy parameters on the base of estimation
of age of globular cluster: for the Hubble key project the best-fit
value of Hubble constant they have obtained \mbox{$w_{de}<-0.4$} and
$\Omega_{m}<0.38$ at 1$\sigma$ C.L.
\index{Hubble constant}%

Therefore, the independent measurements of age of the oldest stars
of our galaxy and Hubble constant from distant galaxies are
important evidence for existence of dark energy in the
Universe.\vspace*{2mm}

\section[\!Constraints on dark energy parameters from
  combined data]{\!\!Constraints on dark\\ \hspace*{-0.95cm}energy parameters from
  combined data\label{ch1-sec5}}

\hspace*{3cm}\index{dark energy parameters}The complex approaches
for establishing the most adequate model of the Universe have been
started in the 90s of the past century. \mbox{Using}
available at that time observational data on large scale structure %
\index{large scale structure|(}%
(galaxies and clusters power spectra, bulk motions, cluster mass and
temperature functions, damped Ly-$\alpha$ systems) and COBE CMB %
\index{CMB anisotropy} %
anisotropy measurements it was stated at 1$\sigma$ confidence level
that $\Omega_{\Lambda}>0$ before indication of dark energy by SN Ia
measurements (see our papers \cite{Kahniashvili1996,Valdarnini1998}
and citing therein). Complementation of these data by SN Ia luminous
distance measurements and CMB acoustic peaks positions from the
ground-based and stratospheric experiments available at the boundary
of millennium enhances the confidence level of existence of dark
energy to 99.99\,\% [135, 137, 215---224].

Simultaneously with progress in accumulation of cosmological data
and increasing their accuracy and quality, the physical
interpretations as well as mathematical and computing methods of
comparison of theoretical predictions with observations and
extracting of the cosmological parameters from them were progressed
too. The highly accurate fast codes for estimation of large number
of parameters become publicly available (CosmoMC, for example).
Together with publications of the results of the digital galaxy sky
surveys and the first year WMAP all-sky CMB survey they originates
the next level of cosmology investigations, dubbed precision
cosmology. It becomes possible to establish the most optimal values
of main cosmological parameters at high confidence level, which is
very important for investigation of the nature of dark energy and
constraining the classes and number of its models. Since its effect
on the predicted values of the characteristics of the Universe (age,
dynamics of expansion, large scale structure) is comparable with
influence of values of rest cosmological parameters ($\Omega_K$,
$\Omega_{cdm}$, $\Omega_b$, $\Omega_{\nu}$, $N_{\nu}$, $H_0$, $A_s$,
$n_s$, $A_t$, $n_t$, $Y_p$, $\tau$), the dark energy parameters must
be determined jointly with all other ones. The analysis of Monte
Carlo Markov
chains %
\index{Markov chain Monte Carlo (MCMC)}%
is widely used for this puprose, in it two functions\vspace*{-2mm}
\begin{equation}
  \mathcal{L}(\mathbf{x};\theta_k)=\exp\left(\!-\frac{1}{2}(x_i-x_i^{th}(\theta_k))C_{ij}(x_j-x_j^{th}(\theta_k))\!\right)\!,
  \label{like_function}
\end{equation}
and
\begin{equation}
  \mathcal{P}(\theta_k;\mathbf{x})=\frac{L(\mathbf{x};\theta_k)p(\theta_k)}{g(\mathbf{x})},
  \label{posterior_function}
\end{equation}

\noindent
the likelihood %
\index{likelihood function} %
and posterior ones correspondingly are under consideration. Here
$\theta_k$ notes all cosmological parameters, $\mathbf{x}$ notes all
observational values, $\mathbf{x}^{th}$ notes their model
predictions, $C_{ij}$ is covariance matrix for all observational
data, $p(\theta_k)$ is prior for $\theta_k$ parameter,
$g(\mathbf{x})$ is probability distribution function of data. If
model is correct, the data are normally distributed and do not
contain dominating systematic errors then normalized to 1 at maximum
dependences of their likelihood and posterior functions on each
parameter marginalized over the rest
ones coincide.%
\index{posterior function|(} %

It was shown (\!\!\cite{Durrer2003,Novosyadlyj2005,WMAP7b} and
citing therein) that contribution of tensor mode of cosmological
perturbations to CMB temperature fluctuations is negligibly small %
\index{CMB temperature fluctuations} %
and, therefore, can be omitted in the problem of determination of
dark energy parameters. Also, there it was shown that the value of
active neutrino density parameter, $\Omega_{\nu}$, is lower than
0.03 at 95.4\,\% confidence level. Its best-fit value is close to
zero. Such low upper limit for active neutrino density parameter
makes it unimportant for dark energy problem. Therefore, two
parameters, $A_t$ and $\Omega_{\nu}$ can be assumed to be zero
without loss of generality in the problem of determination of dark
energy parameters. The spectral index of tensor mode, $n_t$, in this
case is neglected too. The effective number of neutrino species,
$N_{\nu}$, is not crucial parameter too and in our task is commonly
assumed to be equal to effective number from standard model of
particle physics, 3.04.

The similar situation is with primordial helium parameter $Y_p$. Its
value defines the number density of free electrons at decoupling
epoch as $n_e=$\linebreak $=11.31(1+z_{dec})(1-Y_p)\Omega_bh^2$
\cite{Hu1995b} that
influences photon free streaming and suppresses the CMB power spectrum %
\index{power spectrum}%
at small angular scale. The change of primordial helium parameter in
the range of its 95.4\,\% limits, $0.16<Y_p<0.46$ \cite{WMAP7b},
causes the variation of amplitude of 2nd, 3d and 4th acoustic peaks
in the range 1---2\,\%. \index{acoustic peaks|)}So, without loss of
generality one can fix it at standard fiducial value 0.24, which
matches well the Big Bang nucleosynthesis (BBN) const\-raint
\cite{Steigman2007,Wright2007} and incorporates the other
measurements of primordial content of helium
\cite{Izotov2004,Olive2004}. Of course, the exact values of these
parameters are very important for physics and cosmology, but they
correspond to other problems of current cosmology which are beyond
the scope of this book. They become especially actual in the light
of forthcoming experiments.

\begin{figure}
\vskip1mm
  \includegraphics[width=13cm]{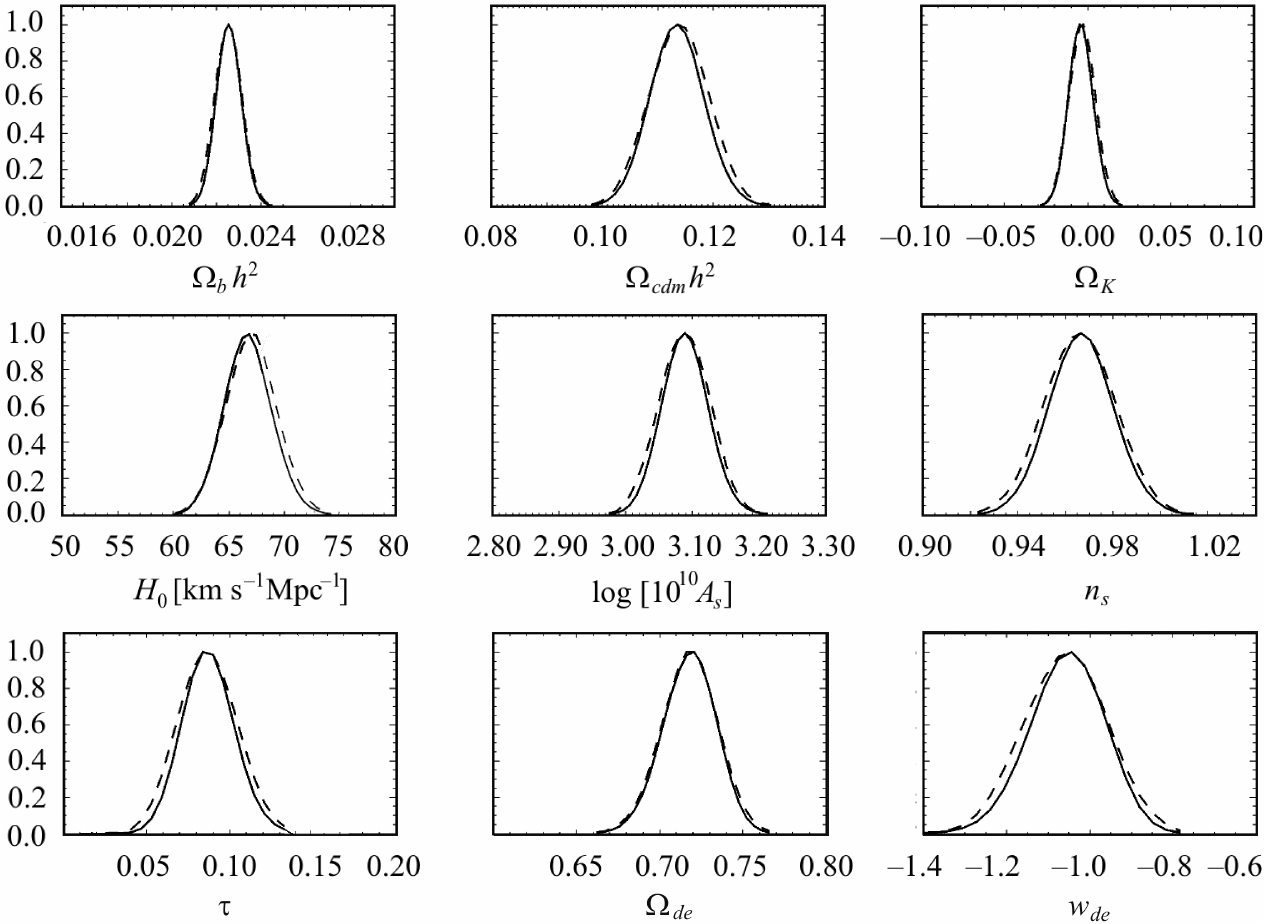}
  \vskip-2mm
  \caption{Posterior (solid line) and likelihood (dashed line)
    functions for main cosmological parameters and combined datasets
    WMAP7 {+} HST {+} BBN {+} SDSS LRG7 {+} SN SDSS SALT2}
  \label{priors_likes_wcdm2}\vspace*{-1.5mm}
\end{figure}

\begin{figure}
\vskip1mm
  \includegraphics[width=13cm]{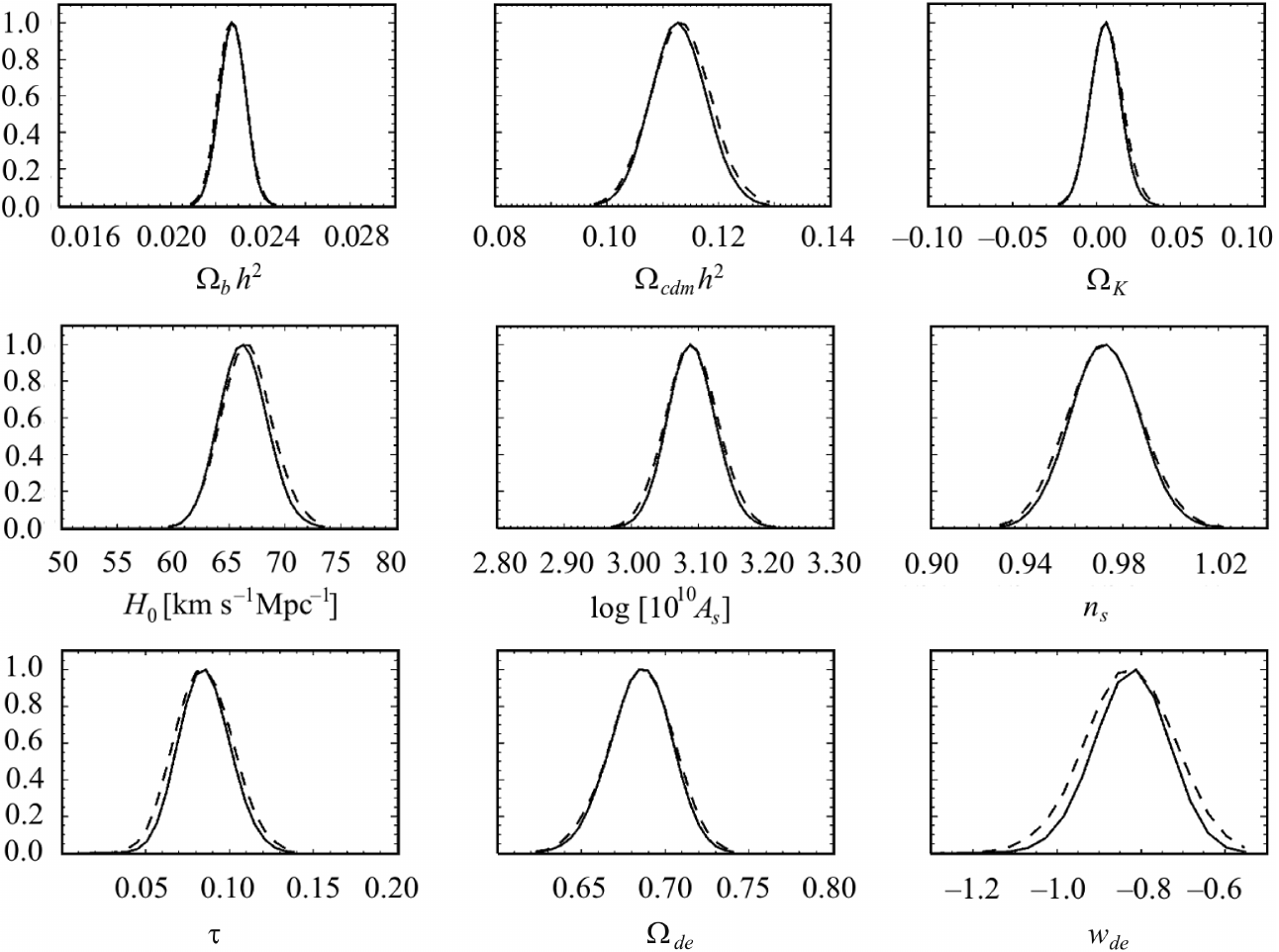}
    \vskip-2mm
  \caption{Posterior (solid line) and likelihood (dashed line)
    functions for main cosmological parameters and combined datasets
    WMAP7 {+} HST {+} BBN {+} SDSS LRG7 {+} SN SDSS MLCS2k2}
  \label{priors_likes_wcdm}\vspace*{-1.5mm}
\end{figure}

Therefore, in the problem of determination of dark energy parameters %
\index{dark energy parameters} %
at current level there are 7 important cosmological parameters
($\Omega_K$, $\Omega_{cdm}$, $\Omega_b$, $H_0$, $A_s$, $n_s$,
$\tau$) apart the dark energy ones $\Omega_{de}$ and $w_{de}$ ones.
In Figs.~1.\ref{priors_likes_wcdm2} and 1.\ref{priors_likes_wcdm}
the marginalized likelihood and posterior functions are presented
for main cosmological parameters and current observational data on
dynamics of
expansion of the Universe (SN SDSS \cite{Kessler2009}, HST %
\index{Hubble Space Telescope (HST)} %
\cite{Riess2009}), its large scale structure (SDSS LRG7 %
\index{large scale structure|)}%
\cite{Reid2010}, BAO \cite{Percival2010}) and CMB anisotropy (WMAP7
[93---95]). %
\index{CMB anisotropy} %
In the first figure SNe Ia from SDSS supernova survey are processed
using SALT2 light-curve fitting %
\index{light-curve fitting}%
\index{best-fit parameters}%
\index{Multicolor Light Curve Shape (MLCS)|(}%
method, in the second one~--- MLCS2k2 method. Maxima of functions
define the best-fit parameters and limits of 68.3\,\% (95.4\,\%)
fraction of area under the curves define 1$\sigma$ (2$\sigma$)
confidence limits of optimal values of corresponding parameters.
They are presented in Table 1.\ref{tab_params} for different
assumptions and SNe Ia compilations. One can see, that best-fit
values of $w_{de}$ are somewhat different for SALT2 and MLCS2k2
light-curve fittings for the same data set: in the first case it is
in the phantom range, in the second one in the quintessence one. But
confidential limits for both are wide and do not exclude each other.
The presented in Fig.~1.\ref{priors_likes_wcdm2} and
1.\ref{priors_likes_wcdm} likelihood and posterior functions, which
are practically superimposed and Gaussian, prove that the theory
matches well observational data and its parameters are determined
surely.  One can see, that current data unambiguously prefer the
models\linebreak with low spatial cur\-va\-tu\-re and high dark
energy density now. Since the admissible values of cur\-va\-tu\-re
parameter are very close to zero and computations\linebreak of
models with
non-zero cur\-va\-tu\-re by CAMB %
\index{CAMB} %
are essentially time con\-su\-ming  most researchers fix it equal to
zero when determine the dark energy parameters.

The remarkable achievement of current cosmology is visible here too:
it is establishing of the existence of dark energy ($\Omega_{de}>0$)
at 99.999\,\% or more confidence level ($\gtrsim$12$\sigma$) with
its content in the 1$\sigma$ range $0.64<\Omega_{de}<0.76$.  In
spite of the relatively small dispersions of parameters shown in
Figs.~1.\ref{priors_likes_wcdm2} and 1.\ref{priors_likes_wcdm} and
Table~1.\ref{tab_params}, a large number of dark energy models can
``belong'' there yet. Really, 1$\sigma$ range of marginalized
likelihood/posterior functions\linebreak for $w_{de}$ includes
$\Lambda$CDM, phantom and quintessence dark energy. \index{phantom
dark energy}\index{quintessence}Its more accurate and assured
determination becomes the key problem of dark energy
\mbox{investigations.}

\begin{table}[b!]
\vspace*{-5mm} \noindent\parbox[b]{13cm}{
\caption{\bf The best-fit values for cosmological parameters\newline
and the \boldmath$1\sigma$ limits from the extremal values of the N-dimensional distribution\newline
determined for the $\Lambda$CDM and $w$CDM models by the MCMC technique\newline
from the combined datasets WMAP7~{+} HST {+} BBN {+} SDSS LRG7 {+}\newline
+ SN SDSS SALT2 (1) and WMAP7~{+} HST {+} BBN {+} SDSS LRG7 {+}\newline
+ SN SDSS MLCS2k2 (2). By the asterisk in brackets the derived\newline 
parameters are noted. The Hubble constant $H_0$ is in units km$\,$s$^{-1}\,$Mpc$^{-1}$\label{tab_params}}
\index{Hubble constant}%
\index{Markov chain Monte Carlo (MCMC)}}\vspace*{2mm} \tabcolsep11.0pt
\noindent{\footnotesize
\begin{tabular}{|c|c|c|c|c|}
    \hline
        \rule{0pt}{5mm}\raisebox{-4mm}[0cm][0cm]{\scriptsize Parameters}&{\scriptsize $\Lambda$CDM}&{\scriptsize $\Lambda$CDM}&{\scriptsize $w$CDM}&{\scriptsize
        $w$CDM}
        \\[2mm] \cline{2-5}
         \rule{0pt}{5mm}&{\scriptsize 1}&{\scriptsize 2}&{\scriptsize 1}&{\scriptsize 2}\\[2mm]
       \hline
\rule{0pt}{6mm}$\Omega_b
h^2$&0.0225$_{-0.0013}^{+0.0016}$&0.0223$_{-0.0013}^{+0.0016}$&
0.0228$_{-0.0012}^{+0.0008}$& 0.0226$_{- 0.0014}^{+
0.0017}$\\[1.5mm]
$\Omega_{cdm} h^2$& 0.113$_{-0.012}^{+0.012}$& 0.117$_{-0.013}^{+0.011}$& 0.115$_{-0.012}^{+0.010}$& 0.115$_{- 0.014}^{+ 0.013}$\\[1.5mm]
$\Omega_{m}^{\,(*)}$& 0.286$_{- 0.043}^{+ 0.042}$& 0.319$_{-0.051}^{+0.047}$& 0.279$_{-0.033}^{+0.048}$& 0.307$_{- 0.049}^{+ 0.057}$\\[1.5mm]
$\Omega_{K}$& --0.001$_{-0.017}^{+0.016}$& --0.004$_{-0.019}^{+0.016}$& --0.002$_{-0.019}^{+0.012}$&0.006$_{-0.023}^{+0.026}$\\[1.5mm]
$H_0$& 68.8$_{-4.9}^{+6.5}$& 66.0$_{-5.0}^{+6.0}$&70.2$_{-6.1}^{+4.3}$&66.8$_{-5.9}^{+5.6}$ \\[1.5mm]
$\log(10^{10}A_s)$& 3.09$_{-0.09}^{+0.09}$& 3.09$_{-0.09}^{+0.08}$&3.09$_{-0.08}^{+0.07}$&3.09$_{-0.09}^{+0.10}$ \\[1.5mm]
$n_s$& 0.970$_{-0.036}^{+0.034}$& 0.962$_{-0.033}^{+0.035}$& 0.969$_{-0.027}^{+0.029}$& 0.973$_{-0.040}^{+0.039}$\\[1.5mm]
$\tau$&0.084$_{-0.033}^{+0.046}$&0.085$_{-0.036}^{+0.041}$&0.085$_{-0.029}^{+0.034}$ &0.082$_{-0.034}^{+0.050}$\\[1.5mm]
$t_0^{\,(*)}$&13.87$_{-0.75}^{+0.78}$&14.04$_{-0.71}^{+0.77}$&13.58$_{-0.58}^{+0.99}$
   &13.62$_{-0.91}^{+0.098}$\\[1.5mm] 
$\Omega_{de}^{\,(*)}$& 0.715$_{-0.037}^{+0.041}$& 0.685$_{-0.043}^{+0.047}$&0.723$_{-0.044}^{+0.032}$&0.687$_{-0.052}^{+0.044}$ \\[1.5mm]
$w_{de}$&--1&--1&--1.04$_{-0.19}^{+0.17}$&--0.84$_{-0.22}^{+0.22}$\\[1mm] 
$-\log L$&3865.11&3859.24&3865.05&3857.32\\[2mm]
\hline
   \end{tabular}
  }
\end{table}

In Figs.~1.\ref{union_constraints}, 1.\ref{snls3_constraints},
1.\ref{w_omm_wmap3}, 1.\ref{w_omm_bao}, 1.\ref{w_omm_xray},
1.\ref{w_omm_isw}, 1.\ref{w_omm_gr_lens} the 1 and 2$\sigma$
contours in $\Omega_m-w_{de}$ plane of likelihood distribution
function marginalized over all rest parameters are presented for the
different determinations. One can see, that practically each method
shows some degeneracy (see expressions (\ref{w_SNIa}),
(\ref{w_CMB}), (\ref{w_BAO})) which makes contours prolate and
convoluted. It is caused by integral character of
``distant---redshift'' relations in the low-$z$ observations or
geometrical projections in the high-$z$ ones.  For example, the
degeneracy in $\Omega_m-w_{de}$ plane of likelihood distribution
function, obtained from CMB data alone, can be understood as
follows. The first peak of the CMB temperature power spectrum is %
\index{power spectrum}%
connected with the sound horizon at decoupling, when the CMB was
last scattered by free electrons. It depends (see equations
(\ref{zdec})---(\ref{rs})) on $\Omega_b$, $\Omega_{m}$, $H_0$, $T_0$
and effective number of relativistic fractions $N_{\rm eff}$.  The
projection of this sound horizon onto the same degree-scale angle on
the sky, as it follows from eqs.~(\ref{dA}) and (\ref{lpm}), can be
realized in cosmologies with different combinations of $H_0$,
$\Omega_b$, $\Omega_K$, $\Omega_m$, $\Omega_{de}$ and $w_{de}$. But
when we take into account the position and amplitude of other peaks
and troughs, which have different dependences on cosmological
parameters, the ranges of possible combination of parameters become
narrow. Their dimensions, amounts and forms depend also on accuracy
of observational data. So, the models with values of parameters
which are inside the contour cannot be surely distinguished using
only the
measurements of primordial CMB power spectrum. %
\index{CMB power spectrum} %
Fortunately, for different techniques of dark energy parameter
determination the contours or surfaces of equal likelihood or
posterior have different orientations, prolateness and convolution
in the parameter space (see Figs.~1.\ref{union_constraints},
1.\ref{snls3_constraints}, 1.\ref{w_omm_wmap3}, 1.\ref{w_omm_bao},
1.\ref{w_omm_xray}, 1.\ref{w_omm_isw}, 1.\ref{w_omm_gr_lens}), that
give the possibility to break the degeneracies by using combined
datasets for determinations of cosmological parameters. It also
narrows the confidence ranges of main cosmological parameters and
establishes the
concordance model of the Universe. %
\index{posterior function|)} %

\begin{figure}
\vskip1mm
\includegraphics[width=13cm]{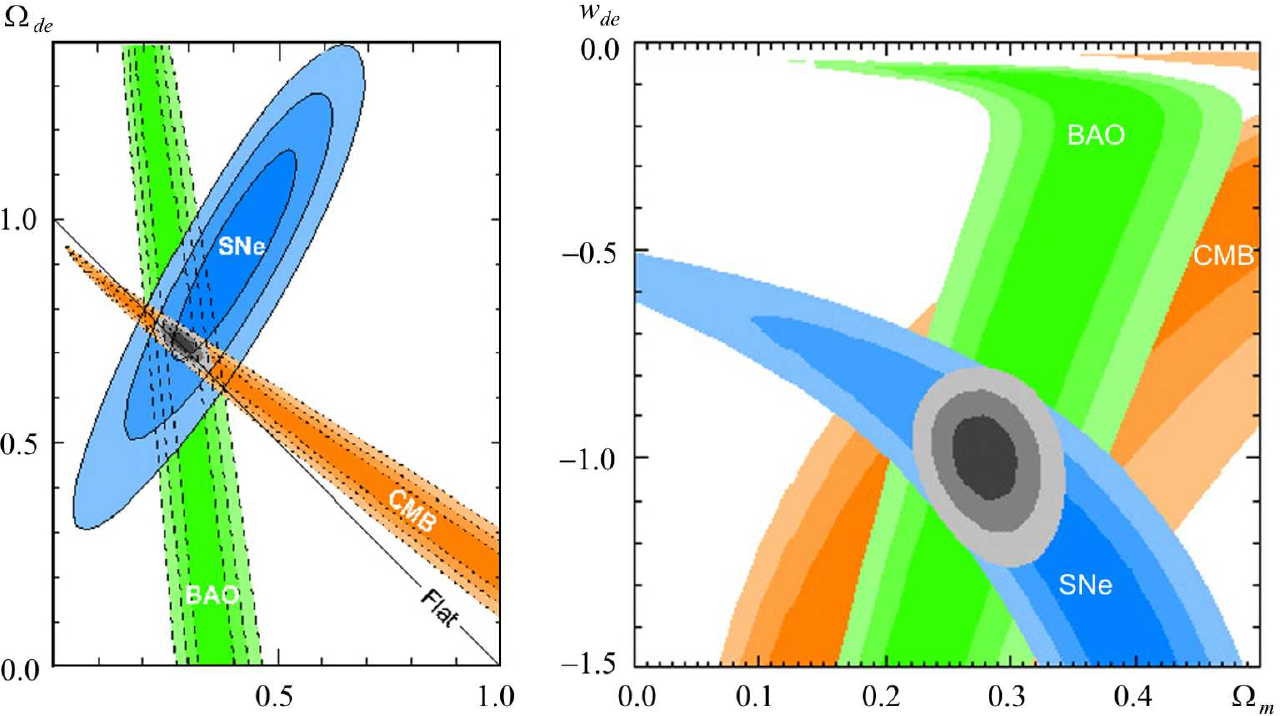}
    \vskip-2mm
  \caption{Confidence level (68.3\,\%, 95.4\,\% and 99.7\,\%) contours in
    $\Omega_m-\Omega_{\Lambda}$ plane (left panel) and in
    $\Omega_m-w_{de}$ plane (right panel) obtained from CMB, BAO and
    SN data alone as well as their combination (From
    \cite{SNUnion})}\vspace*{-1mm}
  \label{w-omde_frieman}
\end{figure}

Fig.~1.\ref{w-omde_frieman} illustrates breaking of the degeneracies
in $\Omega_{de}-\Omega_m$ and \mbox{$w_{de}-\Omega_m$} marginalized
likelihood distributions by using the combined dataset for
determinations of
cosmological parameters. There the Union compilation %
\index{Union sample} %
of SNe Ia \cite{SNUnion}, 5-year WMAP data \cite{WMAP5a} and %
\index{baryon acoustic oscillations (BAO)} %
BAO \cite{Eisenstein2005} have been used. In the left panel the
density parameters $\Omega_{de}$ and $\Omega_m$ are constrained by
the whole
data set for the case of fixed %
\index{EoS parameter} %
EoS parameter $w_{de}=-1$ ($\Lambda$CDM). The best-fit values from
combined dataset are as follows\,\footnote{\,Symmetrized errors
include
  statistical and systematical ones (Table 6 in \cite{SNUnion}) as
  root square from sum of their squares.}:
$\Omega_{de}=0.713 \pm 0.047$, $\Omega_{m}=0.285 \pm 0.022$ and
$\Omega_K=-0.009 \pm 0.0098$. In the right panel the const\-raints
are computed for flat models, there best-fit values of $w_{de}$ and
$\Omega_m$ are \mbox{$-0.969\,\pm\,0.089$} and
\mbox{$0.274\,\pm\,0.020$} correspondingly. These values somewhat
differ from ones presented in Table~1.\ref{tab_params}, where we
used the updated observatio\-nal data
(WMAP7 [93---95], SN SDSS \cite{Kessler2009}, HST %
\index{Hubble Space Telescope (HST)} %
\cite{Riess2009}, SDSS LRG7 \cite{Reid2010}, BAO
\cite{Percival2010}). The main reason for difference of best-fit
values of dark energy parameters consists in different datasets and
different light-curve fitting in SNe Ia compilations, %
\index{light-curve fitting}%
SALT method in Union compilation and MLCS2k2 method in SN SDSS one.
Also, our 1$\sigma$ ranges are wider, since we present the extremal
values of corresponding parameters in N-dimensional distributions.

The main advantage of combined analysis consists in the fact of
existence of concordance model, which matches practically all
cosmological and astrophysical observational data.  This concordance
model is dark energy dominated and close to flat $\Lambda$CDM one at
current epoch. But small departure from standard $\Lambda$CDM allows
existence of large number of alternative models, which include
physical fields, fluids, generalized or modified gravity theories.
In this section we analyzed only $\Lambda$CDM and $w$CDM models as
simplest %
\index{LambdaCDM model ($\Lambda$CDM)} %
ones and showed the incontrovertible observational evidence for
existence of dark energy.

\section{\!Summary\label{ch1-sec9}}

\hspace*{3cm}The state-of-art observational evidence for existence
of dark energy and methods of constraining of its parameters were
discussed in this chapter. The list of independent indicators of
dark energy developed  during last decade in observational, data
processing and theoretical aspects is as follows:

{\footnotesize$\bullet$}\,\,luminosity distance~--- redshift
relation for SNe Ia,

{\footnotesize$\bullet$}\,\,luminosity distance~--- redshift relation for GRBs, %
\index{gamma-ray bursts (GRBs)} %

{\footnotesize$\bullet$}\,\,angular diameter distance~--- redshift
relation for CMB acoustic
  peaks, %
\index{angular diameter distance} %
\index{acoustic peaks} %

{\footnotesize$\bullet$}\,\,angular diameter distance~--- redshift
relation for BAO peaks in
  matter density perturbations,

{\footnotesize$\bullet$}\,\,angular\,diameter\,distance\,---\,redshift\,relation\,for\,X-ray\,clusters\,of\,\mbox{galaxies,}

{\footnotesize$\bullet$}\,\,formation of the large scale structure
of the Universe and its
  elements, %
\index{large scale structure}%

{\footnotesize$\bullet$}\,\,cross-correlation of ISW anisotropy of
CMB with large scale
  structure distribution of galaxies, %
\index{Integrated Sachs---Wolfe effect} %
\index{cosmic microwave background (CMB)|)} %

{\footnotesize$\bullet$}\,\,weak gravitational lensing of CMB,

{\footnotesize$\bullet$}\,\,age of the oldest stars of our Galaxy.

\noindent We have shown that each of from them prefers the dark
energy dominated model for agreement of current physical models of
objects with numerous accurate observational data on their
luminosities, spectra, sizes, distances, ages etc. We suppose that
indirect but important argument for dark energy is existence of
concordance model, in which all heterogeneous observational data are
fitted well simultaneously. The determinations of dark energy
parameter $\Omega_{de}$ from different combinations of observable
data give close values: $\Omega_{de}=$ $=0.69\pm0.05$ for WMAP7 {+}
HST {+} BBN {+} SDSS LRG7 {+}
SN SDSS MLCS2k2 dataset and $\Omega_{de}=0.72\pm0.04$ for WMAP7 {+} %
\index{Multicolor Light Curve Shape (MLCS)|)}%
HST {+} BBN + SDSS LRG7~{+} +~SN SDSS SALT2 dataset. The EoS
parameter
is %
\index{EoS parameter} %
worse determined: its value considerably depends on prior
assumptions and combinations of dataset and most determinations give
its value in the interval (--1.2---0.8).  Its best-fit value is in
the quintessence range ($w_{de}=-0.84\pm0.22$) if dataset contains
the SNe Ia distance
moduli obtained with MLCS2k2 method of light-curve fitting %
\index{light-curve fitting}%
and is in the phantom range ($w_{de}=-1.04\pm0.18$) if used SNe Ia
distance moduli are obtained with SALT2 one. Since they are out of
1$\sigma$-range of each other one can conclude that some
inconsistency or tension exists between fitters SALT2 and MLCS2k2.

In the next chapters we will discuss the different physical models
of dark energy and their agreement with observational data presented
in this chapter.

\setcounter{chapter}{1}
\chapter{\label{s:1}  SCALAR FIELD MODELS\\[1mm]  OF DARK ENERGY}\markboth{CHAPTER 2.\,\,Scalar field models of dark energy}{CHAPTER 2.\,\,Scalar field models of dark energy}
\thispagestyle{empty}\vspace*{-12mm}

\begin{wrapfigure}{l}{2.6cm}
\vspace*{-5.9cm}{\includegraphics[width=3.0cm]{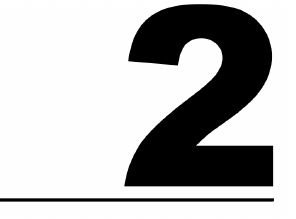}}\vskip17.2cm
\end{wrapfigure}
\vspace*{10mm}

 \setcounter{section}{1} \vspace*{-5mm}
\hspace*{3cm}\section*{\hspace*{-3cm}2.1.\,\,Introduction$_{}$}\label{ch2-Intr}\begin{picture}(10,10)
\put(0,-128){\bfseries\sffamily{72}}
\end{picture}

\vspace*{-1.1cm} \noindent In this and next two chapters we analyze
the different models of dark energy. Since the explanation of
accelerated expansion of the Universe is out of standard models of
matter and gravitation the three ways are to find it: the modifying
of form of matter, the modifying of gravity, or both. The historical
first one is modifying of gravity: Albert
Einstein in 1917 \cite{Einstein1917} added the cosmological constant %
\index{cosmological constant|(} %
to the left part of General Relativity equations. But then the empty
space is curved, that destroys the main conception of General
relativity: the matter-energy causes the space cur\-va\-tu\-re. On
the other hand, then the gravity has two fundamental constants, $G$
and $\Lambda$, which through the cosmological values $H_0$,
$\rho_{cr}$ and $\Omega_{\Lambda}$ can be expressed as
$G=3H_0^2/(8\pi\rho_{cr})$ and $\Lambda=3H_0^2\Omega_{\Lambda}$.
Their numbers in units of $c=1$ are drastically different,
$G=10^{13}$\,cm/g and $\Lambda=10^{-56}$\,cm$^{-2}$, that has not
found the physical interpretation. That is why physicists prefer to
put $\Lambda$ in the right hand side of General Relativity equations
and interpret it as matter-energy component. There it also has not
obtained suitable physical interpretation, but lead to the other
problems, ``fine
tuning'' and ``cosmic coincidence'' %
\index{cosmic coincidence} %
\index{fine tuning} %
ones, which have no satisfactory physical explanations too.  This is
the first reason of existence of large number of alternatives to
cosmological constant, in spite of that the cosmological models with
$\Lambda$ well match the observational data on dynamics of expansion
and large scale structure of the Universe. %
\index{large scale structure}%
The other reason is originated, in our opinion, by successes of
inflation and particle physics theories. Indeed, if scalar field or
inflaton existed in the very early Universe and accelerated its
expansion, then why cannot something like it exist later or now?
Ratra and Peebles (1988) \cite{Ratra1988} and Wetterich (1988)
\cite{Wetterich1988} where first who analyzed the cosmological
consequences of presence of scalar fields %
\index{scalar field|(} %
in the contemporary Universe. The active development of these ideas
in the next years after discovery of accelerated expansion of the
Universe in 1998 (see [37---44] and citing therein) has led to the
current dark energy conception [24---34].\vspace*{-2mm}

\section[\!Cosmological constant as vacuum energy]
{\!Cosmological constant\\ \hspace*{-0.95cm}as vacuum energy: ideas
and
  problems\label{ch2-sec1}}\vspace*{-1mm}

\hspace*{3cm}\index{vacuum energy}The physical interpretation of
cosmological constant has long history. In 1968 Ya.\,Zeldovich
\cite{Zeldovich68} argued that the energy of zero oscillations of
vacuum is Lorentz-invariant $p_{vac}=-\rho_{vac}c^2$, i.e. it is
equivalent to the $\Lambda$-constant $T^{vac}_{\mu\nu} = \Lambda
g_{\mu\nu}$. The importance of such identification can hardly be
underestimated, since the existence of vacuum energy is indisputable
in quantum mechanics and field theory, as it causes the observed
inevitable natural broadening of spectral lines and Lamb
shifts\,\footnote{\,In 1955, Lamb
  was honored with Nobel prize for the discovery of shift in the
  structure of energy spectrum of Hydrogen atom. This shift is
  measured and predicted with accuracy as high as 11 digits. The
  theoretical prediction is based on the interactions of electrons
  with zero oscillations of electromagnetic field.}. On the other
hand, the General theory of relativity states, that all of existing
kinds of energy are the sources of gravitational field and should be
included as proper stress-energy tensors in the right-hand side of
Einstein equations, %
\index{Einstein equations} %
$G_{\mu\nu} = 8\pi G(T_{\mu\nu}+T^{vac}_{\mu\nu})$. This means that
$\Lambda$-constant should be introduced in order to take into
account the gravitational action of the vacuum.

However, the explanation of numerical value of observable
$\Lambda$-constant is complicated. Indeed, the value of energy of
zero oscillations of vacuum can be estimated as follows
\cite{Linde90}:\vspace*{-7mm}
\begin{equation}
  \rho_{vac} \propto \int\limits_0^{k_{cut}} \sqrt{k^2+m^2}k^2dk \propto k_{cut}^4,
\end{equation}\vspace*{-5.5mm}

\noindent where $k_{cut}\gg m$ is the ultraviolet limit, or the
scale of spectrum cut-off, necessary for the finite value of vacuum
energy to be obtained\,\footnote{\,In quantum electrodynamics the
arbitrary large
  value of vacuum energy is eliminated by renormalization, though the
  absolute value (the finiteness) in this theory is not so fundamental
  as in the theory of gravitation.}.  The most plausible scale for the
cut-off is the Planck energy scale $k_{cut}\propto M_{pl}$, for
which the value for the vacuum energy is $\rho_{vac}\propto
M_{pl}^4\approx 10^{96}\,\text{kg}/\text{m}{}^3$.  Such density
exceeds the observable value of $\Lambda$-constant, $\rho_{\Lambda}
= 8\pi G \Lambda = 1.88\,\times$
$\times\,10^{-26}\Omega_{\Lambda}h^2\,\text{kg}/\text{m}^3$, by 123
orders and it is the main problem of such physical interpretation of
$\Lambda$.  Since the appropriate scale for cut-off, $k_{cut}\propto
M_X \approx $ $\approx10^{-3}$\,eV, is impossible to obtain in the
framework of existing particle \mbox{physics} theories, Zeldovich
proposed the idea of ordering of polarized vacuum
\cite{Zeldovich68}. This is a vacuum represented as birth and
annihilation of the same type pairs of particle and antiparticle.
The ordering implies the zero rest mass for particle-antiparticle
pair, so only the energy of their gravitational interaction is left.
For the particles of masses $m$, separated by the distance $\lambda
= \hbar/(mc)$, one can evaluate the energy density of gravitation,
and hence vacuum energy as $\rho_{vac}c^2 \approx (G
m^2/\lambda)/\lambda^3 = G m^6c^4/\hbar^4$.  The estimated value of
vacuum energy is appropriate for the mass of pion, it is
intermediate value between the mass of proton and electron. The
result looks very promising, since the proper order of magnitude for
vacuum energy is obtained, but at the same time the result is vague,
since it has not found further development in particle physics,
hence there is no additional grounding for it. Also the attempts
were made to construct the observed value for $\rho_{\Lambda}$ from
the vacuum energy at Planck scales $\rho_{vac}\propto
10^{96}\,\text{kg}/\text{m}{}^3$ by multiplying by $e^{-2/\alpha}$
\cite{Starobinsky98}, where $\alpha=e^2/(\hbar c)$ is the fine
structure constant, or by $(M_{SUSY}/M_{pl})^8$ \cite{Antoniadis98},
where $M_{SUSY}\simeq 10^3\,\text{GeV}$ is the energy scale of
supersymmetry breaking at electro-weak interaction. Unfortunately,
these attempts have failed to find the grounding too.

Another yet unsolved problem of vacuum energy is connected with the
circumstances of its emergence. It is quite natural to suppose, that
this energy is a remnant of processes taking places in the early
Universe. Since the vacuum energy density does not vary during the
expansion, at the moment of emergence it was by many orders smaller
than the density of any other components of medium. The value of
vacuum energy density should be fine-tuned in the early Universe to
be of the same order as the matter density now. Moreover, the small
variation of the value in the early Universe leads to the crucial
consequences for the formation of its large scale structure in
future.

So, three unresolved problems arise in the way of interpretation of
cosmological constant as vacuum energy: its small value, fine tuning
in the early Universe and strange coincidence of the value of its
density with the matter density now (see also [236---239]). These
problems, on one hand, stipu\-late us to reconsider our point of
view on the nature of vacuum, and, on the other hand, enforce to
search the alternative interpretation of $\Lambda$-constant.  We
suppose, that history of $\Lambda$-cosmology is not completed, since
it is the \mbox{simplest} theory of dark energy from the
mathematical point of view, matching well practically all
observational data mentioned in the
previous chapter. %
\index{cosmological constant|)} \vspace*{-2mm}

\section[\!Scalar fields as dark energy$_{ }$]{\!Scalar fields as dark energy\label{ch2-sec2}\vspace*{-1mm}}

\hspace*{3cm}\index{accelerated expansion}Other radical alternative
of explanation of accelerated expan\-sion of the Universe consists
in abandonment of cosmological constant, zeroing of vacuum
energy\,\footnote{\,There are many hypothetical mechanism to do
  that in supersymmetry and string theories of particle physics.  It
  is easier to build the theory which gives zero for energy of averaged
  vacuum state than such small that corresponds to dark energy
  density.} but introducing the scalar field $\phi(a)$, which smoothly
(does not depend on spatial coordinates in the main order of its
value) fills the Universe and satisfies some conditions. The
simplest variants of such fields assume the minimal coupling with
other matter-energy components of the Universe via gravitation only.
Their physical properties are given by two functions: potential
$U(\phi)$
and Lagrangian density %
\index{Lagrangian density}\vspace*{-2mm}
\begin{equation}
  \mathcal{L}=\mathcal{F}(X,U(\phi)),
  \label{gen_L}
\end{equation}
where $X\equiv \frac{1}{2}\phi_{;\mu}\phi^{;\mu}$ is kinetic term,
which describes the rate of change of the field $\phi$. The
covariant
Euler---Lagrange equation, %
\index{Euler---Lagrange equation} %
or variational principle applied to the action
\begin{equation}
  S=\int\sqrt{-g}\mathcal{L}(X,U)d^4x,
  \label{sf_action}
\end{equation}
gives the field equation of motion in the\index{FRW
metric}\index{Friedmann equations|(}\index{equation of motion} FRW
metric ($g$ is its determinant),\linebreak which can be solved
jointly with Friedmann equations (\ref{Freq1})---(\ref{Freq2}). To
do so, the energy density and stress or energy-momentum tensor of
field \mbox{must be defined.}

\index{energy-momentum tensor} %
The energy-momentum tensor of such field is as follows:
\begin{equation}
  T_{\mu\nu}=\mathcal{L}_{,X}\phi_{,\mu}\phi_{,\nu}-g_{\mu\nu}{\mathcal{L}}.
  \label{sf_Tmunu}
\end{equation}
It can be rewritten in the form of perfect fluid energy-momentum
tensor
\begin{equation*}
T_{\mu\nu}=(\rho_{de}+p_{de})u_{\mu}u_{\nu}-g_{\mu\nu}p_{de}
\end{equation*}
with energy density ($T_0^0$)\vspace*{-2mm}
\begin{equation}
  \rho_{de}=2X\mathcal{L}_{,X}-\mathcal{L},
  \label{rho_de_L}
\end{equation}
pressure ($T_i^i$)\vspace*{-2mm}
\begin{equation}
  p_{de}=\mathcal{L}
  \label{p_de_L}
\end{equation}
and four-velocity $u_{\mu}\equiv\phi_{;\mu}/\sqrt{2X}=(a,0,0,0)$ in
the comoving coordinates.  The value of pressure, as a rule, is
presented in the Friedmann equations by EoS parameter $w_{de}$ %
\index{EoS parameter} %
\index{Friedmann equations} %
\vspace*{-2mm}
\begin{equation}
  p_{de}=w_{de}\rho_{de},
  \label{p_de}
\end{equation} %
which for the scalar field dark energy model should satisfy the
equation
\begin{equation}
  w_{de}(a)\equiv\frac{p_{de}(a)}{\rho_{de}(a)}=\frac{\mathcal{L}}{2X\mathcal{L}_{,X}-\mathcal{L}}.
  \label{w_L}
\end{equation}

It is a function of time and determination of character of its
variation (monotonous, non-monotonous, increasing, decreasing or
oscillating) and its evolution is crucial task for modern cosmology.
Determining of the time-dependence of $w_{de}(a)$ from observations
gives evolution of dark energy density according to equations
(\ref{rho_de})---(\ref{tilde_w_de}). This is enough for description
of dynamics of expansion of the homogeneous Universe using Friedmann
equations (\ref{Freq1})---(\ref{Freq2}), but insufficient for
understanding of field evolution as well as for including the scalar
field in theory of the large scale structure formation.
\index{large scale structure}%

Indeed, applying the Euler---Lagrange equation for action
(\ref{gen_L}), or dif\-ferential energy-momentum conservation law
(\ref{demcl}) for (\ref{sf_Tmunu}), we obtain the equation of motion
for scalar field in the general form:
\begin{equation}
  \left(\ddot{\phi}+2aH\dot{\phi}\right)\mathcal{L}_{,X}-a^2\frac{\partial{U}}{\partial{\phi}}\mathcal{L}_{,U}+\frac
  {\ddot{\phi}\dot{\phi}^2-aH\dot{\phi}^3}{a^2}\mathcal{L}_{,XX}+\frac{\partial{U}}{\partial{\phi}}\dot{\phi}^2\mathcal{L}_{,XU}=0.
  \label{gen_KG}
\end{equation}
It can be solved jointly with Friedmann equations
(\ref{Freq1})---(\ref{Freq2}) for the specified functional form of
Lagrangian $\mathcal{L}(X,U)$ and potential $U(\phi)$.
\index{potential $U(\phi )$} %

On the other hand, Einstein equations and/or differential
energy-momen\-tum conservation law %
\index{conservation law} %
\index{Einstein equations} %
indicate, that such scalar field in contrast to vacuum dark energy
cannot be perfectly smooth, it is perturbed by gravitational
influence of matter-radiation inhomogeneities or has its own ones,
generated in the early Universe. The equations for scalar field
density and velocity perturbations, as we will see below, contain
besides $w_{de}$
two other values, so called effective %
\index{sound speed} %
sound speed $c_s^2=\delta p_{de}/\delta\rho_{de}$ and adiabatic
one\,\footnote{\,The terms ``effective sound speed'' and ``adiabatic
sound
  speed'' of dark energy are used in the literature for designation of
  dark energy intrinsic values which formally correspond to
  thermodynamical ones.}  $c_a^2=\dot p_{de}/\dot\rho_{de}$. Note that
in the case of scalar field dark energy the adiabatic sound speed is
not the true velocity of sound propagation. The perturbed scalar
field has non-negligible entropy and thus non-adiabatic pressure
perturbations. In the dark energy rest frame the total pressure
perturbation can be presented as $\delta
p_{de}=c_s^2\delta\rho_{de}$ and the effective sound speed is
defined for given Lagrangian as
\begin{equation}
  c^2_s\equiv\frac{\delta p}{\delta \rho}=\frac{p_{,X}}{\rho_{,X}}=\frac{\mathcal{L}_{,X}}{\mathcal{L}_{,X}+2X\mathcal{L}_{,XX}}.
  \label{c_s2}
\end{equation}

The adiabatic sound speed is not independent quantity, it is related
to EoS parameter $w_{de}$ by the differential equation %
\index{EoS parameter} %
\begin{equation}
  aw'_{de}=3(1+w_{de})(w_{de}-c_a^2).\label{w'}
\end{equation}
It can be calculated for known $w_{de}(a)$, or, on contrary, used
for solution of equation (\ref{w'}) for $w_{de}(a)$ with defined
$c_a^2$.

Therefore, in cosmology applications of scalar field models of dark
energy the $\Omega_{de}$, $w_{de}(a)$ and $c_s^2$ must be given or
determined. In the previous chapter it was shown that $\Omega_{de}$
and $w_{de}=$~const are determined firmly by available observational
data. The time variable $w_{de}$ is less constrained and the
effective sound speed estimations are very rough now [240---246].
Future observational programs are designed to achieve the 1 percent
accuracy of their determination. But even then, as one can see from
equations (\ref{gen_L})---(\ref{c_s2}), the reconstruction or
reverse engineering of potential and Lagrangian will be an ambiguous
task.

In the case of functional forms of Lagrangian and potential given ad
hoc the EoS parameter $w_{de}$ and the effective sound speed $c_s^2$
are defined by (\ref{w_L}) and (\ref{c_s2}) correspondingly. %
\index{Friedmann equations|)} %
Friedmann equations (\ref{Freq1})---(\ref{Freq2}) together with
(\ref{gen_KG}) form the closed set of equations for determination of
the evolution of $a$ and $\phi$. Together with the set of equations
for evolution of metric, density and velocity perturbations for all
components it gives the possibility to test different scalar fields
as dark energy and determine their intrinsic parameters by
comparison the model-predicted characteristics of the Universe with
observational data. It is one of the ways of intensive
investigations of dark energy since its discovery. Another one
consists in defining of some functional forms for $w_{de}(a)$ and
$c_s^2(a)$ and search for the corresponding forms of Lagrangians and
potentials. The third one is combined~--- definition of functional
forms of $w_{de}(a)$ and $\mathcal{L}(X,U)$, giving the possibility
to calculate $c_s^2(a)$ and reconstruct $U(\phi)$. We will not
discuss the advantages and imperfections of different approaches,
each of them is useful for its specifical aspects, but in the next
sections we will use the combined one to analyze the possibility to
distinguish different scalar field models of dark
energy.\vspace*{-1.5mm}

\section[\!Scalar perturbations of the scalar field]{\!Scalar
  perturbations\\ \hspace*{-0.95cm}of the scalar field and other components\label{ch2-subsec12}}

\hspace*{3cm}In the subsection \ref{ch1-subsec41} it was shown that
the rate of
growth of matter density perturbations %
\index{density perturbations} %
is sensitive to the value of smoothed density of dark energy
($\Omega_{de}$) and tempo of its change in the past via $w_{de}$. It
affects strongly the ratio of amplitudes of matter power spectrum at %
\index{power spectrum}%
different redshifts.  Other possible fingerprints of dark energy in
the matter power spectrum are related to the gravitational interplay
of density perturbations in both components.

Let us consider the two-component model with matter and dark energy
in the terms of their energy densities, pressures and
four-velocities. For derivation of the evolution equations for
scalar linear
perturbations it is convenient to use the conformal Newtonian gauge %
\index{conformal Newtonian gauge} %
with space-time metric\vspace*{-1mm}
\begin{equation*}
  ds^2=a^2(\eta)[(1+2\Psi(\textbf{x},\eta))d\eta^2-(1+2\Phi(\textbf{x},\eta))\delta_{\alpha\beta}dx^{\alpha}dx^{\beta}],
\end{equation*}\vspace*{-6mm}

\noindent where $\Psi(\textbf{x},\eta)$ and $\Phi(\textbf{x},\eta)$
are
gauge-invariant metric perturbations %
\index{metric perturbations}%
called Bardeen's potentials \cite{Bardeen1980}, which in the case of
zero proper anisotropy of medium (as for the dust matter and scalar
fields) have equal absolute values and opposite signs:
$\Psi(\textbf{x},\eta)=-\Phi(\textbf{x},\eta)$ \cite{Kodama1984}.

The perturbations in the energy density $\rho_{de}$, pressure
$p_{de}$ and the four-velocity $u^{\mu}_{(de)}$ of dark energy are
defined in the following way:\vspace*{-1mm}
\[
  \rho_{de}(\eta,\textbf{x})  = \bar{\rho}_{de}(\eta) + \delta \rho_{de}(\eta,\textbf{x})=\bar{\rho}_{de}(\eta)(1+\delta_{de}(\eta,\textbf{x})),
  \]\vspace*{-7mm}
\[
  p_{de}(\eta,\textbf{x})  = \bar{p}_{de}(\eta)+\delta p_{de}(\eta,\textbf{x})= \bar{p}_{de}(\eta)(1 + \pi_{de}(\eta,\textbf{x})),
  \]\vspace*{-6mm}
\[
  u^{\mu}_{(de)}  = \bar{u}^{\mu}_{(de)} + \delta u^{\mu}_{(de)},
\]
where $\bar{\rho}_{de}(\eta)$, $\bar{p}_{de}(\eta)$ and
$\bar{u}_{\mu\,(de)} =(-a,0,0,0)$ are the unperturbed background
values of energy density, pressure and four-velocity in the world
with FRW line element (\ref{ds}), respectively. Since
$u^{\mu}_{(de)}u_{\mu\,(de)} = -1$, it turns out that $a^{-1}\delta
u_{0\,(de)} = a\delta u^{0}_{(de)} =-\Psi $. In the case of scalar
mode of perturbations the spatial part of the four-velocity $\delta
u^{i}_{(de)}$ can be expressed as a gradient of some scalar function
$V(\eta,\textbf{x})$
\begin{equation}
  \delta u^{i}_{(de)} = g^{ij}V_{,j}.
\end{equation}

For scalar fields the entropy perturbations are inherent and cause
in addition to the adiabatic pressure perturbations, which follow
from the variation of (\ref{p_de}), the non-adiabatic pressure ones
$\delta p_{de}^{(nad)}$, so the total perturbation is their sum
[240---254]:
\begin{equation*}
  \delta p_{de}=c_a^2\delta \rho_{de}+\delta p_{de}^{(nad)}.
\end{equation*}
The intrinsic entropy $\Gamma_{de}$ is defined by non-adiabatic part
of pressure as
\begin{equation*}
  \Gamma_{de}=\pi_{de}-\frac{c_a^2}{w_{de}}\delta_{de},
\end{equation*}
and in the variables of conformal-Newtonian gauge equals
\cite{Hu1998,Erickson2002,DeDeo2003,Hannestad2005}
\begin{equation*}
  \Gamma_{de}=\frac{c_s^2-c_a^2}{w_{de}}\left(\!\delta_{de}+3aH(1+w_{de})\frac{V_{de}}{k}\!\right)\!.
\end{equation*}

So, the relative perturbation of dark energy pressure is
\begin{equation}
  \pi_{de}=\frac{c_s^2}{w_{de}}\delta_{de}+3aH(c_s^2-c_a^2)\frac{1+w_{de}}{w_{de}}\frac{V_{de}}{k}.
\end{equation}
In the rest frame of dark energy ($V_{de}=\delta\phi=0$) it can be
presented as
\begin{equation*}
  \pi_{de}= \frac{c_s^2}{w_{de}}\delta_{de},
\end{equation*}
where the effective (rest-frame) sound speed $c_{s}^2$ for the
scalar field with given Lagrangian can be calculated according to
(\ref{c_s2}).

The perturbations are supposed to be small ($|\Phi|\sim|\Psi|\sim
|\delta_{de}|\sim ]\pi_{de}|\sim$ $\sim|V_{de}|\ll1$), henceforth
all following equations are linearized with respect to the perturbed
variables. %
\index{perturbation theory} %
In the linear perturbation theory it is convenient to perform the
Fourier transformation of all spatially-dependent variables, so the
equations are written for the corresponding Fourier amplitudes of
the metric ($\Psi(k,\eta)$), dark energy density
$\delta_{de}(k,\eta)$, pressure $\pi_{de}(k,\eta)$ and velocity
$V_{de}(k,\eta)$ perturbations (here $k$ is wave number). These
variables are gauge-invariant
\cite{Bardeen1980,Kodama1984,Novosyadlyj2007}.

Since we suppose the minimal coupling between dark energy and
matter, then from the differential conversation law
$T^{\mu}_{\nu;\mu\,(de)}=0$ we obtain the equations for evolution of
density and velocity perturbations of dark energy:
\begin{equation}
  \label{deltade-Vde}
  \begin{array}{c}
  \displaystyle  \dot{\delta}_{de}-3aH(w-c_s^2)\delta_{de} +   \\
   \displaystyle+ (1+w_{de})  \left\{\!\left[k^2+9a^2H^2(c_s^2-c_a^2)\right]\frac{V_{de}}{k}-3\dot{\Psi}\!\right\}  =   0,   \\
  \displaystyle  \dot{V}_{de}+aH(1-3c_s^2)V_{de}-  \frac{c_s^2k}{1+w_{de}}\delta_{de}-k\Psi
    = 0.
  \end{array}
\end{equation}

By the same way the equations for evolution of density and velocity
per\-tur\-ba\-ti\-ons can be obtained of %
\index{cold dark matter (CDM)} %
cold dark matter, which is pressureless, $p_{dm}=$ $=\pi_{dm}=0$,
perfect fluid. So, they can be easy deduced from (\ref{deltade-Vde})
by assuming $w_{dm}=c^2_{a\,(dm)}=c^2_{s\,(dm)}=0$ and re-denoting
corresponding values:
\begin{gather}
  \dot{\delta}_{dm}+kV_{dm}-3\dot{\Psi}=0, \label{delta_dm}\\
  \dot{V}_{dm}+aHV_{dm}-k\Psi=0. \label{V_dm}
\end{gather}
The Einstein equations for scalar metric and energy-momentum tensor
per\-tur\-bations, %
\index{Einstein equations} %
\begin{equation*}
  \delta R^{\mu}_{\nu}-\frac{1}{2}\delta^{\mu}_{\nu}\delta R=4\pi G\left(\!\delta T^{\mu}_{\nu\,(dm)}+\delta T^{\mu}_{\nu\,(de)} \right)\!,
\end{equation*}
where $\delta R^{\mu}_{\nu}$ and $\delta R$ are perturbed parts of
Ricci tensor %
\index{tensor Ricci} %
and scalar cur\-va\-tu\-re of four-space correspondingly, complete
the system of equations (\ref{deltade-Vde})---(\ref{V_dm}) by
equations for evolution of gravitational potential. One of them is
as follows:
\begin{equation}
  \dot{\Psi}+aH\Psi-\frac{4\pi Ga^2}{k}\left(\rho_{dm}V_{dm}+\rho_{de}(1+w_{de})V_{de}\right)=0,
  \label{Psi}
\end{equation}

The system of 5 linear differential equations
(\ref{deltade-Vde})---(\ref{Psi}) for 5 unknown functions
$\delta_{de}$, $V_{de}$, $\delta_{dm}$, $V_{dm}$ and $\Psi$ is
closed and can be solved for given initial conditions.  In the first
and this chapters we assume adiabatic initial conditions for
perturbations of dark matter, baryons and relativistic components.
At the beginning of matter dominated era, when density of
relativistic components vanishes and dark energy is subdominant, in
the limit $\Omega_r$ and $\Omega_{de}$ $\rightarrow$ $0$ the growing
solution (growth mode) of equations (\ref{delta_dm})---(\ref{Psi})
is as follows:
$$\Psi={\rm const}, \quad \delta_{dm}=-(2+k^2\eta^2/6)\Psi, \quad V_{dm}=k\eta\Psi/3$$
(for more details see \cite{Novosyadlyj2007}). Putting some small
value $\Psi_{init}$ at $\eta_{init}\ll k^{-1}$ we obtain the
adiabatic initial conditions for the dark matter component:
\begin{equation}
  \delta_{dm}^{init}=-2\Psi_{init}, \quad
  V_{dm}^{init}=k\eta_{init}\Psi_{init}/3. \label{dm_init}
\end{equation}

Assuming that dark energy is subdominant in the early Universe and
$w_{de},$ $c_s^2$, $c_a^2 $ are constant one can solve the equations
(\ref{deltade-Vde}) for $\Psi={\rm const}$ and for superhorizon
scale of perturbations ($k\eta\ll1$) obtaining the initial
conditions for density and velocity perturbations of dark energy:
\begin{equation}
  \delta_{de}^{init}=-6\frac{(1+w_{de})(c_a^2-c_s^2)}{(3-2c_s^2)(w_{de}-c_s^2)}\Psi_{init}, \quad
  V_{de}^{init}=k\eta_{init}\Psi_{init}/(3-2c_s^2). \label{de_init}
\end{equation}

The equations (\ref{delta_dm})---(\ref{Psi}) with initial conditions
(\ref{dm_init})---(\ref{de_init}) have been used for analysis of
mutual
influence of perturbations in two-component Universe %
\index{two-component Universe} %
with scalar field dark energy with different Lagrangians [131,
255---257]. The same equations (\ref{delta_dm})---(\ref{Psi}) have
been used recently in \cite{Haq2011} for analysis of effect of dark
energy perturbations on dark matter ones in scalar field models with
generalization of kinetic term of classical Lagrangian, but,
unfortunately, there the initial conditions for both components are
not specified. \index{classical Lagrangian}In these papers it has
been shown that scalar fields with evolving $w_{de}$ and $c_s^2$
affect distinctly on the power spectrum of matter density
perturbations at \mbox{subhorizon scales.}\index{power spectrum}

Therefore, the dark energy perturbations must be included in
complete theory of cosmological perturbations which is used for
determinations
of %
\index{cosmological parameters} %
cos\-mo\-lo\-gi\-cal parameters from CMB anisotropy %
\index{cosmic microwave background (CMB)}%
\index{CMB anisotropy}%
\index{large scale structure}%
and large scale structure data. The Einstein---Boltzmann equations
for scalar per\-tur\-ba\-ti\-ons in multicomponent Universe (cold
dark matter, baryons, photons and neutrinos), in which all important
physical processes at different epochs have been taken into account,
are presented in the paper \cite{Ma1995}. They are the base for
publicly
available codes CMBFAST \cite{cmbfast96,cmbfast99}, %
\index{CMBFAST} %
CMBEasy \cite{Doran2005}, CAMB %
\index{CAMB} %
\cite{camb,camb_source} and CLASS %
\index{CLASS} %
[122---124], designed for
integration of Einstein---Boltzmann equations %
\index{Einstein---Boltzmann equations} %
and accurate computation of power spectra of CMB anisotropy and %
\index{CMB anisotropy} %
matter density per\-tur\-ba\-ti\-ons for different cosmologies. The
most advanced and widely used in cosmological applications is CAMB
code which gives possibility to include also different models of
dark energy and is  supplied as part of the CosmoMC parameter
estimation package. There the Einstein---Boltzmann equations in
synchronous gauge are integrated. Since this code will be used below
we present here the evolution equations
(\ref{delta_dm})---(\ref{Psi}) and initial conditions
(\ref{dm_init})---(\ref{de_init}) in synchronous gauge too.

\index{synchronous gauge} %
In the synchronous gauge the line element in 4-space with flat
3-space is as follows:\vspace*{-3mm}
\begin{equation}
  ds^2=g_{\mu\nu} dx^{\mu} dx^{\nu} =a^2(\eta)(-d\eta^2+(\delta_{ij}+h_{ij}) dx^{i}dx^{j}),\label{ds_synch}
\end{equation}
where $h_{ij}(\eta,\textbf{x})$ is metric perturbations. %
\index{metric perturbations|(} %
The scalar perturbations of metric $h_{ij}$ can be decomposed into
the trace $h\equiv h_{i}^{i}$ and traceless $\tilde{h}_{ij}$
components as $h_{ij}=h\delta_{ij}/3+\tilde{h}_{ij}$. As above, the
perturbations are supposed to be small ($h\ll1$), so, all following
equations containing $h$ are linearized with respect to the metric
and matter-energy perturbed variables. In the multicomponent fluid
each component moves with a small peculiar velocity
$V^{i}=dx^{i}/d\eta$, defined by its intrinsic properties (density,
pressure, entropy etc.), gravitational potential $h$ and initial
conditions. At the linear stage of evolution of perturbations the
cold dark matter (CDM) component is a pressureless perfect fluid
interacting with other components only via gravity. Therefore, the
synchronous coordinates are usually defined as comoving to the
particles of CDM: $V_{cdm}=0$. The evolution equations for dark
energy perturbations can
be deduced by the same way as in conformal %
\index{conformal Newtonian gauge} %
Newtonian gauge or by gauge transformations
$x^{\mu}_{(con)}\rightarrow x^{\mu}_{(syn)}+\xi^{\mu}$ which
transform $g_{\mu\nu}^{(con)}\rightarrow g_{\mu\nu}^{(syn)}$ keeping
$ds^2$ as invariant.

Both ways lead to the equations for evolution of density and
velocity perturbations of dark energy and cold dark matter as well
as metric perturbations in synchronous gauge as follows:
\begin{gather}%
  \label{d_de_s}%
  \begin{array}{c}%
   \displaystyle \dot{\delta}_{de}  +3(c_s^2-w_{de})aH\delta_{de}  + \\
    \displaystyle + (1+w_{de})\frac{\dot{h}}{2}
    +(1+w_{de})\left[k^2+9a^2H^2(c_s^2-c_a^2)\right]\frac{V_{de}}{k}=0,
  \end{array}\\
  \label{V_de_s}
  \dot{V}_{de}+aH(1-3c_s^2)V_{de}-\frac{c_s^2k}{1+w_{de}}\delta_{de}=0, \\
  \label{d_dm_s}
  \dot{\delta}_{dm}+\frac{\dot{h}}{2}=0,  \\
  \label{V_dm_s} V_{dm}=0,  \\
 \label{h_s}
 \begin{array}{c}
  \displaystyle (aH\dot{h})\dot{ }-  8\pi
   Ga^2\left[\rho_{dm}a\left(a^{-1}\delta_{dm}\right)\dot{ }  + \right.
   \\[2mm]
    \displaystyle \left.  + \rho_{de}\left(\!\dot{\delta}_{de}-aH(1+3w_{de})\delta_{de}-kV_{de}\!\right)\!\right]=0.
  \end{array}%
\end{gather}

The adiabatic initial conditions for dark matter and subdominant
dark energy are:
\begin{equation}
  \label{synch_init}
  \begin{array}{c}
    \displaystyle \delta_{de}^{\;init}  = -\frac{(4-3c_s^2)(1+w_{de})}{8+6c_s^2-12w_{de}+9c_s^2(w_{de}-c_a^2)}
    h_{init},\\[4mm]
  \displaystyle   V_{de}^{\;init}  = -\frac{c_s^2k\eta_{init}}{8+6c_s^2-12w_{de}+9c_s^2(w_{de}-c_a^2)} h_{init},
  \\[4mm]
  \displaystyle   \delta_{dm}^{\;init}  = -\frac{1}{2} h_{init},
  \\[4mm]
   \displaystyle  V_{dm}^{\;init}  = 0.
  \end{array}
\end{equation}

\index{scalar perturbations}%
The character of evolution of scalar field density perturbations
depends on the temporal behavior of EoS parameter, adiabatic and
effective sound speeds, which can be defined or deduced for
specified
$\mathcal{L}(X,U)$ and $U(\phi)$). %
\index{EoS parameter} %

The evolution of dark energy perturbations can be analyzed also in
the
terms of perturbations of field variable %
\index{variable field} %
$$\phi(\eta,\textbf{x})=\bar{\phi}(\eta)+\delta\phi(\eta,\textbf{x}),$$
where it is supposed that the perturbation amplitude is small,
$|\delta\phi|\ll|\phi|$, henceforth all following equations can be
linearized with respect to $\delta\phi$ and its derivatives.
Substituting it into (\ref{sf_Tmunu}), using the
differential conservation law $T^{\mu}_{0\,(de);\mu}=0$ %
\index{conservation law} %
and taking into account (\ref{gen_KG}) one can obtain the second
order linear differential equation for $\delta\phi$, with
coefficients comprising $H$, $w_{de}$, $c_s^2$, $c_a^2$,
$\mathcal{L}{,_X}$, $\mathcal{L}{,_U}U_{,\phi}$ etc.  and free term
with metric perturbations. The density, pressure and velocity
perturbations of dark energy, which are necessary to write the free
term in equations for metric perturbations (\ref{Psi}), are related
with $\delta\phi$ by
relationships:%
\index{metric perturbations}%
\begin{equation}
  \delta\rho_{de}  =\left(\!\dot{\bar{\phi}}\dot{\delta\phi}-\Psi \dot{\bar{\phi}}^2\!\right)\!\left(\!\frac{\partial\mathcal{L}}{\partial X}+2 X\frac{\partial^{2}\mathcal{L}}{\partial X^{2}}\!\right)
  -\left(\!\frac{\partial \mathcal{L}}{\partial U}\frac{\partial U}{\partial \phi}-2X\frac{\partial^{2}\mathcal{L}}{\partial X\partial U}\frac{dU}{d\phi}\!\right)\delta\phi, \label{drho_dphi}
  \end{equation}
 \begin{equation} \delta p_{de}  = \left(\!\dot{\bar{\phi}}\dot{\delta\phi} - \Psi \dot{\bar{\phi}}^2\!\right)\frac{\partial \mathcal{L}}{\partial X}+\frac{\partial \mathcal{L}}{\partial U}\frac{\partial U}{\partial \phi}\delta\phi, \label{dp_dphi}
\end{equation}
  \begin{equation}
  V_{de}  = \frac{k\delta \phi}{\dot{\bar{\phi}}}. \label{v_dphi}
\end{equation}
They can be used also for definition of the initial conditions
$\delta\phi_{init}$ and $\dot{\delta\phi}_{init}$ from
(\ref{de_init}).

Presented in this section evolution equations for homogeneous and
per\-tur\-bed scalar field are used for interpretation of
observational data on dynamics of expansion and large scale
structure formation in
the MD and DED epochs. %
\index{large scale structure}%
\index{cosmic microwave background (CMB)}%
For interpretation of CMB anisotropy data the %
\index{CMB anisotropy}%
equations (\ref{deltade-Vde}) or (\ref{d_de_s})---(\ref{V_de_s}) as
well as corresponding initial conditions must be included in the
Einstein-Boltzmann code, CAMB for example.

\section[\!Specifying the scalar-field models of dark energy]
{\!Specifying the scalar-field models\\ \hspace*{-0.95cm}of dark
energy\label{ch2-sec4}}

\hspace*{3cm}The scalar field model of dark energy should be
specified\linebreak for cosmological applications by definition of
Lagrangian and potential or dependences of EoS parameter and
effective sound speed on scale factor (\mbox{redshift} or time). In
this section we list some of them, most widely used in the
literature.

\subsection{\!Lagrangian\label{ch2-subsec41}}
\hspace*{3cm}The Lagrangians used for specifying the scalar field
dark energy models are as follows:

canonical or classical Lagrangian %
  \index{classical Lagrangian} %
  \index{canonical Lagrangian} %
  [258---268]
  \begin{equation}
    \mathcal{L}=X-U(\phi), \label{lagr_cf}
  \end{equation}

non-canonical Dirac-Born-Infeld one %
  \index{Dirac-Born-Infeld Lagrangian} %
  [269---278],
  which is relativistic generalization of classical Lagrangian,
  \begin{equation}
    \mathcal{L}=-U(\phi)\sqrt{1-2X},  \label{lagr_tf}
  \end{equation}

classical Lagrangian with opposite sign of kinetic term %
\index{classical Lagrangian} %
  [51, 279---284]
  \begin{equation}
    \mathcal{L}=-X-U(\phi), \label{lagr_phantom}
  \end{equation}

 canonical form of Lagrangian with generalization of kinetic term
  in the form some function $F(X)$ \cite{Haq2011}
  \begin{equation}
    \mathcal{L}=F(X)-U(\phi), \label{lagr_fx}
  \end{equation}

 non-canonical form of Lagrangian with non-canonical kinetic term\linebreak  \mbox{[285---291]}
  \begin{equation}
    \mathcal{L}\left(\phi,\phi_{;\nu}\phi^{;\nu},U(\phi)\right), \label{lagr_k_essence}
  \end{equation}

 two-field Lagrangians of canonical or non-canonical form with
  canonical or non-canonical kinetic terms
  \begin{equation}
    \mathcal{L}\left(F(\phi,\phi_{;\nu},\phi_{,\mu}\phi^{,\mu}),U(\phi);F(\psi,\psi_{;\nu},\psi_{,\mu}\psi^{,\mu}),U(\psi)\right). \label{lagr_quintom}
  \end{equation}

In the case of the additional coupling of dark energy with other
com\-po\-nent(s) the Lagrangian contains the additional term(s),
describing this non-gra\-vi\-ta\-tio\-nal interaction(s). The limit
of generalization of functional forms of Lagrangians does not exist,
it is only restricted by fantasia and technical possibilities of
researches.

\subsection{\!Potential\label{ch2-subsec42}}

\hspace*{3cm}\index{potential $U(\phi )$|(}As it follows from
(\ref{w_L}) and (\ref{c_s2}), for definition of $w_{de}$ and $c_s^2$
the potential in Lagrangian must be given too. For different types
of Lagrangians the different potentials are
studied. The scalar fields with %
\index{canonical Lagrangian} %
canonical Lagrangian (\ref{lagr_cf}) have the simplest and best
studied equation of\linebreak \mbox{motion (\ref{gen_KG}),} %
\index{equation of motion} %
\begin{equation}
  \ddot{\phi}+2aH\dot{\phi}+a^2\frac{d{U}}{d{\phi}}=0.
  \label{KG_eq}
\end{equation}
called Klein---Gordon one. %
\index{Klein---Gordon equation} %
They were called by Steinhardt and Caldwell
\cite{Steinhardt1997,Steinhardt1998} the ``quintessence''. %
\index{quintessence} %
The large number of potentials, motivated by particle physics beyond
the standard model, have been used to probe the scalar field as dark
energy. They should satisfy the condition $U>a^{-2}\dot{\phi}^2$,
where $\phi$ is solution of (\ref{KG_eq}), in order to have
$w_{de}<-1/3$ near the current epoch. Other conditions for scalar
field potentials follows from the requirement that the energy
density of the scalar field should be significantly less than that
of radiation and dark matter during RD epoch $U(z<z_{dec})\ll
\rho_{\gamma}-\dot{\phi}^2/2$, provide long enough MD epoch to allow
galaxies to form which requires $U(z< 1)\ll
\rho_{dm}-\dot{\phi}^2/2$ and accelerate the expansion now, that
requires $U(z\approx0)\approx 3.3\rho_{dm}-\dot{\phi}^2/2$ now.

We present here the list of some functional forms of potentials,
which elucidate the main properties of scalar field models of dark
energy with canonical Lagrangian:

 the power-law potential, often used in particle physics
  (\!\!\!\cite{Linde90} and citing therein)
  \begin{equation}
    U= M^{4-n}\phi^n \quad \mathrm{with} \quad n>0,
    \label{U_pl1}
  \end{equation}

 the power-law tracker potential used in SUSY
  \cite{Binetruy1998,Masiero1999} and supergravity
  \cite{Brax1999,Copeland2000} theories
  \begin{equation}
    U= M^{4+n}\phi^{-n}\quad \mathrm{with} \quad n>0,
    \label{U_pl2}
  \end{equation}

 the polynomial form of potential
  \begin{equation}
    U= \Sigma_{n} a_n\phi^n,
    \label{U_poly}
  \end{equation}

 the exponential potential used for moduli or dilaton fields
  \cite{Barreiro1999,Ferreira1997}
  \begin{equation}
    U= M^{4}\exp{(-\beta\phi/M_p}),
    \label{U_exp_moduli}
  \end{equation}

 the exponential tracker field potential
  \begin{equation}
    U= M^{4}\exp{(M_p/\phi)},
    \label{U_exp_track}
  \end{equation}

 the combined power-law and exponential tracker potential used
  also in SUSY \cite{Binetruy1998,Masiero1999} and supergravity
  \cite{Brax1999,Copeland2000} theories
  \begin{equation}
    U= M^{4+n}\phi^{-n}\exp{(\alpha\phi^2/M_p^2)},
    \label{U_exp_pl}
  \end{equation}

the potential used for presenting pseudo-Nambu-Goldstone
  boson (PNGB) \cite{PNGB} and some type of axions
  \begin{equation}
    U= M^{4}\cos^2{(\phi/2f)}
    \label{U_pngb}\mbox{...}
  \end{equation}

Here $M_p$ is Planck mass, other values~--- $n$, $M$, $a_n$,
$\alpha$, $\beta$, $f$~--- are parameters of scalar field which need
the definition or determination to match the observational data on
dynamics of expansion of the Universe.

\begin{figure}
\raisebox{0.0cm}{\parbox[b]{6.2cm}{\caption{Classes of quintessence
scalar fields ($w_{de}(a)>-1$),
    freezing ($w_{de}'<0$) and thawing ($w_{de}'>0$) ones, in the
    phase plane $w_{de}(a)-w_{de}'(a)$ \cite{Caldwell2005}. Black
    solid lines show the boundaries of these classes in the phase
    space, the short-dashed line shows the boundary between field
    evolution accelerating and decelerating down the potential (see
    for details \cite{Caldwell2005}). The solid, dashed, dotted and dash-dotted lines show
    evolutionary tracks of scalar fields with potentials
    (\ref{U_pl1})---(\ref{U_pngb}). The arrows show the direction of
    evolution from beginning ($a=0$: right-most points for freezing
    and left-most points for thawing scalar fields) to current epoch %
    \index{freezing model|)} %
    ($a=1$: left-most points for freezing and right-most points for
    thawing scalar fields) (From \cite{Caldwell2005})\label{Caldwell_Linder}}}}\hspace*{0.3cm}\includegraphics[width=6.5cm]{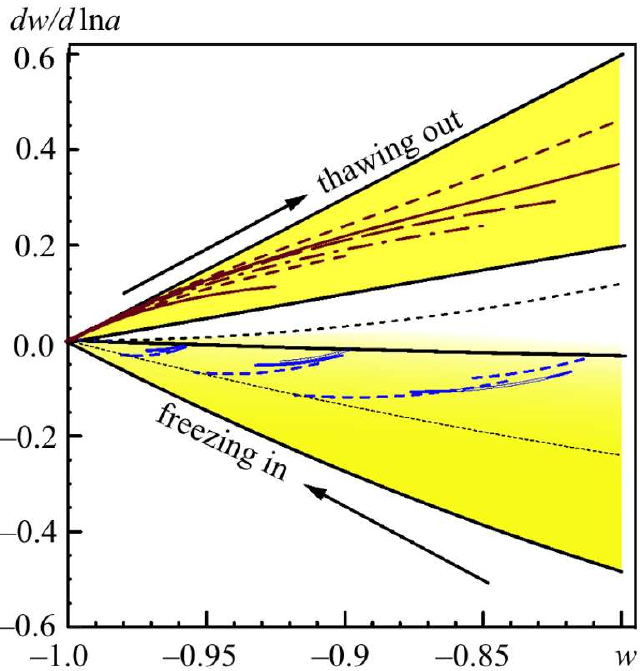}
\end{figure}

Presented list of potentials is far incomplete but spectrum of
scalar
field properties for them is wide enough that to allow the %
\index{freezing model|(} %
\index{thawing model} %
possibility to classify them as ``freezing models'' and ``thawing''
models \cite{Caldwell2005}.  In the class of free\-zing models the
fields were rolling along the potentials in the past, but their
movements gradually slow down after the system enters the phase of
cosmic acceleration and freeze out at the minimum of potential at
finite $\phi$ or at infinity of field variable when minimum of
potential is absent. The EoS parameters $w_{de}$ of such fields can %
\index{EoS parameter} %
start from any value $>$$-1$ and go to $-1$ at the free\-zing stage.
On contrary, in the class of thawing models the fields have been
frozen by Hubble friction (term $2aH\dot{\phi}$ in (\ref{KG_eq}))
until recently and then start to evolve with increasing $w_{de}$. In
this models $w_{de}\approx-1$ at the early epochs and
\mbox{$-1<w_{de}<-1/3$ now.}

These classes of quintessence scalar fields are separated in the
phase plane $w_{de}(a)-w_{de}'(a)$ (shown in
Fig.~2.\ref{Caldwell_Linder}). The evolutionary tracks of scalar
fields with potentials (\ref{U_pl1})---(\ref{U_pngb}) are shown
there too. The freezing models with poten\-tials (\ref{U_pl2}) and
(\ref{U_exp_pl}) are shown there by  solid and dashed lines
correspon\-dingly. The thawing ones in this figure are represented
by potentials (\ref{U_pl1}) and (\ref{U_pngb}). The evolutionary
track for field with potential (\ref{U_pngb}) is shown by  solid
line and for field with potential (\ref{U_pl1}) by short-, dot-, and
long-dashed ones for $n = 1, 2, 4$ correspondingly. All scalar
fields shown there have at current epoch $-1<w_{de}<-0.8$ and
$\Omega_{de}\approx0.7$, so, they can be considered as real
candidates for the dark energy.

Unfortunately, for Lagrangians other than canonical one we have no
such intensive analysis of scalar fields.  Maybe in the nearest
future this lack will \mbox{be removed.}

\subsection{\!EoS parameter\label{ch2-subsec53}}

\hspace*{3cm}\index{EoS parameter}Another way of specifying of
scalar field model of dark energy consists in assumption about the
dependence of EoS parameter $w_{de}$ on the scale factor $a$. It can
be made in \textit{ad hoc} manner or by solution of equations that
implement the defined field properties.

\begin{figure}
\vskip1mm
\includegraphics[width=13cm]{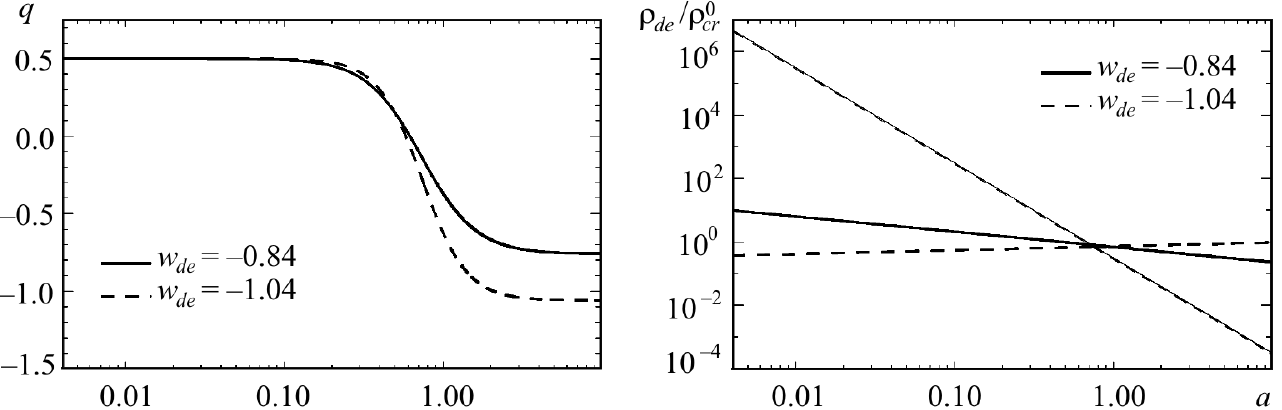}
    \vskip-2mm
  \caption{The evolution of deceleration parameter $q$ (left panel)
    and energy density $\rho_{de}$ in units of critical one at current
    moment (right panel) in the models with parameters of wCDM models
    from Table 1.\ref{tab_params} (dataset 1~--- dashed line, dataset 2~--- solid line). The dark matter density evolution for both models
    is shown for comparison by thin lines (superimposed), which go
    from the upper left to lower right corner}
  \label{q_wcdm}
\end{figure}

{\bf\textit{Ad hoc} setting of \boldmath$w_{de}(a)$}. Traditional
approach is to present $w_{de}$\linebreak \mbox{in the form}
$$w_{de}(a)=\sum_nw_nf_n(a),$$
where $w_n$ is a parameter and $f_n(a)$
some simple function of scale factor $a$. Let us list used in the
literature function forms for $w_{de}(a)$ ordered by the number of
their parameters.

i) \textit{One-parametric EoS}. Appears in the models with
$w_{de}=$~const. They are the simplest and best studied models of
dark energy scalar field. This parameter is constrained by majority
of observational data, though the accuracy of determination of its
best-fit value is too low, $\sim$15---25\,\%, while the density
parameter $\Omega_{de}$ is determined with accuracy $\sim$5\,\% (see
Chapter 1). The best-fit value of constant $w_{de}$ for the most
determinations is in the range (--1.2, --0.8). In these models
$c_a^2=w_{de}$, but $c_s^2$ must be defined additionally by
definition of either Lagrangian or itself in \textit{ad hoc} manner.
Similarly to the simplest models the $w_{de}=$~const-models are very
specific. For
example, in the case of scalar field with classical Lagrangian %
\index{classical Lagrangian} %
(\ref{lagr_cf})
$$
X=\frac{1+w_{de}}{1-w_{de}}U, \quad U=\frac{1-w_{de}}{2}\rho_{de},
\quad c_s^2=1,
$$
in the case of scalar field with Dirac---Born---Infeld one
(\ref{lagr_tf})
$$X=\frac{1+w_{de}}{2}, \quad U=\sqrt{-w_{de}}\rho_{de}, \quad c_s^2=-w_{de},$$
and so on, this can be deduced using equation (\ref{w_L}). The
evolution of energy density and deceleration parameter for $w$CDM
models with parameters from Table 1.\ref{tab_params} are shown in
Fig.~2.\ref{q_wcdm}. The evolution of deceleration parameters in
both models is similar, they differ slightly by value of $q$ at
current epoch $q_0=$ $=\Omega_m+(1+3w_{de})\Omega_{de}$ (in the
model with $w_{de}=-0.84$ $q_0=-0.4$, in the models with
$w_{de}=-1.04$ $q_0=-0.62$) and asymptotic value at
$a\rightarrow\infty$: $q_{\infty}=(1+3w_{de})/2$ (--0.76 for
$w_{de}=-0.84$ and --1.06 for $w_{de}=-1.04$). The energy density
evolution for these models is quite different: it decreases from
$\infty$ at $a=0$ to 0 when $a\rightarrow\infty$ in scalar field
model with $w_{de}=-0.84$ and increases from 0 at $a=0$ to $\infty$
when $a\rightarrow\infty$ in scalar field model with $w_{de}=-1.04$.
The first is quintessence scalar field, the second is phantom
\cite{Caldwell2002} one.

ii) \textit{Two-parametric EoS}. The simplest approximation for
time-variable EoS parameter, which is widely used,
\begin{equation}
  w_{de}(a)=w_0+(1-a)w_a,
  \label{wa_cpl}
\end{equation}
was proposed by Chevallier, Polarski and Linder
\cite{Chevallier2001,Linder2003}, called in the literature the CPL
one. Here, $w_0$ and $w_a$ denote the present values of $w_{de}$ and
its first derivative with respect to $a$ with opposite
sign\,\footnote{\,It is first derivative with respect to $z$ at the
current
  epoch $z=0$.} respectively. The determination of them in
\cite{WMAP7b} on the base of WMAP7 {+} BAO {+} SN data gives
$w_0=-0.93\pm0.12$, $w_a=-0.41\pm0.72$. The dependence
(\ref{wa_cpl}) with these

\begin{wrapfigure}{r}{7.3cm}
\vskip-2mm
\hspace*{2mm}\includegraphics[width=7cm]{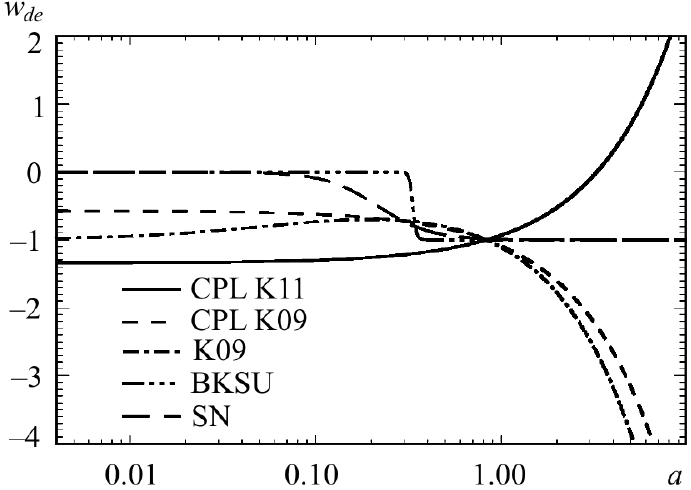}\vspace*{-2mm}\\
\hspace*{2mm}\raisebox{0.2cm}{\parbox[b]{7cm}{\caption{Time dependences of EoS parameter for different %
    \index{EoS parameter|(} %
    parametrizations: CPL with parameters determined in \cite{WMAP7b}
    (CPL K11), CPL with pa\-ra\-me\-ters determined in \cite{WMAP5b} (CPL
    K09) \label{wa_appr}}}}\vskip-7mm
\end{wrapfigure}
\begin{figure}
\vskip1mm
\includegraphics[width=13cm]{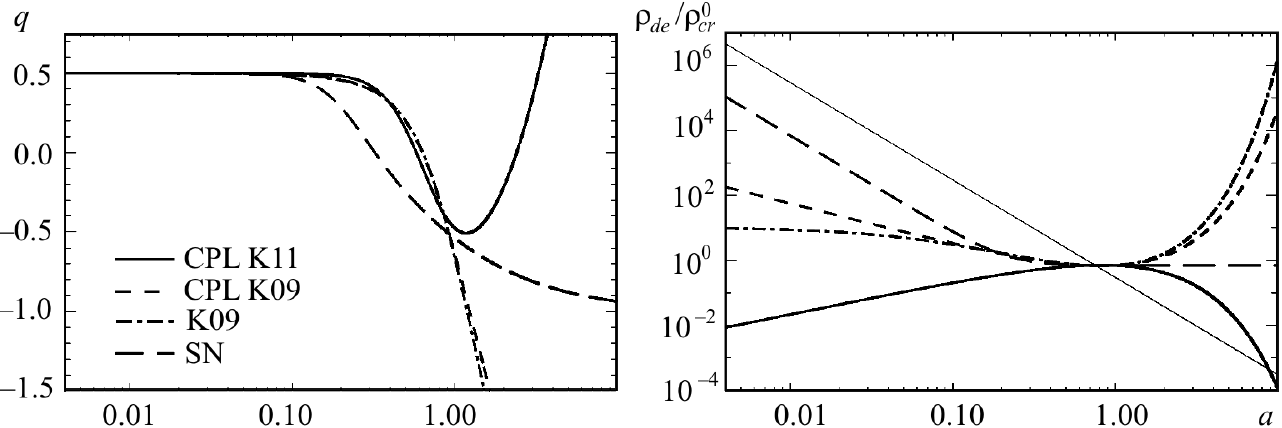}
    \vskip-2mm
  \caption{The evolution of deceleration parameter $q$ (left panel)
    and energy density $\rho_{de}$ in units of critical one at current
    moment (right panel) in the models with different parametrizations
    of EoS parameter: CPL with parameters determined in \cite{WMAP7b}
    (CPL K11), CPL with pa\-ra\-meters determined in \cite{WMAP5b} (CPL
    K09). The dark matter density evolution for all models is shown for
    comparison by thin lines (superimposed), going from the upper left to
    lower right corner}
  \label{q_wa}
\end{figure}


\noindent parameters is shown in Fig.~2.\ref{wa_appr} (CPL K11) in
the range $0.001\le a\le10$. Such dark\linebreak energy evolves from
the field with $w=w_0+w_a=-1.34$ at the early epoch (\mbox{$a=0$})
to \mbox{$w=w_0=-0.93$} at the current one (\mbox{$a=1$}) which will
increase in future. The evolution of deceleration parameter for the
same range of $a$ is shown in Fig.~2.\ref{q_wa} (left panel). Its
energy density increases from zero to $\sim$$1.25\rho^{(0)}_{de}$ at
$a\approx 0.83$ and decreases asymptotically to zero after that
according to (\ref{rho_de}) with effective EoS parameter
\begin{equation*}
\tilde{w}_{de}=w_0+w_a\frac{1-a+\ln{a}}{\ln{a}},
\end{equation*}
which goes to $+\infty$ or $-\infty$ depending on sign of $w_a$ when
$a\rightarrow +\infty$ (Fig.~2.\ref{q_wa}, right panel).

\index{WMAP} %
The previous determination of these parameters by \cite{WMAP5b}
using similar but older datasets gave $w_0=-1.09\pm0.12$,
$w_a=0.52\pm0.46$ (Fig.~2.\ref{wa_appr}, CPL K09). It means that in
this case the dark energy evolves from quintessential field with
$w=-0.57$ at the early epoch to the phantom one at current epoch
with $w=-1.09$. Its density decreases at early epoch, achieves the
minimal value of $\sim$$0.83\rho^{(0)}_{de}$ at the same scale
factor $a\approx 0.83$ and grows later. The evolution of
deceleration parameter and energy density is shown in
Fig.~2.\ref{q_wa} by lines marked as CPL K09.

One can see, that $w_{de}$, $q$ and $\rho_{de}$ diverge
asymptotically at $a>1$, therefore, such models of dark energy are
not usable for prediction of future dynamics of expansion of the
Universe. It is objection to this parametrization.

Another form of $w_{de}$ parametrization,
\begin{equation*}
w_{de}(a)=w_0+w_1z=w_0+w_1\frac{1-a}{a},
\end{equation*}
has been proposed and used in \cite{Weller2002} for analysis of
possibility to discriminate models with constant and time-varying
EoS parameters on the base of SNe data.  It has, however, essential
disadvantage: it can be used only for low-$z$ cosmology, since it
diverges at high $z$.

iii) \textit{Three-parametric EoS}. Recently this form has been
modified by \cite{WMAP5b} in order to bring the behavior of
dynamical dark energy at early epoch closer to that of
$\Lambda$-term:\vspace*{-3mm}
\begin{equation}
  w_{de}(a) = \frac{a}{a+a_{trans}}[w_0+(1-a)w_a] - \frac{a_{trans}}{a+a_{trans}},
  \label{wa_k09}
\end{equation}
This approximation has the additional third parameter $a_{tran}$
which, however, is weakly constrained by observations. One can see,
that at the early epoch when $a\ll a_{tran}$ $w_{de}\approx -1$ and
has explicit asymptotic behavior $a\rightarrow0$
$w_{de}\rightarrow-1$. At $a\gg a_{trans}$ this parametrization
becomes the CPL one.  It also has the analytical form of effective
equation of state, $\tilde{w}_{de}(a)$, which gives the evolution of
dark energy density and deceleration parameter:
$$
\tilde{w}_{de}(a)=-1+\frac{1-a}{\ln a}w_a
+\frac{1+w_0+(1+a_{trans})w_a}{\ln
  a}\ln\frac{a+a_{trans}}{1+a_{trans}}.
$$

The values of parameters in (\ref{wa_k09}) have been determined in
\cite{WMAP5b} for $a_{trans}=$ $=10$ and are as follows:
$w_0=-1.12\pm0.13$, $w_a=0.70\pm0.53$. The $a$-de\-pen\-den\-ces of
$w_{de}$, $q$ and $\rho_{de}$ with them are presented in
Figs.~2.\ref{wa_appr}---2.\ref{q_wa}. One can see, that this
parametrization meets the same objection as CPL one: prediction of
future dynamics of expansion of the Universe is doubtful. The other
objection is restriction of properties of scalar field in the early
Universe: it supposes that scalar field starts from vacuum-like
state ($w_{de}(0)=-1$). But establishing of its true origin from
observations is important for unveiling of nature of dark energy as
well as physics of Very Early Universe and unified theory of
particle physics. One more lack is the weak constraint for the third
parameter from observational data.

iv) \textit{Four-parametric EoS}. The four-parametric EoS in the
form
\begin{equation}
  w(z) = w_0 +  \frac{(w_f - w_0)}{1 + \exp(\frac{z -
      z_t}{\Delta})} \label{w}, \quad z\equiv \frac{1}{a}-1,
\end{equation}
has been proposed in \cite{Bassett2002}. Here $w_0$ is the initial
EoS at $a=0$ and $w_f$ is final one at $a=1$ when the transition
epoch $a_t$ is in the past ($a_t<1$) and $z_t/\Delta\gg 1$. It has
been studied by authors only for the special case of $w_0=0$ (at the
beginning the dark energy is dust-like) and transition is sharp
($z_t/\Delta=$ $=30$). Such time dependence of EoS has physical
motivation: it appears in the models like vacuum metamorphosis where
non-perturbative quantum effects are important at late times
\cite{Parker1999a,Parker1999b,Parker2000}. It was shown
\cite{Bassett2002} that in such case the CMB and SNe Ia data prefer
the model with $z_t=2.0^{+2.2}_{-0.76}$ and $w_f=$\linebreak
$=-1^{+0.2}$
(Fig.~2.\ref{wa_appr}). %
\index{cosmic microwave background (CMB)} %
One can see that the third parameter $z_t$ is poorly constrained
even for fixed $w_0$ and $\Delta$.  It is important to constraint
all four parameters $w_0$, $w_f$, $z_t$ and $\Delta$ jointly, but it
looks impossible at current accuracy level of cosmological
observations. Disadvantage of this simple four-parametric form of
$w(a)$ is absence of analytic solution of integral
(\ref{tilde_w_de}) for $\tilde{w}_{de}$ and, as consequence,
analytic representations for $\rho_{de}(a)$, $H(a)$ and $q(a)$.

The mentioned parametrizations of time dependence of EoS parameter
allow the phantom divide crossing ($w=-1$) and extend the variety of
properties of dark energy and its possible physical interpretations.
The additional degeneracies and uncertainties of parameters related
to the early dark energy density and time variations of EoS
parameter are inherent for them. And vice versa, the value of EoS
$w_{de}$ as well as of energy density $\Omega_{de}$ related to the
late epoch are determined well as a result of their main impact on
the expansion history of the Universe, horizon scale,
distance to CMB last scattering surface and scale-independent %
\index{growth factor} %
growth factor of linear matter density perturbations. These values,
however, give no possibility to constrain essentially the types of
cosmological scalar fields, or, in other words, the forms of their
Lagrangians and potentials.

\begin{figure}
\vskip1mm
\includegraphics[width=13cm]{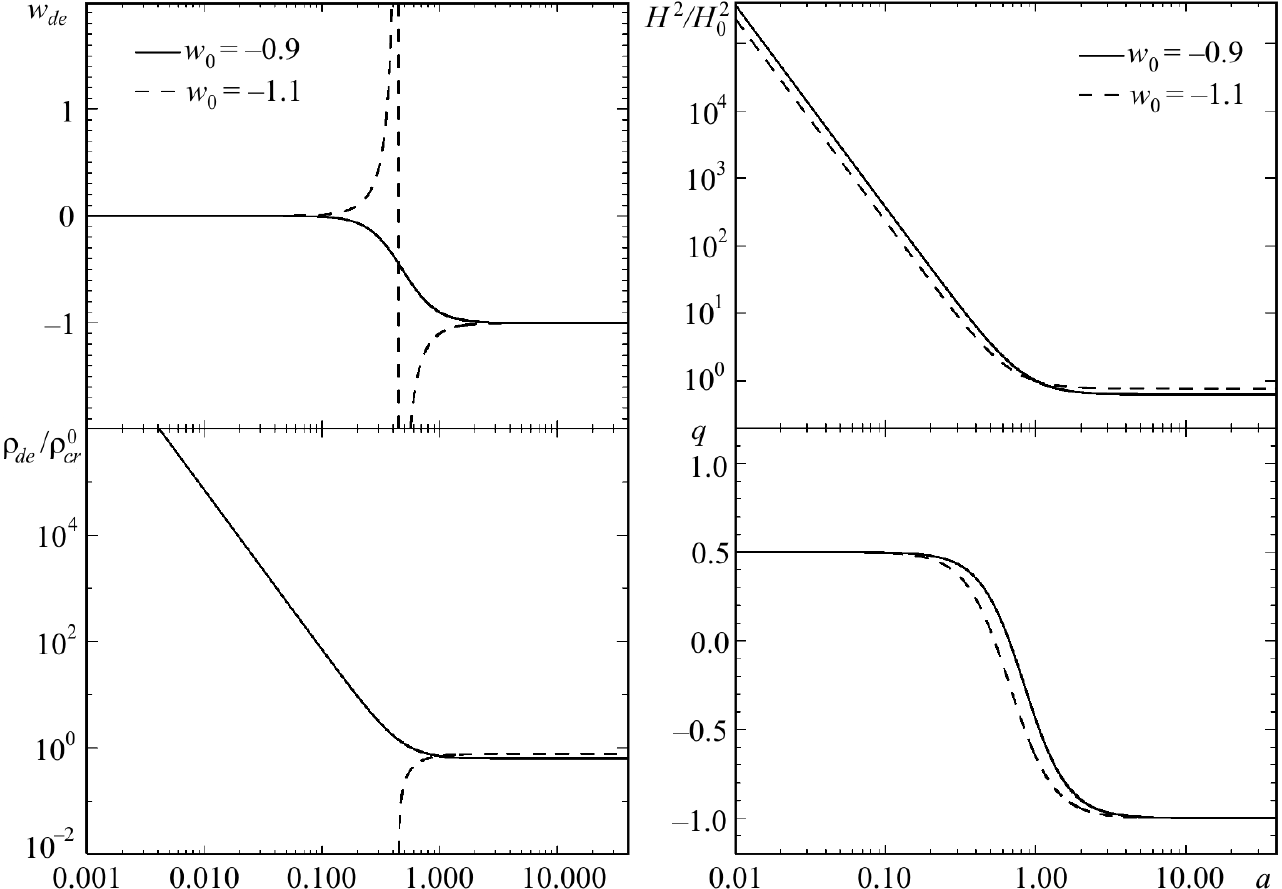}
    \vskip-2mm
  \caption{Left column: top panel~--- the dependences of EoS parameter
    on scale factor for $p_{de}=$~const ($c_a^2=0$) scalar field with
    $w_0=-0.9$ and --1.1; bottom panel~--- the dependences of dark
    energy density (in the units of critical one at the current epoch)
     with EoS parameters from the top panel on scale factor.
    Right column: The dynamics of expansion of the Universe with
    $p_{de}=$~const scalar field ($c_a^2=0$)~--- $H^2(a)$ (top panel) and
    $q(a)$ (bottom one) for the same $w_{de}$ as in left panel}
  \label{wrode_ca0}
\end{figure}

{\bfseries\itshape Some exact solutions of
Eq.\,\,(\ref{w'})}.\label{ch2-subsec43} \index{cosmological
constant}The cosmological constant or va-\linebreak cu\-um-li\-ke
fields as well as $w_{de}$~=~const dark energy are analyzed at
different stages of evolution of the Universe, from Beginning
($a\ll1$) to current epoch ($a=1$), and for prediction of its future
($a\gg1$). In the case of ii)---iv) parametrizations their
application at $a\ll1$ or $a\gg1$ are ambiguous \mbox{since} energy
density and pressure of dark energy can acquire there surprising
values as a consequence of extension of those parametrizations which
are good approximations only at the vicinity of $a=1$.  Instead of
probing the numberless analytical forms of $w_{de}(a)$ one can probe
the scalar field dark energy models assuming their specific
properties. Let us consider the simplest ones.

i) \textit{Constant density, $\rho_{de}=const$.}  If the density of
dark energy is constant in space and time then we have the well
studied model with cosmological constant or vacuum field models,
$\tilde{w}_{de}=-1$, that follows from Eq.~(\ref{rho_de}).

ii) \textit {Constant pressure, $p_{de}=const$.}  Such assumption is
equivalent to the adiabatic sound speed, %
\index{sound speed} %
$c_a^2=\dot{p}_{de}/\dot{\rho}_{de}$, is zero when
$\dot{\rho}_{de}\ne 0$.  Equation (\ref{w'}) in this case has the
simple analytic solution\vspace*{-3mm}
\begin{equation}
  w_{de}(a)=\frac{w_0a^3}{1+w_0(1-a^3)},
\end{equation}\vspace*{-5mm}

\noindent where \mbox{$w_0=w_{de}(a=1)$} is free parameter. Such
one-parametric EoS has in\-te\-res\-ting asymptotic properties: when
$a\rightarrow0$ then $w_{de}\rightarrow0$, and when
$a\rightarrow\infty$ then $w_{de}\rightarrow -1$. Such dark energy
at the Beginning is dust-like and in the future it is similar to
vacuum energy field. It is like to the vacuum metamorphosis model,
mentioned above, but with gradual transition from dust-like state to
va\-cu\-um-li\-ke one
(Fig.~2.\ref{wrode_ca0}).  So, such model has no fine tuning problem. %
\index{fine tuning} %
Its other advantage is that the effective EoS parameter
(\ref{tilde_w_de}) has exact \mbox{analytic form:}\vspace*{-2mm}
\begin{equation*}
  \tilde w_{de}(a)=-\frac{1}{3}\frac{\ln{(1+w_0-w_0a^3)}}{\ln{a}}
\end{equation*}

\noindent and the same asymptotic behavior. The energy density,
accordingly, has simple analytic dependence too\vspace*{-3mm}
\begin{equation}
  \rho_{de}(a)=\rho_{de}^{(0)}[(1+w_0)a^{-3}-w_0]. \nonumber
\end{equation}

In the range $w_0\ge-1$ the EoS parameter $w_{de}$ is monotonic
decreasing function of $a$, repulsive property of scalar field
increases and the Universe with such dark energy will exponentially
expand in far future. If $w_0=-1$, then
$w_{de}(a)=\tilde{w}_{de}(a)=-1$ and we have the vacuum-like field.
When $w_0<-1$, the properties of the field become unusual: at
$a_{d2k}=[(1+w_0)/w_0]^{1/3}<1$ the EoS parameter has discontinuity
of the second kind since energy density of scalar field $\rho_{de}$
becomes zero (left panel of Fig.~2.\ref{wrode_ca0}). This
discontinuity is not physical, since physical values $\rho_{de}$ and
$p_{de}$ are smooth continuous functions. In this case $\rho_{de}$
is negative at $0\le a<a_{d2k}$ and positive later. In spite of
$\rho_{de}\propto a^{-3}$ in the multicomponent medium with matter
and radiation all
energy conditions are always satisfied and Einstein equations for %
\index{Einstein equations} %
evolution of the homogeneous Universe have real solutions (see for
details Chapter~5).  Such dynamics of EoS parameter is interesting
in the scalar field model which supposes negative value of energy
density, that will be discussed later. In the right panel the
evolution of Hubble parameter $H(a)$ %
\index{Hubble parameter} %
and deceleration one $q(a)$ are shown for $w_0=-0.9$ and $-1.1$. One
can see, that dynamical evolution of the Universe filled with
$p_{de}=$~const DE with $w_0=-0.9$ and $-1.1$ in past and in future
is similar though dynamical evolution of fields is quite different.

We have studied this model in
\cite{Novosyadlyj2009,Sergijenko2009b,Novosyadlyj2010} for $w_0\ge
-1$ and shown that it matches all observational data as well as
$\Lambda$CDM and $w$CDM models do. The best-fit parameters of such
dark energy are $\Omega_{de}=0.72^{+0.04}_{-0.05}$,
$w_0=0.99^{+0.03}_{-0.01}$. %
\index{best-fit parameters} %
It was shown also that it is perturbed and causes the appreciable
influence on the matter power spectrum at
subhorizon scales. %
\index{power spectrum}%

iii) \textit {Barotropic EoS or constant adiabatic sound speed
  $c_a^2$.}  This assumption is more general than previous one and
allows other distinct properties of dark energy favorable for
analytic, semianalytic and numerical analysis. In such case the
temporal derivative of $p_{de}(\eta)$ is proportional to the
temporal derivative of $\rho_{de}(\eta)$. The integral form of this
condition is the generalized linear barotropic equation of state
\begin{equation}
  p_{de}=c_a^2\rho_{de}+C,\label{beos}
\end{equation}
where $C$ is a constant. Cosmological scenarios for the Universe
filled with the fluid with such EoS equation\,\footnote{\,Often
called in
  literature ``wet dark fluid''.} have been analyzed in
\cite{Babichev2005,Holman2004}. The solution of the differential
equation (\ref{w'}) for $c_a^2=const$ is following:
\begin{equation}
  w_{de}(a)=\frac{(1+c^2_a)(1+w_0)}{1+w_0-(w_0-c^2_a)a^{3(1+c^2_a)}}-1,\label{w_bar}
\end{equation}
where the integration constant of (\ref{w'}) $w_0$ is chosen as the
current value of $w_{de}$. One can easily find that (\ref{w_bar})
gives (\ref{beos}) with $C=\rho_{de}^{(0)}(w_0-c_a^2)$, where
$\rho_{de}^{(0)}$ is current density of dark energy.  Thus, we have
two values $w_0$ and $c_a^2$ defining the EoS parameter $w_{de}$ at
any redshift $z=a^{-1}-1$.

The effective EoS parameter $\tilde{w}_{de}$ is also analytical
function of scale factor
\begin{equation}
  \tilde w_{de}(a)=-1-\frac{\ln{\left(\!c_a^2-w_0+(1+w_0)a^{-3(1+c_a^2)}\!\right)}-\ln{\left(c_a^2+1\right)}}{3\ln{a}}\label{tilde_w_bar}.
\end{equation}
The differential equation (\ref{rho'}) with $w_{de}$ from
(\ref{w_bar}) has the analytic solu\-tion~too:
\begin{equation}
  \rho_{de}=\rho_{de}^{(0)}\frac{(1+w_0)a^{-3(1+c_a^2)}+c_a^2-w_0}{1+c_a^2}.\label{rho_bar}
\end{equation}
The expressions (\ref{beos}) and (\ref{rho_bar}) can be used for
finding of the allowable values of $c_a^2$. Really, if $c_a^2>0$
then the energy density of scalar field increases with decreasing
$a$ faster then matter density. In the Universe with such scalar
field the
MD epoch, required for large scale structure formation, is absent. %
\index{large scale structure}%
The age of such Universe is lower than age of oldest stars of our
galaxy. Besides, at early epoch $\rho_{de}>\rho_m$ and $p_{de}>0$
that changes drastically the transfer function of matter density
perturbations. So, the range of values for $c_a^2>0$ must be
excluded from consideration. Therefore, the sound range of allowable
values of $c_a^2$ is $<0$ and $w_0<-1/3$. Other constraints for
$c_a^2$ and $w_0$ follow from analysis of dynamics of expansion of
the Universe, but their optimal values one can deduce from
comparison of computed predictions with all set of observational
data, which are mentioned in the Chapter~1.

The dynamical properties of such scalar field depend on the ratio
$c_a^2$ between $w_{0}$ as well as on whether they are $>-1$ or
$<-1$. If any of them equals $-1$, then
$w_{de}(a)=\tilde{w}_{de}(a)=-1$ and we have vacuum-like field
again.  In the case $w_0=c_a^2$ we have the well studied
$w_{de}=$~const model.

\textit {a) Both $c_a^2$ and $w_0$ $>-1$ (quintessential range)}.

\begin{figure}
\vskip1mm
\includegraphics[width=13cm]{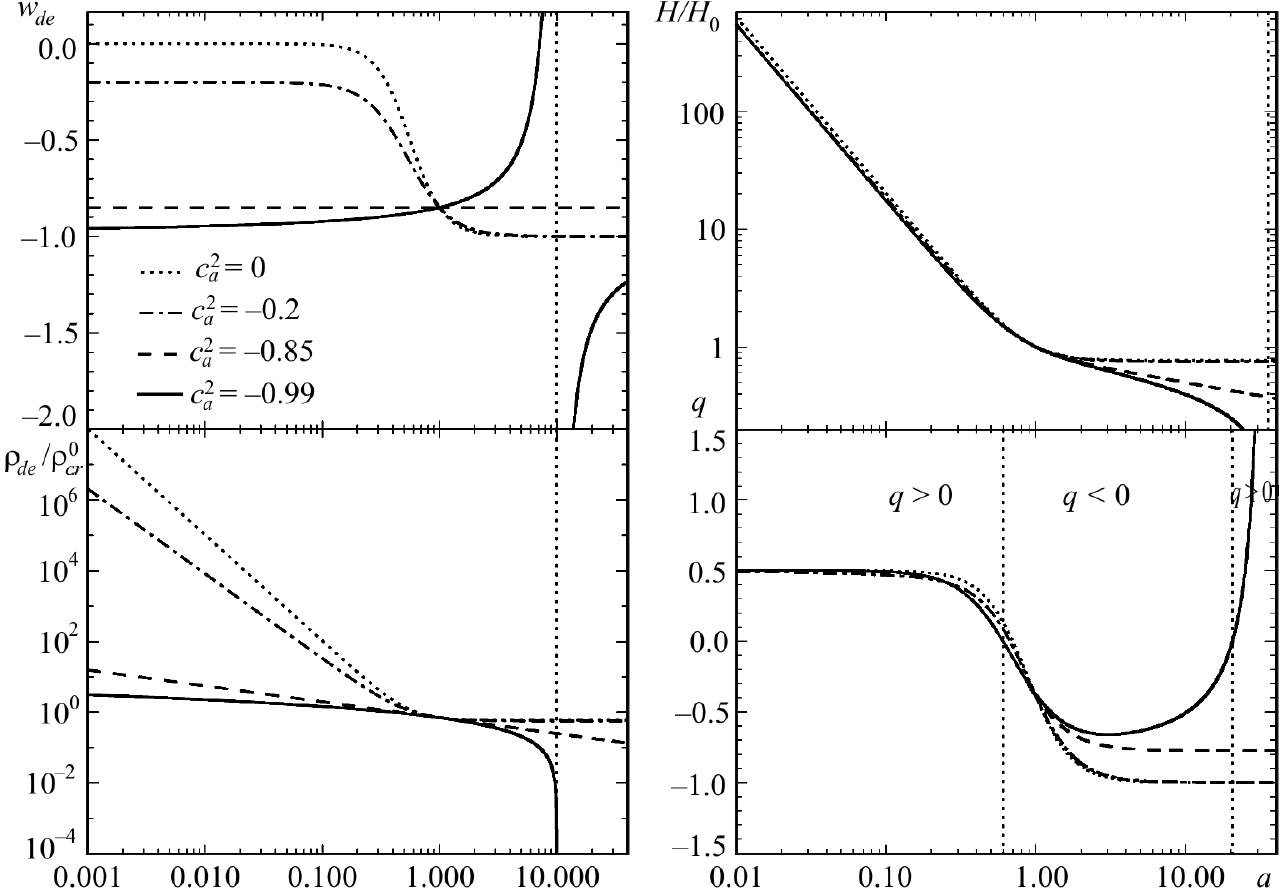}
    \vskip-2mm
  \caption{Left column: top panel~--- the dependences of EoS parameter
    on scale factor for barotropic quintessential scalar field %
    \index{quintessential scalar fields (QSF)} %
    with
    $w_0=-0.85$ and different $c_a^2$ (0, --0.2, --0.85, \mbox{--0.99}); bottom
    panel~--- the dependences of dark energy density (in the units of
    critical one at the current epoch) with EoS parameters presented
    in the top panel on scale factor.  Right column: The dynamics of
    expansion of the Universe with barotropic quintessential scalar
    field~--- $H(a)$ (top panel) and $q(a)$ (bottom one) for the same
    $w_{de}$ as in left panel}
  \label{wrode}
\end{figure}

The time dependences of barotropic EoS parameter for different
values\linebreak of $c_a^2>-1$ are shown in the left top panel of
Fig.~2.\ref{wrode}. As it follows from\linebreak (\ref{w_bar}),
$c_a^2$ corresponds to the EoS parameter at the beginning of
expansion,\linebreak $w_{de}(0)=c_a^2$.

The dependences of dark energy density on scale factor for the same
values of $c_a^2$ are shown in the left bottom panel of
Fig.~2.\ref{wrode}. The dynamics of expansion of the homogeneous
isotropic Universe, described by $H(a)$ and $q(a)$ (Friedmann
equations (\ref{H})---(\ref{q})), with the same scalar field models
is shown in the right panel of Fig.~2.\ref{wrode}.

One can see, that in the case $c_a^2>w_0$ $w_{de}(a)$ is monotonic
decreasing function in the whole range of scale factor variation,
$0<a<\infty$, while in the case $c_a^2<w_0$ $w_{de}(a)$ has
discontinuity of the second kind at \mbox{$a_{d2k}=[(1\,+$}
\mbox{$+\,w_0)/(w_0-c_a^2)]^{1/3(1+c_a^2)}>1$}, where $\rho_{de}$
becomes zero.  After that the energy density of such scalar field
acquires negative values and somewhat later $H$ reaches zero too and
the Universe will start to recollapse (right panel of
Fig.~2.\ref{wrode}). We note here that $\rho_{de}(a)$ and
$p_{de}(a)$ are smooth continuous functions \mbox{at any
$a$.}

\textit {b) Both $c_a^2$ and $w_0$ $<-1$ (phantom range)}.

As it follows from (\ref{w_bar}), in this case $c_a^2$ corresponds
to the EoS parameter at the scale factor infinity,
$w_{de}(a\rightarrow\infty)$. But when $a\rightarrow0$, then
$w_{de}\rightarrow-1$ and $\rho_{de}\rightarrow
\rho_{de}^{(0)}(c_a^2-w_0)/(1+c_a^2)$. The energy density increases
monotonically with increasing of $a$, that follows from
(\ref{rho_bar}). It is always positive for $c_a^2\le w_0$ and
sign-alternating in the case of $c_a^2>w_0$. In the last case the
energy density is negative at $a<a_{d2k}$ and positive at
$a>a_{d2k}$. When $\rho_{de}$ becomes zero at $a=a_{d2k}<1$ the
$w_{de}(a)$ has discontinuity of the second kind in past, but
$\rho_{de}(a)$ and $p_{de}(a)$ are smooth continuous functions.

\begin{figure}
\vskip1mm
\includegraphics[width=13cm]{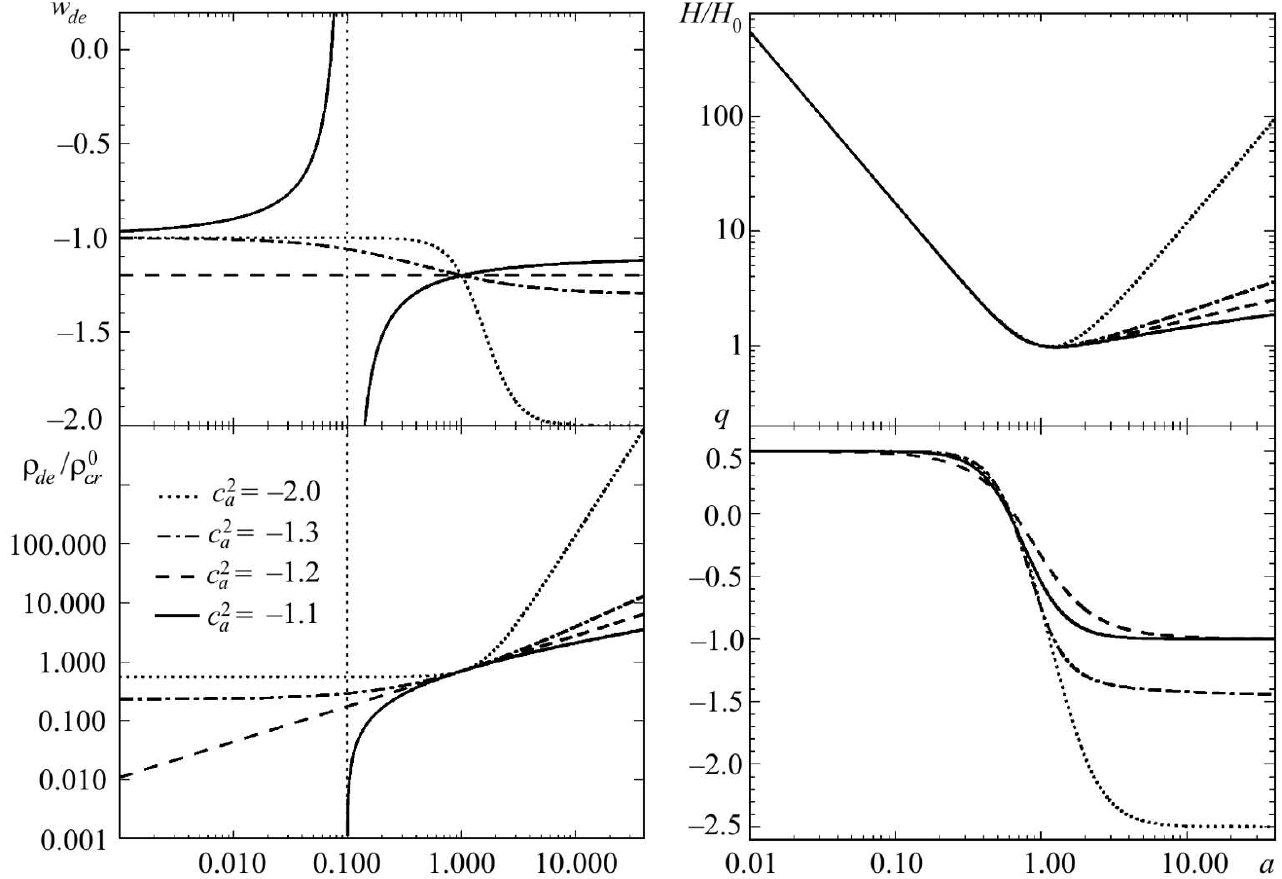}
    \vskip-2mm
  \caption{Left column: top panel~--- the dependences of EoS parameter
    on scale factor for barotropic phantom scalar field with
    $w_0=-1.2$ and different $c_a^2$ (--2.0, --1.3, --1.2, --1.1); bottom
    panel~--- the dependences of dark energy density (in the units of
    critical one at the current epoch) with EoS parameters presented
    in the top panel on scale factor.  Right column: The dynamics of
    expansion of the Universe with barotropic phantom scalar field~---
    $H(a)$ (top panel) and $q(a)$ (bottom one) for the same models
    as in left panel}
  \label{wrode_ph}
\end{figure}

The dependences of $w_{de}$, $\rho_{de}$, $H$ and $q$ on $a$ for
different $c_a^2$ and $w_0=-1.2$ are shown in Fig.~2.\ref{wrode_ph}.
One can see that in spite of the second kind discontinuity of
$w_{de}$ such scalar field practically does not influence the
dynamics of the Universe in the MD and RD epochs (the lines in the
right panels of Fig.~2.\ref{wrode_ph} are superimposed at $a<0.8$.)

\textit {c) Phantom divide crossing}.

\begin{figure}
\vskip1mm
\includegraphics[width=13cm]{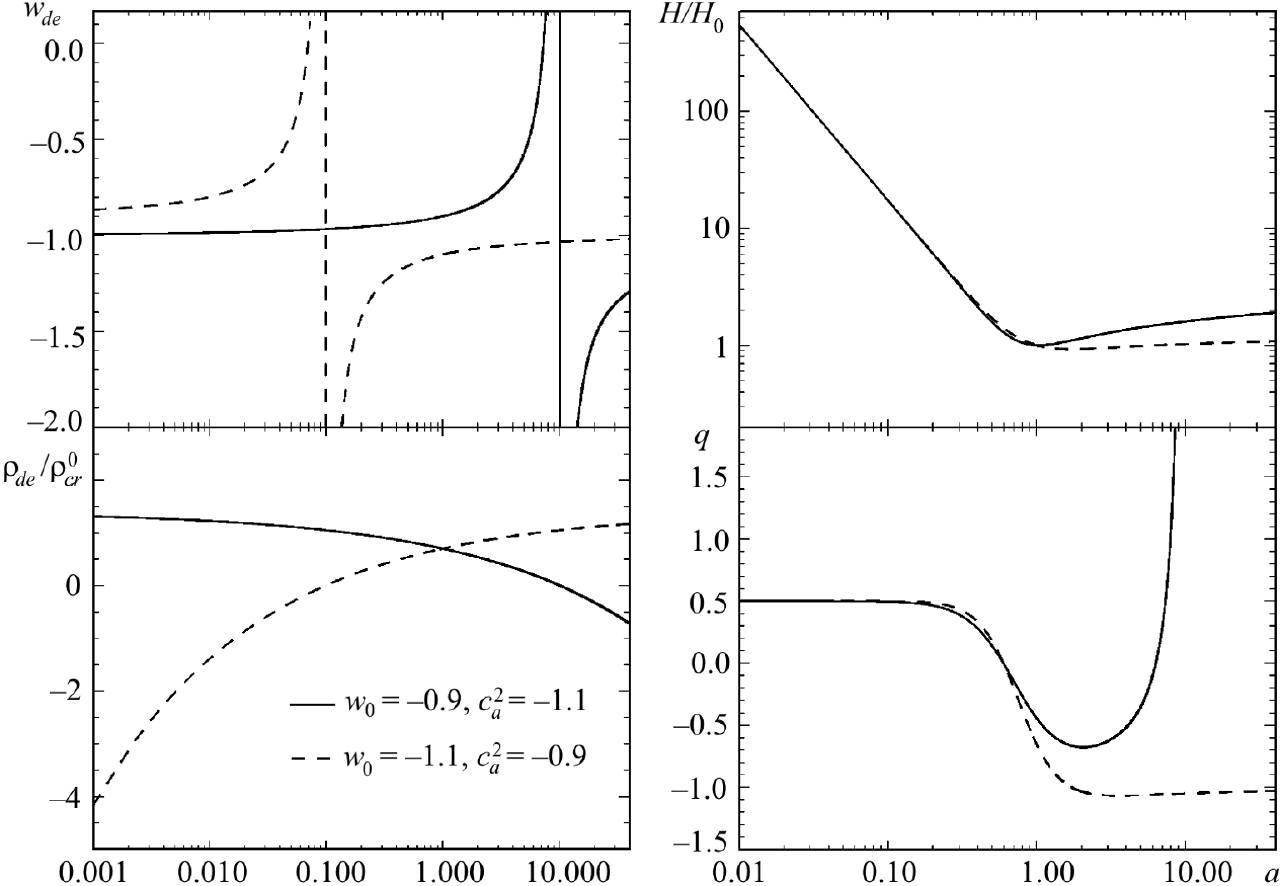}
    \vskip-2mm
  \caption{Left column: top panel~--- the dependences of EoS parameter
    on scale factor for barotropic quintessential scalar field with %
        \index{quintessential scalar field (QSF)} %
    phantom divide crossing EoS parameter ($w_0=$ $=-0.9$, $c_a^2=-1.1$
    and $w_0=-1.1$, $c_a^2=-0.9$); bottom panel~--- the dependences of
    dark energy density (in the units of critical one at the current
    epoch) with EoS parameters presented in the top panel on scale
    factor.  Right column: The dynamics of expansion of the Universe
    with barotropic quintessential scalar field~--- $H(a)$ (top panel)
    and $q(a)$ (bottom one) for the same $w_{de}$ as in left panel}
  \label{wrode_phcr}
\end{figure}

It happens when $c_a^2>-1$ and $w_0<-1$ or when $c_a^2<-1$ and
$w_0>-1$. In both cases the crossing of the line $w_{de}=-1$ passes
as discontinuity of the second kind of $w_{de}(a)$ at $a=a_{d2k}$,
which is in the past in the first case, and in the future in the
last one. In both cases the physical measurable values
$\rho_{de}(a)$ and $p_{de}(a)$ are smooth continuous functions, that
is shown in the left panel \mbox{of Fig.~2.\ref{wrode_phcr}.}

So, the scalar field with barotropic EoS is capable to describe the %
\index{phantom scalar field (PSF)}dif\-ferent possible dynamical
properties of dark energy (like vacuum energy, $w=$ =~const fluid,
quintessence, phantom, transition from quintessence to phantom and
vice-versa), which are defined by two parameters only, $w_0$ and
$c_a^2$. Their determination on the base of observations can unveil
the dynamical properties of dark energy in our Universe. But such
scalar field allows monotonic evolution of energy density which can
acquire negative values in the past or future. Its crossing over
zero leads to second kind discontinuity of $w_{de}(a)$. We have no
other arguments against this possibility besides that the null
energy condition $\sum_N \rho_{N}\ge 0$ must be satisfied always in
the past. This condition can be used for establishing of the limits
for values of $w_0$ for any $c_a^2$. Since it can be violated in the
late DE dominated epoch, the density of relativistic component can
be omitted. So, the null energy condition is satisfied when
\begin{equation}
 w_0\ge -\frac{1+(1+c_a^2)\frac{\Omega_m}{\Omega_{de}}a^{3c_a^2}+c_a^2a^{3(1+c_a^2)}}{1-a^{3(1+c_a^2)}}\label{w0_prior}.
\end{equation}

\begin{figure}
\vskip1mm
\centering
\includegraphics[width=10cm]{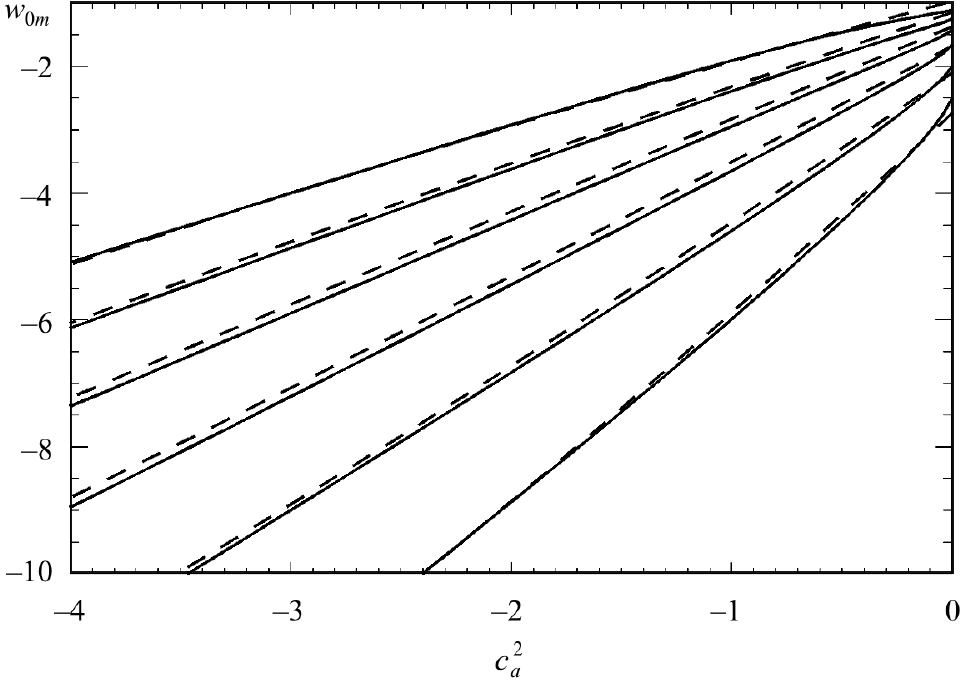}
    \vskip-2mm
  \caption{Minimal $w_0$ as the function of $c_a^2$ for different
    $\Omega_m/\Omega_{de}$ (0.1/0.9, 0.2/0.8, 0.3/0.7, 0.4/0.6,
    0.5/0.5, 0.6/0.4 from top to bottom). For values above the
    corresponding lines the null energy condition $\sum_N \rho_{N}\ge
    0$ is always satisfied in the past. Dashed line shows the analytic
    approximation (\ref{w0m})}
  \label{w0m_fig}
\end{figure}

The right part of inequality as function of $a$ goes to $-\infty$
when $a\rightarrow 0$ or $a\rightarrow 1$ and has maxima at
$0.5<a_m<1$ which we denote by $w_{0m}$. The dependences of $w_{0m}$
on $c_a^2$ for different $\Omega_m/\Omega_{de}$ are shown in
Fig.~2.\ref{w0m_fig}. One can state that the null energy condition
is satisfied in any epoch in the past when $w_0\ge w_{0m}$ for given
$c_a^2$ and $\Omega_m/\Omega_{de}$.  The dependences of $w_{0m}$ on
$c_a^2$ and $\Omega_m/\Omega_{de}$, shown in Fig.~2.\ref{w0m_fig} by
solid lines, can be approximated by simple expression
\[
  w_{0m}=-0.9103-1.272\frac{\Omega_m}{\Omega_{de}}+\left(\!0.7407+1.658\frac{\Omega_m}{\Omega_{de}}\!\right)c_a^2+
\]\vspace*{-3mm}
 \begin{equation}
  +\left(\!-0.03778+0.08091\frac{\Omega_m}{\Omega_{de}}\!\right)c_a^4, \label{w0m}
\end{equation}
the accuracy of which is few percents (dashed lines in
Fig.~2.\ref{w0m_fig}).

Therefore, defining of $w_{de}(a)$ on the whole time axis and
$\Omega_{de}$ at current epoch governs completely the dynamical
properties of homogeneous scalar \mbox{field} and the whole
Universe, gives possibility to describe its history, present and
future. But for analysis of gravitational instability of scalar
field and its
influence on the formation of large scale structure of the Universe %
\index{large scale structure}%
the effective sound speed as parameter of equations
(\ref{deltade-Vde}) must be defined too. %

\subsection{\!The effective sound speed\label{ch2-subsec45}}

\hspace*{3cm}\index{sound speed}If Lagrangian of scalar field is
defined then effective sound speed can be calculated from
(\ref{c_s2}). In the opposite case it can be specified apart. First
of all we must find the allowable range of its values. Analysis of
equations for evolution of scalar field density and velocity
perturbations \linebreak
 shows that $c_s^2$ must be positive or zero,
since in the opposite case the sca\-lar field is strongly
gravitationally unstable and can essentially change the \linebreak
transfer function and power spectrum of matter density perturbations
and, even, the angular power spectrum of CMB temperature
fluctuations. \index{cosmic microwave background (CMB)}\index{CMB
temperature fluctuations}On the other hand, $c_s^2$ cannot exceed
$1$ to retain causality. So, the range of allowable \mbox{values of
$c_s^2$ is $[0,\,1]$.}

\begin{figure}
 \vskip1mm
\includegraphics[width=13cm]{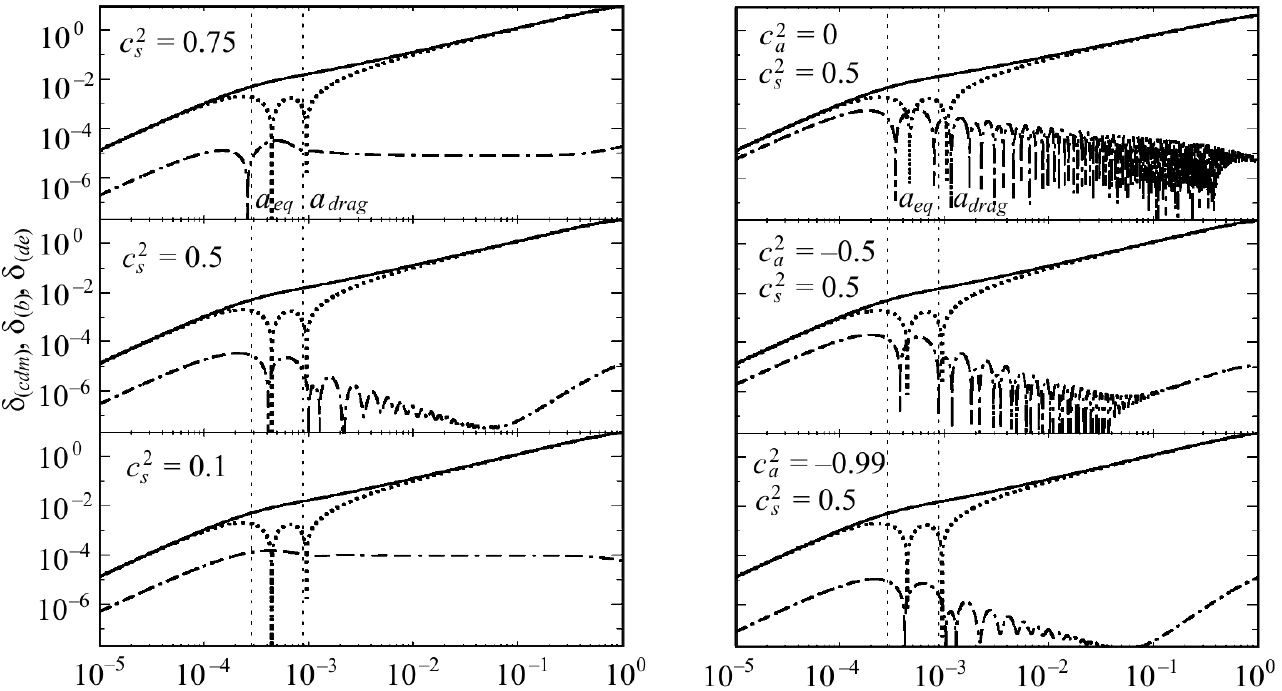}
    \vskip-2mm
  \caption{Evolution of Fourier amplitude ($k=0.05$ Mpc$^{-1}$) of
    density perturbations of cold dark matter (solid line), baryonic
    matter (dotted line) and dark energy (dash-dotted line) computed
    by CAMB for models with constant effective sound speed $c_s^2$.
    Left panel: models with constant EoS parameter ($w_{de}=-0.9$) and
    different $c_s^2$ $(0.75, 0.5, 0.1)$ from top to bottom); right
    panel: models with $w_{de}$ variable according to (\ref{w_bar})
    with $w_0=-0.9$ and $c_a^2=0, -0.5, -0.99$ (from top to
    bottom)}
  \label{ddeb_wa_cs2c}\vspace*{-2mm}
\end{figure}

In most of papers the authors assume some value for $c_s^2$
supposing that it is constant. In Fig.~2.\ref{ddeb_wa_cs2c} the
evolution of Fourier amplitude ($k=0.05$~Mpc$^{-1}$) of linear
density perturbations of dark energy with different values of
constant effective sound speed $c_s^2$ and either constant (left
panel) or variable (right panel) EoS parameter is shown. It is
computed by CAMB for multicomponent Universe in the synchronous
gauge comoving to cold dark matter component. The corresponding
amplitudes for dark matter and baryonic components are presented
there for comparison. A few conclusions can be deduced from their
analysis: a) evolution of energy density perturbations of scalar
field depends on value of effective sound speed; b) the amplitude
increases when scale of perturbation is larger than acoustic horizon
scale ($k^{-1}>c_st$) and decays when it becomes smaller
($k^{-1}<c_st$); c) practically for any $0<c_s^2\le1$ at current
epoch the amplitude of energy density perturbations of scalar field
is essentially lower than amplitudes of dark matter and baryonic
components; d) the value of EoS parameter as well as the character
of its time variation changes the evolution of density (right panel
of Fig.~2.\ref{ddeb_wa_cs2c}) too: for lower initial value of
$w_{de}$~--- lower initial amplitude of scalar field density
perturbations.

The time-variable effective sound speed was considered too. For
example, the authors of \cite{Haq2011} have proposed and analyzed
the simple analytic dependence in form
\begin{equation}
  c_s^2(a)=c_0+c_1\left(\!\frac{a}{1+a}\!\right)^{\!\gamma}\!, \label{cs2a}
\end{equation}
which comprehends infinite number of monotonic evolution tracks
between 0 to 1, defined by constants $c_0$, $c_1$ and $\gamma$. The
first of them equals $c_s^2$ at the Beginning, $a=0$, the second one
is asymptotic value $c_s^2\rightarrow c_1$ when $a\rightarrow\infty$
and $c_0=0$. The power-low index $\gamma$ governs the rate of
change. The increasing

\begin{wrapfigure}{r}{6.8cm}
\hspace*{2mm}\includegraphics[width=6.5cm]{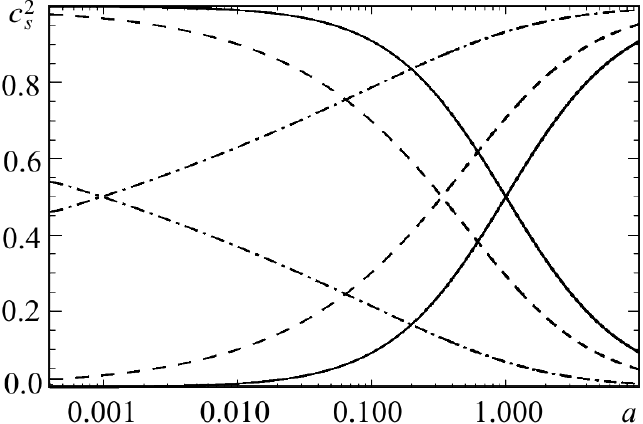}\vspace*{-2mm}\\
\hspace*{2mm}\raisebox{0.2cm}{\parbox[b]{6.5cm}{\caption{Variable\,effective\,sound\,speed\,$c_s^2(a)$
defined by
    (\ref{cs2a}): increasing curves are cal\-culated for $c_0=0$,
    $c_1=1$, $\gamma=1$ (solid li\-ne), 0.5 (dashed line), 0.1
    (dash-dotted line) and decreasing ones for $c_0=1$, $c_1=-1$ and
    the same $\gamma$'s\label{cs2a_fig}}}}\vskip-4mm
\end{wrapfigure}
\begin{figure}[b!]
\vskip-2mm
\includegraphics[width=13cm]{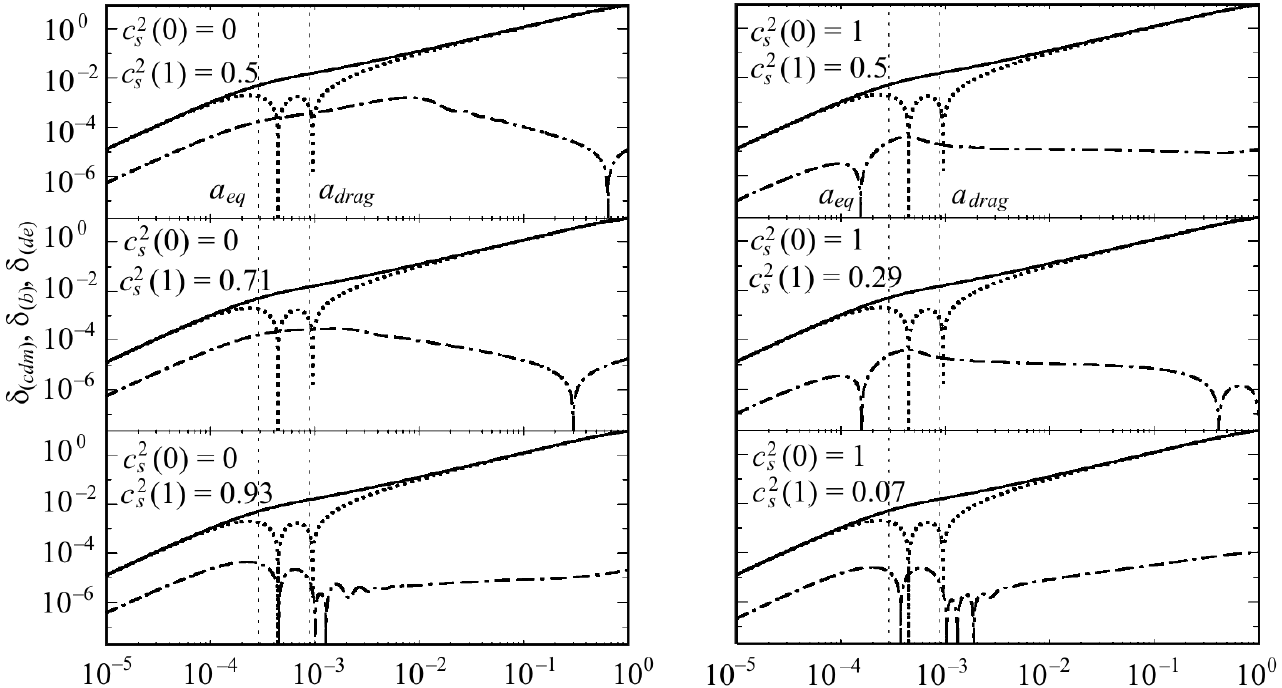}
    \vskip-2mm
  \caption{Evolution of Fourier amplitude ($k=0.05$ Mpc$^{-1}$) of
    density perturbations of cold dark matter (solid line), baryonic
    matter (dotted line) and dark energy (dash-dotted line) for models
    with constant EoS parameter ($w_{de}=-0.9$) and variable effective
    sound speed (\ref{cs2a}).  Left panel: models with increasing
    $c_s^2$ ($c_0=0$, $c_1=1$; $\gamma=1,0.5,0.1$ from top to
    bottom); right panel: models with decreasing $c_s^2$ ($c_0=1$,
    $c_1=-1$; $\gamma=1,0.5,0.1$ from top to bottom)}
  \label{ddeb_w09_cs2z}
\end{figure}


\noindent from 0 to 1 and decreasing from 1 to 0 effective sound
speed for three values of $\gamma=1, 0.5, 0.1$ is shown in
Fig.~2.\ref{cs2a_fig}.  The evolution of density perturbations of
scalar field \mbox{with} these effective sound speeds and
$w_{de}={\rm const}=-0.9$ is presented in Fig.~2.\ref{ddeb_w09_cs2z}
(left panel for increa\-sing, right panel for decreasing $c_s^2$'s).
One can see, that increasing or decreasing of $c_s^2$ as well as
their rates influence the time evolution of energy density
perturbations of scalar fields.

\index{density perturbations} %
In all cases considered here the amplitudes of subhorizon density
perturbations of scalar fields are essentially lower than
corresponding amplitudes of dark matter and baryonic components,
that is caused by decaying of perturbations of scalar fields after
entering into horizon. At super- and near-horizon scales they are
comparable (it depends also on initial conditions for density
perturbations in each component) and their gravitational interaction
can leave
appreciable fingerprints in initial power spectrum of matter density %
\index{power spectrum}%
perturbations. The impact of scalar field perturbations on the
linear power spectrum of matter density ones is expected to be
essentially lower than the growth factor caused by the background
dynamics, but it is scale-dependent and can be appreciable for some
types of scalar fields. Some of them are studied carefully in
\cite{Putter2010,Haq2011}.

At the end of this subsection we would like to note that specifying
of scalar field by defining of $w_{de}$ and $c_s^2$ determines
completely its dynamical and perturbative properties as of energy
component of the Universe but gives no possibility to reconstruct
the Lagrangian, potential and field variable unambiguously, that
follows from Eq.~(\ref{w_L}) and (\ref{c_s2}). So, we can say
\mbox{nothing} about its physical nature and field properties. That
is why we prefer the combined approach to specifying of scalar
field~--- its general property and form of Lagrangian. Below we
consider the scalar fields with generalized barotropic equations of
state and classical, phantom and tachyon Lagrangians and discuss
their properties, possibility to match all set of observational data
as well as possibility of distinguishing between them.
\index{potential $U(\phi )$|)} %

\section[\!Quintessential scalar fields with barotropic EoS]{\!Quintessential scalar
fields\\ \hspace*{-0.95cm}with barotropic EoS}

\hspace*{3cm}\index{quintessential scalar field (QSF)|(}In this
section we consider the scalar field models of dark energy specified
by the barotropic EoS (\ref{beos})---(\ref{rho_bar}) and different
types of Lagrangians to analyze the evolution of fields,
their perturbative properties and influence on the power spectrum of %
\index{power spectrum}%
matter density perturbations. We will determine their parameters
jointly with minimal set of cosmological parameters using current
observational data and discuss the possibility of
\mbox{distinguishing} between different types of scalar field models
of dark energy. We suppose that the Universe is filled with
non-relativistic particles (cold dark matter and baryons),
relativistic ones (thermal electromagnetic radiation and massless
neutrino) and minimally coupled scalar field as dark energy.

\subsection{\!Classical scalar field}

\hspace*{3cm}\index{classical scalar field}The  scalar field with canonical (classical or Klein---\linebreak Gordon) Lagrangian %
  \index{canonical Lagrangian} %
\begin{equation}
  L_{clas}=X-U(\phi),\label{L_clas}
\end{equation}
and positive values of kinetic term $X$ and potential $U(\phi)$ is
called the classical one.  At any time its energy density and
pressure are simply linear combinations of $X$ and $U$,
\begin{equation}
  \rho_{clas}=X+{U}(\phi), \quad  P_{clas}=X-{U}(\phi), \label{rho_XU_clas}
\end{equation} %
and EoS parameter $w_{de}\equiv p_{de}/\rho_{de}$ for such field is
the ratio of these com\-binations,
\begin{equation}
  w_{clas}=\frac{X-U}{X+U}.\label{w_XU_clas}
\end{equation}
One can see, that for positive values of $X$ and $U$ the EoS
parameter always is $\ge$$-1$.  For explanation of accelerated
expansion of the Universe at the current epoch ($q_0<0$) it must
satisfy two conditions:
\begin{equation}
  {\rm a)}\,\, X^{(0)}<U^{(0)}/2, \quad  {\rm b)}\,\, U^{(0)}-2X^{(0)}>\rho_m^{(0)}/2,
\end{equation}
where index ${(0)}$ marks the current values of corresponding
variables.

Using relations (\ref{rho_XU_clas})---(\ref{w_XU_clas}) and
(\ref{w_bar})---(\ref{rho_bar}) the field variable, potential and
kinetic term can be presented in terms of density and EoS parameters
as follows:
\index{EoS parameter|)} %
\begin{equation}%
\label{U_clas}
\begin{array}{c}
 \displaystyle \phi(a)-\phi_0=\pm\sqrt{(1+w_0)\rho_{de}^{(0)}}\int\limits_1^a\frac{da'}{a'^{(\frac{5}{2}+\frac{3}{2}c_a^2)}H(a')},
 \\[7mm]
  \displaystyle U(a)=\frac{(1-c_a^2)(1+w_0)a^{-3(1+c_a^2)}+2(c_a^2-w_0)}{2(1+c_a^2)}\rho_{de}^{(0)},
  \\[5mm]
  \displaystyle X(a)=\frac{1+w_0}{2}a^{-3(1+c_a^2)}\rho_{de}^{(0)}.
\end{array}
\end{equation}
One can see that for cosmological model of real Universe ($H(a)>0$,
$\rho_{de}^{(0)}>0$) the quintessential barotropic scalar field
($w_0>-1$, $c_a^2>-1$) has always real values of field variable and
potential. Its kinetic term $X(a)$ is positive for any $a$, the
potential $U(a)$ is positive for any $c_a^2<1$ at $a\le1$. But its
sign in the future ($a\gg1$) depends on relation of values of
$c_a^2$ and $w_0$. Indeed, when $a\rightarrow\infty$ then
$\displaystyle U\rightarrow\frac{c_a^2-w_0}{1+c_a^2}\rho_{de}^{(0)}$
and is positive for $c_a^2>w_0$ and negative for $c_a^2<w_0$.

\index{expansion dynamics}The dynamics of expansion of the Universe
at late epoch and in the
future depends on the density and EoS parameters of scalar field. %
\index{EoS parameter|(} %
It is shown in Fig.~2.\ref{wrode}. The behavior of field and
dynamics of the Universe expansion can be divided into three types,
defined by the relation between adiabatic sound speed and EoS
parameter or the sign of derivative of EoS parameter with respect to
\mbox{scale factor.}

\begin{figure}
  \vskip1mm
\includegraphics[width=13cm]{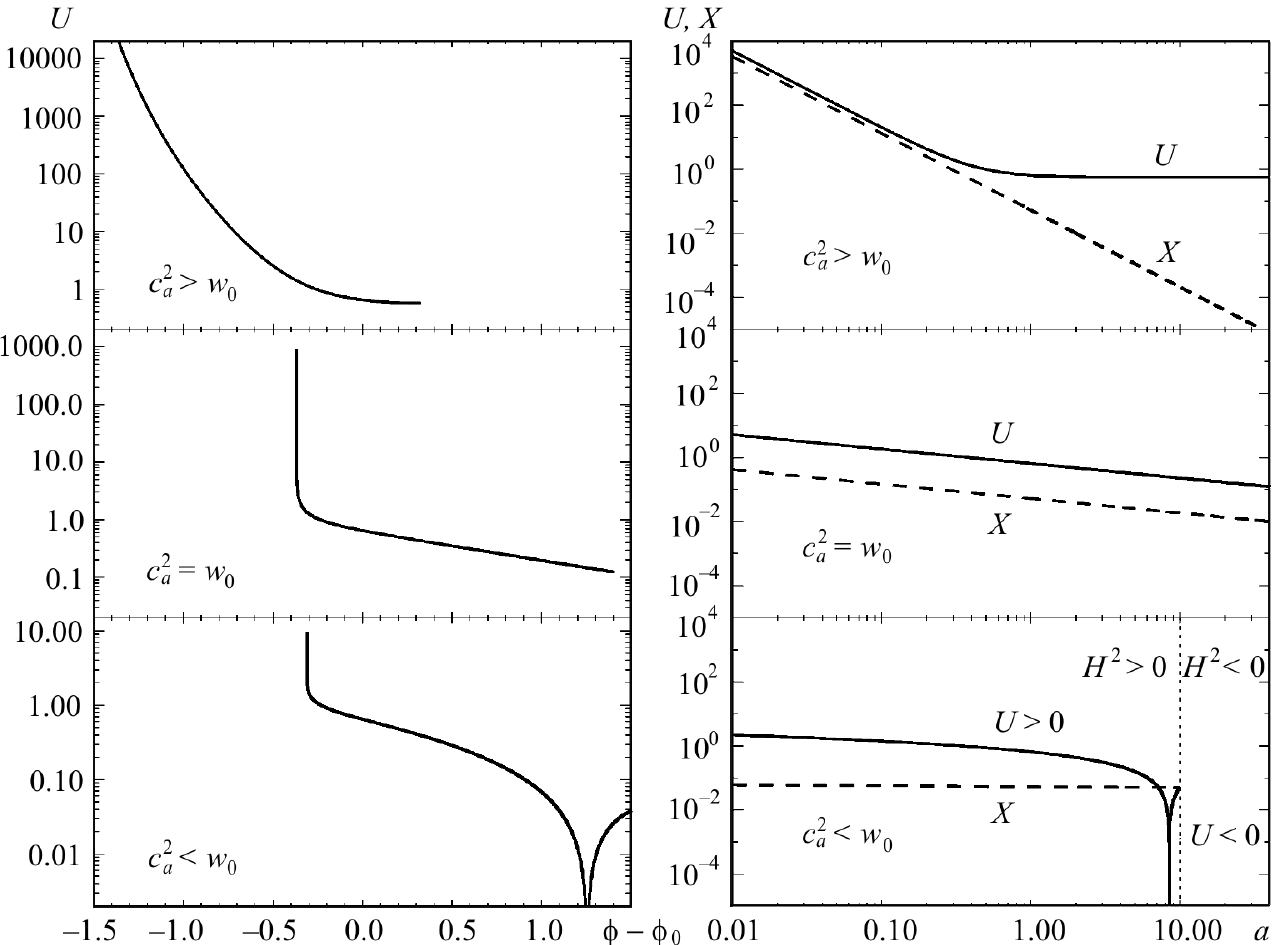}
    \vskip-2mm
  \caption{Potentials $U(\phi-\phi_0)$ (left) and dependences of
    potentials and kinetic terms on scale factor $a$ (right) for
    classical scalar field with barotropic EoS. %
    \index{classical scalar field} %
    In the top panels $c_a^2=-0.2$, in the middle ones $c_a^2=-0.85$
    and in the bottom ones $c_a^2=-0.99$, in all panels
    $w_0=-0.85$. Potential and kinetic term are in the units of
    current critical energy density, $3c^2H_0^2/8\pi G$, the field
    variable in units of $\sqrt{3c^2/8\pi G}$. The current epoch in
    the left panels corresponds to $\phi-\phi_0=0$ and the field
    evolves from left to right}\vspace*{-1mm}
  \label{U_X}
\end{figure}

1) $w'<0$ ($c_a^2>w_0$): As it follows from (\ref{w_bar}), in this
case $w_{de}$ decreases monotonically from $c_a^2$ at the early
epoch to $w_0$ at current one up to --1 at the infinite time. The
constant $C$ in EoS equation (\ref{beos}) is negative. The dark
energy density and pressure tend asymptotically to
$\rho_{de}^{(\infty)}=\rho_{de}^{(0)}(c_a^2-w_0)/(1+c_a^2)$ and
$p_{de}^{(\infty)}=-\rho_{de}^{(\infty)}$. Therefore, in this case
the scalar field rolls down to the minimum of potential (see
Fig.~2.\ref{U_X}) and in far future the Universe will proceed into
de Sitter stage of its expansion with $w^{(\infty)}_{de}=-1$,
$q^{(\infty)}=-1$ and $H^{(\infty)}=$
$=\sqrt{\Omega_{de}(c_a^2-w_0)/(c_a^2+1)}H_0$.  So, the scalar field
of such type has the following general properties (see relations
(\ref{w_XU_clas})---(\ref{U_clas}) and top panels of
Fig.~2.\ref{U_X}): a)~its kinetic term and potential are always real
positive; b)~kinetic term is always lower than potential ($X<U$);
c)~the potential rolls down to minimum ($U_{min}>0$) at the finite
value of the field variable $\phi_{min}$ (left top panel), which is
reached at time infinity (right top panel); d) the kinetic term of
such field tends asymptotically to 0, it means that
$\dot{\phi}\rightarrow 0$ and the field ``freezes''.

2) {$w'=0$ ($c_a^2=w_0$):} It corresponds to the well-studied case
$w_{de}=$~const. In this case $C=0$ and we have usual barotropic EoS
$p_{de}=w_0\rho_{de}$, $\rho_{de}\rightarrow 0$ when $a\rightarrow
\infty$. So, the Universe in the future will experience the
power-law expansion with $a\propto t^{2/3(1+w_0)}$ and deceleration
parameter $q\rightarrow (1+3w_0)/2$. The scalar field of such type
has the following general properties (see relations
(\ref{w_XU_clas})---(\ref{U_clas}) and middle panels of
Fig.~2.\ref{U_X}): a)~its kinetic term and potential have always
real positive values; b)~the ratio of potential to kinetic term is
constant and large than unity, $U/X={\rm const}>1$; c)~its potential
rolls down to minimum $U_{min}=0$ at some finite value of the field
variable and time infinity; d)~the kinetic term of such field tends
asymptotically to 0, it means that $\dot{\phi}\rightarrow 0$ and the
field ``freezes''.

3) {$w'>0$ ($c_a^2<w_0$):} The EoS parameter $w_{de}$ increases %
\index{EoS parameter|)} %
monotonically from $c_a^2$ at the early epoch to $w_0$ at the
current one and still continues to increase after that. The field
will satisfy the strong energy condition $\rho_{de}+3p_{de}\ge 0$
($w_{de}>-1/3$) starting from\vspace*{-5mm}
\begin{equation*}
a_{(w=-\frac{1}{3})}=
\left[\frac{(1+3c_a^2)(1+w_0)}{2(c_a^2-w_0)}\right]^{\frac{1}{3(1+c_a^2)}}
\end{equation*}
and then accelerated expansion of the Universe will be changed by
the decelerated one. The EoS parameter will reach 0 in future
at\vspace*{-1mm}
\begin{equation*}
  a_{(w=0)}=\left[\frac{c_a^2(1+w_0)}{c_a^2-w_0}\right]^{\frac{1}{3(1+c_a^2)}}\!,
\end{equation*}\vspace*{-5mm}

\noindent and 1 at\vspace*{-3mm}
\begin{equation}
  a_{(w=1)}=\left[\frac{(1-c_a^2)(1+w_0)}{2(w_0-c_a^2)}\right]^{\frac{1}{3(1+c_a^2)}}\!,
\end{equation}
when potential becomes zero (relation (\ref{U_clas}) and left bottom
panel of Fig.~2\ref{U_X}). The scalar field energy density at these
$a$ is positive: $\rho_{de}(a_{(w=0)})=\rho_{de}^{(0)}(c_a^2-$
$-w_0)/c_a^2$ and
$\rho_{de}(a_{(w=1)})=\rho_{de}^{(0)}(c_a^2-w_0)/(c_a^2-1)$
correspondingly.  The energy density of scalar field continues
decreasing, reaches 0 at
$$a_{(\rho=0)}=\left[\frac{1+w_0}{w_0-c_a^2}\right]^{\frac{1}{3(1+c_a^2)}}$$
and then becomes negative. The EoS parameter at this moment has
discontinuity of the second kind (Fig.~2.\ref{wrode}). Soon after
that, when $\rho_m+\rho_{de}$ reaches 0, the expansion of the
Universe is
 changed by the contraction since at this moment $\dot{a}=0$,
$\ddot{a}<0$, as it follows from equations (\ref{H}) and (\ref{q}),
which have no solution for larger $a$. Such behavior can be
corrected only slightly by the cur\-va\-tu\-re parameter from the
observationally allowable range.

\subsection{\!Tachyonic scalar field}

\hspace*{3cm}\index{tachyonic scalar field|(}The scalar field $\xi$
with Dirac---Born---Infeld Lagrangian
\index{Dirac---Born---Infeld Lagrangian} %
\begin{equation}
  L_{tach}=-\tilde{U}(\xi)\sqrt{1-2\tilde{X}},\label{L_tach}
\end{equation}
and positive values of kinetic term $0\le\tilde{X}\le1/2$ and
potential $\tilde{U}(\xi)$ can be another good candidate for
quintessential dark energy \cite{Gibbons2002,Padmanabhan2002}. Such
field is often called in the literature tachyonic one.  At any time
the energy density and pressure are functions of $\tilde{X}$ and
$\tilde{U}$:\vspace*{-1mm}
\begin{equation}
  \rho_{tach}=\frac{\tilde{U}(\xi)}{\sqrt{1-2\tilde{X}}}, \quad P_{tach}=-\tilde{U}(\xi)\sqrt{1-2\tilde{X}}. \label{rho_XU_tach}
\end{equation}\vspace*{-4mm}

\begin{figure}
\includegraphics[width=13cm]{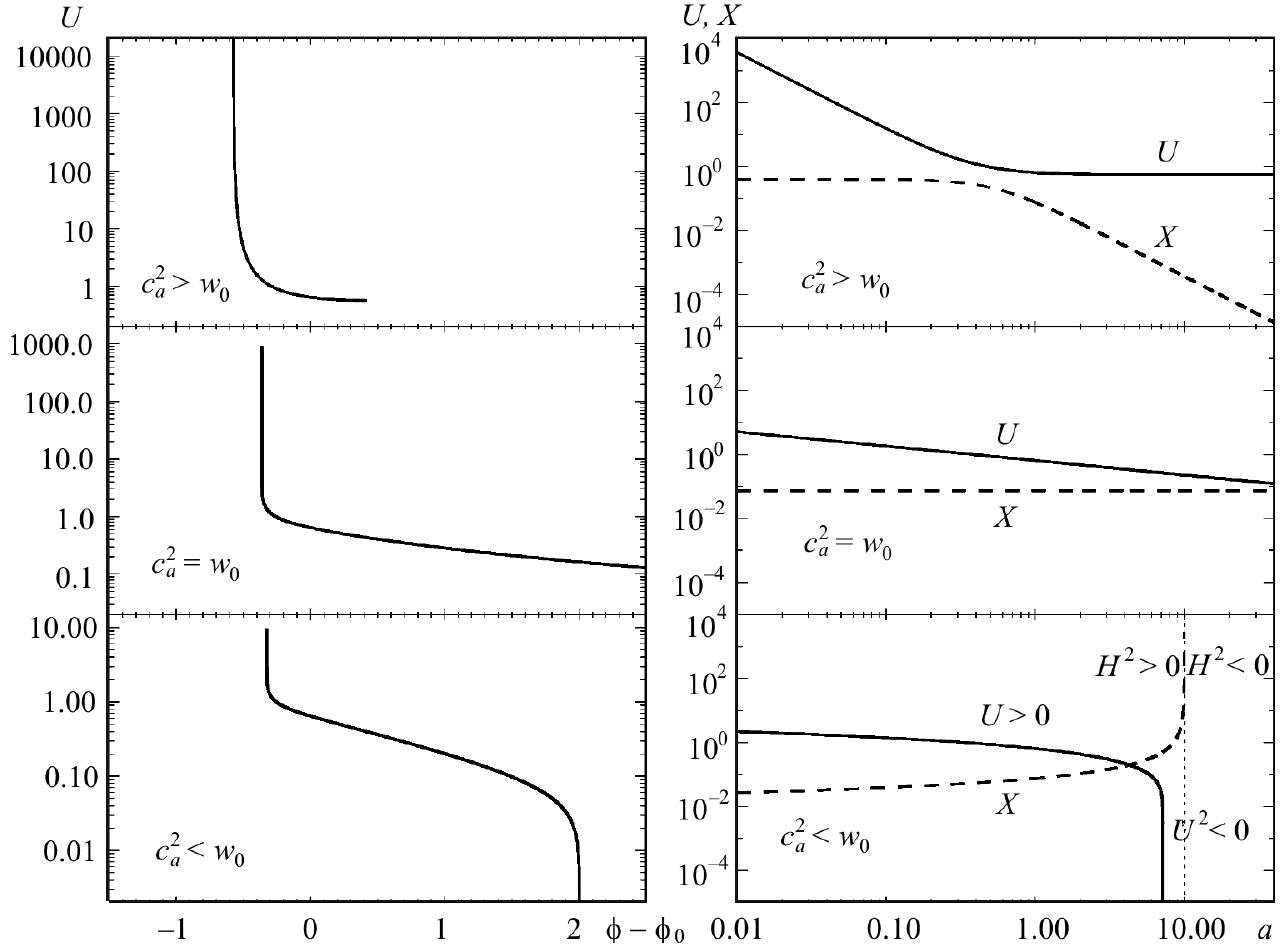}
    \vskip-2mm
  \caption{Potentials $U(\phi-\phi_0)$ (left) and dependences of
    potentials and kinetic terms on scale factor $a$ (right) for
    tachyonic scalar field with barotropic EoS. Dark energy parameters
    and units of variables are the same as in Fig.~2.\ref{U_X}}
  \label{U_X_tach}\vspace*{-1mm}
\end{figure}

\index{EoS parameter} %
The EoS parameter $w_{de}\equiv p_{de}/\rho_{de}$ for this field is
following:\vspace*{-1mm}
\begin{equation}
  w_{tach}=2\tilde{X}-1.\label{w_XU_tach}
\end{equation}\vspace*{-6mm}

One can see, that in the case of tachyonic field the EoS parameter
is always $\ge-1$ for positive values of $\tilde{X}$ independently
on value and sign of $\tilde{U}$.  For explanation of accelerated
expansion of the Universe at the current epoch ($q_0<0$) it must
satisfy two conditions:\vspace*{-4mm}
\begin{equation}
  \text{ a)}\,\, \tilde{X}^{(0)}<1/3, \quad  {\rm b)}\,\, \tilde{U}^{(0)}\frac{1-3\tilde{X}^{(0)}}{\sqrt{1-2\tilde{X}^{(0)}}}>\rho_m^{(0)}/2,
\end{equation}\vspace*{-4mm}

\noindent where index ${(0)}$ marks the current values of
corresponding variables.

The tachyonic field variable, potential and kinetic term can be
presented in terms of dark energy density (\ref{rho_bar}) and EoS
parameter (\ref{w_bar}) as follows:
\begin{equation}
\label{U_tach}
\begin{array}{c}
  \displaystyle\xi(a)-\xi_0
  =\pm\int\limits_1^a\frac{da'\sqrt{1+w_{de}(a')}}{a'H(a')},\\[9mm]
  \displaystyle \tilde{U}(a)  =\rho_{de}(a)\sqrt{-w_{de}(a)}, \\[5mm]
  \displaystyle \tilde{X}(a)  =\frac{1+w_{de}(a)}{2}.
\end{array}
\end{equation}

The potentials $\tilde{U}(\xi-\xi_0)$, evolution of potentials and
kinetic terms for models with decreasing, constant and increasing
EoS parameter are shown in Fig.~2.\ref{U_X_tach}. One can see that
accelerated expansion of the Universe is caused\linebreak by rolling
down of field to minimum of its potential quite similarly as in the
case of classical field. Meanwhile, for the same time dependence of
$\rho_{de}$ and $w_{de}$ (or $p_{de}$) the evolution of $\tilde{U}$
and $\tilde{X}$ for tachyonic field differs essentially from
corresponding evolution for classical one, that follows from
comparison of expressions (\ref{U_clas}) and (\ref{U_tach}) or
Fig.~2.\ref{U_X} and 2.\ref{U_X_tach}.  Moreover, in the case of
increasing EoS parameter the potential of tachyonic field at
$a_{w=0}$ becomes imaginary, while one of classical field is real
 always. Therefore, the same dynamics of expansion of the
homogeneous Universe can be provided by different homogeneous scalar
fields, classical and tachyonic fields are the example of such model
degeneracy. But in the case of these two fields it can be partially
broken if cosmological perturbations are taken into account. That
will be shown below. %
\index{tachyonic scalar field|)} %

\subsection{\!Quintessential scalar fields\\ \hspace*{-1.15cm}in the phase plane}

\begin{figure}
\vskip1mm
\includegraphics[width=13cm]{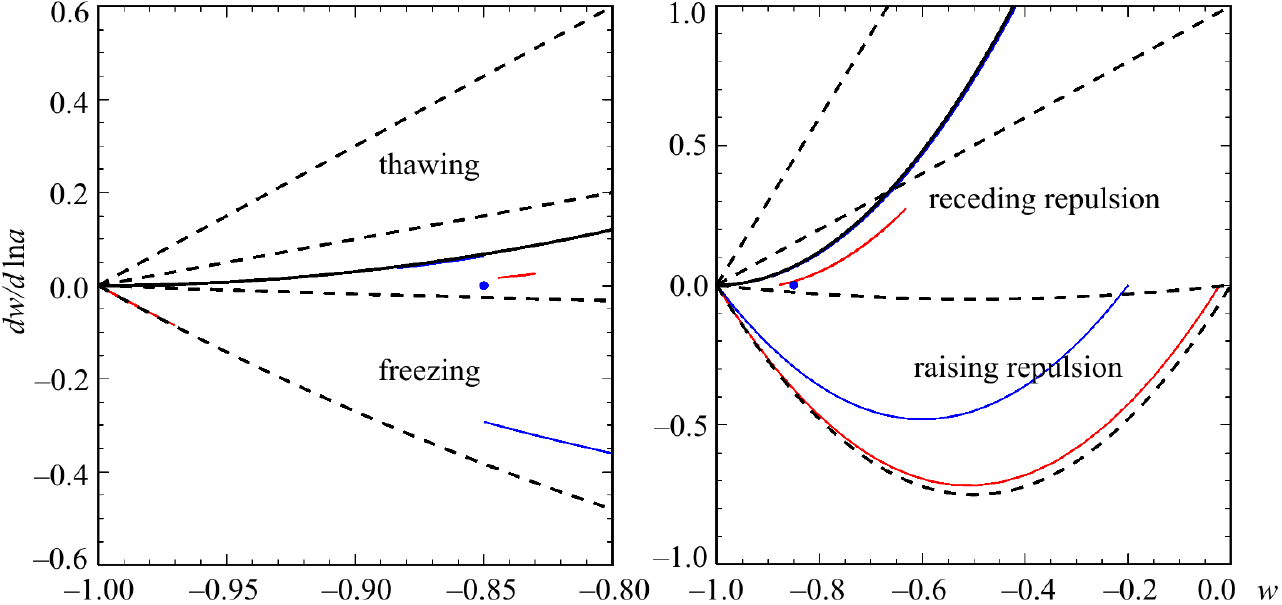}
    \vskip-2mm
  \caption{The $w_{de}-dw_{de}/d\ln{a}$ phase plane for quintessential
    scalar fields with barotropic EoS as models of dynamical dark
    energy (solid lines). If $dw_{de}/d\ln{a}<0$ the fields evolve
    from right to left raising their repulsion properties, if
    $dw_{de}/d\ln{a}>0$ the fields evolve from left to right receding
    them. Thick dashed lines show the ranges occupied by the thawing
    and freezing scalar fields deduced by \cite{Caldwell2005} from the
    analysis of simplest particle physics scalar field models of
    dynamical dark energy. In the left panel the phase plane evolution
    tracks of the scalar fields with barotropic EoS are shown in the
    range $0.5\le a\le 1$ ($0\le z\le 1$) and in the same scale as in
    Fig.~2.\ref{Caldwell_Linder} for easy comparison. In the right panel
    the phase plane evolution tracks of the scalar fields with
    barotropic EoS correspond to the range $0.0001\le a\le 10$
    ($-0.9\le z\le 10000$). Thick solid black lines show the limits
    for such scalar field models: the upper line corresponds to
    $c_a^2=-1$, the lower one to $c_a^2=0$ (superimposed with the
    lower limit for freezing scalar fields from
    \cite{Caldwell2005}). The blue solid lines and dot show the phase
    tracks of models shown in Figs.~2.\ref{U_X} and 2.\ref{U_X_tach}, the
    red solid lines show the phase tracks of the best fitting models
    $\mathbf{q}_1$ and $\mathbf{q}_2$ from Table 2.\ref{tab_qsf}}
  \label{w`w}
\end{figure}

\hspace*{3cm}The general properties of the quintessential scalar
field models of dark energy with barotropic EoS can be deduced also
from the analysis of their occupation of the $w_{de}-dw_{de}/d\ln a$
phase plane. From (\ref{w'}) and the constraint $-1\le c_a^2\le 0$
follows that the scalar field models of dark energy with
$c_a^2=$~const occupy the $w_{de}-dw_{de}/d\ln a$ region limited by
the lines $dw_{de}/d\ln a=3(1+w_{de})^2$ and $dw_{de}/d\ln
a=3w_{de}(1+w_{de})$ (Fig.~2.\ref{w`w}). The last one coincides with
the lower limit for freezing scalar field models of dark energy
deduced by \cite{Caldwell2005} from the analysis of the simplest
particle-physics models of cosmological scalar fields. Below it the
scalar fields have too large density at the early epoch that
contradicts the data on CMB
anisotropy. %
\index{cosmic microwave background (CMB)} %
\index{CMB anisotropy} %
Above the upper limit there is a range of fields that started as
phantom ones, which is excluded for fields with classical Lagrangian
as well as tachyonic one considered above. The scalar fields which
are in the phase plane between the lines $dw_{de}/d\ln a=0$ and
$dw_{de}/d\ln a=3w_{de}(1+w_{de})$ evolve from right to left in
Fig.~2.\ref{w`w} and their repulsion properties are raising with
time. They are unlimited in time and $w_{de}$ %
%
for them tends asymptotically to --1. The scalar fields which are in
the phase plane between the lines $dw_{de}/d\ln a=0$ and
$dw_{de}/d\ln a=3(1+w_{de})^2$ evolve from left to right in
Fig.~2.\ref{w`w} and their repulsion properties are receding with
time ($dw_{de}/d\ln a>0$, $w_{de}$ increases) to change the
accelerated expansion by decelerated one and even collapse. They can
start in the range below the lower limit for thawing scalar fields,
then enter the range of thawing scalar fields limited by
\cite{Caldwell2005}, cross it and go out of upper limit
$dw_{de}/d\ln a=3(1+w_{de})$ when $w_{de}>0$. So, the scalar fields
with $dw_{de}/d\ln a>0$ ($c_a^2<w_0$) can only partially be called
thawing.

%
\begin{table}[t!]
\vspace*{-3mm} \noindent\parbox[b]{13cm}{\caption{\bf The best-fit
values and 1\boldmath$\sigma$ confidential ranges\newline
    of parameters of cosmological model with classical and tachyonic QSF\newline
    determined by the Markov chain Monte Carlo technique using two\newline
    observational datasets:  WMAP7~{+}  HST~{+}~BBN {+} BAO {+} SN SDSS\newline SALT2 ($\mathbf{q}_1$,
    $\mathbf{t}_1$) and WMAP7 {+} HST~{+} BBN~{+} BAO {+} SN SDSS\newline MLCS2k2($\mathbf{q}_2$, $\mathbf{t}_2$).
    The Hubble constant $H_0$ is in units\newline km$\,$s$^{-1}\,$Mpc$^{-1}$. %
    \index{Hubble constant}\index{Markov chain Monte Carlo (MCMC)}\index{Multicolor Light Curve Shape (MLCS)|(}We denote the rescaled energy density\newline  of the component $X$ by $\omega_X \equiv \Omega_Xh^2$\label{tab_qsf}}
  \index{Hubble constant}%
  \index{Markov chain Monte Carlo (MCMC)}}\vspace*{2mm} \tabcolsep13.5pt

\noindent{\footnotesize
\begin{tabular}{|c|c|c|c|c|}
    \hline
        \rule{0pt}{4mm}\raisebox{-3mm}[0cm][0cm]{\scriptsize Parameters}& \multicolumn{2}{c}{\scriptsize Classical QSF}&
        \multicolumn{2}{|c|}{\scriptsize Tachyonic QSF}\\[1.5mm] \cline{2-5}
         \rule{0pt}{3.5mm}&{\scriptsize $\mathbf{q}_1$}&{\scriptsize  $\mathbf{q}_2$}&{\scriptsize $\mathbf{t}_1$}&{\scriptsize $\mathbf{t}_2$}\\[1.5mm]
       \hline
\rule{0pt}{4mm}$\Omega_{de}$&0.73$_{-0.05}^{+0.03}$&0.70$_{-0.05}^{+0.04}$&0.73$_{-0.04}^{+0.03
}$& 0.71$_{-0.05}^{+0.04}$ \\ [1.5mm]
    $w_0$&--0.996$_{-0.004}^{+0.16}$& --0.83$_{-0.17}^{+0.22}$& --0.989$_{-0.011}^{+0.15}$&--0.83$_{-0.17}^{+0.20}$ \\ [1.5mm]
    $c_a^2$& --0.022$_{-0.978}^{+0.022}$& --0.88$_{-0.12}^{+0.88}$& --0.48$_{-0.52}^{+0.48}$&--0.97$_{-0.03}^{+0.96}$ \\ [1.5mm]
    10$\omega_b$& 0.226$_{-0.015}^{+0.015}$&0.226$_{-0.014}^{+0.016}$& 0.226$_{-0.014}^{+0.014}$&0.230$_{-0.017}^{+0.013}$\\ [1.5mm]
    $\omega_{cdm}$& 0.110$_{-0.013}^{+0.011}$&0.108$_{-0.012}^{+0.016}$& 0.111$_{-0.016}^{+0.010}$&0.110$_{-0.013}^{+ 0.014}$\\ [1.5mm]
    $H_0$&70.2$_{-4.3}^{+3.5}$&66.3$_{-3.7}^{+4.3}$&70.2$_{-4.4}^{+3.2}$&67.1$_{-4.9}^{+3.7}$\\ [1.5mm]
    $n_s$& 0.97$_{-0.04}^{+0.04}$&0.97$_{- 0.03}^{+ 0.04}$& 0.97$_{-0.03}^{+0.04}$&0.98$_{-0.04}^{+ 0.04}$\\ [1.5mm]
    $\log(10^{10}A_s)$& 3.09$_{-0.10}^{+0.10}$&3.07$_{- 0.08}^{+ 0.11}$& 3.08$_{-0.08}^{+0.11}$&3.08$_{-0.09}^{+0.11}$\\ [1.5mm]
    $\tau_{rei}$&0.091$_{-0.041}^{+0.040}$&0.089$_{-0.037}^{+0.044}$&0.087$_{-0.037}^{+0.043}$&0.091$_{-0.040}^{+0.042}$\\ [2.0mm]
\hline
    \rule{0pt}{4mm}$-\log L$&3865.01&3857.21&3865.09&3857.23\\ [2mm]
\hline
   \end{tabular}
  }\vspace*{-3mm}
\end{table}%

We propose to call them ``scalar fields receding repulsion'',
reflecting their main properties. Symmetrically, the scalar fields
\mbox{with} $dw_{de}/d\ln a<0$ ($c_a^2>$ $>w_0$), occupying the same
range as freezing scalar fields from \cite{Caldwell2005}, can be
called ``scalar fields raising repulsion''. Most of quin\-tessential
scalar field models of dark energy filling the phase plane fit well
the current observational data and main problem consists now in
distinguishing \mbox{between them.}\vspace*{-2mm}

\subsection{\!Best-fit parameters of QSF}\label{qsf_best-fit}\index{best-fit parameters|(} 

\hspace*{3cm}\index{quintessence}Let us estimate the parameters of
QSF with barotropic EoS ($\Omega_{de}$, $w_0$, $c_a^2$)
simultaneously with other cosmological parameters ($\omega_b$,
$\omega_{cdm}$, $H_0$, $n_s$, $A_s$ and $\tau_{rei}$) using the
following datasets: (1)~\textit{CMB
  temperature fluctuations and polarization angular power spectra}
from the 7-year WMAP observations (hereafter WMAP7)
[93---95]; %
\index{cosmic microwave background (CMB)} %
\index{CMB temperature fluctuations} %
(2)~\textit{Baryon acoustic oscillations} in the space distribution
of galaxies from SDSS DR7 (hereafter BAO) \cite{Percival2010}; (3)~\textit{Hubble constant measurements} from HST (hereafter HST) %
\index{Hubble Space Telescope (HST)} %
\cite{Riess2009}; (4)~\textit{Big Bang Nu\-cleo\-synthesis prior} on
baryon abundance (hereafter BBN) \cite{Steigman2007,Wright2007};
(5)~\textit{supernovae Ia luminousity distances} from SDSS
compilation (hereafter SN SDSS) \cite{Kessler2009}, determined using
SALT2 method of light curve fitting \cite{Guy2007} (hereafter SN
SDSS SALT2) and MLCS2k2 \cite{Jha2007} one (hereafter SN SDSS
MLCS2k2).

\index{Markov chain Monte Carlo (MCMC)}%
In order to find the best-fit values of parameters of cosmological
model with QSF and their confidence limits we perform the Markov
chain Monte Carlo (MCMC) analysis for two combined datasets: WMAP7
{+} HST {+} BBN~{+} +~BAO {+} SN SDSS SALT2 and WMAP7 {+} HST {+}
BBN {+} BAO
{+} SN SDSS MLCS2k2. We use the publicly available package CosmoMC %
\index{CosmoMC}%
\cite{cosmomc,cosmomc_source} including code CAMB %
\index{CAMB}%
\cite{camb,camb_source} for the calculation of model predictions.
This code has been modified to study the dark energy models
discussed here. The parameters $w_0$ and $c_a^2$ are determined
using the priors $-1<w_0<0$ and $-1<c_a^2<0$.

The results of estimations of QSF parameters jointly with the
minimal set of cosmological ones for two sets of observational data
(WMAP7 {+} HST~{+} +~BBN {+} BAO {+} SN SDSS SALT2 and WMAP7 {+} HST
{+} BBN {+} BAO~{+} +~SN SDSS MLCS2k2) are presented in
Table~2.\ref{tab_qsf}. We denoted the sets of best-fit parameters
for them as $\mathbf{q}_1$ and $\mathbf{q}_2$ for classical QSF and
$\mathbf{t}_1$ and $\mathbf{t}_2$ for tachyonic QSF accordingly. The
1$\sigma$ confidential limits are determined from the extremal
values of the N-dimensional distribution. One can see, that
WMAP7~{+} HST~{+} BBN {+} BAO {+} SN SDSS SALT2 dataset prefers the
scalar field model of dark energy with
decreasing EoS parameter: %
\index{EoS parameter|(} %
at the current epoch it is close to --1, at the early epoch it is
--0.02. The acceleration has changed the sign at $z\approx0.75$ and
now the deceleration parameter $q_0$ equals --0.59. In the future
such QSF
will approach the $\Lambda$CDM model %
\index{LambdaCDM model ($\Lambda$CDM)} %
with exponential expansion~--- late eternal inflation. The dataset
WMAP7 {+} HST {+} BBN~{+} +~BAO~{+} SN SDSS MLCS2k2 prefers the
scalar field model of dark energy with slowly increasing EoS
parameter: it started from the value --0.88 at the early epoch and
is --0.83 at current one. In this model the decelerated expan\-sion
has been changed by the accelerated one at $z\approx0.66$ and at
current epoch the deceleration parameter $q_0$ equals --0.38. While
$w_{de}$ continues to increase the deceleration parameter reaches
the minimal value, begins increasing and becomes positive (start of
decelerated expansion) in far future at $a\approx 20.46$ ($z\approx
-0.95$). The turnaround point is at $a\approx 35.5$, when Universe
will 172.5~Gyrs old. Then the redshifts of galaxies will be changed
by \mbox{blueshifts}, the Universe
will start collapsing and will reach the Big Crunch singularity %
\index{singularity} %
in the age of 345~Gyrs. Therefore, the model with parameters
$\mathbf{q}_2$ is limited in time as opposed to the model with
$\mathbf{q}_1$, though both match equally well the observational
dataset corresponding to the past and present of the Universe (see
\mbox{Figs.}~2.\ref{dl_all}---2.\ref{cl_all_te}). The differences
between obtained best-fit parameters of these models are caused by
differences of SNe
Ia distance moduli obtained by SALT2 and MLCS2k2 %
\index{Multicolor Light Curve Shape (MLCS)|)}%
methods of light-curve~fitting.\index{light-curve fitting}

\vspace*{-0.5mm}
 Let us analyze now the possibility of
distinguishing between the quintes\-sential scalar field models with
decreasing and increasing EoS parameters as well as with classical
and tachyonic Lagrangian. Note, that the difference in the
Lagrangian manifests only in the development of cosmological
perturbations due to the different effective sound speed of scalar
fields, so we consider two sets of parameters $\mathbf{q}_1$ and
$\mathbf{q}_2$ for both Lagrangian.

\begin{figure}
\vskip1mm
\includegraphics[width=13cm]{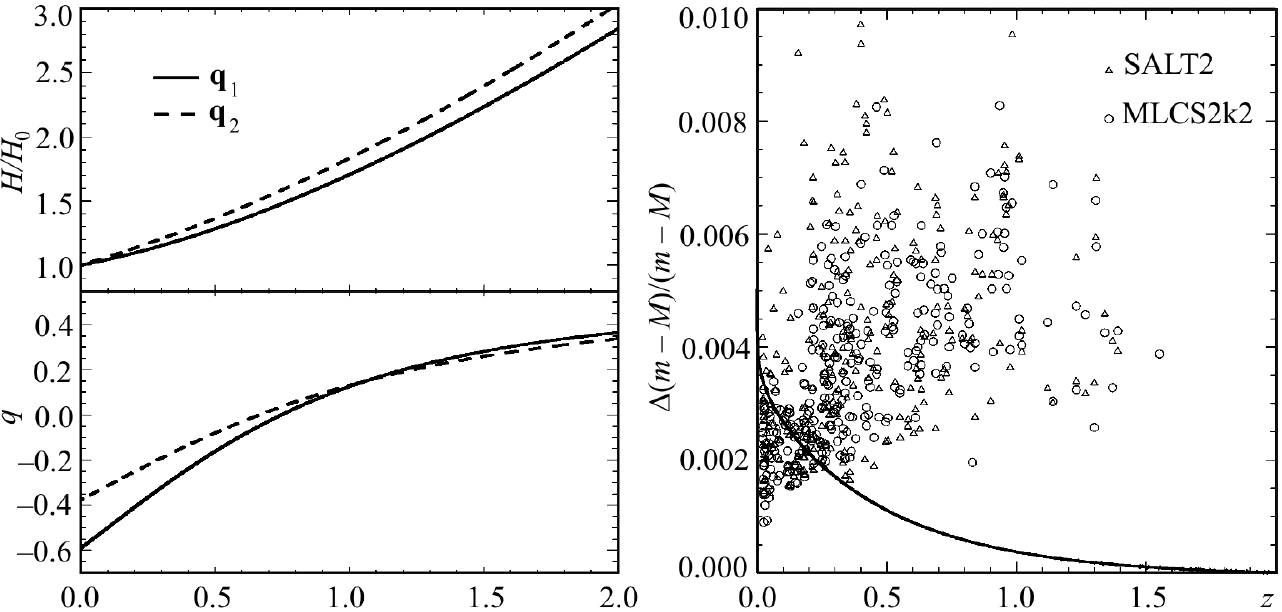}
    \vskip-2mm
  \caption{Dynamics of expansion of the Universe in the redshift range
    $0\le z\le2$ for cosmological models with best-fit parameters
    $\mathbf{q}_1$ and $\mathbf{q}_2$ (left panel) and relative
    differences of distance moduli
    $[(m-M)_{q_2}-(m-M)_{q_1}]/(m-M)_{q_1}$ (solid line in the right
    panel). The symbols show the uncertainties of distance moduli
    determination of SN SDSS SALT2 and SN SDSS MLCS2k2 data}
  \label{ddl_qsf_sn_sdss}\vspace*{-2mm}
\end{figure}

\vspace*{-0.5mm}
 The differences of dynamics of expansion of the
Universe in the cosmological models with best-fit parameters
$\mathbf{q}_1$ and $\mathbf{q}_2$ are shown in the left panel of
Fig.~2.\ref{ddl_qsf_sn_sdss}. The rate of expansion in the model
$\mathbf{q}_1$ increases slower than in the model $\mathbf{q}_2$,
since $q_0$ in it is essentially lower. Both characteristics,
$H(z)/H_0$ and $q(a)$, could be deduced from SNe Ia luminosity
distances, if their number would be sufficient. In their absence the
dependence of SNe Ia luminosity distance on redshift is used. In the
right panel the relative differences of distance moduli
$[(m-M)_{q_2}-(m-M)_{q_1}]/(m-M)_{q_1}$ as well as the statistical
and systematical uncertainties of distance moduli determinations of
SDSS SNe Ia are presented. One can see that observational
uncertainties are comparable with the model differences only at low
redshifts $z<0.3$.

\begin{figure}
\vskip1mm
\includegraphics[width=13cm]{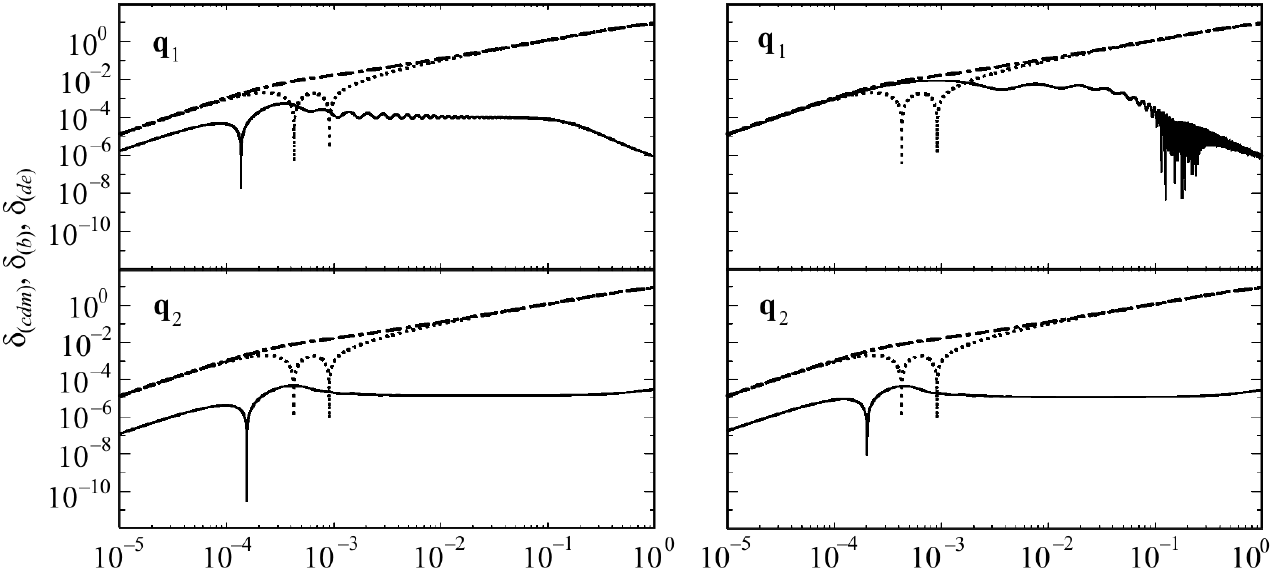}
    \vskip-2mm
  \caption{Evolution of Fourier amplitude ($k=0.05$\,Mpc$^{-1}$) of
    density perturbations for cold dark matter (dashed line), baryonic
    matter (dotted) and QSF (solid).  In the left column the scalar
    field is classical, in the right one tachyonic. The cosmological
    parameters for the computations were taken from Table~2.\ref{tab_qsf} }
  \label{ddeb_qsf}
\end{figure}

\begin{figure}[h!]
\vskip1mm
\includegraphics[width=13cm]{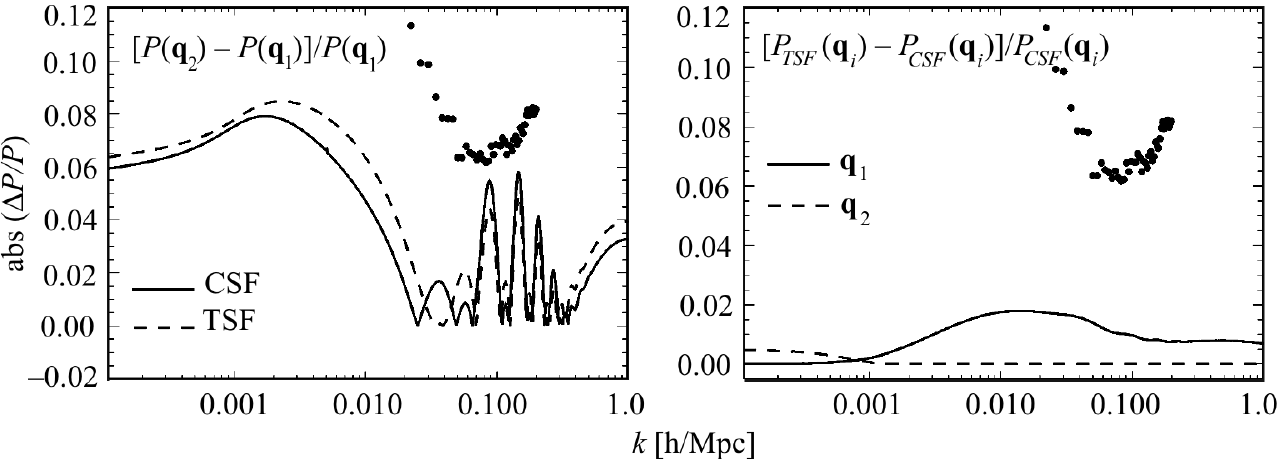}
    \vskip-2mm
  \caption{Left panel: the relative difference of matter density power
    spectra $|\Delta P(k)|/P(k)$ in the models with best fitting
    parameters $\textbf{q}_1$ and $\textbf{q}_2$ for classical scalar
    field (solid line) and tachyonic one (dashed line). %
    \index{classical scalar field} %
    Right panel: the relative difference of matter density power
    spectra $|\Delta P/P|$ in the models with classical and tachyonic
    scalar fields for two sets of the best fitting parameters
    $\textbf{q}_1$ (solid line) and $\textbf{q}_2$ (dashed line). Dots
    show observational uncertainties (1$\sigma$) of SDSS LRG DR7 data
    \cite{Reid2010}}\vskip3mm
  \label{vpkm_qsf}
\end{figure}

\index{scalar perturbations}%
Other tests are based on the evolution of cosmological perturbations
in the multicomponent Universe. The scalar field dark energy affects
the evolution of matter density perturbations via growth factor %
\index{growth factor} %
(dynamics of expansion) and gravitational influence of its own
scalar perturbations \cite{Novosyadlyj2009,Novosyadlyj2010}. The
evolution of density perturbations of main components in the
cosmological models with parameters $\mathbf{q}_1$, $\mathbf{q}_2$,
$\mathbf{t}_1$ and $\mathbf{t}_2$ is shown in Fig.~2.\ref{ddeb_qsf}.
They have been
computed using synchronous gauge, %
\index{synchronous gauge} %
the initial conditions are adiabatic for matter components and
subdominant asymptotic ones for the scalar \mbox{field}
(eqs.~(\ref{synch_init})). The general property is inherent for all
models: for positive matter density perturbation the QSF density
perturbation is positive from initial moment to horizon crossing
one, after that it changes the sign and decays. At the current epoch
the density perturbations of QSF are by $\sim$2---3 orders lower
than matter ones and have opposite sign, so, their imprint in the
large scale
structure of the Universe %
\index{large scale structure}%
is expected to be small. In the left panel of Fig.~2.\ref{vpkm_qsf}
the relative differences of matter power spectra in the models with
$\textbf{q}_1$ and $\textbf{q}_2$ best-fit parameters for classical
(solid line) and tachyonic (dashed line) Lagrangians are shown:
$|P(k;\textbf{q}_2)-P(k;\textbf{q}_1)|/P(k;\textbf{q}_1)$.  The
maximal dif\-ferences $\sim $8\,\% are for large scale perturbations
where observational errors are essentially larger. At the scales
$k\sim 0.1$\,Mpc$^{-1}$, where errors of determinations of matter
power spectrum are minimal $\sim $6---7\,\%, the relative
differences between power spectra for $\textbf{q}_1$ and
$\textbf{q}_2$ best-fit parameters are $\sim$4---6\,\% and we hope
that the current observational program will improve accuracy and
possibility to distinguish this
models. %
\index{best-fit parameters|)} %

In the right panel of Fig.~2.\ref{vpkm_qsf} the relative differences
of matter power spectra in the models with classical and tachyonic
Lagrangians $|P_{TSF}(k;\textbf{q}_i)\,-$
$-\,P_{CSF}(k;\textbf{q}_i)|/P_{CSF}(k;\textbf{q}_i)$ are shown for
$\textbf{q}_1$ (solid line) and $\textbf{q}_2$ (dashed line). Here
differences are caused solely by influence of scalar field density
perturbations on matter density ones. The maximal differences here
are $\sim$1---2\,\% for the model with $\textbf{q}_1$ and $<$1\,\%
for the model with $\textbf{q}_2$. So, distinguishing of classical
scalar field from tachyonic one by the observational data on matter
density power spectra in the nearest future looks problematic.

\begin{figure}[h!]
\includegraphics[width=13cm]{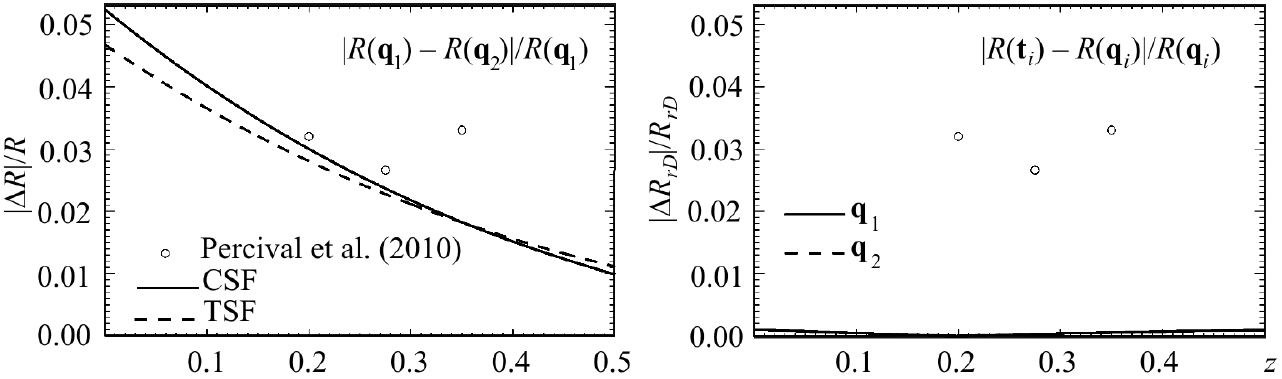}
    \vskip-2mm
  \caption{Left panel: the relative differences of the BAO distance
    measure $R\equiv r_s(z_{drag})/$ $D_V(z)$ in the cosmological models
    with best fitting parameters $\textbf{q}_1$ and $\textbf{q}_2$
    ($|R(\textbf{q}_1)\,-$ $-\,R(\textbf{q}_2)|/R(\textbf{q}_1)$) for
    classical scalar field (solid line) and tachyonic one (dashed
    line). %
    \index{classical scalar field}\mbox{Right} panel: the relative differences of the BAO distance
    measure $|R(\textbf{t}_i)-R(\textbf{q}_i)|/R(\textbf{q}_i)$ in the
    models with best fitting parameters $\textbf{t}_i$ and
    $\textbf{q}_i$ ($i=1, 2$). Dots show the observational 1$\sigma$
    uncertainties of $R$ extracted from SDSS DR7 galaxy redshift
    survey \cite{Percival2010} (symbols)}
  \label{vrbao_qsf}
\end{figure}

\index{cosmological parameters} %
In the procedure of determination of cosmological parameters the
data on the BAO relative distance measure $R\equiv
r_s(z_{drag})/D_V(z)$ \cite{Percival2010} (see subsection
\ref{ch1-sec3}) have been used instead of data on $P(k)$ because
their accuracy is 2--3 times better. In the left panel of
Fig.~2.\ref{vrbao_qsf} the relative differences of the BAO distance
measure $|R(\textbf{q}_1)-R(\textbf{q}_2)|/R(\textbf{q}_1)$ in the
cosmological models with best fitting parameters $\textbf{q}_1$ and
$\textbf{q}_2$ are shown for classical and tachyonic scalar fields.
One can see that increasing of accuracy of measurement of this
parameter will


\noindent give possibility to distinguish between QSF~+~CDM models
with increasing and decreasing EoS parameter. But these data cannot
be used for distingui\-shing between classical and tachyonic scalar
fields since they do not contain the information about evolution of
matter
density perturbations and their power spectrum. %
\index{power spectrum}%
In the right panel of Fig.~2.\ref{vrbao_qsf} the relative
differences of the BAO distance measure
$|R(\textbf{t}_i)-R(\textbf{q}_i)|/R(\textbf{q}_i)$ ($i=1,\,2$) it
is shown, they are essentially smaller than
$|P_{TSF}(k;\textbf{q}_1)-P_{CSF}(k;\textbf{q}_1)|/P_{CSF}(k;\textbf{q}_1)$,
because are caused only by small differences of cosmological
parameters $\textbf{t}_1$ and $\textbf{q}_1$. Therefore, the high
precision power spectrum of matter density perturbations is more
informative about the nature of dark energy than BAO relative
distance measure.

\begin{figure}[t]
\vskip1mm
\includegraphics[width=13cm]{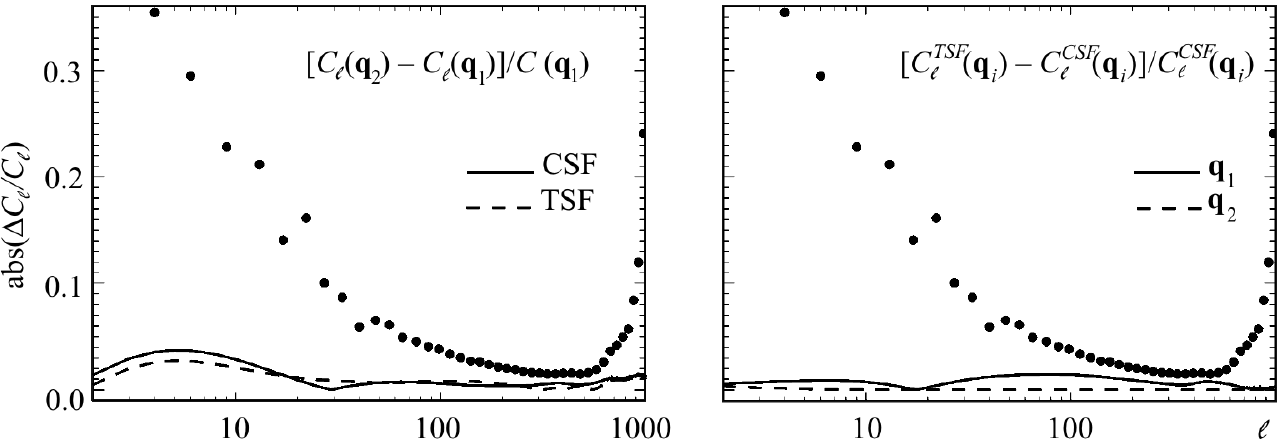}
    \vskip-2mm
  \caption{Left panel: the relative difference of CMB temperature
    fluctuations power spectra $|\Delta C_{\ell}|/C_{\ell}$ in the
    models with best fitting parameters $\textbf{q}_1$ and
    $\textbf{q}_2$ for classical (solid line) and tachyonic (dashed
    line) scalar fields (Table 2.\ref{tab_qsf}). Right panel: the
    relative difference of CMB temperature fluctuations power
    spectra %
    \index{CMB temperature fluctuations} %
    $|\Delta C_{\ell}|/C_{\ell}$ in the models with classical and
    tachyonic scalar fields for two sets of the best fitting
    parameters $\textbf{q}_1$ (solid line) and $\textbf{q}_2$ (dashed
    line). Dots show observational uncertainties (1$\sigma$) of WMAP7
    data}
  \label{vcl_qsf}
\end{figure}

The dynamical properties of scalar field dark energy and its
perturbations leave slight fingerprints in the map of temperature
fluctuations and polarization of CMB radiation %
\index{cosmic microwave background (CMB)|(} %
that have been already discussed above (see also
\cite{Novosyadlyj2010}). Here we show the relative differences of
CMB power spectra
$|C_{\ell}(\textbf{q}_1)-C_{\ell}(\textbf{q}_2)|/C_{\ell}(\textbf{q}_1)$
for decreasing ($\textbf{q}_1$) and increasing ($\textbf{q}_2$) EoS
parameters (left panel of Fig.~2.\ref{vcl_qsf}) as well as
$|C_{\ell}^{CSF}(\textbf{q}_i)-C_{\ell}^{TSF}(\textbf{q}_i)|/C_{\ell}^{CSF}(\textbf{q}_i)$
(right panel) for parameter sets $\textbf{q}_1$ and $\textbf{q}_2$.
In the left panel the differences at low spherical harmonics
($\ell<10$) for both scalar fields are caused mainly by different
contribution of the late Sachs---Wolfe effect ($\sim $2---4\,\%), at
higher ones mainly by different cold dark matter content ($< $2\,\%)
and, partially, by different optical depth ($\sim $0.2\,\%) and
geometrical effect\,\footnote{\,The particle horizon for model with
$\textbf{q}_1$ is
  14430 Mpc, for model with $\textbf{q}_2$ is 14470 Mpc.} ($\sim$0.3\,\%).
  In the right panel the lines show the differences caused
solely by influence of scalar field density perturbations on
formation
of CMB anisotropy. %
\index{CMB anisotropy} %
They do not exceed ($\sim $1---2\,\%) for model \mbox{with}
decreasing EoS, in which the differences between amplitudes of CSF
and TSF density perturbations are most substantial, and are $\ll
$1\,\% for model \mbox{with} \mbox{increasing} EoS parameters
$\textbf{q}_2$. The relative errors of
binned CMB power spectrum $|\Delta C_{\ell}|/C_{\ell}$, %
\index{CMB power spectrum}%
\index{power spectrum}%
obtained in WMAP seven-year experiment, are somewhat larger even in
the range of acoustic peaks, where accuracy is highest. %
\index{acoustic peaks} %
So, the SN SDSS SALT2 data prefer the model with decreasing EoS
parameter, while SN SDSS MLCS2k2 data prefer the model with
increasing
EoS parameter, but the difference of maxima of likelihood functions %
\index{likelihood function}%
\index{Multicolor Light Curve Shape (MLCS)}%
for both models is statistically insignificant. It means, that
current observational data do not distinct these models as well as
the models
with classical and tachyonic Lagrangian. %
\index{quintessential scalar field (QSF)|)} %
\index{cosmic microwave background (CMB)|)} %

\section[\!Phantom scalar fields with barotropic EoS]{\!Phantom scalar fields with barotropic EoS}

\hspace*{3cm}\index{phantom scalar field (PSF)|(}\index{phantom dark
energy}It was mentioned above that current observations allow the
possibility of the equation of state $w_{de}<-1$, which is generally
referred to as phantom dark energy (PDE)
\cite{Caldwell2002,Caldwell2003}. On the other hand PDE emerges
effectively from the gravity sector of brane-world
models~\cite{Sahni2003,Lue2004} (see also section
\ref{brane-sec:phantom} of this book), from superstring
theory~\cite{Neupane2006,Arefeva2008}, from Brans---Dicke
scalar-tensor gravity~\cite{Elizalde2004,Gannouji2006} and quantum
effects that lead to violations of the weak energy condition on
cosmological scales~\cite{Onemli2002,Onemli2004}. Some of these
models have phantom properties only at the current stage of
evolution of the Universe but did not have them at early time or
they lose this feature in the future. Let us analyze the possibility
of modeling of such dark energy by a single minimally coupled scalar
field.

Primarily note, that scalar fields with classical or tachyonic
Lagrangians cannot be PDE since the field variables for them become
imaginary (see eqs.~(\ref{U_clas})---(\ref{U_tach})). So, another
form of Lagrangians must be considered. The simplest one is modified
canonical Lagrangian with altered sign before the kinetic term: %
  \index{canonical Lagrangian} %
\begin{equation}
  L_{de}=-X-U(\phi).\label{L_ph1}
\end{equation}
In this case the energy density and pressure are following linear
combinations of $X$ and $U$:
\begin{equation}
  \rho_{de}=-X+{U}(\phi), \quad  p_{de}=-X-{U}(\phi). \label{rho_XU_ph1}
\end{equation}

The EoS parameter
\begin{equation}
  w_{de}=\frac{-X-U}{-X+U}\label{w_XU_ph1}
\end{equation}
for positive values of $X$ and $U$ is $\le-1$. For explanation of
accelerated expansion of the Universe at the current epoch ($q_0<0$)
the phantom scalar field must satisfy two conditions:
\begin{equation}
  \text{ a)}\,\, 0<X^{(0)}<U^{(0)}, \quad  \text{ b)}\,\, U^{(0)}+2X^{(0)}>\rho_m^{(0)}/2.
\end{equation}

\begin{figure}[t]
\vskip1mm
\includegraphics[width=13cm]{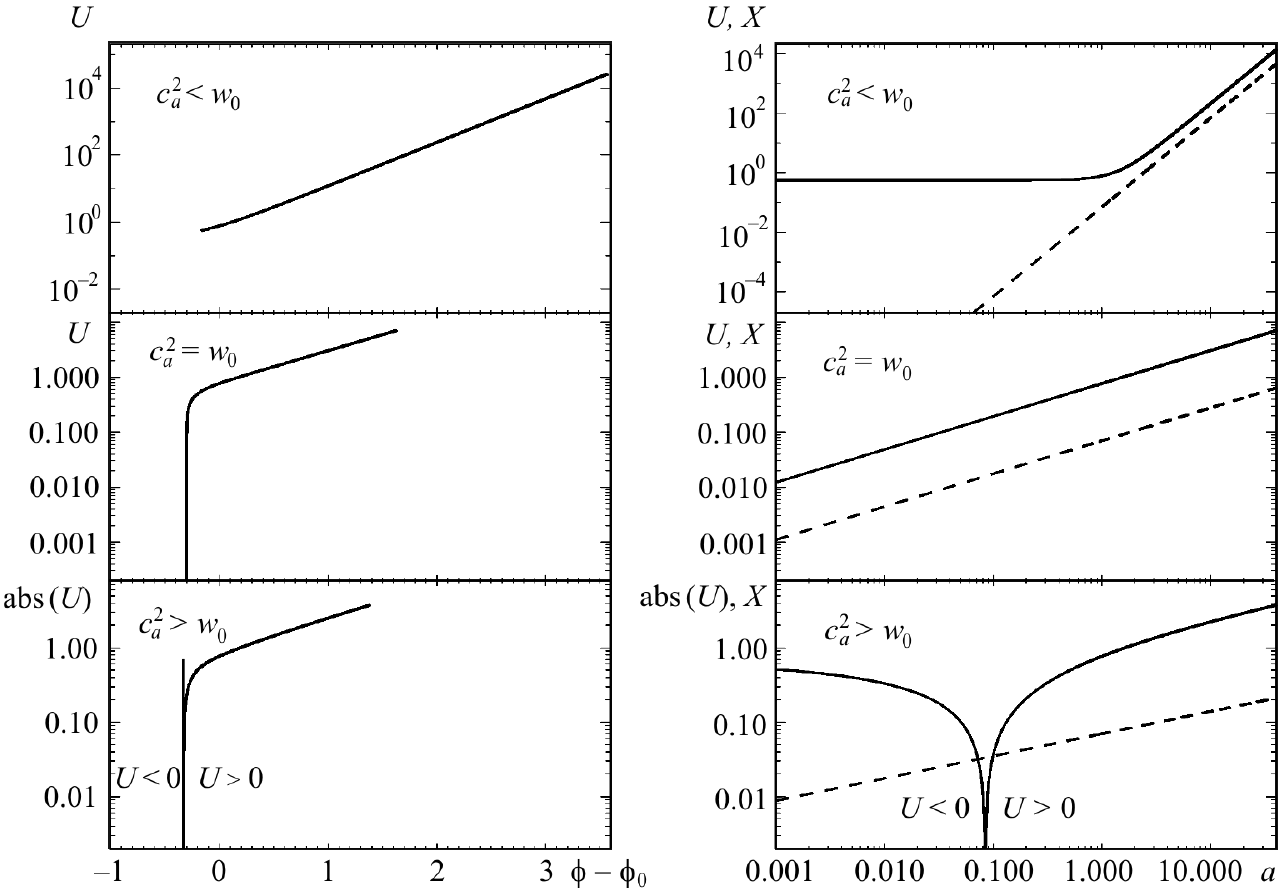}
    \vskip-2mm
  \caption{Potentials $U(\phi-\phi_0)$ (left) and dependences of
    potentials and kinetic terms on scale factor $a$ (right) for
    phantom scalar field with Lagrangian (\ref{L_ph1}) and barotropic
    EoS. In the top panels $c_a^2=-2.0$, in the middle ones
    $c_a^2=-1.2$ and in the bottom ones $c_a^2=-1.1$, in all panels
    $w_0=-1.2$. Potential and kinetic term are in the units of current
    critical energy density, $3c^2H_0^2/8\pi G$, the field variable in
    units of $\sqrt{3c^2/8\pi G}$. The current epoch in the left
    panels corresponds to $\phi-\phi_0=0$ and field evolves from left
    to right}
\label{U_X_ph1}\vspace*{-2mm}
\end{figure}

\index{equation of state (EoS)} %
Assuming barotropic equation of state the field variable, potential
and kinetic term can be presented in terms of density and EoS
parameter as follows:
\begin{equation}
  \begin{array}{c}
    \label{U_ph1}
   \displaystyle\phi(a)-\phi_0=\pm\sqrt{-(1+w_0)\rho_{de}^{(0)}}\int\limits_1^a\frac{da'}{a'^{(\frac{5}{2}+\frac{3}{2}c_a^2)}H(a')},
   \\[7mm]
   \displaystyle U(a)=\frac{(1-c_a^2)(1+w_0)a^{-3(1+c_a^2)}+2(c_a^2-w_0)}{2(1+c_a^2)}\rho_{de}^{(0)},
   \\[5mm]
  \displaystyle X(a)=-\frac{1+w_0}{2}a^{-3(1+c_a^2)}\rho_{de}^{(0)}.
  \end{array}
\end{equation}
One can see that phantom barotropic scalar field ($w_0<-1$,
$c_a^2<-1$) has real values of field variable and potential if
current density of dark energy is positive. Its kinetic term $X(a)$
is positive at any $a$, the potential $U(a)$ is always positive only
in the case $c_a^2\le w_0$. If $w_0<c_a^2<-1$, then $U(a)$ starts
from negative energy density $(c_a^2-w_0)\rho_{de}^{(0)}/(1+c_a^2)$
at $a=0$, changes the sign from ``$-$'' to ``+'' at
$a_{\rho=0}=[2(w_0-c_a^2)/(1-c_a^2)(1+w_0)]^{-\frac{1}{3(1+c_a^2)}}$,
which is always $\le 1$ for phantom case. In any case $U(a)$
increases with $a$, that distinguishes phantom scalar field from the
quintessential one. The potentials $U(\phi-\phi_0)$, evolution of
potentials and kinetic terms for models with $c_a^2<w_0<-1$,
$c_a^2=w_0<-1$\linebreak and $w_0<c_a^2<-1$ are shown in
Fig.~2.\ref{U_X_ph1}. One can see that accelerated expansion of the
Universe is caused by rolling up of field to maximum of its
potential, inversely as in the case of quintessential scalar field.
We must note, that energy density and pressure are smooth monotonous
functions of $a$ for all relations between $c_a^2$ and $w_0$, both
$<$$-1$, while $w_{de}$ has the second kind discontinuity, caused
the passing of scalar field energy density \mbox{over zero (see
Fig.~2.\ref{wrode_ph}).}

The effective sound speed (\ref{c_s2}) for phantom scalar field
(\ref{L_ph1}) is equal to the speed of light ($c_s^2=1$) like in the
case of classical one.

One can construct the Lagrangian like DBI one (\ref{L_tach}) for
phantom range of EoS with positive defined kinetic term. Indeed,
``relativistic'' generalization of (\ref{L_ph1}) is following:
\begin{equation}
 L_{de}=-\tilde{U}(\xi)\sqrt{1+2\tilde{X}}.\label{L_ph2}
\end{equation}
The energy density, pressure and EoS parameter for this field are as
follows:
\begin{equation}
  \rho_{tach}  = \frac{\tilde{U}(\xi)}{\sqrt{1+2\tilde{X}}}, \quad p_{tach}=-\tilde{U}(\xi)\sqrt{1+2\tilde{X}}, \label{rho_XU_ph2} \\
\end{equation}\vspace*{-3mm}
 \begin{equation}
  w_{de}  = -2\tilde{X}-1.\label{w_XU_ph2}
\end{equation}
\index{EoS parameter|)} %
One can see, that in the case of phantom field (\ref{L_ph2}) the EoS
parameter is always $\le-1$ for positive values of $\tilde{X}$
independently on the value and sign of $\tilde{U}$.  For explanation
of accelerated expansion of the Universe at the current epoch
($q_0<0$) such field must satisfy two conditions:
\begin{equation}
  \text{ a)}\,\, \tilde{X}^{(0)},\,\tilde{U}^{(0)}>0, \quad  \text{ b)}\,\, \tilde{U}^{(0)}\frac{1+3\tilde{X}^{(0)}}{\sqrt{1+2\tilde{X}^{(0)}}}>\rho_m^{(0)}/2,
\end{equation}
where index ${(0)}$ marks the current values of corresponding
variables.

In the case of phantom field (\ref{L_ph2}) its field variable,
potential and kinetic term can be presented in terms of dark energy
density (\ref{rho_bar}) and EoS parameter (\ref{w_bar}) as follows:
\begin{equation}
\label{U_ph2}
\begin{array}{c}
 \displaystyle
 \xi(a)-\xi_0=\pm\int\limits_1^a\frac{da'\sqrt{-(1+w_{de}(a'))}}{a'H(a')},\\[7mm]
 \displaystyle \tilde{U}(a)=\rho_{de}(a)\sqrt{-w_{de}(a)}, \\[5mm]
 \displaystyle \tilde{X}(a)=-\frac{1+w_{de}(a)}{2}.
\end{array}
\end{equation}

\begin{figure}[t]
\vskip1mm
\includegraphics[width=13cm]{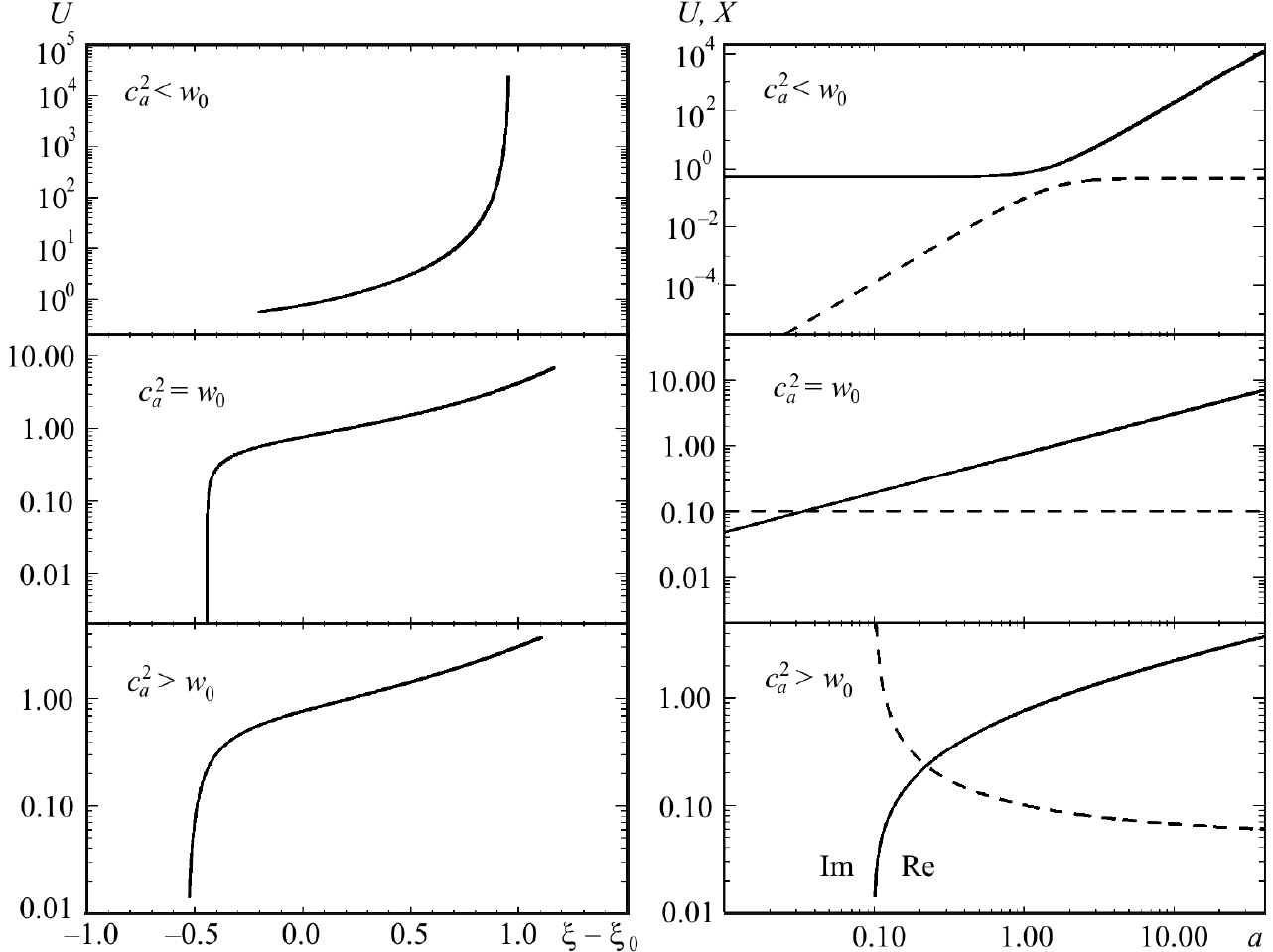}
    \vskip-2mm
  \caption{Potentials $U(\xi-\xi_0)$ (left) and dependences of
    potentials and kinetic terms on scale factor $a$ (right) for
    phantom scalar field with Lagrangian (\ref{L_ph2}) and barotropic
    EoS. In the top panels $c_a^2=-2.0$, in the middle ones
    $c_a^2=-1.2$ and in the bottom ones $c_a^2=-1.1$, in all panels
    $w_0=-1.2$. Potential and kinetic term are in the units of current
    critical energy density, $3c^2H_0^2/8\pi G$, the field variable is
    in units of $\sqrt{3c^2/8\pi G}$. The current epoch in the left
    panels corresponds to $\phi-\phi_0=0$ and field evolves from left
    to right}\vspace*{-2mm}
\label{U_X_ph2}
\end{figure}

The potentials $U(\xi-\xi_0)$, evolution of potentials and kinetic
terms for models with decreasing, constant and increasing EoS
parameters are shown in Fig.~2.\ref{U_X_ph2}. As it is in previous
case, the accelerated expansion of the Universe is caused by rolling
up of field to maximum of its potential. We must note, that energy
density and pressure are smooth monotonous functions of $a$ for all
relations between $c_a^2$ and $w_0$, both $<$$-1$, while $w_{de}$
has discontinuity of second kind, caused by passing of scalar field
energy density over zero (see Fig.~2.\ref{wrode_ph}) in the case
$w_0<c_a^2$. Moreover, as it follows from (\ref{U_ph2}) and
Fig.~2.\ref{wrode_ph}, the field variable $\xi(a)$ and potential
$\tilde{U}(a)$ are imaginary at $a<a_{\rho=0}$, while measurable
values, $\rho_{de}(a)$ and $p_{de}(a)$, are real. Other intriguing
property of such field is effective sound speed: according to
(\ref{c_s2}) it is equal to $-w_{de}$ and for phantom range is
superluminal. Therefore, the phantom scalar field (\ref{L_ph2}) can
be theoretically interesting model of dark energy but unlikely for
realization in our Universe. So, below we will analyse only phantom
\mbox{scalar field (\ref{L_ph1}).}

\begin{figure}
\vskip1mm \raisebox{0.0cm}{\parbox[b]{5.0cm}{\caption{Dependences of
scale factors on time, $a(t)$, for cosmological models with
quinte\-ssen\-ce/phan\-tom scalar fields with $w_0=-1\pm0.2$ and
$c_a^2=-1\pm1$ (dotted line), $-1\,\pm$ $\pm\,0.3$ (dashed line),
$-1\pm0.2$ (dash-dotted line), $-1\pm0.1$ (dash-three-dotted line).
The upper sign is for QSF, lower one is for PSF. For the
$\Lambda$CDM model $a(t)$ is shown by thick solid line. In all
models $\Omega_m=0.3$, $\Omega_{de}=0.7$,
$H_0=70$\,km/s\,$\cdot$\,Mpc\index{LambdaCDM model
($\Lambda$CDM)}\label{ah_ph}}}}\hspace*{0.5cm}\includegraphics[width=7.5cm]{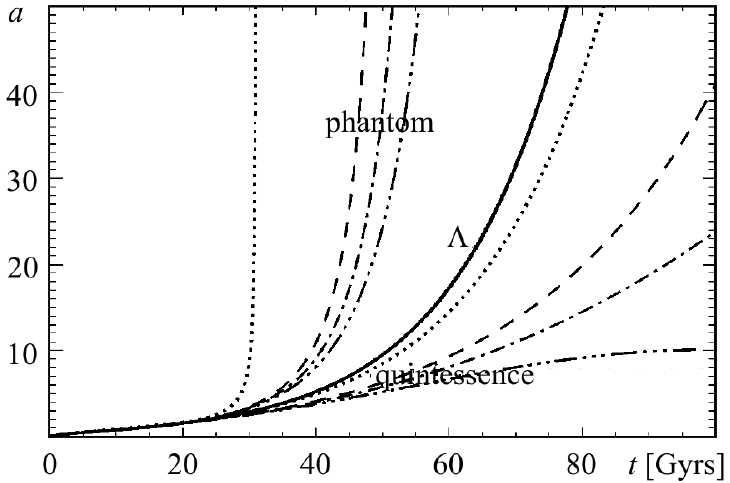}
\end{figure}

Other distinction of PSF from QSF consists in their asymptotic
behavior: PSF mimics cosmological constant at the Beginning for any
$c_a^2<-1$ ($w_{de}$ goes to $-1$ when $a$ goes to 0), while QSF
mimics it at $a$-infinity. So, such PSF always starts as
cosmological
constant with %
\index{cosmological constant} %
$\rho_{de}(a=0)=\rho_{de}^{(0)}(c_a^2\,-\,w_0)/(1\,+\,c_a^2)$, which
is positive for $c_a^2<w_0$ and negative when $w_0<c_a^2<-1$.  This
property distinguishes PSF from ``standard'' PDE
\cite{Caldwell2002,Caldwell2003}, the density of which starts from
zero at $a=0$. But one can see, that PDE is the special case of our
barotropic PSF, when $c_a^2=w_0$.  In far future, when $a\gg1$, its
energy density will increase as $\rho_{de}(a)\propto
(1+w_0)/(1+c_a^2)\rho^{(0)}_{de}a^{-3(1+c_a^2)}$ while $w_{de}$ will
go to $c_a^2$ (see eqs.~(\ref{w_bar}) and (\ref{rho_bar})). So, its
repulsion properties will increase and in finite time reach and
outmatch firstly forces of gravitationally bound objects, then
electrically ones, then strong force ones. All elements of structure
of our Universe~--- galaxies, stars, planets, atoms and protons,
will be ripped in finite time. This moment is dubbed the Big
Rip\,\footnote{\,The
  first name of this singularity was ``Big Smash'', proposed in the
  paper \cite{McInnes2002}.}  \cite{Caldwell2003} and the moment when
it happens can be estimated from the time dependence of scale factor, %
\index{singularity} %
\begin{equation}
  t=\int\limits_0^a\frac{da'}{a'H(a')},
  \label{a(t)}
\end{equation}
which can be calculated numerically using (\ref{H}) for any
cosmological models and parameters of scalar field with barotropic
EoS. In Fig.~2.\ref{ah_ph} we present the time dependences of scale
factors, $a(t)$, for cosmological models with PSF with the same
parameters as in Fig.~2.\ref{wrode_ph}. For comparison we show also
$a(t)$ for $\Lambda$CDM and QSF+CDM models with symmetrical values
of $w_0$ and $c_a^2$ relative to the phantom divide line.  The
phantom range of $a-t$ space is above $a(t)$-line for $\Lambda$CDM
with the same cosmological parameters, the quintessence range is
below one.

At $a\gg1$, when radiation and matter terms in (\ref{H}) can be
neglected, we obtain the approximate analytic formula for $a(t)$:
\begin{equation}
  a(t)\approx\left[\frac{3}{2}H_0(1+c_a^2)\sqrt{\frac{(1+w_0)\Omega_{de}}{1+c_a^2}}(t-t_0)+1\right]^{\frac{2}{3(1+c_a^2)}}\!.\label{a_ph}
\end{equation}
It shows, that $a$-infinity is reached in finite time
\begin{equation}
  t_{BR}-t_0\approx \frac{2}{3}\frac{1}{H_0}\frac{1}{|1+c_a^2|}\sqrt{\frac{1+c_a^2}{(1+w_0)\Omega_{de}}},\label{t_br}
\end{equation}%
\index{Big Rip|(}which is noted as time of Big Rip. Such fast
expansion leads also to freezing \linebreak of the particle horizon
$r_p$ at some $r_p^{max}$ and contraction of the event horizon $r_e$
to point when $t\rightarrow t_{BR}$.  Really, in the co-moving
coordinates they are as follows:
\begin{eqnarray*}
  r_{p}(t)=c\int\limits^{a}_{0}\frac{da'}{a'^{2}H(a')},\quad r_{e}(t)=c\int\limits^{\infty}_{0}\frac{da'}{a'^{2}H(a')}-r_{p}(a),
\end{eqnarray*}
so, starting from $a_f\gg1$, when matter component in (\ref{H}) can
be omitted, the rest of $r_p$-integral from $a_f$ to $a\gg a_f$ has
analytic representation
\begin{equation*}
  I(a_f,a)=\frac{2c}{(1+3c_a^2)H_0}\sqrt{\frac{1+c_a^2}{(1+w_0)\Omega_{de}}}a_f^{\frac{(1+3c_a^2)}{2}}
\end{equation*}
and goes to 0 when $a_f\rightarrow\infty$.

Therefore, the positive energy density of PSF becomes infinite at
finite time (\ref{t_br}), overcoming all other forms of matter. The
phantom scalar field dark energy rips at first the clusters of
galaxies, later Milky Way and other galaxies, then Solar System, a
bit later Earth, Sun and stars and ultimately ``the molecules,
atoms, nuclei, and nucleons of which we are composed, before the
death of the
Universe in a Big Rip'' (see Table 1 in \cite{Caldwell2003}). %
\index{Big Rip|)} %
Will this be the end of Everything? Maybe this will be the beginning
of new worlds~--- if PSF reaches the Planck density, the quantum
fluctuations or interaction of field with the particles (the
phenomenon of confinement) will lead to inflation in some regions of
Planck scales. In the paper \cite{Elizalde2004} it has been
demonstrated that in the case of phantom Big Rip the consideration
of quantum gravity effects might drastically change the future of
our
Universe, removing the singularity in a quite natural way. %
\index{singularity} %

Another feature of phantom dark energy, discussed in the literature,
is its inluence on quantum stability of vacuum.  It was shown
\cite{Carroll2003,Cline2004} that minimally coupled scalar fields
with a linear negative kinetic term may cause a UV quantum
instability of
the vacuum %
\index{quantum instability of the vacuum} %
manifesting itself in the production of pairs of ghosts, photons or
gravitons as a consequence of the violation of the null energy
condition. It can be prevented by introducing the squared kinetic
term in the Lagrangian as in the ghost condensate
model~\cite{Arkani2004,Piazza2004} or by second derivatives of the
scalar field as in the kinetic braiding scalar-tensor
model~\cite{Deffayet2010}. For late type phantom scalar fields the
produced ghosts typically carry low energy, so, their decay rates
are strongly time-dilated. On the other hand, the time scale of this
instability for phantom dark energy can be much larger than the
cosmological one, making this effect unsuitable for constraining the
parameters of the model at the present level of observations. This
is why in this chapter we concentrate our attention on the classical
properties of scalar field models of dark energy and on
possibilities to determine their parameters by comparison of
predictions with available observational data.

\subsection{\!Gravitation instability of PSF\\ \hspace*{-1.2cm}and large scale structure
  formation}\label{psf_instab}

\begin{figure}
\vskip1mm
  \includegraphics[width=8cm]{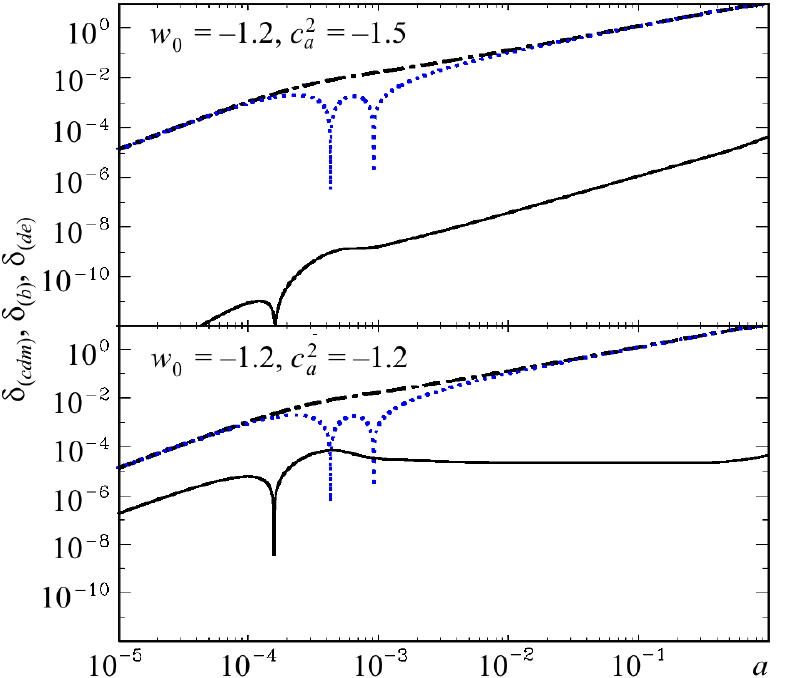}\hspace*{0.5cm}\raisebox{0.0cm}{\parbox[b]{4.5cm}{\caption{Evolution of Fourier amplitude ($k=0.05$~Mpc$^{-1}$) of
    density perturbations of cold dark matter (dashed line), baryonic
    matter (dotted line) and PSF (solid line) with $c_a^2<w_0$ (top
    panel) and $c_a^2=w_0$ (bottom panel)\label{ddeb_p12_05}}}}
\end{figure}

\hspace*{3cm}\index{large scale structure}Before determination of
PSF parameters let us discuss shortly the gravitational instability
of such scalar field and its impact on the matter clustering.  The
complete system of evolution equations for cosmological
per\-tur\-ba\-ti\-ons of cold dark matter, baryons, massless and
massive neutrinos and radiation based on General Relativity,
differential conservation law and Boltzmann equations is presented
in
\cite{Ma1995,Novosyadlyj2007,Durrer2008}. %
\index{conservation law} %
To understand the gravitational instability of PSF and its impact on
the large scale structure formation in the matter and dark energy
dominated epochs it is enough to analyze the subset of differential
equations (\ref{deltade-Vde})---(\ref{Psi}) with initial
condi\-tions (\ref{dm_init})---(\ref{de_init}) or
(\ref{d_de_s})---(\ref{h_s}) with initial conditions
(\ref{synch_init}).  At $\eta_{init}$ for positive matter density
perturbation\,\footnote{\,In the early Universe for
  superhorizon scale it includes also relativistic components, which
  are density dominating.} ($\delta_m>0$) the gravitational potential
$h_{init}$ in %
\index{synchronous gauge} %
synchronous gauge (or $\Psi_{init}$ in conformal Newtonian gauge) %
\index{conformal Newtonian gauge} %
is nega\-tive and the dark energy density perturbation has opposite
sign ($\delta_{de}<0$) for any $w_0,\,c_a^2<-1$ and $c_s^2>0$ (see
equations (\ref{synch_init})). The absolute values of their
amplitudes in synchronous gauge increase $\propto a$ at superhorizon
stage of evolution, but amplitudes of density perturbations of
phantom scalar field change the sign and decay after entering into
horizon at $\eta\approx k^{-1}$. It is shown in
Fig.~2.\ref{ddeb_p12_05}, where the evolution of Fourier mode
$k=0.05$ Mpc$^{-1}$ of density perturbations for dark matter,
baryons and phantom scalar field is presented for two cases:
$c_a^2<w_0$ and $c_a^2=w_0$.

One can see, that in the case of $c_a^2=w_0$ the absolute value of
initial amplitude of $|\delta_{de}|$ is higher than in the case of
$c_a^2<w_0$, but at the epoch of structure formation and at current
epoch they are essentially lower than $\delta_m$. It means that
perturbations of minimally coupled scalar fields with initial
conditions (\ref{synch_init}) practically do not impact structure
formation in matter components.

\begin{figure}
  \raisebox{0.0cm}{\parbox[b]{4.5cm}{\caption{Evolution of matter density perturbations from Dark Ages to
    current epoch in sCDM, $\Lambda$CDM, QSF+CDM (1:~$w_0=-0.8$,
    $c_a^2=-0.8$; 2:~$w_0=-0.8$, $c_a^2=-0.5$) and PSF+CDM (1:~$w_0=-1.2$, $c_a^2=-1.2$; 2:~$w_0=-1.2$,
     $c_a^2=-1.5$)
    models. Amplitudes are normalized to $0.1$ at $z=10$ ($a=0.1$). In
    the models with dark energy $\Omega_m=0.3$,
    $\Omega_{de}=0.7$\label{fig_dm_2}}}}\hspace*{0.5cm}\includegraphics[width=8cm]{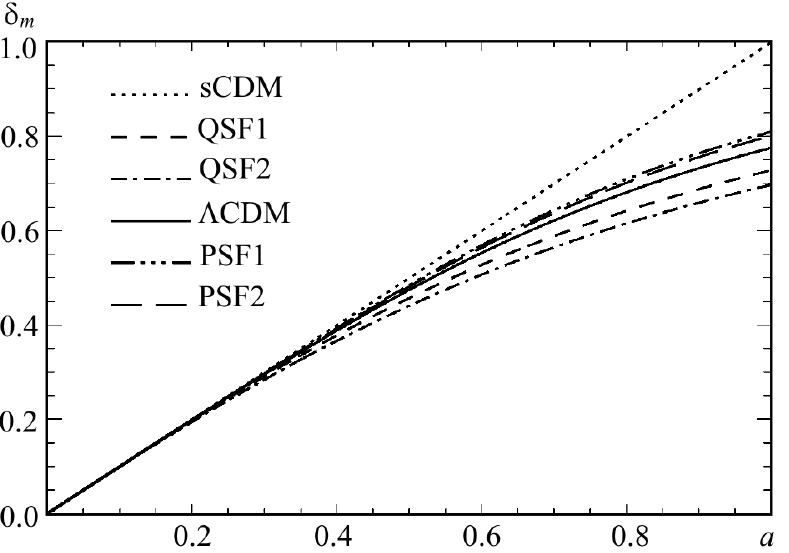}\vspace*{-3mm}
\end{figure}

Nevertheless the parameters of barotropic scalar field can be
constrained by data on the large scale structure of the Universe,
since the rate of increasing of amplitude of matter density
perturbations is enough sensitive to them. It is illustrated in
Fig.~2.\ref{fig_dm_2}, where the matter density evolution,
$\delta_m(a)$, is shown for models with PSF dark energy. In order to
eliminate the $k$-dependence caused by the baryonic component at
small scales and amphasize the influence of dark energy we have
normalized the amplitude of matter density perturbations to 0.1 at
$a=0.1$ (free normalization). Then $\rho_m/\rho_{de}\sim 1000$,
$q\approx0.5$ and amplitudes of all Fourier modes evolve practically
equally. For comparison the same variables for QSF~+~CDM,
$\Lambda$CDM and the standard CDM (sCDM) models are also presented.
So, for the amplitude of large scale structure inhomogeneities the
cosmological model with PSF is distinctive from one with QSF by
$\sim$10\,\% and from $\Lambda$CDM one by few percents at $0\le z\le
1$.


\begin{figure}
\vskip1mm
  \includegraphics[width=13cm]{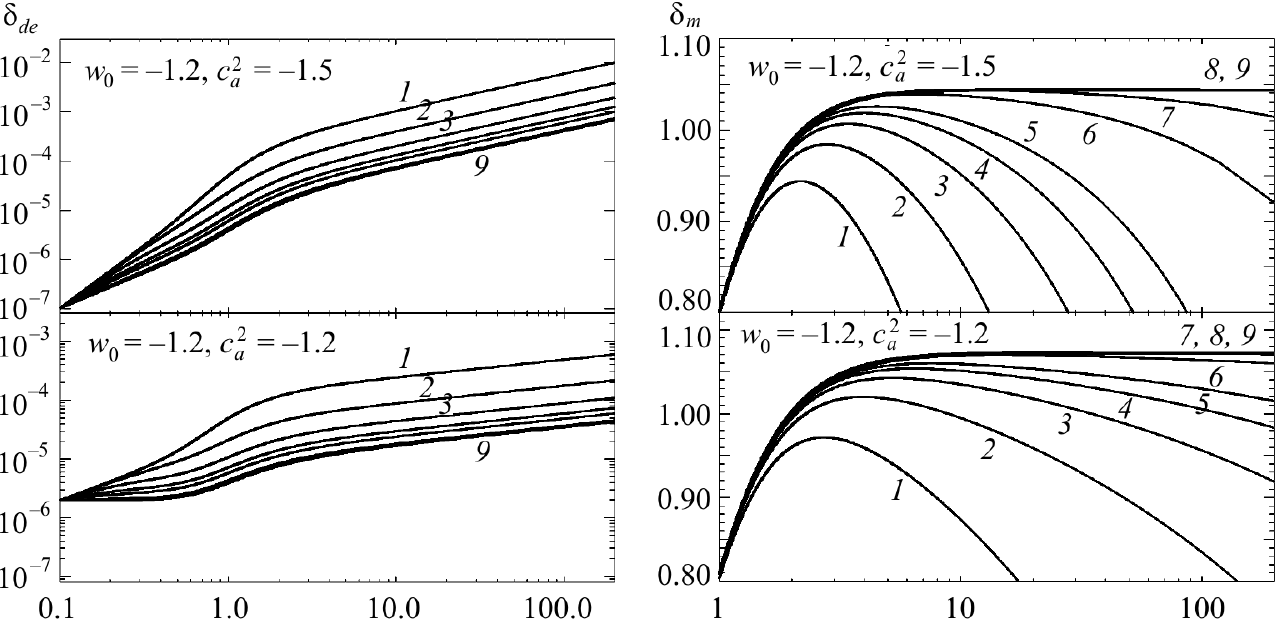}
    \vskip-2mm\caption{Evolution of different Fourier amplitudes of PSF (left
    column) and matter (right column) density perturbations from
    $a=0.1$ to $a=200$ for models with $w_0=-1.2$, $c_a^2=-1.5$ (top
    panels) and $w_0=-1.2$, $c_a^2=-1.2$ (bottom panels). The rest
    parameters are the same as in Figs.2.\ref{ddeb_p12_05} and
   2.\ref{fig_dm_2}. The different lines correspond to wave numbers $k$
    (in Mpc$^{-1}$) as follows: {\it 1}~--- 0.0005, {\it 2}~--- 0.001, {\it 3}~--- 0.0015, {\it 4}~---
    0.002, {\it 5}~--- 0.0025, {\it 6}~--- 0.005, {\it 7}~--- 0.01, {\it 8}~--- 0.05, {\it 9}~--- 0.1~Mpc$^{-1}$. The amplitudes of all $k$-modes of $\delta_{de}$ are
    normalized to
    $0.1\delta_{de}(k=$ $=0.05; a=0.1)/\delta_m(k=0.05; a=0.1)$, the
    amplitudes of all $k$-modes of $\delta_m$ are normalized to 0.1 at
    $a=0.1$}
  \label{dde_p12_k}\vspace*{-2mm}
\end{figure}

Let us analyze the evolution of linear density perturbations in the
future. The first question for elucidation is gravitational
instability of PSF at strongly dark energy dominated epoch. %
\index{dark energy dominated (DED) epoch} %
We have integrated the system of differential equations
(\ref{d_de_s})---(\ref{h_s}) with initial conditions
(\ref{synch_init}) up to $a=200$, when expansion is already
superfast (see Fig.~2.\ref{ah_ph}) and $\rho_{de}/\rho_m\sim
10^{8}$---$10^{10}$.  The results for PSF density perturbations are
shown in Fig.~2.\ref{dde_p12_k} (left column) for different
$k$-modes (0.0005, 0.001, 0.0015, 0.002, 0.0025, 0.005, 0.01, 0.05,
0.1~Mpc$^{-1}$) and two expansion rates, which correspond to models
with $w_0=$ $=-1.2$, $c_a^2=-1.5$ (top panel) and $c_a^2=-1.2$
(bottom panel). One can see, that their amplitudes increase slowly
and the rate depends on background expansion rate as well as on wave
number. In order to visualize this $k$-dependence we remove the
dependence on $k$ caused
by initial power spectrum and transition processes in the early epochs %
\index{power spectrum}%
by renormalization of amplitudes at $a=0.1$ to
$\delta_{de}(k)=0.1\delta_{de}(k=0.05;a=0.1)/\delta_m(k=0.05;a=0.1)$.
So, all $k$-modes of PSF density perturbations in
Fig.~2.\ref{dde_p12_k} have the same amplitudes at $a=$ $=0.1$, but
the ratio of matter to PSF ones is the same as for $k=0.05$
Mpc$^{-1}$ mode at that moment, as computed by CAMB and shown in
Fig.~2.\ref{ddeb_p12_05}. One can see, that rates of increasing of
amplitudes are higher for lower $k$ in the range $a=0.1$---10 and
are practically the same for all $k$-modes at $a>10$:
$\delta_{de}\propto a^{-3(1+c_a^2)/2}$. For the PSF with $w_0=-1.2$
and $c_a^2=-1.5$ the amplitude of $k=0.1$~Mpc$^{-1}$ mode increases
from $a=0.1$ to $a=10$ by 693 times, while the amplitude of
$k=0.0005$~Mpc$^{-1}$ mode increases by 10307 times. For the PSF
with $w_0=c_a^2=-1.2$ these numbers are 9 and 125 correspondingly.
Since the evolution of amplitude of gravitational potential $h$ is
driven by term $\rho_m\delta_m+\rho_{de}(1+3w_{de})\delta_{de}$
(Eq.~\ref{h_s}), shortly after $a=1$ the perturbations of PSF will
become crucial firstly on the largest scales and later on less and
less. They affect the evolution of matter density perturbations,
that is shown in the right panels of Fig.~2.\ref{dde_p12_k} in the
log-norm scales.  At scales with $k\ge 0.05$~Mpc$^{-1}$
(superimposed lines 8, 9 in both panels) the amplitudes of matter
density perturbations in the models with PSF will increase from
$a=1$ to $a=10$ only by $\sim $1.3 times and will freeze at this
value. At these scales in this range of $a$ the influence of PSF
density perturbations on the evolution of matter density ones is
inappreciable. In the $\Lambda$CDM and QSF~+~CDM models all
$k$-modes will evolve similarly as it is shown by line 9. In the
PSF~+~CDM models at scales with $k<0.05$~Mpc$^{-1}$ the effect of
PSF density perturbations on the evolution of matter density ones
becomes important: increasing of amplitude of PSF density
perturbations causes the decaying of matter density ones. The
greater is scale of perturbation, the earlier its amplitude starts
to decay\,\footnote{\,In
  order to visualize this effect in Fig.~2.\ref{dde_p12_k} we have
  normalized all $k$-modes of $\delta_m$ to 0.1 at $a=0.1$.}.

Note, that this decaying of matter density perturbations is caused
solely by influence of phantom scalar field perturbations, not
superfast expansion of background (at $a\sim2$ the rates of
expansion in PSF models are close to ones in $\Lambda$CDM and QSF,
as it can be seen from Figs.~2.\ref{wrode_ph} and 2.\ref{ah_ph}).
Excluding the effect of perturbations, the amplitudes of all
$k$-modes freeze as it is shown by line 9. This is not the beginning
of the Big Rip mentioned
above, but its analog for linear perturbations. %
\index{Big Rip|(} %

\subsection{\!Best-fit parameters of PSF}\label{psf_best-fit}

\hspace*{3cm}\index{best-fit parameters|(}Let us estimate the
parameters of PSF using the same data, method and codes which have
been applied for QSF (see subsection \ref{qsf_best-fit}).  To find
the best-fit values of parameters of cosmological model with PSF and
their confidence limits we perform the
MCMC analysis %
\index{Markov chain Monte Carlo (MCMC)}%
for two combined datasets:\linebreak WMAP7~+ HST + BBN + BAO + SN
SDSS SALT2 and WMAP7~+ HST~+ +~BBN + BAO + SN SDSS MLCS2k2. The
difference in the search procedure consists only in the flat priors
for $w_0$ and $c_a^2$ and starting values for them. In the case of
PSF the priors are as follows: $-2\le w_0\le -1$ and $-2\le c_a^2\le
-1$. Since in
the PSF models with $w_0<c_a^2<-1$ the EoS parameter has second kind %
\index{EoS parameter|(}%
discontinuity at $a<1$, we exclude them from likelihood analysis by
the additional condition $c_a^2\le w_0$.

\begin{table}[t]
\vspace*{-2mm} \noindent\parbox[b]{13cm}{ \caption{\bf The best-fit
values for cosmological parameters of PSF {+} CDM\newline model and their
\boldmath$1\sigma$ limits from the extremal values of the
N-dimensional\newline distribution  determined by the MCMC technique from
the combined\newline datasets WMAP7 {+}HST {+} BBN {+} BAO {+} SN SDSS
SALT2 ($\mathbf{p}_1$)\newline and WMAP7 {+} HST {+} BBN~{+} BAO {+} SN
SDSS MLCS2k2 ($\mathbf{p}_2$).\newline All units and notation are the same
as in Table~2.\ref{tab_qsf}\index{Markov chain Monte Carlo
(MCMC)}\hspace*{2.5cm}\label{tab_phsf}}}\vspace*{2mm}
\tabcolsep7.4pt

\noindent{\footnotesize
\begin{tabular}{|c|c|c||c|c|c|}
    \hline
\rule{0pt}{4mm}{\scriptsize Parameters}&{$\scriptsize
\mathbf{p}_1$}&{\scriptsize $\mathbf{p}_2$} &{\scriptsize
Parameters}&{\scriptsize $\mathbf{p}_1$ }&{\scriptsize
$\mathbf{p}_2$}
\\ [1.0mm]
       \hline
\rule{0pt}{4mm}$\Omega_{de}$&0.72$_{-0.04}^{+0.04}$&0.69$_{-0.04}^{+0.05}$&
$H_0$& 70.4$_{- 3.2}^{+ 4.0}$&67.8$_{- 2.9}^{+4.2}$\\[1.5mm]
$w_0$& --1.043$_{-0.24}^{+0.043}$&--1.002$_{-0.14}^{+0.002}$&$n_s$& 0.96$_{-0.03}^{+0.04}$&0.96$_{-0.04}^{+0.03}$\\[1.5mm]
$c_a^2$& --1.12$_{-0.50}^{+0.12}$&--1.19$_{-0.42}^{+0.19}$& $\log(10^{10}A_s)$& 3.09$_{-0.09}^{+0.09}$&3.11$_{-0.11}^{+0.08}$\\[1.5mm]
10$\omega_b$& 0.223$_{-0.013}^{+0.016}$&0.223$_{-0.013}^{+0.014}$& $\tau_{rei}$&0.085$_{-0.031}^{+ 0.041}$&0.086$_{-0.038}^{+0.036}$\\[1.5mm]
\cline{4-6}
\rule{0pt}{4mm}$\omega_{cdm}$& 0.115$_{-0.010}^{+0.011}$&0.119$_{-0.010}^{+0.009}$&$-\log L$&3864.86&3859.30\\[1.5mm]
    \hline
  \end{tabular}
  }\vspace*{-2mm}
\end{table}

The results of estimation of the PSF parameters jointly with the
minimal set of cosmological parameters for two sets of
observa\-tional data (WMAP7~{+} +~HST {+} BBN {+} BAO {+} SN SDSS
SALT2 and WMAP7 {+}
HST {+} BBN~{+} +~BAO {+} SN SDSS MLCS2k2) are presented in Table~2.\ref{tab_phsf}. %
\index{Multicolor Light Curve Shape (MLCS)}%
We denote the sets of best-fit parameters by $\mathbf{p_1}$ and
$\mathbf{p_2}$. Here $\mathbf{\mathbf{p}_i}=(\Omega_{de}, w_0,
c_a^2, \Omega_b, \Omega_{cdm}, H_0, n_s, A_s, $ $\tau_{rei}$). Both
SN SDSS distance moduli datasets prefer values of $w_0$ slightly
lower than --1. In the past, when $a\rightarrow 0$,
$w_{de}\rightarrow -1$. Hence, the PSFs with parameters
$\mathbf{p}_1$ and $\mathbf{p}_2$ mimic the $\Lambda$-term from the
Big Bang up to the current epoch, but, due to instability of the
value $w_{de}=-1$, even such a small dif\-ference changes
drastically the future fate of the Universe: in $\Lambda$CDM model
the Universe as well as existing structures (in principle) are
time-unlimited, while in the PSF~{+} CDM model it
reaches the Big Rip singularity in finite time, %
\index{singularity} %
\index{Big Rip|)} %
preceded by the destruction of the structure from clusters of
galaxies to elementary particles. More precisely, in the PSF {+} CDM
with parameters $\mathbf{p}_1$ this happens in $\approx$152~Gyrs,
with $\mathbf{p}_2$ in $\approx $594~Gyrs. Long before $t_{BR}$ the
particle horizon\,\footnote{\,At the current epoch $r_p^{(0)}=14260$
Mpc
  in the model with $\mathbf{p}_1$ and 14170~Mpc in the model with
  $\mathbf{p}_2$.} becomes $r_p^{max}\approx18710$~Mpc in model
$\mathbf{p}_1$ and $\approx$19200 in model $\mathbf{p}_2$, just
$\approx$1.3 times larger than the current particle
horizon.\vspace*{-1.0mm}

\section[\!Distinguishing of scalar field models of dark
  energy]{\!Distinguishing of scalar field\\ \hspace*{-0.95cm}models of dark
  energy}\label{disting_sf}

\vspace*{-0.5mm}\hspace*{3cm}The key parameters of barotropic scalar
field are its current density in units of critical one
$\Omega_{de}$, EoS parameter $w_0$, adiabatic sound speed $c_a^2$
and effective sound speed $c_s^2$. Their reliable determination will
unveil the inherent properties of dark energy:
whether it is the cosmological constant (\mbox{$w_0=c_a^2=-1$}) %
\index{cosmological constant} %
or dynamical dark energy in the form of QSF or PSF. The best-fit
values and confidential ranges of cosmological parameters for flat
QSF~{+}~CDM and PSF~{+} CDM models determined by MCMC method on the
base of WMAP7~+~HST {+} BBN {+} BAO {+} SN SDSS SALT2 and WMAP7~{+}
HST {+} BBN {+} BAO {+} SN SDSS MLCS2k2 datasets are presented in
Tables~2.\ref{tab_qsf} and 2.\ref{tab_phsf} correspondingly. Similar
computations have been carried out for $\Lambda$CDM model and the
best-fit values and confidential ranges of cosmological parameters
$\mathbf{l}_1$ and
$\mathbf{l}_2$ are presented in Table~2.\ref{tab_lcdm}. %
\index{Multicolor Light Curve Shape (MLCS)} %

\begin{table}[b!]
\vspace*{-7mm} \noindent\parbox[b]{13cm}{\caption{\bf The best-fit
values for cosmological parameters of \boldmath$\Lambda$CDM\newline model
and their
$1\sigma$ limits from the extremal values of the N-dimensional\newline
 distribution determined by the MCMC technique from the combined\newline datasets WMAP7~{+} HST {+} BBN {+} BAO {+} SN SDSS SALT2
 ($\mathbf{l}_1$)\newline and WMAP7 {+} HST {+} BBN~{+} BAO {+} SN SDSS MLCS2k2 ($\mathbf{l}_2$). %
\index{cosmic microwave background (CMB)}%
    \index{Markov chain Monte Carlo (MCMC)}%
  \newline  All units and notation are the same as in Tables 2.\ref{tab_qsf} and 2.\ref{tab_phsf}. %
  \index{LambdaCDM model ($\Lambda$CDM)}\hspace*{2.5cm}\label{tab_lcdm}}}\vspace*{2mm}
\tabcolsep8.2pt

\noindent{\footnotesize
\begin{tabular}{|c|c|c||c|c|c|}
    \hline
\rule{0pt}{4mm}{\scriptsize Parameters}&{\scriptsize
$\mathbf{l}_1$}&{\scriptsize $\mathbf{l}_2$}&{\scriptsize
Parameters}&{\scriptsize $\mathbf{l}_1$ }&{\scriptsize
$\mathbf{l}_2$ }
\\ [1.0mm]
       \hline
\rule{0pt}{4mm}$\Omega_{\Lambda}$&0.73$_{-0.04}^{+0.03}$&0.70$_{-0.04}^{+0.04}$&$n_s$& 0.97$_{-0.03}^{+0.03}$&0.96$_{-0.03}^{+0.03}$\\[1.5mm]
    10$\omega_b$& 0.226$_{-0.013}^{+0.012}$&0.224$_{-0.013}^{+0.013}$&$\log(10^{10}A_s)$& 3.08$_{-0.09}^{+0.09}$&3.10$_{-0.08}^{+0.08}$\\[1.5mm]
    $\omega_{cdm}$& 0.112$_{-0.008}^{+0.009}$&0.118$_{-0.009}^{+0.008}$& $\tau_{rei}$&0.087$_{-0.031}^{+ 0.041}$&0.082$_{-0.030}^{+0.038}$\\[1.5mm]
\cline{4-6}
\rule{0pt}{4mm}$H_0$& 70.4$_{- 3.4}^{+2.9}$&68.2$_{-
3.1}^{+3.3}$&$-\log
    L$&3864.96&3859.15\\[1.5mm]
    \hline
  \end{tabular}
  }
\end{table}

\begin{figure}
\vskip1mm
  \includegraphics[width=13cm]{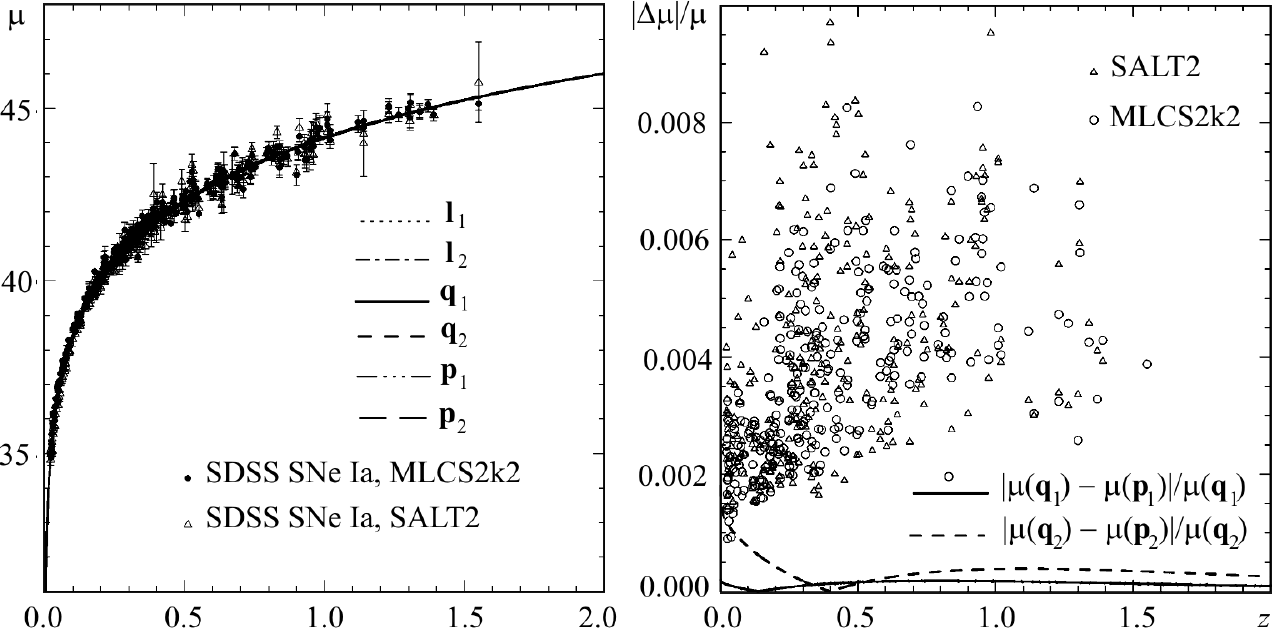}
   \vskip-2mm
   \caption{Left panel: the dependences of distance modulus $\mu\equiv
    m-M$ on redshift $z$ for SNe Ia in the models with best-fit
    parameters $\mathbf{l}_1$, $\mathbf{l}_2$, $\mathbf{q}_1$,
    $\mathbf{q}_2$, $\mathbf{p}_1$ and $\mathbf{p}_2$ (superimposed
    lines) and observational data SDSS SNe Ia (signs). Right panel:
    the relative differences of distance moduli between QSF~+~CDM and
    PSF~+~CDM models with best-fit parameters determined for the same
    datasets (lines) and observational uncertainties for SDSS SNe Ia
    (signs)}
\label{dl_all}
\end{figure}

\index{likelihood function}%
If we compare the maxima of likelihood functions
\mbox{$\chi^2=-\log\,(L_{max})$}, presented in the last rows of
Tables~2.\ref{tab_qsf}---2.\ref{tab_lcdm}, we see that the current
observa\-tional dataset including SN SDSS SALT2 prefers phantom
scalar field model of dark energy, since
$\chi^2_{PSF}<\chi^2_{\Lambda
  CDM}<\chi^2_{QSF}$. The opposite trend is for the dataset including
SN SDSS MLCS2k2, $\chi^2_{QSF}<\chi^2_{\Lambda CDM}<\chi^2_{PSF}$,
so, it prefers quintessential scalar field model of dark energy. The
differences between all $\chi^2$ in the tables are not statistically
significant, but a clear trend supports that.  In the paper
\cite{Kessler2009} the differences between 2 methods of light-curve
\mbox{fitting}, SALT2 and MLCS2k2, are thoroughly analyzed but
convincing arguments for one or the other are not given. So, we can
conclude only that WMAP7\,{+} +~HST~{+}~BBN~{+}~BAO {+} SN SDSS
SALT2 dataset prefers slightly the phantom scalar field model of
dark energy, while WMAP7~{+} HST {+} BBN~{+}~BAO\,+ {+}~SN SDSS
MLCS2k2 dataset prefers slightly the quintessence one, but none of
considered QSF~{+} CDM, $\Lambda$CDM and PSF~+~CDM models with
best-fit parameters has statically significant advantages by current
datasets, used for their determination. All these models match well
the observational data on SNe Ia distance moduli
(Fig.~2.\ref{dl_all}), BAO relative distance measure
(Fig.~2.\ref{rbao_all}), power spectrum of matter density perturbations %
\index{power spectrum}%
(Fig.~2.\ref{pk_all}), CMB temperature fluctuations
(Fig.~2.\ref{cl_all}) and temperature-polarization
(Fig.~2.\ref{cl_all_te}) power spectra. Let us analyze the
possibility to distinguish QSF~+~CDM and PSF~+~CDM by each type of
observational data.

The lines corresponding to different DE models, QSF, PSF and
$\Lambda$, in the left panel of Fig.~2.\ref{dl_all}) look as
perfectly superimposed, that means the complete model degeneracy.
The maximal values of relative differences of SNe Ia distance moduli
$\Delta(m-M)/(m-M)\sim 0.4\,\%$ at low redshifts are for QSF models
with $\mathbf{q}_1$ and $\mathbf{q}_2$ shown in
Fig.~2.\ref{ddl_qsf_sn_sdss}. The similar relative differences
between these dependences for QSF {+} CDM and PSF~+~CDM models with
best-fit parameters determined with the same fitters
($|\mu(\mathbf{q}_i)-\mu(\mathbf{p}_i)|/\mu(\mathbf{q}_i)$) are
$\le$0.1\,\% (right panel of Fig.~2.\ref{dl_all}), that makes these
data still quite useless to set the type of scalar field as dark
energy.

\begin{figure}
 \vskip1mm
  \includegraphics[width=13cm]{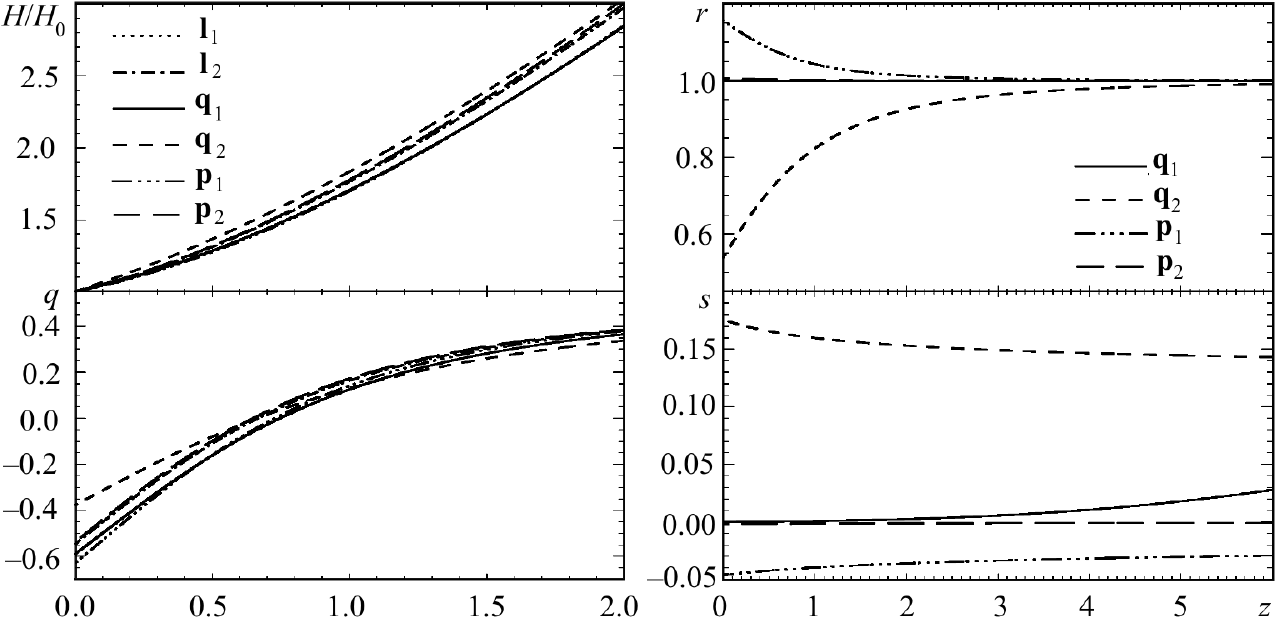}
   \vskip-2mm
  \caption{The dependences of dimensionless parameters describing the
    dynamics of expansion of the Universe on redshift in the models
    with best-fit parameters $\mathbf{l}_1$, $\mathbf{l}_2$,
    $\mathbf{q}_1$, $\mathbf{q}_2$, $\mathbf{p}_1$ and $\mathbf{p}_2$:
    the rate of expansion $H/H_0$ (left panel, top), the deceleration
    parameter $q=-\dot{H}/(aH^2)-1$ (left panel, bottom) and the
    statefinder parameters $r= \ddot{H}/(a^2H^3)+2\dot{H}/(aH^2)+1$
    and $s\equiv (r-1)/3(q-1/2)$ (right panel). %
    For the $\Lambda$CDM model the last two parameters equal 1 and 0
    correspondingly\index{LambdaCDM model ($\Lambda$CDM)}}
  \label{qh_all}\vskip3mm
  \end{figure}

\begin{figure}[h!]
  \includegraphics[width=13cm]{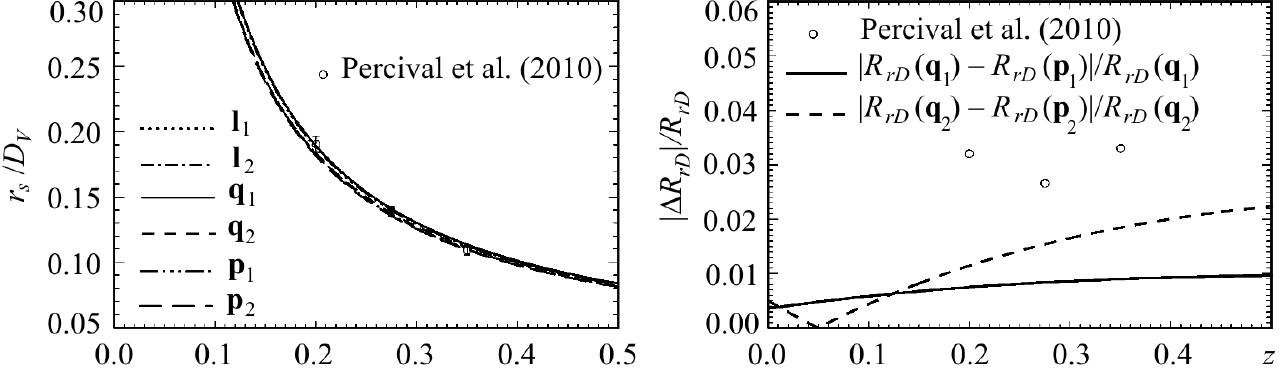}
   \vskip-2mm
  \caption{Left panel: the BAO relative distance measure $R_{rD}\equiv
    r_s(z_{drag})/D_V(z)$ in the cosmological models with best-fit
    parameters $\mathbf{l}_1$, $\mathbf{l}_2$, $\mathbf{q}_1$,
    $\mathbf{q}_2$, $\mathbf{p}_1$ and $\mathbf{p}_2$ (lines) and
    observational data extracted from SDSS DR7 galaxy redshift survey
    \cite{Percival2010} (symbols).  Right panel: the relative
    difference of the BAO distance measure $|\Delta R_{rD}|/ R_{rD}$
    in the models with best fitting parameters $\textbf{q}_i$ and
    $\textbf{p}_i$. Dots show observational 1$\sigma$ relative errors
    of data in the left panel}
  \label{rbao_all}
\end{figure}

\begin{figure}
 \vskip1mm
  \includegraphics[width=13cm]{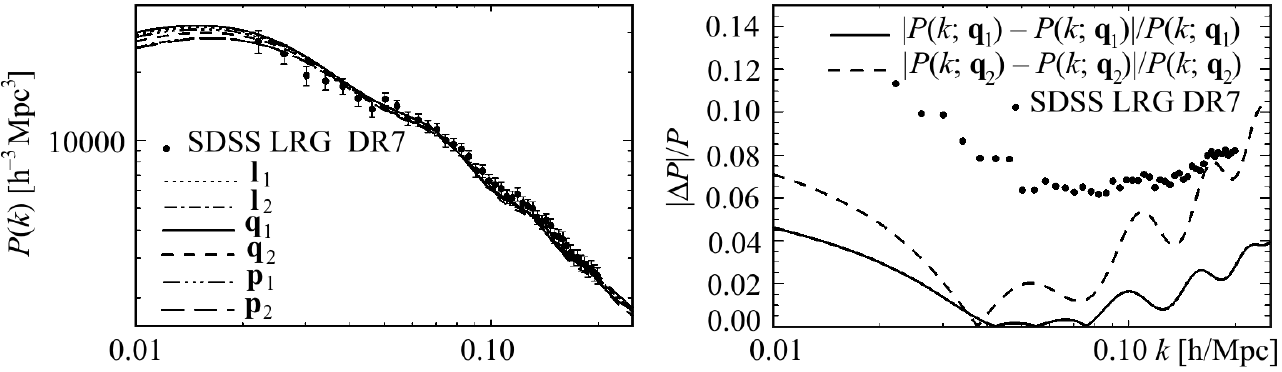}
   \vskip-2mm
  \caption{Left panel: the power spectrum of matter density
    perturbations in the cosmological models with best-fit parameters
    $\mathbf{l}_1$, $\mathbf{l}_2$, $\mathbf{q}_1$, $\mathbf{q}_2$,
    $\mathbf{p}_1$ and $\mathbf{p}_2$. Dots show observational SDSS
    LRG DR7 power spectrum \cite{Reid2010}.  Right panel: the relative
    difference of matter density power spectra $|\Delta P(k)|/P(k)$ in
    the models with best fitting parameters $\textbf{q}_i$ and
    $\textbf{p}_i$. Dots show observational uncertainties (1$\sigma$)
    of SDSS LRG DR7 data \cite{Reid2010}}\vskip3mm
  \index{power spectrum|(}%
  \label{pk_all}
\end{figure}

\begin{figure}
  \includegraphics[width=13cm]{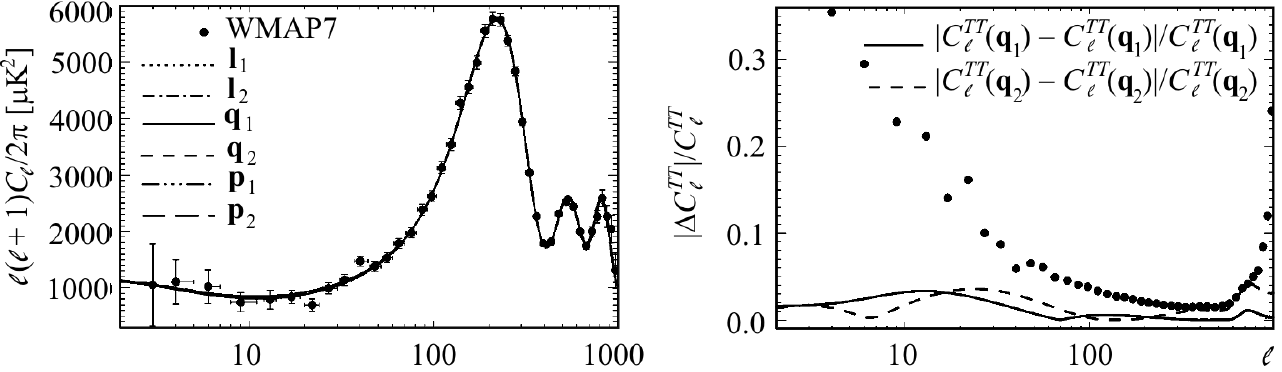}
   \vskip-2mm
  \caption{Left panel: the power spectra of CMB temperature
    fluctuations $\ell(\ell+1)C^{TT}_{\ell}/2\pi$ in the cosmological
    models with best-fit parameters $\mathbf{l}_1$, $\mathbf{l}_2$,
    $\mathbf{q}_1$, $\mathbf{q}_2$, $\mathbf{p}_1$ and $\mathbf{p}_2$
    (superimposed lines) and observational one from WMAP7
    \cite{WMAP7a} (dots). %
    \index{cosmic microwave background (CMB)|(} %
    \index{CMB temperature fluctuations} %
    Right panel: the relative differences of CMB temperature
    fluctuations power spectra $|\Delta C^{TT}_{\ell}|/C^{TT}_{\ell}$
    in the models with best fitting parameters $\textbf{q}_i$ and
    $\textbf{p}_i$ (Tables 2.\ref{tab_qsf} and 2.\ref{tab_phsf}). Dots
    show observational uncertainties (1$\sigma$) of WMAP7
    $\ell(\ell+1)C^{TT}_{\ell}/2\pi$}\vspace*{-1mm}
  \label{cl_all}
\end{figure}

\begin{figure}
 \vskip1mm
  \includegraphics[width=13cm]{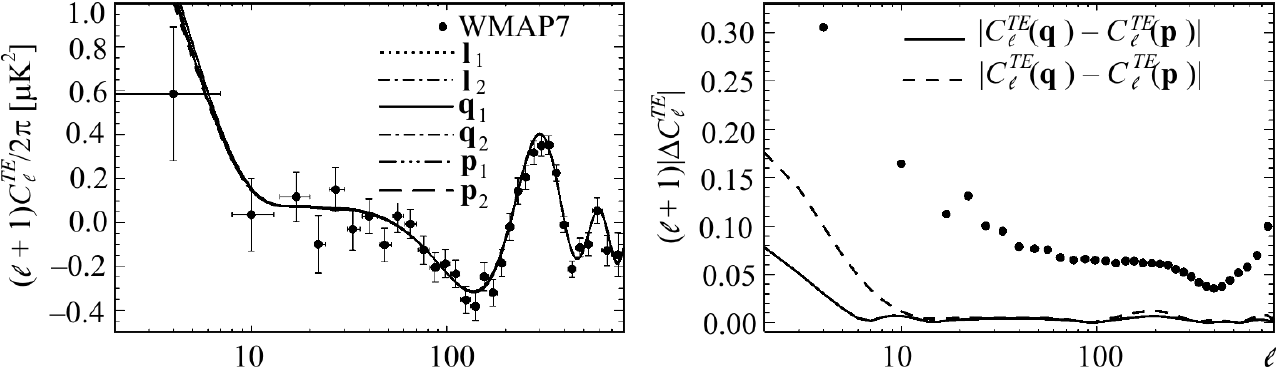}
   \vskip-2.5mm
  \caption{Left panel: the temperature-polarization power spectra of
    CMB $(\ell+1)C^{TE}_{\ell}/2\pi$ in the cosmological models with
    best-fit parameters $\mathbf{l}_1$, $\mathbf{l}_2$,
    $\mathbf{q}_1$, $\mathbf{q}_2$, $\mathbf{p}_1$ and $\mathbf{p}_2$
    (superimposed lines) and observational one from WMAP7
    \cite{WMAP7a} (dots). Right panel: the relative differences of CMB
    temperature-polarization power spectra $|\Delta
    C^{TE}_{\ell}|/C^{TE}_{\ell}$ in the models with best fitting
    parameters $\textbf{q}_i$ and $\textbf{p}_i$ (Tables 2.\ref{tab_qsf}
    and 2.\ref{tab_phsf}). Dots show observational uncertainties
    (1$\sigma$) of WMAP7 $(\ell+1)C^{TE}_{\ell}/2\pi$}
  \label{cl_all_te}\vspace*{-4mm}
\end{figure}

Fortunately, other characteristics of dynamics of expansion of the
Universe, based on the measurements of the first and second time
derivatives of Hubble parameter\,\footnote{\,$\mu(z)$ dependence is
  integral of $1/H$ over redshift (Eq.~\ref{dl}).}  $H(\eta)$, are
essentially more sensitive to the value and time dependence of EoS
parameter. %
\index{EoS parameter|)} %
The dependences of dimensionless parameters describing the dynamics
of expansion of the Universe on redshift, such as the rate of
expansion $H/H_0$, the deceleration parameter $q=-\dot{H}/(aH^2)-1$
and the statefinder parameters \cite{Sahni2003a} $r=
\ddot{H}/(a^2H^3)+2\dot{H}/(aH^2)+1$ and $s\equiv (r-1)/3(q-1/2)$,
are shown in Fig.~2.\ref{qh_all} for the models with best-fit
parameters $\mathbf{l}_1$, $\mathbf{l}_2$, $\mathbf{q}_1$,
$\mathbf{q}_2$, $\mathbf{p}_1$ and $\mathbf{p}_2$. Here the
$r$-parameter is {\it
  jerk} (\ref{j_z}), %
\index{jerk} %
the $s$-parameter is linear combination of $q$ and $r$, not snap
one, discussed in the section \ref{ch1-sec2}. For matter plus dark
energy
dominated epoch they can be presented in our parametrization %
\index{dark energy dominated (DED) epoch} %
(\ref{w_bar}) as follows:
\begin{equation*}
  r=1+4.5(1+w_{de})c_a^2\Omega_{de}(a), \quad s=(1+w_{de})c^2_a/w_{de},
\end{equation*}
where $\Omega_{de}(a)\equiv 8\pi G\rho_{de}(a)/3H^2$.  One can see
that the differences between $s(\textbf{q}_i)$ and $s(\textbf{p}_i)$
at high $z$ as well as between $r(\textbf{q}_i)$ and
$r(\textbf{p}_i)$ at low $z$ are essentially larger than for
parameters $H$ and $q$. Maybe the future high-precision measurements
of dynamics of expansion of the Universe will give possibility to
distinguish the QSF and PSF models of DE.

%
%

The BAO relative distance measure $R_{rD}(z)\equiv
r_s(z_{drag})/D_V(z)$ (see sub\-sec\-tion \ref{ch1-sec3}) extracted
from SDSS DR7 galaxy redshift survey \cite{Percival2010} is matched
well by QSF+CDM, PSF+CDM and $\Lambda$CDM models with best-fit
parameters $\mathbf{q}_1$, $\mathbf{q}_2$, $\mathbf{p}_1$,
$\mathbf{p}_2$, $\mathbf{l}_1$ and $\mathbf{l}_2$ correspondingly.
(Fig.~2.\ref{rbao_all}, left panel). The relative differences of
$R_{rD}(z)$ for models with QSF and PSF are $\le2\,\%$ at $z\le0.4$,
\mbox{while} observational errors are $\sim $3\,\% at $0.2\le z\le
0.35$ (right panel of Fig.~2.\ref{rbao_all}). It means that future
extensive measurements of galaxy space inhomogeneities in principle
can distinguish these scalar fields of dark energy. The power
spectrum of matter density perturbations extracted from luminous red
galaxies of SDSS DR7 catalogue by Reid et al. (2010) \cite{Reid2010}
has not been used for determination of best-fit parameters of
QSF~+~CDM, PSF~+~CDM and $\Lambda$CDM models, however the computed
for them power spectra match it perfectly too. The experimental
errors of its determination are still too large (8---12\,\%) to
distin\-guish between different scalar field models of dark energy.
In addition, at the small scales ($k\ge 0.1$~hMpc$^{-1}$) there are
uncertainties in the computation of the power spectrum associated
with the non-linear evolution of perturbations and unknown type of
dark matter, cold or warm, that is being actively discussed in the
literature.

Above we have discussed the importance of data on CMB anisotropy for %
\index{CMB anisotropy} %
determination of cosmological parameters and, particularly, the dark
energy ones. The key cosmological data now are WMAP all sky maps,
which contain information about position and amplitude of acoustic
peaks at decoupling epoch as well as the amplitudes of large scale
matter density perturbations at the late epoch causing the
integrated
Sachs---Wolfe effect. %
\index{acoustic peaks} %
In Fig.~2.\ref{cl_all} (left panel) the binned power spectrum of
temperature fluctuations extracted from the 7-year WMAP all sky
measurements [93---95] is shown. Its accuracy is maximal (minimal
errors $\sim $2---4\,\%) in the range of acoustic peaks ($\ell\sim
200$---600), this allows the accurate determination of main
cosmological parameters. The maximal accuracy of determination of
parameters of dark energy is achieved when these data are used
together with SNe Ia and BAO relative distance measure or matter
density power spectrum. The computed power spectra
$\ell(\ell+1)C^{TT}_{\ell}/2\pi$ for cosmological models
$\Lambda$CDM, QSF~+~CDM and PSF~+~CDM with best-fit parameters
$\mathbf{l}_1$, $\mathbf{l}_2$, $\mathbf{q}_1$, $\mathbf{q}_2$,
$\mathbf{p}_1$ and
$\mathbf{p}_2$ match well the observational one. %
\index{best-fit parameters|)} %
The relative differences between power spectra in QSF+CDM and
PSF+CDM models, shown in the right panel of Fig.~2.\ref{cl_all}, do
not exceed the relative observational uncertainties of
$\ell(\ell+1)C^{TT}_{\ell}/2\pi$.  Additional constraints on the
cosmological parameters are obtained when CMB polarization data are
included. For the illustration of agreement between theory and
observations in the left panel of Fig.~2.\ref{cl_all_te} the power
spectra of CMB temperature-polarization $(\ell+1)C^{TE}_{\ell}/2\pi$
in the cosmological models with best-fit parameters $\mathbf{l}_1$,
$\mathbf{l}_2$, $\mathbf{q}_1$, $\mathbf{q}_2$, $\mathbf{p}_1$ and
$\mathbf{p}_2$ as well as WMAP7 \cite{WMAP7a} one are presented. All
lines are superimposed at $\ell>10$ with sub-percent accuracy, while
minimal errors of observational power spectrum at high spherical
harmonics are $\sim $4---6\,\% (right \mbox{panel of Fig.~2.\ref{cl_all_te})}. %
\index{power spectrum}%

In the paper \cite{Novosyadlyj2012} we have used the newer data on
SNe Ia distance moduli from SNLS3 compilation (hereafter SNLS3)
\cite{snls3} and Union2.1 compilation (hereafter Union2.1)
\cite{union} together with data on BAO from the WiggleZ Dark Energy
Survey (hereafter WiggleZ) \cite{wigglez}.  The results for the
combined datasets WMAP7 {+} HST {+} BBN {+} BAO {+} WiggleZ {+}
SNLS3 and WMAP7 {+} HST~{+} +~BBN {+} BAO {+} WiggleZ {+} Union2.1
are presented in Tables 2.\ref{tab_snls} and 2.\ref{tab_union}
correspondingly.

Both these combined datasets prefer phantom fields,
$\chi^2_{PSF}<\chi^2_{\Lambda CDM}<$ $<\chi^2_{QSF}$, and for both
the differences of maximum of likelihoods between PSF~{+}~CDM,
$\Lambda$CDM and QSF {+} CDM are still statistically insignificant
too.

The results of determination of cosmological parameters, especially
$H_0$, $\Omega_{de}$, $w_{de}$ and $c_a^2$, presented in Tables
2.\ref{tab_qsf}---2.\ref{tab_lcdm}, also indicate certain
inconsistency or tension between fitters SALT2 and MLCS2k2 applied
to the same SNe
Ia. %
\index{Multicolor Light Curve Shape (MLCS)}%
It was clearly highlighted and analyzed in the papers
\cite{Kessler2009,Bengochea2011}, but up to now we have no decisive
arguments for favor of one of them.

Therefore, any of the used observational data at the current level
of accuracy cannot prefer QSF {+} CDM, PSF {+} CDM or $\Lambda$CDM
at statistically significant level.  In the framework of each of
them the model with best-fit parameters exists, it matches well each
type of
data and all together with close goodness. %
\index{cosmic microwave background (CMB)|)} %
The increasing of accuracy of observational CMB power spectra
\mbox{jointly} with high precision matter density one and SNe Ia
luminosity distance %
\begin{table}[h!]
\vspace*{-4mm} \noindent\parbox[b]{13cm}{\caption{\bf The best-fit
values and 1\boldmath$\sigma$ confidence ranges of the
N-dimensional\newline distribution for the dark energy parameters in
QSF\,{+}\,CDM, $\Lambda$CDM and PSF\,{+}\newline +\,CDM models determined
by the Markov chain Monte Carlo technique\newline using the dataset
WMAP7\,{+}\,HST\,{+}\,BBN\,{+}\,BAO\,{+}\,WiggleZ\,{+}\,SNLS3. The\newline
current Hubble parameter $H_0$ is in units km$\,$s$^{-1}\,
$Mpc$^{-1}$. (From \cite{Novosyadlyj2012})\index{Markov chain Monte
Carlo (MCMC)}}\label{tab_snls}}\vspace*{2mm} \tabcolsep23.9pt

\noindent{\footnotesize
\begin{tabular}{|c|c|c|c|}
    \hline
\rule{0pt}{4mm}{\scriptsize Parameters}&{\scriptsize QSF +
CDM}&{\scriptsize $\Lambda$CDM}&{\scriptsize PSF + CDM}\\ [1.5mm]
       \hline
\rule{0pt}{5mm}$\Omega_{de}$&0.72$_{-0.04}^{+0.04}$&0.73$_{-0.04}^{+0.04}$&0.73$_{-0.04}^{+0.04}$\\[1.5mm]
    $w_0$&--0.994$_{-0.006}^{+0.14}$&--1& --1.10$_{-0.27}^{+0.10}$\\[1.5mm]
    $c_a^2$&--0.72$_{-0.28}^{+0.72}$&--1& --1.29$_{-0.33}^{+0.29}$ \\[1.5mm]
    $H_0$&70.1$_{-4.6}^{+3.6}$&70.3$_{-3.4}^{+3.5}$&71.5$_{-4.1}^{+5.1}$\\[2mm]
     $-\log L$&3947.00&3946.75&3945.98\\[2mm]
    \hline
  \end{tabular}
  }
\end{table}%
\begin{table}[h!]
\vspace*{-8mm} \noindent\parbox[b]{13cm}{\caption{\bf The best-fit
values and
1\boldmath$\sigma$ confidence ranges of the N-dimensional\newline
distribution
   for the dark energy parameters in QSF\,+\,CDM, $\Lambda$CDM and PSF\,+\newline +\,CDM determined by the
    Markov chain Monte Carlo technique using the\newline observational dataset
     WMAP7\,{+}\,HST\,{+}\,BBN\,{+}\,BAO\,{+}\,WiggleZ\,{+}\,Union2.1.\newline The current Hubble parameter $H_0$ is
      in units km\,s$^{-1}$\,Mpc$^{-1}$. (From \cite{Novosyadlyj2012})\label{tab_union}}}\vspace*{2mm}
\tabcolsep23.9pt

\noindent{\footnotesize
\begin{tabular}{|c|c|c|c|}
    \hline
\rule{0pt}{4mm}{\scriptsize Parameters}&{\scriptsize QSF +
CDM}&{\scriptsize $\Lambda$CDM}&{\scriptsize PSF + CDM}\\ [1.5mm]
       \hline
\rule{0pt}{5mm}$\Omega_{de}$&0.72$_{-0.04}^{+0.03}$&0.72$_{-0.04}^{+0.04}$&0.73$_{-0.04}^{+0.03}$ \\[1.5mm]
    $w_0$&--0.995$_{-0.005}^{+0.17}$&--1& --1.13$_{-0.23}^{+0.13}$\\[1.5mm]
    $c_a^2$&--0.55$_{-0.45}^{+0.55}$&--1& --1.54$_{-0.09}^{+0.54}$ \\[1.5mm]
    $H_0$&69.7$_{-4.5}^{+3.1}$&69.8$_{-3.2}^{+3.2}$&71.4$_{-4.4}^{+4.7}$\\[2mm]
       $-\log L$&3800.89&3800.76&3800.48\\[2mm]
    \hline
  \end{tabular}
  }
\end{table}%
measurements will give possibility to establish the dynamical
properties of dark energy and, maybe, its nature. Indeed, in the
papers \cite{Novosyadlyj2011,Sergijenko2011} it was shown that
accuracy of the power spectra that are expected in the experiment
Planck will significantly narrow the allowable range of values of
parameters for the scalar field models of dark
energy.\vspace*{-1.5mm}

\section{\!Summary}
\vspace*{-0.5mm}

\hspace*{3cm}In the chapter the different methods of modeling of
dark energy by single scalar field are analyzed. The main attention
was paid to the scalar field with barotropic equation of state and
classical or tachyonic (DBI) Lagrangians. Such field has only three
free parameters, $\Omega_{de}$, $w_0$ and $c_a^2$, which define
completely its dynamical behavior from the Big Bang to the current
epoch and in the future as well as the dynamics of expansion of the
Universe.  The different combinations of values of $w_0$ and $c_a^2$
from the range [--2, 0] correspond to different types of dark
energy, quintessence and phantom, to different character of
evolution of EoS parameter, \index{EoS parameter}decreasing,
increasing or constant, and, accordingly, different repulsion
properties, \mbox{raising}, re\-ce\-ding or stable. The well studied
$\Lambda$CDM and $w$CDM models of dark energy are partial cases of
this barotropic scalar field model when $w_0=-1$ and $w_0=c_a^2$
correspondingly. If $c_a^2=0$ and $-1<w_0<-1/3$ then the scalar
field \mbox{with} tachyon Lagrangian at the Beginning mimics
dust-like matter, but later becomes quintessential dark energy with
decreasing EoS parameter, \mbox{which} becomes $w_0$ at current
epoch and goes asymptotically to --1, mimicking $\Lambda$-term in
far future. When Lagrangian of scalar field is classical then dark
energy is never dust-like with the same time evolution of $w_{de}$
since its effective sound speed is always $c_s^2=1$.  If
$-1<w_0<c_a^2<0$ then we have quintessential scalar field model of
dark energy with decreasing EoS parameter and raising repul\-sion:
$w_{de}$ monotonically decreases from $c_a^2$ at the Big Bang to
$w_0$ now and will continue decreasing to --1 up to time infinite.
For the other order of inequality, $-1<c_a^2<w_0<0$, we have
opposite evolution of $w_{de}$: it monotonically increases from
$c_a^2$ in the Big Bang to $w_0$ now and will continue increase in
the future, will become zero, then positive and will go to
discontinuity of second kind since the energy density of scalar
field, $\rho_{de}$, during decreasing will pass zero. In this model,
in contrast to previous ones, the accelerated expansion will not
last forever~--- it will be changed by the decelerated one, will
reach the turnaround point, then will start to collapse and finally
it will end in the Big Cranch singularity. The accelerated expansion
in all these cases is caused by slow roll of scalar field to minimum
of its potential. The energy density of quintessence scalar field
decreases at the stage of expansion of the Universe much more slowly
than for any other component, that explains its crucial role in
the later epochs. %
\index{singularity} %

If $w_0$ and $c_a^2 <-1$, then the scalar field with barotropic EoS
has phan\-tom properties~--- its density monotonically increases,
that causes the su\-per\-ex\-po\-nen\-tial expansion of the Universe
and reaching of
the Big Rip singularity in finite time. %
\index{Big Rip|(} %
The field variable $\phi$ and potential $U$ are always real when the
Lagrangian, classical or tachyonic, has the opposite sign before
kinetic term. Moreover, the potential and kinetic term are always
positive also if $c_a^2<w_0<-1$. The superfast expansion in this
model is caused by roll up of the phantom scalar field to maximum of
its potential which is $+\infty$, but is reached in finite time. The
repulsion properties of PSF increase and in finite time they reach
and outmatch firstly the gravitational force, then electromagnetic
forces and finally strong interactions. All bound structures in the
Universe~--- galaxies, stars, planets, atoms and protons~--- will be
ripped in finite time. A distinctive feature of this class of
phantom dark energy is that in a
general case it mimics the positive or negative cosmological constant %
\index{cosmological constant} %
($w_{de}(a)\rightarrow-1$) in the early epoch ($a\rightarrow0$),
that is defined by inequality $c_a^2<w_0$ or $w_0<c_a^2$
correspondingly. The inherent ``phantom'' property, when field
starts from zero value of energy density, occurs only in the
particular case $w_0=c_a^2$. In the opposite asymptotic range,
$a\rightarrow +\infty$, the equation of state parameter $w_{de}(a)$
asymptotically goes to $c_a^2$, that defines the physical meaning of
the term ``adiabatic sound speed'' in the case of phantom scalar
field with barotropic EoS.

Here it has been shown also that dynamics of expansion of the
Universe depends on values of parameters $\Omega_{de}$, $w_0$ and
$c_a^2$ of scalar field with barotropic EoS and this fact allows to
estimate them using the cosmological test based on relations
``luminosity distance~--- redshift'' and ``angular diameter
distance~--- redshift'' for different
classes of astrophysical objects. %
\index{angular diameter distance} %
The ``strength'' of scalar field repulsion and its time dependence,
defined by these parameters, affect also the formation of large
scale structure of the Universe via growth factor of linear matter
density perturbations.
\index{large scale structure}%
\index{growth factor}%

The important feature of the scalar fields is their gravitational
instability, which was carefully analyzed in the chapter. It was
shown that in the case of adiabatic initial conditions for matter
density perturbations and subdominant asymptotic ones for dark
energy the amplitudes of scalar field density perturbations in the
past and current epochs are essentially lower than matter density
ones. However, they leave their subtle ``fingerprints'' via
gravitational interaction between dark energy and dark matter scalar
density perturbations. It can be used as additional source of
information about the nature of dark energy since it is sensitive to
another important parameter of scalar field model of DE~--- the
effective sound speed $c_s^2$. It was shown that the effect is more
noticeable for smaller values of $c_s^2$.  In the case of phantom
scalar field %
\index{phantom scalar field (PSF)|)} %
the distinctive property is that the perturbation of energy density
and gravitational potential have the same sign. It causes the decay
of large scale linear density perturbations of matter long before
the Big Rip singularity.
\index{Big Rip|)} %

So, the determination of values of $\Omega_{de}$, $w_0$ and $c_a^2$
for barotropic scalar field with given Lagrangian by comparison of
theoretical predictions with observational data gives possibility to
define the type and main dynamical properties of dark energy. %
\index{best-fit parameters} %
The best-fit parameters of quintessence and phantom scalar fields
with
barotropic EoS have been determined jointly with all relevant %
\index{Markov chain Monte Carlo (MCMC)}%
cosmological parameters by MCMC method using the datasets WMAP7 {+}
HST {+} +~BBN~{+} BAO {+} SN SDSS SALT2 and WMAP7 {+} HST {+} BBN
{+}
BAO {+} +~SN SDSS MLCS2k2. %
\index{Multicolor Light Curve Shape (MLCS)}%
It was shown that the dataset including SNe Ia distance moduli
obtained with SALT2 fitter prefers slightly the phantom model of
dark energy ($\Omega_{de}=0.72\pm0.04$,
$w_0=-1.043^{+0.043}_{-0.24}$, $c_a^2=-1.12_{-0.50}^{+0.12}$), while
the dataset with the same SNe Ia but obtained with MLCS2k2 prefers
slightly the quintessence model ($\Omega_{de}=0.70\pm0.05$,
$w_0=-0.83^{+0.22}_{-0.17}$, $c_a^2=-0.88_{-0.12}^{+0.0.88}$).
However, the difference of the maximum likelihoods between them is
statistically insignificant. The same conclusions apply to the
datasets WMAP7 {+}\linebreak +~HST~{+} BBN {+} BAO {+} WiggleZ {+}
SNLS3 and WMAP7 {+} HST {+} BBN~{+} +~BAO {+} WiggleZ {+} Union2.1.
The possibility of distinguishing between quintessence and phantom
scalar fields by current and expected datasets is analyzed and it is
concluded that more accurate future observations will enable
us to do that.  \index{scalar field|)} %

\newpage


\setcounter{chapter}{2}
\chapter{\label{chap:KK}KALUZA---KLEIN MODELS}\markboth{CHAPTER 3.\,\,Kaluza---Klein models$_{ }$}{3.\,\,Kaluza---Klein models$_{ }$}
\thispagestyle{empty}\vspace*{-12mm}

\begin{wrapfigure}{l}{2.6cm}
\vspace*{-5.5cm}{\includegraphics[width=3.0cm]{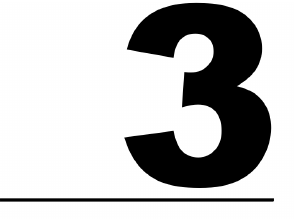}}\vskip17.2cm
\end{wrapfigure}
\vspace*{10mm}

 \setcounter{section}{1} \vspace*{-3mm}
\hspace*{3cm}\section*{\hspace*{-3cm}3.1.\,\,Introduction$_{}$}\label{zh-Intr}\begin{picture}(10,10)
\put(123,-135){\bfseries\sffamily{133}}
\end{picture}

\vspace*{-1.1cm} \noindent \index{LambdaCDM model ($\Lambda$CDM)}As
we have already pointed out in the previous chapters, the
lar\-ge-sca\-le dynamics of the observable part of our pre\-sent
time Universe is well described by the $\Lambda$CDM model with the
four-dimensional Friedmann---Robertson---Walker (FRW)
metric.\index{FRW metric} However, it is possible that space-time at
short (Fermi or Planck) distances might have a dimen\-sio\-na\-lity
of more than four and possess a rather complex topology \cite{zh1}.
This idea takes its origin from the pio\-nee\-ring papers by Kaluza
and Klein (KK) \cite{zhKK,zhKK_1}.\index{Kaluza---Klein models} They
were \mbox{first} who indicated how to unify gravity with
elect\-ro\-mag\-ne\-tism. Moreover, Klein introduced the idea of
com\-pac\-ti\-fi\-ca\-tion of the extra (\mbox{fifth}) dimension
which provides the natural explanation of the extra dimension
unob\-ser\-va\-bi\-li\-ty (see [332---334] for review of the KK
models).

String theory \cite{zh1a,zh1a_1} and its recent generalizations~---
p-bra\-ne, M- and F-theory \cite{zh1b,zh1c}~--- widely use this
con-\linebreak cept and give it a new foundation.  The most
con\-sis\-tent for\-mu\-la\-ti\-ons of these theories are possible
in space-ti\-mes \mbox{with} critical dimensions \mbox{$ D_c>4 $},
for example, in \mbox{string} theory there are \mbox{$D_c=26$} or 10
for the bo\-so\-nic and su\-per\-sym\-met\-ric version,
respectively. In KK models, it is supposed that a
$D$-di\-men\-sio\-nal manifold $M$ undergoes a ``spon\-ta\-neous
com\-pac\-ti\-fi\-ca\-tion'' [339---342]: \mbox{$M\to
M^4\times B^{D-4}$}, where $%
M^4$ is the four-di\-men\-sio\-nal external spa\-ce-ti\-me and
$B^{D-4}$ is a compact internal space. So it is natural to consider
cosmological consequences of such com\-pac\-ti\-fi\-ca\-tions.
\mbox{With} this in mind, we shall investigate
mul\-ti\-di\-men\-sio\-nal cos\-mo\-lo\-gi\-cal models (MCM) with
the topology\index{multidimensional cosmological models}
\begin{equation}
  \label{zh1.1}M=M_0\times M_1\times \mbox{...} \times M_n ,
\end{equation}
where $M_0$ denotes the $(D_0=d_0+1)$-dimensional (usually $d_0=3$)
external space-time and $M_i$ $(i=1,\mbox{...}, n)$ are
$d_i$-dimensional internal spaces. To\linebreak make the extra
dimensions unobservable at the present time these internal spaces
$M_i$ have to be compact and reduced to scales near the Fermi length
$L_{\rm F} \sim 10^{-17}$\,cm.

This scale is dictated by the level of energies achieved up to now
in the accelerators and which were less than 1\,TeV.  In KK models,
each type of particles of the Standard Model has infinite number of
partners (the Kaluza---Klein tower of massive states commonly
referred
to KK particles (see e.g. \cite{zhCambridge})) %
\index{Kaluza---Klein particles} %
\mbox{with} masses inversely proportional to the size of the
internal space. KK particles were not detected in all previous
experiments. This result can be interpreted in such a way that
energies up to 1\,TeV were not enough to excite them. In other
words, the size of the internal spaces should be of the order or
less than ${(1\,\text{TeV})}^{-1}\sim 10^{-17}$\,cm. On the other
hand, it was not also registered any reaction where usual particles
disappear in the extra dimensions. For considered models, it occurs
if wavelength of these particles are bigger than the size of the
extra dimensions. In the case of TeV energy particles, their
wavelength is of order of $L_{\mathrm{F}}$. We again arrive at the
indicated above restriction on the size of the internal space.  To
discover the extra dimensions, it is necessary to increase energies
of accelerators.

Large Hadronic Collider in CERN should reach 14\,TeV. Therefore, if
the internal spaces have the Fermi length size, we shall detect them
in this experiment. Thus, in subsequent sections, we assume that
scale factors $a_i$ of the internal spaces should be of order or
less of $L_{\rm F}$.

There is no problem in constructing compact spaces with a positive
cur\-va\-tu\-re \cite{zhp1, zhp2}.  (For example, every Einstein
manifold with constant positive cur\-va\-tu\-re is necessarily
compact \cite{zhp3}.) However, Ricci-flat spaces and negative
cur\-va\-tu\-re spaces also can be compact. This can be achieved by
appropriate periodicity conditions for the coordinates [347---351]
or, equivalently, through the action of discrete groups $\Gamma$ of
isometries related to face pairings and to the manifold's topology.
For example, three-dimensional spaces of constant negative
cur\-va\-tu\-re are isometric to the open, simply connected,
infinite hyperbolic (Lobachevsky) space $H^3$ \cite{zhp1, zhp2}.
However, there exist also an infinite number of compact, multiply
connected, hyperbolic coset manifolds $H^3/\Gamma $, which can be
used for the construction of FRW metrics with negative
cur\-va\-tu\-re [347, 349]. These manifolds are built from a
fundamental polyhedron (FP) in $H^3$ with faces pairwise identified.
The FP determines a tessellation of $H^3$ into cells that are
replicas of the FP, through the action of the discrete group
$\Gamma$ of isometries \cite{zh4}. The simplest example of
Ricci-flat compact spaces is given by $D$-dimensional tori $T^D
={\mathbb{R}}^D/\Gamma$. Thus internal spaces may have non-trivial
global topology, being compact (i.e. closed and bounded) for any
sign of \mbox{spatial cur\-va\-tu\-re.}

In the cosmological context, internal spaces can be called
compactified, when they are obtained by a compactification in the
usual mathematical under\-standing (e.g. by replacements of the type
$\mathbb{R} ^D\rightarrow S^D$, $\mathbb{R} ^D\rightarrow \mathbb{R}
^D/\Gamma $ or $H^D\rightarrow$ $\rightarrow H^D/\Gamma $) with
additional contraction of the sizes to Fermi scale.  There are a
number of interesting cosmological and astrophysical exact solutions
(see, e.g., Refs.~[352---367]). For the most of exact cosmological
solutions, the internal spaces have dynamical behavior (see section
\ref{zh-Sp} below). However, the physical constants that appear in
the effective four-dimensional theory after dimensional reduction of
an originally higher-dimensional model are the result of integration
over the extra dimensions. If the volumes of the internal spaces
would change, so would the observed constants (see
Eq.~(\ref{zh4.39}) below). Because of limitation on the variability
of these constants (see the relevant discussion in the subsequent
sections), the internal spaces should be static or at least slowly
variable since the time of primordial nucleosynthesis and, as we
mentio\-ned above, their sizes are of the order of the Fermi length.
Obviously, such compactifications have to be stable against small
fluctuations of the sizes (the scale factors $a_i$) of the internal
spaces.  This means that the effective potential of the model
obtained under dimensional reduction to a four-dimensional effective
theory should have minima at $a_i\sim L_{\rm F}$
$(i=1,\mbox{...},n)$. These minima play the role of the cosmological
constant (dark energy!) in our
effective four- \mbox{dimensional Universe.} %
\index{cosmological constant} %

Additionally, small excitations of a system near a minimum can be
obser\-ved as massive scalar fields in the external space-time.
These scalar \mbox{fields}\index{scalar field} very weakly interact
with the Standard Model (SM) particles. Therefore, they belong to a
class of the dark matter particles.


In the next section, we consider in detail the procedure of the
dimensio\-nal reduction of KK models. Before that, it worthy to note
that the idea of the multidimensionality of our Universe has
received a great deal of renewed attention over the last few years
within the ``brane-world'' description of the Universe. In this
approach the $SU(3)\times SU(2)\times U(1)$ Standard Model fields,
related to usual four-dimensional physics, are localized on a
three-dimensional space-like hy\-per\-sur\-fa\-ce (brane) whereas
the gravitational field propagates in the whole (bulk) space-time.
This approach is different from KK one and will be the subject of
the third part of our book.

\newpage

\section[\!Dimensional reduction$_{ }$]{\!Dimensional reduction, stable\\
  \hspace*{-0.95cm}compactification, gravitational excitons,\\ \hspace*{-0.95cm}effective cosmological
  constant}\vspace*{-1mm} \label{zh-Exci}



\subsection{\!General setup\label{zh-general}}\vspace*{-1mm}

\hspace*{3cm}\index{gravitational excitons}In this section we
present a sketchy outline of the basics of dimensional reduction and
gravitational excitons. A more detailed description can be found,
e.g., in the papers [368---370].

\index{space-time manifold} %
Let us consider a multidimensional space-time manifold ${ M}$ with
warped product topology (\ref{zh1.1}) and metric\vspace*{-4mm}
\begin{equation}
  \label{zh2.1}g=g_{MN}(X)dX^M\otimes dX^N=g^{(0)}+\sum_{i=1}^ne^{2\beta
    ^i(x)}g^{(i)},
\end{equation}\vspace*{-3mm}

\noindent where $x$ are some coordinates of the
$(D_0=4)$-dimensional manifold ${ M}_0 $ and
\begin{equation}
  \label{zh2.2}g^{(0)}=g_{\mu \nu }^{(0)}(x)dx^\mu \otimes dx^\nu .
\end{equation}\vspace*{-4mm}

Let further the internal factor manifolds ${ M}_i$ be
$d_i$-dimensional warped Einstein spaces with warp factors
$e^{\beta^i
  (x)}$ and metrics $g^{(i)}=g^{(i)}_{m_in_i}(y_i)\, dy_i^{m_i}\otimes
$ $\otimes\, dy_i^{n_i}$, i.e.,\vspace*{-3mm}
\begin{equation}
  \label{zh2.3}R_{m_in_i}\left[ g^{(i)}\right] =\lambda
  ^ig_{m_in_i}^{(i)},\quad m_i,n_i=1,\mbox{...},d_i
\end{equation}\vspace*{-5mm}

\noindent and\vspace*{-3mm}
\begin{equation}
\label{zh2.4}R\left[ g^{(i)}\right] =\lambda ^id_i\equiv R_i.
\end{equation}\vspace*{-4mm}

In the case of constant cur\-va\-tu\-re spaces parameter $\lambda^i$
are normalized as $\lambda^i=k_i(d_i-1)$ with $k_i = 0,\pm 1$.
Let $b_i \equiv e^{\beta^i}$ and $b_{(0)i} \equiv e^{\beta^i_0}$
denote the scales factors of the internal spaces $M_i$ at arbitrary
and at present time. (Obviously, to get scale factors in dimensional
units, we must multiply them by the Planck length $L_{\rm Pl} \sim
10^{-33}$cm, e.g. $b_{(0)i} = L_{\rm Pl} e^{\beta^i_0}$.)  Then the
total volume of the internal spaces at the present time is given
by\vspace*{-1mm}
\begin{equation}
  \label{zh2.5}
  V_{D'}  \equiv V_I\times v_0 \equiv \prod_{i=1}^n\int\limits_{M_i}d^{d_i}y
  \sqrt{|g^{(i)}|} \times \left( \prod_{i=1}^n e^{d_i\beta^i_0} \right) = V_I \times
  \prod_{i=1}^n b_{(0)i}^{d_i} ,
\end{equation}\vspace*{-3mm}

\noindent where $D' =\sum^n_{i=1}d_i$ is the total number of the
extra dimensions. The factor $V_I$ is dimensionless and defined by
geometry and topology of the internal spaces. We also denote the
deviations of the internal scale factors from their present day
values:\vspace*{-3mm}
\begin{equation}
  \label{zh2.6}
  \tilde \beta^i = \beta^i - \beta^i_0 .
\end{equation}\vspace*{-5mm}

For the demonstration of the dimensional reduction %
\index{dimensional reduction} %
we consider a multidimensional action with a bare $D$-dimensional
cosmological constant $\Lambda$ and a minimal scalar field $\Phi$: %
\index{cosmological constant} \vspace*{-1mm}
\begin{equation}
\label{zh2.7}
\begin{array}{c}
 \displaystyle S  = \frac 1{2\kappa_D
    ^2}\int\limits_Md^Dx\sqrt{|g|}\left\{ R[g]-2\Lambda \right\}+S_m + S_{YGH} ,
    \\[6mm]
 \displaystyle S_m  = -\frac12 \int\limits_Md^Dx\sqrt{|g|}\left[ g^{MN}\partial_M
    \Phi
    \partial_N \Phi + 2 U(\Phi) \right]\!,
\end{array}
\end{equation}
where $\kappa^2_D$ is a $D$-dimensional gravitational
constant\,\footnote{\,$\kappa^2_D$ is connected with the
multidimensional
  fundamental mass scale $ M_{\rm Pl(D)}$ and the surface area $S_{D-1}=2\pi
  ^{(D-1)/2}/\Gamma [(D-1)/2]$ of a unit sphere in $D-1$ dimensions by
  the relation $$ \kappa^2_D = 2S_{D-1} / M_{\rm Pl(D)}^{D'+2}.$$\label{kappa D}}
and $S_{YGH}$ is the standard York---Gibbons---Hawking boundary term
\cite{zhYork,zhGH}. The field $\Phi$ itself can be considered in its
zero-mode approximation. This means that $\Phi$ and its potential
$U(\Phi)$ depend only on the coordinates of the external space, and
the dimen\-sio\-nal reduction of the model can be performed by a
simple integration over the coordinates of the internal spaces.
Moreover, we usually assume that $\Phi$ depends only on time to be
in concordance with the homogeneity and isotropy \mbox{of the
Universe.}

Scalar cur\-va\-tu\-re of the metric (\ref{zh2.1}) reads
\[
R[g] = R[g^{(0)}]+\sum_{i=1}^{n} e^{-2\beta^i} R[g^{(i)}] -
\]\vspace*{-3mm}
\begin{equation}
  \label{zh2.8}
   -
 \sum_{i,j=1}^{n} (d_i\delta_{ij}+d_i d_j) \ g^{(0)\mu\nu}
  \frac{\partial\beta^i}{\partial
    x^\mu} \frac{\partial\beta^j}{\partial x^\nu} -2 \sum_{i=1}^{n} d_i \Delta[ g^{(0)}] \beta^i .
\end{equation}
where $\Delta[\bar g^{(0)}]$ is the Laplace---Beltrami operator on
$M_0$:
\begin{equation}
  \label{2.9}
  \Delta[g^{(0)}]= \frac{1}{\sqrt{|g^{(0)}|}} {\frac{\partial}{\partial x^{\mu}}} \left(\! \sqrt{|g^{(0)}|}\,
    g^{(0)\mu\nu} {\frac{\partial}{\partial x^{\nu}}}\! \right)\! ,
\end{equation}

\noindent Taking into account the relation
\[
     { \frac{1}{\kappa_D^2} \int\limits_{M} {\rm d}^Dx \sqrt{|g|}
      \sum_{i=1}^{n} d_i \Delta[g^{(0)}] \beta^i =
    }  \]\vspace*{-3mm}
\[
     = \frac{\mu}{\kappa_D^2} \sum_{i=1}^{n} d_i \int\limits_{
      M_0}{\rm d}^{D_0}x \sqrt{| g^{(0)}|} \prod_{l=1}^n e^{d_l\beta^l}
    \frac{1}{\sqrt{| g^{(0)}|}}
    \frac{\partial}{\partial x^{\lambda}} \left(\! \sqrt{| g^{(0)}|}\, g^{(0)\lambda\nu}
      \frac{\partial}{\partial x^{\nu}}\beta^i \!\right) =  \]
\[
     = \frac{1}{\kappa^2_0} \sum_{i=1}^{n} d_i \int\limits_{ M_0}{\rm d}^{D_0}x
    \Bigg[
      \frac{\partial}{\partial x^{\lambda}} \left(\! \sqrt{|g^{(0)}|}\,  g^{(0)\lambda\nu} \prod_{l=1}^n e^{d_l\beta^l} \frac{\partial}{\partial x^{\nu}}\beta^i \!\right)  -
      \]
\[
     -  \sqrt{|g^{(0)}|}\, g^{(0)\lambda\nu} \frac{\partial
        \beta^i}{\partial x^\nu} \prod_{l=1}^n e^{d_l\beta^l}
      \sum_{j=1}^{n} d_j \frac{\partial \beta^j}{\partial x^\lambda}
    \Bigg] =  \]
\begin{equation}     = S_{YGH} -\frac{1}{\kappa^2_0} \int\limits_{ M_0}{\rm d}^{D_0}x \sqrt{|
      g^{(0)}|}\, \prod_{l=1}^n e^{d_l\beta^l} \sum_{i,j=1}^{n} d_i
    d_j g^{(0)\lambda\nu} \frac{\partial \beta^i}{\partial x^\lambda}
    \frac{\partial \beta^j}{\partial x^\nu}  , \label{zh2.10}
\end{equation}

\noindent the action (\ref{zh2.7}) can be easily reduced to the
following expression
\[
\displaystyle  S= \frac 1{2\kappa _0^2}\int\limits_{M_0}d^{D_0}x\sqrt{|g^{(0)}|}%
  \prod_{i=1}^ne^{d_i\tilde \beta ^i}\Bigg\{\! R\left[ g^{(0)}\right]
    -G_{ij}g^{(0)\mu\nu }\partial _\mu \tilde \beta ^i\,\partial _\nu
    \tilde \beta ^j+
\]\vspace*{-3mm}
\begin{equation}
\label{zh2.11} \displaystyle  +   \sum_{i=1}^n \tilde R_i
e^{-2\tilde \beta^i}-2\Lambda -
    g^{(0)\mu \nu}\kappa^2_D
    \partial_{\mu} \Phi \partial_{\nu} \Phi -2\kappa^2_D
    U(\Phi)\!\Bigg\}\!  ,
\end{equation}
where the notations $\tilde R_i := R_i e^{-2\beta^i_0}$,
$G_{ij}=\delta_{ij}d_i-d_i d_j$ are used. Here, $D_0$-dimensional
gravitational constant $\kappa^2_0 \equiv 8\pi G_N =8\pi/M_{\rm
Pl(4)}^2$ ($G_N$ is the Newton gravitational constant and $M_{\rm
Pl(4)}$ is the four-dimensional Planck mass)\linebreak \mbox{is
given as}
\begin{equation}
  \label{zh2.12}
  \kappa^2_0 = \kappa^2_D/V_{D'} .
\end{equation}

Action (\ref{zh2.11}) of the four-dimensional effective model is
written in Brans---Dicke frame, i.e., it has the form of a
generalized Brans---Dicke theory. As next step, we remove the
explicit dilatonic coupling term in (\ref{zh2.11}) by conformal
transformation\vspace*{-3mm}
\begin{equation}
  \label{zh2.13}
  g_{\mu \nu }^{(0)}= \Omega^2 \tilde
  g_{\mu \nu }^{(0)} := {\left( \prod_{i=1}^ne^{d_i\tilde \beta
        ^i}\!\right) } ^{-2/(D_0-2)} \tilde g_{\mu \nu }^{(0)}
\end{equation}
and obtain the effective action in the Einstein frame
\[
S=\frac 1{2\kappa _0^2}\int\limits_{M_0}d^{D_0}x\sqrt{|\tilde
    g^{(0)}|}
    \bigg\{\! \tilde R\left[ \tilde g^{(0)}\right] -\bar G_{ij}
      \tilde g^{(0)\mu \nu }\partial _\mu \tilde \beta ^i\,
      \partial _\nu \tilde \beta ^j -
\]\vspace*{-3mm}
\begin{equation}
  \label{zh2.14}
 - \tilde g^{(0)\mu \nu}\kappa^2_D
      \partial_{\mu} \Phi \partial_{\nu} \Phi - 2U_\mathrm{eff}
    \!\bigg\}\!.
 \end{equation}
\index{midisuperspace metric} %
The tensor components of the midisuperspace metric (target space
metric on $\mathbb{R} _T^n$) $\bar G_{ij}\ (i,j=1,\mbox{...} ,n)$,
its inverse metric $\bar G^{ij} $ and the effective
potential\linebreak are given as\vspace*{-3mm}
\begin{gather}
  \label{zh2.15}\bar G_{ij}=d_i\delta _{ij}+\frac 1{D_0-2}d_id_j,
  \quad \bar G^{ij}=\frac{\delta ^{ij}}{d_i}+\frac 1{2-D}
  \intertext{ and }
  \label{zh2.16}
  U_\mathrm{ eff}[\tilde \beta ,\Phi ] = {\left(
      \prod_{i=1}^ne^{d_i\tilde \beta ^i}\!\right) }^{- \frac{2}{(D_0-2)}}\left[
    -\frac 12\sum_{i=1}^n\tilde R_ie^{-2\tilde \beta ^i}+\Lambda
    +\kappa_D^2 U(\Phi )\right]\!.
\end{gather}

It is obvious that the internal spaces can stabilize if the
effective potential (\ref{zh2.16}) has at least one minimum with
respect to the scale factors $\tilde \beta^i$.  Because the
conformal transformation (\ref{zh2.13}) was performed only with
respect to the external metric $g^{(0)}$, the stability of the
internal space configurations does not depend on the concrete choice
of the frame (Einstein or Brans---Dicke). In general, the effective
potential $U_{\mathrm{eff}}$ can have more than one minimum so that
transitions between these minima should be possible. In the
following we consider models with the internal space scale factors
localized at their present day values: ${\tilde \beta^i} =0,\;
\left. \partial U_\mathrm{
    eff}/\partial \tilde \beta ^i\right| _{\tilde \beta^i =0}=0$.

With the help of a regular coordinate transformation $\psi =Q\beta
,\; \beta =Q^{-1}\psi$ midisuperspace metric (target space metric)
$\bar G$ can be transformed to a pure Euclidean form:
$\kappa_0^{-2}\bar G_{ij}d\beta ^i\otimes d\beta ^j=\sigma
_{ij}d\psi ^i\otimes d\psi ^j=\sum_{i=1}^nd\psi ^i\otimes d\psi ^i,$
$ \sigma =$ $=\mathrm{ diag\ }(+1,+1,\mbox{...} ,+1)$.  An
appropriate transformation to normal modes $ \psi ^j=$ $=Q_i^j\beta
^i$ can be found e.g. in \cite{zhGZ1}. In the special case of only
one internal space ($n=1$), this procedure reduces to a simple
rescaling\,\footnote{\,\label{rescaling}The relation between $\tilde
  \beta^1$ and $\psi^1$ is determined up to sign. For definiteness, we
  chose the minus sign.}
\begin{equation}
  \label{zh2.17}
  \tilde \beta^1 = - \kappa_0 \sqrt{\frac{D_0-2}{d_1(D-2)}}\psi^1  ,
\end{equation}

It is usually assumed that metric $\tilde g^{(0)}$ in action
(\ref{zh2.14}) is the Friedmann---Robertson---Walker one. Then, the
dynamics
of scale factor and scalar fields %
\index{scalar field|(} %
for arbitrary form of potential $U_\mathrm{ eff}$ can be described
with the help of numerical calculation of the system of the first
order ordinary differential equations (ODEs) (A.13)---(A.16).

\index{stable compactification} %
Below, we show that the stabilization of the internal spaces in
model with the minimal coupled scalar field takes place if scalar
field is in its minimum position $\Phi_0$ too.  For small
fluctuations of the normal modes and scalar field in the vicinity of
the minimum of the effective potential, action (3.15) reads
\[
S = \frac{1}{2\kappa _0^2}\int \limits_{M_0}d^{D_0}x \sqrt
    {|\tilde g^{(0)}|}\left\{\!\tilde R\left[\tilde g^{(0)}\right] -
      2\Lambda
      _\mathrm{ eff}\!\right\} -
\]\vspace*{-3mm}
  \begin{equation}
    \label{zh2.18}
    - \frac{1}{2}\int \limits_{M_0}d^{D_0}x \sqrt {|\tilde
      g^{(0)}|}\left\{\sum_{i=1}^n \left( \tilde g^{(0)\mu \nu} \psi
        ^i_{,\mu}\psi^i_{,\nu} + m_i^2\psi ^i\psi ^i\right) + \tilde
      g^{(0)\mu \nu} \phi_{,\mu} \phi_{,\nu} + m_{\phi}^2 \phi \phi
    \!\right\}\!,
  \end{equation}
where $\Lambda_\mathrm{ eff}\equiv U_\mathrm{ eff}(\tilde
\beta^i=0,\Phi=\Phi_0)$ plays the role of a $D_0$-dimensional
effective cosmological constant, %
\index{cosmological constant|(} %
$m_i^2$ and $m_{\phi}^2$ are mass squared of the normal modes and
scalar field, respectively, and for convenience we use the
normalizations for scalar field fluctuations: $\sqrt{V_{D^{\prime
    }}}(\Phi -\Phi_0) \equiv \phi$. In the case of one internal space
\begin{equation}
  \label{zh2.19}
  m_1^2 = \frac{D_0-2}{d_1(D-2)}\left. \frac{\partial^2 U_\mathrm{ eff}}{\partial
      {(\tilde
 \beta^1)}^2}\right|_{\tilde \beta^1 =0,\Phi=\Phi_0}\!.
\end{equation}

Summarizing this section, we conclude that conformal zero-mode
excitations of the internal factor spaces $M_i$ have the form of
massive scalar \mbox{fields} developing on the background of the
external space-time $M_0$. In analogy\linebreak \mbox{with} excitons
in solid state physics (excitations of the electronic subsystem of a
crystal), we called these conformal excitations of the internal
spaces {\itshape
  gravitational excitons} \cite{zhGZ1}. %
\index{gravitational excitons} %
Later, since Refs.~\cite{zhsub-mill3,zhsub-mill2} these geometrical
mo-\linebreak duli excitations are also known as radions. Within the
framework of multidimensional cosmological models such excitations
were investigated in [375---378]. Obviously, positive
$\Lambda_\mathrm{ eff}$ plays the role of the {\bfseries dark
  energy} %
\index{dark energy in KK models} %
in our Universe and weakly interacting gravexcitons form the
{\bfseries dark matter} (see the\linebreak \mbox{following sections}). %
\index{dark matter in KK models} %

Now we consider some specific examples of stable compactification.

\subsection{\!Stable compactification\\ \hspace*{-1.2cm}with minimal
  scalar fields\label{zh-scalar}}

\hspace*{3cm}\index{effective potential}This model is described by
effective action (\ref{zh2.14}). It is clear now that stabilization
of the internal spaces can be achieved if the effective potential
$U_\mathrm{ eff}$ (\ref{zh2.16}) has a minimum with respect to
fields $\tilde \beta^i$ (or fields $\psi^i$). Let us find conditions
which ensure a minimum at $\tilde \beta^i = 0$.

The extremum condition yields:
\begin{equation}
  \label{zh2.20}
  \left.\frac{\partial U_\mathrm{
        eff}}{\partial \tilde \beta^k}\right|_{\tilde \beta^i =0} =0
  \Longrightarrow \tilde R_k = -\frac{d_k}{D_0-2}\left( \sum_{i=1}^n
    \tilde R_i -2(\Lambda + \kappa^2_D U(\Phi) )\!\right)\! .
\end{equation}

The left-hand side of this equation is a constant but the right-hand
side is a dynamical function. Thus, stabilization of the internal
spaces in such type of models is possible only when the effective
potential has also a minimum with respect to the scalar field $\Phi$
(in Ref.~\cite{zhBZ} it was proved that for this model the only
possible solutions with static internal spaces correspond to the
case when the minimal coupled scalar field is in its extremum
position too). Let $\Phi_0$ be the minimum position for field
$\Phi$.  {}From the structure of the effective potential
(\ref{zh2.16}) it is clear that minimum positions of the potentials
$U_\mathrm{ eff}[\tilde \beta ,\Phi]$ and $U(\Phi)$ with respect to
field $\Phi$ coincide with each other:
\begin{equation}
  \label{zh2.21}
  \left. \frac{\partial U_\mathrm{ eff}}{\partial \Phi}\right|_{\Phi_0} = 0
  \Longleftrightarrow \left. \frac{\partial U(\Phi )}{\partial \Phi }\right|_{\Phi_0} = 0 .
\end{equation}
Hence, we should look for parameters which ensure a minimum of
$U_\mathrm{ eff}$ at the point ${\tilde \beta^i = 0, \Phi =
  \Phi_0}$. Eqs.~\ref{zh2.20} show that there exists a fine tuning %
\index{fine tuning} %
condition for the scalar cur\-va\-tu\-res of the internal spaces:
\begin{equation}
  \label{zh2.22}
  \frac{\tilde R_k}{d_k} = \frac{\tilde R_i}{d_i} , \quad (i,k = 1,\mbox{...} ,n) .
\end{equation}

\noindent Introducing the auxiliary quantity
\begin{equation}
  \label{zh2.23}
  \tilde \Lambda \equiv \Lambda + \left. \kappa^2_D U(\Phi)\right|_{\Phi_0} \! ,
\end{equation}
we get the useful relations
\begin{equation}
  \label{zh2.24}
  \Lambda_\mathrm{ eff} :=\left. U_\mathrm{ eff}\vphantom{\int} \right|_ {\tilde \beta^i  =0, \Phi = \Phi_0} = \frac{D_0-2}{D-2}\, \tilde \Lambda = \frac{D_0-2}{2}\, \frac{\tilde R_k}{d_k} ,
\end{equation}
which show that $\mbox{sign}\, \Lambda_\mathrm{ eff} =\mbox{sign}\,
\tilde \Lambda =\mbox{sign}\, R_k$. As we already mentioned above,
$\Lambda_\mathrm{ eff}$ plays the role of an effective cosmological
constant in the external space-time. For the masses of the normal
mode
excitations of the internal spaces (gravitational excitons) %
\index{gravitational excitons} %
and of the scalar field near the extremum position we obtain
respectively \cite{zhGZ1}:
\begin{equation}
\label{zh2.25}
\begin{array}{c}
 \displaystyle m_1^2  = \mbox{...} =m_n^2 = -\frac{4\Lambda_\mathrm{
      eff}}{D_0-2}=-2\frac{\tilde
    R_k} {d_k} > 0 ,  \\[5mm]
 \displaystyle m_{\Phi}^2  := \left. \frac{d^2 U(\Phi )}{d \Phi^2}
  \right|_{\Phi_0}\! .
\end{array}
\end{equation}

These equations show that for our specific model a global minimum
can only exist in the case of compact internal spaces with negative
cur\-va\-tu\-re $R_k <0$ $ (k=1,\mbox{...} ,n)$. The effective
cosmological constant is negative also: $\Lambda_\mathrm{ eff} <0$.
Obviously, in this model it is impossible to trap the internal
spaces at a minimum of $U_\mathrm{ eff}$ if they are tori ($\tilde
R_i = 0$) because for Ricci-flat internal spaces the effective
potential has no minimum at all. Equations (\ref{zh2.24}) and
(\ref{zh2.25}) show also that a stabilization by trapping takes
place only for $\tilde \Lambda <0$. This means that the minimum of
the scalar field potential should
be negative $U(\Phi_0)<0$ for non-negative bare cosmological constant %
\index{cosmological constant|)} %
$\Lambda \ge 0$ or it should satisfy inequality $\kappa^2_D
U(\Phi_0)<|\Lambda |$ for $\Lambda <0$. In paper \cite{zhGZ(PRD2)},
it is shown a possibility of the early inflation in this model.
However, because of the negative sign of $\Lambda_\mathrm{ eff}$,
the configurations with stabilized extra dimensions do not provide a
late-time acceleration of our Universe in the model with minimal
scalar field.  This configurations are asymptotically
anti-De~Sitter. Therefore, there is no dark energy in this model. To
get dark energy, it is necessary to include additional matter. It
may shift a minimum of the effective potential from negative to
positive values. Different forms of matter can be described with the
help of a \mbox{perfect fluid.}


\subsection{\!Perfect fluid: no-go theorem\label{zh-perfect}}

\hspace*{3cm}\index{no-go theorem}In conventional cosmology matter
fields are taken into ac\-count in a phenomenological way as a
perfect fluid with equal pressure in all three spatial directions.
It provides homogeneous (if energy density and pressure depends only
on time) and isotropic picture of the Universe. In
mul\-ti\-di\-men\-sio\-nal case we generalize this approach to a
$m$-component perfect fluid with ener\-gy-mo\-men\-tum tensor
\cite{zhKZ,zhCQG1996,zhIv2,zhIv4}
  \begin{gather}
    \label{zh2.26}
    T_N^M=\sum_{c=1}^m{T^{(c)}}^M_N , \\
    \label{zh2.27} {T^{(c)}}^M_N = \mathrm{ diag\ } (\, -\rho
    ^{(c)}(\tau ), \underbrace{P_0^{(c)}(\tau ),\mbox{...},P_0^{(c)}(\tau
      )}_\text{$d_0$ times}%
    ,\mbox{...} , \underbrace{P_n^{(c)}(\tau ),\mbox{...} ,P_n^{(c)}(\tau
      )}_\text{$d_n$ times} ) .
  \end{gather}
The conservation equations we impose on each component separately
\begin{equation}
  \label{zh2.28} {T^{(c)}}^M_{N;M} = 0.
\end{equation}

To investigate dynamical behavior of our Universe and internal
spaces for a model with such perfect fluid, the metric (\ref{zh2.1})
should be also written in homogeneous form:
\[
g= g^{\, (0)}(x)+\sum_{i=1}^n e^{2\beta ^i(\tau )}g^{(i)}(y)\equiv
\]\vspace*{-3mm}
\begin{equation}
\label{zh2.29}
   \equiv  -e^{2\gamma (\tau )}d\tau \otimes d\tau + e^{2\beta
    ^0(\tau )}q^{(0)}(\vec{x}) + \sum_{i=1}^n e^{2\beta ^i(\tau)}g^{(i)}(y)  ,
\end{equation}
where $q^{(0)}$ is a metric of the constant cur\-va\-tu\-re space:
$R\left[
  q^{(0)}\right] =kd_0(d_0-1)$ with $k = 0,\pm 1$. The choice of the
function $\gamma (\tau)$ defines different gauges, e.g. the
synchronous time gauge %
\index{synchronous time gauge} %
$\gamma =0$ or the conformal time gauge %
\index{conformal time gauge} %
$\gamma (\tau) =\beta^0(\tau)$, etc. In what follows, we use the
notations $a \equiv e^{\beta^0}$ and $b_i \equiv e^{\beta^i}$ $ (i
=1,\mbox{...} ,n)$ to describe scale factors of the external and
internal spaces, respectively.

Denoting by an overdot differentiation with respect to time $\tau$,
the conservation equations (\ref{zh2.28}) for the tensors
(\ref{zh2.27}) read
\begin{equation}
  \label{zh2.30}
  \dot\rho ^{(c)}+\sum_{i=0}^nd_i\dot \beta ^i\left( \rho ^{(c)}+P_i^{(c)}\!\right) =0 .
\end{equation}

\noindent If the pressures and energy density are related via
equations of state
\begin{equation}
\label{zh2.31} P_i^{(c)}=\left(\! \alpha _i^{(c)}-1\!\right) \rho
^{(c)},\quad i=0,\mbox{...} ,n,\quad c=1,\mbox{...} ,m ,
\end{equation}
then Eq.~(\ref{zh2.30}) has the simple integral
\begin{equation}
  \label{zh2.32}
  \rho ^{(c)}(\tau )=A^{(c)}a^{-d_0 \alpha_0^{(c)}}\times \prod_{i=1}^n b_i^{-d_i\alpha
    _i^{(c)}} \!,
\end{equation}
where $A^{(c)}$ is the constant of integration.

To investigate the problem of the stable compactification, it is
helpful to use the equivalence between the Einstein equations %
\index{Einstein equations} %
and the Euler---Lagrange equations for Lagrangian obtained by
dimension reduction of the action (\ref{zh2.7}) with
\begin{equation}
  \label{zh2.33}
  S_m = - \int\limits_Md^Dx\sqrt{|g|} \rho ,
\end{equation}
where $\rho $ is given by Eq.~(\ref{zh2.32}) (see
\cite{zhKZ,zhCQG1996,zhIv2,zhIv4} for details). This equivalence
takes place for homogeneous model (\ref{zh2.29}).  However, we can
generalize it to the inhomogeneous case allowing inhomogeneous
fluctuations $\tilde \beta^i (x)$ $ (i=1,\mbox{...} ,n)$ over stably
compactified background $\beta_0^i = \mbox{const}$ (see
Eq.~(\ref{zh2.6})). It can be easily seen that the dimensional
reduction of action (\ref{zh2.7}) with the matter term
(\ref{zh2.33}) results in effective theory (\ref{zh2.14}) (where we
should drop the scalar field) with the effective potential
\begin{equation}
  \label{zh2.34}
  U_\mathrm{ eff}={\left( \prod_{i=1}^ne^{d_i\tilde \beta ^i}\!\right) }^{\!- 2/(D_0-2)}\left[
    -\frac 12\sum_{i=1}^n\tilde R_ie^{-2\tilde \beta ^i}+\Lambda_D +\kappa
    ^2_D\sum_{c=1}^m\rho ^{(c)}\right]\! ,
\end{equation}
where $\rho^{(c)}$ is defined by Eq.~(\ref{zh2.32}). If we suppose
that
the external space-time metric in the Einstein frame has also FRW form: %
\index{FRW metric} %
\begin{equation}
  \label{zh2.35}
  \tilde g^{(0)}=\Omega ^{-2} g^{(0)}=\tilde g_{\mu \nu }^{(0)}dx^\mu \otimes dx^\nu
  :=-e^{2\widehat{ \gamma }}d\widehat{ \tau} \otimes d\widehat{ \tau} + e^{2\widehat{\beta} ^0(x)}q^{(0)} ,
\end{equation}
which results in the following connection between the external scale
factors in the Brans---Dicke frame $a\equiv e^{\beta ^0}$ and in the
Einstein frame $\tilde a \equiv e^{\widehat{ \beta }^0}$:
\begin{equation}
  \label{zh2.36}
  a={\left( \prod_{i=1}^ne^{d_i\tilde \beta ^i}\!\right) }^{\!- 1/(D_0-2)}\tilde a,
\end{equation}
then, expression (\ref{zh2.32}) for $\rho^{(c)}$ can be rewritten in
the form:
\begin{equation}
  \label{zh2.37}
  \kappa^2_D \rho ^{(c)}= \kappa^2_0 \rho^{(c)}_{(d_0)}\prod_{i=1}^n
  e^{-\xi_i^{(c)}\tilde
    \beta^i} ,
\end{equation}
where\vspace*{-3mm}
\begin{equation}
  \label{zh2.38}
  \rho^{(c)}_{(d_0)} = \tilde A^{(c)} \tilde a^{-d_0\alpha_0^{(c)}} , \quad \tilde A^{(c)} = A^{(c)} V_I \prod_{i=1}^n b_{(0)i}^{d_i(1-\alpha_i^{(c)})}
\end{equation}
and\vspace*{-3mm}
\begin{equation}
  \label{zh2.39}
  \xi_i^{(c)} = d_i\left(\! \alpha _i^{(c)}-\frac{\alpha _0^{(c)}d_0}{%
      d_0-1}\!\right)\! .
\end{equation}\vspace*{-3mm}

\noindent It can be easily verified that $\tilde A^{(c)}$ has
dimension $\mbox{cm}^{d_0\alpha_0^{(c)}-D_0}$.

Thus, the problem of stabilization of the extra dimensions is
reduced now to search of minima of the effective potential
$U_\mathrm{ eff}$ with respect to the fluc\-tuations $\tilde
\beta^i$:\vspace*{-1mm}
\begin{equation*}
  \left.\frac{\partial U_\mathrm{ eff}}{\partial \tilde \beta^k}\right|_{\tilde
    \beta =0} = 0  ,
\end{equation*}\vspace*{-5mm}

\noindent implying\vspace*{-3mm}
\begin{equation}
\label{zh2.40} \tilde R_k = -\frac{d_k}{D_0-2}\left[
  \sum_{i=1}^n \tilde R_i -2\Lambda_D \right]\! + \kappa^2_0
\sum_{c=1}^m \rho^{(c)}_{(d_0)}\left(\!\xi^{(c)}_k +
  \frac{2d_k}{D_0-2}\!\right)\!, \; k = 1,\mbox{...} ,n .
\end{equation}
The left-hand side of this equation is a constant but the right-hand
side is a dynamical function because of dynamical behavior of the
effective $d_0$-dimensional energy density $\rho^{(c)}_{(d_0)}$.
Thus, we arrived at the following {\itshape no-go theorem}
\cite{zhZhuk-no-go}: \vspace*{2mm}

{\itshape Multidimensional cosmological Kaluza---Klein models with
the
  perfect fluid as a matter source do not admit stable
  compactification of the internal spaces with exception of two
  special cases:} %
\index{Kaluza---Klein models} \vspace*{-2mm}
\begin{align}
 \hspace*{-0.4cm} &\mbox{\,\,I.}  \quad \alpha^{(c)}_0 = 0, \quad \forall \;
  \alpha_i^{(c)} ,  \quad i=1,\mbox{...} ,n,\quad c=1,\mbox{...} ,m . \label{zh2.41} \\
  \hspace*{-0.4cm}&\mbox{II.}  \quad \xi_i^{(c)} = -\frac{2d_i}{d_0-1} \Longrightarrow
  \begin{cases}
    \alpha _0^{(c)} = \displaystyle\frac 2{d_0}+ \frac{d_0-1}{d_0}\alpha ^{(c)} ,\\[2ex]
    \alpha _i^{(c)} = \alpha ^{(c)},\quad i=1,\mbox{...} ,n, \quad
    c=1,\mbox{...} ,m \label{zh2.42} .
  \end{cases}\!\!\!\!
\end{align}\vspace*{-3mm}

\noindent First case corresponds to vacuum in the external space
$\rho^{(c)}_{(d_0)} = \tilde A^{(c)}=\mbox{const}$ and arbitrary
equations of state in the internal spaces.  Some bulk matter can
mimic such behavior, e.g. vacuum fluctuations of quantum fields
(Casimir effect) \cite{zhGZ1,zhGKZ}, monopole form fields
\cite{zhGZ1,zhGMZ2} and gas of branes \cite{zhKaya}.

In the second case, the energy density in the external space is not
a constant but a dynamical function with the following
behavior:\vspace*{-1mm}
\begin{equation}
  \label{zh2.43}
  \rho _{(d_0)}^{(c)}(\tilde a)=\tilde A^{(c)}\frac 1{\tilde a^{2+(d_0-1)\alpha ^{(c)}}}
  \Longrightarrow  \rho _{(3)}^{(c)} = \tilde A^{(c)}\frac 1{\tilde a^{2(1 +\alpha^{(c)})}} .
\end{equation}\vspace*{-5mm}

\noindent\index{equation of state (EoS)}The corresponding equation
of state is:\vspace*{-1mm}
\begin{equation}
  \label{zh2.44}
  P_{(d_0)}^{(c)} = (1/3)(2\alpha^{(c)}-1)\rho _{(d_0)}^{(c)} \Longrightarrow
  P_{(3)}^{(c)}= (\alpha_0^{(c)}-1)\rho _{(3)}^{(c)} ,
\end{equation}\vspace*{-5mm}

It can be easily seen from Eq.~(\ref{zh2.37}) that in the case of
stabilized internal spaces (i.e. $\tilde \beta^i =0$) $\rho
_{(d_0)}^{(c)} = \rho ^{(c)}V_{D'}$ where the internal space volume
$V_{D'}$ is defined in Eq.~(\ref{zh2.5}). Similar relation takes
place for $P_{(d_0)}^{(c)}$ and $P_{0}^{(c)}$: $ P_{(d_0)}^{(c)} =$
$= P_{0}^{(c)}V_{D'}$.  Therefore, second case corresponds to
ordinary matter in our three-dimensional space. For example, for
$d_0=3$, besides the exotic case of gas of cosmic strings with
$\alpha^{(c)}=0$, the choice $\alpha ^{(c)}=1/2$ provides dust in
our space: $\alpha_0^{(c)}=1  \rightarrow  P_{(3)}^{(c)}=0$ and
equation of state $\alpha_i^{(c)}=1/2  \rightarrow
P_i^{(c)}=-(1/2)\rho^{(c)}$ in the internal spaces.  It is worth
noting that these equations are exactly the black string/branes
equations of state (see section \ref{zh-KK problem}).

It is clear that the cases I and II are the necessary but not
sufficient conditions for stabilization. As we shall show below,
stability is ensured by the matter from the first case with a proper
choice of the parameters of models. The matter related to the second
case provides the standard evolution of the Universe and does not
spoil the stabilization.

The no-go theorem (the case II) clearly shows that the condition of
the internal space stabilization requires the violation of symmetry
(in terms of equations of state) between our three dimensions and
the extra dimensions. The need for such a violation is especially
seen in the example of radiation. It is well known that radiation
satisfies the equation of state $P=(1/3)\rho$. If we assume equality
of all dimensions and allow light to move around all
multidimensional space, then equation of state will be
$P=(1/D)\rho$, which apparently contradicts the observations for
$D>3$. Therefore, radiation should not move in the extra dimensions.
Exactly this situation we have in case II. If we take
$\alpha^{(c)}=1$, then we obtain the usual equation of state for
radiation in our Universe $\alpha_0^{(c)}=4/3  \rightarrow
P_{(3)}^{(c)}=(1/3)\rho _{(3)}^{(c)}$ and dust in the internal
space: $\alpha_i^{(c)}=1  \rightarrow  P_i^{(c)}=0$. The latter
means that the light does not move in the extra dimensions. Such
situation is realized if light is localized on a brane
\cite{zhRubakov}.


\subsection{\!Conventional cosmology\\ \hspace*{-1.2cm}from multidimensional models\label{zh-conventional}}

\hspace*{3cm}Now, we want to present toy models with stabilized
internal spaces and the standard Friedmann-like behavior of the
external space (our Universe).  These models give a possibility to
demonstrate the typical problems of KK multidimensional cosmological
models.


\label{zh-fine tuning}{\bf\itshape Dark energy from extra
  dimensions: fine tuning problem.}
\index{fine tuning} %
\index{dark energy in KK models} %
Let us consider a model where multicomponent perfect fluid is the
combination of the cases I and II. To be more precise, additionally
to the the perfect fluid of the type II, we endow our model with the
monopole form fields \cite{zhGZ1,zhGMZ2}:\vspace*{-2mm}
\begin{equation}
  \label{zh2.45}
  S_m = -\frac12 \int\limits_M d^Dx \sqrt{|g|}\sum_{i=1}^n
  \frac{1}{d_i!}  \left(\!F^{(i)} \!\right)^2 = - \int\limits_M d^Dx \sqrt{|g|}\sum_{i=1}^n
  \frac{f_i^2}{b_i^{2d_i}} ,
\end{equation}
where $f_i \equiv \mbox{const}$ are arbitrary constants of
integration (free parameters of the model) and for real form fields
$f^2_i > 0$
(see, e.g., Eqs.~(\ref{zh5.57})---(\ref{zh5.60})). %
\index{fields of forms} %
Comparison of this expression with Eqs.~(\ref{zh2.33}) and
(\ref{zh2.32}) shows that such monopole form fields are equivalent
to $n$-component perfect fluid with $\alpha_0^{(c)} =0 ,$
$\alpha^{(c)}_i = 2\delta^c_i $, $c,i = 1,\mbox{...} ,n$, i.e.
belong to the case I.

Without loss of generality, we can perform our analysis in the case
of one internal space $n=1$.  Then, the effective potential
(\ref{zh2.34}) for such combined model undergoes the following
separation:
\begin{equation}
  \label{zh2.46}
 U_\mathrm{ eff} =  \underbrace{{\left(\! e^{d_1\tilde
            \beta^1}\!\right)}^{-2/(D_0-2)}\left[ -\frac 12\tilde
        R_1e^{-2\tilde \beta^1}+\Lambda_D + \tilde
        f^{2}_1e^{-2d_1\tilde \beta^1} \right] }_{U_{int}(\tilde
    \beta^{1})} +
      \underbrace{\kappa_0^2\sum_{c=1}^m\rho _{(d_0)}^{(c)}(\tilde a)}_{U_{ext}(\tilde a)},
\end{equation}\vspace*{-1mm}

\noindent where $\rho _{(d_0)}^{(c)}(\tilde a)$ is defined by
Eq.~(\ref{zh2.43}) and $\tilde f^2_1 \equiv \kappa^2_D f^2_1 /
b_{(0)1}^{2d_1}$.  These two terms $U_{int}(\tilde \beta{\, ^1})$
and $U_{ext}(\tilde a)$ depend only on scale factors of the internal
and external spaces, respectively.  Therefore, the first term is
responsible for the internal space stabilization due to its minimum
and the second term provides the Friedmann-like behavior of the
external space.

First, we investigate the problem of stable compactification of the
internal space. It is clear that such stabilization for our model
takes place if potential $U_{int}$ has a minimum with respect to the
fluctuation field $\tilde \beta^{ 1}$:
\begin{equation}
  \label{zh2.47}
  \left.\frac{\partial U_{int}}{\partial \tilde
      \beta^1}\right|_{\tilde \beta^1 =0} =0 \Longrightarrow
  \frac{D-2}{2d_1}\, \tilde R_1 = \Lambda_D +d_0\tilde f^{2}_1 .
\end{equation}\vspace*{-1mm}

The value of this potential at the minimum plays the role of the
effective four-dimensional cosmological constant: %
\index{cosmological constant|(} %
\begin{equation}
  \label{zh2.48}
  \Lambda_\mathrm{ eff} := \left. U_{int}\right|_{\tilde
    \beta^{1}=0} = -\frac12 \tilde R_1 +\Lambda_D + \tilde f^2_1 .
\end{equation}
With the help of the extremum condition (\ref{zh2.47}),
$\Lambda_\mathrm{eff}$ can be written in the form
\begin{align}
  \Lambda_\mathrm{ eff}  = \frac{D_0-2}{2d_1}\, \tilde R_1 - (D_0-2)
  \tilde f^2_1= \label{zh2.49} \\[2mm]
   = \frac{D_0-2}{D-2} \Lambda_D - \left(\! \frac{d_0 d_1}{D-2} - 1
  \!\right)\tilde f^2_1= \label{zh2.50} \\[2mm]
   = \frac{d_0-1}{d_0}\Lambda_D - \frac12 \left(\!1 -
    \frac{D-2}{d_0d_1}\!\right)\tilde R_1 \label{zh2.51} .
\end{align}

Second derivative of $U_{int}$ in the extremum position reads
  \begin{equation}
    \label{zh2.52}  \left.\frac{\partial^2 U_{int}}{\partial {\tilde
          {\beta^{1}}}^2}\right|_{\tilde \beta^1 =0} =
    -2\left(\!\frac{D-2}{D_0-2}\!\right)^{\!\!2}\tilde R_1 + \left(\!\frac{2d_1}{D_0-2}\!\right)^{\!\!2}\Lambda_D + \left(\!\frac{2d_0d_1}{D_0-2}\!\right)^{\!\!2}\tilde f^2_1=
\end{equation}\vspace*{-4.0mm}
\begin{equation}
     =  \frac{4}{D_0-2}\left[-\frac12 (D-2)\tilde R_1 + 4d_0d_1^2\tilde
    f^2_1\right]=\label{zh2.53}
\end{equation}\vspace*{-1.5mm}
\begin{equation}
     = \frac{4d_1}{(D_0-2)^2}\left[-(D_0-2)\Lambda_D + (D-2)d_0
      \left(\!\frac{d_0d_1}{D-2}-1\!\right)\tilde f^2_1
      \right]=\label{zh2.54}
\end{equation}\vspace*{-1mm}
     \begin{equation}
     = \frac{4}{(D_0-2)^2}\left[-d_1^2 (d_0-1)\Lambda_D +
      \frac12(D-2)^2\left(\!\frac{d_0d_1}{D-2}-1\!\right)\tilde
      R_1\right]\!.\label{zh2.55}
\end{equation}

For stable compactification, this extremum should be a minimum.
Then, small fluctuations above it describe minimal scalar field
(gravitational excitons \cite{zhGZ1}) %
\index{gravitational excitons} %
propagated in the external space with the mass squared
\begin{equation}
  \label{zh2.56} m^2_{1} \equiv
  \frac{D_0-2}{d_1(D-2)}\left.\frac{\partial^2 U_{int}}{\partial {\tilde
        {\beta^{1}}}^2}\right|_{\tilde \beta^1 =0} >0.
\end{equation}
Additionally, the effective four-dimensional cosmological constant
should be positive $\Lambda_\mathrm{ eff} > 0$. This is the
necessary condition for the late time acceleration of the Universe
in considered model. Both of these conditions (the positiveness of
$\Lambda_\mathrm{
  eff}$ and $m^2_{1}$) lead to the following inequalities
  \begin{gather}
    \label{zh2.57}
    d_1 \tilde f^2_1 < \frac12 \tilde R_1 < \frac{4d_0d_1}{D-2}\times d_1\tilde f^2_1 , \\
    \label{zh2.58}
    \frac{D-2}{D_0-2}\left(\!\frac{d_0d_1}{D-2}-1\!\right)\tilde f^2_1 <
    \Lambda_D < d_0\times
    \frac{D-2}{D_0-2}\left(\!\frac{d_0d_1}{D-2}-1\!\right)\tilde f^2_1 , \\
    \label{zh2.59}
    \frac12
    \frac{D-2}{d_1(d_0-1)}\left(\!\frac{d_0d_1}{D-2}-1\!\right)\tilde R_1
    < \Lambda_D < \frac{D-2}{d_1}\times \frac12
    \frac{D-2}{d_1(d_0-1)}\left(\!\frac{d_0d_1}{D-2}-1\!\right)\tilde R_1
    .
  \end{gather}

The inequality on the left-hand side follow from the condition
$\Lambda_{\mathrm{eff}} > 0$ applied to
Eqs.~(\ref{zh2.49})---(\ref{zh2.51}), whereas the inequalities on
the right-hand side follow from the minimum condition $m^2_{1} >0$
applied to Eqs.~(\ref{zh2.53})---(\ref{zh2.55}).  Thus, for the most
reasonable case $\tilde f^2_1 >0 $ of the real form field we get
$\tilde R_1$, $\Lambda_D > 0$. It can be easily seen that the case
$\tilde f^2_1 =0$ results in negative effective cosmological
constant
\cite{zhCQG(1998)}. %
\index{cosmological constant} %
To shift it to non-negative values we introduced the real form
fields.

According to the present day observations, our Universe undergoes
the late time accelerating expansion due to dark energy. The origin
of the dark energy is the great challenge of the modern theoretical
physics and cosmology. The cosmological constant is one of the most
probable candidate for it. The observations give $\Lambda_{DE} \sim
10^{-123}\Lambda_{\rm Pl(4)} \sim 10^{-57}\,\text{cm}^{-2}$. Let us
estimate now a possibility for our effective cosmological constant
to admit this quantity: $\Lambda_{\mathrm{eff}} \sim \Lambda_{DE}
\sim 10^{-57}\,\text{cm}^{-2}$.

As we noted in section (3.1), in KK models the size of the extra
dimensions at present time should be of the order or less than
\mbox{$b_{(0)1} \sim 10^{-17}\,\mbox{cm} \sim 1\,\mbox{TeV}^{-1}$}.
In the case of the upper limit \mbox{$b_{(0)1} \sim
10^{-17}\,\mbox{cm}$}, we get \mbox{$\tilde R_1 \sim b_{(0)1}^{-2}
\sim 10^{34}\,\mbox{cm}^{-2}$}. On the other hand, inequalities
(\ref{zh2.57})---(\ref{zh2.59}) show that $\tilde R_1$, $\Lambda_D$
and $\tilde f^2_1$ are of the same order of magnitude, i.e.
\mbox{$\tilde R_1 \sim \Lambda_D \sim \tilde f^2_1 \sim
10^{34}\,\text{cm}^{-2}$}, and have the same sign. Thus, these
parameters should be extremely fine tuned (in Eq.~(\ref{zh2.48})) to
compensate each other in such a way that to leave only
$10^{-57}\,\text{cm}^{-2}$. One of possibilities to avoid this
problem consists in inclusion of different form fields/fluxes which
may result in a big number of minima (landscape) \cite{zhlandscape}
with sufficient large probability to find oneself in a dark energy
minimum.

\index{fine tuning} %
It is clear that extreme fine tuning has arisen because of two
extremely different scales present in the model, namely
\mbox{$\Lambda_{\mathrm{eff}} \sim 10^{-57}\,\text{cm}^{-2}$} and
\mbox{$\tilde R_1 \sim$} \mbox{$\sim 10^{34}\,\text{cm}^{-2}$}. We
can avoid this problem by removing one of the scales (or both of
them). Let us consider these possibilities.

(i)\,\,$\tilde R_1 = \Lambda_{\mathrm{eff}}=0$ This case does not
  work because Eqs.~(\ref{zh2.47}) and (\ref{zh2.48}) contradict each other
  for $d_0>1$.

(ii)\,\,$\tilde R_1 \neq 0$,  $\Lambda_{\mathrm{eff}}=0$ Here,
  $\tilde R_1 =2d_1 \tilde f^2_1$ and $\Lambda_D = (d_1-1) \tilde
  f^2_1$. Hence, $\tilde R_1 \sim$ $\sim\Lambda_D \sim \tilde f^2_1
  $. Additionally, condition $m^2_{1} >0$ requires that these
  parameters are positive. Therefore, if \mbox{$b_{(0)1} \sim
  10^{-17}\,\text{cm} \sim 1\,\text{TeV}^{-1}$}, then we obtain the
  physically reasonable result that all parameters in this model are
  of the TeV scale, e.g. $m_{1} \sim$~1\,TeV.  Unfortunately, the
  condition $\Lambda_{\mathrm{eff}}=0$ makes it impossible to solve
  the problem of dark energy.

(iii)\,\,$\tilde R_1 = 0, \; \Lambda_{\mathrm{eff}} >0$ Here,
  $\Lambda_D =-2d_0 \tilde f^2_1$ and $\Lambda_{\mathrm{eff}} =
  -(d_0-1) \tilde f^2_1=$ $=[(d_0-1)/d_0]\Lambda_D$ which result in
  conditions $\Lambda_D>0,\; \tilde f^2_1<0$ for
  $\Lambda_{\mathrm{eff}}>0$. However, the condition $m^2_{1}>0$
  requires the opposite inequalities: $\Lambda_D<0$, $\tilde
  f^2_1>0$. Therefore, we arrive at a contradiction. Nevertheless,
  models with Ricci-flat ($\tilde R_1 = 0$) internal spaces are not
  excluded. For example, this space can be an orbifold with branes in
  fixed points \cite{zhUED}. This model is investigated in the next
  section.

Now, we turn to the dynamical behavior of the external factor space.
We consider zero order approximation, when all excitations are
frozen out (or heavy enough to decay long before the present
time\,\footnote{\,Gravexcitons with masses $m_{1} \sim
1\,\text{TeV}$
  decay (e.g. due to reaction $\psi \rightarrow 2\gamma$) before
  primordial nucleosynthesis and they do not contradict the
  observational data \cite{zhGSZ,zhiwara} (see section
  \ref{zh-heavy}). Therefore, in this case the effective potential
  contributes in the form of the effective cosmological constant
  (\ref{zh2.48}).}). Because our Universe (external space) is homogeneous
and isotropic, functions $\widehat \gamma $ and $\widehat \beta^0$
depends only on time: $\widehat \gamma =\widehat \gamma (\widehat
\tau )$ and $\widehat \beta ^0=\widehat \beta^0 (\widehat \tau )$.
Then, the action functional (\ref{zh2.14}) (without scalar field and
for the effective potential (\ref{zh2.46})) after dimensional
reduction reads:
\[
 S  = \frac 1{2\kappa _0^2}\int\limits_{M_0}d^{D_0}x \sqrt{|\tilde
    g^{(0)}|}\left\{\! \tilde R\left[ \tilde g^{(0)}\right]
    -2U_{\mathrm{eff}}\!\right\} =
\]\vspace*{-3mm}
\[
= \frac{V_0}{2\kappa _0^2} \int d\widehat \tau \Bigg\{\!\!
e^{\widehat
      \gamma +d_0\widehat \beta^0}e^{-2\widehat \beta^0}
    R[q^{(0)}]+d_0(1-d_0)e^{-\widehat \gamma +d_0\widehat \beta^0}
    \left(\! \frac{d\widehat \beta^0}{d\widehat \tau
    }\!\right)^{\!\!2}-
  \]\vspace*{-1mm}
\begin{equation}
\label{zh2.60}%
 -2e^{\widehat \gamma +d_0\widehat \beta^0}\left(\! \Lambda
      _\mathrm{ eff}+ \kappa_0 ^2\sum_{c=1}^m\rho
      _{(d_0)}^{(c)}\!\right)\!\! \Bigg\} + \frac{V_0}{2\kappa _0^2}d_0\int
  d\widehat \tau \frac d{d\widehat \tau }
 \left(\! e^{-\widehat \gamma +d_0\widehat \beta^0} \frac{d\widehat \beta^0 } {d\widehat \tau }
  \!\right)\!,
\end{equation}%
where $V_0 := \int_{\mathcal{M}_0}d^{d_0}x \sqrt{|q^{(0)}|}$ is the
volume of the external space, $\rho _{(d_0)}^{(c)}$ is defined by
Eq.~(\ref{zh2.43}) and usually $R[q^{(0)}]=kd_0(d_0-1)$, $k=\pm
1,0$. The
constraint equation $\partial L/\partial \widehat \gamma =0$ in the %
\index{synchronous time gauge} %
synchronous time gauge $\widehat \gamma =0$ yields
\begin{equation}
  \label{zh2.61}%
  \left(\! \frac 1{\tilde a}\frac{d\tilde a}{d\tilde t}\!\right) ^{\!\!2}=-\frac
  k{\tilde a^2}+\frac 2{d_0(d_0-1)}\left(\! \Lambda _{\mathrm{eff}}+
    \kappa_0 ^2\sum_{c=1}^m\rho _{(d_0)}^{(c)}(\tilde a)\!\right)\!,
   \end{equation}
which results in\vspace*{-3mm}
\[
\tilde t + \mathrm{ const} = \int \dfrac{d\tilde a}{\left[
        -k+\dfrac{2\Lambda _{\mathrm{eff}}}{d_0(d_0-1)}\tilde a^2+
        \dfrac{2\kappa_0 ^2}{d_0(d_0-1)}\sum\limits_{c=1}^m\dfrac{\tilde
          A^{(c)}}{\tilde
          a^{(d_0-1)\alpha ^{(c)}}}\right]^{1/2}} ,
\]\vspace*{-1mm}
   \begin{equation}
  \label{zh2.62}%
     \hspace*{-1.6cm}= \int \dfrac{d\tilde a}{\left[ -k +\dfrac{\Lambda_{\mathrm{ eff}}}
        3\tilde a^2+\dfrac{\kappa_0^2}3\sum\limits_{c=1}^m \dfrac{\tilde
          A^{(c)}}{\tilde a^{2\alpha ^{(c)}}}\right] ^{1/2}},
 \end{equation}
where in the last line we put $d_0=3$.

\index{cosmological constant} %
Thus in zero order approximation we arrived at a Friedmann model in
the presence of positive cosmological constant $\Lambda
_{\mathrm{eff}}>0$ and a multicomponent perfect fluid.  It is
assumed that $\Lambda_{\mathrm{eff}}$ defines dark energy observed
now.  The perfect fluid has the form of a
dust for $\alpha ^{(c)}=1/2$ and radiation for $\alpha ^{(c)}=1$.

There is also a possibility for a primordial inflation. For this
purpose we can consider one component perfect fluid with
$\alpha^{(1)} <0$, e.g. $\, \alpha^{(1)} = -1/2 \Rightarrow
\alpha_0^{(1)}=$ $= 1/3$ which describes a frustrated network of
domain walls in the external space. It is well known that such
perfect fluid results in
acceleration of the Universe. %
\index{accelerated expansion} %
The flat Universe ($k=0$) in the case $\alpha_0^{(1)}= 1/3$
undergoes the power law inflation at early times: $\tilde a \sim
\tilde t^2$. If domain walls decay into ordinary matter, then the
described above Friedmann-like scenario follows the inflation.


\label{zh-orbifolds}{\bf\itshape Dark energy in the universal extra
  dimension models.}
In this section, to avoid the fine tuning problem, we consider
Ricci-flat internal spaces. In this case scalar cur\-va\-tu\-res of
the internal spaces are absent and there is no need for extreme fine
tuning of the parameters to get the observable dark energy.

\index{Universal extra dimension model}Among such models, Universal
extra dimension models (UED) are of special interest \cite{zhUED}.
Here, the internal spaces are orbifolds\,\footnote{\,For example,
$S^1/Z_2$ and $T^2/Z_2$ which
  represent circle and square folded onto them\-sel\-ves due to $Z_2$
  symmetry.} with branes in fixed points. The compactification of the
extra dimensions on orbifolds has a number of very interesting and
useful properties, e.g. breaking (super)symmetry and obtaining
chiral fermions in four dimensions (see, e.g., paper by H.-C. Cheng
at al in \cite{zhUED}). The latter property gives a possibility to
avoid famous no-go theorem of KK models (see e.g. \cite{zhno-go}).

In UED models, the Standard Model fields are not localized on the
brane but can move in the bulk. Branes in fixed points contribute in
action functional (\ref{zh2.7}) in the form:
\begin{equation}
  \label{zh2.63}
  \sum_{^\mathrm{ fixed}_\mathrm{points}}\left. \int d^4x \sqrt{ g^{(0)}(x)}\, L_b
  \right|_{^\mathrm{ fixed}_\mathrm{ point}}\!,
\end{equation}
where $ g^{(0)}(x)$ is induced metric (which for our geometry
(\ref{zh2.1}) coincides with the metric of the external space-time
in the Brans---Dicke frame) and $L_b$ is the matter Lagrangian on
the brane. In what follows, we consider the case where branes are
characterized by their tensions $L_{b(k)} = -\tau_{(k)},\,
k=1,2,\mbox{...} ,m$ and $m$ is the number of branes.

After conformal transformation (\ref{zh2.13}), the action
(\ref{zh2.63}) reads
\begin{equation}
  \label{zh2.64}
  \frac{1}{2\kappa^2_0} \int d^4x \sqrt{\tilde g^{(0)}(x)} \left(\prod_{i=1}^n e^{d_i  \tilde \beta^{\, i}}\! \right)^{\!\!-2/(D_0-2)}
  \!\left[-2\kappa^2_0 \sum_{k=1}^m \tau_{(k)}  \prod_{i=1}^ne^{-d_i\tilde \beta^{\, i}}\right]\!.
\end{equation}

The comparison of this expression with Eqs.~(\ref{zh2.32}) and
(\ref{zh2.34}) shows that branes contribute in the effective
potential in the form of one component perfect fluid ($c=1$) with
equations of state: $\alpha^{(1)}_0 =0,\, \alpha^{(1)}_i = 1, \,
i=1,\mbox{...} ,n$, i.e., from the case I of the no-go theorem. It
means that they contribute only to the $U_\mathrm{ int}$:
\begin{equation}
  \label{zh2.65}%
  U_{\mathrm{int}} = {\left(\! e^{d_1\tilde \beta{\, ^1}}\!\right) }^{\!-
  2/(D_0-2)}\!
  \left[ \Lambda_D + \tilde f^{2}_1e^{-2d_1\tilde \beta^{\, 1}} - \lambda e^{-d_1\tilde \beta^{\, 1}}\right]\!,
\end{equation}
where we consider the case of one internal space $i=1$ and introduce
notation $\lambda \equiv - \kappa^2_0 \sum_{k=1}^m \tau_{(k)}$.

Obviously, the internal space is stabilized if potential
(\ref{zh2.65}) has a minimum with respect to $\tilde \beta^{\, 1}$.
The extremum condition reads:
\begin{equation}
  \label{zh2.66}
  \left.\frac{ \partial U_{\mathrm{int}}}{\partial \tilde \beta^{\, 1 }}\right|_{\tilde \beta^{\, 1} =0} =0 \Longrightarrow
  \frac{d_1D_0}{D_0-2} \lambda =  \frac{2d_1}{D_0-2}\Lambda_D + \frac{2d_1(D_0-1)}{D_0-2}\tilde f^{2}_1 .
\end{equation}

The value of this potential at the minimum plays the role of
effective
four-dimensional cosmological constant: %
\index{cosmological constant|)} %
\begin{equation}
  \label{zh2.67}
  \Lambda_{\mathrm{eff}} := \left. U_{\mathrm{int}}
  \right|_{\tilde \beta^{\, 1}=0} =
  \Lambda_D + \tilde f^2_1 -\lambda  >0,
\end{equation}
which we assume to be positive. For the mass of gravexcitons we
obtain:
\[
     m_{1}^2 \sim \left.\frac{\partial^2 U_{\mathrm{int}}}{\partial
        {\tilde {\beta^{1}}}^2}\right|_{\tilde \beta^{\, 1} =0}  =
     \left(\!\frac{2d_1}{D_0-2}\!\right)^{\!2}\Lambda_D+
\]\vspace*{-1mm}
\begin{equation}
  \label{zh2.68}
    +\left(\!\frac{2d_1(D_0-1)}{D_0-2}\!\right)^{\!2}\tilde f^{2}_1
    -\left(\!\frac{d_1D_0}{D_0-2}\!\right)^{\!2}\lambda >0.
 \end{equation}

It can be easily seen from condition (\ref{zh2.66}) and inequality
(\ref{zh2.67}) that all three parameters are positive: $\tilde
f^{2}_1, \Lambda_D, \lambda >0$. Thus, from $\lambda >0$ follows
that summarized tensions of branes should be negative. Taking into
account inequality (\ref{zh2.68}), it can be easily verified that
all these parameters have the same order of magnitude:
\begin{equation}
  \label{zh2.69}
  \tilde f^{2}_1 \sim \Lambda_D \sim \lambda \sim \Lambda_\mathrm{eff} \sim m_{1}^2.
\end{equation}
Therefore, there is no need for fine tuning of parameters to obtain
the observable value of dark energy. To get it, it is sufficient to
suppose that all these parameters, including $\Lambda_\mathrm{
eff}$, are of the order of $\Lambda_{\mathrm{DE}} \sim
10^{-123}\Lambda_{\rm Pl(4)}\sim $ $\sim 10^{-57}\,\text{cm}^{-2}$.
However, it is natural to assume that parameters of the model have
the same order of magnitude. On the other hand, our model does not
answer why this value is equal to $10^{-123}L_{\rm Pl}$. According
to the anthropic prin\-ciple, it takes place because human life is
possible only at this value of \mbox{dark energy.}

If we assume that our parameters are defined by
$\Lambda_{\mathrm{DE}} \sim 10^{-123}\,\Lambda_{\rm Pl(4)}$, then we
get for the gravexciton masses \mbox{$m_{1} \sim 10^{-33}\,\text{eV}
\sim 10^{-61}\,M_{\rm Pl(4)}$}. These ultra-light particles have a
period of oscillations \mbox{$t \sim 1/m_{1} \sim
10^{18}\,\text{sec}$} (see section \ref{zh-light exci}) which is of
the order of the Universe age. So, up to now these cosmological
gravexcitons did not start to oscillate but slowly move to the
position of minimum of the effective potential. In this case it is
hardly possible to speak about stabilization of the internal space
(the effective potential $U_{\mathrm{int}}$ is too flat) and we
arrive at the problem of the fundamental constant variations (see
section \ref{zh-fine str}).

Of course, this problem is absent for gravexcitons with sufficiently
large masses. For example, we can put $m_{1}\sim 1$\,TeV. Then, as
it follows from (\ref{zh2.68}), $\Lambda_D,\tilde f^{2}_1, \lambda
\sim m_1^2 \sim 1\,\mbox{TeV}^2$ and to get observable dark energy
\mbox{$\Lambda_\mathrm{eff} \sim \Lambda_\mathrm{ DE}\sim $}
\mbox{$\sim10^{-123}M_{\rm Pl(4)}^2 \sim 10^{-91}\,\mbox{TeV}^2$}
the parameters $\Lambda_D,\tilde f^{2}_1, \lambda$ should be
extremely fine tuned each other.

Similarly to the previous section, we can avoid this problem, if we
give up the dark energy in this model: $\Lambda_{\mathrm{eff}}=0$.
In this case we get $\Lambda_D= \tilde f^{2}_1 = \lambda/2$ and
$m^2_{1}=2d_1^2\Lambda_D$. Thus, if we naturally assume that
$\Lambda_D\sim 1$\,TeV, then all parameters of the model (including
$m_{1}$) are of the TeV scale, which is physi\-cally~reasonable.

Thus, the two previous sections demonstrate the typical problems of
the KK models, when we want to both stabilize the internal spaces
and to solve the problem of dark energy.  Either we are faced with
an extremely fine tuning of parameters or the effective potential is
very flat, which leads to the problem of variation of fundamental
constants.

\section[\!Abelian gauge fields in KK models, dimensional reduction]{\!Abelian gauge fields\index{Abelian gauge fields}\\ \hspace*{-0.95cm}in KK
  models, dimensional reduction} \label{zh-em}

\hspace*{3cm}In the section \ref{zh-general} we have shown that
conformal zero-mode excitations of the internal factor spaces $M_i$
have the form of massive scalar fields (gravexcitons/radions)
developing on the background of the external space-time $M_0$ i.e.
in our Universe. It is of interest to consider the interaction of
these scalar particles with fields from the Standard Model.  In this
section, we study the
interaction of gravitational excitons %
\index{gravitational excitons} %
with Abelian gauge fields, and in particular with the standard
electromagnetic field of $U(1)_{EM}$ symmetry. Strictly speaking,
the photon will not exist as a separate gauge boson at temperatures
higher than the electroweak scale $M_{\rm EW} \sim 246\,\text{GeV}$
where the full electroweak $SU(2)\times U(1)$ model should be
considered. Nevertheless, our results should reproduce the correct
coupling term between the gravexciton sector and the EM gauge field
sector of the theory. In the next section we will use this
coup\-ling term for estimating the strength of cosmological and
astrophysical effects related to the corresponding interaction
channel.

In order to derive the concrete form of the coupling term in the
dimensionally reduced, four-dimensional effective theory, we start
from the simplified toy model ansatz\vspace*{-3mm}
\begin{equation}
  \label{zh3.1}
  S_{EM} = -\frac12 \int\limits_M d^DX\sqrt {|g|}F_{MN}F^{MN},
\end{equation}
where the gauge field is assumed Abelian also in the
higher-dimensional setup.  Additionally, we work in the zero mode
approximation for these fields, i.e. we keep only the zero modes of
the harmonic expansion in mass eigenstates of the higher-dimensional
fields\,\footnote{\,The excitation of Kaluza---Klein modes of
Abelian gauge
  fields was considered, e.g., in Ref.~\cite{zhMUM}.},
\cite{zh18,zh21}. In this case, the Abelian vector potential depends
only on the external coordinates, $A_M=A_M(x),\, (M=$ $=1,\mbox{...}
,D)$, and the corresponding non-zero components of the field
strength tensor are \mbox{$F_{\mu \nu }=\partial _\mu A_\nu
-\partial _\nu A_\mu$},  $(\mu ,\nu =1,\mbox{...} ,D_0)$ and $F_{\mu
m_i}=\partial_\mu A_{m_i}-
\partial _{m_i}A_\mu =$ $=\partial _\mu A_{m_i},\; (m_i=1,\mbox{...} ,d_i;\;
i=1,\mbox{...} ,n)$.

Dimensional reduction of the gauge field action (\ref{zh3.1}) yields
\[
S_{EM} =  -\frac12 \int\limits_{M_0} d^{D_0}x\sqrt {|g^{(0)}|}
  \prod_{i=1}^n e^{d_i\tilde \beta^i} \times
\]\vspace*{-2mm}
\begin{equation}
\label{zh3.2} \times \left\{\! F_{\mu \nu }F^{\mu \nu} + 2g^{(0)\mu
\nu }
    \sum_{i=1}^n e^{-2\tilde \beta^i (x)} \bar
    g^{(i)m_in_i} \partial_{\mu } A_{m_i}\partial_{\nu } A_{n_i}
 \! \right\}\! ,
\end{equation}
where we introduced the metric integral
\begin{equation}
  \label{zh3.3} \bar g^{(i)m_in_i}:= \frac
  1{V_{d_i}}\int\limits_{M_i}d^{d_i}y\sqrt{%
    |g^{(i)}|}g^{(i)m_in_i}(y^i)
\end{equation}
and included the factor $\sqrt{V_{D'}}$ into $A_M$ for convenience:
$\sqrt{V_{D'}} A_M\rightarrow A_M$. Due to this redefinition, the
field strength tensor $F_{\mu \nu}$ acquires the usual
dimensionality $\mbox{cm}^{-D_0/2}$ (in geometrical units $\hbar = c
=1$). In Eq.~(\ref{zh3.2}) we assumed $F^{\mu \nu }=g^{(0)\mu \kappa
}g^{(0)\nu
  \delta }F_{\kappa \delta }$.

It is easily seen that the $A_{m_i}$ components play the role of
scalar fields in the $D_0$-dimensional space-time. In what follows,
we will not investigate the dynamics of these fields. Instead, we
will concentrate on the interaction between gravexcitons and the
2-form field strength $F=dA,\; A=A_\mu dx^\mu $ which is described
by the first term of the action functional (\ref{zh3.2}). The
corresponding truncated action (without $A_{m_i}$ terms) will be
denoted by $\bar {S}_{EM}$.

The exact field strength 2-form $F=dA$ with components $F_{\mu \nu
}$ is invariant under gauge transformations $A\mapsto A^f=A+df$,
$F^f=dA+d^2f=dA=F$, with $f(x)$ any smooth and single-valued
function. Accordingly, $\bar {S}_{EM} $ is gauge invariant too (see
Eq.~(\ref{zh3.2})).

The action functional (\ref{zh3.2}) is written in a Brans---Dicke
frame. For passing by the conformal transformation (\ref{zh2.13}) to
the Einstein frame we choose an ansatz
\begin{equation}
  \label{zh3.4}A =\Omega ^k\tilde A
\end{equation}
for the vector potential and introduce the auxiliary field strength
$\bar F$ by the relation
\begin{equation}
  \label{zh3.5}
  \begin{array}{c}
    F  = dA = d(\Omega^k \tilde A)=\Omega^k \bar F,\\[2mm]
    \bar F  = d(\ln \Omega^k )\wedge \tilde A + d \tilde A  .
  \end{array}
\end{equation}

The conformally transformed effective action reads then
\begin{equation}
\label{zh3.6}%
  \bar{S}_{EM}=-\frac12 \int\limits_{M_0} d^{D_0}x \sqrt{|\tilde g^{(0)}|}
  \left\{\! \Omega^{2(k-1)} \bar F_{\mu \nu }\bar F^{\mu \nu }
  \!\right\}\! ,
\end{equation}
where the external space indices are raised and lowered by the
metric $\tilde g^{(0)}$.  With $\tilde F=d\tilde A$, we have in
Eq.~(\ref{zh3.6}) explicitly
\[
\bar F_{\mu \nu }\bar F^{\mu \nu } =  \tilde F_{\mu \nu }\tilde
  F^{\mu \nu } - 2 \tilde F^{\mu \nu } \left[ \tilde A_{\mu }
    \partial_{\nu }(\ln \Omega ^k) - \tilde A_{\nu } \partial_{\mu }
    (\ln \Omega ^k) \right] +
\]
\begin{equation}
\label{zh3.7}
 + 2\left[ \tilde g^{(0)\mu \kappa }\partial_{\mu } (\ln \Omega
    ^k)\partial_{\kappa }(\ln \Omega ^k) \tilde A^{\nu }\tilde A_{\nu
    } - \left(\! \tilde A^{\mu }\partial_{\mu }(\ln \Omega
    ^k)\!\right)^{\!2}
  \right]\!.
\end{equation}

In order to fix the conformal weight $k$ of the vector potential in
Eq.~(\ref{zh3.4}), we require the effective external field strength
tensor $\bar F_{\mu \nu }$ in Eq.~(\ref{zh3.6}) to be
gauge-invariant, i.e. to be invariant under $\tilde A \mapsto \tilde
A^f=\tilde A+df$. From Eq.~(\ref{zh3.5}) we have for this
transformation
\begin{equation}
\label{zh3.8} \bar F \mapsto \bar F^f = d\tilde A + d^2f + d(\ln
\Omega ^k)\wedge (\tilde A\ +df) = \bar F + d(\ln \Omega ^k)\wedge
df ,
\end{equation}
so that for non-trivial $\Omega \neq 1$ the gauge invariance $\bar
F=\bar F^f$ is only achieved for zero conformal weight $k=0$. The
same result follows also directly from the gauge invariance of the
field strength tensor $F$ in Eq.~(\ref{zh3.2}) and the ansatz
(\ref{zh3.4}): One checks immediately that $\bar F$ is invariant
under a transformation $\tilde A\mapsto \check A=\tilde
A+\Omega^{-k}df,$ which only for $k=0$ is a gauge transformation.

This means that in order to preserve the gauge invariance of the
action functional, when passing from the Brans---Dicke frame to the
Einstein frame, we have to keep the vector potential unchanged, i.e.
we have to fix the conformal weight at $k=0$. As a result, we arrive
at the action functional
\begin{equation}
  \label{zh3.9}
  \bar{S}_{EM}= -\frac12 \int\limits_{M_0} d^{D_0}x \sqrt{|\tilde g^{(0)}|} \left\{\! e^{\frac{2}{D_0-2}\sum ^n_{i=1}d_i\tilde \beta^i(x)} F_{\mu \nu } F^{\mu \nu }\! \right\}
\end{equation}
with dilatonic coupling of the Abelian gauge fields to the
gravitational exci-\linebreak tons \cite{zhGSZ}.

For completeness, we note that for $k=1$, according to
Eqs.~(\ref{zh3.6}) and (\ref{zh3.7}), we obtain a theory with a pure
free action term $\tilde F_{\mu \nu }\tilde F^{\mu \nu }$ without
any prefactor $(\Omega ^{2(k-1)}=1)$ but with explicitly destroyed
gauge invariance. The corresponding effective action reads
 \[
\bar{S}_{EM}  = -\frac12 \int\limits_{M_0} d^{D_0}x \sqrt{|\tilde
    g^{(0)}|} \bigg\{\! \tilde F_{\mu \nu }\tilde F^{\mu \nu }
    - 4\tilde F^{\mu \nu} \tilde A_{[\mu } \partial_{\nu ]} ( \ln \Omega ) +
 \]\vspace*{-2mm}
\begin{equation}
\label{zh3.10}
 +  2 \left[ \tilde g^{(0)\mu \kappa }\partial_{\mu } \left(
        \ln \Omega \right) \partial_{\kappa } \left( \ln \Omega
      \right) \tilde A^{\nu }\tilde A_{\nu } - \left(\! \tilde A^{\mu
        }\partial_{\mu } \ln \Omega \!\right)^{\!2} \right]\!\! \bigg\}\! .
\end{equation}

Obviously, the localization of the internal space scale factors at
their present values implies $\tilde \beta^i =0$ which yields
$\Omega \equiv 1$.  Then, both approaches (\ref{zh3.9}) and
(\ref{zh3.10}) coincide with each other. However, the presence of
small scale factor fluctuations above this background will restore
the dilatonic coupling of Eq.~(\ref{zh3.9}) (see also the next
section).

\section[\!Gravitational excitons]{\!Gravitational excitons and
their\\
  \hspace*{-0.95cm}cosmological and astrophysical implications.\\
  \hspace*{-0.95cm}Dark matter from extra  dimensions} \label{astro}

\hspace*{3cm}\index{gravitational excitons|(}\index{dark matter in
KK models}In this section, we discuss some cosmological and
astrophysical implications related to the possible existence of
gravitational excitons. We suppose that the scale factor background
of the internal spaces is localized in one of the minima of the
effective potential and that gravexcitons are present as small
fluctuations above this static background. Our analysis is based on
the dilatonic coupling (\ref{zh3.9}) which describes the interaction
between gravexcitons and zero mode photons\,\footnote{\,Brane-world
models with on-brane dilatonic
  coupling terms have been considered, e.g., in
  Refs.~\cite{zhboehm,zhIJMPD}. In a rough approximation, the results
  of the present section will also hold for these models.}.
Hereafter, we treat these KK zero mode photons as our usual SM
matter photons.  In particular, the vector potential ${A_{\mu}}(x)$
of the
previous section corresponds to our 4D photons. %
\index{gravitational excitons} %

In the following we consider the simplest example~--- the
interaction between gravitational excitons and photons in a system
with only one internal space $(n=1)$ with its scale factor $\beta^1$
localized in one of the minima of the effective potential (e.g.
potential (\ref{zh2.34})). Then, for small scale factor fluctuations
$\tilde \beta^1 \ll 1$ action (\ref{zh2.18}) (without minimal scalar
field) together with (\ref{zh3.9}) reads
\[
S  =  \frac{1}{2\kappa _0^2}\int \limits_{M_0}d^{D_0}x \sqrt
  {|\tilde g^{(0)}|}\left\{\!\tilde R\left[\tilde g^{(0)}\right] -
    2\Lambda _\mathrm{ eff}\!\right\} +
\]\vspace*{-2mm}
\[
 + \frac{1}{2}\int \limits_{M_0}d^{D_0}x \sqrt {|\tilde
    g^{(0)}|}\left\{\!-\tilde g^{(0)\mu \nu}\psi _{,\mu}\psi _{,\nu} -
    m_{\psi}^2\psi \psi \!\right\} -
\]\vspace*{-2mm}
\begin{equation}
 \label{zh4.1} - \frac{1}{2}\int \limits_{M_0}d^{D_0}x \sqrt {|\tilde
    g^{(0)}|}\left\{\! F_{\mu \nu}F^{\mu \nu} -
    2\sqrt{\frac{d_1}{(D_0-2)(D-2)}}\kappa_0 \psi F_{\mu \nu} F^{\mu
      \nu}\!\right\} + \mbox{...}
\end{equation}

\begin{wrapfigure}{l}{5.0cm}
\includegraphics[width=5cm]{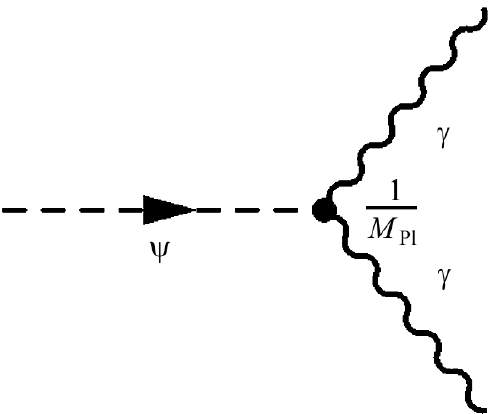} \vspace*{-5mm}\\
\raisebox{0.0cm}{\parbox[b]{5cm}{\caption{\footnotesize Planck-scale
suppressed gravexciton decay:
    $\psi \longrightarrow 2 \gamma$\label{zh-fig1}}}}\vspace*{3mm}
\end{wrapfigure}

We have used the notation of the action (\ref{zh2.18}), $m_{\psi} :=
m_{1}$ and relation (\ref{zh2.17}) between $\tilde \beta^1$ and the
rescaled fluctuational component $\psi\equiv \psi^1$. As mentioned
in section \ref{zh-general}, $\kappa^2_0 = 8\pi / M_{\rm Pl(4)}^2 $
is the $D_0$-dimensional (usually $D_0 = 4$) gravitational constant.
The last term in Eq.~(\ref{zh4.1}) describes the interaction between
gravitational excitons and photons. In the lowest-order tree-level
approximation, this term corresponds to the vertex $1/M_{\rm Pl(4)}$
of Fig.~3.\ref{zh-fig1} \cite{zhGSZ}
 and describes the decay of a gravitational exciton into
two photons. The probability of this decay is easily estimated
as\,\footnote{\,Exact calculations give $\Gamma =
[2d_1/(d_1+2)](m_{\psi}^3/M_{\rm Pl(4)}^2)$.}
\begin{equation}
\label{zh4.2}
  \Gamma \sim \left(\!\frac{1}{M_{\rm Pl(4)}}\!\right)^{\!\!2} m_{\psi}^3 =
  \left(\!\frac{m_{\psi}}{M_{\rm Pl(4)}}\!\right)^{\!\!3}\frac{1}{t_{\rm Pl}} \ll m_{\psi}
   ,
\end{equation}

\noindent which results in a life-time $\tau$ of the gravitational
excitons with respect to this decay channel of\vspace*{-4mm}
\begin{equation}
\label{zh4.3} \tau = \frac{1}{\Gamma} \sim \left(\!\frac{M_{\rm
Pl(4)}}{m_{\psi}}\!\right)^{\!\!3} t_{\rm Pl} .
\end{equation}\vspace*{-3mm}

For example, for light excitons with\linebreak masses
\mbox{$m_{\psi} \le 10^{-21} M_{\rm Pl} \sim 10^{-2}\,\mbox{GeV}
\sim20 m_e$} (where $m_e$ is the electron mass) we get that their
life-time $\tau $ is greater than the age of the Universe $\tau \ge
10^{19}\, \mbox{sec}
> t_{univ} \sim10^{18}$\,sec (see also section \ref{zh-light
exci}).

\index{scalaron}Similar to Polonyi particles in spon\-ta\-neous\-ly
broken supergravity \cite{zhCFKRR,zhENQ}, scalarons in the $(R
+R^2)$ fourth order theory of gravity \cite{zhStar1980} or moduli
fields in the hidden sector of SUSY [398---401], gravitational
excitons are \mbox{WIMPs} (Weakly-Interacting Massive Particles
\cite{zhKT}) because their coupling to the observable matter is
suppressed by powers of the Planck scale. Thus, the
gravexcitons/radions can contribute to {\bfseries dark
matter}\,\footnote{\,The type of the DM depends on the
  masses of the particles which constitute it. It is hot for
  $m_\mathrm{DM} \le 50$---$100$\,eV, warm for $ 100\,{\rm eV} \le
  m_\mathrm{DM} \le 10$\,KeV and cold for $m_\mathrm{DM} \ge 10$---50\,KeV\,$\sim$ $\sim 10^{-5}$\,GeV.\label{DM}} (DM).
  As we have shown above, light
gravexcitons with masses $m_{\psi} \le$\linebreak $ \le 10^{-21}
M_{\rm Pl(4)} \sim 10^{-2}\,\mbox{GeV} $  have life-time $\tau $
greater than the age of the Universe. \index{age of
Universe}However, there are strong cosmological restrictions on
masses of gravexcitons.  To show it, we investigate now the role of
gravexcitons in different cosmological scenarios
\cite{zhiwara}.\vspace*{3mm}


\subsection{\!Effective equation\\ \hspace*{-1.2cm}of motion for
  gravexcitons\label{zh-eff eq}}

\hspace*{3cm}The effective equation of motion for massive
gravexcitons in FRW metric (\ref{zh2.35}) is\index{FRW
metric}\index{equation of motion}\index{gravitational
excitons|)}\vspace*{-2mm}
\begin{equation}
  \label{zh4.4}
  \ddot \psi + (3H+\Gamma)\dot \psi + m_\psi^2 \psi= 0,
\end{equation}
where $H=\dot {\tilde a}/\tilde a$ is the Hubble constant, %
\index{Hubble constant}%
$\tilde a$ is the scale factor of the external space in the Einstein
frame, and dots denote the derivatives with respect to the
synchronous time $\tilde t$ in the Einstein frame. In sections
\ref{zh-eff
  eq}---\ref{zh-liv}, for \mbox{simplicity} of notations we shall write $a$ and
$t$ without symbol tilde above them. In Eq.~(\ref{zh4.4}) (by
analogy with Ref.~\cite{zhKLS1994}) we took into account the
effective decay of gravexcitons into ordinary matter, e.g. into 4D
photons (\ref{zh4.2}). In Refs.~\cite{zhGSZ,zhkolb2} it was shown
that the gravexciton production due to interactions with matter
fields is negligible for the models under consideration.\linebreak A
corresponding source term on the right-hand side of
Eq.~(\ref{zh4.4}) can, hence, be omitted.

The investigation of  Eq.~(\ref{zh4.4}) is most conveniently started
by
 a sub\-stitution\vspace*{-2mm}
\begin{equation}
\label{zh4.5}
  \psi := B(t) u(t) := M_{\rm Pl(4)} e^{- \frac 12 \int
     (3H+\Gamma)dt}u(t),
\end{equation}
which, for constant $\Gamma$, leads to the following differential
equation for the auxiliary function $u(t)$:\vspace*{-3mm}
\begin{equation}
  \label{zh4.6} \ddot u +\left[m^2_\psi-\frac 14 (3H+\Gamma)^2 - \frac 32
    \dot H \right]u=0.
\end{equation}
\index{Hubble parameter} %
This equation shows that at times when the Hubble parameter $H = s/t
\sim 1/t$ is less than the mass\vspace*{-3mm}
\begin{equation}
  \label{zh4.7}
  H < m_{\psi} \Longrightarrow t> t_{in} \sim
  H^{-1}_{in} \sim \frac{1}{m_{\psi}},
\end{equation}\vspace*{-5mm}

\noindent the scalar field oscillates\vspace*{-3mm}
\begin{equation}
\label{zh4.8} \psi \approx C B(t) \cos (m_{\psi}t + \delta).
\end{equation}

The time $t_{in}$ roughly indicates the beginning of the
oscillations. Substituting the Hubble parameter $H = s/t$ into the
definition of $B(t)$ we obtain %
\index{Hubble parameter} %
\begin{equation}
  \label{zh4.9}
  B(t) = M_{\rm Pl(4)}\; e^{- \frac 12 \Gamma t}\frac{1}{(M_{\rm Pl(4)}\;
    t)^{3s/2}},
\end{equation}
where $s=1/2, 2/3$ for radiation dominated (RD) %
and matter dominated
(MD) stages, respectively. %
\index{radiation dominated (RD) epoch} %
\index{matter dominated (MD) epoch} %
The parameter $C$ in Eq.~(\ref{zh4.8}) can be obtained from the
amplitude of the initial oscillation $\psi_{in}$:
\begin{equation}
  \label{zh4.10}
  \psi_{in} \sim C B(t_{in}) \Longrightarrow
  C_r \sim \frac{\psi_{in}}{M_{\rm Pl(4)}}\left(\!\frac{M_{\rm Pl(4)}}{m_{\psi}}\!\right)^{\!\!3/4}\!, \quad C_m  \sim \frac{\psi_{in}}{m_{\psi}}.
\end{equation}
$C_r$ and $C_m$ correspond to particles which start to oscillate
during the RD and MD stages. Additionally, we took into account that
$\Gamma t_{in} \sim \Gamma/m_{\psi} \ll1$.

Further useful differential relations are those for $B(t)$ in
(\ref{zh4.5}), (\ref{zh4.9}) and for the energy density
$\rho_{\psi}$ and the number density $n_{\psi}$ of the gravexcitons.
It can be easily seen from the definition of $B(t)$ that this
function satisfies the differential equation\vspace*{-3mm}
\begin{equation}
  \label{zh4.11}
  \frac{d}{d t}\left(  a^3B^2 \right) = -\Gamma   a^3B^2 ,
\end{equation}
with $a(t) \sim t^{s}$ as scale factor of the Friedmann Universe.
The energy density of the gravexciton field and the corresponding
number density, which can be approximated as
\begin{equation}
  \label{zh4.12}
  \rho_{\psi} = \frac12 \dot \psi^2 + \frac12 m^2_{\psi}\, \psi^2
  \approx \frac12 C^2 B^2 m^2_{\psi}, \quad n_{\psi}\approx \frac12
  C^2 B^2 m_{\psi},
\end{equation}
satisfy the DEs\vspace*{-3mm}
\begin{equation}
  \label{zh4.13}
  \frac{d}{d t}( a^3\rho _{\psi}) =-\Gamma a^3\rho _{\psi} \text{ and
  } \frac{d}{d t}( a^3n_{\psi}) =-\Gamma a^3n_{\psi}
\end{equation}
with solutions\vspace*{-3mm}
\begin{equation}
  \label{zh4.14}
  \rho _{\psi} \sim e^{-\Gamma t} a^{-3} \text{ and } n_{\psi}\sim
  e^{-\Gamma t} a^{-3}.
\end{equation}

This means that during the stage $ m_{\psi}> H \gg \Gamma $ the
gravexcitons perform damped oscillations and their energy density
behaves like a red-shifted dust-like perfect fluid $(\rho
_{\psi}\sim a^{-3})$ with slow decay $e^{-\Gamma t}\sim
1$:\vspace*{-1mm}
\begin{equation}
  \label{zh4.15}
  \rho_{\psi} \approx \psi_{in}^2 m_{\psi}^2
  \left(\!\frac{T}{T_{in}}\!\right)^{\!\!3}\!.
\end{equation}\vspace*{-4mm}

$T_{in}$ denotes the temperature of the Universe when the
gravexcitons started to oscillate. According to the Friedmann
equation (the 00-component of the Einstein equation), the Hubble
parameter and the energy density (which defines the dynamics of the
Universe) are connected (for flat spatial sections) by the
relation\vspace*{-3mm}
\begin{equation}
  \label{zh4.16}
  H(t) M_{\rm Pl(4)} = \sqrt{\frac{8\pi}{3}\, \rho (t)} \sim \sqrt{\rho (t)}.
\end{equation}
During the RD stage it holds $\rho (t)\sim T^4$ and, hence, $H^2
\sim T^4/M_{\rm Pl(4)}^2$. For gravexcitons which start their
oscillations during this stage, the temperature $T_{in}$ is now
easily estimated as\vspace*{-3mm}
\begin{equation}
  \label{zh4.17}
  H_{in} \sim m_{\psi} \sim \frac{T_{in}^2}{M_{\rm Pl(4)}} \Longrightarrow T_{in} \sim \sqrt{m_{\psi}M_{\rm Pl(4)}}.
\end{equation}\vspace*{-4mm}

If there is no broad parametric resonance (``preheating'')
\cite{zhKLS1994}, then the decay plays the essential role when
$H\lesssim \Gamma$ and the evolution of the energy density of the
gravexcitons is dominated by an exponential decrease. The most
effective decay takes place at times\vspace*{-4mm}
\begin{equation}
  \label{zh4.18}
  t_D \sim H_D^{-1} \sim \Gamma^{-1} \sim
  \left(\!\frac{M_{\rm Pl(4)}}{m_{\psi}}\!\right)^{\!\!2}m_{\psi}^{-1}.
\end{equation}

\subsection{\!Light and ultra-light\\
  \hspace*{-1.2cm}gravexcitons: \boldmath$m_{\psi} \leq 10^{-2}$\,GeV\label{zh-light exci}}

\hspace*{3cm}\index{gravitational excitons}If the decay time $t_D$
of the gravexcitons exceeds the age of the
Universe $t_{univ} \sim 10^{18}$\,sec then the decay can be neglected. %
\index{age of Universe} %
Eq.~(\ref{zh4.18}) shows that this is the case for particles with
masses $m_{\psi} \leq 10^{21}M_{\rm Pl(4)} \sim10^{-2}$\,GeV~$\sim$
$\sim 20 m_e$ (where $m_e$ is the electron mass).

Subsequently, we split our analysis, considering separately
particles \mbox{which} start to oscillate before matter/radiation
equality $t_{eq} \sim H_{eq}^{-1}$ (i.e. during the RD stage) and
after $t_{eq}$ (i.e. during the MD stage). According to present WMAP
data for the
$\Lambda$CDM model, the following holds: %
\index{LambdaCDM model ($\Lambda$CDM)} %
\begin{equation}
  \label{zh4.19}
  H_{eq} \equiv m_{eq} \sim 10^{-56}M_{\rm Pl(4)} \sim
  10^{-28}~\mbox{eV} .
\end{equation}

An obvious requirement is that gravexciton should not over-close the
observable Universe.  This means that for particles with masses
$m_{\psi} > m_{eq}$ the energy density at the time $t_{eq}$ should
not exceed the critical density\,\footnote{\,It is clear that the
ratio
  between the energy densities of gravexciton and matter becomes fixed
  after $t_{eq}$. If $\rho_{\psi}$ is less than the critical density
  at this moment then it \mbox{will} remain under-critical forever.}:
\begin{equation}
  \label{zh4.20}
  \left. \sqrt{\rho_{\psi}}\; \right|_{t_{eq}\sim
    H^{-1}_{eq}} \lesssim H_{eq} M_{\rm Pl(4)} \Longrightarrow m_{\psi}\lesssim
  m_{eq}\left(\!\frac{M_{\rm Pl(4)}}{\psi_{in}}\!\right)^{\!\!4}\!.
\end{equation}

Here we used the estimate $\rho_{\psi} \sim
(\psi_{in}/t)^2(m_{\psi}t)^{1/2}$ which follows from
Eqs.~(\ref{zh4.9}), (\ref{zh4.10}) and (\ref{zh4.12}). For particles
with masses $m_{\psi} \gtrsim m_{eq}$, relation (\ref{zh4.20})
implies the additional consistency condition $\psi_{in} \lesssim
M_{\rm Pl(4)}$, or more exactly
\begin{equation}
  \label{zh4.21}
  \psi_{in} \lesssim
  \left(\!\frac{m_{eq}}{m_{\psi}}\!\right)^{\!\!1/4}M_{\rm Pl(4)} \lesssim M_{\rm Pl(4)}.
\end{equation}

Let us now consider particles with masses $m_{\psi} \lesssim m_{eq}$
which start to oscilla\-te during the MD stage. From
Eqs.~(\ref{zh4.9}), (\ref{zh4.10}) and (\ref{zh4.12}) one finds for
these particles $\rho_{\psi} \sim (\psi_{in}/t)^2$ so that the
inequality
\begin{equation}
  \label{zh4.22}
  \left.\sqrt{\rho_{\psi}}\; \right|_{t_{in} \sim
    H_{in}^{-1}} \lesssim H_{in}M_{\rm Pl(4)} \Longrightarrow \psi_{in} \lesssim M_{\rm Pl(4)}
\end{equation}
ensures under-criticality of the energy density with respect to
over-closure of the Universe.

It is worth noting that
the combination $-(9/4)H^2 - (3/2)\dot H $ in (\ref{zh4.6}) vanishes
for the MD stage because of $H = 2/(3t)$. Hence, for times $t \geq
t_{eq} \sim $ $\sim1/m_{eq}$ the solutions of equation (\ref{zh4.6})
have an oscillating behavior (provided that $m_{\psi} > \Gamma$)
with a period of oscillations $t_{osc} \sim 1/m_{\psi}$. For light
particles with masses $m_{\psi} \leq m_{eq}$ this implies that the
initial oscillations start at $t_{in} \sim 1/m_{\psi} $.

Particles with masses \mbox{$m_{\psi} \sim 10^{-33}\, \text{eV} \sim
10^{-61}\,M_{\rm Pl(4)}$} are of special interest because via
$\Lambda_{\mathrm{eff}} \sim m_{\psi}^{2}$ (see section
\ref{zh-orbifolds}) they are related to the recently observed value
of
the effective cosmological constant (dark energy) %
\index{cosmological constant} %
\mbox{$\Lambda_{\mathrm{eff}} \sim$} \mbox{$\sim
10^{-123}\Lambda_{\rm Pl(4)} \sim 10^{-57}\,\text{cm}^{-2}$}. These
ultra-light particles begin to oscillate at characteristic times $t
\sim 10^{18}$\,sec which is of order of the Universe age. Thus,
these particles did not oscillate coherently up to the present time
and a split\-ting of the scale factor of the internal space into a
background component and gravexcitons makes no sense. A more
adequate interpretation of the scale factor dynamics would be in
terms of a slowly varying background in the sense of a quintessence
scenario \cite{Sahni2000, zhWCOS}.
However, such light gravexcitons will lead to a temporal variability
of the fine structure constant above the experimentally established
value \cite{zhGSZ} (see section \ref{zh-fine str}).

Finally, it should be noted that light gravexcitons can lead to the
ap\-pea\-ran\-ce of a fifth force with characteristic length scale
$\lambda \sim 1/m_{\psi}$. Recent gravitational (Cavendish-type)
experiments (see e.g. \cite{zhexperiment}) exclude fifth force
particles with masses $m_{\psi}\lesssim 1/(10^{-2}\,\text{cm})\sim
10^{-3}\,\text{eV}$. This sets an additional restriction on the
allowed mass region of gravexcitons. Thus, physically sensible
models should allow for parameter configurations which exclude such
ultra-light gravexcitons.


\subsection{\!Heavy gravexcitons: \boldmath$m_{\psi} \geq
  10^{-2}$\,GeV\label{zh-heavy}}
\hspace*{3cm}This section is devoted to the investigation of
gravexcitons/ radions with masses $m_{\psi} \geq
10^{-2}\,\text{GeV}$ for which the decay plays an impor\-tant role.
Because of $m_{\psi} \gg m_{eq}$, the corresponding modes begin to
oscillate during the RD stage.  We consider two scenarios
separately. The first one con\-tains an evolutionary stage with
transient gravexciton dominance ($\psi$-dominance), whereas in the
second one gravexcitons remain always sub-dominant.

\label{zh-heavy trans} {\bf\itshape The transiently
\boldmath$\psi$-dominated Universe}. In this subsection we consider
a scenario where the Universe is already at the RD stage when the
gravexcitons begin their oscillations. The initial heating could be
induced, e.g., by the decay of some additional very massive
(inflaton) scalar field. We assume that the Hubble parameter at this
stage is defined by the energy density of the radiation. The
gravexcitons start their oscillations when the radiation cools down
to the temperature $T_{in} \sim \sqrt{m_{\psi}M_{\rm Pl(4)}}$ (see
Eq.~(\ref{zh4.17})). From the dust-like red-shifting of the energy
density $\rho_{\psi}$ (see Eq.~(\ref{zh4.15})) follows that the
ratio $\rho_{\psi}/\rho_{rad}$ increases like $ 1/T $ when $T$
decreases. At some critical temperature $T_{crit}$ this ratio
reaches $\sim $1 and the Universe becomes
$\psi$-dominated:\vspace*{-1mm}
\begin{equation}
  \label{zh4.23}%
  m_{\psi}^2 \psi_{in}^2 \left(\!\frac{T}{T_{in}}\!\right)^{\!\!3}
  \sim T^4 \Longrightarrow T_{crit} \sim T_{in}
  \left(\!\frac{\psi_{in}}{M_{\rm Pl(4)}}\!\right)^{\!\!2}\!.
\end{equation}

\index{Hubble parameter} %
After that the Hubble parameter is defined by the energy density of
the gravexcitons: $H^2M_{\rm Pl(4)}^2 \sim \rho_{\psi}$ (with
$\rho_{\psi}$ from Eq.~\ref{zh4.15}). This stage is transient and
ends when the gravexcitons decay at the temperature $T_D$:
\[
 H_D^2 M_{\rm Pl(4)}^2 \sim \Gamma^2 M_{\rm Pl(4)}^2 \sim m_{\psi}^2
  \psi_{in}^2 \left(\!\frac{T_D}{T_{in}}\!\right)^{\!\!3} \Longrightarrow
\]\vspace*{-3mm}
\begin{equation}
  \label{zh4.24}
   \Longrightarrow T_D \sim T_{in}
   \left(\!\frac{M_{\rm Pl(4)}}{\psi_{in}}\!\right)^{\!\!2/3}\!\left(\!\frac{m_{\psi}}{M_{\rm Pl(4)}}\!\right)^{\!\!4/3}\! .
\end{equation}

We assume that, due to the decay, all the energy of the gravexcitons
is converted into radiation and that a reheating occurs. The
corresponding reheating temperature can be estimated
as:\vspace*{-3mm}
\begin{equation}
  \label{zh4.25} H_D^2 M_{\rm Pl(4)}^2 \sim \Gamma^2 M_{\rm Pl(4)}^2 \sim
  T_{RH}^4 \Longrightarrow T_{RH} \sim \sqrt{\frac{m_{\psi}^3}{M_{\rm Pl(4)}}}.
\end{equation}
Because the Universe before the gravexciton decay was
gravexciton-dominated, it is clear that the reheating temperature
$T_{RH}$ should be higher than the decay temperature $T_D$. This
provides a lower bound for $\psi_{in}$:
\begin{equation}
  \label{zh4.26}
  T_{RH} \geq T_D \Longrightarrow \psi_{in}
  \geq \sqrt{m_{\psi}M_{\rm Pl(4)}} \sim T_{in}.
\end{equation}
Substitution of this estimate into Eq.~(\ref{zh4.23}) shows that the
minimal critical temperature (at which the Universe becomes
$\psi$-dominated) is equal to the reheating temperature: $T_{crit
  (min)} \sim T_{RH}$.

If we additionally assume the natural initial condition $\psi_{in}
\sim M_{\rm Pl(4)}$, then it holds $T_{crit} \sim T_{in}$ and the
Universe will be $\psi$-dominated from the very begin\-ning of the
gravexciton oscillations. The upper bound on $\psi_{in}$ is set by
the exclusion of quantum gravity effects: $m_{\psi}^2\psi_{in}^2
\leq M_{\rm Pl(4)}qu$. Hence, in the considered scenario it should
hold
\begin{equation}
  \label{zh4.27}
  T_{in} \leq \psi_{in} \leq M_{\rm Pl(4)}
  \left(\!\frac{M_{\rm Pl(4)}}{m_{\psi}}\!\right)\!.
\end{equation}

A successful nucleosynthesis requires a temperature $T \gtrsim
1$\,MeV during the RD stage.  If we assume that this lower bound is
fulfilled for the reheating temperature (\ref{zh4.25}), then we find
the lower bound on the gravexciton mass
\begin{equation}
  \label{zh4.28}
  m_{\psi} \gtrsim 10^{4}\,\text{GeV}.
\end{equation}

It is also possible to consider a scenario where the $\psi-$field
acts as inflaton itself. In such a scenario, the Universe is
$\psi$-dominated from the very beginning and for the amplitude of
the initial oscillations one obtains:
  \begin{equation}
  \label{zh4.29}
  H M_{\rm Pl(4)} \sim \sqrt{\rho_{\psi}} \Longrightarrow H_{in}M_{\rm Pl(4)} \sim m_{\psi}M_{\rm Pl(4)}
  \sim \psi_{in}m_{\psi} \Longrightarrow \psi_{in} \sim M_{\rm Pl(4)}.
\end{equation}

\noindent The reheating temperature is then again given by the
estimate (\ref{zh4.25}) and the gravexciton masses should also
fulfill the requirement (\ref{zh4.28}).

\label{zh-heavy sub}{\bfseries\itshape Sub-dominant gravexcitons}.
In this subsection we consider a scenario where the $\psi$-field
undergoes a decay, but the gravexcitons never dominate the dynamics
of the Universe. The Hubble parameter of the Universe is then
defined by the energy density of other (matter) fields which behave
as radiation for $t \leq t_{eq}$ and as dust for $t \geq t_{eq}$.
The energy density $\rho_{\psi}$ is always much less than the total
energy density of the other fields.

\noindent i.~\textit{Decay during the RD stage}

Here, we analyze the behavior of $\psi$-particles that decay during
the RD stage, when $H^2 \sim T^4/M_{\rm Pl(4)}^2$.  Again we will
clarify for which masses $m_{\psi}$ and initial oscillation
amplitudes $\psi_{in}$ such a scenario can hold. A decay during RD
implies that the decay temperature $T_D$, estimated as
\begin{equation}
  \label{zh4.30} \Gamma \sim H_D \sim \frac{T_D^2}{M_{\rm Pl(4)}}
  \Longrightarrow T_D \sim \sqrt{\frac{m_{\psi}^3}{M_{\rm Pl(4)}}},
  \end{equation}
should be higher than the temperature
  \begin{equation}
  \label{zh4.31} T_{eq} \sim \sqrt{H_{eq}M_{\rm Pl(4)}} \sim 1\, \mbox{eV}\,
\end{equation}
of the matter/radiation equality. This yields the following
restriction on the gravexciton masses:\vspace*{-3mm}
\begin{equation}
  \label{zh4.32}
  T_D \gtrsim T_{eq} \Longrightarrow m_{\psi}
  \gtrsim M_{\rm Pl(4)}\left(\!\frac{T_{eq}}{M_{\rm Pl(4)}}\!\right)^{\!\!2/3} \equiv m_d \sim 1\, \mbox{GeV}.
\end{equation}

The mass parameter $m_d$ corresponds to particles which decay at the
moment $t_{eq}$:  $\Gamma \sim H_{eq}$. The bound on the initial
oscillation amplitude $\psi_{in}$ can be found from the energy
sub-dominance condition for the gravexcitons at the moment of their
decay $t_D$\vspace*{-3mm}
\begin{equation}
  \label{zh4.33}
  \left. \rho_{\psi}\right|_{t_D} \approx \psi_{in}^2 m_{\psi}^2
  \left(\!\frac{T_D}{T_{in}}\!\right)^{\!\!3} < T^4_D.
\end{equation}
It reads
\begin{equation}
  \label{zh4.34} \psi_{in} < \sqrt{m_{\psi}M_{\rm Pl(4)}} \sim T_{in} ,
\end{equation}
where $T_{in}$ is defined by Eq.~(\ref{zh4.17}). It can be easily
seen that condition (\ref{zh4.34}) is supplementary to the condition
(\ref{zh4.26}). During the decay, the energy of the gravexcitons is
converted into radiation: $\rho_{\psi} \rightarrow \rho_{r,2}$ with
temperature $T_r^4 \sim$ $\sim \left. \rho_{\psi}\right|_{t_D}$. For
a scenario with $\left. \rho_{\psi}\right|_{t_D}\ll T_D^4$, and
hence $T_r \ll T_D$, the energy density $\rho_{r,2}$ contributes
only negligibly to the total energy density and the gravexciton
decay does not spoil the standard picture of a hot Universe with
successful big bang nucleosynthesis (BBN).

\noindent ii.~\textit{Gravexciton decay during the MD stage}

\index{Hubble parameter} %
At the MD stage (for $t>t_{eq}$) the Hubble parameter reads (see
e.g. \cite{zhKT}, page 504)\vspace*{-3mm}
\begin{equation}
  \label{zh4.35}
  t \sim H^{-1} \sim \frac{M_{\rm Pl(4)}}{T^{3/2}T_{eq}^{1/2}}
  \Longrightarrow  H M_{\rm Pl(4)} \sim T^{3/2}T_{eq}^{1/2},
\end{equation}
and the decay temperature $T_D$ of the gravexcitons can be estimated
as
\[
\Gamma \sim H_D \sim
  \frac{T_D^{3/2}T_{eq}^{1/2}}{M_{\rm Pl(4)}} \Longrightarrow
\]\vspace*{-3mm}
\begin{equation}
\label{zh4.36}
  \Longrightarrow T_D^{3} \sim \frac{(\Gamma M_{\rm Pl(4)})^2}{T_{eq}} \sim
  T_{eq}^3
  \left(\!\frac{m_{\psi}}{M_{\rm
  Pl(4)}}\!\right)^{\!\!4}\!\left(\!\frac{m_{\psi}}{m_{eq}}\!\right)^{\!\!2}\!
  .
\end{equation}

For a decay during MD this decay temperature should be less then
$T_{eq}$, and as implication a restriction on the mass of the
$\psi$-field can be obtained\vspace*{-1mm}
\begin{equation}
  \label{zh4.37}
  T_D < T_{eq} \Longrightarrow m_{\psi} <
  M_{\rm Pl(4)}\left(\!\frac{T_{eq}}{M_{\rm Pl(4)}}\!\right)^{\!\!2/3}\!\! = m_d ,
\end{equation}
which is supplementary to the inequality (\ref{zh4.32}). The
restriction on the initial amplitude $\psi_{in}$ can be found from
the condition of matter dominance and the fact that heavy
gravexcitons begin to oscillate at the RD stage when \mbox{$T_{in}
\sim \sqrt{m_{\psi}M_{\rm Pl(4)}}$}\vspace*{-3mm}
\[
\psi_{in}^2m_{\psi}^2
  \left(\!\frac{T_D}{T_{in}}\!\right)^{\!\!3} \!\sim T_r^4 < H_D^2M_{\rm Pl(4)}^2 \sim
  \frac{T_D^3T_{eq}}{M_{\rm Pl(4)}^2}M_{\rm Pl(4)}^2\Longrightarrow
\]\vspace*{-2mm}
\begin{equation}
  \label{zh4.38}
  \Longrightarrow
  \psi_{in} < M_{\rm Pl(4)}\left(\!\frac{T_{eq}}{T_{in}}\!\right)^{\!\!1/2} \!\sim
  M_{\rm Pl(4)}\left(\!\frac{H_{eq}}{m_{\psi}}\!\right)^{\!\!1/4} \!\ll M_{\rm Pl(4)}.
\end{equation}
Condition (\ref{zh4.38}) guarantees that there is no additional
reheating and the BBN is not spoiled.

In this section, we discussed different cosmological scenarios
affected by the dynamics of gravitational excitons/radions. These
massive moduli fields describe the conformal excitations of the
internal spaces in higher dimensional models and are WIMPs in the
external space-time. We demonstrated that observable cosmological
data set strong constraints on the gravexciton masses and the
amplitudes of their initial oscillations.


\subsection{\!Variation\\
  \hspace*{-1.2cm}of the fine-structure
  constant\label{zh-fine str}}

\hspace*{3cm}\index{fine-structure constant}As we have seen in section (\ref{zh-light exci}),
light gravexcitons %
\index{gravitational excitons} %
with masses $m_{\psi} \leq 10^{-2}\,\text{GeV}$ have a decay time
greater than the age of the Universe. %
\index{age of Universe} %
Moreover, ultra-light particles with masses $m_{\psi} \leq
10^{-33}\,\text{eV}$ have not yet started to oscillate.  It means
that, in the case of such ultra-light particles, the internal spaces
are not stabilized up to present time in the minimum of the
effective potential. However, there is very strong restrictions on
dynamical behavior of the internal spaces following from experiments
on the time variation of the fine-structure constant
$\alpha=e^2/\hbar c$. In KK models the effective fine-structure
constant is inversely proportional to the volume of the internal
spaces: $\alpha \sim $ $\sim V^{-1}_{D'}(t)$, where $V_{D'}(t) =\exp
(\sum_{i=1}^n d_i\tilde \beta^i) V_{D'}$ is the internal space
volume at the moment $t$ and $V_{D'}$ is defined in
Eq.~(\ref{zh2.5}). The origin of such dependence can be easily seen
from action (\ref{zh3.9}) where the Lagrangian density can be
rewritten in the form\,\footnote{\,It is well known that the theory
of electromagnetic
  field can be described with the help of the Lagrangian density of
  the form $\mathcal{L} = -(1/2) \sqrt{|g|}(1/e^2)F^2$ where $e$ is
  the electron charge.}  $\mathcal{L} = -(1/2) \sqrt{|\tilde g^{(0)}|}
(V_{D'}(t)/V_{D'})(1/e^2)F^2$ and we chose $D_0=4$. Therefore, we
can introduce an effective electron charge
$e^2_{\mathrm{eff}}(t)=e^2V_{D'}/V_{D'}(t)$, which defines the
written above dependence of the fine-structure constant on the
internal space volume. Thus, for the variation of $\alpha$ with time
we obtain the following equation\,\footnote{\,We remind that dots
denote the derivatives
  with respect to the synchronous time in the Einstein frame. We omit
  here the symbol ``tilde'' for $t$.}:
\begin{equation}
  \label{zh4.39} \left|\frac{\dot \alpha}{\alpha}\right| =\left |\frac{\dot
      V_{D'}(t)}{V_{D'}(t)}\right | \!.
\end{equation}\vspace*{-1mm}

In the case of one internal space $n=1$ this formula
yields\vspace*{1mm}
\begin{equation}
  \label{zh4.40}
  \left|\frac{\dot \alpha}{\alpha}\right|=
  d_1|\dot{\tilde{\beta}}^1| =\frac{\sqrt{8\pi}}{M_{\rm Pl(4)}}\, \sqrt{\frac{2d_1}{D-2}}\; |\dot
  \psi^1| =\frac{\sqrt{8\pi}}{M_{\rm Pl(4)}}\, \sqrt{\frac{2d_1}{D-2}}\; \left|\frac{\Delta
      \psi^1}{\Delta
      t}\right|\!,
\end{equation}\vspace*{-2mm}

\noindent where we used the relation (\ref{zh2.17}).

The experimental bounds on the time variation of $\alpha$ have been
considerably refined during recent years (see, e.g., [407---411] and
references therein). Different experiments give different bounds on
$|\dot{\alpha}/\alpha|$ (see Table II in \cite{zhUzan}), from
$\lesssim 10^{-12}\, \mbox{yr}^{-1}$ (following from the data
analysis of the observed cosmic microwave background
\cite{zhHannestad}) to $\lesssim 10^{-17}\,\text{yr}^{-1}$
(following from the Oklo experiment \cite{zhDD}). Estimates on
primordial nucleosynthesis require $|\Delta \alpha/\alpha| \lesssim
10^{-4}$ at a redshift of order $z = 10^9$---$ 10^{10}$
\cite{zhKPW}, i.e. $|\dot{\alpha}/\alpha| \lesssim 10^{-14}\,
\mbox{yr}^{-1}$. The WMAP data
analysis \cite{zhMMRTAV} gives upper bounds on the variation of
$\alpha$ during the time from re-ionization/recombination ($z \sim
1100$) until today: $|\Delta \alpha/\alpha| \lesssim 2\,\times$
$\times 10^{-2}$---$ 6\times 10^{-2}$, i.e. $|\dot \alpha/\alpha |
\lesssim 2\times 10^{-12} $---$ 6\times 10^{-12}\, \mbox{yr}^{-1}$.
In all these estimates $\dot{\alpha} = \Delta \alpha /\Delta t$ is
the average rate of change of $\alpha$ during the time interval
$\Delta t$ (corresponding to a redshift $z$). For our estimates, we
use the bound $|\dot{\alpha}/\alpha| \lesssim 10^{-15}\,
\mbox{yr}^{-1}$ \cite{zhWebb}, which follows from observations of
the spectra of quasars at a Hubble time scale $\Delta t \sim H^{-1}
\sim 10^{10}$ years.

Coming back to Eq.~(\ref{zh4.40}) for the fine-structure constant
variation, we obtain in our model
\begin{equation}
  \label{zh4.41}
  \left|\frac{\dot \alpha}{\alpha}\right| \sim \sqrt{8\pi}\,
  \sqrt{\frac{2d_1}{D-2}}\; 10^{-10} {\mbox {yr}}^{-1} ,
\end{equation}\vspace*{-2mm}

\noindent where for the Hubble time scale $\Delta t \sim H^{-1} \sim
10^{10}$ we used $|\Delta \psi^1| \sim \psi_{in} \sim$ $\sim M_{\rm
Pl(4)}$ yr which is usual assumption for light gravexcitons (see
section \ref{zh-light
  exci}). It is obvious that this value is many orders of magnitude
larger than the experimental bounds $10^{-15}\,\mbox{yr}^{-1}$. This
contradiction means that light gravexcitons with masses $m_{\psi}
\lesssim 10^{-2}\,\text{GeV}$ and initial amplitude $\psi_{in} \sim
M_{\rm Pl(4)}$ should have decayed at sufficiently early times of
the evolution of the Universe in order not to contradict the
experimental bounds on the variation of the fine structure constant.
From this point of view, the presence of such light gravexcitons is
unacceptable for the time after the end of primordial
nucleosynthesis. This restriction can be circumvented in the case
$\psi_{in} \ll M_{\rm Pl(4)}$ (see the next \mbox{section
\ref{zh-liv}}). %
\index{scalar field|)} %

The above estimate shows that too much variation of the
fine-structure constant is a typical problem of models with
dynamical behavior of the\linebreak internal spaces if they are not
stabilized at sufficiently early times (see, e.g., section
\ref{zh-Sp}).\vspace*{-3mm}


\subsection{\!Lorentz invariance violation\label{zh-liv}}

\hspace*{3cm}\index{Lorentz invariance violation}Obviously, the
term, describing interaction between gravexcitons and photons in
action (\ref{zh4.1}), modifies the Maxwell equations, and,
con\-se\-quent\-ly, results in a modified dispersion relation for
photon \cite{zhLIV}. In this section we demonstrate that this
modification has a rather specific form. For simplicity we assume
that $\tilde
g^{0}$ in action (\ref{zh4.1}) is the flat FRW metric %
\index{FRW metric} %
and for the scale factors $\tilde a$ and synchronous time $\tilde
t$, we omit the tilde symbol.  It is worth noting that action
(\ref{zh4.1}) (the third line) is conformally invariant in the case
when $D_0=4$, $D_0$-dimensional field strength tensor, $F_{\mu
\nu}$, is gauge invariant and the electromagnetic field is
antisymmetric as usual,
$F_{\mu\nu}=\partial_{\mu}A_{\nu}-\partial_{\nu}A_{\mu}$.  Varying
(\ref{zh4.1}) with respect to the electromagnetic vector potential,
\begin{equation}
  \label{zh4.42}
  \partial_{\nu}\left[ \sqrt{-g} \left(1-\mathcal G\kappa_0 \psi
    \right)F^{\mu \nu} \right]=0,
\end{equation}
where $\mathcal G := 2 \sqrt{d_1/[(D_0-2)(D-2)]}$.  The second term
in the round brackets $\mathcal G\kappa_0 \psi F^{\mu\nu}$ reflects
the interaction between photons and the scalar field $\psi$, and as
we show below, it is responsible for Lorentz invariance violation
(LV). In particular, coupling between photons and the scalar field
$\psi$ makes the speed of photons different from the standard speed
of light. Eq.~(\ref{zh4.42}) together with Bianchi identity,
$F_{(\mu\nu,\lambda)}=0,$ (which is preserved in the considered
model due to gauge-invariance of the tensor $F_{\mu\nu}$
\cite{zhGSZ}) defines a complete set of the generalized Maxwell
equations. As we noted, the electromagnetic part of action
(\ref{zh4.1}) is conformally invariant in the $4D$ dimensional
space-time. So, it is convenient to present the flat FRW metric
$\tilde g^{0}$ in the conformally flat form: $\tilde
g^{0}_{\mu\nu}=a^2\eta_{\mu \nu}$, where $\eta_{\mu\nu}$ is the
Minkowski metric.

\index{tensor of electromagnetic field} %
Using the standard definition
of the electromagnetic field tensor $F_{\mu \nu}$:%
\begin{equation}
\label{zh4.43} F_{\mu\nu}= \left(\!\!\!\begin{array}{cccc} 0 &\!\!
-E_1 &\!\! -E_2 &\!\!
    -E_3 \\ E_1 &\!\! 0 &\!\! B_3 &\!\! -B_2 \\ E_2 &\!\! -B_3 &\!\! 0 &\!\! B_1 \\ E_3 &\!\! B_2
    &\!\! -B_1 & 0
  \end{array}\!\!\!\!\right)\!,~~  F^{\mu\nu}= \left(\!\!\!\begin{array}{cccc} 0 &\!\! E_1 &\!\! E_2 &\!\! E_3
    \\ -E_1 &\!\! 0&\!\!
    B_3 &\!\! -B_2
    \\ -E_2 &\!\! -B_3 &\!\! 0 &\!\! B_1 \\ -E_3 &\!\! B_2 &\!\! -B_1 &\!\! 0
  \end{array}\!\!\!\!\right)\!,
\end{equation}%
we obtain the complete set of the Maxwell equations in
vacuum\,\footnote{\,The electric ($\mathbf{E}$) and magnetic
  ($\mathbf{B}$) fields are related to the vector potentials (scalar
  $\phi$ and vector $\mathbf{A}$) as\vspace*{-2mm}
$$
\mathbf{B}= \mathbf{\nabla} \times \mathbf{A},~
\mathbf{E}=-\mathbf{\nabla} \phi -
\frac{\partial\mathbf{A}}{\partial
  \eta},
$$
where the operator $\mathbf{\nabla}$ denotes the spatial derivatives
in $3$-dimensional flat space.},
\begin{equation}
  \mathbf{\nabla} \cdot \mathbf{B}  = 0, \label{zh4.44}
  \end{equation}
  \begin{equation}
  \mathbf{\nabla} \cdot \mathbf{E}  = \frac{{\mathcal G}\kappa_0}{1-
    {\mathcal G}\kappa_0 \psi} (\mathbf{\nabla} \psi \cdot
  \mathbf{E}),
  \label{zh4.45}
  \end{equation}
  \begin{equation}
  \mathbf{\nabla} \times \mathbf{B}  = \frac{\partial
    \mathbf{E}}{\partial \eta} - \frac{{\mathcal G}
    {\kappa_0\dot\psi}}{1- {\mathcal G}\kappa_0 \psi} \mathbf{E} +
  \frac{{\mathcal G}\kappa_0}{1- {\mathcal G}\kappa_0 \psi}
  [\mathbf{\nabla} \psi \times \mathbf{B}],
  \label{zh4.46}
  \end{equation}
  \begin{equation}
  \mathbf{\nabla} \times \mathbf{E}  = -\frac{\partial
    \mathbf{B}}{\partial \eta}, \label{zh4.47}
\end{equation}
where all operations are performed in the Minkowski space-time,
$\eta$
denotes conformal time %
\index{conformal time} %
related to physical time $t$ as $dt = a(\eta)d\eta$, and an overdot
represents a derivative with respect to conformal time $\eta$.

Eqs.~(\ref{zh4.44}) and (\ref{zh4.47}) correspond to Bianchi
identity, and since it is preserved, Eqs.~(\ref{zh4.44}) and
(\ref{zh4.47}) keep their usual forms. Eqs.~(\ref{zh4.45}) and
(\ref{zh4.46}) are modified due to interactions between photons and
gravexcitons ($\propto \mathcal \kappa_0 \psi$). These modifications
have simple physical meaning: the interaction between photons and
the scalar field $\psi$ acts in Eq.~(\ref{zh4.45}) as an effective
electric charge $e_\mathrm{
  eff}$. This effective charge is proportional to the scalar product
of the $\psi$ field gradient and the $\mathbf{E}$ field, and it
vanishes for an homogeneous $\psi$ field. The modification of
Eq.~(\ref{zh4.46}) corresponds to an effective current
$\mathbf{J}_\mathrm{ eff}$, which depends on both electric and
magnetic fields. This effective current is determined by variations
of the $\psi$ field over the time ($\dot\psi$) and space ($\nabla
\psi$). For the case of a homogeneous $\psi$ field the effective
current is still present and LV takes place. The modified Maxwell
equations are conformally invariant. To account for the expansion of
the Universe we rescale the field components as
$\mathbf{B},\mathbf{E} \rightarrow \mathbf{B}a^2, \mathbf{E}a^2$
\cite{zhgrasso}. Thus, the components of the physical electric and
magnetic field are diluted as $1/a^2$.

To obtain a dispersion relation for photons, we use the Fourier
transform between position and wave number spaces as,
\begin{equation}
  \label{zh4.48}
  \begin{array}{c}
   \displaystyle \mathbf{F}(\mathbf{k}, \omega)  =
    \int \int d \eta ~  d^3\!x \,  e^{-i(\omega \eta - \mathbf{k} \cdot \mathbf{x})} \mathbf{F}(\mathbf{x}, \eta
    ),\\[3mm]
  \displaystyle  \mathbf{F}(\mathbf{x}, \eta )  = \frac{1}{(2\pi)^4}\int \int
    {d\omega}~ {d^3\!k} e^{i(\omega \eta - \mathbf{k}\cdot
      \mathbf{x})} \mathbf{F}(\mathbf{k}, \omega ).
  \end{array}
\end{equation}

Here, $\mathbf{F}$ is a vector function describing either the
electric or the magnetic field, $\omega$ is the angular frequency of
the electromagnetic wave measured today, and $\mathbf{k}$ is the
wave vector. We assume that the field $\psi$ is an oscillatory field
with the frequency $\omega_{\psi}$ and the momentum $\mathbf{q}$, so
$ \psi(\mathbf{x}, \eta )=C\exp[{i(\omega_\psi \eta - \mathbf{q}
\cdot
  \mathbf{x})}]$, $C={\rm const}$.  Eq.~(\ref{zh4.44}) implies
$\mathbf{B} \perp \mathbf{k}$. Without loosing of generality, and
for simplicity of description we assume that the wave vector
$\mathbf{k}$ is oriented along the $\mathbf{z}$ axis. Using
Eq.~(\ref{zh4.47}) we get $\mathbf{E} \perp \mathbf{B}$.  Therefore,
Eqs.~(\ref{zh4.45}) and (\ref{zh4.47}) in the component form read
\begin{equation}
  (1-{\mathcal G}\kappa_0 \psi) k E_z = {\mathcal G}\kappa_0 \psi
  (q_{x} E_x + q_{y} E_y + q_{z} E_z) \label{zh4.49}
\end{equation}
and\vspace*{-3mm}
\begin{equation}
    kE_y  = \omega B_x \label{zh4.50},
\end{equation}\vspace*{-5mm}
\begin{equation}
    kE_x  = -\omega B_y \label{zh4.51}.
\end{equation}
Eq.~(\ref{zh4.46}) in components can be rewritten as
\begin{equation}
  (1-{\mathcal G}\kappa_0 \psi)kB_y  = -(1-{\mathcal G}\kappa_0
  \psi)\omega E_x + {\mathcal G}\kappa_0 \psi \, (\omega_\psi E_x +
  q_{z} B_y ),
  \label{zh4.52}
  \end{equation}\vspace*{-6mm}
\begin{equation}
  (1-{\mathcal G}\kappa_0 \psi)kB_x  = (1-{\mathcal G}\kappa_0
  \psi)\omega E_y -{\mathcal G}\kappa_0 \psi \, (\omega_\psi E_y -
  q_{z} B_x ),
  \label{zh4.53}
  \end{equation}\vspace*{-5mm}
\begin{equation}
  0  = (1-{\mathcal G}\kappa_0 \psi)\omega E_z -{\mathcal G}\kappa_0\psi \, (\omega_\psi E_z + q_{y} B_x
  - q_{x} B_y).
  \label{zh4.54}
\end{equation}
A linearly polarized wave can be expressed as a superposition of
left (L, $-$) and right (R, $+$) circularly polarized (LCP and RCP)
waves. Using the polarization basis of Sec.~1.1.3 of
Ref.~\cite{zhvarshalovich89}, we derive $E^\pm =(E_x \pm i
E_y)/\sqrt{2}$.  From Eqs.~(\ref{zh4.50})---(\ref{zh4.53}), for LCP
and RCP waves we get
\begin{equation}
  (1-n_{+}^2)E^{+}=0,  \quad (1-n_{(-)}^2)E^{-}=0,
  \label{zh4.55}
\end{equation}
where $n_{+}$ and $n_-$ are refractive indices for RCP and LCP
electromagnetic waves
\begin{equation}
  n_+^2= \frac{k^2  \left[1-{\mathcal G} \kappa_0 \psi
      (1+q_{z}/k)\right]}{\omega^2\left[1-{\mathcal G} \kappa_0\psi(1+
      \omega_\psi /\omega)\right]}= n_-^2. \label{zh4.56}
\end{equation}

In the case when Lorentz invariance is preserved, the
electromagnetic waves propagating in vacuum have $n_+=n_-=n=k/\omega
\equiv 1$. For the electromagnetic waves propagating in the
magnetized plasma, $k/\omega \neq 1$, and the difference between the
LCP and RCP refractive indices describes the Faraday rotation
effect, $\alpha \propto \omega (n_+ - n_-)$ \cite{zhkrall}. In the
considered model, since $n_+=n_-$ the rotation effect is absent, but
the speed of electromagnetic waves propagation in vacuum differs
from the speed of light $c$ (see also Ref. \cite{zhCantcheff:2004dn}
for LV induced by electromagnetic field coupling to other generic
field). This difference implies the propagation time delay effect
\begin{equation}
  \label{zh4.57}
  \Delta t = \Delta l \left(\!1-\frac{\partial k}{\partial
      \omega}\!\right)\!,
\end{equation}
where $\Delta l$ is a propagation distance, $\Delta t$ is the
difference between the photon travel time and that for a ``photon''
which travels at the speed of light $c$. Here, $t$ is physical
synchronous time. This formula does not take into account the
evolution of the Universe.  However, it is easy to show that the
effect of the Universe expansion is negligibly small. To get
$\partial k/\partial \omega$, we should note that we have defined
the system of 6 equations (\ref{zh4.49})---(\ref{zh4.54}) with
respect to 6 components of the vectors $\mathbf{E}$ and
$\mathbf{B}$. This system has non-trivial solutions only if its
determinant is zero. From this condition we get the dispersion
relation:
\begin{equation}
  \label{zh4.58} \omega^2\left(\!1-{\mathcal G} \kappa_0\psi\left(\!1+
    \frac{\omega_\psi}{\omega}\!\right)\!\right) = k^2 \left(\!1-{\mathcal G}
    \kappa_0 \psi \left(\!1+\frac{q_{z}}{k}\!\right)\!\right)\!.
\end{equation}

Solving the dispersion relation as a square equation with respect to
$k$, we can obtain
\begin{equation}
  \label{zh4.59} \frac{\partial k}{\partial \omega}\simeq
  \pm\left\{\!1+\frac{1}{2}\left[\frac{\omega_{\psi}^2-q_z^2}{4\omega^2}\right]({\mathcal G} \kappa_0\psi)^2\!\right\}\!\!,
\end{equation}
where $\pm$ signs correspond to photons forward and backward
directions res- \mbox{pectively.}

The modified inverse group velocity (\ref{zh4.58}) shows that the LV
effect can be measured if we know the gravexciton frequency
$\omega_{\psi}$, $z$-component of the momentum $q_z$ and its
amplitude $\psi$. For our estimates, we assume that $\psi$ is the
oscillatory field, satisfying (in local Lorentz frame) the
dispersion relation, $\omega_\psi^2 = m_\psi^2 + \mathbf{q}^2$,
where $m_{\psi}$ is the mass of gravexcitons\,\footnote{\,To get
physical values of the
  corresponding parameters we should rescale them by the scale factor
  $a$.}. Unfortunately, we do not have any information concerning
parameters of gravexcitons (some estimates are given in sections
\ref{zh-light exci} and \ref{zh-heavy}). Thus, we intend to use
possible LV effects (supposing it is caused by interaction between
photons and gravexcitons) to set limits on gravexciton parameters.
For example, from Eqs.~(\ref{zh4.57}) and (\ref{zh4.59}) we can
easily get the following estimate for the upper limit of the
amplitude of gravexciton oscillations:
\begin{equation}
  \label{zh4.60}
  |\psi| \approx \frac{1}{\sqrt{\pi}\, \mathcal G}\,
  \sqrt{\left|\frac{\Delta t}{\Delta l}\right|}\, \frac{\omega}{m_{\psi}}\, M_{\rm Pl(4)},
\end{equation}
where for $\omega$ and $m_{\psi}$ we can use their physical values.
In
the case of gamma-ray burst (GRB) %
\index{gamma-ray bursts (GRBs)|(} %
with \mbox{$\omega \sim 10^{21}\div 10^{22}\,\text{Hz} \sim
10^{-4}\div 10^{-3}\,\text{GeV}$} and \mbox{$\Delta l \sim 3\div
5\times$} \mbox{$\times 10^{9}\,\,\text{y} \sim 10^{17}$}\,\,sec,
the typical upper limit for the time delay is \mbox{$\Delta t
\sim$}\linebreak $\sim 10^{-4}\,\text{sec}$ \cite{zhMP06}.

For these values the upper limit on gravexciton amplitude of
\mbox{oscillations is}
\begin{equation}
  \label{zh4.61}
  \left|\kappa_0\psi \right| \approx
  \frac{10^{-13}\,\text{GeV}}{m_{\psi}}.
\end{equation}

This estimate shows that our approximation $\kappa_0 \psi <1$ works
for gravexciton masses $m_{\psi}>10^{-13}\,\text{GeV}$. Future
measurements of the time-delay effect for GRBs at frequencies
$\omega \sim 1$---$10\,\text{GeV}$ would increase significantly the
limit up to $m_{\psi} > 10^{-9}\,\text{GeV}$. On the other hand, as
we wrote in section \ref{zh-light exci}, the Cavendish-type
experiments exclude fifth force particles with masses $m_{\psi}
\lesssim$ $\lesssim 1/(10^{-2}\,\text{cm}) \sim
10^{-12}\,\text{GeV}$, which is rather close to our lower bound for
$\psi$ field masses. Respectively, we slightly shift the considered
mass lower limit to be $m_{\psi} \geq 10^{-12}\,\text{GeV}$.  These
masses are considerably higher than the mass corresponding to the
equality between the energy densities of the matter and radiation
(matter/radiation equality), \mbox{$m_{eq}\sim H_{eq}\sim
10^{-37}\,\text{GeV}$}, where $H_{eq}$ is the Hubble ``constant'' at
matter/radiation equality. It means that such $\psi$-particles start
to oscillate during the radiation dominated epoch \index{radiation
dominated (RD) epoch}(see section \ref{zh-light exci}).  Another
bound on the $\psi$-particles masses comes from the condition of
their stability. With respect to decay $\psi \to \gamma \gamma$ the
life-time of $\psi$-particles is $\tau \sim (M_{\rm
Pl(4)}/m_{\psi})^3t_{\rm Pl}$ (see Eq.~(\ref{zh4.3})), and the
stability conditions requires that the
decay time should be greater than the age of the Universe. %
\index{age of Universe} %
According to this we consider light gravexcitons with masses
$m_{\psi} \le 10^{-21} M_{\rm Pl(4)} \sim 10^{-2}\,\text{GeV} \sim$
$\sim 20 m_e$ (where $m_e$ is the electron mass).

\index{cosmological gravexcitons}An additional restriction arises
from the condition that such cosmological gravexcitons should not
overclose the observable Universe. This results in conditions
(\ref{zh4.20}) and (\ref{zh4.21}) for $m_{\psi}$ and $\psi_{in}$,
respectively. Thus, for the range of
masses\,\footnote{\,Gravexcitons with such masses are either warm or
cold
  {\bfseries dark matter} (see footnote \ref{DM}).}
$10^{-12}\,\text{GeV}\leq m_{\psi}\leq 10^{-2}\,\text{GeV}$, we
obtain respectively $\psi_{in}\lesssim $ $\lesssim10^{-6}M_{\rm
Pl(4)}$ and $\psi_{in}\lesssim 10^{-9}M_{\rm Pl(4)}$. Now, we want
to estimate the amplitude of oscillations of the considered
gravexciton at the present time. To perform it, we should mention
that prefactors $C_r$ and $B(t)$ in Eq.  (\ref{zh4.8}) are given by
(Eqs.~\ref{zh4.10}) and (\ref{zh4.9}), respectively. We are
interested in the gravexciton oscil\-lations at the present time
$t=t_{univ}$. In this case $s=2/3$ and for $B(t_{univ})$ we obtain:
$B(t_{univ})\sim t^{-1}_{univ}\approx 10^{-61}M_{\rm Pl(4)}$, where
we took into account that for considered masses $\Gamma
t_{univ}\lesssim 1$. Thus, the amplitude of the light gravexciton
oscillations at the present time reads:
\begin{equation}
  \label{zh4.62}
  |\kappa_0\psi| \sim 10^{-60} \frac{ \psi_{in}}{M_{\rm Pl(4)}}
  \left(\!\frac{M_{\rm Pl(4)}}{m_{\psi}}\!\right)^{\!\!3/4}\!.
\end{equation}

Together with the nonovercloseness condition, we obtain from this
expression that \mbox{$|\kappa_0\psi| \sim 10^{-43}$ for
$m_{\psi}\sim 10^{-12}\,\text{GeV}$} and $\psi_{in}\sim
10^{-6}\,M_{\rm Pl(4)}$ and $|\kappa_0\psi| \sim$ $\sim 10^{-53}$
for $m_{\psi}\sim 10^{-2}\,\text{GeV}$ and $\psi_{in}\sim
10^{-9}\,M_{\rm Pl(4)}$. Obviously, it is much less than the upper
limit (\ref{zh4.61}). Note, as we mentioned above, gravexcitons
\mbox{with} masses $m_{\psi}\gtrsim 10^{-2}\,\text{GeV}$ can start
to decay at the present epoch. However, taking into account the
estimate $|\kappa_0\psi| \sim 10^{-53}$, we can easily get that
their energy density $\rho_{\psi} \sim (|\kappa_0\psi|^2/8\pi)M_{\rm
Pl(4)}^2m_{\psi}^2\sim 10^{-55}\,\mathrm{g}/\mathrm{cm}^3$ is much
less than the present energy density of the radiation
$\rho_{\gamma}\sim 10^{-34}\,\mathrm{g}/\mathrm{cm}^3$. Thus,
$\rho_{\psi}$ contributes negligibly in $\rho_{\gamma}$. Otherwise,
the gravexcitons with masses $m_{\psi} \gtrsim 10^{-2}\,\text{GeV}$
should be observed at the present time, which, obviously, is not the
case.


Additionally, it follows from section \ref{zh-fine str} that to
avoid the problem of the fine structure constant variation, the
amplitude of the initial oscillations should satisfy the condition:
$\psi_{in} \lesssim 10^{-5}\,M_{\rm Pl(4)}$ which, obviously,
completely agrees with our upper bound $\psi_{in} \lesssim 10^{-6}\,
\text{GeV}$.

In summary, we shown that LV effects can give additional
restrictions on parameters of gravexcitons. First, gravexcitons
should not be lighter than $10^{-13}\,\text{GeV}$. It is very close
to the limit following from the fifth-force experiment.  Moreover,
experiments for GRB at frequencies $\omega > 1\,\text{GeV}$ can
result in significant shift of this lower limit making it much
stronger than the fifth-force estimates. Together with the
nonovercloseness condition, this estimate leads to the upper limit
on the amplitude of the gravexciton initial oscillations. It should
not exceed $\psi_{in}\lesssim 10^{-6}\,\text{GeV}$. Thus, the bound
on the initial amplitude obtained from the fine structure constant
variation is one magnitude weaker than the bound found from LV even
for the limiting case of the gravexciton masses. This limit becomes
stronger for heavier gravexcitons. Our estimates for the present-day
amplitude of the gravexciton oscillations, following from the above
obtained limitations, show that we cannot use the LV effect for the
direct detections of the gravexcitons.  Nevertheless, the obtained
bounds can be useful for astrophysical and cosmological
applications. For example, let us suppose that gravexcitons with
masses $m_{\psi}>10^{-2}\,\text{GeV}$ are produced during late
stages of the Universe expansion in some regions and GRB photons
travel to us through these regions. Then, Eq.~(\ref{zh4.62}) is not
valid for such gravexcitons having astrophysical origin and the only
upper limit on the amplitude of their oscillations (in these
regions) follows from Eq.~(\ref{zh4.61}). In the case of TeV masses
we get $|\kappa_0\psi|\sim 10^{-16}$. If GRB photons have
frequencies up to 1 TeV, $\omega \sim 1\,\text{TeV}$, then this
estimate is increased by six orders of magnitude.\vspace*{-1.5mm}

\section[\!Dark energy in multidimensional models] {\!Dark energy in
  cur\-va\-tu\-re-non-linear \boldmath$f(R)$\\ \hspace*{-0.95cm}multidimensional cosmological models\label{zh-nonlin}}
\vspace*{-0.5mm}

\hspace*{3cm}\index{dark energy in multidimensional models}Starting
from the pioneering paper \cite{zhStar1980}, the non-linear
(\mbox{with} respect to the scalar cur\-va\-tu\-re $R$) theories of
gravity $f(R)$ have attracted the great deal of interest because
these models can provide a natural mecha\-nism of the early
inflation. Non-linear models may arise either due to quantum
fluctuations of matter fields including gravity \cite{zhBirrDav}, or
as a result of compactification of extra spatial dimensions
\cite{zhNOcompact}. Compared, e.g., to others higher-order gravity
theories, $f(R)$ theories are free of ghosts and of Ostrogradski
instabilities \cite{zhWoodard}. Recently, it was realized that these
models can also explain the late-time acceleration of the Universe.
This fact resulted in a new wave of papers devoted to this topic
(see e.g., recent reviews \cite{zhreviews,zhCLF}).

\index{cur\-va\-tu\-re-non-linear models} %
The most simple, and, consequently, the most studied models are
po\-ly\-no\-mials of $R$: $f(R)=\sum_{n=0}^k C_n R^n \,$ $(k >1)$,
e.g., quadratic $R+R^2$ and quar\-tic $R+R^4$ ones. Active
investigation of these models, which started in 80-th years of the
last century
\cite{zh80-th,zhMaeda}, continues up to now. 
Obviously, the correction terms (to the action of Hilbert---Einstein
type) with $n>1$ give the main contribution in the case of large
$R$, e.g., in the early stages of the Universe's evolution. As it
was shown first in \cite{zhStar1980} for the quadratic model, such
modification of gravity results in early inflation. On the other
hand, function $f(R)$ may also contain negative degrees of $R$. For
example, the simplest model is $R+R^{-n},\, n\geq 1$. In this case
the correction term plays the main role for small $R$, e.g., at the
late stage of the Universe's evolution (see, e.g., numerous
references in \cite{zhGZBR,zhSZPRD2007},). Such modification of
gravity may result in the late-time acceleration of our Universe
\cite{zhCarrolletal}. Non-linear models with polynomial as well as
$R^{-n}$-type correction terms have also been generalized to the
multidimensional case (see, e.g., [377, 378, 428, 429, 431---435]).
Special emphasis was laid on finding parameter regions (regions in
moduli space) which ensure the existence of at least one minimum of
the effective potential for the volume moduli of the internal spaces
and which in this way allow for their stabilization. Additionally,
positive minimum of the effective potential plays the role of the
positive cosmological constant which gives the possibility to
resolve the dark energy problem. It is well known that non-linear
models are equivalent to linear-cur\-va\-tu\-re models with
additional minimal scalar \mbox{field} $\phi$ (dubbed scalaron in
\cite{zhStar1980}). This scalar \mbox{field} corresponds to
additional degree of freedom of non-linear models. This equivalence
is very useful tool for the investigation of the problems of the
internal space stabilization and the external space acceleration.
Let us show it in more detail.
\index{gamma-ray bursts (GRBs)|)} %

\subsection{Internal space stabilization\\ \hspace*{-1.2cm}for
  pure geometrical \boldmath$f(R)$ models\label{zh-general f(R)}}

\hspace*{3cm}We consider a $D= (4+D^{\prime})$~--- dimensional
non-linear pure gravitational theory with action
\begin{equation}
  \label{zh5.1}
  S = \frac {1}{2\kappa^2_D}\int\limits_M d^Dx \sqrt{|\overline g|}
  f(\overline R) ,
\end{equation}
where $f(\overline R)$ is an arbitrary smooth function with mass
dimension $\mathcal{O}(m^2)$ \ ($m$ has the unit of mass) of a
scalar cur\-va\-tu\-re $\overline R = R[\overline g]$ constructed
from the D-dimensional metric $\overline g_{ab}\; (a,b =
1,\mbox{...},D)$.

\index{equation of motion} %
The equation of motion for this theory reads
\begin{equation}
  \label{zh5.2}
  f^{\prime }\overline R_{ab} -\frac12 f\, \overline g_{ab} - \overline
  \nabla_a \overline \nabla_b f^{\prime } + \overline g_{ab} \overline{\square } f^{\prime } = 0 ,
\end{equation}
where $ f^{\prime } =df/d\overline R$, $\; \overline R_{ab} =
R_{ab}[\overline g]$.  $ \overline \nabla_a$ is the covariant
derivative with respect to the metric $\overline g_{ab}$; and the
corresponding Laplacian is denoted by
\begin{equation}
  \label{zh5.3}
  \overline{\square}  = \square [\overline g] = \overline g^{ab}\overline
  \nabla_a \overline \nabla_b = \frac{1}{\sqrt{|\overline g|}}
  \partial_a \left(\! \sqrt{|\overline g|}\,\overline g^{ab}
    \partial_b\! \right)\!\!.
\end{equation}
Eq.~(\ref{zh5.2}) can be rewritten in the form
\begin{equation}
  \label{zh5.4}
  f^{\prime }\overline G_{ab} +\frac12 \overline g_{ab} \left( \overline R
    f^{\prime} - f\right) - \overline \nabla_a \overline \nabla_b f^{\prime } + \overline g_{ab}
  \overline{\square } f^{\prime } = 0,
\end{equation}
where $\overline G_{ab} = \overline R_{ab} -\frac12 \overline R \;
\overline g_{ab}$. The trace of Eq.~(\ref{zh5.2}) is
\begin{equation}
  \label{zh5.5} (D-1)\overline{\square } f^{\prime } = \frac{D}{2} f
  -f^{\prime }\overline R
\end{equation}
and can be considered as a connection between $\overline R$ and $f$.

It is well known, that for\,\footnote{\,We consider the positive
branch
  $f'(\bar{R})>0$.  Although the negative $f'<0$ branch can be
  considered as well (see e.g. Refs.~\cite{zhGMZ2,zhMaeda,zhGZBR}). However, negative values of $f'(\bar
  R)$ result in negative effective gravitational ``constant''
  $G_{\mathrm{eff}}=\kappa^2_D/f'$. Thus $f'$ should be positive for
  the graviton to carry positive kinetic energy (see, e.g.,
  \cite{zhCLF}).\label{zh-f'>0}} $f'(\overline R) >0$ the conformal
transformation
\begin{equation}
  \label{zh5.6} g_{ab} = \Omega^2 \overline g_{ab},
\end{equation}
with
\begin{equation}
\label{zh5.7} \Omega = \left[ f'(\overline R)\right]^{1/(D-2)}\!,
\end{equation}
reduces the non-linear theory (\ref{zh5.1}) to a linear one with an
additional scalar field. The equivalence of the theories can be
easily proven with the help of the following auxiliary formulas:
\begin{equation}
    \label{zh5.8}
    \square = \Omega^{-2}\left[\overline{\square } + (D-2)\overline
      g^{ab}\Omega^{-1}\Omega_{,a}\partial_b\right]
    \Longleftrightarrow \overline {\square} = \Omega^{2}\square -
    (D-2) g^{ab}\Omega \;
    \Omega_{,a}\partial_b ,
\end{equation}
\begin{equation}
    \label{zh5.9}
    R_{ab} = \overline R_{ab} +\frac{D-1}{D-2}( f')^{-2} \overline {\nabla}_a f'
    \overline {\nabla}_b f' -(f')^{-1} \overline {\nabla}_a \overline {\nabla}_b f'-
    \frac{1}{D-2} \overline g_{ab} (f')^{-1} \overline {\square} f'
\end{equation}
and\vspace*{-3mm}
\begin{equation}
    \label{zh5.10} R = (f')^{2/(2-D)}\left\{\! \overline R +\frac{D-1}{D-2}
      (f')^{-2} \overline g^{ab}\partial_a f'\partial_b f'-
      2\frac{D-1}{D-2}(f')^{-1} \overline {\square} f'\!\right\}\!.
\end{equation}

\noindent Thus, Eqs.~(\ref{zh5.4}) and (\ref{zh5.5}) can be
rewritten as
\begin{equation}
  \label{zh5.11}
  G_{ab} = \phi_{,a}\phi_{,b} -\frac12
  g_{ab}g^{mn}\phi_{,m}\phi_{,n} - \frac12 g_{ab}\; e^{\frac {-D}{\sqrt{(D-2)(D-1)}}\phi} \left(\overline R f'- f\right)
\end{equation}
 and
\begin{equation}
  \label{zh5.12}
  \square \phi = \frac {1}{\sqrt{(D-2)(D-1)}}\;
  e^{\frac {-D}{\sqrt{(D-2)(D-1)}}\phi} \left(\!\frac {D}{2} f-f'\overline R \!\right)\!,
\end{equation}
 where\vspace*{-3mm}
\begin{equation}
  \label{zh5.13}
  f' = \frac {df}{d \overline R} := e^{A \phi} > 0,\quad A :=
  \sqrt{\frac{D-2}{D-1}}.
\end{equation}
Eq.~(\ref{zh5.13}) can be used to express $\overline R$ as a
function of the
dimensionless field (scalaron) $\phi$: $\overline R = \overline R( \phi )$. %
\index{scalar perturbations}%

It is easily seen that Eqs.~(\ref{zh5.11}) and (\ref{zh5.12}) are
the equations of motion for the action\vspace*{-1mm}
\begin{equation}
    \label{zh5.14}
    S = \frac{1}{2\kappa^2_D} \int\limits_M d^D x \sqrt{|g|}
    \left(\!
      R[g] - g^{ab} \phi_{,a} \phi_{,b} - 2 U(\phi )\!\right)\!,
\end{equation}\vspace*{-5mm}

\noindent where\vspace*{-3mm}
\begin{equation}
    \label{zh5.15}
    U(\phi ) = \frac12 e^{- B \phi} \left[ \overline R (\phi )e^{A \phi }
      - f\left( \overline R (\phi )\right) \right]\!, \quad B := \frac
    {D}{\sqrt{(D-2)(D-1)}}
\end{equation}
and they can be written as follows:\vspace*{-1mm}
\begin{equation}
  \label{zh5.16} G_{ab} = T_{ab}\left[ \phi , g \right]\!,
\end{equation}
\begin{equation}
  \label{zh5.17}
  \square \phi = \frac {\partial U(\phi)}{\partial \phi}.
\end{equation}

\index{energy-momentum tensor} %
Here, $T_{ab}\left[ \phi , g \right]$ is the standard expression of
the energy--momentum tensor for the minimally coupled scalar field
with potential (\ref{zh5.15}). Eq.~(\ref{zh5.17}) can be considered
as a constraint equation following from the reduction of the
non-linear theory (\ref{zh5.1}) to the linear one (\ref{zh5.14}).
\index{scalar field} %

Let us consider what will happen if, in some way, the scalar field
$\phi$ tends asymptotically to a constant: $\phi \to \phi_{0} $.
From Eq.~(\ref{zh5.13}) we see that in this limit the non-linearity
disappears and (\ref{zh5.1}) becomes a linear theory $f(\overline R)
\sim$\linebreak \mbox{$\sim c_1 \overline R + c_2$} with \mbox{$c_1
= f' = \exp(A \phi_{0})$} and a cosmological constant $-c_2/(2c_1)$.
In the case of homogeneous and isotropic space-time manifolds,
linear purely geometrical theories with constant $\Lambda $-term
necessarily imply an (A)dS geometry. Thus, in the limit $\phi \to
\phi_{0}$ the D-dimensional theory (\ref{zh5.1}) can asymptotically
lead to an (A)dS with scalar cur\-va\-tu\-re:\vspace*{-3mm}
\begin{equation}
  \label{zh5.18}
  \overline R \rightarrow -\frac{D}{D-2}\, \frac{c_2}{c_1}.
\end{equation}

\noindent Clearly, the linear theory (\ref{zh5.14}) would reproduce
this asymptotic (A)dS-limit for $\phi \to \phi_{0}$:\vspace*{-1mm}
\begin{equation}
  \label{zh5.19} R \rightarrow  2\frac{D}{D-2}\, U(\phi_0) = -
  \frac{D}{D-2}\; c_2\,  c_1^{-\frac{D}{D-2}}.
\end{equation}
Hence, in this limit $\overline R / R \to c_1^{\frac{D}{D-2}}$ in
accordance with Eq.~(\ref{zh5.10}) and $f'= c_1$.

Next, we assume that the D-dimensional bulk space-time $M$ undergoes
a spontaneous compactification to a warped product manifold
\begin{equation}
  \label{zh5.20}
  M = M_0 \times M_1 \times \mbox{...} \times M_n
\end{equation}\vspace*{-5mm}

\noindent with metric\vspace*{-3mm}
\begin{equation}
  \label{zh5.21}
  \bar g=\bar g_{ab}(X)dX^a\otimes dX^b=\bar
  g^{(0)}+\sum_{i=1}^ne^{2\bar {\beta} ^i(x)}g^{(i)}.
\end{equation}

The coordinates on the $(D_0=d_0+1)$-dimensional manifold $M_0 $
(usually interpreted as our observable $(D_0=4)$-dimensional
Universe) are denoted by $x$ and the corresponding metric by
\begin{equation}
  \label{zh5.22}
  \bar g^{(0)}=\bar g_{\mu \nu }^{(0)}(x)dx^\mu \otimes
  dx^\nu.
\end{equation}

The internal factor manifolds $M_i$ are taken in the form of
$d_i$-dimensional Einstein spaces (\ref{zh2.3}) and (\ref{zh2.4}).
The specific metric ansatz (\ref{zh5.21}) leads to a scalar
cur\-va\-tu\-re $\bar R$ which depends only on the coordinates $x$
of the external space: $\bar R[\bar g] = \bar R(x)$.
Correspondingly, also the non-linearity field $\phi$ depends on $x$
only: $\phi = \phi (x)$.

Passing from the $\bar R$-non-linear theory (\ref{zh5.1}) to the
equivalent $R$-linear theory (\ref{zh5.14}) the metric
(\ref{zh5.21}) undergoes the conformal transformation \mbox{$\bar g
\mapsto g$} [see relation (\ref{zh5.6})]
\begin{equation}
  \label{zh5.23}
  g = \Omega^2 \bar g = \left(\! e^{A \phi }\!\right)^{2/(D-2)} \bar g\:
  := g^{(0)}+\sum_{i=1}^ne^{2
    \beta^i(x)}g^{(i)}
\end{equation}\vspace*{-5mm}

\noindent with\vspace*{-1mm}
\begin{equation}
  \label{zh5.24}
  g^{(0)}_{\mu \nu} := \left(\! e^{A \phi}\!\right)^{2/(D-2)} \bar
  g^{(0)}_{\mu \nu}, \quad \beta^i := \bar {\beta} ^i +
  \frac{A}{D-2} \phi.
\end{equation}

Therefore, the problem of the internal spaces stabilization can be
solved in full analogy with sections \ref{zh-general} and
\ref{zh-scalar} where in formulas of the latter section we should
put $\Lambda=0$ (for simplicity of notations, in the present section
\ref{zh-nonlin} we use $\phi$ instead of $\Phi$ of section
\ref{zh-scalar}.). In section \ref{zh-scalar}, we have shown that in
the case $\Lambda =0$, the stabilization\,\footnote{\,It is worth
noting
  that despite the existence of a negative minimum of the effective
  potential,
  the internal spaces are not fully stabilized in the case of flat
  external space. This follows easily from the equations of
  Appendix~A. Indeed, for the Hubble parameter $H>0$
  (expanding external space), the friction term in (\ref{zha9}) results
  in decrease of the amplitude of $\varphi^i$ and its velocity $\dot
  \varphi^i$ with time (see also section \ref{zh-eff eq}). This
  decrease continues until $H$ in (\ref{zha4}) becomes equal to 0. Then,
  the scale factor of the Universe reaches its maximum and $H$ changes
  the sign (respectively, the friction term in (\ref{zha9}) changes its
  sign). After that, scalar fields $\varphi^i $ start to oscillate
  around the minimum position with increasing
  amplitude. 
  Therefore, at the present stage of expanding Universe, the internal
  spaces are not fully stabilized at the minimum position but
  oscillate with decreasing amplitude. The frequency of oscillations
  is equal to the gravexciton masses (see Eq.  (\ref{zh4.8})) and the
  period of oscillations $T_{osc}$ is quite short for sufficiently
  large masses. For example, $T_{osc}\sim 10^{-12}$\,sec for
  $m_{\psi}\sim 10^{-2}\,\text{GeV}$.  Obviously, for time intervals
  $t \gg T_{osc}$, with a high degree of accuracy, the internal space
  scale factors are equal to their values at the equilibrium
  positions.} of the extra dimensions automatically results in
condition $\phi \to \phi_{0}$ with $U(\phi_0) < 0$. Thus, the
D-dimensional space-time (bulk) becomes asymptotically
$\text{AdS}_D$ (see Eq.~(\ref{zh5.19})) and there is no dark energy
in these pure geometrical theories. The main difference from section
\ref{zh-scalar} is that in non-linear $f(R)$ models the potential of
the scalar field is not arbitrary, but completely determined by the
form of the scalar cur\-va\-tu\-re non-linearity. Let us consider a
few examples.

{\bfseries\itshape i. \boldmath$1/R$ non-linearity:}
\begin{equation}
  \label{zh5.25}
  f(\bar R) = \bar R - \mu / \bar R, \quad \mu >0 .
\end{equation}
In front of the $\bar R^{-1}$-term, the minus sign is chosen,
because otherwise the potential $U(\phi)$ will have no extremum.
With the help of definition (\ref{zh5.13}), we express the scalar
cur\-va\-tu\-re $\bar R$ in terms of scalaron $\phi$ and obtain two
real-valued solution
branches %
\index{scalaron|(}\vspace*{-2mm}
\begin{equation}
  \label{zh5.26} \bar R_\pm = \pm \sqrt{\mu} \left(\! e^{A\phi} -
    1\!\right)^{-1/2} \Longrightarrow  \phi >0
\end{equation}
of the quadratic equation $f'(\bar R)=e^{A\phi}$.

The corresponding potentials (\ref{zh5.15})
\begin{equation}
  \label{zh5.27}
  U_\pm(\phi) = \pm \sqrt{\mu}\, e^{-B\phi }\,
  \sqrt{e^{A\phi }-1}
  \end{equation}
have extrema for cur\-va\-tu\-res
\begin{equation}
  \label{zh5.28} \bar R_{0,\pm} =\pm \sqrt{\mu}\sqrt{\frac{D+2}{D-2}}
  \, \Longrightarrow \, e^{A\phi_0} = \frac{2B}{2B-A} = \frac{2D}{D+2}
  > 1 \, \text{for} \, D \geq 3\,
\end{equation}
and take for these cur\-va\-tu\-res the values
\begin{equation}
  \label{zh5.29}
  U_\pm(\phi_0) = \pm
  \sqrt{\mu}\sqrt{\frac{D-2}{D+2}}\; e^{-B\phi_0} = \pm
  \sqrt{\mu}\sqrt{\frac{D-2}{D+2}}\left(\! \frac{2D}{D+2} \!\right)^{\!\!-D/(D-2)}\!.
\end{equation}

The stability defining second derivatives at the extrema
(\ref{zh5.28}),
\[
\left.  \partial^2_\phi U_\pm\right|_{\phi_{0}} =
    \mp \sqrt{\mu} \frac{D}{D-1}\sqrt{\frac{D+2}{D-2}}\; e^{B\phi_{0}} =
\]\vspace*{-3mm}
  \begin{equation}
    \label{zh5.30}
     =    \mp \sqrt{\mu} \frac{D}{D-1}\sqrt{\frac{D+2}{D-2}}
     \left(\!
      \frac{2D}{D+2} \!\right)^{\!\!-D/(D-2)}\!,
  \end{equation}
shows that only the negative cur\-va\-tu\-re branch $\bar R_-$
yields a minimum with stable internal space components. The positive
branch has a maximum with $U_+(\phi_0)>0$.  According to
(\ref{zh2.24}) (where $\Lambda =0$) it can provide an effective
{\bfseries dark energy} contribution with
$\Lambda_{\mathrm{eff}}>0$, but due to its tachyonic behavior with
$\partial_\phi^2U(\phi_0)<0$ it cannot give stably frozen internal
dimensions \cite{zhGZBR}.

{\bfseries\itshape ii. cur\-va\-tu\-re-squared non-linearity:}
\begin{equation}
  \label{zh5.31}
  f(\overline R ) = \overline R + \alpha \overline R^{\; 2} - 2\Lambda _D,
\end{equation}

For this theory we obtain
\begin{equation}
  \label{zh5.32}
  1 + 2\alpha \overline R = e^{A \phi} \Longleftrightarrow \overline R =
  \frac{1}{2\alpha } \left(\!e^{A \phi } - 1\!\right)\!,~~ -\infty <
  \phi < \infty,
\end{equation}
and
\begin{equation}
  \label{zh5.33}
  U(\phi ) = \frac12 e^{-B \phi }\left[ \frac{1}{4\alpha }\left(\! e^{A
        \phi } - 1\!\right)^2 + 2 \Lambda _D \right]\!.
\end{equation}

The parameter region which ensures the stabilization of the internal
space is described in \cite{zhGMZ}. In this region, the effective
potential has a negative global minimum.  Thus, the D-dimensional
space-time becomes asymptotically AdS.

{\bfseries\itshape iii. cur\-va\-tu\-re-quartic non-linearity:}
\begin{equation}
  \label{zh5.34}
  f(\bar R) = \bar R +\gamma \bar R^{4} -2\Lambda_D.
\end{equation}

\noindent For this model, $\bar R$ and scalaron $\phi$ are related
as
\begin{equation}
  \label{zh5.35}
  \bar R = (4\gamma)^{-1/3}\left(\! e^{A\phi} -1\!\right)^{1/3}\!,~~ -\infty < \phi <  \infty,
\end{equation}
and potential $U(\phi)$ reads
\begin{equation}
  \label{zh5.36}
  U(\phi ) = \frac{1}{2}e^{-B\phi} \left[
    \frac{3}{4} (4\gamma)^{-1/3}\left(\!e^{A\phi}-1 \!\right)^{4/3} + 2\Lambda_D\right]\!.
\end{equation}

The internal space stability region in parameter space is described
in \cite{zhGZBR}. It is shown that this stability region depends on
the total dimension $D=\dim(M)$ of the higher dimensional space-time
$M$. For $D>8$ the stability region consists of a single (absolutely
stable) sector which is shielded from a conformal singularity %
\index{singularity} %
(and an antigravity sector beyond it) by a potential barrier of
infinite height and width. This sector is smoothly connected
\mbox{with} the stability region of a cur\-va\-tu\-re-linear model.
For $D<8$ an additional (meta-stable) sector exists which is
separated from the conformal singularity by a potential barrier of
finite height and width so that systems in this sector are prone to
collapse into the conformal singularity. This second sector is not
smoothly connected with the first (absolutely stable) one. As we
mentioned above, the external space-time in this model is necessary
AdS and the corresponding negative effective cosmological constant,
$\Lambda_{\mathrm{eff}}<0$,
forbids a late-time acceleration. %
\index{cosmological constant} %

{\bfseries\itshape iv. cur\-va\-tu\-re-squared and
cur\-va\-tu\-re-quartic non-linearities:}
\begin{equation}
  \label{zh5.37}
  f(\bar{R})=\bar{R}+\alpha\bar{R}^{2}+\gamma\bar{R}^{4}-2\Lambda_{D}.
\end{equation}
The relation between the scalar cur\-va\-tu\-re $\bar{R}$ and the
scalaron field $\phi$ is
\begin{equation}
  \label{zh5.38}
  f' = e^{A\phi} = 1 +2\alpha \bar{R} + 4\gamma \bar{R}^3.
\end{equation}
This equation has three solutions $\bar R_{1,2,3}(\phi )$, where one
or three of them are real-valued. Therefore, in general, the
potential
\begin{equation}
  \label{zh5.39}
  U(\phi ) = (1/2)e^{-B\phi}\left(\alpha\bar{R}^2+3\gamma\bar{R}^4+2\Lambda_D\right)
\end{equation}
is multivalued and consists of a number of branches. The case of one
real-valued solution for $(D=8)$-dimensional space-time was
investigated in \cite{zhSZGC2006} where the parameter region for the
freezing stabilization of the internal spaces was described. The
external space-time is asymptotically AdS.

Very interesting case of multivalued solutions for
$(D=4)$-dimensional space-time was considered in \cite{zhSZPRD2010}.
Here, the branches of the potential $U(\phi )$ are fitted with each
other in the branching and monotonic points.  It was shown that the
monotonic points are penetrable for the scalaron, while in the
vicinity of the branching points, the scalaron has the bouncing
behavior and cannot cross these
points. Moreover, there are branching points where the scalaron %
\index{scalaron|)}%
bounces an infinite number of times with decreasing amplitude, and
the Universe asymptotically approaches the de Sitter stage. Such
accelerating behavior was called bouncing inflation.  For this
accelerating expansion, there is no need for original potential
$U(\phi)$ to have a minimum or to check the slow-roll conditions. A
necessary condition for such inflation is the existence of the
branching points. This is a new type of inflation. Such bouncing
inflation takes place both in the Einstein and Brans---Dicke frames.
This type of inflation was found for the model with the
cur\-va\-tu\-re-squared and cur\-va\-tu\-re-quartic correction terms
which play an important role during the early stages of the Universe
evolution. However, the branching points take also place in models
with $\bar R^{-1}$-type correction terms \cite{zhFrolov}. These
terms play an important role at late times of the evolution of the
Universe. %
\index{accelerated expansion} %
Therefore, bouncing inflation may be responsible for the late-time
accelerating expansion of the Universe ({\bfseries dark energy}).

\subsection{\!Dark energy in \boldmath$f(R)$ models with form fields\label{zh-form f(R)}}

\hspace*{3cm}As we saw above, in the pure geometrical non-linear
models, the internal space freezing stabilization is achieved due to
negative minimum of the effective potential. Thus, these models are
asymptotically AdS without accelerating behavior of our Universe
(see, however, comments with respect to the bouncing inflation).
However, the inclusion of matter can uplift potential to the
positive values. In this section, we shall demonstrate such
uplifting for
non-linear models with form fields. %
\index{fields of forms} %

To start with, let us present general theory of such models
\cite{zhGMZ2}. We consider a $D=(4+D')$-dimensional non-linear
gravitational theory with action
\begin{equation}
  \label{zh5.40}
  S = \frac {1}{2\kappa_D^2}\int\limits_M d^Dx \sqrt{|\overline g|} f(\overline R) -
  \frac12 \int\limits_M d^Dx \sqrt{|\overline g|}\sum_{i=1}^n \frac{1}{d_i!} \left(\!F^{(i)}
  \!\right)^2\!,
\end{equation}
where $f(\overline R)$ is an arbitrary smooth function with mass
dimension $\mathcal{O}(m^2)$ \ ($m$ has the unit of mass) of the
scalar cur\-va\-tu\-re $\overline R = R[\overline g]$ constructed
from the D-dimensional metric $\overline g_{ab}\; (a,b =
1,\mbox{...},D)$. In action (\ref{zh5.40}), $F^{(i)} = F^{(i)}_{m_i
n_i \dots q_i}$, $i=1, \mbox{...} ,n$ is an antisymmetric tensor
field of rank $d_i$ (a $d_i$-form field strength) with indices from
an index set $s_{(i)}=\{m_i: \ \max (m_i) - \min (m_i) =d_i\}$,
where $ m_i, n_i, \mbox{...} , q_i \in s_{(i)}$.  For simplicity, we
suppose that the index sets $s_{(i)}$, $s_{(j)}$ of tensors
$F^{(i)}$, $F^{(j)}$ with $i\neq j$ contain no common elements as
well as no indices corresponding to the coordinates of the
$D_0$-dimensional external space-time (usually $D_0 =4$).
Additionally, we assume that for the sum of the ranks holds
$\sum_{i=1}^n d_i = D-D_0 := D^{\prime}$. Obviously, this model can
be generalized to tensor configurations $F^{(i)}$, $F^{(j)}$ with
intersecting (overlapping) index sets. In this case explicit field
configuration can be obtained, e.g., when the indices satisfy
special overlapping rules \cite{zhIv6}. Such a generalization is
beyond the scope of our consideration. Furthermore, we assume in our
subsequent considerations that the index sets $m_i,n_i,\mbox{...}
,q_i \ne 0$ do not contain the coordinates of the external
space-time $M_0$ and, hence, the field strengths $F^{(i)}$ can be
associated with a magnetic (solitonic) $p$-brane system located in
the extra dimensions as discussed, e.g., in Refs.
\cite{zhIv6,zhStelle,zhLOW}.

The equation of motion for the gravitational sector of (\ref{zh5.40}) reads %
\index{equation of motion} %
\begin{equation}
  \label{zh5.41}%
  f^{\prime }\overline R_{ab} -\frac12 f \overline g_{ab} - \overline
  \nabla_a \overline \nabla_b f^{\prime } + \overline g_{ab} \overline{\square } f^{\prime } =
  \kappa_D^2T_{ab} \left[ F,\overline g\right] \!,
\end{equation}
where $a,b = 1,\mbox{...},D$, $f^{\prime } =df/d\overline R$,
$\overline R_{ab} = R_{ab}[\overline g]$, $ \overline R =
R[\overline g]$. $\overline \nabla_a$ and $\overline{\square}$
denote the covariant derivative and the Laplacian with respect to
the metric $\overline g_{ab}$ (see equation (\ref{zh5.3})).
Eq.~(\ref{zh5.41}) can be rewritten in the form
\begin{equation}
  \label{zh5.42}
  f^{\prime }\overline G_{ab} +\frac12 \overline g_{ab} \left( \overline R
    f^{\prime} - f\right) - \overline \nabla_a \overline \nabla_b f^{\prime } + \overline g_{ab}
  \overline{\square } f^{\prime } = \kappa_D^2T_{ab} \left[ F,\overline g\right]\!,
\end{equation}
where $\overline G_{ab} = \overline R_{ab} -\frac12 \overline R \;
\overline g_{ab}$, with its trace
\begin{equation}
  \label{zh5.43}
  (D-1)\overline{\square } f^{\prime } = \frac{D}{2} f
  -f^{\prime }\overline R + \kappa_D^2T\left[ F,\overline g\right]\!.
\end{equation}

The energy momentum tensor (EMT) $T_{ab}\left[ F,\overline g\right]$
is defined in the standard way as
\begin{equation}
  \label{zh5.44} T_{ab}\left[ F,\overline g\right] \equiv \frac1{\sqrt{|\overline
      g|}}\frac{\delta \left(\!\sqrt{|\overline g|} \sum_{i=1}^n
      \frac{1}{d_i!}\left(F^{(i)} \right)^2\right)} {\delta \overline
    g^{ab}} = \sum_{i=1}^n T_{ab}\left[ F^{(i)},\overline g\right]\!,
\end{equation}
where
  \begin{equation}
  \label{zh5.45}
  \hspace*{-0.3cm}T_{ab}\left[ F^{(i)},\overline g\right] =
  \frac1{d_i!}\left(\!-\frac12\overline g_{ab} F^{(i)}_{m_i n_i \dots
      q_i}F^{(i)\, m_i n_i \dots q_i} + d_i F^{(i)}_{a n_i\dots
      q_i}F_b^{(i)\, n_i\dots q_i}\!\right)\!.\!\!\!
\end{equation}
For the trace of this tensor we obtain
  \begin{equation}
  \label{zh5.46} T\left[ F,\overline g\right] = \sum_{i=1}^n T\left[
    F^{(i)},\overline g\right]
\end{equation}
 with
  \begin{equation}
  \label{zh5.47}
  T\left[ F^{(i)},\overline g\right] = \frac{2d_i-D}{2(d_i!)}F^{(i)}_{m_i
    n_i\dots q_i} F^{(i)\, m_i n_i\dots q_i}\!.
\end{equation}

The field strengths  $F^{(i)}$ satisfy the equations of motion
  \begin{equation}
    \label{zh5.48} {F^{(i)\, m_i n_i \dots q_i}}_{;\, q_i} = 0
    \Longleftrightarrow \frac1{\sqrt{|\overline g|}} \left(\! \sqrt{|\overline g|}\; F^{(i)\, m_i n_i \dots q_i}\!\right)_{\!, q_i}=0,~ i=1,\mbox{...} ,n.
  \end{equation}
and the Bianchi identities
\begin{equation}\label{zh5.49} F^{(i)}_{[m_i n_i \dots q_i ,\; a]} =
0,~
i=1,\mbox{...} ,n.
\end{equation}

Similar to the previous section, we perform the conformal
transformation (\ref{zh5.6}) and reduce the non-linear gravitational
theory to a linear one with additional scalar field $\phi$
(scalaron). %
\index{scalaron|(}%
This transformation is well defined for $f'(\overline R)>0$ (see
footnote \ref{zh-f'>0}). The equivalence of the theories can be
easily proven with the help of the auxiliary formulas
(\ref{zh5.8})---(\ref{zh5.10}).

Defining the scalar $\phi $ by the relation (\ref{zh5.13}), and
making use of (\ref{zh5.8})---(\ref{zh5.10}), equations
(\ref{zh5.42}) and (\ref{zh5.43}) can be rewritten as
\begin{equation}
  \label{zh5.50}
  G_{ab} = \kappa^2_D T_{ab}\left[ F,\phi, g\right] +
  T_{ab}\left[ \phi ,g\right]
\end{equation}
and
\[
  \square \phi = \frac {1}{\sqrt{(D-1)(D-2)}}\; e^{\frac
    {-D}{\sqrt{(D-1)(D-2)}}\phi} \left(\! \frac {D}{2} f - f'\overline R
    \!\right)+
\]
\begin{equation} \label{zh5.51}
  + \frac1{\sqrt{(D-1)(D-2)}}\kappa^2_D T\left[ F,\phi ,g\right]\!.
\end{equation}

\noindent The EMTs read
\begin{equation}
  \label{zh5.52} T_{ab}\left[ \phi,g\right] = \phi_{,a}\phi_{,b}
  -\frac12 g_{ab}g^{mn}\phi_{,m}\phi_{,n} - \frac12 g_{ab}\; e^{\frac
    {-D}{\sqrt{(D-1)(D-2)}}\phi} \left(\overline R f'- f\right)\!,
\end{equation}
\begin{equation}
  \label{zh5.53} T_{ab}\left[ F,\phi ,g\right] = \sum_{i=1}^n
  e^{\frac{2d_i-D}{\sqrt{(D-1)(D-2)}}\phi} T_{ab}\left[
    F^{(i)},g\right]
\end{equation}
and
\begin{equation}
  \label{zh5.54}
  T\left[ F,\phi,g\right] = \sum_{i=1}^n
  e^{\frac{2d_i-D}{\sqrt{(D-1)(D-2)}}\phi} T\left[ F^{(i)},g\right]\!,
\end{equation}
where $T_{ab}\left[ F^{(i)},g\right]$, $T\left[ F^{(i)},g\right]$
are given by replacing $\overline g \to g$ in equations
(\ref{zh5.45}), (\ref{zh5.47}). The indices of the field strengths
$F^{(i)}$ are now raised and lowered with the metric $g$.

The equations of motion (\ref{zh5.48}) for $F^{(i)}$ transform to
\begin{equation}
\label{zh5.55} \frac1{\sqrt{|g|}} \left( \sqrt{|g|}\;
e^{\frac{2d_i-D}{\sqrt{(D-1)(D-2)}}\phi}\; F^{(i)\, m_i n_i \dots
q_i}\right)_{\!,\, q_i}=0,\quad i=1,\mbox{...},s ,n,
\end{equation}
whereas the Bianchi identities (\ref{zh5.49}) do not change.

It can be easily checked that Eqs.~(\ref{zh5.50}), (\ref{zh5.51})
and (\ref{zh5.55}) are the equations of motion for the action
\[
S = \frac{1}{2\kappa_D^2} \int\limits_M d^D x \sqrt{|g|} \Bigg\{\!
R[g] -
    g^{ab}
    \phi_{,a} \phi_{,b} - 2 U(\phi )
    -
\]\vspace*{-3mm}
\begin{equation}
  \label{zh5.56}
   - \kappa^2_D\sum_{i=1}^n \frac1{d_i!} \;
    e^{\frac{2d_i-D}{\sqrt{(D-1)(D-2)}}\phi}\;F^{(i)}_{m_i n_i \dots
      q_i} F^{(i)\, m_i n_i \dots q_i}\!\Bigg\}\!,
\end{equation}%
\index{potential $U(\phi )$} %
where potential $U(\phi )$ is defined by formula (\ref{zh5.15}). The
scalaron $\phi$ is the result and the carrier of the cur\-va\-tu\-re
non-linearity of the original theory (\ref{zh5.40}).
Correspondingly, Eq.~(\ref{zh5.51}) has a two-fold interpretation.
It is the equation of motion for the field $\phi$ and at the same
time it can be considered as constraint equation following from the
reduction of the non-linear theory (\ref{zh5.40}) to the linear one
(\ref{zh5.56}). Furthermore, we note that in the linear theory
(\ref{zh5.56}) the form fields are non-minimally coupled with the
scalaron $\phi$. (A minimal coupling occurs only for a model with
$n=1$, $d_1=D_0$, where according to (\ref{zh5.47}) the trace of the
form field EMT vanishes.) A comparison of the action functional with
(\ref{zh5.53}) shows that the last term in (\ref{zh5.56}) coincides
with the expression for the energy density $-T^0_0[F,\phi,g]$ of the
solitonic form field (due to $F^{(i)}_{0n_i\dot q_i}\equiv 0$ by the
definition of $F^{(i)}$).

Now, we assume that the multidimensional space-time manifold
undergoes a spontaneous compactification (\ref{zh5.20}) in
accordance with the block-or\-tho\-go\-nal structure of the field
strength $F$, and that the form fields $F^{(i)}$, each nested in its
own $d_i-$dimensional factor space $M_i$, respect a generalized
Freund---Rubin ansatz \cite{zhFR} (see also
\cite{zhIv6,zhLOW,zhwiltshire,zhGC1}). The factor spaces $M_i$ are
then Einstein spaces with metrics $\hat g^{(i)} \equiv
e^{2\beta^i(x)} g^{(i)}$ which depend only through the warp factors
$a_i(x) :=e^{2\beta^i(x)}$ on the coordinates $x$ of the external
space-time $M_0$ (see Eqs.~(\ref{zh5.21})---(\ref{zh5.24})). This
allows us to perform a dimensional reduction of our model and the
internal space stabilization along the lines of section
\ref{zh-Exci}. Additionally, it is not difficult to show that in the
case of the freezing
stabilization\,\footnote{\,\label{frizstabil}Without loss of
generality,
  we can choose the stability position $\beta^i_0=0$. Then, the
  internal space fluctuations $\tilde \beta^i$ in section
  \ref{zh-Exci} coincide with $\beta^i$.} $(\beta^i=0,\phi =\phi_0)$,
the asymptotic multidimensional space-time is built up from
Einstein-space blocks, but is itself a non-Einsteinian space due to
an additional terms which depend on form fields \cite{zhGMZ2}.

Similar to the previous section, it can be easily seen that for
considered product manifold ansatz, the scalar cur\-va\-tu\-re
$\overline R$ depends only on the coordinate $x$ of the
$D_0-$dimensional external space-time $M_0$: $\overline R[\overline
g] = \overline R(x)$. This implies that the scalaron field $\phi$ is
also a function only of $x$: $\phi = \phi (x)$.

We choose the form-field components in the generalized
Freund---Rubin ansatz as
\begin{equation}
\label{zh5.57}
\begin{array}{c}
  \displaystyle F^{(i)}_{m_i n_i\dots q_i} = \sqrt{2}\; \sqrt{|\hat g^{(i)}|}\;
  \varepsilon_{m_i n_i\dots q_i} f^{(i)}(x),  \\[3mm]
 \displaystyle  F^{(i)\; m_i n_i\dots q_i} = \left(\!\sqrt{2} / \sqrt{|\hat
      g^{(i)}|}\!\right) \varepsilon^{m_i n_i\dots q_i} f^{(i)}(x).
\end{array}
\end{equation}

\index{Levi---Civita symbol} %
For the Levi---Civita symbol $\varepsilon_{m_i n_i\dots q_i}$ we use
conventions where for Rie\-mann spaces holds $\varepsilon_{m_i
n_i\dots
  q_i} = \varepsilon^{m_i n_i\dots q_i}$ and $\varepsilon_{m_i
  n_i\dots q_i}\varepsilon^{m_i n_i\dots q_i} =d_i\,!$. It can be
easily seen that the ansatz (\ref{zh5.57}) satisfies
Eq.~(\ref{zh5.55}) (because $\phi$ and $f$ depend only on $x$ and
the $\sqrt{|g^{(i)}|}$ factors cancel). The Bianchi identities
(\ref{zh5.49}) reduce to the equations\vspace*{1mm}
\begin{equation}
\label{zh5.58} \frac{\partial\left(\! a^{d_i}_i(x)
f^{(i)}(x)\!\right)}{\partial x^{\mu}} = 0
\end{equation}
 with solutions
\begin{equation}
\label{zh5.59}
 f^{(i)}(x) = \frac{f_i}{a^{d_i}_i}
\end{equation}
and $f_i\equiv {\rm const}$. With (\ref{zh5.59}) the energy density
of the solitonic form field, and correspondingly the last term in
action (\ref{zh5.56}), reads
\[
 -T^0_0[F,\phi,g] =  \frac12\sum_{i=1}^n \frac1{d_i!} \;
    e^{\frac{2d_i-D}{\sqrt{(D-1)(D-2)}}\phi}\;F^{(i)}_{m_i n_i \dots q_i} F^{(i)\, m_i n_i \dots
    q_i}=
\]
\begin{equation}
\label{zh5.60}
      =  \sum_{i=1}^n \, e^{\frac{2d_i-D}{\sqrt{(D-1)(D-2)}}\phi}\;
    \frac{f_i^2}{a_i^{2d_i}} :=\rho (\beta , \phi) ,
\end{equation}

\noindent where for real form fields $f_i^2 \ge 0$. Again we see
that for models with $n=1$ and $d_1=D_0$ this energy density
decouples from the scalaron field~$\phi$: \mbox{$\rho(\beta^1, \phi
) \!\rightarrow\! \rho(\beta^1\!)$.}

Let us consider now a model with only one $d_1$-dimensional internal
space. After dimensional reduction and subsequent conformal
transformation to the Einstein frame (along the lines of section
\ref{zh-Exci}), the action functional (\ref{zh5.56})
reads
\[
S=\frac 1{2\kappa_0^2}\int\limits_{M_0}d^{D_0}x\sqrt{|\tilde
    g^{(0)}|}\bigg\{\! R\left[ \tilde g^{(0)}\right] -
    \tilde g^{(0) \mu \nu} \partial_{\mu}\varphi \partial_{\nu}
    \varphi\,
    -
\]
\begin{equation}
  \label{zh5.61}
    -\, \tilde g^{(0) \mu \nu} \partial_{\mu}\phi \partial_{\nu} \phi
    -2U_\mathrm{ eff} (\varphi ,\phi ) \!\bigg\}\!,
\end{equation}\vspace*{-1mm}

\noindent where field $\varphi$ is defined by equation
(\ref{zh2.17}):
\begin{equation}
  \label{zh5.62} \varphi = -\sqrt{\frac{d_1(D-2)}{D_0-2}}\beta^1 .
\end{equation}
A stable compactification of the internal space $M_1$ is ensured
when its scale factor $\varphi$ is frozen at one of the minima of
the effective potential
\[
U_{\mathrm{eff}}(\varphi ,\phi )    =  e^{2\varphi
\sqrt{\frac{d_1}{(D-2)(D_0-2)}}} \,\times
\]
\begin{equation}
  \label{zh5.63}
  \times
  \left[ -\frac12 R_1e^{2\varphi
      \sqrt{\frac{D_0-2}{d_1(D-2)}}}+ U (\phi ) + \kappa^2_D\, \rho (\varphi ,\phi ) \right]\! ,
\end{equation}
where $R\left[ g^{(1)}\right] \equiv R_1$ is the scalar
cur\-va\-tu\-re of the factor-space $M_1$ (see Eq.~(\ref{zh2.4}))
and the energy density (\ref{zh5.60}) of the solitonic form field
reads
\begin{equation}
  \label{zh5.64} \kappa^2_D\, \rho (\varphi ,\phi )\, =\,
  \kappa^2_D\,f^2_1\, e^{\frac{2d_1-D} {\sqrt{(D-1)(D-2)}}\phi }\, e^{2\varphi
    \sqrt{\frac{d_1(D_0-2)}{D-2}}}.
\end{equation}
The value of the effective potential at the minimum plays the role
of
the effective $D_0$-dimensional cosmological constant: %
\index{cosmological constant} %
$\left. U_{\mathrm{eff}}\right|_{min} \equiv \Lambda_\mathrm{ eff}$.
It can be {\bfseries dark energy} in the case of positive
$\Lambda_{\mathrm{eff}}>0$.

The potential $U(\phi)$ of the scalaron field is given by
Eq.~(\ref{zh5.15}) and its exact expression depends on the form of
non-linearity $f(R)$. The $1/R$ and $R^2$ non-linearities were
considered in detail in papers \cite{zhGMZ2} and \cite{zhSZPRD2007}.
It was shown that for all these models, there exist parameter
configurations that can provide positive values of the effective
four-dimensional cosmological constant.  Thus, an accelerated
expansion of the Universe (dark energy) %
\index{accelerated expansion} %
is possible in accordance with observational data. However, the
observational value of dark energy is achieved with the help of fine
tuning of parameters, similar to how it happens for the linear model
in section \ref{zh-fine tuning}.

To conclude this section, we would like to note that there is also a
possibility to consider non-linear models with form fields where the
action functional reads \cite{zhSZPRD2009}:\vspace*{-5mm}
\[
S =\frac{1}{2\kappa^2_D}\int_M d^Dx\sqrt{|\overline g|}f(\overline
R) -\frac12
  \int\limits_M
  d^Dx\sqrt{|g|}\frac{1}{d_1!}\left(\!F^{(1)}\!\right)^2 -
\]\vspace*{-5mm}
\begin{equation}
  \label{zh5.65}
   -\sum_{k=1}^m \int\limits_{M_0} d^4x \sqrt{ |g^{(0)}(x)}|\; \tau_{(k)}.
\end{equation}\vspace*{-3mm}

The main difference between this expression and the action
functional (\ref{zh5.40}) is that the form field $F^{(1)}$ is
originally coupled with metric $g$ but not with $\overline g$.
Additionally, we suppose that the internal space $M_1$ is a flat
orbifold with $m$ branes in fixed points and branes are uniquely
characterized by their tensions $\tau_{k}$ (see section
\ref{zh-orbifolds}). Then, the effective
potential for this model reads\vspace*{-2mm}%
\[
 U_{\mathrm{eff}}(\varphi ,\phi )
    e^{2\varphi \sqrt{\frac{d_1}{(D-2)(D_0-2)}}}\times
\]\vspace*{-6mm}
\begin{equation}
  \label{zh5.66}
     \times \Big[U(\phi ) +
      \kappa^2_D f_1^2e^{2\varphi \sqrt{\frac{d_1(D_0-2)}{D-2}}}
      -\lambda e^{\varphi \sqrt{\frac{d_1(D_0-2)}{D-2}}\; } \Big]\!,
\end{equation}\vspace*{-4mm}

Therefore, in this approach the energy density of the solitonic form
\mbox{field} decouples from the scalaron field $\phi$ for any number
of \index{scalaron}dimensions $d_1$: $\rho(\varphi, \phi
)\rightarrow$ $\rightarrow \rho(\varphi)$, and this greatly
simplifies the calculations. This model was investigated in
\cite{zhSZPRD2009} for $R^2$ and $R^4$ non-linearities. To avoid the
fine-tuning problem, the main attention was paid to the case of zero
effective cosmological constant \index{cosmological
constant}$\Lambda_{\mathrm{eff}}=0$. Conditions, that ensure stable
compactification of the internal space in zero minimum of the
effective potentials, were defined. Such effective potentials have
interesting and rather complicated form with a number of local
minima, maxima and saddle points. It was shown (with the help of
numerical calculation of equations in Appendix~A) that the $R^2$-
and $R^4$ models can produce up to 10 and 22 e-foldings,
respectively. These values are not sufficient to solve the
homogeneity and isotropy problem but big enough to explain the
recent CMB data. \index{cosmic microwave background
(CMB)|)}Additionally, the $R^4$ model, with saddle points of the
effective potential, can provide conditions for eternal topological
inflation. The main drawback of the obtained inflationary models
consists in a spectral index $n_s$ that is less than the presently
observed $n_s\approx 1$. For the $R^4$ model, \mbox{e.g., $n_s
\approx 0.61$.}\vspace*{-2mm}


\section[\!S$p$-branes]{\!S\boldmath$p$-branes. Dynamical\\
\hspace*{-1.15cm} dark energy from extra   dimensions}\label{zh-Sp}

\hspace*{3cm}\index{extra dimensions}\index{brane}As we have
repeatedly noted above, recent astronomical ob\-ser\-va\-tions
abundantly evidence that our Universe underwent stages of
accelera\-ting expansion during its evolution. There are at least
two of such stages: early inflation and late-time acceleration. The
latter began approximately at the redshift $z \sim 0.35$ (see, e.g.,
\cite{zhFTH}) and continues until now. Thus, the construction and
investigation of models with stages of acceleration is one of the
main challenge of the modern cosmology. In previous sections
(\ref{zh-Exci})---(\ref{zh-nonlin}), we demonstrated theories where
dark energy appears in multidimensional models due to minima of the
effective potentials. Such dark energy is time independent. There
are also models with dynamical dark energy. Among such models, the
models originating from fundamental theories (e.g. string/M-theory)
are of the most of interest. For example, it was shown that some of
space-like brane (S-brane) solutions have a stage of the
accelerating expansion.
In $D$-dimensional manifold, S$p$-branes are time dependent
solutions with ($p+1$)-dimensional Euclidean world volume and, apart
from time, they have $(D-p-2)$-dimensional hyperbolic, flat or
spherical spaces as transverse/additional dimensions
\cite{zhGS}:\vspace*{-1mm}
\begin{equation}
  \label{zh6.1}
  ds_D^2 = -e^{2\gamma(\tau )} d\tau^2 +
  a_0^2(\tau ) \left( dx_1^2 + \mbox{...} + dx_{p+1}^2 \right) + a^2_1(\tau )
  d\Sigma^2_{(D-p-2),\, \sigma} ,
\end{equation}\vspace*{-5mm}

\noindent where $\gamma (\tau )$ fixes the gauge of time, $a_0 (\tau
)$ and $a_1 (\tau )$ are time dependent scale factors, and $\sigma =
-1,0,+1 $ for hyperbolic, flat or spherical spaces respectively.
Obviously, $p=2$ if brane describes our 3-dimensional space. These
branes are known as SM2-branes if original theory is 11-dimensional
M-theory and SD2-branes in the case of 10-dimensional Dirichlet
strings. For this choice of $p$, the evolution of our Universe is
described by the scale factor $a_0$. In general, the scale factor
$a_1$ can also determine the behavior of our 3-dimensional Universe.
Hence, $D-p-2=3$ and we arrive to SM6-brane in the case of the
M-theory and SD5-brane for the Dirichlet string. Usually, S$p$-brane
models include form fields (fluxes) and massless scalar fields
(dilatons) as a matter sources. If SD$p$-branes are obtained by
dimensional reduction of 11-dimensional M-theory, then the dilaton
is associated with the scale factor of a compactified 11-th
dimension.

\index{S-brane}Starting from \cite{zhGS}, the S-brane solutions were
also found, e.g., in Refs. [444---447]. It was quite natural to test
these models for the accelerating expansion of our Universe. Really,
it was shown in \cite{zhOh2} that the SM2-brane as well as the
SD2-brane have stages of the accelerating behavior. This result
generalizes conclusions of \cite{zhTW} for models with hyperbolic
compact internal spaces. Here, the cosmic acceleration (in Einstein
frame) is possible due to a negative cur\-va\-tu\-re of the internal
space that gives a positive contribution to an effective potential.
This acceleration is not eternal but has a short period and the
mechanism of such short acceleration was explained in \cite{zhEG}.
It was indicated in \cite{zhOh2} that the solution of \cite{zhTW} is
the vacuum case (the zero flux limit) of the S-branes. It was
natural to suppose that if the acceleration takes place in the
vacuum case, it may also happen in the presence of fluxes. Indeed,
it was confirmed for the case of the compact hyperbolic internal
space. Even more, it was found that periods of the acceleration
occur in the cases of flat and spherical internal spaces due to the
positive contributions of fluxes into the effective potential.
Similar effect of uplifting of the effective potential due to the
form field was already considered in section \ref{zh-form f(R)} for
non-linear models.

Along with Ref.~\cite{zhOh2} mentioned above, the accelerating
S-brane cosmolo\-gies (in the Einstein frame) were obtained and
investigated, e.g., in Refs. [451---455]. Accelerating solutions
closely related to them were also found in
Refs.~\cite{zhCHNW,zhIv3a}. It should be noted that some of these
solutions are not new ones but either rediscovered or written in
different parametrization (see correspon\-ding comments in Refs.
\cite{zhIv2a,zhIv3a}). For example, the first vacuum solution for a
product manifold (consisting of $(n-1)$ Ricci-flat spaces and one
Einstein space with non-zero constant cur\-va\-tu\-re) was found in
\cite{zhIv1a}. This solution was generalized to the case of a
massless scalar field in Refs. \cite{zhBZ1,zhBZ2}. Obviously,
solutions in Refs. \cite{zhBZ1,zhBZ2,zhIv1a} are the zero flux limit
of the S$p$-branes and the result of \cite{zhTW} is a particular
case of \cite{zhBZ2}. Some of solutions in \cite{zhCHNOW,zhCHNW}
coincides with corresponding solutions in
Refs.~\cite{zhBZ1,zhBZ2,zhIv1a,zhBZ3}. An elegant minisuperspace
approach for the investigation of the product space manifolds
consisting of Einstein spaces was proposed in \cite{zhIMZ}. Here, it
was shown that the equations of motion have the most simple form in
a
harmonic time %
\index{harmonic time gauge} %
gauge\,\footnote{\,For Eq.~(\ref{zh6.1}), it reads $\gamma =
(p+1)\ln a_0 +
  (D-p-2)\ln a_1 $. In the harmonic time gauge, time satisfies
  equation $ \Delta [g] \tau = 0$ \cite{zhIMZ}.} because the
minisuperspace metric is flat in this gauge. Even if the authors of
the above-mentioned papers were not aware of it, they intuitively
used this gauge to get exact solutions. New solutions also can be
generated (from the known solutions) with the help of a topological
splitting when the Einstein space with non-zero cur\-va\-tu\-re is
split into a number of Einstein spaces of the same sign of the
cur\-va\-tu\-re (see Refs.~\cite{zhGIM,zhsplit}). This kind of
solutions was found, e.g., in Refs.~\cite{zhCHNOW,zhCHNW}.

\index{S$p$-brane}Now, to show the main characteristic properties of
S$p$-brane solutions, we consider some particular solutions from
\cite{zhBZ1} and \cite{zhBZ2}. To start with, we derive the
connection between different quantities in the Einstein and
Brans---Dicke frames. The dimensionally reduced actions in these
frames have, e.g., forms of equations (\ref{zh2.11}) and
(\ref{zh2.14}). The conformal transformation between the external
space-time metrics in the Einstein and Brans---Dicke frames is given
by Eq.~(\ref{zh2.13}). For simplicity, we consider the case of one
internal space, i.e. $n=1$ in (\ref{zh2.13}). Additionally, for the
model with dynamical internal spaces, there is no sense to split the
internal space scale factors into background and fluctuations.
Therefore, $\tilde \beta^i \equiv \beta^i$ in (\ref{zh2.13}) and
\mbox{$\Omega^2 =
\left(\!e^{d_1\beta^1}\!\right)^{-2/(d_0-1)}=\left(\!a_1^{d_1}\!\right)^{-2/(d_0-1)}$}.
Thus, the metric (\ref{zh6.1}) in different gauges reads (see also
Eqs.~(\ref{zh2.35}) and~(\ref{zh2.36}))
\[
 g = - e^{2\gamma _0 }d\tau \otimes d\tau + a_{0}^2 q^{(0)} +
  a_{1}^2  g^{(1)}
  = - dt\otimes dt + a_{0}^2 q^{(0)} + a_{1}^2 g^{(1)} =
\]\vspace*{-5mm}
\begin{equation}
  \label{zh6.2}
    = \Omega ^2\left(\! - d\tilde {t}\otimes d\tilde {t} + \tilde
    {a}_{0}^2 q^{(0)} \!\right) + a_{1}^2 g^{(1)} ,
\end{equation}
where the first equality is the metric in the harmonic time gauge %
\index{harmonic time gauge} %
$(\gamma=\gamma_0=$\linebreak $=d_0\beta^0+d_1\beta^1)$ in the
Brans---Dicke frame,
the second equality is the metric in the synchronous time gauge %
\index{synchronous time gauge} %
in the Brans---Dicke frame, and the third equality is the metric in
the synchronous time gauge in the Einstein frame.
Equations~(\ref{zh6.2}) show that the external scale factors in the
Einstein and Brans---Dicke frames are related as in
Eq.~(\ref{zh2.36}):\vspace*{-1mm}
\begin{equation}
  \label{zh6.3}
  \tilde {a}_{0} = \Omega ^{ - 1}a_{0}
\end{equation}\vspace*{-6mm}

\noindent and there exists the following correspondence between
different times\,\footnote{\,To have the same directions of the
arrows of time, we
  choose the plus sign for the square root.}:
\begin{equation}
  \label{zh6.4}%
  dt = e^{\gamma _0 (\tau )}d\tau \Longrightarrow t = \int {e^{\gamma
      _0 (\tau )}d\tau  } + \text{const},
\end{equation}\vspace*{-5mm}
\begin{equation}
  \label{zh6.5}%
  d\tilde {t} = \Omega ^{ - 1}e^{\gamma _0 (\tau )}d\tau
  \Longrightarrow \tilde {t} = \int {\Omega ^{ - 1}e^{\gamma _0 (\tau
      )}d\tau } + \text{const}.
\end{equation}

For two component cosmological model with $R[q^{(0)}]= 0, \;
R[g^{(1)}]\equiv R_1\ne 0$ and minimally coupled free scalar field,
the explicit expressions for the scale factors (in the Brans---Dicke
frame) and scalar field as functions of harmonic time read
\cite{zhBZ1,zhBZ2}:\vspace*{-4mm}
\begin{equation}
  \label{zh6.6}
  a_0 (\tau )  = \exp( \beta^0(\tau )) =A_0
  \exp \left(\!\frac{\xi _1 }{d_0 }\tau \!\right)\!,
\end{equation}\vspace*{-3mm}
\begin{equation}
  \label{zh6.7} a_1 (\tau ) = \exp \left(\beta^1(\tau )\right)
  =a_{(c)1} \exp \left(\! - \frac{\xi _1 }{d_1 - 1}\tau\! \right)
  \times\frac{1}{g_\pm (\tau )} ,
\end{equation}\vspace*{-3mm}
\begin{equation}
  \label{zh6.8} \varphi (\tau )  = p^2\tau + q,
\end{equation}\vspace*{-6mm}

\noindent where
\begin{equation}
  \label{zh6.9}
  g_ + = \cosh^{1/(d_1 - 1)}\left( {\xi _2 \tau } \right)\!, \quad
  (-\infty < \tau < +\infty ),
\end{equation}\vspace*{-5mm}

\noindent for $R_1 > 0\,$ and
\begin{equation}
  \label{zh6.10}
  g_ - = \sinh^{1/(d_1 - 1)}\left( {\xi _2 |\tau |} \right)\!, \quad (|\tau |>0),
\end{equation}
for $R_1 < 0$. Here,
$a_{(c)1}=A_1(2\varepsilon/|R_1|)^{1/2(d_1-1)}$, $\xi_1 =
[d_0(d_1-1)/(D-2)]^{1/2} p^1$, $\xi _2=[(d_1 -
1)/d_1]^{1/2}(2\varepsilon )^{1/2}$ and
$2\varepsilon=(p^1)^2+(p^2)^2$. Parameters $A_0 , A_1, p^1, p^2$ and
$q$ are the constants of integration and $A_0, A_1$ satisfy the
following const\-raint: $A_0^{d_0}A_1^{d_1}=A_0$. It was shown in
\cite{zhZhukquant} that $p^1$ and $p^2$ are the momenta in the
minisuperspace ($p^1$ is related to the momenta of the scale factors
and $p^2$ is responsible for the momentum of the scalar field) and
$\varepsilon$ plays the role of energy. Obviously, these solutions
are the zero flux limit of the S$p$-branes.

This model was investigated in detail in \cite{zhBZ-Sp1}. Here, both
the Ricci-flat space and non-zero cur\-va\-tu\-re space may play the
role of our Universe (with corresponding changes in
Eqs.~(\ref{zh6.2})---(\ref{zh6.5})). The analysis was performed in
the Brans---Dicke and Einstein frames.
It was shown that in the context of the considered models, the
Brans---Dicke gravity provides more possibilities for accelerating
cosmologies than the Einsteinian one. Such different behavior of the
external space scale factors in both of these frames is not
surprising because these scale factors are described by different
variables connected with each other via the conformal transformation
in Eqs.~(\ref{zh6.2})---(\ref{zh6.5}). Moreover, the synchronous
times in both of these frames are also different. As a consequence
of these discrepancies, the scale factors of the external space in
both frames behave differently. In the Brans---Dicke frame, stages
of the accelerating expansion exist for all types of the external
space (flat, spherical and hyperbolic). However, in the Einstein
frame, the model with flat external space and hyperbolic
compactification of the internal space is the only one with the
stage of the accelerating expansion. The presence of a minimally
coupled free scalar field does not help the acceleration because
this field does not contribute to the potential. Nevertheless, it
make sense to include such field in the model because it results in
more reach and interesting dynamical behavior. For example, it was
shown that scalar field can prevent the acceleration in the Einstein
frame.

As it was shown in section \ref{zh-fine str}, the dynamical behavior
of the internal spaces results in the variation of the
fine-structure constant (see Eq.~(\ref{zh4.39})).  Thus, any
multidimensional cosmological models with time dependent internal
spaces should be tested from this point of view. It was demonstrated
that the examined model runs into significant problems related to
the too large variations of the fine-structure constant.  The case
of the hyperbolic external space in the Brans---Dicke frame is the
only possibility to avoid this problem.

\subsection{\!Dark energy in pure geometrical
  S\boldmath$p$-brane\\ \hspace*{-1.2cm}model with hyperbolic internal space\label{zh-Sp hyper}}

\hspace{3cm}The considered above S$p$-brane model was carefully
investigated in \cite{zhBZ-Sp1} for an arbitrary range of
parameters. Is it possible to fix these parameters with the help of
the modern observational data? This interesting problem was
investigated in paper \cite{zhBZ-Sp2} where the metric (\ref{zh6.2})
is defined on the manifold with product topology
\begin{equation}
  \label{zh6.11}
  M = \mathbb{R} \times \mathbb{R}^{d_0} \times \mathbb{H}^{d_1}/\Gamma,
\end{equation}
where $\mathbb{R}^{d_0}$ is $d_0$-dimensional Ricci-flat external
(our) space with metric $q^{(0)}$: $R[q^{(0)}]=0$ and scale factor
$a_0$, and $\mathbb{H}^{d_1}/\Gamma$ is $d_1$-dimensional hyperbolic
(com\-pact) internal space with metric $g^{(1)}$:
$R[g^{(1)}]=-d_1(d_1-1)$ and scale factor $a_1$. Both $a_0$ and
$a_1$ depend only on time.  As we already mentioned, the first
equality in (\ref{zh6.2}) is the metric in
the Brans---Dicke frame in the harmonic time gauge %
\index{harmonic time gauge} %
where $e^{\gamma_0}=a_{0}^{d_0}a_1^{d_1}$. The third equality in
(\ref{zh6.2}) is the metric in the Einstein frame in the synchronous
time gauge.  According to formulas (\ref{zh2.36}) and (\ref{zh6.3}),
the scale factors $a_{0}$ of the external space in the Brans---Dicke
frame is connected with the scale factor $\tilde a_0$ in the
Einstein frame as follows: $\tilde a_0=\Omega^{-1}a_{0}$, where
conformal factor
$\Omega=a_1^{-d_1/(d_0-1)}$. Harmonic time $\tau$ %
\index{harmonic time} %
is related to
synchronous time $\tilde t$ as $d\tilde t=f(\tau)d\tau$, %
\index{synchronous time gauge} %
where $f(\tau)=\Omega^{-1}a_{0}^{d_0}a_1^{d_1} =\tilde a_0^{d_0}$.
Hereafter we consider 3-dimensional external space: $d_0=3$. Taking
into account these relations, the solutions (\ref{zh6.6}) and
(\ref{zh6.7}) (in the case of the absence of scalar field:
$p^2\equiv 0$) in the Einstein frame can be rewritten as follows:
\begin{equation}
  \label{zh6.12}
  \tilde a_0(\tau)=A_1^{\frac{d_1+2}{6}}\!
  \left(\!\sqrt{\dfrac{2\varepsilon}{|R_1|}}\!\right)^{\frac{d_1}{2(d_1-1)}}
  \frac{\exp\left(\!-\sqrt{\dfrac{d_1+2}{12(d_1-1)}{2\varepsilon}}\;
      \tau\!\right)} {\sinh^{d_1/[2(d_1-1]}\left(\!-\sqrt{\dfrac{d_1-1}{d_1}
      2\varepsilon}\; \tau\!\right)},
   \end{equation}

\noindent and
\begin{equation}
  \label{zh6.13}
  a_1(\tau)=A_1\!\left(\sqrt{\frac{2\varepsilon}{|R_1|}}\right)^{\frac{1}{d_1-1}}
  \frac{\exp\left(\!-\sqrt{\dfrac{3}{(d_1-1)(d_1+2)}2\varepsilon}{\;
        \tau}\!\right)} {\sinh^{1/(d_1-1)}\left(\!-\sqrt{\dfrac{d_1-1}{d_1}
      2\varepsilon}\; \tau\!\right)},
\end{equation}
where $A_1$ and $\varepsilon$ are the constants of integration. The
function $f(\tau)$ can be easily obtained from Eq.~(\ref{zh6.12})
via expression $f(\tau) = \tilde a_0^3(\tau)$.

Solutions (\ref{zh6.12}) and (\ref{zh6.13}) for the metric
(\ref{zh6.2}) is a particular case of the S$p$-branes with
$(p+1=d_0)$-dimensional Ricci-flat external space. In the case
$d_0=3$ we obtain $p=2$. Therefore, if underlying model is
$(D=11)$-dimensional M-theory, we arrive at M2-branes where the
number of internal dimensions is equal to 7. As we already mentioned
above, such models with hyperbolic internal space undergo the stage
of accelerating expansions \cite{zhBZ-Sp1}. However, the parameters
of the model in (\ref{zh6.12}) and (\ref{zh6.13}) are still not
connected with observational data. So, now we want to use the modern
cos\-mo\-lo\-gi\-cal data (the present
day value for the Hubble parameter %
\index{Hubble parameter|(} %
and the red\-shift when our external space transits from
deceleration to acceleration) to fix all arbitrary parameters of the
considered model and obtain corresponding dynamical behavior for the
scale factors, the Hubble parameter, the deceleration parameter and
the fine-structure ``constant''.

Besides the external $\tilde a_0$ and internal $a_1$ scale factors
described by Eqs.\,(\ref{zh6.12}) and (\ref{zh6.13}), we also
consider the Hubble parameter for each of the factor
spaces\vspace*{2mm}
\[
H_0 = \frac{1}{\tilde a_0}\frac{d\tilde a_0}{d\tilde
t}=\frac{1}{\tilde
    a_0f(\tau ) }\frac{d\tilde a_0}{d\tau}=
\]
\begin{equation}
\label{zh6.14}%
 =
  -\frac{\sqrt{2\varepsilon}}{f(\tau)}\left(\!\sqrt{\frac{d_1+2}{12(d_1-1)}}+
    \sqrt{\frac{d_1}{4(d_1-1)}}\coth\left(\!\sqrt{\frac{d_1-1}{d_1}2\varepsilon}\;
      \tau\!\right)\!\right)\!,
\end{equation}
\[
H_1 =\frac{1}{a_1}
    \frac{da_1}{d\tilde t}=\frac{1}{a_1f(\tau ) }\frac{da_1}{d\tau} =
\]
\begin{equation}
  \label{zh6.15}%
    =
    -\frac{\sqrt{2\varepsilon}}{f(\tau)}\sqrt{\frac{3}{(d_1-1)(d_1+2)}}
    \left(1+\sqrt{\frac{d_1+2}{3d_1}}
      \coth\left(\!\sqrt{\frac{d_1-1}{d_1}2\varepsilon}\;
        \tau\!\right)\!\!\right)\!,
\end{equation}
%
\noindent the external space deceleration
parameter\,\footnote{\,Note that
  overdots in definition of Hubble and deceleration parameters after
  Eqs.~(\ref{H}) and (\ref{q}) denote the derivatives with respect to the
  conformal time $\eta$.} %
\index{conformal time} %
\[
  q_0 = -\frac{d^2\tilde a_0}{d\tilde t^2}\frac{1}{H^2_0\tilde a_0} =
  -\frac{1}{f(\tau)}\frac{d}{d\tau}\left(\!\frac{1}{f(\tau)}\frac{d\tilde
      a_0}{d\tau}\!\right)\frac{1}{H^2_0\tilde a_0}=
\]\vspace*{-3mm}
\[
  = -2\sinh^{-2} \left(\!\sqrt{\frac{d_1-1}{d_1}2\varepsilon}\; \tau
  \!\right) \times \]\vspace*{-2mm}
\begin{equation}
  \label{zh6.16}%
  \times \left[ \sqrt{\frac{d_1+2}{3(d_1-1)}}+\sqrt{\frac{d_1}{d_1-1}}
    \coth\left(\!\sqrt{\frac{d_1-1}{d_1}2\varepsilon}\;
      \tau\!\right)\!\right]^{-2} \!+2
\end{equation}
\noindent and the variation of the fine-structure constant (as a
function of redshift $z$)
\begin{equation}\label{zh6.17}
\Delta \alpha=\frac{\alpha(z) -\alpha(0)}{\alpha(0)}=
\frac{a_1^{d_1}(0)}{a_1^{d_1}(z)}-1,
\end{equation}
where we took into account that for $d_0=3$ the fine-structure
constant $\alpha \sim a_1^{-d_1}$ (see section \ref{zh-fine str}).
We also assume that the solution (\ref{zh6.12}), (\ref{zh6.13})
describes the M2-brane, that is $d_1=7$.

According to the recent observational data (see, e.g., [442, 464]),
the pre\-sent acceleration stage began at redshift $z\approx 0.35$
and the Hubble parameter now is $H_0 (\tilde t_p) \equiv H_p \approx
72\,\mbox{km}/\mbox{sec}/\mbox{Mpc} = 2.33 \times
10^{-18}\,\text{sec}^{-1}$. Hereafter, the letter $p$ denotes the
present day values. Additionally, at the present time the value of
the external space scale factor can be estimated as $\tilde
a_0(\tilde t_p)\approx cH_p^{-1} \approx$ $\approx 1.29\times
10^{28}\,\text{cm}$.  We shall use these observational conditions to
fix the free parameters of the model $A_1$ and $\varepsilon$ (the
constants of integration) and to define the present
time\,\footnote{\,It is obvious that our model
  cannot pretend to describe the full history of the Universe. We try
  to apply this model to explain the late time acceleration of the
  Universe which starts at the redshift $z\approx 0.35$. Before this
  time, the Universe evolution is described by the standard Big Bang
  cosmology. Therefore, in our model $\tilde t=0$ corresponds to
  $z=0.35$ (i.e.  $q_0(z=0.35)=0$) and $\tilde t_p$ is the time from
  this moment to the present day.} $\tilde t_p$.  Observational data
also show that for different redshifts the fine structure constant
variation does not exceed $10^{-5}$: $|\Delta \alpha|<10^{-5}$
\cite{zhUzan}.

Below, all quantities are measured in the Hubble units. For example,
the scale factors are measured in $cH_p^{-1}$ and synchronous time
$t$ is measured in $H_p^{-1}$.  Therefore, $\tilde a_0(\tilde
t_p)=1$ and $H_p=1$.

To fix all free parameters of the model, we use the following logic
chain. First, from the equation $q_0(\tau)=0$ we obtain the harmonic
time $\tau_{in}$ %
\index{harmonic time} %
of the beginning of the stage of acceleration. We find that this
equation has two roots which describe the beginning and end of the
acceleration. Second, we define the constant of integration $A_1$
from the equation $z=0.35=1/\tilde a_0(\tau_{in})-1$ where we use
the condition that acceleration starts at $z=0.35$ and that $\tilde
a_0(\tau_p)=1$. Third, we find the present harmonic time $\tau_p$
from the condition $\tilde a_0(\tau_p)=1$. It is worth noting that
$\tau_{in}, \tau_{p}$ and $A_1$ are the functions of $\varepsilon$.
To fix this parameter, we can use the condition $H_0(\tau_p)=1$.
Finally, to find the value of the present synchronous time, we use
the equation $\tilde t_p=\int_{\tau_{in}}^{\tau_p}f(\tau)d\tau$
where $f(\tau)=\tilde a_0^3(\tau)$. In the case $d_1=7$, direct
calculations give for the constants of integration $A_1=1.23468$ and
$\varepsilon=1.53097$. It results in $\tilde t_{p}=0.296\sim
4\mbox{Gyr}$, $q_0(\tilde t_{p})=$ $=-0.960572$ and for the internal
space $a_1(\tilde t_{p})=1.24319$, $H_1(t_{p})=0.0500333$.

\begin{figure}
\vskip1mm \includegraphics[width=13cm]{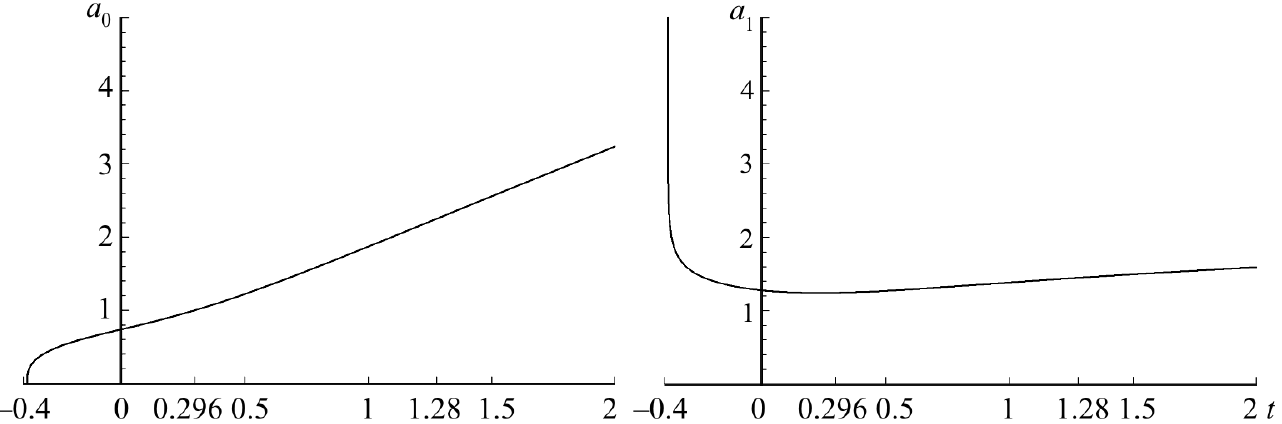} \vskip-1mm
  \caption{\label{zh-Sp1}The scale factors (in the Einstein frame) of
    the external space (left panel) and in\-ter\-nal space (right panel)
    versus synchronous time $\tilde t$}\vspace*{-1mm}
\end{figure}

Dynamical behavior of the considered model is depicted in
Figs.~3.\ref{zh-Sp1}---3.\ref{zh-Sp3} \cite{zhBZ-Sp2}.
Fig.~3.\ref{zh-Sp1} shows the dynamics of the external space scale
factor $\tilde a_0(\tilde t)$ (left panel) and the internal space
scale factor $a_1(\tilde t)$( right panel). Here, $\tilde t=0.296$
is the present time, and $\tilde t=0$ and $\tilde t=1.28$ correspond
to the beginning and the end of the stage of acceleration,
respectively. It follows that the internal space is the same order
of magnitude as the external one at the
present time. However, for the standard Kaluza---Klein models %
\index{Kaluza---Klein models|(} %
there is the experimental restriction on sizes of the extra
dimensions: $l_{extra}\leq 10^{-17}\,\text{cm}$. That is $\tilde
a_0/a_1 \geq 10^{45}$. Obviously, our model does not satisfy this
condition. One of the possible way to avoid this problem consists in
proposal that the Standard Model matter is localized on a brane. In
this case the extra dimensions can be much bigger that
$10^{-17}\,\text{cm}$ (even an infinite). However, such model
requires the generalization of our metric (\ref{zh6.2}) to the
non-factorisable case and this investigation is out of the scope of
this section.

\begin{figure}
\vskip1mm
  \includegraphics[width=13cm]{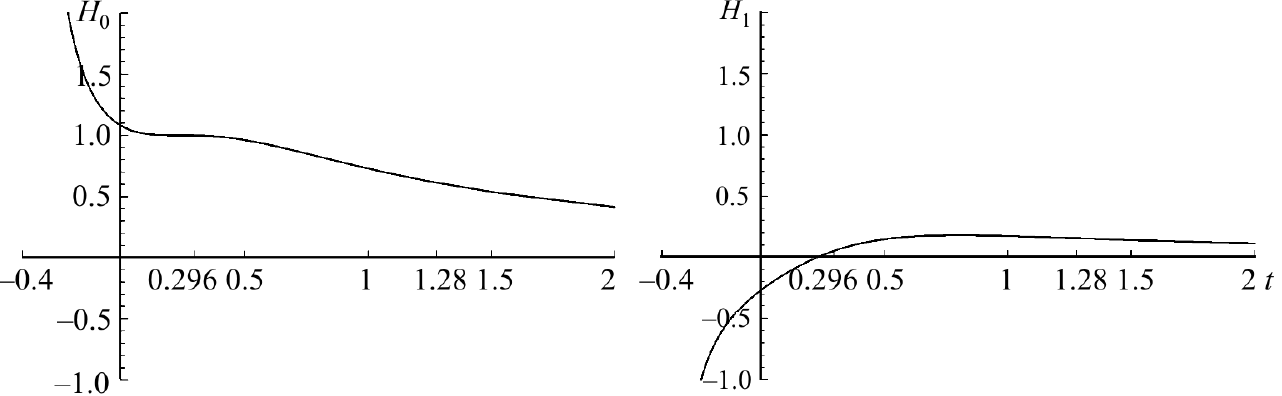}
\vskip-1mm
  \caption{\label{zh-Sp2}The Hubble parameters of the external space %
\index{Hubble parameter|)} %
    (left panel) and internal space (right panel) versus synchronous
    time $\tilde t$}\vskip3mm
\end{figure}

\begin{figure}
  \includegraphics[width=13cm]{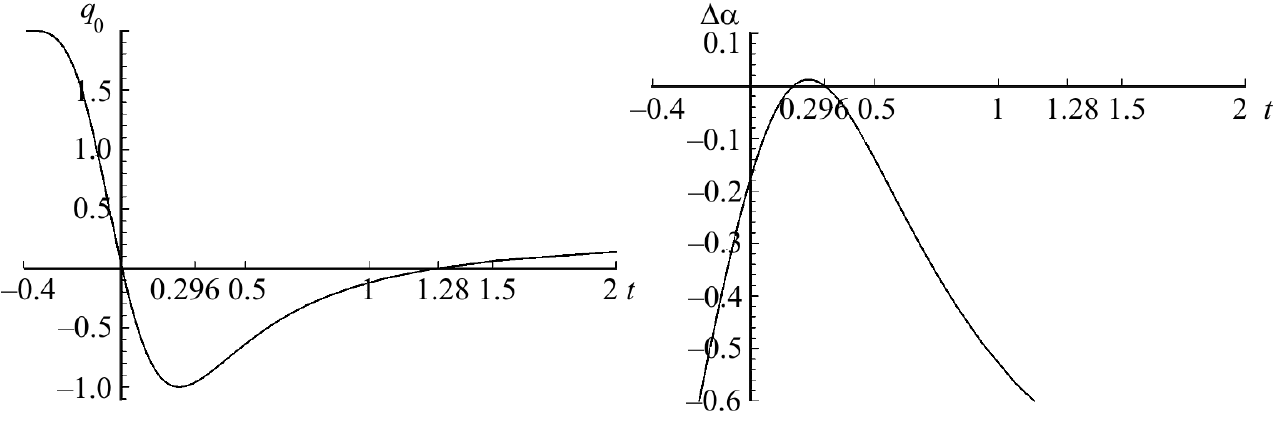}
\vskip-2mm
  \caption{\label{zh-Sp3}The deceleration parameter of the external
    space (left panel) and variation of the fine structure constant
    (right panel) versus synchronous time $\tilde t$}
\end{figure}

We plot in Fig.~3.\ref{zh-Sp2} the evolution of the Hubble
parameters $H_0(\tilde t)$ (left panel) and $H_1(\tilde t)$ (right
panel). We can see that their values are comparable with each other.
Thus, the internal space is rather dynamical and this fact is the
main reason of too large variations of the fine structure constant
\mbox{(see Fig.~3.\ref{zh-Sp3}).}

We present in Fig.~3.\ref{zh-Sp3} the evolution of the deceleration
parameter $q_0(\tilde t)$ (left panel) and the variation of the fine
structure constant $\Delta \alpha (\tilde t)$ (right panel). Left
picture clearly shows that the acceleration stage has the final
period for the considered model. It starts at $\tilde t=0$ and
finishes at $\tilde t=1.28$. The right picture demonstrates that
$\Delta \alpha$ does not satisfy the observable restrictions
$|\Delta \alpha|<10^{-5}$. There is the only very narrow region in
the vicinity of $z=0.13$ (or equivalently $\tilde t=0.17$ in
synchronous time) where $\Delta \alpha$ changes its sign.  However,
it is the exceptional region but restriction $|\Delta
\alpha|<10^{-5}$ works for very large diapason of redshifts $z$
\cite{zhUzan}.


\index{accelerated expansion} %
Therefore, despite the satisfactory description of the accelerated
expansion of our Universe at the late stages of its evolution, this
model has two significant drawbacks.  On the one hand, the internal
space is too big with respect to the standard Kaluza---Klein
restrictions $a_{internal}\leq 10^{-17}$cm and, on the other hand,
this space is not sufficiently constant to satisfy the observable
limits on the fine-structure constant variations. These are typical
problems for the Ka\-luza---Klein multidimensional cosmological
models with dynamical \mbox{internal spaces.}


\section{\!Problematic aspects of Kaluza---Klein models} \label{zh-KK
  problem}

\hspace*{3cm}\index{Kaluza---Klein models}In this section we want to
point out some problems of Kalu\-za---Klein models. We first
consider a model with a matter source in the form of a point-like
mass.  This approximation seems physically reasonable at
sufficiently large distances from the compact material sources.
Moreover, this approach works very well in General Relativity for
calculation in a weak-field limit of the formulas for the well
known gravitational experiments: perihelion shift, %
\index{perihelion shift} %
deflection of light and time delay of radar echoes \cite{zhLandau}.
Thus, we expect that such approach will be also applicable to
Kaluza---Klein models. However, it is not the case.

To prove it, we can use asymptotic expression for the metric
coefficients\,\footnote{\,\label{zh-sign}In this section \ref{zh-KK
    problem}, we use the sign convention of the book
  \cite{zhLandau}. We also indicate explicitly the speed of light $c$,
  because in weak-field approximation we expand quantities in powers
  of $1/c$. It is also convenient to use the letter $D$ to indicate
  the total number of spatial dimensions and the dimension of
  space-time is denoted by the letter $\mathcal{D}$, i.e.
  $\mathcal{D}=1+D$.} in $(\mathcal{D}=1+D)$-dimensional space-time
with toroidal extra dimensions. For a point-like mass $m$ at rest
the line element is \cite{zhEZ3}:
\[
ds^2  \approx \left(\!1\!-\frac{r_g}{r_3}+\frac{r_g^2}{2r_3^2}
  \!\right)c^2dt^2
  -\left(\!1+\frac{1}{D-2}\,
  \frac{r_g}{r_3}\!\right)\left(dr_3^2+r_3^2d\Omega^2_2\right)-
\]
\begin{equation}
\label{zh7.1}
  - \left(\!1+\frac{1}{D-2}\,
    \frac{r_g}{r_3}\!\right)\sum\limits_{i=1}^{N}ds^2_{i},
\end{equation}
where $r_3$ is the length of a radius vector in three-dimensional
space, $r_g=$ $=2G_Nm/c^2$ is three-dimensional Schwarzschild
radius, $G_N$ is the Newtonian gravitational constant,
$ds^2_{i}=\sum_{j=1}^{d_i}d\xi^2_{(i)j}$ is a metric of
$d_i$-dimensional torus, and we used three-dimensional isotropic
(with respect to our three-dimensional space) coordinates. We
suppose that the $(D=3+D')$-dimensional space has the factorisable
geometry of a product manifold $M_D=\mathbb{R}^3\times
T_1^{d_1}\times \mbox{...} \times T_N^{d_N}$\!. $\mathbb{R}^3$
describes the three-dimensional asymptotically flat external (our)
space and the internal space consists of $N$ $d_i$-dimensional tori
with the total number of the extra dimensions $D'=\sum_{i=1}^N d_i$.

We would like to mention that metrics (\ref{zh7.1}) is written for
the distances from gravitating mass $m$ which much larger than
periods of tori. In this case we can restrict ourselves to the zero
Kaluza---Klein mode (see \cite{zhEZ1,zhEZ2}). For example, this
approximation is very well satisfied for the planets of the solar
system because the inverse-square law experiments show that the
extra dimensions in Kaluza---Klein models should not exceed
submillimeter scales \cite{zhexperiment}. Then, the gravitational
potential reads \mbox{$\varphi(\mathbf{r})\approx -G_N m/r_3 =
-r_gc^2/(2r_3)$}. Mo\-re\-over, in the case of gravitating mass
uniformly smeared over the extra dimensions, the Newton's law
preserves its shape for arbitrary distances and this approximate
formula for $\varphi(\mathbf{r})$ becomes the exact equality
\cite{zhEZ1,zhEZ2}.

Now, we can use the metric (\ref{zh7.1}) to calculate the perihelion
shift \index{perihelion shift} and deflection of light, and to
compare the results with the observations. Such calculations were
performed in \cite{zhEZ3} where it was shown that considered model
significantly contradicts the experimental data. However, there is
more short way to show it with
the help of parameterized post-Newtonian (PPN) parameters. %
\index{parametrized post-Newtonian (PPN) parameters} %
According to PPN formalism (see, e.g.,
Refs.~\cite{zhWill,zhStraumann}), the static spherically symmetric
metrics in isotropic coordinates is parameterized as
follows:\vspace*{-3mm}
\begin{equation}\label{zh7.2}
  ds^2=\left(\!\!1-\frac{r_g}{r_3}+\beta\frac{r_g^2}{2r_3^2} \!\right)c^2dt^2 - \left(\!1+\gamma
    \frac{r_g}{r_3}\!\right)\left(dr_3^2+r_3^2d\Omega^2_2\right)\!.
\end{equation}
In General Relativity we have $\beta=\gamma=1$. To get $\beta$ and
$\gamma$ in the case of a point-like mass, it is sufficient to
compare the metric coefficients in (\ref{zh7.2}) with the
corresponding asymptotic expression (\ref{zh7.1}) what immediately
gives the PPN parameters for the point-like mass
\begin{equation}\label{zh7.3}
\beta =1, \quad \gamma=\frac{1}{1+D'}.
\end{equation}

The latter expression shows that parameter $\gamma$ coincides with
the correspon\-ding value of General Relativity if the number of
extra dimensions $D'=0$. Only in this case $\gamma =1$. According to
the experimental data, $\gamma $ should be very close to 1. The
tightest
constraint on $\gamma$ comes from the Shapiro time-delay experiment %
\index{Shapiro time-delay effect} %
using the Cassini spacecraft: $\gamma-1 =(2.1\pm 2.3)\times 10^{-5}$
[469---471]. On the other hand, for the point-like mass we get
$\gamma -1 =-D'/(1+D') \sim \mathcal{O}(1)$ what is very far from
the experimental data.

After this negative (and, to some extent unexpected) result with
respect to point-like masses, it is of interest to find metrics
which are in good agreement with observations.  From this point the
soliton solutions play important role. These solutions belong to a
class of metrics of the form
\begin{equation}\label{zh7.4}
  ds^2 = A(r_3)c^2dt^2+B(r_3)(dr_3^2
  +r_3^2d\Omega^2_2)+\sum\limits_{i=1}^{N}C_{i}(r_3)ds^2_{i}.
\end{equation}

\index{Einstein equations} %
They are vacuum solutions of Einstein equations with the proper
boundary conditions. The dependence of the metric coefficients in
(\ref{zh7.4}) only on $r_3$ means that the matter source for such
metrics is uniformly smeared over the extra dimensions
\cite{zhEZ1,zhEZ2}. It is clear that in this case the
non-relativistic gravitational potential depends only on $r_3$ and
exactly coincides with the Newtonian one. However, as we shall see
below, in general case this coincidence is not sufficient to be
agreement with observations.

In 5-dimensional space-time, soliton solutions %
\index{soliton solutions} %
were found in papers [474---476]. Then, they have been generalized
to an arbitrary number of dimensions in
\cite{zhIv1,zhIv0,zhLeon1,zhLeon2}. To our knowledge, the most
general form of these solutions was given in \cite{zhLeon1} and in
isotropic coordinates it reads\vspace*{-1mm}
\[
ds^2 =\left(\!\frac{ar_3 -
1}{ar_3+1}\!\right)^{\!\!2\alpha}\!c^2dt^2 -
  \left(\!1 -
    \frac{1}{a^2r_3^2}\!\right)^{\!\!2}\!\left(\frac{ar_3+1}{ar_3-1}\!\right)^{\!\!2\alpha(1-\tau)}\!
  \left(dr_3^2 +r_3^2d\Omega^2_2 \right)-
\]\vspace*{-3mm}
\begin{equation}
  \label{zh7.5}
  -\sum\limits_{i=1}^{N}\left(\!\frac{ar_3+1}{ar_3-1}\!\right)^{\!\!2\alpha\gamma_{i}}ds^2_{i}
  ,
\end{equation}
where parameters $\alpha $ and $\gamma_i$ satisfy the
condition\vspace*{-2mm}
\begin{equation}\label{zh7.6}
  \alpha^2[(\tau - 1)^2 +\sigma +1] = 2,\quad
  \tau\equiv\sum\limits_{i=1}^{N}d_{i}\gamma_{i}, \quad \sigma \equiv
  \sum\limits_{i=1}^{N}d_{i}\gamma^2_{i}.
\end{equation}\vspace*{-3mm}

In the weak-field limit $1/(ar_3)\ll 1$, the metric coefficients
are\vspace*{-1mm}
\begin{equation}
  \label{zh7.7} A(r_3)  \approx 1-\frac{4\alpha}{ar_3}+\frac{16\alpha^2}{a^2}\frac{1}{2 r_3^2},
  \end{equation}\vspace*{-4mm}
\begin{equation}
  B(r_3)  \approx -1-\frac{4\alpha(1-\tau)}{ar_3}\label{zh7.8},
  \end{equation}\vspace*{-3mm}
\begin{equation}
  C_i(r_3) \approx -1-\frac{4\alpha \gamma_i}{ar_3}\label{zh7.9}.
\end{equation}
The comparison of these asymptotes with the metric coefficients in
Eq.~(\ref{zh7.1}) gives a possibility to single out the soliton
solution which corresponds to the point-like mass. To get such
correspondence, first, the following relation must
hold:\vspace*{-2mm}
\begin{equation}\label{zh7.10}
  \frac{4\alpha}{a}=r_g.
\end{equation}\vspace*{-3mm}

It follows that ${\rm sign}\, a = {\rm sign}\, \alpha$. Because the
solution (\ref{zh7.5}) is invariant under the simultaneous change
$a\to -a, \alpha \to -\alpha$, we can choose $a,\alpha >0$.  Second,
the parameters $\gamma_i$ should take the same value for all
internal spaces:\vspace*{-2mm}
\begin{equation}\label{zh7.11}
  \gamma_1=\gamma_2=\mbox{...} =\gamma_N =\frac{1}{1+D'},
\end{equation}\vspace*{-7mm}

\noindent and, third, the parameters $\alpha$ and $a$
are\vspace*{-1mm}
\begin{equation}\label{zh7.12}
  \alpha = \sqrt{\frac{2(1+D')}{2+D'}}, \quad a =
  \frac{4}{r_g}\sqrt{\frac{2(1+D')}{2+D'}},
\end{equation}\vspace*{-5mm}

\noindent where we also took into account the constraint
(\ref{zh7.6}) and the relation (\ref{zh7.10}).

Therefore, Eqs.~(\ref{zh7.10})---(\ref{zh7.12}) completely define
the point-mass soliton, i.e.  the solution where delta-shaped
$T_{00}$ is the only non-zero component of the energy-momentum
tensor. To demonstrate it, we derive in the next subsection
equations of state for general soliton solution (\ref{zh7.5}). In
5-dimensional space-time, the experimental bounds as well as
equations of state for soliton solutions were investigated in
\cite{zhEZ4}. In the present section we consider the general case
(\ref{zh7.5}) of arbitrary number of dimensions and perform our
investigations following Ref.~\cite{zhEZ5}.

\subsection{\!\label{zh-7.1}Equations of state in general case}

\hspace*{3cm}\index{equation of state (EoS)}As we noted above, the
dependence of the metric coefficients in (\ref{zh7.5}) only on $r_3$
means that the matter source for such metrics is uniformly
``smeared'' over the extra dimensions.  It is clear that in this
case the non-relativistic gravitational potential depends only on
$r_3$ and exactly coin\-cides \mbox{with} the Newtonian one
\cite{zhEZ1,zhEZ2}. Because the function $A(r_3)$ is the metric
coefficient $g_{00}$ (which in the weak-field limit defines the
non-relativistic potential) this demand leads to the condition of
the form of (\ref{zh7.10}): $4\alpha /a=$ $=r_g=2G_Nm/c^2$. Then,
the expansions (\ref{zh7.7})---(\ref{zh7.9}) become
\begin{equation}
  A(r_3)  \approx 1-\frac{r_g}{r_3}+\frac{1}{2}\frac{r_g^2}{r_3^2} \label{zh7.13} ,
  \end{equation}\vspace*{-3mm}
  \begin{equation}
  B(r_3)  \approx -1-(1-\tau)\frac{r_g}{r_3}\label{zh7.14},
  \end{equation}\vspace*{-3mm}
  \begin{equation}
  C_i(r_3)  \approx -1-\gamma_i\frac{r_g}{r_3}\label{zh7.15}.
\end{equation}
From these expressions, we can easily get the perturbations
$h_{00}=-r_g/r_3$, $h_{\alpha\alpha}=-(1-\tau)r_g/r_3$ and
$h_{\mu_i\mu_i}=-\gamma_ir_g/r_3$ of the order of $1/c^2$ over the
flat space-time, that gives us the possibility to find components of
Ricci tensor up to the same order: %
\index{tensor Ricci} %
\begin{equation}
  \label{zh7.16}
  R_{00}  \approx \frac{1}{2}\triangle h_{00}= \frac{1}{2}\kappa_0^2m\delta(\mathbf{r}_3)c^2=\frac{1}{2}\kappa_0^2\rho_{(3)} c^2,
\end{equation}\vspace*{-3mm}
  \begin{equation}
  R_{\alpha\alpha}  \approx \frac{1}{2}\triangle h_{\alpha\alpha} =\frac{1}{2}(1-\tau)\kappa_0^2\rho_{(3)} c^2,~ \alpha=1,2,3,\label{zh7.17}
\end{equation}\vspace*{-3mm}
  \begin{equation}
  R_{\mu_i\mu_i}  \approx \frac{1}{2}\triangle h_{\mu_i\mu_i} =\frac{1}{2}\gamma_{i}\kappa_0^2\rho_{(3)} c^2,\label{zh7.18}
\end{equation}\vspace*{-3mm}
  \begin{equation*}
  \mu_i  = 1+ \sum_{j=0}^{i-1} d_j,\; \mbox{...} , d_i +
  \sum_{j=0}^{i-1} d_j  ;~ i=1,\mbox{...},N,
\end{equation*}
where $d_0=3$, $\kappa_0^2\equiv 8\pi G_N/c^4$ and $\triangle =
\delta^{ik}\partial^2/\partial x^{i}\partial x^{k}$ is the
$D$-dimensional Laplace operator (see \cite{zhEZ3} for details). We
also introduced the non-relativistic three-dimensional mass density
$\rho_{(3)} =m\delta(\mathbf{r}_3)$, which is connected with
$D$-dimensional mass density $\rho=\rho_{(3)}/V_{D'}$. Here,
$V_{D'}$ is the total volume of the internal spaces. For example, if
i-th torus has periods $a_{(i)j}$, then $V_{D'}=$\linebreak
$=\prod_{i=1}^N\prod_{j=1}^{d_i} a_{(i)j}$.

Now, we want to define the components of the energy-momentum tensor
with the help of Einstein equation in $(1+D)$-dimensional space-time: %
\index{Einstein equations} %
\begin{equation}\label{zh7.19}
  R_{ik}=\frac{2S_D\tilde G_{\mathcal{D}}}{c^4}\left(\!T_{ik}-\frac{1}{D-1}g_{ik}T\!\right)\!,
\end{equation}
where $S_D=2\pi^{D/2}/\Gamma (D/2)$ is the total solid angle
(surface area of the \mbox{$(D-1)$}-dimensional sphere of unit
radius) and $\tilde G_{\mathcal{D}}$ is the gravitational constant
in the $(\mathcal{D}=1+D)$-dimensional space-time. Introducing the
quantity $\kappa^2_{\mathcal{D}}\equiv$\linebreak $\equiv2S_D\tilde
G_{\mathcal{D}}/c^4$ (see also footnote \ref{kappa D} where we
should replace $D$ by $\mathcal{D}=1+D$) and keeping in mind that we
consider compact astrophysical object at rest in our
three-dimensional space (it results in $T_{11}=T_{22}=T_{33}=0$), we
arrive at the following Einstein equations:
\begin{gather}
  \frac{1}{2}\kappa_0^2\rho_{(3)} c^2\approx \kappa^2_{\mathcal{D}}\left(\!T_{00}-\frac{1}{D-1}Tg_{00}\!\right)\!,  \label{zh7.20}\\
  \frac{1}{2}(1-\tau)\kappa_0^2\rho_{(3)} c^2\approx \kappa^2_{\mathcal{D}}\left(\!-\frac{1}{D-1}Tg_{\alpha\alpha}\!\right)\!,\label{zh7.21}\\
  \frac{1}{2}\gamma_{i}\kappa_0^2\rho_{(3)} c^2\approx
  \kappa^2_{\mathcal{D}}\left(\!T_{\mu_i\mu_i}-\frac{1}{D-1}Tg_{\mu_i\mu_i}\!\right)\!.\label{zh7.22}
\end{gather}

Therefore, the required components of the energy-momentum are
\begin{equation}
  T_{00}\approx \frac{\kappa_0^2
    V_{D'}}{\kappa^2_{\mathcal{D}}}\left(\!1-\frac{\tau}{2}\!\right)\rho
  c^2,\quad T_{\alpha\alpha}=0, \label{zh7.23}
\end{equation}\vspace*{-3mm}
\begin{equation}
  T_{\mu_i\mu_i}  \approx \frac{\kappa_0^2
    V_{D'}(\gamma_{i}-1+\tau)}{2\kappa^2_{\mathcal{D}}}\rho
  c^2 \label{zh7.24}.
\end{equation}

The equation for the 00-component shows that the parameter $\tau$
cannot be equal to 2 because for $\tau =2$ we get $T_{00}=0$, what
corresponds to uninteresting case of absence of matter. Moreover,
$T^0_0=\varepsilon $ is the energy density of matter (remind that in
this section we use the sign convention of the book
\cite{zhLandau}). Therefore, up to the terms $1/c^2$, we have
$T_{00}\approx \varepsilon \approx \rho c^2$. It requires the
following relation between Newtonian and multidimensional
gravitational constants:
\begin{equation}
  \label{zh7.25}
  \kappa_0^2 =\frac{2}{2-\tau}\kappa^2_{\mathcal{D}}/V_{D'} \; \Longrightarrow \; 4\pi G_N
  = \frac{2}{2-\tau}S_D \tilde G_{\mathcal{D}}/V_{D'}.
\end{equation}
In the particular case of a point-like mass source with
$\tau=D'/(1+D')$, this relation was given in \cite{zhEZ2,zhEZ4}.
From Eqs.~(\ref{zh7.23}) and (\ref{zh7.24}) we also obtain the
relation
\begin{equation}
  \label{zh7.26}
  T_{\mu_i\mu_i}\approx \frac{\gamma_{i}-1+\tau}{2-\tau}\; T_{00}.
\end{equation}
Taking into account that up to the terms $1/c^2$, components
$T_{\mu_i\mu_i}$ define pressure in $i$-th internal space:
$T_{\mu_i\mu_i}\approx P_i$, we get from Eq.~(\ref{zh7.26}) the
following equations of state in these spaces:
\begin{equation}
  \label{zh7.27}
  P_i = \frac{\gamma_{i}-1+\tau}{2-\tau}\, \varepsilon ,\quad  i=1,\mbox{...} ,N.
\end{equation}

Since $T_{11}=T_{22}=T_{33}=0$, in our three-dimensional space we
have dust-like equation of state: $P_0=0$. In the case of a
point-like mass, the parameters $\gamma_i$ satisfy the condition
(\ref{zh7.11}). It can be easily seen that for these values of
$\gamma_i$, all $T_{\mu_i\mu_i}$ are equal to zero. Therefore, in
this case, $T_{00}$ is the only non-zero component and in the
external/our space, as well as in all internal spaces, we have the
same dust-like equations of state: $P_i=0$, $i=0,\mbox{...} ,N.$


\subsection{\!\label{zh-7.2}Latent solitons}

\hspace*{3cm}\index{latent solitons|(}Asymptotic expressions
(\ref{zh7.13}) and (\ref{zh7.14}) also enable to get the PPN
parameters in general case. Comparing these equations with the
corresponding metric coefficients in (\ref{zh7.2}), we immediately
find for solitons:
\begin{equation}\label{zh7.28}
  \beta_s =1, \quad \gamma_s=1-\tau.
\end{equation}
With the help of these PPN parameters, we can easily get formulas
for the famous gravitational experiments \cite{zhWill,zhWill2}:

{\bfseries\itshape (i) Perihelion shift}
\[
\delta\psi=\frac{6\pi mG_N}{\lambda
    \left(1-e^2\right)c^2}\frac{1}{3}
  (2+2\gamma_s-\beta_s)=
\]\vspace*{-3mm}
\begin{equation}
  \label{zh7.29}
    = \frac{6\pi mG_N}{\lambda \left(1-e^2\right)c^2}\frac{3-2\tau}{3}=
  \frac{\pi r_g}{\lambda \left(1-e^2\right)}(3-2\tau),
\end{equation} %
\index{perihelion shift}where $\lambda$ is the semi-major axis of
the ellipse and $e$ is its eccentricity.

{\bfseries\itshape (ii) Deflection of light}
\begin{equation}
  \label{zh7.30}
  \delta\psi=(1+\gamma_{s})\frac{r_g}{\rho}=(2-\tau)\frac{r_g}{\rho},
\end{equation} %
\index{deflection of light}where $\rho$ is the distance of closest
approach (impact parameter) of the ray's path to the gravitating
mass $m$.

{\bfseries\itshape (iii) Time delay of radar echoes (Shapiro
time-delay
  effect)}
\[
\delta t=(1+\gamma_{s})\frac{r_g}{c}\ln\left(\!\frac{4r_\mathrm{
        Earth}r_\mathrm{ planet}} {R_\mathrm{ Sun}^2}\!\right)=
\]\vspace*{-3mm}
\begin{equation}
  \label{zh7.31}
   = (2-\tau)\frac{r_g}{c}\ln\left(\!\frac{4r_\mathrm{ Earth}r_\mathrm{
        planet}}{R_\mathrm{Sun}^2}\!\right)\!.
\end{equation} %
\index{Shapiro time-delay effect} %

Comparison of the formulas (\ref{zh7.29})---(\ref{zh7.31}) with the
experimental data gives the possibility to restrict parameters of
the soliton solutions. In fact, as in the case of the point-like
mass, we can also get it directly from experimental restriction on
PPN parameter $\gamma $: $\gamma-1 =(2.1\,\pm\,2.3)\times 10^{-5}$.
Thus, from (\ref{zh7.28}) we find that solitonic parameter $\tau$
should satisfy the condition
\begin{equation}
  \label{zh7.32}
  \tau = -(2.1\pm 2.3)\times 10^{-5}.
\end{equation}

In the case of the point-like massive soliton described by
Eqs.~(\ref{zh7.10})---(\ref{zh7.12}), we have $\tau =D'/(1+D')\sim
\mathcal{O}(1)$, what obviously contradicts to Eq.~(\ref{zh7.32}).

Equation (\ref{zh7.28}) shows that there is very interesting class
of solitons which are defined by the condition\vspace*{-3mm}
\begin{equation}
  \label{zh7.33}
  \tau =\sum_{i=1}^N d_i \gamma_i = 0.
\end{equation}

With the help of the gravitational experiments mentioned above, it
is impossible to differ these Kaluza---Klein solitons from general
relativity because they have $\gamma_s=1$ as in general
relativity\,\footnote{\,It can be easily seen from equations
  (\ref{zh7.13})---(\ref{zh7.15}) that the parameter $\tau$ defines also the
  difference between perturbations $h_{00}$ and $h_{\alpha\alpha}$:
  $\; h_{00}-h_{\alpha\alpha}=-\tau r_g/r_3$. Precisely because of
  this difference gravitational experiments in KK models and in
  General Relativity lead to different results. When $\tau \to 0$,
  this difference disappears. The additional limit $\gamma_i \to 0
  \Longrightarrow$ $\Longrightarrow h_{\mu_i\mu_i}\to 0$ provides stabilization of the
  internal spaces \cite{zhZhuk-no-go,zhEZ4}.}. For this reason we
called these solutions {\textit{latent solitons}} \cite{zhEZ5}. %
For these latent solitons, equations of state (\ref{zh7.27}) in the
internal spaces are
\begin{equation}\label{zh7.34}
  P_i = \frac{\gamma_{i}-1}{2}\, \varepsilon ,\quad  i=1,\mbox{...} ,N.
\end{equation}

Black strings ($N=1,\, d_1=1$) and black branes ($N>1$) are %
\index{black strings} %
\index{black branes} %
characterized by the condition that all $\gamma_i =0,\, i\geq 1$.
Obviously, they belong to the class of latent solitons and they have
the equations of state
\begin{equation}
  \label{zh7.35}
  P_i = -\frac{1}{2}\, \varepsilon ,\quad  i=1,\mbox{...} ,N.
\end{equation}

It is known (see section (\ref{zh-perfect}) and
Refs.~\cite{zhZhuk-no-go,zhEZ4}) that in the case of
three-dimensional external/our space such equations of state are the
only ones which do not spoil the condition of the internal space
stabilization for the compact astrophysical objects with the
dust-like equation of state $P_0=0$ in the external space.
Therefore, it is tempting to treat non-zero parameters $\gamma_i$
 as a measure of the latent soliton
destabilization\,\footnote{\,\label{destabilization}  If we rewrite
equations of state in the form of Eq.~(\ref{zh2.31})
  $P_i=(\alpha_i-1)\varepsilon, \; i=0,\mbox{...} ,N $, then for the
  latent solitons we have $\alpha_0=1,\, \alpha_i=(1+\gamma_i)/2$,
  $i=1,\mbox{...} ,N$. For these values of $\alpha_0, \,\alpha_i$, we get
  on the right-hand side of the equation (\ref{zh2.40}) (for $d_0=3$) the
  terms $(\gamma_id_i/2)\kappa_N\rho_{(3)}$.  These terms are
  dynamical functions because of dynamical behavior of the energy
  density $\rho_{(3)}$. This results in violation of the necessary
  condition for the internal space stabilization.}.
  However, a careful analysis (see [483])shows that the variation of
  the total volume of the internal space is equal to zero.
  Consequently, in the case of latent solitons variation of
  fundamental constants are also absent.

We would like to stress the following: It is well known that black
strings (branes) have the topology
$$\text{(4-dimensional
  Schwarzschild space-time)}\times\text{(flat internal spaces)}.$$%
In this case, it does not seem surprising that gravitational
experiments lead to the same results as for general relativity.
However, the latent solitons, in general case, do not have either
Schwarzschildian metrics for 4-dimensional part of space-time nor
flat metrics for the extra dimensions. Nevertheless, within the
considered accuracy, it is also impossible to distinguish them from
General Relativity. This is really surprising.

To conclude this section, we would like to mention that the relation
between Newtonian and multidimensional gravitational constants for
latent solitons is reduced to the equation
(\ref{zh2.12}):\vspace*{-2mm}
\begin{equation}
  \label{zh7.36}
  4\pi G_N = S_D \tilde G_{\mathcal{D}}/V_{D'}.
\end{equation}


\subsection{\!Experimental restrictions on the equations\\
  \hspace*{-1.2cm}of state of a multidimensional perfect fluid\label{zh-7.3}}

\hspace*{3cm}Now, we want to show in general case that for static
spherically symmetric perfect fluid with dust-like equation of state
in our space, the condition $h_{00}=h_{\alpha\alpha}$ (which ensures
the agreement with the gravitational experiments at the same level
of accuracy as General Relativity) results in the latent soliton
condition (\ref{zh7.33}) and equations of state (\ref{zh7.34}), and
additional condition $R_{\mu_i\mu_i}=0 \Longrightarrow
h_{\mu_i\mu_i}=0$ reduces (\ref{zh7.34}) to (\ref{zh7.35}) (which is
necessary for the internal space stability) and singles out $d_0=3$
for the number of the external dimensions.

Let us consider a static spherically symmetric perfect fluid with
energy-momentum tensor (see footnote \ref{zh-sign} about the sign
convention in this section):
\begin{equation}
  \label{zh7.37}
  {T^i}_k = \mathrm{ diag\ } ( \varepsilon, \underbrace{-P_0,\mbox{...} ,-P_0}_{\text{$d_0$ times} }, \mbox{...} , \underbrace{-P_N,\mbox{...} ,-P_N}_{ \text{$d_N$ times}} ).
\end{equation}%
We recall that we are using the notations: $i,k = 0,1, \mbox{...},
D$; $a,b = 1,\mbox{...} , D$; $\alpha,\beta =1, \mbox{...}, d_0$ and
$\mu_i = 1+ \sum_{j=0}^{i-1} d_j, \mbox{...} , d_i +
\sum_{j=0}^{i-1} d_j $, $i=1,\mbox{...},N$. For static spherically
symmetric configurations we have $g_{0a}=0$ and $g_{ab}=0,\, a\neq
b$. Since we want to apply this model to ordinary astrophysical
objects where the condition ${T^0}_0 \gg |{T^{\alpha}}_{\alpha}|$
usually holds, we assume the dust-like equation of state in
$d_0$-dimensional external space: $P_0=0$, but the equations of
state are arbitrary ones in i-th internal space: $P_i=\omega_i
\varepsilon$. Obviously, $\varepsilon$ is equal to zero outside the
compact astrophysical objects.

Moreover, we consider the weak-field approximation where the metric
coefficients can be expressed in the form%
\begin{equation}
  \label{zh7.38}
  g_{00}\approx 1+h_{00},\quad g_{aa}\approx -1+h_{aa}, \quad h_{00},h_{aa} \sim
  O(1/c^2).
\end{equation}%

As an additional requirement, we impose that the considered
configuration does not contradict the observations. It will be so if
the following conditions hold: $h_{00}=h_{\alpha\alpha}$ and
$h_{\mu_i\mu_i}=0$ (see Ref.~\cite{zhEZ4}). In what follows, we
define which equations of state are obtained as a result of these
restrictions.  Taking into account that $T=\sum_{i=0}^D
{T^i}_i=\varepsilon (1-\sum_{i=1}^N\omega_i d_i)$,
$T_{\alpha\alpha}=0$, $\varepsilon \sim O(c^2)$ and, up to terms
$c^2$, that $T_{00}\approx {T^0}_0$, $T_{\mu_i\mu_i}\approx
-{T^{\mu_i}}_{\mu_i}$, we get from the Einstein equation
(\ref{zh7.19})
the non-zero components of Ricci tensor (up to $1/c^2$): %
\index{tensor Ricci} %
\begin{equation}
  \label{zh7.39}
  R_{00}  \approx \frac{\varepsilon \kappa_{\mathcal{D}}}{D-1}\;
  \left[d_0-2+\sum_{i=1}^{N}d_{i}(1+\omega_{i})\right]\!,
\end{equation}\vspace*{-3mm}
\begin{equation}
  R_{\alpha\alpha}  \approx \frac{\varepsilon \kappa_{\mathcal{D}}}{D-1}\; \left[1-\sum_{i=1}^{N}d_{i}\omega_{i}\right]\!,\label{zh7.40}
\end{equation}\vspace*{-3mm}
\begin{equation}
  R_{\mu_i\mu_i}  \approx \frac{\varepsilon
    \kappa_{\mathcal{D}}}{D-1}\times
  \left[\omega_i\left(\sum_{j=0}^{N}{'}d_{j}-1\!\right)+1-\sum_{j=1}^{N}{'}d_{j}\omega_{j}\right]\!, \label{zh7.41}
\end{equation}
where $\kappa_{\mathcal{D}}\sim O(1/c^4)$ is defined in section
\ref{zh-7.1} and the prime in the summation of Eq.~(\ref{zh7.41})
means that we must not take into account the $i$-th term.
Eqs.~(\ref{zh7.39}) and (\ref{zh7.40}) shows that $R_{00}$ and
$R_{\alpha\alpha}$ components are related as follows:
\begin{equation}
  \label{zh7.42}
  R_{\alpha\alpha} = \frac{1-\sum_{i=1}^{N}d_{i}\omega_{i}}{d_0 - 2 +
    \sum_{i=1}^{N}d_{i}(1+\omega_i)}R_{00}.
\end{equation}

On the other hand, in weak-field limit the components of Ricci tensor read %
\index{tensor Ricci} %
\begin{equation}
  \label{zh7.43}
  R_{00}\approx \frac12 \triangle h_{00},\quad  R_{aa}\approx \frac12 \triangle h_{aa},\quad  a=1,\mbox{...} ,D,
\end{equation}
where as usual we can put $h_{00}\equiv 2\varphi/c^2$ and
$\triangle$ is $D$-dimensional Laplace operator defined in
Eqs.~(\ref{zh7.16})---(\ref{zh7.18}). Therefore, from equations
(\ref{zh7.42}) and (\ref{zh7.43}) we obtain\vspace*{-3mm}
\begin{equation}\label{zh7.44}
  h_{\alpha\alpha}=\frac{1-\sum_{i=1}^{N}d_{i}\omega_{i}}{d_0 - 2 +
    \sum_{i=1}^{N}d_{i}(1+\omega_i)}h_{00},~ \alpha =1,\mbox{...} ,d_0.
\end{equation}
As we have mentioned above, to be in agreement with experiment at
the same level of accuracy as General Relativity we should demand
$h_{\alpha\alpha}=h_{00}$, what leads to the restriction on the
parameters $\omega_i$ of the equations of state:
\begin{equation}
  \label{zh7.45}
  3-d_0 - \sum_{i=1}^{N}d_{i} = 2\sum_{i=1}^{N}d_{i}\omega_{i}.
\end{equation}
In the case of three-dimensional external space $(d_0=3)$, this
constraint is reduced to\vspace*{-2mm}
\begin{equation}
  \label{zh7.46}
  \sum_{i=1}^N d_i\left(\!\omega_i+\frac12\!\right)=0.
\end{equation}

If we parameterize\vspace*{-3mm}
\begin{equation}\label{zh7.47}
  \omega_i = \frac{\gamma_i-1}{2},~ i=1,\mbox{...} ,N,
\end{equation}%
\index{latent solitons} %
then we arrive at the latent soliton condition (\ref{zh7.33}).
Therefore, the demand that multidimensional perfect fluid (with
dust-like equation of state in the external space $P_0=0$) provides
the same results for gravitational experiments as General
Relativity, leads to the latent soliton equations of state
(\ref{zh7.34}) in the internal spaces.  However, it is known (see
section \ref{zh-perfect}) that the internal spaces can be stabilized
if multidimensional perfect fluid (with $P_0=0$) has the same
equations of state $\omega_i=-1/2$ in all internal spaces and the
external space is three-dimensional $d_0=3$. In other words, it
takes place if all $\gamma_i=0$ in (\ref{zh7.47}). Let us show that
the additional requirement $R_{\mu_i\mu_i}=0$ ensures the
fulfillment of these conditions. Indeed, from Eq.~(\ref{zh7.41}) we
get
\begin{equation}
  \label{zh7.48}
  R_{\mu_i\mu_i}=0~ \Longrightarrow~ \omega_i =-\frac12,~ i=1,\mbox{...} ,N,
\end{equation}
where we used the constraint
(\ref{zh7.45})\,\footnote{\,\label{additional}It
  can be also easily seen that $R_{\mu_i\mu_i}=0\; \Longrightarrow \;
  \triangle h_{\mu_i\mu_i}=0$, what, together with the boundary
  conditions (finiteness of $h_{\mu_i\mu_i}$ at $r_3=0$ and
  $h_{\mu_i\mu_i}\to 0$ for $r_3\to +\infty$), gives
  $h_{\mu_i\mu_i}=0$.}. Now, substitution $\omega_i=-1/2$ in
(\ref{zh7.45}) singles out $d_0=3$.  Therefore, the demand of the
internal space stabilization leads, for multidimensional perfect
fluid (with $P_0=0$), to the black string/brane equations of state
(\ref{zh7.35}) in the internal spaces and, additionally, it selects
uniquely the number of the external spaces to be $d_0=3$.

To conclude the consideration of this perfect fluid, we want to get
the metric coefficients up to $\mathcal{O}(1/c^2)$ (see
Eq.~(\ref{zh7.38})). To do so, it is sufficient to define the
function $\varphi$. It can be easily seen from (\ref{zh7.39}) and
(\ref{zh7.43}) that this function satisfies the equation
\begin{equation}
  \label{zh7.49}
  \triangle\varphi =\frac{c^2}{2} \triangle h_{00}  \approx c^2 R_{00} \approx S_D \tilde
  G_{\mathcal{D}}\rho,
\end{equation}
where we use the constraint (\ref{zh7.45}) for arbitrary $d_0$ and
relation $\varepsilon \approx \rho c^2$. Therefore, to get the
metric coefficients we need to solve this equation \mbox{with}
proper boundary conditions. We want to reduce this equation to
ordinary \mbox{Poisson} equation in three-dimensional external space
$d_0=3$. To do so, we consider the case in which matter is uniformly
smeared over the extra dimensions, then $\rho=\rho_{(3)}/V_{D'}$
(see section \ref{zh-7.1}).

In this case the non-relativistic potential $\varphi$ depends only
on our external coordinates and $\triangle$ is reduced to
three-dimensional Laplace operator $\triangle_3$. The\-re\-fore,
Eq.~(\ref{zh7.49}) is reduced to
\begin{equation}
  \label{zh7.50}
  \triangle_3 \varphi \approx  (S_D \tilde G_{\mathcal{D}}/V_{D'})\rho_{(3)}=4\pi G_N
  \rho_{(3)},
\end{equation}
where we use the relation (\ref{zh7.36}) between Newtonian and
multidimensional gravitational constants. This is usual Poisson
equation. It is worth noting that $\rho_{(3)}=0$ outside the compact
astrophysical object and it is necessary to solve (\ref{zh7.50})
inside and outside of the object, followed by matching these
solutions at the boundary.

We can summarize the main conclusion of this section as follows. For
compact astrophysical objects with dust-like equation of state in
the external space ($P_0=0$), the demand of the agreement with the
gravitational experiments requires the condition (\ref{zh7.32}),
namely: \mbox{$\tau = -(2.1\,\pm\,2.3)$\,$\times$\,$10^{-5}$}.
However, to be at the same level of accuracy as General Relativity,
we must have $\tau=0$. In other words, we should consider the latent
solitons with equations of state (\ref{zh7.34}) in the internal
spaces (in the case $d_0=3$). Moreover, the condition of stability
of the internal spaces singles out black \mbox{strings}/bra\-nes
from the latent solitons and leads uniquely to
$P_i=-(1/2)\varepsilon$ as the black \mbox{string}/bra\-ne equations
of state in the internal spaces, and to the number of the external
dimensions $d_0=3$.
The main problem with the black \mbox{strings}/bra\-nes is to find a
physically reasonable mechanism which can explain how the ordinary
particles forming the astrophysical objects can acquire rather
specific equations of state ($P_i=-\varepsilon/2$) in the internal
spaces.

As we have seen above, to be in agreement with the observations, it
is necessary to break the symmetry (in terms of equations of state)
between the external/our and internal spaces. In our opinion,
braneworld models are the most promising alternative to the KK
models because they naturally break the symmetry between our
three-dimensional Universe and the extra dimensions, and the
following chapter is devoted to the problems of dark energy and dark
matter in the braneworld models.\vspace*{2mm}



\section{\!Summary}

\hspace*{3cm}In this chapter, we have considered a possibility for
our space-time to have extra spatial dimensions, and what
observational con\-se\-quen\-ces follow from this. In particular,
can we explain dark matter and dark energy due to the existence of
the extra dimensions? We supposed that fields from the Standard
Model of particle physics are not localized on a three-dimensional
hy\-per\-sur\-fa\-ce but can propagate throughout the
multidimensional space-time. Such models are called Kaluza---Klein
ones. The case of localization will be considered in the following
chapter 4.

We first investigated an important problem of stable
compactification of the internal space. The point is that in
Kaluza---Klein models, to make the extra dimensions unobservable at
the present time, the internal spaces have to be compact and reduced
to very small scales (of the order of or less than the Fermi
length). Therefore, we proposed general mechanism of dimension
reduction of multidimensional models. We have shown that after such
reduction the considered models take the form of an effective
four-dimensional Brans---Dicke or Einstein (after conformal
transformation) theories with scalar fields. These scalar fields are
defined by scale factors of the internal spaces. The form of a
potential energy of these fields depends on topology of the internal
spaces and matter content of considered models. Moreover, to avoid a
problem of too large variations of the fine structure constant, the
compactification have to be stable against fluctuations of these
fields (i.e. small fluctuations of the scale factors of the internal
spaces). This means that the effective potential of the model
obtained under dimensional reduction to a four-dimensional effective
theory should have minima with respect to these fluctuations. These
minima play the role of the cosmological constant in our observable
Universe and can be dark energy in the case of positive values.  We
have also shown that small excitations of a system near a minimum of
the effective potential can be observed as massive scalar fields in
the external (our) space-time.  These scalar fields very weakly
interact with the Standard Model particles. Therefore, they belong
to a class of the dark matter
particles. We called these particles gravitational excitons. %
\index{gravitational excitons} %
They may play an important role during the Universe evolution.
Therefore, we have investigated the dynamical behavior of
gravexcitons depending on the value of their mass. We have also
considered effects of Lorentz invariance violation due to
interaction between gravexcitons and four-dimensional photons. It
was shown that experimental limitations on such violation can
restrict parameters (e.g. masses of gravexcitons) of the models.

In conventional cosmology matter fields are taken into account in a
phe\-no\-me\-no\-lo\-gi\-cal way as a perfect fluid. Therefore, we
have proven an important theorem (the no-go theorem) which defines
the classes of the perfect fluid allowing the stable
compactification of the internal spaces.

Non-linear gravitational $f(R)$ models have attracted the great deal
of interest from the eighties of last century because these models
can provide a natural mechanism of the early inflation.  Recently,
it was realized that these models can also explain the late-time
acceleration of the Universe (dark energy). This fact resulted in a
new wave of papers devoted to this topic. Therefore, we have
generalized these theories to the case of multidimensional
space-time. We have shown that these models are reach enough to
explain both early inflation and accelerated expansion of the
Universe at a late stage of its evolution. However, there are still
problems with fine-tuning of parameters of the models as well as
with sufficient number of e-foldings.

We have also considered multidimensional cosmological models which
can mimic dynamical dark energy. They are the so-called S$p$-branes.
In this case, there are no minima of the effective potential for
gravexcitons and acceleration has different origin.  For example,
the negative cur\-va\-tu\-re of the internal space can lead to such
acceleration. However, we have shown that, despite the satisfactory
description of the accelerated expansion of our Universe at the late %
\index{accelerated expansion} %
stages of its evolution, this model has two significant drawbacks.
On the one hand, the internal space is too big with respect to the
standard Kaluza---Klein restrictions and, on the other hand, this
space is not sufficiently invariable to satisfy the observable
limits on the fine-structure constant variations. These are typical
problems for the Kaluza---Klein multidimensional cosmological models
with dynamical
internal spaces. %
\index{Kaluza---Klein models|)} %

\index{latent solitons|)} %
It is well known that General Relativity in four-dimensional
space-time is in good agreement with gravitational experiments such
as
perihelion shift, %
\index{perihelion shift} %
deflection of light, time delay of radar echoes and %
\index{parametrized post-Newtonian (PPN) parameters} %
PPN parameters. Therefore, it is important to verify Kaluza---Klein
models as to their conformity with these experiments. We first have
considered models with toroidal compactification of the extra
dimensions. A matter source was taken in the form of a point-like
mass with a dust-like $p=0$ equation of state in all (external and
internal) spatial dimensions. This approach works very well in
General Relativity for calculation in a weak field limit of the
formulas for the gravitational experiments.  However, in the case of
considered Kaluza---Klein models, we found that PPN parameters
demonstrate good agreement with the experimental data only in the
case of ordinary three-dimensional space. Therefore, the point-like
gravitational source with dust-like equations of state strongly
contradicts the observations. It is important to note that the
result does not depend on the point-like approximation. Instead of
the delta-shaped form, we can consider a compact object in the form
of a perfect fluid with the dust-like equation of state in all
spatial dimensions, and we obtain the same negative result. It
turned out that to satisfy the experimental data, the matter source
should have negative equations of state (tension) in the internal
spaces. For example, latent solitons (which are exact solutions
(\ref{zh7.5}) with condition (\ref{zh7.33})) have such tension and
they satisfy the gravitational tests at the same level of accuracy
as General Relativity. The uniform black strings and black branes
are particular examples of the latent solitons. It is of interest to
understand why some models meet the classical gravitational tests,
while others do not. In our recent paper \cite{zhEZ6}, we have shown
that the variation of the total volume of the internal spaces
generates the fifth force in the case of toroidal models with the
dust-like equations of state in all spatial dimensions. This is the
main reason of the problem. However, in the case of the latent
solitons, tension of the gravitating source fixes the internal space
volume, eliminating the fifth force contribution and resulting in
agreement with the observations. Therefore, tension plays a crucial
role here. In the case of spherical compactification of the internal
space, the fifth force is replaced by the Yukawa interaction for
models with the stabilized internal space [484---486]. For large
Yukawa masses (gravexciton masses), the effect of this interaction
is negligibly small, and considered models satisfy the gravitational
tests at the same level of accuracy as General Relativity. It
happens for an arbitrary equation of state (including the dust-like
$p=0$) in the internal space. However, we have shown
\cite{zhEFZ,zhEZ7} for this model that gravitating masses acquire
effective relativistic pressure in the external space. Such pressure
contradicts the observations of compact astrophysical objects (e.g.,
the Sun). The tension (with the parameter of equation of state
$\omega=-1/2$) in the internal space is the only possibility to
preserve the dust-like equation of state in the external space.
Therefore, in spite of agreement with the gravitational experiments
for an arbitrary value of $\omega$, tension ($\omega=-1/2$) also
plays a crucial role for the models with spherical compactification.
The problematic aspect of all these models with tension consists in
physically reasonable explanation of the origin of tension for
ordinary astrophysical objects.

\newpage

\setcounter{chapter}{3}
\chapter{\label{chap:brane}  BRANEWORLD MODELS}\markboth{CHAPTER 4.\,\,Braneworld models}{CHAPTER 4.\,\,Braneworld models}
\thispagestyle{empty}\vspace*{-12mm}

\begin{wrapfigure}{l}{2.6cm}
\vspace*{-5.5cm}{\includegraphics[width=3.0cm]{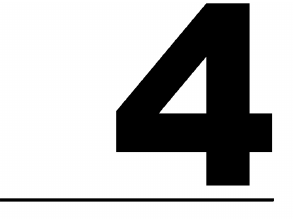}}\vskip17.2cm
\end{wrapfigure}
\vspace*{10mm}

 \setcounter{section}{1} \vspace*{-3mm}
\hspace*{3cm}\section*{\hspace*{-3cm}4.1.\,\,Introduction$_{}$}\label{brane-sec:intro}\begin{picture}(10,10)
\put(0,-135){\bfseries\sffamily{206}}
\end{picture}

\vspace*{-1.1cm}\noindent \index{braneworld
model}Cosmological\,braneworld\,models\,constitute\,the\,branch\,of\,phy-
sics with non-compact (``large'') extra-dimensions in \mbox{which}
our world is a four-dimensional hypersurface (the brane) embedded in
a higher-dimensional manifold. This represents an alternative to the
Kaluza---Klein compactification of ext\-ra dimensions that was
described in Sec.~3.1 of \mbox{chapter \ref{chap:KK}.}

One of the first models with localization of matter on the brane was
constructed in \cite{RubSha} (see also \cite{Akama:1982jy} and
reviews in \cite{zhRubakov,Barvinsky}).  In this model, the brane
represented a domain wall (kink) in a five-dimensional space-time,
constructed of a real scalar field $\varphi$ with spontaneous
breaking of the discrete $Z_2$ symmetry with respect to the
transformation $\varphi \to - \varphi$.  A Dirac fermion field
coupled to this scalar has a zero Kaluza---Klein mode (i.e., with
zero four-dimensional invariant mass) which is localized in the
neighborhood of the brane and has left chirality.  Such massless
fermions can imitate matter localized on the brane. At high
interaction energies, such zero modes can create excitations from
the continuum of the Kaluza---Klein spectrum, which then would
correspond to the process of particles leaving the brane and going
into higher dimension [492---493].  It is also possible (but more
difficult) to confine vector gauge fields on
the brane in the field-theoretical setup \cite{Dvali:1996xe}. %
\index{scalar field} %

More recent motivation for alternative compactification of extra
dimensions is the discovery of $p$-branes~--- extended dynamical
submanifolds in the multidimensional space~--- in string theory
\cite{zh1b,zh1c}.  Some types of $p$-branes can confine matter
fields; for instance, gauge fields can live on the so-called
Dirichlet-branes, or D-branes (see a review in
\cite{Polchinski:1996na}).  D$p$-branes are $(p + 1)$-dimensional\,
time-like\, hy\-per\-sur\-fa\-ces

\begin{wrapfigure}{r}{5.1cm}
\includegraphics[width=5cm]{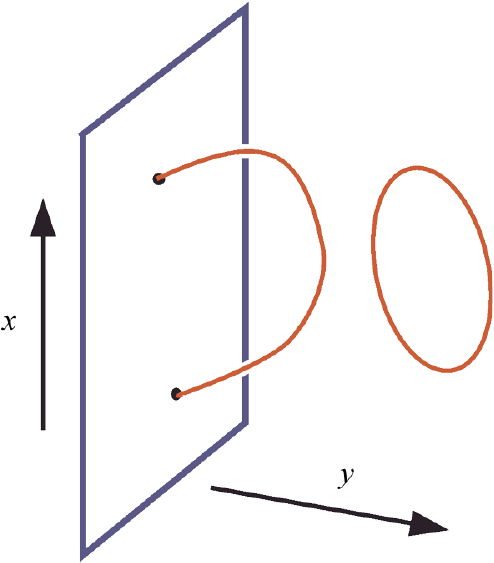}
\raisebox{0.2cm}{\parbox[b]{5cm}{\caption{\footnotesize An open
string whose ends are attached to a D-brane, and a
    closed string that can propagate free\-ly in a higher-dimensional
    space \label{brane-fig:string}}}}\vspace*{-3mm}
\end{wrapfigure}

\noindent  at which the ends of open strings can be localized (see
Fig.~4.\ref{brane-fig:string}). Since the ends of open strings carry
gauge fields, these fields on a fundamental level are $(p +
1)$-di\-men\-sio\-nal objects residing entirely on the brane and
having no Kaluza---Klein coun\-ter\-parts.  On the other hand,
closed strings, descri\-bing excitations of spin two (i.e.,
gravitons), can freely propagate in the mul\-ti\-di\-men\-sio\-nal
volume. This explains the basic structural features of the new
picture of ext\-ra dimen\-sions, in which matter fields and gravity
play different role.

\index{M-theory} %
The idea of a ``braneworld'' arises na\-tu\-ral\-ly in the context
of the fundamental M-theo\-ry (see \cite{Kiritsis} as a review).
Although the phenomenological models usually under consideration in
cosmology are quite simplistic, one can hope that the analogs of at
least some of their properties can be found in a realistic
fundamental theory. This relates, in particular, to the properties
of gravity and dark energy in the braneworld model, which is the
subject of the present chapter.

In the end of 1990ies, the braneworld theory gave a fresh view of
the problem of the Planck hierarchy, which served as one of the
motivations for its subsequent development.  In the model proposed
by Arkani-Hamed, Di\-mo\-pou\-los and Dvali (the ADD model)
\cite{ADD} (see also \cite{AADD}), a multidimensional gravitational
action of Hilbert---Einstein type
\begin{equation} \label{add-action} S = M^{D-2} \int {\mathcal R}
  \sqrt{-g}\, d^D x ,
\end{equation}
was under consideration, where $M$ is the Planck mass in a
$D$-dimensional space-time, and ${\mathcal R}$ is the curvature
scalar of the corresponding metric $g_{ab}$.

In the presence of a three-brane, under compactification of the
extra $d =$ $= D - 4$ dimensions and under the assumption of
independence of the graviton wave function on the extra coordinates,
action (\ref{add-action}) reduces to a four-dimensional effective
action on the brane of the form
\begin{equation}
  S_{\rm brane} = M_{\rm P} \int R \sqrt{-h}\, d^4 x,
\end{equation}
where $R$ is the curvature scalar of the induced metric $h_{ab}$ on
the brane, and the Planck mass $M_{\rm P}$ is expressed through the
fundamental mass $M$ and characteristic compactification radius of
extra dimensions $L$ as follows:\vspace*{-1mm}
\begin{equation}
  M_{\rm P} = M \left( M L \right)^{d/2}\!.
\end{equation}

By setting $M \sim 1$~TeV, one can estimate the value of
$L$:\vspace*{-1mm}
\begin{equation}
L \sim M^{-1} \left(\! \frac{M_{\rm P}}{M} \!\right)^{\!2/d} \!\sim
10^{32/d - 17}~\mbox{cm} .
\end{equation}
Of interest is the case $d = 2$, for which we have $L \sim 1$~mm.
The analysis of the processes of graviton production during
supernovae explosions leads to a somewhat stronger constraint on the
quantity $M$, namely \cite{zhRubakov,ADD1}, $M > 30$~TeV, which
gives $L \sim 1$$-$$10$~$\mu$m for $d = 2$. Thus, compactification
of extra dimensions on very large (from particle-physics viewpoint)
spatial scale, in principle, could solve the problem of the Planck
hierarchy, at the same time leading to modifications of the law of
gravity at small distances.  This observation stimulated search of
the deviation of the gravitational law from Newtonian at
submillimeter distances.

\index{ADD model} %
The ADD model did not address the problem of hierarchy connected
\mbox{with} a large spatial compactification scale $L$.  Moreover,
this model neglected the curvature of the multidimensional space, in
particular, caused by possible gravitational effect of the brane in
this space. Both shortcomings were removed in the model due to
Randall and Sundrum (the RS model) \cite{RS1,RS2}.  Inspired by the
eleven-dimensional Ho\v{r}ava--Witten model \cite{HW1,HW2} on the
orbifold $R^{10} \times S^1/Z_2$, this version of the theory had
only one large extra spatial dimension. The arising effective
five-dimensional space-time was curved, and the four-dimensional
character of the laws of gravity on the brane was achieved by
localization of the massless gravitational mode in the neighborhood
of the brane.  The flat character of the brane (i.e., the absence of
a large effective cosmological constant on the brane) was achieved
by fine tuning the brane tension $\sigma$ and the gravitational and
cosmological constants in the fife-dimensional space-time. %
\index{cosmological constant|(} %

The next important step in the construction of braneworld models was
made in \cite{DGP,CH} (and, independently, in \cite{Shtanov1}),
where it was pointed out that the quantum character of the fields
localized on the brane, in general, leads to the appearance of the
term with the curvature scalar of the induced metric on the brane.
(This mechanism of generating an effective action for gravity was,
in fact, first considered by Sakharov \cite{Sakharov}.)  It is this
version of the theory that will be the focus of our discussion in
this chapter.

\section{\!General setup and notation} \label{brane-sec:setup}

\hspace*{3cm}Throughout this chapter, our main object of
investigation will be a time-like hy\-per\-sur\-fa\-ce (called
brane) in a five-dimensional manifold $\mathcal{B}$ (the bulk).  In
this case it is convenient to use the conventions of \cite{Wald}.
Specifically, the vectors and tensors tangent to the brane are
regarded as vectors also tangent to the five-dimensional manifold,
and thus all tensors carry the five-dimensional abstract indices
$a$, $b$, $c$, $\mbox{...}$ By a tensor field tangent to the brane,
we mean any tensor field $T^{a
  \cdots}{}_{b \cdots}$ defined at the position of the brane and such
that\vspace*{-1.5mm}
\begin{equation}
  n_a T^{a \cdots}{}_{b \cdots} = n^b T^{a \cdots}{}_{b \cdots} = \mbox{...} = 0
\end{equation}
for any vector field $n^a$ normal to the brane.

The metric in the bulk is denoted by $g_{ab}$, while the induced
metric on the brane is $h_{ab} = g_{ab} - n_a n_b$, where $n^a$ is
the {\em unit\/} vector field normal to the brane, which we always
set to be the inner normal, pointed {\em from\/} the brane in the
direction of the bulk space.  Everywhere we employ the space-time
signature $(-,+,+,+,+)$.

\index{brane} %
The induced metric on the brane defines the unique covariant
derivative on the brane compatible with this metric, which we denote
by $D_a$, to distinguish it from the covariant derivative in the
five-dimensional bulk space $\nabla_a$.  The curvature tensor for
$g_{ab}$ in the bulk is denoted by $\mathcal{R}^a{}_{bcd}$.  The
curvature tensor of $h_{ab}$ on the brane is denoted as
$R^a{}_{bcd}$; regarded as a tensor in $\mathcal{B}$, it is tangent
to the brane by construction.  The tensor of extrinsic curvature of
the brane in $\mathcal{B}$ is defined as\vspace*{-1.5mm}
\begin{equation}
  K_{ab} = h^c{}_a \nabla_c n_b \equiv - \frac12 \mathcal{L}_n h_{ab} ,
\end{equation}
where $\mathcal{L}_n$ denotes the lie derivative along the vector
field $n^a$ (arbitrarily but smoothly extended from the brane to the
bulk space). It is also tangent to the brane by definition, as can
easily be verified.

\index{bulk} %
We consider a braneworld model described by the following simple yet
generic action, which includes gravitational and cosmological
constants in the bulk ($\mathcal{B}$) and on the
brane:\vspace*{-3mm}
\[
S = \sum_{i = 1}^N M_i^3 \left[
    \int\limits_{{\mathcal B}_i} \left({\mathcal R}_i - 2 \Lambda_i \right) -
    2 \int\limits_\mathrm{ brane} K_i \right] +
\]\vspace*{-3mm}
\begin{equation}
\label{brane-action} + \int\limits_\mathrm{ brane} \left(
    m^2 R - 2 \sigma \right)
    + \int\limits_\mathrm{ brane} L (h_{ab}, \phi)   .
\end{equation}

\begin{figure}
\vskip1mm
    \includegraphics[width=13cm]{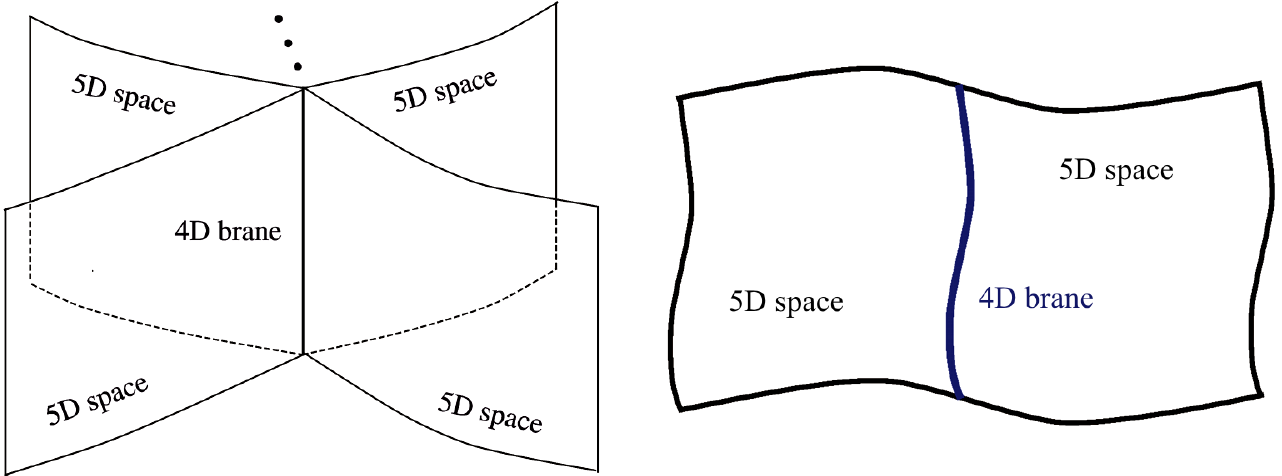}
\vskip-1mm
  \caption{On the left: A brane as a boundary of several volume
    spaces. The field theory of this configuration is studied in
    \cite{Fomin:2000tv}. On the right: A special, physical, case of
    two volume spaces \label{brane-fig:space}}\vspace*{-3mm}
\end{figure}

Here, ${\mathcal R}_i$ is the scalar curvature of the
five-dimensional metric $g^i_{ab}$ on ${\mathcal B}_i$, $i = 1,
\mbox{...}, N$, the $N$ bulk spaces for which the brane is a
boundary (see the left panel in Fig.~4.\ref{brane-fig:space}), and
$R$ is the scalar curvature of the induced metric $h_{ab}$ on the
brane. The quantity $K_i = K^i_{ab} h^{ab}$ is the trace of the
symmetric tensor of extrinsic curvature $K^i_{ab}$ of the brane in
the space ${\mathcal B}_i$. The symbol $L (h_{ab}, \phi)$ denotes
the Lagrangian density of the four-dimensional matter fields $\phi$
the dynamics of which is restricted to the brane so that they
interact only with the induced metric $h_{ab}$.  All integrations
over ${\mathcal B}_i$ and over the brane are taken with the
corresponding natural volume elements. The symbols $M_i$, $i =
1,\mbox{...},N,$ and $m$ denote the Planck masses of the
corresponding spaces, $\Lambda_i$, $i = 1,\mbox{...},N,$ are the
five-dimensional cosmological constants on each side of the brane,
and $\sigma$ is the brane tension.

Action (\ref{brane-action}) can be regarded as the simplest local
action for gravity in the braneworld setup under consideration.
Indeed, it is the lowest-order non-trivial action in the number of
derivatives acting on the metric. The braneworld model described by
(\ref{brane-action}) can be classified according to the number of
the bulk spaces $N$ for which the brane is a boundary, and according
to the values taken by the fundamental constants $M_i$, $\Lambda_i$,
$m$, and $\sigma$.  We do it as follows:

{\footnotesize$\bullet$}\,\,The case $N = 1$ is the simplest one and
most frequently
  discussed in the literature.  It described a brane which is a
  boundary of a five-dimensional bulk space.  Alternatively, it can be
  obtained from the physically more natural case $N = 2$ (see the
  right panel in Fig.~4.\ref{brane-fig:space}) by making identification
  between the two sides of the brane, or by imposing the so-called
  mirror symmetry of reflection of the bulk space with respect to the
  brane.  The cases with $N \geq 2$ need not have this mirror
  symmetry.

  In most of this chapter, we consider the simple case $N = 1$ (or
  brane with mirror symmetry). The case of $N = 2$ (without the mirror
  symmetry) will be considered in Sec.~\ref{brane-sec:asym}.

{\footnotesize$\bullet$}\,\,A special case where $m = 0$ will be
referred to as the
  Randall---Sundrum (RS) type of model since it is this form of action
  that was employed in the original papers \cite{RS1,RS2}.

{\footnotesize$\bullet$}\,\,A more general case where $m \ne 0$ will
be referred to as
  models with induced gravity on the brane, and a special case of such
  models with vanishing cosmological constants on the brane and in
  the %
  \index{cosmological constant|)} %
  bulk, i.e., $\sigma = 0$ and $\Lambda_i = 0$, will be referred to as
  the Dvali---Gabadadze---Porrati (DGP) type of model owing to its first
  appearance in \cite{DGP}.

Variation of action (\ref{brane-action}) with respect to the metric
$g_{ab}$ gives rise to the Einstein equations in each of the $i$
bulk spaces: \index{Einstein equations}\vspace*{-2mm}
\begin{equation}
  \label{brane-bulk} \mathcal{G}_{ab} + \Lambda_i g_{ab} = 0
\end{equation}
and the junction conditions on the brane
\begin{equation}
  \label{brane-brane}
  m^2 G_{ab} + \sigma h_{ab} = \sum_i M_i^3 \left( K_{ab}^i - h_{ab} K_i \right) + T_{ab} ,
\end{equation}
where $G_{ab}$ and $\mathcal{G}_{ab}$ are the Einstein tensors on
the brane and in the bulk, respectively, and $T_{ab}$ is the
stress-energy tensor on the brane~--- the variation of the matter
action with respect to $h_{ab}$:
\begin{equation}
  T_{ab} = \frac{1}{\sqrt{h}} \frac{\delta}{\delta h^{ab}} \int\limits_\mathrm{ brane} L \left(
    h_{ab}, \phi \right)\!.
\end{equation}
The York---Gibbons---Hawking \cite{zhYork,zhGH} boundary terms $- 2
\int_\mathrm{ brane} K_i $ are required for consistency of this
variational problem.  Treatment of the variational problem in
(\ref{brane-action}) with the presence of boundary can be found in
Appendix~B.1.

\section{\!Cosmological solutions} \label{brane-sec:cosmo}

\hspace*{3cm}\index{FRW metric}A brane with FRW metric
\begin{equation}
  \label{brane-FRW}
  ds^2 = - dt^2 + a^2(t) \gamma_{ij} (x) dx^i dx^j ,
\end{equation}
where $\gamma_{ij} (x)$ is the Euclidean metric of the homogeneous
and isotropic three-space, is embedded into the
Schwarzschild---(anti)-de~Sitter solution of the bulk equations
(\ref{brane-bulk}):
\begin{equation}
  \label{brane-AdS} ds_\mathrm{ bulk}^2 = - f(r)
  d\tau^2 + \frac{dr^2}{f(r)} + r^2 \gamma_{ij} (x) dx^i dx^j ,
\end{equation}
where\,\footnote{\,The variables $(\tau, r)$ denote the coordinates
of the
  bulk space in this section.  Where only one bulk space is under
  consideration, we omit the index $i$ and denote the bulk
  cosmological constant by $\Lambda_\mathrm{b}$.}
\begin{equation}
  \label{brane-fR}
  f(r) = \kappa - \frac{C}{r^2} -
  \frac{\Lambda_\mathrm{b}}{6} r^2,
\end{equation}
$\kappa = 0, \pm 1$ in this chapter denotes the spatial curvature of
the metric $\gamma_{ij}$, and $C$ is an arbitrary constant of
integration describing the presence of a black hole in the bulk. The
embedding is described by the trajectory
\begin{equation}
  \label{brane-trajec}
  r = a (\tau) ,
\end{equation}
which is the scale factor of the FRW metric (\ref{brane-FRW}).  The
cosmological time $t$ on the brane in (\ref{brane-FRW}) is then
related to the time $\tau$ in the bulk by
\begin{equation}
\label{brane-propert} \frac{dt}{d\tau} = \sqrt{ f \left( a(\tau)
\right) -
  \frac{a'(\tau)}{f\left(a(\tau)\right)} } .
\end{equation}

The coordinate $r$ is related to the Gaussian normal coordinate $y$
in the bulk by
\begin{equation}
  y = \int\limits^r \frac{dx}{\sqrt{f(x)}} .
\end{equation}

The extrinsic curvature of the hy\-per\-sur\-fa\-ce
(\ref{brane-trajec}) with unit normal $n^a$ pointing in the
direction of increasing $r$ is calculated straightforwardly, so
that, in the $t, x^i$ coordinates on the brane, we have
\begin{equation}
  \label{brane-extr}
  K^\alpha{}_\beta -
  h^\alpha{}_\beta K = \left\{\! - \frac{3 \sqrt{ f(a) - \dot a^2}}{a} \,
    , \, \, - \delta^i{}_j \frac{1}{a^2 \dot a} \frac{d}{dt} \left(\! a^2
      \sqrt{ f(a) + \dot a^2} \right)\! \right\}\!,
\end{equation}
where the dot denotes the derivative with respect to $t$.

After embedding the brane in the bulk via (\ref{brane-trajec}), we
can make it a boundary by leaving only one of the two sides of the
bulk solution, $r > a(\tau)$ or $r < a(\tau)$. This gives rise to
one of two possible branches in each of the bulk spaces
$\mathcal{B}_i$, $i = 1, \mbox{...}, N$.  Using this fact and
substituting (\ref{brane-extr}) into the brane equation
(\ref{brane-brane}), we obtain the following equation for the
cosmological evolution of the scale factor $a(t)$:
\begin{equation}
\label{brane-cosmol} H^2 + \frac{\kappa}{a^2} = \frac{\rho +
\sigma}{3 m^2} + \frac{1}{m^2} \sum_{i = 1}^N \zeta^{}_i M_i^3
\sqrt{H^2 + \frac{\kappa}{a^2} -
  \frac{\Lambda_i}{6} - \frac{C_i}{a^4} } ,
\end{equation}
where $\rho$ is the total energy density of matter on the brane, and
$\zeta_i = \pm 1$ correspond to the two possible ways of bounding
each of the bulk spaces $\mathcal{B}_i$, $i = 1, \mbox{...}, N$, by
the brane \cite{CH,Shtanov1,Deffayet}.  The branch with the bulk
space $r > a(\tau)$ corresponds to $\zeta = 1$. All the richness of
the braneworld homogeneous and isotropic cosmology is encoded in
equation (\ref{brane-cosmol}).

The simplest case in braneworld cosmology is, of course, $N = 1$ in
(\ref{brane-action}), where the brane is a boundary of a single bulk
space.  It is equivalent to the case of arbitrary $N$ with identical
bulk spaces (as was noted in Sec.~\ref{brane-sec:setup}, in the case
$N = 2$, this is called mirror symmetry of reflection of the bulk
space with respect to the brane).

For $N = 1$, omitting the index $i$, we have the cosmological
equation on the brane
\begin{equation}
  \label{brane-cosmol1}
  H^2 + \frac{\kappa}{a^2} =
  \frac{\rho + \sigma}{3 m^2} \pm \frac{M^3}{m^2} \sqrt{H^2 +
    \frac{\kappa}{a^2} - \frac{\Lambda_\mathrm{b}}{6} - \frac{C}{a^4}
  } ,
\end{equation}
where the $\pm$ sign corresponds to the two branches arising from
two possible ways of bounding the bulk space by the brane, as
described in the previous section.

\index{Randall---Sundrum model} %
The cosmological solution based on the Randall---Sundrum model
\cite{RS1,RS2}, which has $m = 0$, is obtained as the lower branch
of (\ref{brane-cosmol1}) in this limit [511---514]:\vspace*{-3mm}
\begin{equation}
  \label{brane-cosmolim}
  H^2 + \frac{\kappa}{a^2} = \left( \!\frac{\rho + \sigma}{3 M^3} \!\right)^{\!\!2} + \frac{\Lambda_\mathrm{b}}{6} + \frac{C}{a^4} .
\end{equation}
Since the last term enters the modified Friedmann equation and has
de\-pen\-den\-ce on the scale factor exactly like radiation, it was
called ``dark radiation'' in the literature. \index{dark
radiation}The name extends to the term $C / a^4$ under the square
\mbox{root of (\ref{brane-cosmol1}).}

The cosmological solution based on the DGP model \cite{DGP} is
obtained by setting $\sigma = 0$, $C = 0$ and $\Lambda_\mathrm{b} =
0$ in (\ref{brane-cosmol1}) \cite{DDG,DLRZA}:\vspace*{-1mm}
\begin{equation}
\label{brane-DGP} H^2 + \frac{\kappa}{a^2} = \frac{\rho}{3 m^2} \pm
\frac{M^3}{m^2} \sqrt{H^2 +
  \frac{\kappa}{a^2}} .
\end{equation}\vspace*{-4mm}

\noindent A remarkable property of this model is that it can
describe an
accelerating Universe even in the absence of cosmological constants %
\index{cosmological constant} %
either on the brane or in the bulk. Indeed, the branch with the
upper sign in (\ref{brane-DGP}) in the asymptotic future tends to
the
de~Sitter regime with the Hubble constant %
\index{Hubble constant}\vspace*{-1mm}
\begin{equation}
  H_{\mathrm{DGP}} = \frac{M^3}{m^2} .
\end{equation}\vspace*{-5mm}

This is the reason why it received the name ``self-accelerating
branch'' in the literature.  This simple alternative model of the
cosmic acceleration at present appears to be ruled out both
observationally and theoretically, since it has a ghost perturbation
in its spectrum [517---521]. The branch with the lower sign in
(\ref{brane-DGP}) can self-accelerate only in the presence of
cosmological constants; it is called ``normal branch'' in the
literature.  The ``normal branch'' is free from ghosts and is in
agreement with all current observations.

The case of $N = 1$ can be treated in an alternative way.  By
contracting the Gauss identity\vspace*{-2mm}
\begin{equation}
  R_{abc}{}^d = h_a{}^f h_b{}^g h_c{}^k h^d{}_j \mathcal{R}_{fgk}{}^j + K_{ac}
  K_b{}^d - K_{bc} K_a{}^d
\end{equation}\vspace*{-5mm}

\noindent on the brane and using Eq.~(\ref{brane-bulk}), one obtains
the constraint equation\vspace*{-1mm}
\begin{equation} \label{brane-constraint0} R - 2 \Lambda_\mathrm{b} +
  K_{ab} K^{ab} - K^2 = 0 ,
\end{equation}\vspace*{-5mm}

\noindent which, together with (\ref{brane-brane}) taken in the case
$N = 1$, implies the following closed scalar equation on the
brane:\vspace*{-2mm}
\[
M^6 \left( R - 2 \Lambda_\mathrm{b} \right) + \left( m^2 G_{ab} +
    \sigma h_{ab} -
    T_{ab} \right) \left( m^2 G^{ab} + \sigma h^{ab} - T^{ab}
    \right)-
\]\vspace*{-5mm}
\begin{equation}
  \label{brane-closed0}
   - \frac{1}{3} \left( m^2 R - 4 \sigma + T \right)^2 = 0,
\end{equation}\vspace*{-5mm}

\noindent where $T = h^{ab} T_{ab}$.

One method for obtaining solutions of the theory consists in first
\mbox{solving} the scalar equation (\ref{brane-closed0}) on the
brane together with the stress-energy conservation equation, and
then integrating the Einstein equations in the bulk with the given
data on \index{Einstein equations|(}the brane [522---524].  The
gravitational equations in the bulk can be integrated by using, for
example, Gaussian normal coordinates. Specifically, in the Gaussian
normal coordinates $(x, y)$, where $x = \{x^\alpha\}$ are the
coordinates on the brane and $y$ is the fifth coordinate in the
bulk, the metric in the bulk is written as
\begin{equation}
  d s^2_5 = dy^2 + h_{\alpha\beta} (x, y) dx^\alpha dx^\beta \!.
\end{equation} %
\index{tensor of extrinsic curvature} %
Introducing also the tensor of extrinsic curvature $K_{ab}$ of every
hy\-per\-sur\-fa\-ce $y = \mathrm{ const}$ in the bulk, one can
obtain the following system of differential equations for the
components $h_{\alpha\beta}$ and $K^\alpha{}_\beta$:
\[
\frac{\partial K^\alpha{}_\beta}{\partial y}  =  R^\alpha{}_\beta -
K K^\alpha{}_\beta -  \frac16 \delta^\alpha{}_\beta \left( R + 2
\Lambda_\mathrm{b} + K^\mu{}_\nu K^\nu{}_\mu -  K^2
  \right)=
\]\vspace*{-3mm}
\begin{equation}
 = R^\alpha{}_\beta - K K^\alpha{}_\beta - \frac23
  \delta^\alpha{}_\beta \Lambda_\mathrm{b} , \label{brane-s}
\end{equation}\vspace*{-3mm}
\begin{equation}
  \frac{\partial h_{\alpha\beta}}{\partial y}  = 2 h_{\alpha\gamma} K^\gamma{}_\beta , \label{brane-metric}
\end{equation}
where $R^\alpha{}_\beta$ are the components of the Ricci tensor %
\index{tensor Ricci} %
of the metric $h_{\alpha\beta}$ induced on the hy\-per\-sur\-fa\-ce
$y = \mathrm{ const}$, $R = R^\alpha{}_\alpha$ is its scalar
curvature, and $K = K^\alpha{}_\alpha$ is the trace of the tensor of
extrinsic curvature. The second equality in (\ref{brane-s}) is true
by virtue of the constraint equation (\ref{brane-constraint0}).
Equations (\ref{brane-s}) and (\ref{brane-metric}) together with the
constraint equation (\ref{brane-constraint0}) represent the $4 + 1$
splitting of the Einstein equations in the Gaussian normal
coordinates. The initial conditions for these equations are defined
on the brane through Eq.~(\ref{brane-brane}).  We emphasize that, to
obtain a complete braneworld theory in the general case (including a
stability analysis), one must also specify additional conditions in
the bulk such as the presence of other branes or certain regularity
conditions. In this section, we deal only with the homogeneous and
isotropic cosmology on the brane, so this issue does not arise. In
this sense, we are studying here the cosmological features common to
the whole class of braneworld models described by action
(\ref{brane-action}) with arbitrary boundary conditions in the bulk.

\section{\!Vacuum and static branes} \label{brane-sec:vacua}

\hspace*{3cm}\index{vacuum brane}In this section, we discuss the
situation pertaining to a va\-cuum brane, i.e., when the matter
stress-energy tensor $T_{ab} = 0$.  It is interesting that the brane
approaches this condition during the course of cosmological
evolution provided it expands forever and its matter density
asymptotically declines to zero. In the simplifying case of $N = 1$
(brane as a boundary of a single bulk space),
Eq.~(\ref{brane-closed0}) takes the form
\begin{equation}
  \label{brane-vacuum}
  \left(\!M^6 + \frac23 \sigma m^2
  \!\right) R + m^4 \left( \!R_{ab} R^{ab} - \frac13 R^2 \!\right) - 4 M^6
  \Lambda_\mathrm{ RS} = 0 ,
  \end{equation}\vspace*{-5mm}

\noindent where\vspace*{-3mm}
\begin{equation}
  \label{brane-LRS}
  \Lambda_\mathrm{ RS} = \frac{\Lambda_\mathrm{b}}{2} +
  \frac{\sigma^2}{3 M^6}
\end{equation}\vspace*{-2mm}

\noindent is the expression arising in the Randall---Sundrum model
\cite{RS1,RS2}.

It is important to note that the second term in
Eq.~(\ref{brane-vacuum}) has {\em precisely\/} the form of one of
the terms in the expression for the conformal anomaly, which
describes the vacuum polarization at the one-loop level in curved
space-time (see, e.g., \cite{zhBirrDav})\,\footnote{\,It is
interesting that, while the
  conformal anomaly term $R_{ab} R^{ab} - \dfrac13R^2$ cannot be
  obtained by the variation of a local four-dimensional Lagrangian,
  the very same term is obtained via the variation of a local
  Lagrangian in the five-dimensional braneworld theory under
  investigation.}.  It therefore immediately follows that all
\emph{symmetric spaces} are solutions of Eq.~(\ref{brane-vacuum})
with appropriate $\Lambda_{\mathrm{RS}}$, just as they are solutions
of the Einstein equations with one-loop quantum-gravitational
corrections \cite{SK}. Symmetric spaces satisfy the condition $D_a
R_{bcde} = 0$, which implies that geometrical invariants such as
$R_{abcd}R^{abcd}, R_{ab}R^{ab}$, and $R$ are constants so that
Eq.~(\ref{brane-vacuum}) becomes an algebraic equation. Prominent
members of this family include:

 {\footnotesize$\bullet$}\,\,the homogeneous and isotropic de~Sitter space-time
  \begin{equation}
   \label{brane-desitter}
   ds^2 = - dt^2 +
   \frac{1}{H^2} \cosh^2{Ht} \left[ d\chi^2 + \sin^2 \chi \left( d
       \theta^2 + \sin^2 \theta d \phi^2 \right)\right]\!,
  \end{equation}
  where $-\infty < t < \infty$, $0 \leq \chi,\theta \leq \pi$, $0 \leq \phi \leq
  2\pi$.  The four-dimensional metric (\ref{brane-desitter}) has the property
  $R^a{}_b = 3 H^2 h^a{}_b$.  It formed the basis for Starobinsky's first
  inflationary model sustained by the quantum conformal anomaly
  \cite{zhStar1980};

 {\footnotesize$\bullet$}\,\,the homogeneous and anisotropic Nariai metric
  \cite{nariai1,nariai2,kss} %
  \index{Nariai metric}
  \begin{equation}
    ds^2 = k^2 \left( - dt^2 + \cosh^2{t} dr^2 + d\theta^2 + \sin^2\theta d\phi^2
    \right)\!,
  \end{equation}
  where $k$ = constant, $-\infty < t < \infty$, $0 \leq \theta \leq
  \pi$, $0 \leq \phi \leq 2\pi$ and for which $R^a{}_b =
  h^a{}_b/k^2$. In fact, it is easy to show that any metric for which
  $R$ and $R_{ab} R^{ab}$ are constants will automatically be a
  solution to Eq.~(\ref{brane-vacuum}) with an appropriate choice of
  $\Lambda_\mathrm{ RS}$.

Both de~Sitter space and the Nariai metric belong to the class of
space-times which satisfy the vacuum Einstein equations with a
cosmological constant\index{Einstein equations|)}\index{cosmological
constant}\vspace*{-2mm}
\begin{equation} \label{brane-ricci} R_{ab} = \Lambda_\mathrm{ eff}
  h_{ab} .
\end{equation}
Such space-times also satisfy Eq.~(\ref{brane-vacuum}) if
\begin{equation}
  \label{brane-lambdaeff}
  \Lambda_{\mathrm{eff}} = \frac{1}{m^2} \left[ \!\left(\! \frac{3 M^6}{2 m^2} + \sigma \!\right) \pm
    \sqrt{ \left(\! \frac{3 M^6}{2 m^2} + \sigma \!\right)^{\!\!2} - 3 M^6
      \Lambda_{\mathrm{RS}}}\, \right] \!.
\end{equation}
Equation (\ref{brane-lambdaeff}) expresses the resulting
cosmological constant on the brane in terms of the coupling
constants of the theory. For the Randall---Sundrum model ($m = 0$),
one obtains $\Lambda_{\mathrm{eff}} = \Lambda_{\mathrm{RS}}$. The
two signs in (\ref{brane-lambdaeff}) again correspond to the two
different ways in which the lower-dimensional brane can form the
boundary of the higher-dimensional bulk.

The condition $\Lambda_{\mathrm{RS}} = 0$ is the well-known
fine-tuning condition of Randall and Sundrum \cite{RS1,RS2} and
leads to the vanishing of the cosmological constant on an empty
brane if we set $m=0$ in (\ref{brane-action}). Note that, under the
Randall---Sundrum condition, expression (\ref{brane-lambdaeff}) with
the sign opposite to the sign of the quantity $3 M^6 / 2 m^2 +
\sigma$ also gives a zero value for the resulting cosmological
constant on the brane, but the other sign usually leads to
$\Lambda_{\mathrm{eff}} \neq 0$.

We would like to draw the reader's attention to the fact that
Eq.~(\ref{brane-lambdaeff}) is meaningful only when the expression
under the square root is non-negative. When it is negative,
solutions describing the corresponding empty Universe simply do not
exist. This leads to the following important conclusion: a Universe
which contains matter and satisfies
\begin{equation}
  \frac{3 M^6 \Lambda_\mathrm{ RS}}{\left( 3 M^6 / 2
      m^2 + \sigma \right)^2 }  > 1 ,
\end{equation}
\emph{cannot expand forever}.

For the special case $3 M^6 / 2 m^2 + \sigma = 0$, the expression
for $\Lambda_{\mathrm{eff}}$ on the brane is given by
\begin{equation}
  \Lambda_{\mathrm{eff}} = \pm \frac{M^3}{m^2} \sqrt{- 3 \Lambda_{\mathrm{RS}}} .
\end{equation}
In this case, both $\sigma$ and $\Lambda_{\mathrm{RS}}$ must be
negative in order that the corresponding empty Universe exist, but
the resulting cosmological constant on the brane can be of any sign.

Another interesting example is that of a static empty Universe. The
radius (scale factor) $a$ of such a Universe is easily determined
from (\ref{brane-vacuum}) to be
\begin{equation}
  \label{brane-static}
  a^2 = \frac{\kappa}{\Lambda_{\mathrm{RS}} } \left(\! \frac32 + \frac{\sigma m^2}{M^6}
 \! \right)\! ,
\end{equation}
where $\kappa = \pm 1$ is the sign of the spatial curvature.  One
can see that the radius of the Universe can be arbitrarily large. In
the general case, the development of this solution to the
five-dimensional bulk leads to a Schwarzschild--anti-de~Sitter
metric. It was shown in \cite{Shtanov1} that, for $\kappa = 1$, this
metric is purely anti-de~Sitter (with zero Schwarzschild mass) if
the constants of the theory satisfy the condition\vspace*{-1mm}
\begin{equation} \frac{\sigma}{m^2} - \frac{\Lambda_\mathrm{b}}{2} +
  \frac{3 M^6}{4 m^4} = 0 ,
\end{equation}
which implies negative brane tension $\sigma$. It should be pointed
out that the static and empty braneworld solution described by
(\ref{brane-static}) does not possess a general-relativistic analog,
since, in General Relativity, a static cosmological model (the
`static Einstein Universe') {\em cannot\/} be empty (see, for
instance, \cite{Sahni2000}). Furthermore, from (\ref{brane-static})
we find that the static empty Universe can be spatially open
($\kappa = -1$)~--- for example, in the case $\Lambda_\mathrm{ RS} <
0$ and $\sigma > - 3 M^6 / 2 m^2$,~--- again a situation without an
analog in General Relativity.

For static homogeneous and isotropic braneworlds filled with
matter,\linebreak Eq.~(\ref{brane-closed0}) gives the following
relation:
\begin{equation}
  a^2 \left[ \rho_\mathrm{ tot} (\rho_\mathrm{ tot} + 3 p_\mathrm{ tot}) - 3 \Lambda_\mathrm{b}
    M^6 \right] = 3 \kappa \left[ m^2 (\rho_\mathrm{ tot} + 3 p_\mathrm{ tot}) - 3 M^6
  \right]\! ,
\end{equation}
where the total energy density $\rho_\mathrm{ tot}$ and pressure
$p_\mathrm{ tot}$ include the contribution from the brane tension,
i.e.,\vspace*{-1mm}
\begin{equation}
  \rho_\mathrm{ tot} = \rho + \sigma , \quad p_\mathrm{ tot} = p - \sigma ,
\end{equation}
and $\kappa = 0, \pm1$ corresponds to the sign of the spatial
curvature.  This relation reduces to (\ref{brane-static}) for $\rho
= p = 0$.

Having obtained all these solutions on the brane, one can find the
cor\-res\-pon\-ding solutions in the bulk by integrating
Eqs.~(\ref{brane-s}) and (\ref{brane-metric}) with the initial
conditions on the brane given by Eq.~(\ref{brane-brane}).  In doing
this, one can consider various additional conditions in the bulk,
for example, the existence of other branes, or one can impose
certain regularity conditions.  It is worth noting that one and the
same cosmological solution on the given brane can correspond to
different global solutions in the bulk, for example, other branes
may be present or absent, be static or evolving, close or far away
from our brane, etc.  In the most general case (for instance in the
absence of special symmetries on the brane) integration on the brane
needs to be performed in conjunction with dynamical integration in
the bulk. All such situations must be separately studied and issues
such as their stability to linearized perturbations must be examined
on a case-by-case basis.

Consider any solution to (\ref{brane-ricci}) on the brane with
effective cosmological constant $\Lambda_\mathrm{ eff}$ given by
(\ref{brane-lambdaeff}).  It is obvious that the solution of system
(\ref{brane-s}), (\ref{brane-metric}), describing the metric in the
Gaussian normal coordinates $y \geq 0$ with the brane situated at $y
= 0$, can be sought for in the form
\begin{equation}
  h_{\alpha\beta} (x, y) = f(y) h_{\alpha\beta} (x, 0) , \quad  K^\alpha{}_\beta =
  \frac14 K (y) \delta^\alpha{}_\beta ,
\end{equation}
with $f (0) = 1$.  For this metric, we have
\begin{equation}
  R^\alpha{}_\beta (x, y) = \frac{\Lambda_\mathrm{ eff}}{f (y)} \delta^\alpha{}_\beta ,~~ G^\alpha{}_\beta (x, y) = - \frac{\Lambda_\mathrm{ eff}}{f (y)} \delta^\alpha{}_\beta
  .
\end{equation}
The brane junction condition (\ref{brane-brane}) for the case $N =
1$ under consideration then gives the initial condition for $K (y)$:
\begin{equation}
K (0) = \frac{4}{3 M^3} \left( m^2 \Lambda_\mathrm{ eff} - \sigma
\right) \!.
\end{equation}

Equations (\ref{brane-s}), (\ref{brane-metric}) then lead to a
system of differential equations for $K (y)$ and $f (y)$:
\begin{equation}
  K' =  \frac{4 \Lambda_\mathrm{ eff}}{f} - K^2 - \frac83 \Lambda_\mathrm{b} ,
\end{equation}\vspace*{-3mm}
\begin{equation}
  f' = \frac12 K f .
\end{equation}

In the simplifying case of a Ricci-flat brane with $\Lambda_\mathrm{
  eff} = \Lambda_\mathrm{ RS} = 0$, we have $K ^2 (0) = - 8
\Lambda_\mathrm{b} / 3$, and the solution is
\begin{equation}
  K (y)
  \equiv K(0) = - \frac{ 4 \sigma}{3 M^3} ,
\end{equation}\vspace*{-3mm}
\begin{equation} f (y) = \exp
  \left(\! - \frac{2 \sigma}{3 M^3} y \!\right) = \exp \left(\!  \pm \sqrt{
      - \frac23 \Lambda_\mathrm{b}}\, y \!\right)\! ,
\end{equation}
where the sign in the last exponent depends on the sign of the brane
tension.  For the flat metric $h_{\alpha\beta} = \eta_{\alpha\beta}$
on the brane, this gives the famous Randall---Sundrum solution
\cite{RS1,RS2}.

\section{\!Properties of braneworld gravity}\label{brane-sec:gravity}

\hspace{3cm}In this section, we discuss some generic properties of
bra\-ne\-world gravity. We will restrict ourselves to the simplest
case $N = 1$ in (\ref{brane-action}).

Action (\ref{brane-action}) in the case $N = 1$ has two important
scales, namely, the length scale\vspace*{-3mm}
\begin{equation}
  \label{brane-ell}
  \ell = \frac{2 m^2}{M^3} ,
\end{equation}
which describes the interplay between the bulk and brane gravity,
and the energy scale\vspace*{-3mm}
\begin{equation}
  \label{brane-k}
  k_\sigma = \frac{\sigma}{3 M^3} ,
\end{equation}
which determines the role of the brane tension in the dynamics of
the brane. In a model characterized by the Randall---Sundrum
constraint \cite{RS1,RS2}
\begin{equation}
  \label{brane-RS}
  \Lambda_\mathrm{ RS} \equiv \frac{\Lambda_\mathrm{b}}{2} + \frac{\sigma^2}{3 M^6} = 0 ,
\end{equation}
the absolute value of $k_\sigma$ is equal to the inverse curvature
$\ell_{\Lambda_\mathrm{b}} = \sqrt{- 6 / \Lambda_\mathrm{b}}$ of the
bulk space. Note that $k_\sigma$ is negative in the case $\sigma <
0$. The quantity $M$ has to be taken positive since, in the opposite
case, the massive Kaluza---Klein gravitons become ghosts
\cite{Padilla,SVi}.

Following the procedure first employed in \cite{SMS} for the
Randall---Sundrum (RS) model \cite{RS1,RS2} and subsequently applied
in \cite{MMT} to the more general model under consideration, we make
one contraction of indices in the Gauss identity
\begin{equation}\label{brane-Gauss}
  R_{abc}{}^d = h_a{}^f h_b{}^g h_c{}^k h^d{}_j \mathcal{R}_{fgk}{}^j + K_{ac}
  K_b{}^d - K_{bc} K_a{}^d
\end{equation}
on the brane and, using Eq.~(\ref{brane-bulk}), obtain the equation
\begin{equation}
\label{brane-effective}
  G_{ab} + \Lambda_\mathrm{ eff} h_{ab} = 8 \pi G_\mathrm{ eff} T_{ab} + \frac{1}{
    \beta + 1} \left(\! \frac{1}{M^6} Q_{ab} - C_{ab} \!\right)\!,
\end{equation}\vspace*{-5mm}

\noindent where\vspace*{-3mm}
\begin{equation}
  \label{brane-beta}
  \beta = \frac{2 \sigma m^2}{3 M^6} = k_\sigma \ell
\end{equation}\vspace*{-5mm}

\noindent is a dimensionless parameter,\vspace*{-3mm}
\begin{equation}
  \label{brane-lambda-eff}
  \Lambda_\mathrm{ eff} = \frac{\Lambda_\mathrm{ RS} }{ \beta + 1}
\end{equation}\vspace*{-5mm}

\noindent is the effective cosmological constant,\vspace*{-3mm}
\begin{equation}
 \label{brane-g-eff}
 8 \pi G_\mathrm{ eff} = \frac{\beta }{ \beta + 1} \frac{1 }{
   m^2}
\end{equation}\vspace*{-5mm}

\noindent is the effective gravitational constant,\vspace*{-3mm}
\begin{equation}
\label{brane-q} Q_{ab} = \frac13 E E_{ab} - E_{ac} E^{c}{}_b +
\frac12 \left(\!E_{cd}
  E^{cd} - \frac13 E^2 \!\right) h_{ab}
\end{equation}
is a quadratic expression with respect to the ``bare'' Einstein
equation $E_{ab} \equiv $ $\equiv m^2 G_{ab} - T_{ab}$ on the brane,
and $E = h^{ab} E_{ab}$. The symmetric traceless tensor $C_{ab}
\equiv n^c n^d C_{acbd}$ in (\ref{brane-effective}) is a projection
of the bulk Weyl tensor $C_{abcd}$.  It is related to the tensor
$Q_{ab}$ through the covariant conservation equation on the brane
\begin{equation}\label{brane-conserv}
  D_a \left( Q^a{}_b - M^6 C^a{}_b \right) = 0 .
\end{equation}

It is important to note that all couplings in
Eq.~(\ref{brane-effective}), including the effective cosmological and %
\index{cosmological constant} %
gravitational constants, are inversely proportional to $\beta + 1$,
which indicates that the theory becomes singular for the special
case $\beta = - 1$ (see \cite{SVi,Smolyakov}).  We are not going to
study this degenerate case here.

In the absence of the curvature term on the brane ($m = 0$), we
obtain Eq.~(\ref{brane-effective}) in which $8 \pi G_\mathrm{ eff} =
2 \sigma / 3 M^6$ is the gravitational constant in the RS model
\cite{RS1,RS2}, and $\beta = 0$; in this form,
Eq.~(\ref{brane-effective}) was first derived in \cite{SMS}. The
conditions $\sigma = 0$ and $\Lambda_\mathrm{b} = 0$ are
characteristic of the DGP model \cite{DGP}, which also has $\beta =
0$. In this model, the effective gravitational constant
(\ref{brane-g-eff}) turns to zero, i.e., the term linear in the
stress--energy tensor on the brane vanishes in
Eq.~(\ref{brane-effective}).

Equation (\ref{brane-effective}) is not closed on the brane in the
sense that it contains the symmetric traceless tensor $C_{ab}$ whose
dynamics on the brane is not determined by the dynamics of matter
alone.  Some additional information from the bulk is needed to solve
the braneworld equations completely, e.g., some boundary conditions
in the bulk are to be specified.  However, for the homogeneous and
isotropic cosmology, this ambiguity manifests itself only in the
appearance of the dark-radiation term $C / a^4$ in
(\ref{brane-cosmol1}), characterized by one constant.

Consider now some properties of braneworld gravity. The expression
for $Q_{ab}$ in Eq.~(\ref{brane-q}) is quadratic in the curvature as
well as in the stress-energy tensor. On the other hand, the tensor
$C_{ab}$ is related to $Q_{ab}$ through the conservation equation
(\ref{brane-conserv}). One might, therefore, expect that the term in
the parentheses on the right-hand side of
Eq.~(\ref{brane-effective}), namely $Q_{ab}/M^6 - C_{ab}$, will be
insignificant on sufficiently large length scales, and that the
braneworld theory on those scales should reduce to Einstein gravity
with the effective constants given by (\ref{brane-lambda-eff}) and
(\ref{brane-g-eff}). This expectation is borne out by a detailed
analysis \cite{SVi} carried out for a positive-tension brane
($\sigma > 0$) in the specific case when the braneworld satisfies
the RS constraint (\ref{brane-RS}). In this case, the gravitational
potential of a unit mass on large scales (on the positive-tension
brane) has the Newtonian form with a small RS correction \cite{SVi}:
\begin{equation}
  \label{brane-large-scale}
  V(r) = - \frac{G_\mathrm{ eff} }{ r} \left[ 1 + \frac{2 }{ 3 (\beta + 1) (k_\sigma r)^2}
  \right]\!, \, k_\sigma r \gg 1 ,
\end{equation}
where $G_\mathrm{ eff}$ is given by (\ref{brane-g-eff}).

On smaller spatial scales, $k_\sigma r \ll 1, \beta$, the potential
in linear theory again has the Newtonian form with a small
logarithmic correction:
\begin{equation}
  \label{brane-small-scale}
  V (r) = - \frac{\widetilde G_\mathrm{ eff} }{ r} - \left(\!\frac{15 }{ 8} +
    \frac{2}{\beta} \!\right) \frac{k_\sigma }{ 3 \pi^2 m^2} \log \left[\!\left(\!\frac{15 }{ 8} +
      \frac2\beta \!\right) k_\sigma r \right]\!, \, k_\sigma r \ll 1, \beta ,
\end{equation}
but with a different expression for the effective gravitational
constant \cite{SVi}
\begin{equation}
  \label{brane-g-tilde}
  \widetilde G_\mathrm{ eff} = \left[ 1 + \frac{1 }{ 3 ( 1 + \beta )} \right] \frac{1 }{
    8\pi m^2} = \left(\!1 + \frac{4 }{ 3 \beta} \!\right)  G_\mathrm{ eff} .
\end{equation}
For $k_\sigma \to 0$ (hence, also $\beta \to 0$), this reproduces
the result obtained for the DGP model in
\cite{Gruzinov,Porrati,DGZal} on scales $r \ll \ell$.

It is worth noting that gravity on these smaller scales $k_\sigma r
\ll 1, \beta$, in principle, involves the massless scalar radion,
i.e., it is of scalar--tensor type. As a consequence, for the
spherically symmetric solution, it violates the property $h_{00} (r)
= {} - h_{rr}^{-1} (r) $, or, in the linear approximation,
$\gamma_{00} (r) = \gamma_{rr} (r)$, where $\gamma_{\alpha\beta}
(r)$ are the components of metric perturbation in the spherically
symmetric coordinate system. Specifically, in the model with the RS
constraint $\Lambda_\mathrm{ RS} = 0$, one can obtain the relation:
\begin{equation}
  \label{brane-discrep}
  \frac{\Delta \gamma }{ \gamma_{00}} \equiv \frac{\gamma_{00} - \gamma_{rr} }{
    \gamma_{00}} = \frac{1 }{ 1 + 3 \beta / 4} .
\end{equation}
Since there are stringent experimental upper bounds
\cite{zhBertotti,zhWill} on the left-hand side of
(\ref{brane-discrep}) in the neighborhood of the solar system (it
should not exceed $10^{-5}$ by order of magnitude), if solution
(\ref{brane-small-scale}) were applicable in this domain, it would
imply that only very large values of $\beta$ are permissible in the
braneworld theory under consideration [namely, the braneworld model
(\ref{brane-action}) with $N = 1$ and with the RS constraint
(\ref{brane-RS})].

We should stress, however, that the applicability region of the
linear approximation (\ref{brane-small-scale}) is bounded from below
by a length scale which depends upon the mass of the central source,
as has been demonstrated for the DGP model in [533---535].
Specifically, the dynamics of the radion develops strong non-linear
corrections on sufficiently small scales, leading to the breakdown
of linearized theory. (This also creates the so-called
strong-coupling problem in the DGP model \cite{LPR,Rubakov}.) In
this case, in order to study gravity at small distances from the
source, one should turn to the fully non-linear theory.

To determine the distances at which the linearized theory breaks
down and to establish the correct behavior of the potential on such
scales, we turn to the effective equation (\ref{brane-effective}).
Taking the trace of Eq.~(\ref{brane-effective}), we get the
following closed scalar equation on the brane:
\begin{equation}
  \label{brane-trace}
  {} - R + 4 \Lambda_\mathrm{ eff} - 8 \pi G_\mathrm{ eff} T = \frac{Q }{ (\beta + 1)
    M^6} ,
\end{equation}
where the left-hand side contains terms which are linear in the
curvature and in the stress--energy tensor while the right-hand side
contains the quadratic term $Q = h^{ab} Q_{ab}$.

Suppose that we are interested in the behavior of gravity in the
neigh\-bor\-hood of a spherically symmetric source with density
$\rho_s$, total mass $\mathcal{M}_s$, and radius $r_s$.  First of
all, we assume that one can neglect the tensor $C_{ab}$ and the
effective
cosmological constant in the neighborhood of the source. %
\index{cosmological constant} %
As regards the effective cosmological constant, this assumption is
natural since the observed cosmological constant is small.
Concerning the tensor $C_{ab}$, its smallness in the neighborhood of
the source represents some additional condition on the spherically
symmetric solution. A condition of this sort is likely to arise in
any consistent and viable braneworld theory as, without it, one has
a large number of spherically symmetric solutions on the brane, many
of them non-physical (see \cite{VW} for a comprehensive treatment in
the framework of the RS model).

Within the source itself, we have two qualitatively different
options: an approximate solution can be sought either neglecting the
quadratic part or linear part of Eqs.~(\ref{brane-effective}) and
(\ref{brane-trace}).  We should choose the option that gives smaller
error of approximation in Eq.~(\ref{brane-trace}). In the first
case, neglecting the quadratic part and the effective cosmological
constant, we have\vspace*{-1mm}
\begin{equation}\label{brane-lin}
G_{ab} - 8 \pi G_\mathrm{ eff} T_{ab} \approx 0  \Rightarrow \frac{Q
}{ (\beta + 1) M^6} \sim \frac{\rho_s^2 }{ (\beta + 1)^3 M^6} .
\end{equation}\vspace*{-3mm}

\noindent In the second case, we neglect the linear part, so
that\vspace*{-1mm}
\begin{equation}\label{brane-quadrat}
Q_{ab} \approx 0  \Rightarrow  E_{ab} \approx 0 \Rightarrow   R + 8
\pi G_\mathrm{ eff} T \sim \frac{\rho_s }{ (\beta + 1) m^2 } .
\end{equation}\vspace*{-3mm}

The final expression on the right-hand side of (\ref{brane-quadrat})
is smaller than the corresponding expression in (\ref{brane-lin})
if\vspace*{-1mm}
\begin{equation}\label{brane-scale}
\rho_s > (\beta + 1)^2 \frac{M^6 }{ m^2}  \Rightarrow  r_s^3 < r_*^3
\sim \frac{\mathcal{M}_s \ell^2 }{ (\beta + 1)^2 m^2 } ,
\end{equation}\vspace*{-3mm}

\noindent where we have used the relation $\mathcal{M}_s \sim \rho_s
r_s^3$. Thus, we can expect that, in the neighborhood of the source,
on distances smaller than $r_*$ given by (\ref{brane-scale}), the
solution is determined mainly by the quadratic part $Q_{ab}$ in
Eq.~(\ref{brane-effective}), which means that it respects the
``bare'' Einstein equation $m^2 G_{ab} = T_{ab}$ to a high
precision.  This effect is sometimes described as the ``gravity
filter'' of the DGP model \cite{DGP}, which screens the scalar
graviton in the neighborhood of the source making the gravity
effectively Einsteinian. Some aspects of this interesting phenomenon
are discussed in \cite{Kaloper1,Kaloper2}.

Expression (\ref{brane-scale}) generalizes the length scale
[533---535] of the DGP model, below which non-linear effects become
important, to the case of non-zero brane tension
(non-zero $\beta$) and bulk cosmological constant satisfying the RS %
\index{cosmological constant} %
const\-raint (\ref{brane-RS}). The observable gravitational constant
on scales much smaller than $r_*$ is then given by\vspace*{-2mm}
\begin{equation}\label{brane-g-obs}
  8 \pi G_\mathrm{ obs} = \frac{1 }{ m^2} .
\end{equation}
For the Sun, the scale $r_*$ is estimated as\vspace*{-1mm}
\begin{equation}
  r_{*}^\odot \sim \frac{10^{16}\, \mbox{km} }{ (\beta + 1)^{2/3}\,
    \Omega_\ell^{1/3} } ,
\end{equation}\vspace*{-3mm}

\noindent where $\Omega_\ell = 1 / \ell^2 H_0^2$. For interesting
values of $\beta \sim 1$ and $\Omega_\ell \sim 1$, this distance
will be very large.  The corresponding radius for the Earth is
smaller only by two orders of magnitude.

This, however, is not the full story.  As argued in [541,\,542], the
gravitational potential of a spherically symmetric body on scales
$r_* \lesssim r \ll \ell$ is corrected by the cosmological
expansion. Moreover, the critical scale $r_*$ becomes dependent on
the
value of the Hubble parameter, %
\index{Hubble parameter} %
and can be different from (\ref{brane-scale}) for values of $\ell$
of the order of the Hubble length.  One also should note that
gravity on scales $r \ll r_*\,$, although close to Einstein gravity,
is not exactly Einsteinian, and these deviations can be used to test
braneworld theory on solar-system scales [533---535, 541---543].
Specifically, in the case of the DGP model ($m = 0$, $\sigma = 0$
and $\Lambda_\mathrm{b} = 0$), the authors of \cite{LS,LSS} obtain a
small correction to the locally static metric:
\begin{equation}
  ds^2 = - \left[ 1 + 2 n(r) \right] dt^2 + \left[ 1 + 2 a (r)
  \right] dr^2 + r^2 \left[ 1 + 2b(r) \right] d \Omega^2 .
\end{equation}\vspace*{-7mm}

\noindent Here,\vspace*{-3mm}
\begin{equation}
  a (r) = \frac{r_g}{2 r} \left[ 1 - \delta (r) \right] +
  \frac12 H^2 r^2 ,
\end{equation}
and $n (r)$ satisfies the differential equation
\begin{equation}
  r n' (r) = \frac{r_g}{2 r} \left[ 1 + \delta (r) \right] - H^2 r^2
  ,
\end{equation}\vspace*{-7mm}

\noindent where\vspace*{-1mm}
\begin{equation}\delta (r) = \frac{3 r^3}{\ell^2 r_g}
  \left( 1 \pm \ell H \right) \left[ \sqrt{ 1 + \frac{2 \ell^2 r_g}{9
        r^3 \left( 1 \pm \ell H \right)^2 }} - 1 \right]\! ,
\end{equation}
and $r_g = \mathcal{M} / 4 \pi m^2$ is the gravitational radius of a
central body.  For
\begin{equation}
r  \ll r_* = \left(\! \frac{r_g}{H^2} \!\right)^{\!\!1/3} \!,
\end{equation}\vspace*{-6mm}

\noindent one obtains\vspace*{-1mm}
\begin{equation}
  n (r) =  - \frac{r_g}{2 r} \pm \sqrt{\frac{2 r_g r}{\ell^2}} ,~~
  a (r) = \frac{r_g}{2 r} \mp \sqrt{\frac{r_g r}{4 \ell^2}} .
\end{equation}

The last terms in these expressions, in particular, lead to orbit
precession with constant rate\vspace*{-1mm}
\begin{equation}
  \frac{d}{dt} \Delta
  \phi_{\mathrm{DGP}} = {} \mp \frac{3}{4 \ell} = {} \mp 5~
  \frac{\mu \text{as}}{\text{year}}
\end{equation}%
for the value of $\ell \simeq 10\,\text{Gpc}$, which is a best-fit
value for the DGP cosmological model by supernovae Ia.

Although the preceding reasoning is applicable to both the
positive-tension and the negative-tension brane, the current
understanding of the braneworld gravitational physics supports only
the positive-tension case. From Eq.~(\ref{brane-effective}), one
might expect that a negative-tension brane will show reasonable
physical behavior in the case $|\beta| > 1$ (note that $\beta < 0$
for a negative-tension brane), in which the gravitational constant
(\ref{brane-g-eff}) is positive. However, direct calculation (along
the lines of \cite{SVi}) in the two-brane model with the RS
constraint (\ref{brane-RS}) shows that, in this case, the
gravitational interaction between material bodies on large scales is
dominated by the ghost-like radion, with the effective gravitational
coupling\vspace*{-3mm}
\begin{equation}
  \label{brane-g-radion}
  G_\mathrm{ radion} = {} - \frac13\, G_\mathrm{ eff} ,~~ k_\sigma r \gg 1 ,
\end{equation}
where $G_\mathrm{ eff}$ is given by the same expression
(\ref{brane-g-eff}).  The radion-dominated gravity on these scales
is formally attractive in the case $G_\mathrm{ eff} < 0$, and is
repulsive for $G_\mathrm{ eff} > 0$.  However, on smaller spatial
scales $k_\sigma r_* \lesssim k_\sigma r \ll 1, |\beta|$, Newton's
law similar to (\ref{brane-small-scale}) is reproduced with the
gravitational constant given by (\ref{brane-g-tilde}), which is
positive if $|\beta| > 4/3$. The gravity on these scales is of
scalar--tensor character. On still smaller distances from the
central source, $r < r_*$, the theory may approach Einstein gravity
with the effective gravitational constant (\ref{brane-g-obs}).

\section[\!Phantom property of
  braneworld dark energy]{\!Phantom property of
  braneworld dark energy} \label{brane-sec:phantom}

\hspace{3cm}\index{phantom dark energy}We proceed in this section to
a more detailed investigation of the specific features of braneworld
cosmology.  The first such interesting feature is the phantom-like
behavior of the dark energy in a braneworld, which is a generic
property of one of the branches in (\ref{brane-cosmol1}).  This
equation can be solved with respect to the Hubble
parameter:\vspace*{-1mm}
\begin{equation}
\label{brane-cosmol2} H^2 + \frac{\kappa }{ a^2} = \frac{\rho +
\sigma }{ 3 m^2} + \frac{2 }{ \ell^2} \left[1 \pm \sqrt{1 +
    \ell^2 \left(\!\frac{\rho + \sigma}{3 m^2} - \frac{\Lambda_\mathrm{ b}}{ 6}
      - \frac{C}{a^4} \!\right)} \right]\! ,
\end{equation}\vspace*{-5mm}

\noindent where $\ell$ is given by (\ref{brane-ell}).

\index{cosmological parameters} %
It is convenient to introduce the dimensionless cosmological
parameters\vspace*{-1mm}
\begin{equation}
\label{brane-omegas}
\begin{array}{c}
  \displaystyle\Omega_\mathrm{ m} =  \frac{\rho_0}{3 m^2 H_0^2} , \quad
  \Omega_\kappa  =  - \frac{\kappa}{a_0^2 H_0^2} , \quad
  \Omega_\sigma  =  \frac{\sigma}{3  m^2 H_0^2} ,  \\[5mm]
 \displaystyle \Omega_\ell =  \frac{1}{\ell^2 H_0^2} , \quad
  \Omega_{\Lambda_\mathrm{ b}}  = - \frac{\Lambda_\mathrm{ b}}{6
    H_0^2} , \quad  \Omega_C  = - \frac{C}{a_0^4 H_0^2} ,
\end{array}
\end{equation}
where the subscript ``{\small 0}'' refers to the current values of
cosmological quantities. The cosmological equation
(\ref{brane-cosmol2}) with the energy density $\rho$ dominated by
dust-like matter can now be written in the form:\vspace*{-1mm}
\[
\frac{H^2(z)}{H_0^2} = \Omega_\mathrm{ m} (1 + z)^3 + \Omega_\kappa
  (1 + z)^2 + \Omega_\sigma+
\]\vspace*{-3mm}
\begin{equation}
  \label{brane-hubble0}
   + 2 \Omega_\ell \pm 2
  \sqrt{\Omega_\ell}\, \sqrt{\Omega_\mathrm{ m} (1 + z )^3 +
    \Omega_\sigma + \Omega_\ell + \Omega_{\Lambda_\mathrm{ b}} +
    \Omega_C (1 + z)^4} .
\end{equation}

The model satisfies the constraint equation\vspace*{-1mm}
\begin{equation}
\label{brane-omega-r1} \Omega_\mathrm{ m} + \Omega_\kappa +
\Omega_\sigma \pm 2 \sqrt{\Omega_\ell}\, \sqrt{1 -
  \Omega_\kappa + \Omega_{\Lambda_\mathrm{ b}} + \Omega_C} = 1
\end{equation}
reducing the number of independent $\Omega$ parameters.  The sign
choices in\linebreak Eqs.~(\ref{brane-hubble0}) and
(\ref{brane-omega-r1}) always correspond to each other if $1 -
\Omega_\kappa + \Omega_{\Lambda_\mathrm{ b}} + \Omega_C >$ $>
\Omega_\ell\,$.  This condition is necessary for the model to have
physical meaning and we assume it to be valid in what follows.  The
signs in (\ref{brane-omega-r1}) correspond to the two possible ways
of bounding the Schwarzschild---(anti)-de~Sitter bulk space by the
brane \cite{CH,Shtanov1,Deffayet}, as discussed in
Sec.~\ref{brane-sec:cosmo}.

In what follows, we consider a spatially flat Universe ($\kappa =
0$) without dark radiation ($C = 0$).  In the physical case $1 +
\Omega_{\Lambda_\mathrm{ b}} > \Omega_\ell\,$, substituting
$\Omega_\sigma$ from (\ref{brane-omega-r1}) into
(\ref{brane-hubble0}), we get
\[
\frac{H^2(z)}{H_0^2} = \Omega_\mathrm{ m} (1 + z)^3 + 1 -
  \Omega_\mathrm{ m} +
  2 \Omega_\ell \mp 2\sqrt{\Omega_\ell}\, \sqrt{1+\Omega_{\Lambda_\mathrm{
  b}}}\pm
\]\vspace*{-3mm}
\begin{equation}
  \label{brane-hubble1}
  \\
  \pm 2 \sqrt{\Omega_\ell}\, \sqrt{\Omega_\mathrm{ m} (1 + z)^3 -
    \Omega_\mathrm{ m} + \left(\! \sqrt{1+\Omega_{\Lambda_\mathrm{ b}}}
      \mp \sqrt{\Omega_\ell} \!\right)^{\!\!2}}\! .
\end{equation}

The cosmological models with the lower  and upper sign were called
BRANE1 and BRANE2 models in \cite{Sahni2003}, respectively.  In the
physical region of parameters, they are equivalent to the previously
introduced ``normal'' and ``self-accelerating'' branches,
respectively. As we noted already in Sec.~\ref{brane-sec:cosmo}, the
latter name is obtained due to the property of the branch with the
upper sign that,
even in the absence of cosmological constants %
\index{cosmological constant} %
on the brane ($\sigma = 0$) and in the bulk space
($\Lambda_\mathrm{b} = 0$), one passes to the de~Sitter space in the
future with the
asymptotic Hubble parameter that follows from (\ref{brane-hubble0}): %
\index{Hubble parameter} %
\begin{equation}
  H_\mathrm{ DGP}^2 = 4 H_0^2 \Omega_\ell .
\end{equation}

Typical values of the $\Omega$ parameters (\ref{brane-omegas}) that
we consider in this section are of order $\Omega \sim 1$.  For such
values, the fundamental constants of the theory have the following
orders of magnitude:
  \begin{equation}
    m^2 \simeq M_\mathrm{ P}^2 \sim 10^{19}~\mbox{GeV}, \,\, M \sim 100\,\mbox{MeV}  , \, \Lambda_\mathrm{ b} \sim
    \frac{\sigma}{m^2} \sim H_0^2 \sim 10^{-
      56}~\mbox{cm}^{-2} .
  \end{equation}

The smallness of the bare cosmological constants in the bulk and on
the brane represents a fine-tuning similar to what is the case for
the
cosmological constant in the standard LCDM (or $\Lambda$CDM) model. %
\index{cosmological constant} %
However, even with such small values of the bare cosmological
constants in action (\ref{brane-action}), the braneworld model of
dark energy exhibits qualitatively new properties when compared to
the case where these constants are set to zero.

Observations of high-redshift type Ia supernovae indicate that these
ob\-jects are fainter than they would be in a standard cold dark
matter cosmology (SCDM) with $\Omega_\mathrm{ m} = 1$
\cite{Riess1998}. This observation is taken as support for a
Universe which is accelerating, fueled by a form of energy with
negative pressure (dark energy). In standard FRW cosmology the
acceleration of the Universe is described by the equation \index{FRW
metric} \vspace*{-2mm}
\begin{equation}
  \label{brane-eq:acc}
  \frac{\ddot a}{a} = -\frac{4\pi G}{3}\sum_i (\rho_i + 3p_i),
\end{equation}\vspace*{-4mm}

\noindent where the summation is over all matter fields contributing
to the dynamics of the Universe. It is easy to show that a necessary
(but not sufficient) condition for acceleration (${\ddot a} > 0$) is
that the strong energy condition is violated for {\em at least one}
of the matter fields in (\ref{brane-eq:acc}), so that $\rho + 3p <
0$. In the case of the popular LCDM model, this requirement is
clearly satisfied since $p_\mathrm{m} = 0$ in pressureless (cold)
matter, while $p_\Lambda = -\rho_\Lambda \equiv -\Lambda/8\pi G$ in
the cosmolo\-gical~constant.

The situation with respect to braneworld models is different since
the braneworld evolution is distinct from FRW evolution at late
times. However it is easy to show that braneworld models can
accelerate. We demonstrate this by noting that a completely general
expression for the deceleration parameter $q = -{\ddot a}/aH^2$ is
provided by\vspace*{-3mm}
\begin{equation}
  \label{brane-decel}
  q(z) = \frac{H'(z)}{H(z)} (1+z) -
  1 ,
\end{equation}
where the derivative is with respect to $z$. In our case, $H(z)$ is
given by (\ref{brane-hubble0}) or (\ref{brane-hubble1}), and the
current value of the deceleration parameter is easily calculated to be %
\index{deceleration parameter}\vspace*{-2mm}
\begin{equation}
  q_0 = \frac32 \Omega_\mathrm{ m} \left(\! 1 - \frac{\sqrt{\Omega_\ell}}{\sqrt{\Omega_\ell} \mp
      \sqrt{ 1 + \Omega_{\Lambda_\mathrm{b}}} } \!\right) - 1 ,
\end{equation}
where the lower and upper signs correspond to BRANE1 and BRANE2
models, respectively. The present Universe will accelerate for brane
parameter values which satisfy $q_0 < 0$.

Observationally, a pivotal role in the case for an accelerating
Universe is played by the \emph{luminosity distance} $d_L(z)$, since
the flux of light received from a distant source varies inversely to
the square of the luminosity distance, $F \propto d_L^{-2}$.  This
effect is quantitatively described by the magnitude---luminosity
relation: $m_B = M_0 + 25 + 5\log_{10}{d_L}$, where $m_B$ is the
corrected apparent peak $B$ magnitude and $M_0$ is the absolute peak
luminosity of the supernova. A supernova will therefore appear
fainter in a Universe which possesses a larger value of the
luminosity distance to a given redshift.

In a FRW Universe, the luminosity distance (\ref{dl}) is determined
by the Hubble parameter and three-dimensional spatial curvature as
\cite{Sahni2000}
\begin{equation}
  d_L(z) = \frac{1 + z}{H_0 \sqrt{\left| \Omega_\mathrm{ total} - 1
      \right|}}\, S \left( \eta_0 - \eta \right) \,
  , \label{brane-age11d}
\end{equation}\vspace*{-5mm}

\noindent where\vspace*{-1mm}
\begin{equation}
   \eta_0 - \eta = H_0
  \sqrt{ \left| \Omega_\mathrm{ total} - 1 \right|} \int\limits_0^z
  \frac{dz'}{H(z')} , \label{brane-lumdis1}
\end{equation}
and $S(x)$ is defined as follows: $S(x) = \sin x$ if $\kappa = 1$
($\Omega_\mathrm{ total} > 1$), $S(x) = \sinh x$ if $\kappa = -1$
($\Omega_\mathrm{ total} < 1$), and $S(x) = x$ if $\kappa = 0$
($\Omega_\mathrm{ total} = 1$).  For a spatially flat Universe under
consideration in this section, Eq.~(\ref{dl}) or
(\ref{brane-age11d}) simplifies to\vspace*{-1mm}
\begin{equation} \label{brane-lumdis2} d_L(z) = (1 + z) \int\limits_0^z
  \frac{dz'}{ H(z')} .
\end{equation}

\begin{figure}
\vskip1mm \raisebox{0.0cm}{\parbox[b]{6.0cm}{\caption{The luminosity
distance is shown as a function of redshift
    for the two bra\-ne\-world models BRANE1 \& BRANE2, LCDM, SCDM, and
    `phantom energy'. All models, with the exception of SCDM, have
    \mbox{$\Omega_\mathrm{m} = 0.3$}. SCDM has \mbox{$\Omega_\mathrm{ m} = 1$}. The
    BRANE1 \& BRANE2 models have $\Omega_\ell =$ =~$ 0.3$ and vanishing
    cosmological constant %
    \index{cosmological constant} %
    in the bulk. LCDM and the phantom model have the same dark energy
    density $\Omega_\Lambda = \Omega_X = 0.7$. The equation of state
    for dark energy is $w_\Lambda = -1$ for LCDM and $w = p_X/\rho_X =
    -1.5$ for phantom. The luminosity distance is greatest for BRANE1
    \& phantom, and least for SCDM\@. BRANE1 \& BRANE2 lie on either
    side of LCDM \label{brane-fig:lum}}}}\hspace*{0.5cm}\includegraphics[width=6.5cm]{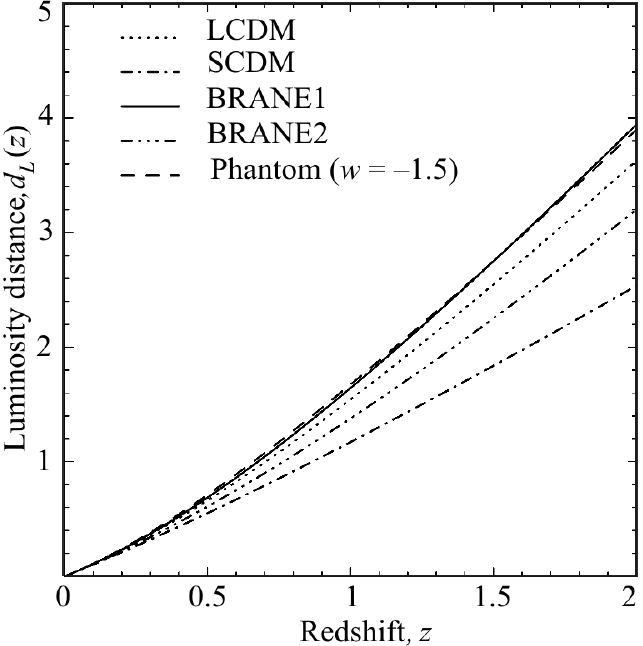}
\end{figure}

In Fig.~4.\ref{brane-fig:lum} we show the luminosity distances for
the BRANE1 \& BRANE2 models. Also shown for comparison is the value
of $d_L(z)$ in a spatially-flat two-component FRW Universe with the
Hubble parameter
\begin{equation} H(z) = H_0\left[ \Omega_\mathrm{ m}
    (1 + z)^3 + \Omega_X (1 + z)^{3(1 + w)} \right]^{1/2}\!,
  \label{brane-hubble3}
\end{equation}
where $\Omega_X$ describes dark energy with equation of state $w =
p_X/\rho_X$. Three cosmological models will be of interest to us in
connection with (\ref{brane-hubble3}):

{\bf(i) SCDM:} The standard cold dark matter Universe with
$\Omega_\mathrm{ m} = 1$ and  $\Omega_X = 0$.

{\bf(ii) LCDM:} Cold dark matter $+$ a cosmological constant
with $w = -1$. %
\index{cosmological constant} %

{\bf(iii) Phantom models:} Cold dark matter $+$ `phantom energy'
satisfying $w <
    -1$ \cite{Caldwell2002}.

We find from Fig.~\ref{brane-fig:lum} that the luminosity distance
in both braneworld models exceeds that in SCDM\@. In fact, BRANE1
models have the unusual feature that their luminosity distance can
even exceed that in LCDM (for a fixed value of $\Omega_\mathrm{
m}$).  In fact it can easily be shown that
\begin{equation}
  \label{brane-lum1}
  d_L^{~\mathrm{dS}}(z) \geq d_L^{~\mathrm{BRANE1}}(z) \geq d_L^{~\mathrm{LCDM}}(z) ,
\end{equation}
where $d_L^{~\mathrm{dS}}(z)$ refers to the luminosity distance in
the spatially flat coordinatization of de~Sitter space
(equivalently, the steady state Universe). The second inequality
presumes a fixed value of $\Omega_\mathrm{ m}$. BRANE2 models show
complementary behavior
\begin{equation}
  \label{brane-lum2}
  d_L^{~\mathrm{LCDM}}(z) \geq
  d_L^{~\mathrm{BRANE2}}(z) \geq d_L^{~\mathrm{SCDM}}(z) ,
\end{equation}
where the first inequality is valid for a fixed value of
$\Omega_\mathrm{ m}$.  In the case $\Omega_\ell = 0$, the equations
of the braneworld theory formally reduce to those of General
Relativity, and we have $d_L^{~\mathrm{BRANE1}}(z) =
d_L^{~\mathrm{BRANE2}}(z) = d_L^{~\mathrm{LCDM}}(z)$.

One might add that the behavior of BRANE1 is mimicked by FRW models
with $w \leq -1$, whereas BRANE2 resembles dark energy with $-1 \leq
w \leq 0$ \cite{Sahni2000}. In fact, from Fig.~4.\ref{brane-fig:lum}
we see that the luminosity distance in the BRANE1 model is quite
close to what one gets from `phantom energy' described by
(\ref{brane-hubble3}) with $w = -1.5$. (The parameters for this
BRANE1 model are $\Omega_\mathrm{ m} = \Omega_\ell =$ $= 0.3$,
$\Omega_{\Lambda_\mathrm{
    b}} = 0$, and $\Omega_\sigma = 1 - \Omega_\mathrm{ m} + 2
\sqrt{\Omega_\ell} \approx 1.8$.)  It should be pointed out that
phantom energy models were introduced by Caldwell
\cite{Caldwell2002}, who made the observation that dark energy with
$w < -1$ appeared to give a better fit to the current supernova
observations than LCDM (which has $w = -1$). However, the models
with phantom energy have several bizarre properties, some of which
are summarized below (see also \cite{Caldwell2002,innes}):

{\bf (i)} A negative equation of state suggests that the effective
  velocity of sound in the medium $v = \sqrt{|dp/d\rho|}$ can become
  larger than the velocity of light.

{\bf (ii)} The expansion factor of a Universe dominated by phantom
energy grows as
  \begin{equation}
    a(t) \simeq a\left( t_\mathrm{ eq} \right) \left[(1 + w) \frac{t}{t_\mathrm{ eq}} - w
    \right]^{2 \big/ 3(1+w)} , ~~ w < -1 ~,
  \end{equation}
  where $t_\mathrm{ eq}$ marks the epoch when the densities in matter
  and phantom energy are equal: $\rho_\mathrm{m}(t_\mathrm{ eq})
  \simeq \rho_X(t_\mathrm{ eq})$. It immediately follows that the
  scale factor diverges in a \emph{finite} amount of cosmic time
  \begin{equation}
    a(t) \to \infty \ \ \mbox{as} \ \ t \to t_* = \left(\! \frac{w}{1+w} \!\right)
    t_\mathrm{ eq} .
  \end{equation}

Substitution of $z \to -1$ and $w < -1$ in (\ref{brane-hubble3})
shows
that the Hubble parameter %
\index{Hubble parameter} %
also diverges as $t \to t_*$, implying that an infinitely rapid
expansion rate for the Universe has been reached in the
\emph{finite} future.

As the Universe expands, the density of phantom energy ($w < -1$)
\emph{grows} instead of decreasing ($w > -1$) or remaining constant
($w = -1$),
\begin{equation}
  \rho(t) \propto \left[(1 + w) \frac{t}{t_{\mathrm{eq}}} - w \right]^{-2} \!,
\end{equation}%
reaching a singular value in a finite interval of time $\rho(t) \to
\infty$, $t \to t_*$. This behavior should be contrasted with the
density of ordinary matter which drops to zero: $\rho_\mathrm{m} \to
0$ as $t \to t_*$. A Universe dominated by phantom energy is thus
doomed to \emph{expand towards a physical singularity} which is
reached in a finite amount of proper time. (An exact expression for
the time of occurrence of the phantom singularity %
\index{singularity} %
can be found in \cite{star99}, which also contains an interesting
discussion of related issues.)

\begin{figure}
\vskip1mm
     \includegraphics[width=13cm]{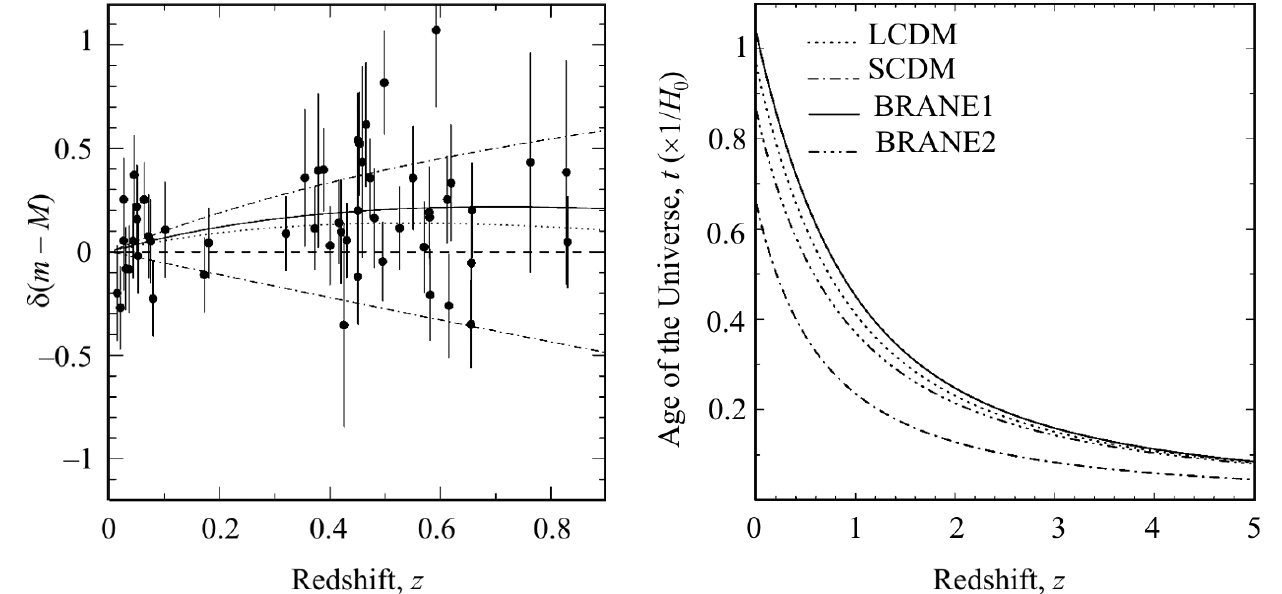}
\vskip-2mm
  \caption{The distance modulus ($m-M$) of Type Ia supernovae (the
    primary fit of the Supernova cosmology project) %
    \index{Supernova Cosmology Project (SCP)} %
    \index{angular diameter distance} %
    is shown relative to an empty $\Omega_\mathrm{ m} \to 0$ Milne
    Universe (dashed line). The solid line refers to the distance
    modulus in BRANE1 with $\Omega_\ell = \Omega_\mathrm{ m} = 0.3$,
    and vanishing cosmological constant in the bulk. The dotted line
    (below the solid) is LCDM with $(\Omega_\Lambda, \Omega_\mathrm{
      m}) = (0.7,0.3)$. The uppermost and lowermost (dot-dashed) lines
    correspond to de~Sitter space $(\Omega_\Lambda, \Omega_\mathrm{
      m}) = (1,0)$ and SCDM $(\Omega_\Lambda, \Omega_\mathrm{ m}) =
    (0,1)$, respectively.  Figure taken from
    \cite{Sahni2003} \label{brane-fig:sn}} \vspace*{-2mm}
    \caption{The age of the Universe (in units of the inverse Hubble
    parameter) %
    \index{Hubble parameter} %
    is plotted as a function of the cosmological redshift for the
    models discussed in Fig.~4.\ref{brane-fig:lum}\@. (The phantom model
    is not shown.) BRANE1 models have the oldest age while SCDM is
    youngest. %
    \index{age of Universe} %
    Figure taken from \cite{Sahni2003}
    \label{brane-fig:age}}\vspace*{-2mm}
\end{figure}

At this stage one must emphasize that, although the BRANE1 model has
several features in common with phantom energy (which is the reason
why it also provides a good fit to supernova data; see
Fig.~4.\ref{brane-fig:sn} as an illustration), it is not necessarily
afflicted with phantom's pathologies. Indeed, in a broad range of
parameters, both BRANE1 and BRANE2 are physically well motivated and
remain {\em well behaved during all times\/}.  The model safely
passes both the geometrical tests using supernovae type Ia and
baryon acoustic oscillations \cite{brane_obs1,brane_obs2} and the
tests connected with the integrated Sachs---Wolfe effect
\cite{Giannantonio:2008qr}. The development of braneworld cosmology,
therefore, added an important new dimension to the debate about the
acceleration of the Universe by showing that cosmological models
with $d_L(z) > d_L^{~\mathrm{LCDM}}(z)$ are possible to construct
within the framework of the braneworld scenario and should be taken
seriously.

\begin{figure}
\vskip1mm
     \includegraphics[width=13cm]{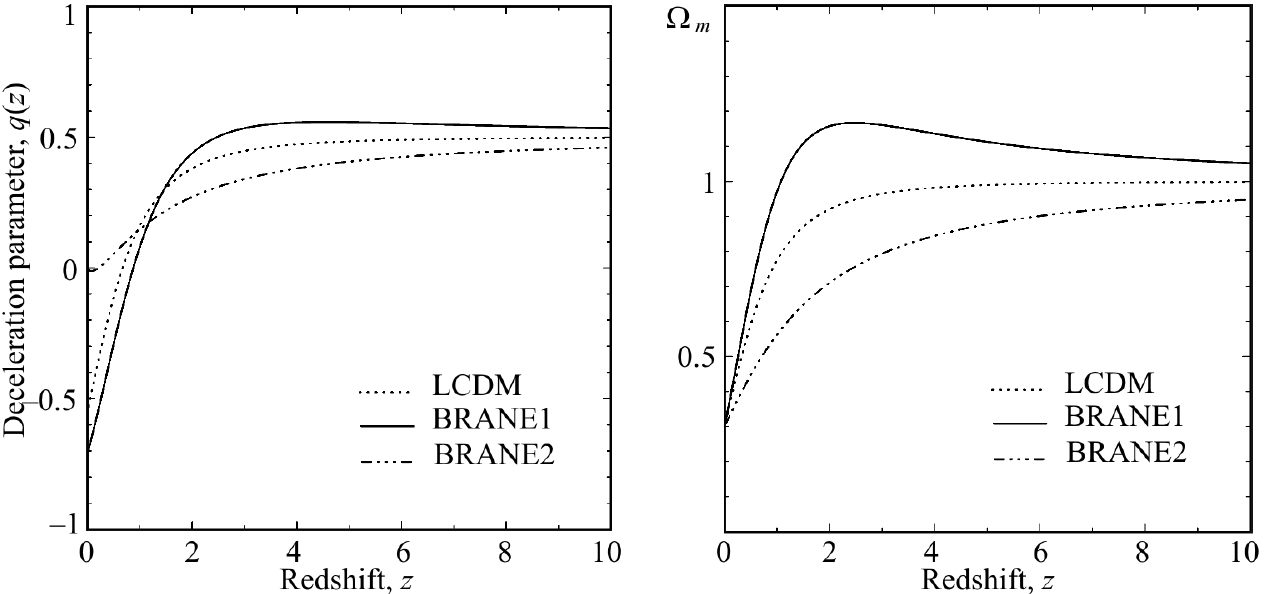}
\vskip-2mm
      \caption{The deceleration parameter $q(z)$ is shown for BRANE1,
    BRANE2 and LCDM\@. The model parameters are as in
    Fig.~4.\ref{brane-fig:lum}\@. For reference it should be noted that
    $q = 0.5$ for SCDM while de~Sitter space has $q = -1$.  Figure
    taken from \cite{Sahni2003} \label{brane-fig:decel}}
    \vspace*{-2mm}
     \caption{The dimensionless matter density $\Omega_\mathrm{ m}(z)$ is
    shown for the two braneworld models and LCDM\@. ($\Omega_\mathrm{
      m} = 1$ in SCDM\@.)  Parameter values are the same as in
    previous figures. BRANE1 has the interesting feature that
    $\Omega_\mathrm{ m}(z)$ {\em exceeds unity\/} for $z \gtrsim
    1$. Figure taken from \cite{Sahni2003} \label{brane-fig:omega}} \vspace*{-2mm}
\end{figure}

The angular diameter distance $d_A$ is related to the luminosity
distance $d_L$ through the equation (see
Sec.~\ref{ch1-sec3})\vspace*{-3mm}
\begin{equation} \label{brane-eq:angdis}
d_A(z) = \frac{d_L (z)}{(1 + z)^2} .
\end{equation}
Therefore, much of the above analysis carries over when one
discusses properties of the angular diameter distance within the
framework of braneworld models. Some cosmological features of
braneworld models are shown in
Figs.~4.\ref{brane-fig:age}---4.\ref{brane-fig:omega}. In
Fig.~4.\ref{brane-fig:age}, the age of the Universe at a given
cosmological redshift
\begin{equation} \label{brane-age} t(z) = \int\limits_z^\infty
  \frac{dz'}{(1+z') H(z')}
\end{equation}
is shown for the two braneworld models and for LCDM \& SCDM\@. We
find that the age of the Universe in BRANE1 (BRANE2) is larger
(smaller) than in LCDM for identical values of the cosmological
density parameter $\Omega_\mathrm{ m}$. This is a direct consequence
of the fact that the Hubble parameter in BRANE1 (BRANE2) is smaller
(larger) than in LCDM\@.  Both braneworld models are significantly
older than SCDM.

\index{deceleration parameter} %
Considerable insight into the dynamics of the Universe is provided
by the cosmological deceleration parameter (\ref{brane-decel}). The
results obtained in \cite{Sahni2003}, shown in
Fig.~4.\ref{brane-fig:decel}, indicate that at late times the BRANE1
(BRANE2) Universe accelerates at a faster (slower) rate than LCDM
(with identical $\Omega_{\mathrm{m}}$).  Curiously, the BRANE1
Universe shows an earlier transition from deceleration to
acceleration than any of the other models. (For the given choice of
parameters this transition takes place at $z \simeq 1$ for BRANE1
and at $z \simeq 0.7$ for LCDM\@. The BRANE2 model begins
accelerating near the present epoch at $z \simeq 0$.) A related
point of interest is that at $z \gtrsim 2$ the deceleration
parameter in BRANE1 marginally exceeds that in SCDM indicating that
the BRANE1 model is decelerating at a faster rate than SCDM ($q =
0.5$). In conventional models of dark matter this behavior can occur
only if the equation of state of the dark component is stiffer than
dust, implying $w > 0$ in (\ref{brane-hubble3}), or if the Universe
is spatially closed. On the other hand, the current
\emph{acceleration} rate of BRANE1 in our example ($q_0 \simeq
-0.7$) significantly exceeds that of LCDM ($q_0 \simeq -0.55$) with
an identical value of $\Omega_\mathrm{ m} = 0.3$ in both models.
Within the framework of four-dimensional Einstein gravity, this
situation can only arise if the equation of state of dark energy is
strongly negative: $w < -1$ in (\ref{brane-hubble3}).

The unusual high-$z$ behavior of the deceleration parameter in
BRANE1 can be better understood if we consider the cosmological
density parameter
\begin{equation}
  \Omega_\mathrm{ m}(z) = \left[ \frac{H_0}{H(z)} \right]^2 \Omega_\mathrm{ m}(0) (1+z)^3
  . \label{brane-eq:omega}
\end{equation}

Here $H(z)$ again should be given by (\ref{brane-hubble1}) for
braneworld models. From Fig.~4.\ref{brane-fig:omega} we notice that,
for $z \gtrsim 1$, the value of $\Omega_\mathrm{ m}(z)$ in BRANE1
{\em
  exceeds\/} its value in SCDM ($\Omega_\mathrm{ m} = 1$). This is
precisely the redshift range during which $q(z)_\mathrm{ BRANE1} >
q(z)_\mathrm{ SCDM}$. Thus, the rapid deceleration of BRANE1 at high
redshifts can be partly attributed to the larger value of the matter
density $\Omega_\mathrm{ m}(z)$ at those redshifts, relative to
SCDM.

Having established partial similarity of BRANE1 with phantom models
at low redshifts, we can investigate the analogy further and
calculate the effective equation of state of dark
energy\vspace*{-2mm}
\begin{equation} \label{brane-state} w(z) = \frac{2 q(z) - 1 }{3 \left[
      1 - \Omega_\mathrm{ m}(z) \right] } ,
\end{equation}
where $\Omega_\mathrm{ m}(z)$ is given by (\ref{brane-eq:omega}).
One
notes that $w(z)$ has a pole-like singularity %
\index{singularity} %
at $z \simeq 1$ for BRANE1, which arises because $\Omega_\mathrm{ m}
(z)$ crosses the value of unity at $z \simeq 1$ (see
Fig.~4.\ref{brane-fig:omega}). This demonstrates that the notion of
`effective equation of state' is of limited utility for this model.
Equations (\ref{brane-hubble1}), (\ref{brane-decel}), and
(\ref{brane-state}) also illustrate the important fact that dark
energy in braneworld models, though similar to phantom energy in
some respects, differ from it in others. For instance, in both
braneworld models, $w(z) \to -0.5$ at $z \gg 1$ and $w(z) \to -1$ as
$z \to -1$, whereas phantom energy has $w(z) < -1$ at {\em all\/}
times.

A useful quantity is the {\em current value\/} of the effective
equation of state of dark energy in braneworld
theories:\vspace*{-2mm}
\begin{equation}
  w_0 = \frac{2 q_0 - 1 }{3 \left( 1 - \Omega_\mathrm{ m} \right)} = - 1 -
  \left(\! \frac{\Omega_\mathrm{ m} }{ 1 - \Omega_\mathrm{ m}} \!\right) \frac{\sqrt{\Omega_\ell} }
  { \sqrt{\Omega_\ell} \mp
    \sqrt{ 1 + \Omega_{\Lambda_\mathrm{b}}} }  ,
\end{equation}
where the lower and upper signs, as usual, correspond to BRANE1 and
BRANE2 models, respectively.  We easily see that $w_0 < - 1$ for
BRANE1, whe\-reas $w_0 > - 1$ for BRANE2.

The reason for the effect of phantom-like behavior of the BRANE1
model can be seen directly from equation (\ref{brane-hubble0}) or
(\ref{brane-hubble1}), which can be written in the
form\vspace*{-1mm}
\begin{equation} \frac{H^2(z) }{ H_0^2} = \Omega_\mathrm{ m} (1 + z)^3 +
  \Omega_{\Lambda_\mathrm{ eff}} (z) ,
\end{equation}\vspace*{-5mm}

\noindent where\vspace*{-3mm}
\[
\Omega_{\Lambda_\mathrm{ eff}} (z)  = \Omega_\sigma + 2
    \Omega_\ell \pm 2
    \sqrt{\Omega_\ell}\, \sqrt{\Omega_\mathrm{ m} (1 + z )^3 + \Omega_\sigma + \Omega_\ell +  \Omega_{\Lambda_\mathrm{
    b}}}=
\]\vspace*{-3mm}
\[
= 1 - \Omega_\mathrm{ m} + 2 \Omega_\ell \mp
    2\sqrt{\Omega_\ell}\,
    \sqrt{1+\Omega_{\Lambda_\mathrm{ b}}} \,\pm
\]\vspace*{-3mm}
\begin{equation}
   \pm\, 2 \sqrt{\Omega_\ell}\, \sqrt{\Omega_\mathrm{ m} (1 + z)^3
      - \Omega_\mathrm{ m} + \left(\!
        \sqrt{1+\Omega_{\Lambda_\mathrm{b}}} \mp \sqrt{\Omega_\ell}
      \right)^{\!2}}
  \end{equation}
is the omega parameter for the effective time-dependent cosmological
constant. %
\index{cosmological constant} %
One can see that, for the branch with the lower sign (BRANE1),
$\Omega_{\Lambda_\mathrm{ eff}} (z)$ increases with time, while, for
the branch with the upper sign (BRANE2), it decreases.  This
explains the properties (\ref{brane-lum1}) and (\ref{brane-lum2}).

\section{\!Disappearing dark energy} \label{brane-sec:dde}

\hspace{3cm}\index{disappearing dark energy}The braneworld models
admit an intriguing possibility that the current acceleration of the
Universe may not be a lasting feature. It may be recalled that most
models of dark energy, including the cosmological constant, have the
property that, once the Universe begins to accelerate, it
accelerates forever. Although this is not a problem from the
viewpoint of cosmology; nevertheless, as shown in a number of
papers, an eternally accelerating Universe is endowed with a
cosmological event horizon which prevents the construction of a
conventional S-matrix describing particle interactions within the
framework of string or M-theory [549---551]. In this section we show
that, provided the Randall---Sundrum constraint relation
(\ref{brane-RS}) is satisfied, the accelera\-tion of the Universe
can be a transient phenomenon in braneworld models.\linebreak An
anisotropic solution of Bianchi~V class with the same feature was
descri-\linebreak bed in \cite{Kofinas}.

From Eq.~(\ref{brane-hubble1}) we obtain the following asymptotic
expressions for the Hubble parameter $H_\infty$ as $z \to -1$, %
\index{Hubble parameter} %
assuming that the Universe expands forever:
\begin{equation}
  \left(\!\frac{H_\infty }{ H_0}\!\right)^{\!\!2} = \Omega_\sigma + 2 \Omega_\ell \pm 2
  \sqrt{\Omega_\ell} \, \sqrt{\Omega_\sigma + \Omega_\ell + \Omega_{\Lambda_\mathrm{
        b}}} ,
\end{equation}
where the lower and upper signs correspond to BRANE1 and BRANE2
models, respectively.  In applying the Randall---Sundrum constraint
(\ref{brane-RS}), we first consider the case where $\Omega_\sigma >
0$. Then
\begin{equation}
  \Omega_\sigma = 2 \sqrt{\Omega_\ell \Omega_{\Lambda_\mathrm{ b}}}
\end{equation}
and
\begin{equation} \label{brane-positive} \left(\!\frac{H_\infty }{
      H_0}\!\right)^{\!\!2} = 2 \sqrt{\Omega_\ell} \left[ \sqrt{\Omega_\ell} +
    \sqrt{\Omega_{\Lambda_\mathrm{ b}}} \pm \left(\! \sqrt{\Omega_\ell}
      + \sqrt{\Omega_{\Lambda_\mathrm{ b}}} \right) \!\right]\! .
\end{equation}
One can see that this expression vanishes for the lower sign.  Thus,
for positive $\Omega_\sigma$, it is the BRANE1 model that leads to
vanishing effective cosmological constant in the future.  However,
in this case, the constraint equation (\ref{brane-omega-r1}) becomes
\begin{equation}
  \Omega_\mathrm{ m} - 2 \sqrt{\Omega_\ell} \left(\! \sqrt{1 + \Omega_{\Lambda_\mathrm{b}}} - \sqrt{\Omega_{\Lambda_\mathrm{ b}}} \right) = 1
\end{equation}
and implies $\Omega_\mathrm{ m} > 1$, which is hardly compatible
with the observations.

In the case of $\Omega_\sigma < 0$, we have
\begin{equation}
  \label{brane-eq:omsigma}%
  \Omega_\sigma = - 2 \sqrt{\Omega_\ell\, \Omega_{\Lambda_\mathrm{
        b}}}
\end{equation}\vspace*{-5mm}

\noindent and\vspace*{-2mm}
\begin{equation}
  \label{brane-negative}%
  \left(\!\frac{H_\infty }{ H_0}\!\right)^{\!\!2} = 2 \sqrt{\Omega_\ell}
  \left(\!
    \sqrt{\Omega_\ell} - \sqrt{\Omega_{\Lambda_\mathrm{ b}}} \pm
    \left|\sqrt{\Omega_\ell} - \sqrt{\Omega_{\Lambda_\mathrm{ b}}}
    \right| \right)\! .
\end{equation}

If $\Omega_\ell > \Omega_{\Lambda_\mathrm{ b}}$, then this
expression vanishes for the lower sign, which \mbox{brings} us back
to the non-physical BRANE1 models with $\Omega_\mathrm{ m} > 1$. In
the case $\Omega_\ell \le \Omega_{\Lambda_\mathrm{ b}}$, expression
(\ref{brane-negative}) vanishes for the upper sign, which
corresponds to BRANE2 models. The constraint equation
(\ref{brane-omega-r1}) now reads
\begin{equation}
\label{brane-eq:con-rs} \Omega_\mathrm{ m} + 2 \sqrt{\Omega_\ell}
\left(\! \sqrt{1 + \Omega_{\Lambda_\mathrm{ b}}} -
  \sqrt{\Omega_{\Lambda_\mathrm{ b}}} \right) = 1
\end{equation}
and implies $\Omega_\mathrm{ m} < 1$.

\begin{figure}
\vskip1mm
     \includegraphics[width=13cm]{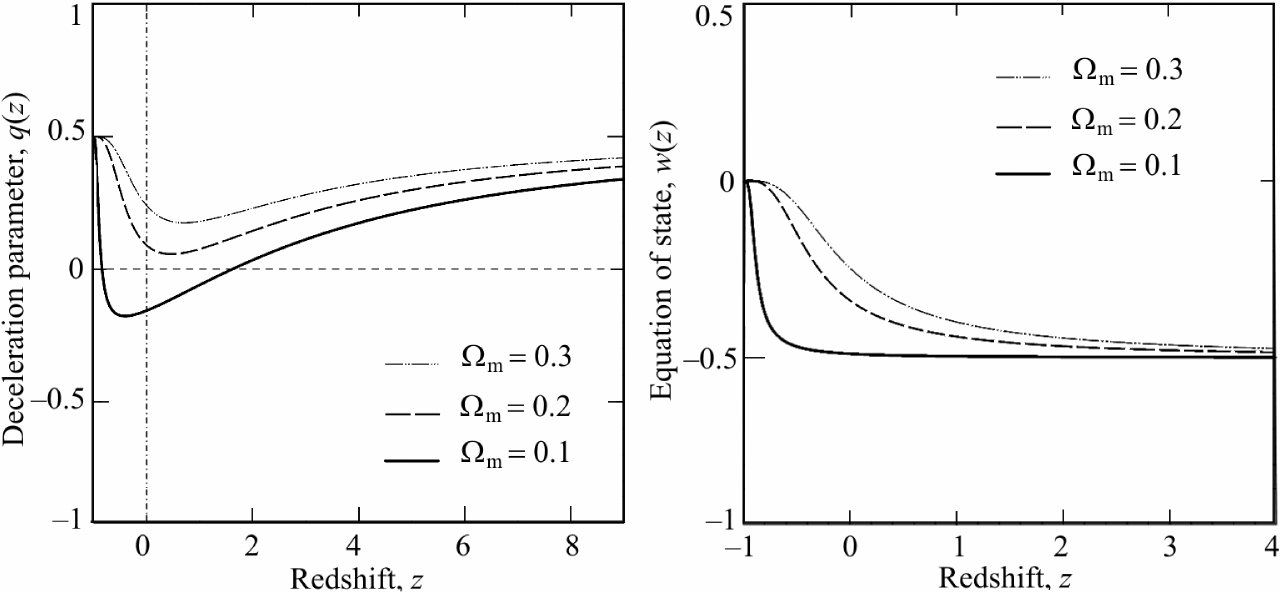}
\vskip-2mm
  \caption{The deceleration parameter is plotted as a function of
    redshift for the BRANE2 model with the Randall---Sundrum constraint
    (\ref{brane-RS}), $\Omega_{\Lambda_\mathrm{ b}} = 2$, and
    $\left(\Omega_\mathrm{ m}, \Omega_\ell \right) = (0.3, 1.2)$,
    $(0.2, 1.6)$, $(0.1, 1.98)$ (top to bottom). The vertical
    (dot-dashed) line at $z = 0$ marks the present epoch, while the
    horizontal (dashed) line at $q = 0$ corresponds to a Milne
    Universe [$a(t) \propto t$] which neither accelerates nor
    decelerates. Note that the Universe {\em ceases to accelerate\/}
    and becomes matter dominated in the future.  Figure taken from
    \cite{Sahni2003} \label{brane-fig:decel_plot}}
\vspace*{-2mm} \caption{The effective equation of state for dark
energy in the
    BRANE2 model is shown as a function of redshift. Model parameters
    are as in the previous figure. Note that the past and future
    asymptotes of $w(z)$ are quite different: $w(z) \to -1/2$ for $z
    \gg 0$, while $w(z) \to 0$ for $z \to -1$.  Braneworld dark energy
    therefore effectively disappears in the future, giving rise to a
    matter-dominated Universe. Figure taken from
    \cite{Sahni2003}\label{brane-fig:state_future}}\vspace*{-2mm}
\end{figure}

Therefore, BRANE2 with $\Omega_\sigma < 0$ and $\Omega_\ell \le
\Omega_{\Lambda_\mathrm{ b}}$, provides us \mbox{with} an
interesting example of a physically meaningful cosmological model in
\mbox{which} the current acceleration of the Universe is a {\em
transient phenomenon\/}. An example of this behavior as probed by
the deceleration parameter is shown in
Fig.~4.\ref{brane-fig:decel_plot}, which demonstrates that the
current period of cosmic acceleration takes place between two
matter-dominated epochs. We emphasize that these models require
negative brane tension $\sigma$.  Since an observer in this model
resides on a negative-tension brane, one must ponder over the issue
of whether such a braneworld will be perturbatively stable and hence
physically viable. We consider this to be an open question for
future investigations. Remarks made at the end of
Sec.~\ref{brane-sec:vacua} are relevant, however, since one and the
same cosmolo\-gical solution on the `visible' (negative tension)
brane can correspond to many dif\-ferent global conditions in the
bulk, for instance, other (`hidden') branes may be present or
absent, static or evolving, close to or far away from \mbox{our
brane, etc.}\looseness=1

\index{equation of state (EoS)} %
Useful insight into the BRANE2 model is also provided by the
effective equation of state of dark energy (\ref{brane-state}). Our
results \cite{Sahni2003}, shown in
Fig.~4.\ref{brane-fig:state_future}, indicate that the past and
future behavior of dark energy in the braneworld Universe can be
very different. The past behavior $w(z) \to -0.5$ for $z \gg 1$
arises because, in a spatially flat braneworld, the second most
important contribution to braneworld expansion at high redshifts is
caused by the $(1+z)^{3/2}$ term in (\ref{brane-hubble1}); see also
\cite{DDG,DLRZA}. The future behavior $w(z) \to 0$ as $z \to -1$, on
the other hand, reflects the decreasing importance of dark energy as
the Universe expands. The acceleration of the Universe is therefore
a transient phenomenon which ends once the Universe settles back
into the matter-dominated regime.

Finally, we should mention that a transiently accelerating regime
also arises in a class of BRANE2 models which do not satisfy the
Randall---Sundrum constraint (\ref{brane-RS}). In these models the
current epoch of acceleration is succeeded by an epoch during which
the deceleration parameter grows without bound. This unusual `future
singularity' %
\index{singularity} %
is reached in a {\em finite\/} interval of expansion time and is
characterized by the fact that both the matter density and the
Hubble
parameter %
\index{Hubble parameter} %
remain finite, while ${\ddot a} \to \infty$ (a feature that
distinguishes it from the phantom singularities discussed in
Sec.~4.6). A detailed discussion of the `new' singularities which
occur in braneworld models will be made in
Sec.~\ref{brane-sec:singular}.

\section{\!Cosmic mimicry} \label{brane-sec:mimicry}

\hspace{3cm}\index{cosmic mimicry}It was noted \cite{MMT} that the
cosmological evolu\-tion in braneworld theory, from the viewpoint of
the Friedmannian Universe, proceeds with a time-dependent
gravitational constant.  Indeed, the cosmological equation on the
brane (\ref{brane-cosmol2}) can be
written in the form of one of the Friedmann equations %
\index{Friedmann equations}
\begin{equation}
  H^2 + \frac{\kappa }{ a^2} = \frac{8 \pi G_\mathrm{ eff} (\rho)}{3} \rho ,
\end{equation}
where $G_\mathrm{ eff} (\rho)$ is the effective time-dependent
gravitational constant that can be read-off from
(\ref{brane-cosmol2}).

In this section, we describe another interesting feature of
braneworld cos\-mo\-lo\-gy, which we call ``cosmic mimicry''
\cite{mimicry}. It turns out that, for a broad range of parameter
values, the braneworld model behaves {\em exactly as a LCDM
($\Lambda$ + Cold Dark Matter)
  Universe\/} with different values of the effective cos\-mo\-lo\-gi\-cal
density parameter $\Omega_\mathrm{ m}$ at different epochs.
Moreover, the cos\-mo\-lo\-gi\-cal density parameter inferred from
the observations of the large-sale structure and CMB and that
determined from neoclassical cosmological tests such as observations
of supernovae (SN) can
potentially have different values. %
\index{cosmic microwave background (CMB)} %

An important feature of this model is that, although it is very
similar to LCDM at the present epoch, its departure from
``concordance cosmology'' can be significant at intermediate
redshifts, leading to new possibilities for the Universe at the end
of the ``dark ages'' which may be worth exploring.

We relate the ``mimicry'' properties of the braneworld cosmology
with the properties of gravity in braneworld theories.  In
particular, we show that the change in the cosmological density
parameter $\Omega_\mathrm{ m}$ as the Universe evolves can be
related to the spatial scale dependence of the effective
gravitational constant in braneworld theory [533---535]. This can
have important consequences for cosmological models based on the
braneworld theory and calls for more extensive analysis of their
cosmological history.

The basic equation describing the cosmological evolution is
(\ref{brane-hubble1}).  For sufficiently high redshifts, the first
term on the right-hand side of this equation dominates, and the
model reproduces the matter-dominated Friedmann Universe with the
density parameter $\Omega_\mathrm{m}$.  Now we note that, for the
values of $z$ and parameters $\Omega_{\Lambda_\mathrm{ b}}$ and
$\Omega_\ell$ which satisfy
\begin{equation}
  \Omega_\mathrm{ m} (1 + z)^3 \ll \left(\! \sqrt{1+\Omega_{\Lambda_\mathrm{ b}}} \mp
    \sqrt{\Omega_\ell} \right)^{\!2}\! , \label{brane-eq:mimic0}
\end{equation}
Eq.~(\ref{brane-hubble1}) can be well approximated as
\begin{equation}
  \frac{H^2(z) }{ H_0^2} \simeq \Omega_\mathrm{ m} (1 +
  z)^3 + 1 - \Omega_\mathrm{ m} -
  \frac{\sqrt{\Omega_\ell}}{\sqrt{\Omega_\ell} \mp \sqrt{1 +
      \Omega_{\Lambda_\mathrm{ b}}}} \left[ \Omega_\mathrm{ m} (1 +
    z)^3 - \Omega_\mathrm{ m} \right]\!.
\end{equation}

We introduce the positive parameter $\alpha$ by the equation
\begin{equation} \label{brane-alpha}
\alpha = \frac{\sqrt{1 +
\Omega_{\Lambda_\mathrm{b}}}}{\sqrt{\Omega_\ell}} .
\end{equation}
Then, defining a new density parameter by the relation
\begin{equation}
\label{brane-lcdm} \Omega^\mathrm{ LCDM}_\mathrm{ m} =
\frac{\alpha}{\alpha \mp 1}\, \Omega_\mathrm{ m} ,
\end{equation}\vspace*{-7mm}

\noindent we get
\begin{equation}
\frac{H^2(z)}{H_0^2} \simeq \Omega^\mathrm{ LCDM}_\mathrm{ m} (1 +
z)^3 + 1 - \Omega^\mathrm{ LCDM}_\mathrm{ m}, \label{brane-eq:lcdm}
\end{equation}
\index{Hubble parameter|(} %
which is precisely the Hubble parameter describing a LCDM Universe.
[Note that the braneworld parameters $\Omega_\ell$ and
$\Omega_{\Lambda_\mathrm{ b}}$ have been effectively absorbed into a
``renormalization'' of the matter density $\Omega_\mathrm{ m} \to
\Omega^\mathrm{ LCDM}_\mathrm{ m}$, defined by (\ref{brane-lcdm}).]

Thus, our braneworld displays the following remarkable behavior
which we refer to as \emph{``cosmic mimicry''}:

{\footnotesize$\bullet$}\,\,A BRANE1 model, which at high redshifts
expands with density pa\-ra\-me\-ter $\Omega_\mathrm{m}$, at lower
redshifts {\em masquerades
    as a LCDM Universe\/} with a {\em smaller value\/} of the density
  parameter. In other words, at low redshifts, the BRANE1 Universe
  expands as the LCDM model (\ref{brane-eq:lcdm}) with
  $\Omega^\mathrm{ LCDM}_\mathrm{ m} < \Omega_\mathrm{ m}$ [where
  $\Omega^\mathrm{ LCDM}_\mathrm{ m}$ is determined by
  (\ref{brane-lcdm}) with the lower (``$+$'') sign].

{\footnotesize$\bullet$}\,\,A BRANE2 model at low redshifts also
masquerades as LCDM but
  with a {\em larger value\/} of the density parameter. In this case,
  $\Omega^\mathrm{ LCDM}_\mathrm{ m} > \Omega_\mathrm{ m}$ with
  $\Omega^\mathrm{ LCDM}_\mathrm{ m}$ being determined by
  (\ref{brane-lcdm}) with the upper (``$-$'') sign\,\footnote{\,For
    $\alpha < 1$, the BRANE2 model behaves like that with {\em
      negative\/} matter density and demonstrates unwanted bouncing at
    low redshifts.}.

The range of redshifts over which this cosmic mimicry occurs is
given by $0 \leq z \ll z_{\mathrm{m}}$, with $z_{\mathrm{m}}$
determined by (\ref{brane-eq:mimic0}). Specifically,
\begin{equation}
  z_{\mathrm{m}} = \frac{ \left(\!\sqrt{1 + \Omega_{\Lambda_\mathrm{ b}}} \mp \sqrt{\Omega_\ell}
    \right)^{2/3}} {\Omega_{\mathrm{m}}^{1/3}} - 1 , \label{brane-eq:mimic1}
\end{equation}
which can also be written as\vspace*{-3mm}
\begin{equation}
  (1+z_{\mathrm{m}})^3 = \frac{\Omega_\mathrm{m} \left(1 + \Omega_{\Lambda_\mathrm{ b}} \right)}{
    \left(\Omega_{\mathrm{m}}^\mathrm{ LCDM} \right)^2} \label{brane-eq:new1}
\end{equation}\vspace*{-5mm}

\noindent for both braneworld models.

\begin{figure}
\vskip1mm
     \includegraphics[width=13cm]{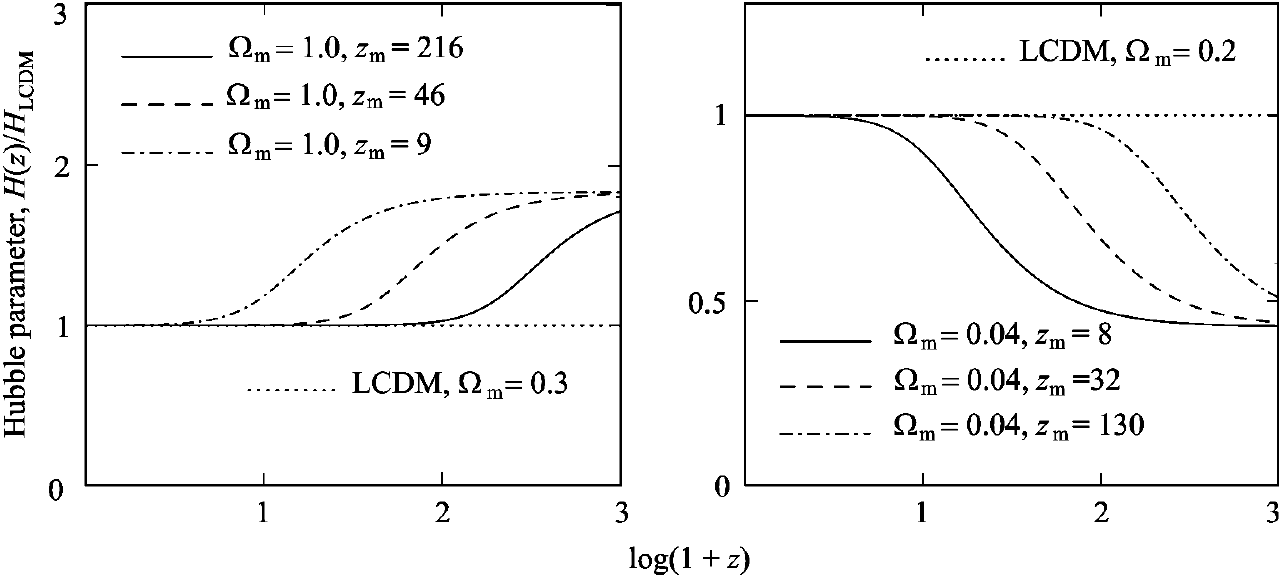}
\vskip-2mm
   \caption{\label{brane-fig:mimic} An illustration of cosmic mimicry
for
  the BRANE1 (left) and BRANE2 (right) models. The Hubble parameter in
  three high-density BRANE1 models with $\Omega_\mathrm{m} = 1$ and
  low-density BRANE2 models with $\Omega_\mathrm{m} = 0.04$ is
  shown. Also shown is the Hubble parameter in the LCDM model (dotted
  line) which closely mimics this braneworld but has a lower mass
  density $\Omega_{\mathrm{m}}^\mathrm{LCDM} = 0.3$ ($\Omega_\Lambda =
  0.7$) for BRANE1 and a higher mass density
  $\Omega_{\mathrm{m}}^\mathrm{ LCDM} = $ $=0.2$ ($\Omega_\Lambda = 0.8$)
  for BRANE2. The brane matter density ($\Omega_\mathrm{m}$) and the
  matter density in LCDM are related through $\Omega_\mathrm{m} =
  \Omega_\mathrm{m}^\mathrm{ LCDM} \left[ 1 \pm
    \sqrt{\Omega_\ell/\left(1 + \Omega_{\Lambda_b} \right)} \right] $,
  so that $\Omega_\mathrm{m} \gtrsim \Omega_\mathrm{m}^\mathrm{ LCDM}$
  for BRANE1 (lower sign) and $\Omega_\mathrm{m} \lesssim
  \Omega_\mathrm{m}^\mathrm{ LCDM}$ for BRANE2 (upper sign). The
  braneworld model masquerades as LCDM for $z \ll z_\mathrm{m}$, where
  $z_\mathrm{m} = 9$, $46$, $216$ for BRANE1 and $z_\mathrm{m} = 8$,
  $32$, $130$ for BRANE2 (left to right) for three different choices
  of parameters. Figure taken from \cite{mimicry}}\vspace*{-1mm}
\end{figure}

Examples of cosmic mimicry are shown in Fig.~4.\ref{brane-fig:mimic}
for the BRANE1 model (left) and BRANE2 model (right). One striking
consequence of the BRANE2 model in Fig.~4.\ref{brane-fig:mimic} is
that a low-density ($\Omega_\mathrm{m} = 0.04$) Universe consisting
{\em
  entirely\/} of baryons mimics a higher-density LCDM model
($\Omega^\mathrm{ LCDM}_\mathrm{m} = 0.3$) and can therefore be in
excellent agreement with the SN data.

In view of relation (\ref{brane-eq:new1}), it is interesting to note
that we can use the equations derived in this section to relate the
three free parameters in the braneworld model: $\left\lbrace
  \Omega_\ell, \Omega_{\Lambda_\mathrm{b}},\Omega_\mathrm{m}
\right\rbrace$ to $\left\lbrace \Omega_\mathrm{m}, z_\mathrm{m},
  \Omega_\mathrm{m}^\mathrm{ LCDM} \right\rbrace$. These relations
(which turn out to be the same for BRANE1 and BRANE2 models) are:
\begin{equation}
  \label{brane-eq:relation}
  \frac{1 + \Omega_{\Lambda_\mathrm{b}}}{\Omega_\mathrm{m}^\mathrm{
      LCDM}}  = \displaystyle \frac{\Omega_\mathrm{m}^\mathrm{
      LCDM}}{\Omega_\mathrm{m}} (1+z_\mathrm{m})^3 ,
\end{equation}\vspace*{-3mm}
\begin{equation}
  \frac{\Omega_\ell}{\Omega_\mathrm{m}^\mathrm{LCDM}}  = \left[
    \sqrt{\frac{\Omega_\mathrm{m}^\mathrm{LCDM}}{\Omega_\mathrm{m}}} -
    \sqrt{\frac{\Omega_\mathrm{m}}{\Omega_\mathrm{m}^\mathrm{LCDM}}}
  \right]^2 (1+z_\mathrm{m})^3 .
\end{equation}

Furthermore, if we assume that the value of
$\Omega_{\mathrm{m}}^\mathrm{ LCDM}$ is known (say, from the
analysis of SN data), then the two braneworld parameters
$\Omega_\ell$ and $\Omega_{\Lambda_\mathrm{b}}$ can be related to
the two parameters $\Omega_\mathrm{m}$ and $z_\mathrm{m}$ using
(\ref{brane-eq:relation}), so it might be more convenient to analyze
the model in terms of $\Omega_\mathrm{m}$ and $z_{\rm m}$ (instead
of \mbox{$\Omega_\ell$ and $\Omega_{\Lambda_\mathrm{b}}$)}.

We also note that, under condition (\ref{brane-eq:mimic0}), the
brane tension $\sigma$, determined by (\ref{brane-omega-r1}), is
positive for BRANE1 model, and negative for BRANE2 model.

Since the Hubble parameter in braneworld models departs from that in
LCDM at {\em intermediate\/} redshifts ($z > z_\mathrm{ m}$), this
could leave behind an important cosmological signature especially
since several key cosmological observables depend upon the Hubble
parameter either differentially or integrally.\linebreak \mbox{Examples include:}%
\index{Hubble parameter|)} %

{\footnotesize$\bullet$}\,\,the luminosity distance $d_L(z)$, given
by (\ref{brane-lumdis2}),

{\footnotesize$\bullet$}\,\,the angular diameter distance (\ref{brane-eq:angdis}), %
\index{angular diameter distance} %

{\footnotesize$\bullet$}\,\,the product $d_A(z)H(z)$, which plays a
key role in the Alcock---Paczynski
    anisotropy test \cite{alcock,KLSW},

{\footnotesize$\bullet$}\,\,the product $d_A^2(z)H^{-1}(z)$, which
is used in the volume-redshift test
    \cite{davis},

{\footnotesize$\bullet$}\,\,the deceleration parameter
(\ref{brane-decel}),

{\footnotesize$\bullet$}\,\,the effective equation of state of dark
energy (\ref{brane-state}),

{\footnotesize$\bullet$}\,\,the age of the Universe (\ref{brane-age}), %
\index{age of Universe|(} %

{\footnotesize$\bullet$}\,\,the ``statefinder pair''
\cite{Sahni2003a,ASSS}:
\begin{equation}
  \begin{array}{c}
   \displaystyle r =  \frac{\stackrel{...}{a}}{a H^3} \equiv 1 + \left\lbrack
      \frac{H''}{H} + \left (\!\frac{H'}{H}\!\right )^{\!2} \right\rbrack
    (1+z)^2 - 2\frac{H'}{H}(1+z) , \\[5mm]
 \displaystyle   s =  \frac{r - 1}{3(q - 1/2)} , \label{brane-eq:state}
  \end{array}
\end{equation}

{\footnotesize$\bullet$}\,\,the electron-scattering optical depth to
a redshift $z_\mathrm{ reion}$
    \cite{peebles,dodelson}
\begin{equation}
  \tau(z_\mathrm{ reion}) = c\int\limits_0^{z_\mathrm{ reion}}\frac{n_e(z)\sigma_T ~dz}{(1+z)H(z)} \,
  , \label{brane-reion}
\end{equation}
where $n_e$ is the electron density and $\sigma_T$ is the Thomson
cross-section describing scattering between electrons and CMB
photons.
\index{cosmic microwave background (CMB)} %

A degree of caution should be exercised when comparing the late-time
LCDM behavior (\ref{brane-eq:lcdm}) of the model under consideration
with different sets of observations, since the parameter
$\Omega^\mathrm{ LCDM}_{\rm m}$, residing in (\ref{brane-eq:lcdm}),
which is effectively used in determinations of the luminosity
distance (\ref{brane-lumdis2}) and angular diameter distance
(\ref{brane-eq:angdis}), may very well be different from the value
of $\Omega_\mathrm{m}$ inferred from observations of gravitational
clustering. These issues should be kept in mind when performing a
maximum-likelihood analysis using data belonging to different
observational streams.

\begin{figure}
\vskip1mm
    \includegraphics[width=13cm]{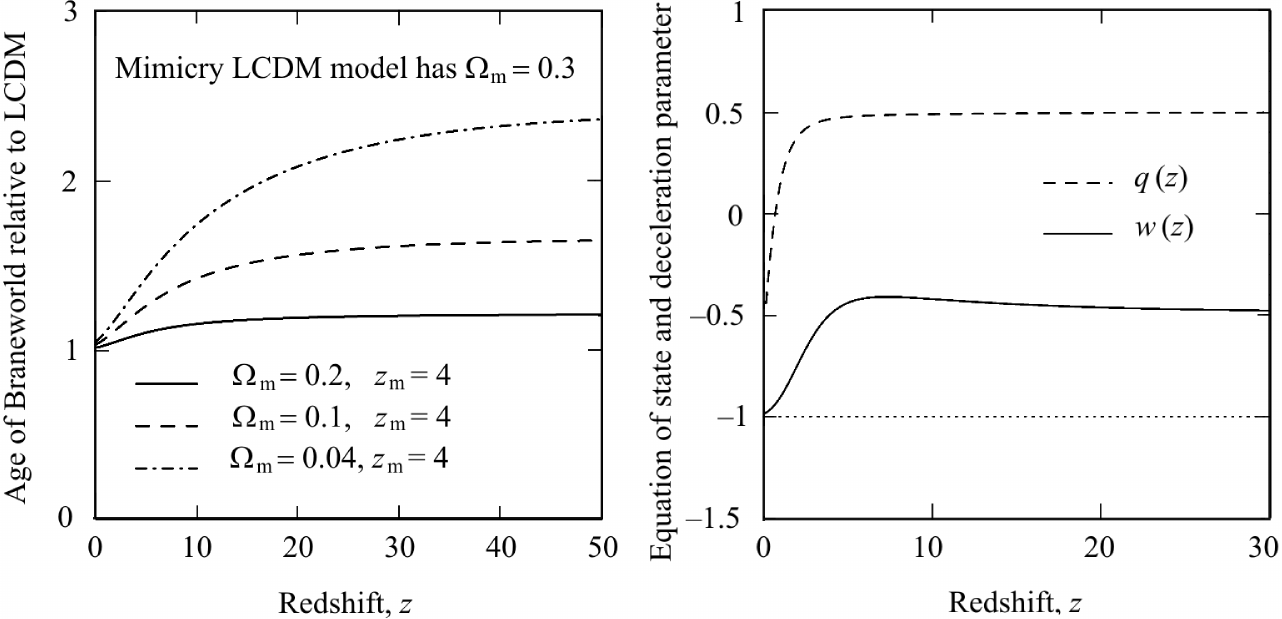}
\vskip-2mm
  \caption{\label{brane-fig:age-mimic}The age of the Universe in the
    BRANE2 model is shown with respect to the LCDM value. %
    \index{age of Universe|)} %
    The mimicry redshift (\ref{brane-eq:new1}) is $z_\mathrm{m} = 4$
    so that $H_\mathrm{ brane}(z) \simeq H_\mathrm{ LCDM}(z)$ at $ z
    \ll 4$. The braneworld models have $\Omega_\mathrm{m} = 0.2$,
    $0.1$, $0.04$ (bottom to top) whereas $\Omega_\mathrm{m}^\mathrm{
      LCDM} = 0.3$. Note that the braneworld models are older than
    LCDM. Figure taken from \cite{mimicry}}
    \vspace*{-2mm}
\caption{\label{brane-fig:state-mimic}The effective equation of
    state (solid line) and the deceleration parameter (dashed line) of
    the BRANE2 model are shown. (The dotted line shows $w=-1$ which
    describes the LCDM model.) The mimicry redshift
    (\ref{brane-eq:new1}) is $z_\mathrm{m} = 4$ so that $H_\mathrm{
      brane}(z) \simeq H_\mathrm{ LCDM}(z)$ at $ z \lesssim 4$. The
    braneworld has $\Omega_\mathrm{m} = 0.2$ whereas
    $\Omega_\mathrm{m}^\mathrm{ LCDM} = 0.3$. Figure taken from
    \cite{mimicry}}\vskip1mm
\end{figure}

Cosmological tests based on the luminosity distance and angular
diameter distance typically probe lower redshifts $z \lesssim 2$. %
\index{angular diameter distance} %
Therefore, if the mimicry redshift is $z_\mathrm{m} \geq 2$, the
braneworld model will, for all practical purposes, be
indistinguishable from the LCDM cosmology on the basis of these
tests alone. However, tests which probe higher redshifts should be
able to distinguish between these models.  For instance, since $H(z)
< H_\mathrm{ LCDM}(z)$ in BRANE2 at redshifts larger than the
mimicry redshift, it follows that the age of the Universe will be
greater in this model than in the LCDM cosmology. This is
illustrated in Fig.~4.\ref{brane-fig:age-mimic} for three distinct
values of the cosmological density parameter: $\Omega_\mathrm{m} =
0.2$, $0.1$, $0.04$, all of which are lower than $\Omega_{\rm
m}^\mathrm{ LCDM} = 0.3$. Since the late-time evolution of the
Universe is
\begin{equation}
  t(z) \simeq \frac{2}{3H_0 \sqrt{\Omega_\mathrm{ m}} } (1+z)^{-3/2} ,
\end{equation}
one finds, for $z \gg 1$,
\begin{equation}
  \frac{t_\mathrm{ brane}}{t_\mathrm{ LCDM}}(z) \simeq \sqrt{\frac{\Omega_{\rm m}^\mathrm{LCDM}}{\Omega_{\rm m}}} .
\end{equation}
Since $\Omega_\mathrm{m} < \Omega_{\rm m}^\mathrm{ LCDM}$ in the
BRANE2 model, we find that the age of a BRANE2 Universe is greater
than that of a LCDM Universe. (The reverse is true for the BRANE1
model, for which $\Omega_\mathrm{ m} >
\Omega_{\mathrm{m}}^\mathrm{LCDM}$.)

The altered rate of expansion in the braneworld model at late times
($z >$ $> z_{\rm m}$) also affects other cosmological quantities
including the redshift of reionization which, for the BRANE2 model,
becomes smaller than that in the LCDM cosmology. This is because the
lower value of $H(z)$ in the BRANE2 model (relative to the LCDM
model), when substituted to (\ref{brane-reion}), gives a
correspondingly lower value for $z_\mathrm{ reion}$ for an identical
value of the optical depth $\tau$ in both models. (In fact, it is
easy to see that, for the BRANE2 model, the value of $z_\mathrm{
reion}$ decreases with decreasing $z_{\rm m}$ and
$\Omega_\mathrm{m}$.)

Both an increased age of the Universe and a lower redshift of
reionization are attractive properties of the braneworld model
which, as we have seen, mimics the LCDM cosmology at lower redshifts
$z < z_{\rm m}$\,\footnote{\,Note that the decreased redshift of
reionization and
  the increased age of the Universe are properties that the BRANE2
  model shares with the loitering Universe %
  \index{loitering Universe}%
  \cite{loiter} discussed in Sec.~\ref{brane-sec:loiter}.}. It is
important to note that the presence of high-redshift quasi-stellar
objects (QSO's) and galaxies at redshifts $z \gtrsim 6$ indicates
that the process of structure formation was already in full swing at
that early epoch when the LCDM Universe was less than a billion
years old. Most models of QSO's rely on a central supermassive black
hole ($M_\mathrm{ BH} \gtrsim 10^9M_\odot$) to power the quasar
luminosity via accretion. Since structure forms hierarchically in
the cold dark matter scenario, the presence of such supermassive
black holes at high redshift suggest that they formed through an
assembly mechanism involving either accretion or mergers or both. It
is not clear whether either of these processes is efficient enough
to assemble a large number of high-redshift QSO's in a LCDM
cosmology \cite{richards03,bh}.  We would like to note in this
section that braneworld cosmology may successfully alleviate some of
the tension currently existing between theory and observations at
moderate redshifts, while allowing the Universe to be ``LCDM-like''
at the present epoch.

The effective equation of state and the deceleration parameter of
the \mbox{BRANE2} model are shown in
Fig.~4.\ref{brane-fig:state-mimic}. The braneworld has
$\Omega_\mathrm{m}= 0.2$ and, at $z \lesssim 4$, masquerades as a
higher-density LCDM model with $\Omega_\mathrm{m}^\mathrm{ LCDM} =
0.3$.  Note that the {\em
  effective\/} equation of state (\ref{brane-state}) is a {\em
  model-dependent\/} quantity, involving the model-dependent
cosmological parameter $\Omega_\mathrm{m}$ in its definition.  In
our case, we use the braneworld theory as our model with
$\Omega_\mathrm{m}$ defined in (\ref{brane-omegas}), and the
effective equation of state (\ref{brane-state}) is then {\em
  redshift-dependent\/} even du\-ring the mimicry period when
$H_\mathrm{ brane} \bigl( \Omega_\mathrm{m}, z \bigr) \simeq
H_\mathrm{ LCDM} \bigl( \Omega_\mathrm{m}^\mathrm{ LCDM},z \bigr)$.
A theorist who is unaware of the possibility of cosmic mimicry, when
reconstructing the cosmic equation of state from (\ref{brane-state})
with $\Omega_\mathrm{m}^\mathrm{ LCDM} = 0.3$ in the place of
$\Omega_\mathrm{m}$, will arrive at a different conclusion that $w =
-1$. This example demonstrates some of the pitfalls associated with
the cosmological reconstruction of the equation of state, which
depends on the underlying theoretical model and for which an
accurate knowledge of $\Omega_\mathrm{m}$ is essential; see [322,
557, 563---566] for a discussion of related issues.

The ``cosmic mimicry'' exhibited by braneworld cosmology can be
related to the gravitational properties of braneworld theory
described in Sec.~\ref{brane-sec:gravity}.  In this context, it is
remarkable that the parameter $\beta$ introduced in
(\ref{brane-beta}) is very close (in absolute terms) to the
parameter $\alpha$ introduced in (\ref{brane-alpha}) in our
discussion of mimicry models. Specifically,
\begin{equation}
  \label{brane-albeta}
  \beta = \frac{1 - \Omega_\mathrm{m}}{2 \Omega_\ell} \mp \alpha ,
\end{equation}
so that\vspace*{-2mm}
\begin{equation}
  \label{brane-alphabeta}
  |\beta| \approx \alpha
\end{equation}
when $|1 - \Omega_\mathrm{m}| \ll \Omega_\ell\,$. This last
inequality follows naturally from condition (\ref{brane-eq:mimic0})
for values of $\alpha$ of order unity, which are of interest to us.
As a consequence, the term which appears in the ``renormalization''
of the cosmological mass density $\Omega_\mathrm{m}$ in
(\ref{brane-lcdm}) is almost identical to the term which redefines
the gravitational constant in (\ref{brane-g-eff}).  This coincidence
can be explained by inspecting the brane equation
(\ref{brane-effective}).  First, we recall that cosmological
solutions without dark radiation are embeddable in the
anti-de~Sitter bulk space-time (see Sec.~\ref{brane-sec:setup}), so
that $C_{ab} = 0$ for these solutions. For high cosmological matter
densities, the quadratic expression (\ref{brane-q}) dominates in
Eq.~(\ref{brane-effective}), and the Universe is described by the
``bare'' Einstein equation $m^2 G_{ab} - T_{ab} = 0$, with the
effective gravitational coupling being equal to $1 / m^2$. As the
matter density decreases, the role of this quadratic term becomes
less and less important, and the effective gravitational coupling
eventually is determined by the linear part of
Eq.~(\ref{brane-effective}), i.e., by the gravitational constant
(\ref{brane-g-eff}).

Thus, comparing (\ref{brane-g-eff}) and (\ref{brane-lcdm}), one has
the following natural relation, valid to a high precision in view of
(\ref{brane-alphabeta}):
\begin{equation}
  \label{brane-omega-m}
  \Omega_\mathrm{m}^\mathrm{ LCDM} = \frac{8 \pi G_\mathrm{ eff} \rho_0}{3 H_0^2 } .
\end{equation}

Since, in our case, interesting values of the parameters
$\Omega_\ell$ and $\Omega_{\Lambda_\mathrm{b}}$ are
con\-si\-de\-ra\-bly greater than unity, the RS constraint
(\ref{brane-RS}), which can also be writ\-ten in the form
\begin{equation}
  \Omega_\sigma \pm 2 \sqrt{\Omega_\ell \Omega_{\Lambda_\mathrm{b}}} = 0 ,
\end{equation}
is satisfied to a good precision in view of
Eq.~(\ref{brane-omega-r1}). Note that $\Omega^\mathrm{
  LCDM}_\mathrm{m}\approx 1$ in the case of the RS constraint.  A
slight violation of the RS constraint causes the appearance of a
small
effective cosmological constant (\ref{brane-lambda-eff}) on the brane, %
\index{cosmological constant} %
which can be thought to be inessential for the discussion of the
Newtonian part of the gravitational physics performed in
Sec.~\ref{brane-sec:gravity}.

The cosmological model under consideration appears to safely pass
the existing constraints on the variation of the gravitational
constant from primordial abundances of light elements synthesized in
the big-bang nucleosynthesis (BBN) and from CMB anisotropy
\cite{UIY}. The value of the gravitational constant at the BBN epoch
in our model coincides with the value measured on small scales
(\ref{brane-g-obs}), and the effective gravitational constant
(\ref{brane-g-tilde}) or (\ref{brane-g-eff}) that \mbox{might}
affect the large-scale dynamics of the Universe responsible for the
CMB fluctuations is within the
uncertainties estimated in \cite{UIY}. %
\index{cosmic microwave background (CMB)} %

Cosmic mimicry in braneworld models is most efficient in the case of
parameter $\alpha \sim 1$, which, according to (\ref{brane-alpha}),
implies that the two spatial scales, namely, the brane length scale
given by (\ref{brane-ell}) and the curvature scale of the bulk are
of the same order: $\ell \sim \ell_{\rm warp} = \sqrt{- 6 /
\Lambda_{\rm
    b}}\,$.  This coincidence of the orders of magnitude of completely
independent scales can be regarded as some tuning of parameters,
although it is obviously a mild one.\vspace*{1mm}


\section[\!Loitering$_{ }$]{\!Loitering\vspace*{-1mm}} \label{brane-sec:loiter}

\subsection{\!Loitering Universe}

\hspace*{3cm}\index{loitering Universe}In the models of dark energy,
the deceleration parameter at some point
changes sign while the Hubble parameter is usually assumed to be a %
\index{Hubble parameter} %
monotonically decreasing function of the cosmic
time\,\footnote{\,Phantom
  models may provide an exception to this rule; see \cite{sahni04} and
  references therein.}. In the present section, we show that this need
not necessarily be the case in braneworld models, and that
compelling dark-energy models can be constructed in which $H(z)$
dips in value at high redshifts. In these models, $dH(z)/dz \simeq
0$ at $z_\mathrm{
  loit} \gg 1$, which is called the ``loitering redshift''.  (A
Universe which loiters has also been called a ``hesitating''
Universe, since, if $H(z_{\mathrm{loit}}) \simeq 0$, the Universe
hesitates at the redshift $z_\mathrm{ loit}$ for a lengthy period of
time --- before either collapsing or re-expanding.) Loitering
increases the age of the Universe at high $z$ and also provides a
boost to the growth of density inhomogeneities, thereby endowing a
dark-energy model with
compelling new properties. %
\index{age of Universe} %

Before discussing loitering in braneworld models \cite{loiter}, we
briefly review the status of loitering in standard General
Relativity in this subsection. Within a FRW setting, loitering can
only arise in a Universe which is spatially closed and which is
filled with matter
and a cosmological constant %
\index{cosmological constant} %
(or some other form of dark energy --- see \cite{sfs92}). The
evolution of such a Universe is described by the
equation\vspace*{-2mm}
\begin{equation}
  H^2 = \frac{8\pi G}{3}\frac{\rho_{0} a_0^3}{a^3} + \frac{\Lambda}{3} - \frac{\kappa}{a^2}
  , \qquad \kappa = 1 , \label{brane-eq:frw}
\end{equation}
where $\rho_{0}$ is the present matter density. Loitering in
(\ref{brane-eq:frw}) arises if the cur\-va\-tu\-re term ($1/a^2$) is
large enough to substantially offset the dark-matter + +~dark-energy
terms but not so large that the Universe collapses. The redshift at
which the Universe loitered can be determined by rewriting
(\ref{brane-eq:frw}) in the form
\begin{equation}
  h^2(z) \equiv \frac{H^2(z)}{H_0^2} = \Omega_\mathrm{ m} (1 + z)^3  + \Omega_{\Lambda} +
  \Omega_{\kappa} (1 + z)^2  , \label{brane-eq:frw1}
\end{equation}
where $\Omega_\kappa = -\kappa/a_0^2H_0^2$, $\Omega_\mathrm{ m} =
8\pi G\rho_{0}/3H_0^2$, $\Omega_{\Lambda} = \Lambda/3H_0^2$, the
subscript ``{\small 0}'' refers to present epoch, and the constraint
equation requires $\Omega_\kappa = 1 - \Omega_\mathrm{ m} -
\Omega_{\Lambda}$.

The loitering condition $dh/dz = 0$ gives
\begin{equation}
1 + z_\mathrm{ loit} = \frac{2\vert\Omega_\kappa\vert}{3
\Omega_\mathrm{ m}} ,
\end{equation}
and it is easy to show that $z_\mathrm{ loit} \leq 2$ for
$\Omega_\mathrm{ m} \geq 0.1$ \cite{sfs92}. (Note that a large value
of $\vert\Omega_\kappa\vert$ can cause the Universe to recollapse.)
The value of the Hubble parameter at loitering can be determined by
substituting $z_\mathrm{ loit}$ into (\ref{brane-eq:frw1}). Note
that, since ${\ddot a}/a = {\dot H} + H^2$, it follows that $({\ddot
  a}/a)\big\vert_{z=z_\mathrm{ loit}} = H^2(z_\mathrm{ loit})$ at
loitering. (The special case ${\dot a} = 0$, ${\ddot a} = 0$
corresponds to the static Einstein Universe \cite{Sahni2000}. For a
detailed discussion of loitering in FRW models with dark energy see
\cite{sfs92}. Loitering in more general contexts has been discussed
in \cite{robert1,robert2}.)

Interest in loitering FRW models has waxed and waned ever since the
original discovery of a loitering cosmology by Lema\^{\i}tre over
seventy years ago \cite{Lemaitre1927}. Among the reasons why the
interest in loitering appears to have declined in more recent times
are the following: (i)~even though loitering models can accommodate
an accelerating Universe, the loitering redshift is usually small:
$z_\mathrm{ loit} \leq 2$ in LCDM; (ii)~loitering models require a
large spatial curvature, which is at variance with inflationary
predictions and CMB observations both of which support a flat
Universe. %
\index{cosmic microwave background (CMB)} %
As we shall show, in marked contrast with the above scenario,
loitering in braneworld models can take place in a spatially flat
Universe and at high redshifts ($z \gtrsim 6$). At late times, the
loitering braneworld model has properties similar to those of LCDM.

\subsection{\!Loitering in braneworld models}

\hspace*{3cm}Loitering can be realized in a braneworld model
described by action (\ref{brane-action}) with $N = 1$ and
cosmological equation (\ref{brane-hubble1}), which, for a spatially
flat Universe ($\kappa = 0$) can be written in the form
\begin{equation}
  \label{brane-hubble}
  H^2 (a)  = \frac{A}{a^3} + B + \frac{2}{\ell^2} \left[ 1 \pm \sqrt{1 + \ell^2
      \left(\!\frac{A}{a^3} + B  - \frac{\Lambda_\mathrm{ b}}{6} - \frac{C}{a^4} \!\right)} \right]\!,
\end{equation}
where
\begin{equation}\label{brane-ab}
  A = \frac{\rho_{0} a_0^3}{3 m^2} , \quad B = \frac{\sigma}{3 m^2} , \quad
  \ell = \frac{2 m^2}{M^3} .
\end{equation}

Of crucial importance to the present analysis is the dark-radiation
term $C/a^4$ in (\ref{brane-hubble}) whose presence is a generic
feature in braneworld models and which describes the projection of
the bulk degrees of freedom onto the brane. [As was noted in the
remark following Eq.~(\ref{brane-fR}), it corresponds to the
presence of the bulk black hole.] An interesting situation arises
when $C< 0$ and $\ell^2 {|C|/a^4} \gg 1$. In this case, if $\ell^2
{|C|/a^4}$ is larger than the remaining terms under the square root
in (\ref{brane-hubble}), then that equation reduces
to\,\footnote{\,The
  negative value of the dark-radiation term implies the presence of
  black hole with negative mass~--- hence, naked singularity~--- in
  the complete extension of the bulk geometry.  In principle, this
  singularity could be ``closed from our view'' by another (invisible)
  brane.} %
\index{singularity} %
\begin{equation}
H^2(a) \approx \frac{A}{a^3} + B \pm \frac{2 \sqrt{- C}}{\ell a^2} .
\label{brane-eq:loit1}
\end{equation}

Equation (\ref{brane-eq:loit1}) bears a close formal resemblance to
(\ref{brane-eq:frw}), which gave rise to loitering solutions in
standard FRW geometry for $\kappa = 1$. Indeed, the role of the
spatial curvature in (\ref{brane-eq:loit1}) is played by the
dark-radiation term; consequently, a spatially open Universe is
mimicked by the BRANE2 model [the upper sign in
(\ref{brane-hubble})] while a closed Universe is mimicked by BRANE1
[the lower sign in (\ref{brane-hubble})]. In analogy with standard
cosmology, one might expect the braneworld model
(\ref{brane-hubble}) to show loitering behavior in the BRANE1 case.
This is indeed the case, and stronly loitering solutions to
(\ref{brane-hubble}) and (\ref{brane-eq:loit1}) can be found by
requiring $H'(a)=0$.

Although this is the general procedure which we follow, for
practical pur\-po\-ses it is more suitable to rewrite
(\ref{brane-hubble}) in the form (\ref{brane-hubble0}).  When the
dark-radiation term is strongly dominating, Eq.~(\ref{brane-hubble})
or, for that matter (\ref{brane-hubble0}) with $\kappa = 0$, then
reduces to
\begin{equation}
  \frac{H^2(z)}{H_0^2} \simeq \Omega_\mathrm{ m}(1 +
  z)^3 + \Omega_\sigma - 2 \sqrt{\Omega_\ell \Omega_C}(1 + z)^2 \,
  , \label{brane-eq:hubble2}
\end{equation}
which is the braneworld analog of (\ref{brane-eq:frw1}). The
loitering redshift in this case can be defined by the condition
$H'(z_\mathrm{
  loit}) = 0$; as a result, one gets
\begin{equation}
  1 + z_\mathrm{ loit} \simeq \frac{4}{3} \, \frac{\sqrt{\Omega_C\Omega_\ell}}{\Omega_\mathrm{ m}}
  . \label{brane-eq:loit}
\end{equation}
From this expression we find that the Universe will loiter at a
large redshift $z_\mathrm{ loit} \gg 1$ provided $\Omega_C
\Omega_\ell \gg \Omega_\mathrm{ m}^2$. Since $\Omega_\mathrm{ m}^2
\ll 1$, this is not difficult to achieve in practice. Successful
loitering of this type requires the following two conditions to be
satisfied:
\begin{equation}
\begin{array}{c}
  \Omega_C(1 + z_\mathrm{ loit})^4 \gg \Omega_\mathrm{ m}(1 + z_\mathrm{ loit})^3 + \Omega_\sigma + \Omega_\ell + \Omega_{\Lambda_\mathrm{ b}} ,
  \\[3mm]
  \Omega_\sigma \sim \sqrt{\Omega_\ell \Omega_C}(1 + z_\mathrm{
    loit})^2 .
\end{array}
\label{brane-eq:constraint}
\end{equation}

The first inequality ensures that the dark-radiation term dominates
over the remaining terms under the square root of
(\ref{brane-hubble1}) during loitering, while the second makes sure
that this term is never so large as to cause the Universe to
recollapse.

Substituting the value for $1 + z_\mathrm{ loit}$ from
(\ref{brane-eq:loit}) into (\ref{brane-eq:constraint}), we
obtain\vspace*{-1mm}
\begin{equation}
  \Omega_\sigma \sim \frac{(\Omega_C\Omega_\ell)^{3/2}}{\Omega_\mathrm{ m}^2} \gg \Omega_\ell
  , \label{brane-eq:constraint1}
\end{equation}\vspace*{-3mm}

\noindent which is a necessary condition for loitering in our
braneworld model.

Finally, the Hubble parameter at loitering is given by the
approximate expression\vspace*{-1mm}
\begin{equation}
  \label{brane-eq:hub_loiter}
  \frac{H^2(z_\mathrm{ loit})}{H_0^2} \simeq \Omega_\sigma -
  \frac{32}{27}\frac{(\Omega_C\Omega_\ell)^{3/2}}{\Omega_\mathrm{
      m}^2} .
\end{equation}

\index{loitering Universe}%
Note that conventional loitering is usually associated with a
vanishingly small value for the Hubble parameter at the loitering
redshift \cite{sfs92}. The Hubble parameter at loitering can be set
as close to zero as possible; however, we do not require it to be
{\em
  very\/} close to zero. A small ``dip'' in the value of $H(z)$, which
is sufficient for our purposes, arises for a far larger class of
parameter values than the more demanding condition $H(z_\mathrm{
  loit}) \simeq 0$.

Moreover, in a wide range of parameters, the Universe evolution may
not exhibit a minimum of the Hubble parameter $H(z)$.  In this case,
the definition of the loitering redshift by the condition
$H'(z_\mathrm{ loit}) = 0$ is not appropriate and can be generalized
in several different ways, one of which is described in
Sec.~\ref{brane-sec:loiter-param}.

\begin{figure}
\vskip1mm
    \includegraphics[width=13cm]{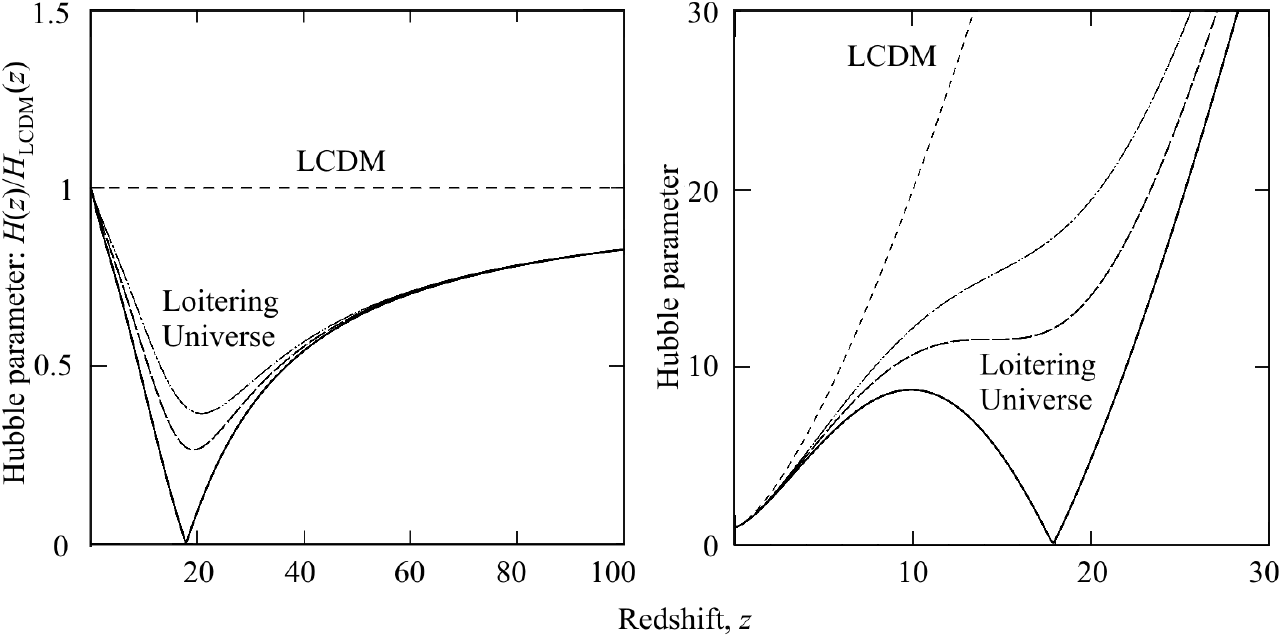}
    \vskip-2mm
  \caption{The Hubble parameter for a Universe that loiters at %
    \index{Hubble parameter} %
    $z_\mathrm{ loit} \simeq 18$. Parameter values are
    $\Omega_\mathrm{ m} = 0.3$, $\Omega_C = 8.0$, $\Omega_\ell = 3.0$,
    and $\Omega_{\Lambda_\mathrm{ b}}/10^5 = 6,\, 4.5,\, 3.4$ (solid
    lines, from top to bottom).  The left panel shows the Hubble
    parameter with respect to the LCDM value while, in the right
    panel, the LCDM (dashed) and loitering (solid) Hubble parameters
    are shown separately. Figure taken from
    \cite{loiter} \label{brane-fig:hubble}}
\end{figure}

An example of a loitering model is shown in
Fig.~4.\ref{brane-fig:hubble}, where the Hubble parameter of a
Universe which loitered at $z \simeq 18$ is plotted against the
redshift, keeping $\Omega_\mathrm{ m}$, $\Omega_\ell$, and
$\Omega_C$ fixed and varying the value of $\Omega_{\Lambda_\mathrm{
b}}$. The right-hand panel of Fig.~4.\ref{brane-fig:hubble}
illustrates the fact that the loitering Universe can show a variety
of interesting behavior: (i)~top curve, $H(z)$ is monotonically
increa\-sing and $H'(z) \simeq \mathrm{
  constant}$ in the loitering interval; (ii)~middle curve, $H(z)$
appears to have an inflexion point ($H' \simeq 0$, $H'' \simeq 0$)
during loitering; (iii)~lower curve, $H(z)$ has both a maximum and a
minimum, the latter occurring in the \mbox{loitering regime.}

\index{Randall---Sundrum model} %
At this point, we would like to stress an important difference
existing between the Randall---Sundrum braneworld
(\ref{brane-cosmolim}) and our Universe (\ref{brane-hubble}) due to
which the latter can accommodate a large value of dark radiation
without violating nucleosynthesis constraints whereas the former
cannot. In the Randall---Sundrum braneworld (\ref{brane-cosmolim}),
the dark-radiation term ($C/a^4$) affects cosmological expansion in
{\em
  exactly the same way\/} as the usual radiation density
$\rho_\mathrm{ r}$, so that this model comes into serious conflict
with the predictions of the big-bang nucleosynthesis if $|C|$ is
very large \cite{ichiki}. In the loitering braneworld, on the other
hand, the dark-radiation term resides under the square root in
(\ref{brane-hubble}); due to this circumstance its effect on the
cosmological expansion is less severe and, more importantly, {\em
  transient\/}. Indeed, even if the dark-radiation term is very large
($|C|/a^4 > \rho_\mathrm{ m},\,\rho_\mathrm{ r}$), its influence on
expansion can only be $\propto$$ 1/a^2$, which does not pose a
serious
threat to the standard predictions of the big-bang nucleosynthesis.%

\begin{figure}
\vskip1mm
    \includegraphics[width=13cm]{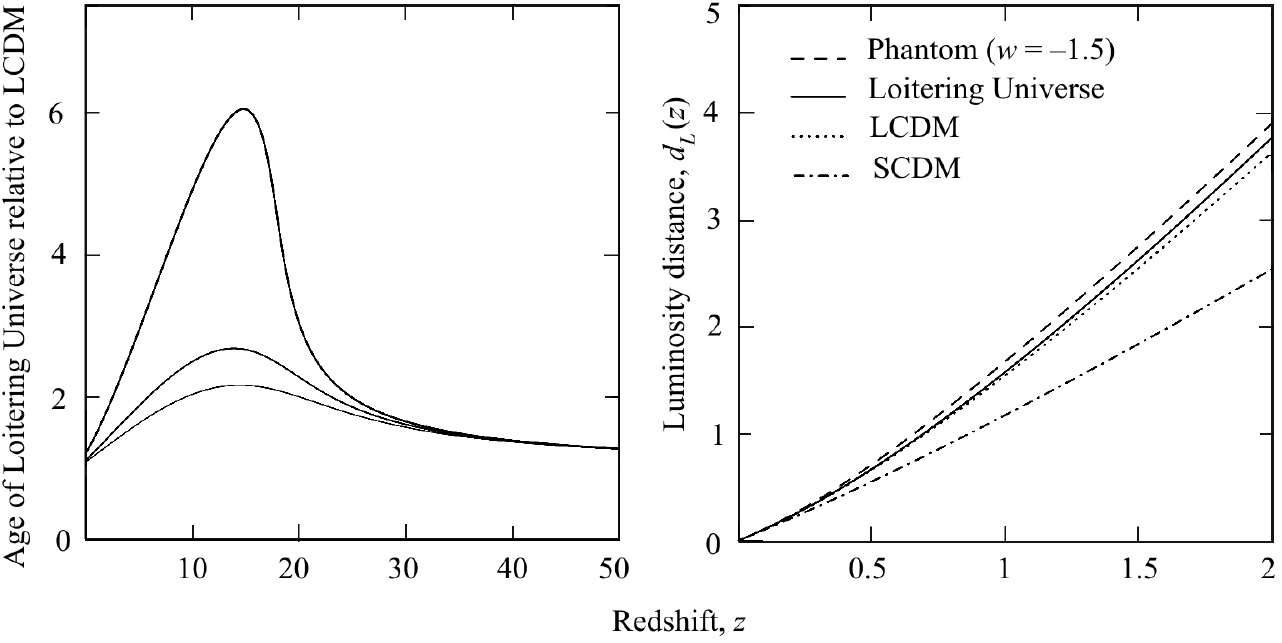}
    \vskip-2mm
      \caption{\label{brane-fig:age-loiter} In the left panel, the age of
    three loitering models is shown relative to the age in LCDM (model
    parameters are the same as in Fig.~4.\ref{brane-fig:hubble}). Note
    that the age of the Universe near loitering ($z_\mathrm{ loit}
    \sim 18$) is significantly greater than that in LCDM although, at
    the present epoch, the difference in ages between the two models
    is relatively small. In the right panel, the luminosity distance
    in a Universe that loiters at $z_\mathrm{ loit} \simeq 18$ is
    shown in comparison with other models. Note that the luminosity
    distance in the loitering model is only slightly larger than that
    in LCDM and smaller than that in a phantom model with $w =
    -1.5$. Figure taken from \cite{loiter}}%
  \index{loitering Universe}%
  \index{age of Universe} %
\end{figure}

 A loitering Universe could have several important
cosmological con-\linebreak sequences:


(i)  \index{age of Universe} %
  The age of the Universe during loitering {\em increases\/}, as shown
  in Fig.~4.\ref{brane-fig:age-loiter}. The reason for this can be seen
  immediately from expression (\ref{brane-age}). Clearly, a lower
  value of $H(z)$ close to loitering will boost the age of the
  Universe at that epoch. In Fig.~4.\ref{brane-fig:age-loiter}, the age
  of the Universe is plotted with reference to a LCDM Universe, which
  has been chosen as our fiducial model. It is interesting to note
  that, while the {\em age at loitering\/} can be significantly larger
  in the loitering model than in LCDM $\left[t_\mathrm{
      loit}(z_\mathrm{ loit}) \sim \mathrm{ few}\times t_\mathrm{
      LCDM}(z_\mathrm{ loit})\right]$, the present age of the Universe
  in both models is comparable $\left[t_\mathrm{ loit}(0) \lesssim
    1.2\times t_\mathrm{ LCDM}(0)\right]$\,\footnote{\,The age of a LCDM
    Universe at $z \gg 1$ is ~$t(z) \simeq
    (2/3H_0\sqrt{\Omega_\mathrm{ m}}) (1+z)^{-3/2} = 5.38\, \times$ $\times\, 10^8
    (1+z/10)^{-3/2} $ years for $\Omega_\mathrm{ m} = 0.3$ and $h =
    0.7$.}. An important consequence of having a larger age of the
  Universe at $z \sim 20$ (or so) is that astrophysical processes at
  these redshifts have more time in which to develop. This is
  especially important for gravitational instability which forms
  gravitationally bound systems from the extremely tiny fluctuations
  existing at the epoch of last scattering. Thus, an early loitering
  epoch may be conducive to the formation of Population~III stars and
  low-mass black holes at $z \sim 17$ and also of $\sim 10^9 M_\odot$
  black holes at lower redshifts ($z \sim 6$).

(ii) In Fig.~4.\ref{brane-fig:age-loiter}, the luminosity
  distance (\ref{brane-lumdis2}) for the loitering model is shown,
  again with LCDM as the base model.  One finds from
  Fig.~4.\ref{brane-fig:age-loiter} that the luminosity distance in the
  loitering model, although somewhat larger than in LCDM, is smaller
  than in a phantom model with $w=-1.5$. Since both phantom and LCDM
  models provide excellent fits to type~Ia supernova data
  \cite{alam03,alam04,Riess1998,Caldwell2003}, we expect our family of
  ``high redshift loitering models'' to also be in good agreement with
  observations. (A detailed comparison of loitering models with
  observations lies outside of the scope of the present book.)

  The reason why both the luminosity distance and the current age of
  the Universe have values which are close to those in the LCDM model
  is clear from Fig.~4.\ref{brane-fig:hubble}, where we see that the
  difference between the Hubble parameters for the loitering models %
  \index{Hubble parameter} %
  and LCDM model is small at low redshifts. Since both $d_L(z)$ and
  $t(z)$ probe $H^{-1}(z)$, and since the value of the Hubble
  parameter at low $z$ is much smaller than its value at high $z$
  (unless parameter values are chosen to give $H(z_\mathrm{ loit})
  \simeq 0$ with a high precision), it follows that $|d_L^\mathrm{
    loit}(z)\, -$ $- \,d_L^\mathrm{ LCDM}(z)| \ll d_L^\mathrm{ LCDM}(z)$ and
  $|t^\mathrm{ loit}(z) - t^\mathrm{ LCDM}(z)| \ll t^\mathrm{
    LCDM}(z)$ for $z \ll 1$.

(iii) The growth of density perturbations depends sensitively
  upon the behavior of the Hubble parameter, as can be seen from the
  following equation describing the growth of linearized density
  perturbations $\delta = (\rho-{\bar\rho})/{\bar\rho}$ in a FRW
  Universe (ignoring the effects of pressure):
  \begin{equation} {\ddot \delta} + 2H{\dot\delta} - 4\pi
    G{\bar\rho}\,\delta = 0 .
    \label{brane-eq:delta}
  \end{equation}
  In Eq.~(\ref{brane-eq:delta}), the second term $2H{\dot\delta}$
  damps the growth of perturbations; consequently, a lower value of
  $H(z)$ during loitering will boost the growth rate in density
  perturbations, as originally demonstrated in \cite{sfs92}.%
\index{loitering Universe}%

  Here we should note that Eq.~(\ref{brane-eq:delta}) for
  perturbations is perfectly valid only in General Relativity and, in
  principle, may be corrected or modified in the braneworld theory
  under consideration. Thus, for the DGP braneworld model \cite{DGP}
  (which corresponds to setting $\sigma = 0$, $\Lambda_\mathrm{ b} =
  0$ and $C =0$ in Eq.~(\ref{brane-hubble})), the linearized equation
  \begin{equation} {\ddot \delta} + 2H{\dot\delta} - 4\pi
    G{\bar\rho}\left(\!1 + \frac{1}{3 \beta_\mathrm{ DGP}}\!\right) \delta
    = 0
    \label{brane-eq:delta1}
  \end{equation}
  was derived in \cite{LS,LSS}, where
  \begin{equation}
    \beta_\mathrm{ DGP} = - \frac{1 + \Omega_\mathrm{ m}^2 (t)}{1 - \Omega_\mathrm{ m}^2 (t)} ,~~ \Omega_\mathrm{ m} (t) \equiv \frac{8 \pi G \bar \rho (t)}{3 H^2 (t)} .
  \end{equation}
  It is important to note the similarities as well as differences
  between (\ref{brane-eq:delta}) and (\ref{brane-eq:delta1}). Thus,
  cosmological expansion works in the same way for both models and
  introduces the damping term $2H{\dot\delta}$ in
  (\ref{brane-eq:delta}) as well as in (\ref{brane-eq:delta1}).
  However, in contrast to (\ref{brane-eq:delta}), the braneworld
  perturbation Eq.~(\ref{brane-eq:delta1}) has a time-dependent
  (decreasing) effective gravitational constant
  \begin{equation}
    G_\mathrm{ eff} = G \left(\!1 + \frac{1}{3\beta_\mathrm{ DGP}}\!\right)\!,
  \end{equation}
  which is expected to affect the growth rate of linearized density
  perturbations in this model. For the generic braneworld models which
  we study in this book [which has non-zero brane and bulk
  cosmological constants %
  \index{cosmological constant} %
  and especially non-zero dark radiation: $C \ne 0$ in
  Eq.~(\ref{brane-hubble})], the corresponding equation for
  cosmological perturbations remains to be derived. We expect the form
  of this equation to be dependent on the additional boundary
  conditions in the bulk or on the brane. However, we anticipate that
  such an equation will contain the damping term $2H{\dot\delta}$
  which serves to enhance the growth of perturbations in the case of
  loitering.  At the same time, braneworld-specific effects may act in
  the opposite direction leading to the suppression of the growth of
  perturbations relative to the FRW model, as is the case, for
  instance, with the last term in (\ref{brane-eq:delta1}) for the DGP
  model \cite{LS,LSS}.

(iv) The deceleration parameter $q$ and the effective equation
  of state $w$ in our loitering model are given by the expressions
  (\ref{brane-decel}) and (\ref{brane-state}), respectively, in which
  $H(z)$ is to be determined from (\ref{brane-hubble1}) and
  (\ref{brane-omega-r1}). The current values of these quantities are
  \begin{equation}
    q_0 = \frac{3}{2}\Omega_\mathrm{ m}\left [1 -
      \frac{\sqrt{\Omega_\ell}}{\sqrt{\Omega_\ell} + \sqrt{1 + \Omega_{\Lambda_\mathrm{ b}} +
          \Omega_C}}
      \left(\!1 + \frac{4}{3}\frac{\Omega_C}{\Omega_\mathrm{ m}}\!\right)\!\right ] - 1 ,
  \end{equation}
  \begin{equation} \label{brane-eq:w_0} w_0 = -1 -
    \frac{\Omega_\mathrm{ m}}{(1-\Omega_\mathrm{ m})} \cdot
    \frac{\sqrt{\Omega_\ell}} {\sqrt{\Omega_\ell} + \sqrt{1 +
        \Omega_{\Lambda_\mathrm{ b}} + \Omega_C}} \left(\!1 +
      \frac{4}{3}\frac{\Omega_C}{\Omega_\mathrm{ m}} \!\right)\!.
  \end{equation}
  From Eq.~(\ref{brane-eq:w_0}) we find that $w_0 < -1$ if $\Omega_C
  \geq 0$; in other words, our loitering Universe has a phantom-like
  effective equation of state.  (In particular, for the loitering
  models shown in Fig.~4.\ref{brane-fig:hubble}, we have $w_0 =
  -1.035$, $-1.04$, $-1.047$ (top to bottom), all of which are in
  excellent agreement with observations \cite{Seljak:2004xh}.)
  However, in contrast to phantom models, the Hubble parameter in a
  loitering Universe (\ref{brane-hubble1}) does not encounter a future
  singularity %
  \index{singularity} %
  since $\Omega_C,\,\Omega_\sigma > 0$ is always satisfied in models
  which loitered in the past. (Future singularities can arise in
  braneworld models if $\Omega_C,\, \Omega_\sigma < 0$~--- see
  \cite{SS1} for a comprehensive discussion of this issue and
  \cite{NO,ANO} for related ideas.)

  \index{loitering Universe}%
  An interesting consequence of the loitering braneworld is that the
  time-dependent density parameter $\Omega_\mathrm{ m} (z) = 8\pi G
  \rho_\mathrm{ m} (z) /3H^2 (z)$ {\em exceeds\/} unity at some time
  in the past. This follows immediately from the fact that, since the
  value of $H(z)$ in the loitering braneworld model is {\em smaller\/}
  than its counterpart in LCDM, the value of $\Omega_\mathrm{ m}(z)$
  is larger than its counterpart in LCDM. One important consequence of
  this behavior is that, as expected from (\ref{brane-eq:w_0}), the
  effective equation of state blows up precisely when $\Omega_\mathrm{
    m} (z) =1$. In Fig.~4.15, we show that, in
  contrast to the singular behavior of the equation of state, the
  deceleration parameter remains finite and well behaved even as $w
  \to \infty$. Note that the finite behavior of $q(z)$ reflects the
  fact that the equation of state for the braneworld is an {\em
    effective\/} quantity and not a real physical property of the
  theory~--- see \cite{Sahni2003a,ASSS} for a related discussion of
  this issue and \cite{linder04} for an example of a different
  dark-energy model displaying similar behavior. (The deceleration
  parameter experiences near-singular behavior at the higher,
  loitering redshift, as $H \to 0$ so that $q \to \infty$.)

  \begin{figure}[h!]
\vskip1mm
\includegraphics[width=6.5cm]{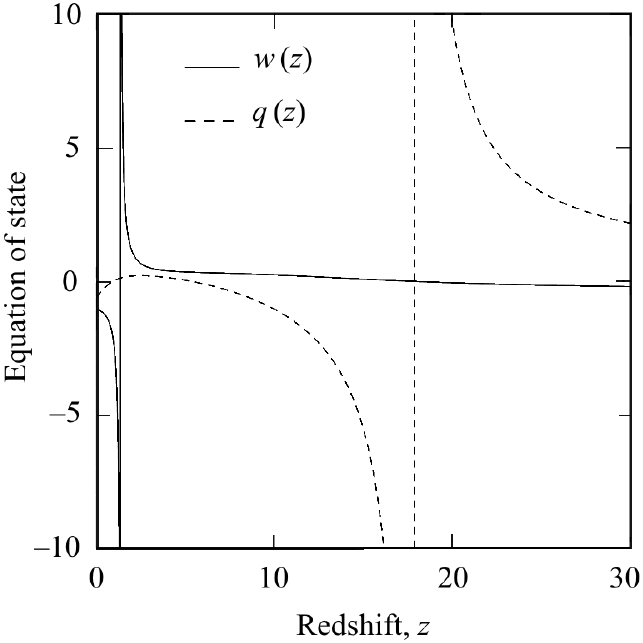}\hspace*{0.5cm}\raisebox{0.0cm}{\parbox[b]{6.0cm}{\caption{\label{brane-fig:state-loiter}
The effective equation of
      state of dark energy (solid) and the deceleration parameter
      (dashed) are shown for a Universe which loitered at $z \simeq
      18$. Note that the effective equation of state of dark energy
      becomes infinite at low redshifts when $\Omega_\mathrm{ m}(z) =
      1$. However, this behavior is not reflected in the deceleration
      parameter, which becomes large only near the loitering redshift.
      Figure taken from \cite{loiter}}}}\vspace*{-2mm}
\end{figure}

(v) Finally, we draw attention to the fact that a loitering
  epoch at $z_\mathrm{ loit}$ can significantly alter the reionization
  properties of the Universe at lower redshifts. The electron
  scattering optical depth to a redshift $z_\mathrm{ reion}$ is given
  by Eq.~(\ref{brane-reion}). Clearly, were $H(z)$ to drop below its
  value in LCDM it would imply a lower value for $z_\mathrm{
    reion}$. Since this is precisely what happens in a loitering
  cosmology, one expects ${z_\mathrm{ reion} \vert}_\mathrm{
    loitering} < {z_\mathrm{ reion}\vert}_\mathrm{ LCDM}$ if
  $z_\mathrm{ loit} \lesssim 20$. As an example, consider the
  loitering models shown for illustrative purposes in
  Fig.~4.\ref{brane-fig:hubble}. Not surprisingly, the redshift of
  reionization drops to $z_\mathrm{ reion} \leq 12$ (from the value
  $z_\mathrm{ reion} \simeq 17$) for the loitering models shown in
  Fig.~4.\ref{brane-fig:hubble}.  By decreasing the redshift of
  reionization as well as increasing the age of the Universe, the
  loitering braneworld may help in alleviating the possible tension
  between the high-redshift Universe and dark-energy cosmology. %
\index{age of Universe} %

\subsection{\!The parameter space\\ \hspace*{-1.2cm}in loitering
  models} \label{brane-sec:loiter-param}

\hspace*{3cm}\index{parameter space}As pointed out earlier, while
not all loitering models pass through a
minimum of the Hubble parameter, %
\index{Hubble parameter} %
a minimum value of the ratio $H (z) / H_\mathrm{ LCDM} (z)$ is
generic and is exhibited by all models. It is therefore useful to
supplement the definition of loitering given in
(\ref{brane-eq:loit}) by defining the loite\-ring redshift
$z_\mathrm{ loit}$ as the epoch associated with the minimum of $H
(z) / H_\mathrm{ LCDM} (z)$ (both models are assumed to have the
same value of $\Omega_\mathrm{ m}$). In order to quantify the degree
of loitering, it is useful to introduce the function\vspace*{-1mm}
\begin{equation}
  \label{brane-eq:f}
  f(z) \equiv 1 - \frac{H^2 (z)}{H^2_\mathrm{ LCDM} (z) },
\end{equation}
where $0 \leq f < 1$. Small values $0 \leq f \leq 1/2$ imply {\em
weak
  loitering}, whereas larger values $1/2 < f < 1$ correspond to {\em
  strong loitering}. It is straightforward to derive expressions for
the loitering redshift $z_\mathrm{ loit}$ and the degree of
loitering $f(z_\mathrm{ loit})$:\vspace*{-3mm}
\begin{equation}
  \label{brane-xextr}
  (1 + z_\mathrm{ loit})^4 \approx \frac{3 \left(\! \sqrt{1 + \Omega_{\Lambda_\mathrm{ b}} +
        \Omega_C} + \sqrt{\Omega_\ell} \right)^2}{\Omega_C} ,
\end{equation}\vspace*{-1mm}
\begin{equation}
  \label{brane-fextr}
  f(z_\mathrm{ loit}) \approx \frac{2 \sqrt{\Omega_\ell} \left(\!\sqrt{1 +
        \Omega_{\Lambda_\mathrm{ b}} + \Omega_C} + \sqrt{\Omega_\ell} \right)}{
    \Omega_\mathrm{ m} (1 + z_\mathrm{ loit})^3 } ,
\end{equation}
which are valid under the single assumption $\Omega_\mathrm{ m} (1 +
z_\mathrm{ loit})^3 \ll \Omega_C (1 + z_\mathrm{ loit})^4$,
or\vspace*{-1mm}
\begin{equation}
  \Omega_\mathrm{ m} \ll \Omega_C^{3/4} \left(\! \sqrt{1 + \Omega_{\Lambda_\mathrm{ b}} +
      \Omega_C} + \sqrt{\Omega_\ell} \right)^{1/2}\!.
\end{equation}

From (\ref{brane-xextr}) and (\ref{brane-fextr}) one has the useful
approximate conditions\vspace*{-1mm}
\begin{equation}
  \label{brane-A}
  2 \sqrt{\Omega_\ell} \left(\!\sqrt{1 + \Omega_{\Lambda_\mathrm{ b}} + \Omega_C} +
    \sqrt{\Omega_\ell} \right) \approx \Omega_\mathrm{ m} f(z_\mathrm{ loit}) (1 + z_\mathrm{
    loit})^3 ,
\end{equation}\vspace*{-5mm}
\begin{equation}
  \label{brane-B}
  \Omega_C \Omega_\ell \approx \frac{3}{4}\Bigl[ \Omega_\mathrm{ m}f(z_\mathrm{ loit})
  (1 + z_\mathrm{ loit})\Bigr]^2\!.
\end{equation}

\begin{figure}
\vskip1mm
\raisebox{0.0cm}{\parbox[b]{6.0cm}{\caption{\label{brane-fig:state-loiter}
\label{brane-fig:param_space} The parameter space $\lbrace
    \Omega_\ell, \Omega_C\rbrace$ is shown for models which exhibit
    (i)~weak loitering: \mbox{$f (z_\mathrm{ loit}) \leq 1/2$} in
    (\ref{brane-eq:f}) (lower left corner); (ii)~strong loitering:
    $1/2 <$ $< f (z_\mathrm{ loit}) < 1$ in (\ref{brane-eq:f}) (shaded
    region). The prohibited region corresponding to braneworld models
    which recollapse {\em before\/} reaching the present epoch is
    shown on the far right. The dashed lines show contours of $\lbrace
    \Omega_\ell, \Omega_C\rbrace$ with current values of the
    effective equation of state: $w_0 =$ =~$ -1.01$, $-1.015$, $-1.02$, $-1.025$,
    $-1.03$, $-1.035$ (from left to right). All models loiter at
    $z_\mathrm{ loit} = 20$ and have $\Omega_\mathrm{ m} =
    0.3$. Figure taken from \cite{loiter}}}}\hspace*{0.5cm}\includegraphics[width=6.5cm]{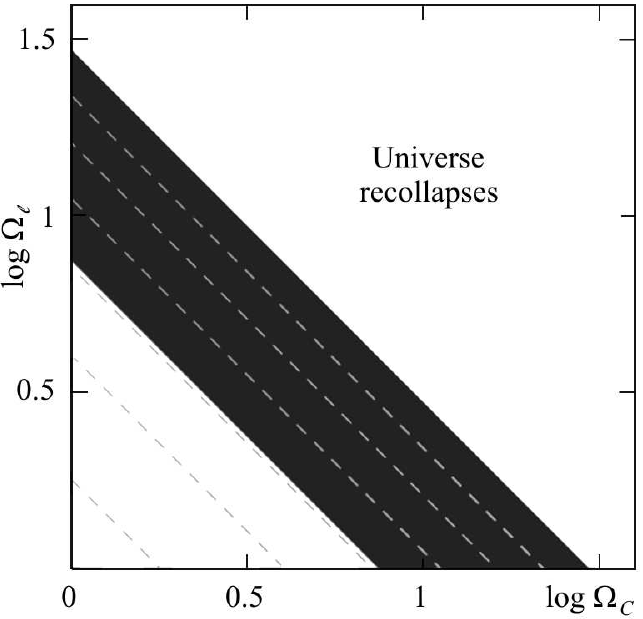}
    \vspace*{-4mm}
\end{figure}

In practice, it is often convenient to take the values of
$\Omega_\mathrm{ m}$, $(1 + z_\mathrm{ loit})$, and $f(z_\mathrm{
  loit})$ as control parameters and to determine the approximate
ranges of $\Omega_\ell$, $\Omega_C$, and $\Omega_{\Lambda_\mathrm{
    b}}$ from equations (\ref{brane-xextr})---(\ref{brane-B}). In
Fig.~4.\ref{brane-fig:param_space}, we show, as an example, the
range of allowed values for the parameter pair $\lbrace \Omega_\ell,
\Omega_C\rbrace$ for a model which loiters at $z_\mathrm{ loit} =
20$ and has $\Omega_\mathrm{ m} = 0.3$.

It is necessary to draw the reader's attention to the fact that not
every set of parameter values gives rise to a ``realistic''
cosmology. For some of them, the Universe recollapses before
reaching the present epoch. (The loitering braneworld shares this
property with a closed FRW Universe, and the reader is referred to
\cite{felten86} for an extensive discussion of this issue.) It is
obvious that the model approaches a recollapsing Universe as the
\mbox{loitering} parameter \mbox{$ f (z_\mathrm{ loit}) \to 1$}.
Thus, setting \mbox{$f (z_\mathrm{ loit}) = 1$} in estimate
(\ref{brane-B}), we obtain the approximate boundary of the region of
recollapsing universes in the parameter space $\lbrace \Omega_\ell,
\Omega_C\rbrace$:\vspace*{-3mm}
\begin{equation}
  \Omega_C \Omega_\ell \gtrsim \frac34 \Omega_\mathrm{ m}^2 (1 + z_\mathrm{ loit})^2\!,
\end{equation}
which corresponds to the ``prohibited'' region in
Fig.~4.\ref{brane-fig:param_space} for the particular choice of
$z_\mathrm{ loit} = 20$ and $\Omega_\mathrm{ m} =
0.3$.\vspace*{-2mm}

\subsection{\!Inflation in braneworld models with loitering}

\hspace*{3cm}The loitering braneworld models considered in the
previous section place certain constraint on the duration of the
inflationary stage, as we are going to show.  First, we note that,
during the inflationary
stage, the Hubble parameter %
\index{Hubble parameter} %
as a function of the scale factor can be approximated with a great
precision as follows [cf.\@ with (\ref{brane-eq:loit1})]:
\begin{equation} \label{brane-eq:inflat} H^2 (a) = \frac{\rho_\mathrm{ i}
    (a)}{3 m^2} - 2 \frac{\sqrt{-C}}{\ell a^2} ,
\end{equation}
where $\rho_\mathrm{ i} (a)$ is the energy density during inflation,
which typically changes very slowly with the scale factor $a$.
Since, on the contrary, the last term in (\ref{brane-eq:inflat})
changes rapidly during inflation, one can easily see that inflation
{\em
  should have a beginning\/} in this model at the scale factor roughly
given by the estimate
\begin{equation}
  a_\mathrm{ i}^2 \simeq \frac{6 m^2 \sqrt{-C}}{\ell \rho_\mathrm{ i}} ,  \mbox{~or~}
   \left(\!\frac{a_\mathrm{ i}}{a_0}\!\right)^{\!\!2} \simeq \frac{2 \rho_0}{\rho_\mathrm{ i}}
  \sqrt{\Omega_\ell \Omega_C} .
\end{equation}
Using (\ref{brane-B}), one can write the following estimate for the
redshift $z_\mathrm{ i}$ at the beginning of
inflation:\vspace*{-3mm}
\begin{equation}
  \label{brane-eq:zinfl}
  (1 + z_\mathrm{ i})^2 \simeq \frac{\rho_\mathrm{ i}}{\rho_0 } \left[\sqrt{3}
    \Omega_\mathrm{ m} f(z_\mathrm{ loit}) \left(1 +  z_\mathrm{ loit} \right)
    \right]^{-1}\!
  ,
\end{equation}
where the loitering redshift $z_\mathrm{ loit}$ and the quantity
$f(z_\mathrm{ loit})$, which quantifies the degree of loitering and
takes values in the range between zero and unity, are defined in
Sec.~\ref{brane-sec:loiter-param}.

To estimate the {\em total number\/} of the inflationary
$e$-foldings, we consider a simple model of inflation based on the
inflaton $\phi$ with potential $V(\phi) = \displaystyle \frac12
m_\phi^2 \phi^2$.  In this case, as can be shown, inflation proceeds
at the values of the scalar field $\phi_\mathrm{ i} \simeq
M_\mathrm{ P} \equiv \sqrt{8
  \pi}\,m$ and ends approximately at $\phi_\mathrm{ f} \simeq
M_\mathrm{ P}^2 / \sqrt{12 \pi}$.  This leads to the following
relation between the typical energy density during inflation and at
its end:
\begin{equation}
\frac{\rho_\mathrm{ i}}{\rho_\mathrm{ f}} \simeq 12 \pi .
\end{equation}
Then using (\ref{brane-eq:zinfl}) and the estimate for the redshift
at the end of inflation
\begin{equation}
  1 + z_\mathrm{ f} = \frac{a_0}{a_\mathrm{ f}}  \simeq \frac{T_\mathrm{ rh}}{T_0} \simeq
  \left(\!\frac{\rho_\mathrm{ f}}{\Omega_\mathrm{ r} \rho_0} \!\right)^{\!\!1/4}
\end{equation}
which assumes that preheating takes place instantaneously with
effective tem\-pe\-ra\-tu\-re $T_\mathrm{ rh}$, we can estimate the
redshift ratio
\begin{equation}
  \begin{aligned}
    \frac{z_\mathrm{ i}}{z_\mathrm{ f}} &\simeq \left[ \frac{4
        \rho_\mathrm{ i}^2}{27\rho_0 \rho_\mathrm{ f}}
      \frac{\Omega_\mathrm{ r}}{\Omega_\mathrm{ m}^2 f^2(z_\mathrm{
          loit})
        (1 + z_\mathrm{ loit})^2 } \right]^{1/4}\!\simeq \\
    &\simeq \left[ \frac{16 \pi \rho_\mathrm{ i}}{9 \rho_0 }
      \frac{\Omega_\mathrm{ r}}{\Omega_\mathrm{ m}^2 f^2(z_\mathrm{
          loit}) (1 + z_\mathrm{ loit})^2 } \right]^{1/4} \!.
  \end{aligned}
\end{equation}
Here, $\Omega_\mathrm{ r} \simeq 10^{-5}$ is the current value of
the radiation density parameter.

For our typical loitering redshift $z_\mathrm{ loit} \approx 18$,
for the degree of loitering $f(z_\mathrm{ loit}) \sim 1$, and for
the estimate of the inflationary energy density in agreement with
the CMB fluctuations spectrum as \cite{lyth-liddle} $\rho_\mathrm{
i} / \rho_0 \sim 10^{112}$, this will restrict the total number of
inflationary $e$-foldings $N$ by
\begin{equation}
  e^N \equiv \frac{z_\mathrm{ i}}{z_\mathrm{ f}} \lesssim 10^{26} \simeq e^{60} .
\end{equation}
It is interesting that the {\em total\/} number of inflationary
$e$-foldings in the loite\-ring braneworld is close to the expected
number of $e$-foldings associated with horizon crossing in
inflationary models \cite{lyth-liddle}. The exact upper bound on the
number of inflationary $e$-foldings depends on a concrete model of
braneworld inflation in the presence of loitering, and we propose to
study this issue in greater detail in a future work.

\index{Hubble parameter} %
Returning to (\ref{brane-eq:inflat}), we would like to draw the
reader's attention to the fact that, depending upon the form of the
inflaton potential, the evolution of the Hubble parameter at very
early times could have proceeded in two fundamentally different and
complementary ways:

(i) If the shape of the inflaton potential $V(\phi)$ is
  sufficiently flat, then, for a field rolling slowly, $\rho_\mathrm{
    i} = \rho_\phi$ behaves like a slowly varying $\Lambda$ term. As a
  result, the $1/a^2$ term is expected to dominate at early times
  giving rise to a cosmological ``bounce'' ($H \simeq 0$) when the two
  terms in (\ref{brane-eq:inflat}) become comparable.

(ii) Alternatively, it might well be that the potential
  $V(\phi)$ is not uniformly flat, but changes its form and becomes
  steep for large values of $\phi$ (within the context of chaotic
  inflation). In this case, the bounce will be avoided if, for small
  values of $a$, $\rho_\mathrm{ i} (a)$ increases faster than the
  $1/a^2$ term in (\ref{brane-eq:inflat}). Such a rapid change in
  $\rho_\mathrm{ i} (a)$ at early times will be accompanied by the
  fast rolling of the inflaton field until the latter evolves to
  values where the potential is sufficiently flat for inflation to
  commence.

Interestingly, both (i) and (ii) lead to departures from scale
invariance of the primordial fluctuation spectrum on very large
scales, and have been discussed in \cite{piao03} and
\cite{contaldi03}, respectively, as providing a means of
suppres\-sing
power on very large angular scales in the CMB fluctuation spectrum. %
\index{cosmic microwave background (CMB)} %
In analogy with the discussion in these papers, we expect that the
present loitering scenario too may give rise to a smaller amplitude
for scalar perturbations %
\index{scalar perturbations}%
on the largest scales, thereby providing better agreement with the
CMB anisotropy results obtained by COBE \cite{COBE1992,cobe} and
WMAP
\cite{WMAP7b}. %
\index{CMB anisotropy} %


\section{\!Quiescent singularities} \label{brane-sec:singular}
\vspace*{-1mm}
\subsection{\!Homogeneous case}
\vspace*{-0.5mm}

\hspace*{3cm}In this section, we describe another interesting
property of the braneworld theory, namely, that it admits
cosmological singularities of very unusual form and nature
\cite{SS1}.\index{singularity|(} Let us first consider the case with
$N = 1$ in action (\ref{brane-action}). The corresponding
cosmological equation of the theory is (\ref{brane-cosmol2}), where
the integration constant $C$ corresponds to the presence of a black
hole in the five-dimensional bulk solution, and the term $C/a^4$
(occasionally referred to as `dark radiation') arises due to the
projection of the bulk gravitational degrees of freedom \mbox{onto
the brane.}

The new singularities that we are going to discuss in this section
are connected with the fact that the expression under the square
root of (\ref{brane-cosmol2}) {\em turns to zero\/} at some point
during evolution, so that solutions of the cosmological equations
{\em cannot
  be continued\/} beyond this point.  There are essentially two types
of `quiescent' singularities displaying this behavior:

A \textbf{type 1} singularity (S1) is essentially induced by the
presence of the `dark radiation' term under the square root of
(\ref{brane-cosmol2}) and arises in either of the following two
cases:

{\footnotesize$\bullet$}\,\,$C > 0$ and the density of matter
increases {\em slower\/} than
  $a^{-4}$ as $a \to 0$. Such singularities occur if the Universe is
  filled with matter having equation of state $p/\rho < 1/3$, an
  example is provided by pressureless matter (dust) for which $\rho
  \propto a^{-3}$.  A special case is an empty Universe ($\rho = 0$).

{\footnotesize$\bullet$}\,\,The energy density of the Universe is
radiation-dominated so that $\rho =$ $=
    \rho_0 / a^4$ and $C > \rho_0 $.

The singularities discussed above can take place either in the past
of an expanding Universe or in the future of a collapsing one.

A {\bfseries type 2} singularity (S2) arises if\vspace*{-1mm}
\begin{equation} \label{brane-ii} \ell^2 \left(\!\frac{\sigma}{3 m^2} -
    \frac{\Lambda_\mathrm{b}}{6} \!\right) < - 1 .
\end{equation}\vspace*{-3mm}

In this case, it is important to note that the combination $\rho / 3
m^2 - C /a^4$ decreases monotonically as the Universe expands. The
expression under the square root of (\ref{brane-cosmol2}) can
therefore become zero at suitably late times, in which case the
cosmological solution cannot be extended beyond this time.
Singularity S2 is even more interesting than S1 since: (i) it can
occur during the late time expansion of the Universe; (ii) it can
occur {\em even if\/} dark radiation is entirely absent ($C = 0$).

\begin{figure}
 \vskip1mm
    \centering{\includegraphics[width=6cm]{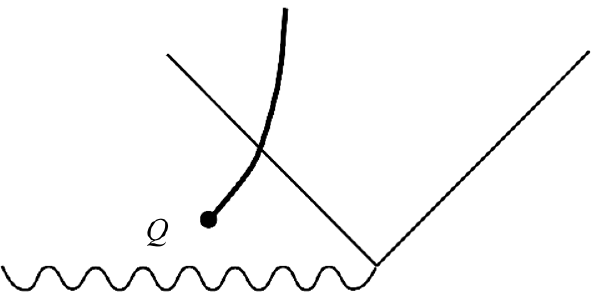}\hspace*{0.5cm}\includegraphics[width=6cm]{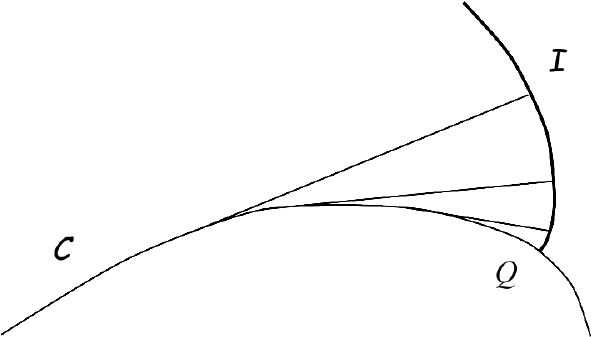}}
\vskip-1mm
  \caption{Conformal diagram showing the trajectory of a
    spatially spherical braneworld embedded in the five-dimensional
    Schwarzschild space-time.  The trajectory is not smoothly
    extendable beyond the point $Q$ \label{brane-fig:brane0}}
    \vspace*{-2mm}
\caption{The involute $\mathcal{I}$ of a planar curve $\mathcal{C}$
    is not smoothly extendable beyond the starting point
    $Q$ \label{brane-fig:involute}} \vspace*{-1mm}
    \end{figure}

For both S1 and S2, the scale factor $a(t)$ and its first time
derivative remain finite, while all the higher time derivatives of
$a(t)$ tend to infinity as the singularity is approached. As an
example consider a type 2 singularity with $C = 0$, for
which\vspace*{1mm}
\begin{equation}
  \frac{d^na}{dt^n} = \mathcal{O} \left(\! \left[ \rho(t) - \alpha \right]^{3/2-n}
  \!\right)\!, \, n \geq 2 ,
\end{equation}\vspace*{-1mm}

\noindent as $\rho(t) \to \alpha = \Lambda_\mathrm{b} m^2/2 - \sigma
- 3m^2/\ell^2$. We therefore find that the scalar curvature $R \to
\infty$ near the singularity, while the energy density and pressure
{\em remain finite}. Although this situation is quite unusual from
the viewpoint of the intrinsic dynamics on the brane, it becomes
comprehensible when one considers the embedding of the brane in the
bulk. As we already know, the cosmological braneworld under
consideration can be isometrically embedded in the five-dimensional
solution of the vacuum Einstein equations described by metric
(\ref{brane-AdS}), (\ref{brane-fR}). The embedding of the brane is
defined by the function (\ref{brane-trajec}), and one can then
proceed to define evolution in terms of the proper cosmological time
$t$ of the induced metric on the brane, given by
(\ref{brane-propert}).  The cosmological singularity under
consideration is connected with the fact that the brane embedding is
not extendable beyond some moment of time $T$ because the function
$a(T)$ that defines the embedding cannot be smoothly continued
beyond this point (see Fig.~4.\ref{brane-fig:brane0}).

This specific feature of the brane embedding can be compared to the
behavior of the involute of a planar curve.  The involute
$\mathcal{I}$ of a convex planar curve $\mathcal{C}$ is a line which
intersects the tangent lines of $\mathcal{C}$ orthogonally
\cite{lip}. $\mathcal{I}$ can be visualized as the trajectory
described by the end of a strained thread winding up from
$\mathcal{C}$ (see Fig.~4.\ref{brane-fig:involute}).  The involute
of a typical curve is sharp at the starting point $Q$ so that it is
not smoothly extendable beyond the \mbox{point $Q$.}\looseness=1

This analogy can be traced further.  Note that the evolution of the
brane in theory (\ref{brane-action}) is described by equation
(\ref{brane-brane}) which, for the case $N = 1$, reads
\begin{equation} \label{brane-brane1}
m^2 G_{ab} + \sigma h_{ab} = M^3 \left( K_{ab} - h_{ab} K \right) +
T_{ab} .
\end{equation}

\index{extrinsic curvature} %
One can see that it is the influence of the extrinsic curvature on
the right-hand side that leads to the singularities under
investigation, so that the singularity of the Einstein tensor
$G_{ab}$ is accompanied by the singularity of the extrinsic
curvature $K_{ab}$, while the induced metric $h_{ab}$ and the
stress-energy tensor $T_{ab}$ on the brane remain finite. Quite
similarly, the involute of a curve is defined through the extrinsic
curvature of its embedding in the plane, as is clear from
Fig.~4.\ref{brane-fig:involute}, and its singularity at the point
$Q$ is connected with the fact that the extrinsic curvature diverges
at this point. Specifically, the parametric equation for the
involute $\mathbf{x}_* (s)$, $s \ge 0$, in Cartesian coordinates on
the plane can be written as \cite{lip}
\begin{equation}
  \mathbf{x}_* (s) = \mathbf{x} (q - s) + s \cdot \mathbf{x}' (q - s) ,
\end{equation}
where $\mathbf{x} (s)$ is the curve $\mathcal{C}$ parametrized by
the natural parameter $s$, and $\mathbf{x} (q) =$ $= \mathbf{x}_*
(0)$ is the coordinate of the starting point $Q$ of the involute.
The extrinsic curvature of the involute is
\begin{equation}
  k (s) = \frac{1}{s} ,
\end{equation}
which diverges at the starting point $Q$ corresponding to $s = 0$.

One should also highlight an important difference between the 1D and
4D embeddings: the involute being one-dimensional, a singularity in
its extrinsic curvature does not lead to a singularity in its
intrinsic geometry. As we have seen, this is not the case with the
brane for which the extrinsic and intrinsic curvatures are related
through (\ref{brane-brane}), so that a singularity in $K_{ab}$ is
reflected in a singularity in $G_{ab}$.

\begin{figure}
\vskip1mm \raisebox{0.0cm}{\parbox[b]{6.0cm}{\caption{The
deceleration parameter (solid line) is shown for a
    braneworld mo\-del with $\Omega_\mathrm{ m} = 0.2$, $\Omega_\ell
    =
    0.4$, $\Omega_{\Lambda_\mathrm{b}} =\Omega_\kappa =$ $=  0$, and
    $\Omega_\sigma$ determined from (\ref{brane-omega-r1}). We find
    that $q\,(z) \to 0.5$ for $z \gg 1$ while $q\,(z) \to\infty$ as $z
    \to -0.312779$... Cur\-rent\-ly $q_0 < 0$, which indicates that the
    Uni\-ver\-se is accelerating. Also shown is the di\-men\-sion\-less
    \index{Hubble parameter}Hubble parameter $h(z) = 0.1\,\times$ $\times\, H(z)/H_0$ (dashed line) for
    this model. The vertical line at $z = 0$ shows the present
    epoch. Figure taken from \cite{SS1}
    \label{brane-fig:decel-sing}}}}\hspace*{0.5cm}\includegraphics[width=6.5cm]{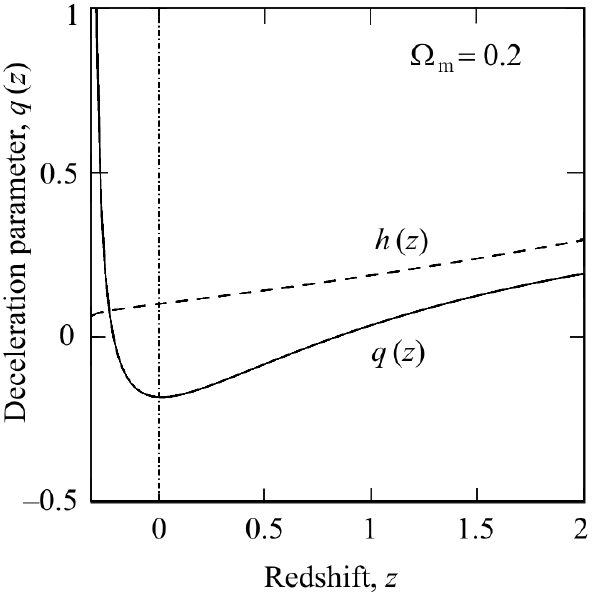}
    \vspace*{-4mm}
\end{figure}

Interestingly, an S2 singularity can arise in the distant future of
a Universe resembling our own.  To illustrate this we can consider
Eq.~(\ref{brane-hubble0}) with $\Omega_C = 0$. For simplicity, we
shall only discuss the solution corresponding to the upper sign in
(\ref{brane-hubble1}) (called BRANE2 in Sec.~\ref{brane-sec:phantom}
and in \cite{Sahni2003}).  Our model satisfies the constraint
equation (\ref{brane-omega-r1}) with $\Omega_C = 0$.  Inequality
(\ref{brane-ii}) now becomes
\begin{equation}
  \label{brane-inequa}
  \Omega_\sigma + \Omega_\ell + \Omega_{\Lambda_\mathrm{b}}  < 0 ,
\end{equation}
and the limiting redshift, $z_s = a_0/a(z_s) - 1$, at which the
braneworld becomes singular is given by\vspace*{-3mm}
\begin{equation}
  z_s = \left (\!-\frac{\Omega_\sigma + \Omega_\ell + \Omega_{\Lambda_\mathrm{b}}} {\Omega_\mathrm{
        m}}\!\right )^{\!\!1/3} - 1 .
\end{equation}
The time of occurrence of the singularity (measured from the present
moment) can easily be determined from\vspace*{-3mm}
\begin{equation}
  \Delta t_s = t(z=z_s) - t(z=0) = \int\limits_{z_s}^0\frac{dz}{(1+z)H(z)},
\end{equation}
where $H(z)$ is given by (\ref{brane-hubble1}) (see also
\cite{star99}). In Fig.~4.\ref{brane-fig:decel-sing} we show a
specific braneworld model having $\Omega_\mathrm{ m} = 0.2$,
$\Omega_\ell = 0.4$, $\Omega_{\Lambda_\mathrm{b}} = \Omega_\kappa =
0$. In keeping with observations of high-redshift supernovae, our
model Universe is currently accelerating \cite{Riess1998}, but will
become singular at $z_s \simeq -0.3 \Rightarrow a(z_s) \simeq
1.4\times a_0$, i.e. after $\Delta t_s \simeq 4.5 \, h^{-1}$~Gyr ($h
= H_0/100$ km/sec/Mpc). Figure~4.\ref{brane-fig:decel-sing}
demonstrates that the deceleration parameter (\ref{brane-decel})
becomes singular as $z_s$ is approached: $\lim\limits_{z \to z_s}
q(z) \to \infty$, while the
Hubble parameter remains finite: %
\index{Hubble parameter} %
\begin{equation} \frac{H^2(z_s)}{H_0^2} = \Omega_\ell -
  \Omega_{\Lambda_\mathrm{b}} .
\end{equation}

It should be noted that, for a subset of parameter values,
inequality (\ref{brane-inequa}) can be satisfied simultaneously with
$\Omega_{\Lambda_\mathrm{b}} > \Omega_\ell$. In these models, the
Universe will recollapse (under the influence of the negative brane
tension) before the S2 singularity is reached. (The marginal case
$\Omega_{\Lambda_\mathrm{b}} = \Omega_\ell$ corresponds to the
Hubble parameter vanishing {\em at\/} the singularity.)

\newpage

\subsection{\!Inhomogeneous case}
\vspace*{-0.5mm}

\hspace*{3cm}The preceding discussion focused on a homogeneous and
isotropic Universe whose expansion was governed by the brane
equations of motion. Since the real Universe is quite inhomogeneous
on spatial scales\linebreak $\lesssim $100~Mpc, it is worthwhile to
ask whether any of our previous results may be generalized to this
case.

Although we are not yet able to provide a self-consistent treatment
of the brane equations for this important case, still, some aspects
of the problem can be discussed at the phenomenological level.
Consider, for instance, the expansion law (\ref{brane-cosmol2}) with
$C = 0$. A necessary condition for the existence of a quiescent
singularity is that the matter density $\rho$ drops to a value which
is small enough for the square root on the right-hand side of
(\ref{brane-cosmol2}) to \mbox{vanish}.  When this happens, the
Universe encounters the quiescent singularity at \mbox{which} $\rho$
and $H$ remain finite, but ${\ddot a}$ and higher derivatives of the
scale factor diverge.  Note, however, that, according to
(\ref{brane-cosmol2}), the Universe encounters the quiescent
singularity {\em homogeneously\/}, i.e., every part of the
(spatially infinite) Universe encounters the singularity at {\em one
  and the same\/} instant of time. This follows from the fact that the
density in (\ref{brane-cosmol2}) depends only upon the cosmic time
and nothing else. In practice, however, the Universe is anything but
homogeneous: its density varies from place to place. For instance,
it is well known that the density of matter in galaxies is
$\gtrsim$$ 10^6$ times the average value while, in voids, it drops
to only a small fraction of the average value. This immediately
suggests that the brane should encounter the quiescent sin\-gularity
in a very inhomogeneous fashion. Underdense regions (voids) will be
the first to encounter the singularity. Even in this case, since the
density in individual voids is inhomogeneously distributed, more
underdense regions \mbox{lying} closer to the void center will be
the first to experience the singularity. It therefore follows that
the quiescent singularity will first form near the centers of very
underdense regions. As the void expands, its density at larger radii
will drop below $\rho_s$, where\vspace*{-4.0mm}
\begin{equation}
\sqrt{1 + \ell^2 \left(\!\frac{\rho_s + \sigma}{3 m^2} -
\frac{\Lambda_\mathrm{ b}}{6} \!\right)} = 0 ;
\end{equation}\vspace*{-4mm}

\noindent consequently, the singularity will propagate outward from
the void center in the form of a quasi-spherical singular front.
(For simplicity, we have assumed that all voids have a spherical
density profile; this assumption may need to be modified for more
realistic cases; see \cite{sss94,SaCo,SFHH,sheth,shandarin} and
references therein.)

The above approach provides us with a very different perspective of
the quiescent singularity than that adopted in the previous
subsection (and in \cite{SS1}). For one thing, the singularity may
be present in certain regions of the Universe {\em right now\/}, so
it concerns us directly (as astrophysicists) and not as some
abstract point to which we may (or may not) evolve in the distant
future. The second issue is related to the first, since the
singularity could already exist within several voids (there are as
many as a million voids in the visible Universe in at least some of
which the condition $\rho \simeq \rho_s$ could be satisfied), a
practical observational strategy needs to be adopted to search for
singularities in voids. (Similar strategies combined with strenuous
observational efforts have led to the discovery of dozens of black
holes in the centers of galaxies \cite{narayan}.)

A number of important issues therefore need to be addressed:

1.\,\,Since $R_{iklm}R^{iklm} \to \infty$ within a finite region at
  the very center of a void, it follows that, unless this region is
  contained within an event horizon, we will find ourselves staring at
  a naked singularity. (As shown in \cite{quantum}, quantum effects do
  soften the singularity so that $R_{iklm}R^{iklm}$ may remain finite
  if these effects are included.)

2.\,\,The moment we drop the homogeneity assumption, the issue of
  particle production immediately crops up, and we must take it into
  account if our treatment is to be at all complete
  \cite{Zeld-72,Zeld-77}. (In a related context, the quantum creation
  of gravitons takes place even in a homogeneous and isotropic
  Universe, since these fields couple minimally, and not conformally,
  to gravity \cite{grish75,grish77}.)

Let us discuss the possible effect of particle production in more
detail. First, we consider the model of homogeneous Universe taking
it as an approximation to the situation inside an underdensity
region (void). Gravitational quantum particle production occurs as
the singularity is approached. Since the local value of the Hubble
parameter %
\index{Hubble parameter} %
remains finite at the singularity, production of the conformally
coupled particles (like photons) is expected to be negligible.
However, particles that are non-conformally coupled to gravity
(which could be, for example, Higgs bosons in the Standard Model)
will be copiously produced as the acceleration of the Universe
$\ddot a$ rapidly increases.  The rate of particle production
depends not only on their coupling to gravity but also on their
coupling between themselves. Gravitationally created primary
particles will decay into conformally coupled secondaries
(electrons, photons, neutrino, etc.\@), which will influence the
rate of production of the primaries by causing decoherence in their
quantum state. The whole process is thus not easy to calculate in
detail. However, from very general arguments it can be seen that
creation of matter due to quantum particle production is important
for the dynamics of the Universe during its later
stages\,\footnote{\,Effects of particle
  production are negligible in the neighborhood of the usual
  cosmological singularity of the Friedmann Universe because the
  energy density of ordinary matter and radiation strongly diverges
  and thus dominates at this singularity
  \cite{Zeld-72,Zeld-77,zhBirrDav}.  In our case, the energy density
  of ordinary matter remains finite during the classical approach to
  the quiescent singularity, hence, particle production effects are of
  crucial significance.}.

For the sake of physical simplicity, we restrict ourselves to the
case
of vanishing bulk cosmological constant $\Lambda_\mathrm{b}$ %
\index{cosmological constant} %
and write Eq.~(\ref{brane-cosmol2}) in the form
\begin{equation} \label{brane-cosmo1}
H = \frac{1}{\ell} + \sqrt{\frac{\Delta \rho}{3 m^2}} ,
\end{equation}
where\vspace*{-3mm}
\begin{equation}
\Delta \rho = \rho - \rho_s , \quad \rho_s = - \sigma - \frac{3
m^2}{\ell^2} ,
\end{equation}
and where we have chosen the physically interesting ``$+$'' sign in
Eq.~(\ref{brane-cosmol2}). Thus, we have two free parameters in our
theory, namely, $\ell$ and $\rho_s$. The value of $m$ is assumed to
be of the order of the Planck mass.  In this case the early-time
behavior of the Universe follows the standard Friedmann model, as
can be seen from (\ref{brane-cosmol2}) or (\ref{brane-cosmo1}).

Let the average particle energy density production rate be $\dot
\rho_\mathrm{ prod}$. Then, differentiating
Eq.~(\ref{brane-cosmo1}), we obtain
\begin{equation} \label{brane-rate}
\dot H = \frac{\dot {\rho}_\mathrm{ prod} - \gamma H \rho}{2 \left(3
m^2 \Delta \rho \right)^{1/2} } ,
\end{equation}
where $\gamma > 0$ corresponds to the effective equation of state of
matter in the Universe: if $p = w \rho$, then $\gamma = 3 (1 + w)$.
The second term in the numerator of (\ref{brane-rate}) follows from
the conservation law and describes the effect of the Universe
expansion on the matter density. (Note that $\rho$ includes
contributions from quantum and classical matter.)
\index{conservation law} %

In order to qualitatively assess the effects of particle production,
let us examine two fundamentally distinct possibilities.

1.\,\, Suppose that, in the course of evolution, $\Delta \rho \to 0$
is
  reached in a finite interval of time.  Since the Hubble parameter
  is %
  \index{Hubble parameter|(} %
  a unique function of the energy density, given by
  (\ref{brane-cosmo1}), and since the singularity value $\rho_s$ of
  the energy density is approached from above, it follows that $\dot H
  \le 0$ in the neighborhood of the singular point.  In the purely
  classical case we find, after setting $\dot \rho_\mathrm{ prod}$ to
  zero in (\ref{brane-rate}), that $\dot H \to - \infty$ as the
  quiescent singularity is approached. It is well known, however, that
  particle production effects are sensitive to the change in the rate
  of expansion \cite{zhBirrDav}, and it is expected that $\dot
  \rho_\mathrm{ prod}$ will go to infinity as $\dot H \to -
  \infty$. Since $\dot {\rho}_\mathrm{ prod} \gg \gamma H \rho$, this
  will result in $\dot H$ becoming positive, which contradicts the
  assumption that $\dot H \le 0$.

  Therefore, under the assumption that the critical density $\rho_s$
  is reached in a finite time, the only possibility for $\dot H$ is to
  {\em remain bounded}. In other words, the rate of particle
  production should exactly balance the decrease in the matter density
  due to expansion, turning the numerator in (\ref{brane-rate}) to
  zero:
  \begin{equation}
    \label{brane-prodrate}
    \dot \rho_\mathrm{ prod} -
    \gamma H \rho \to 0\,  \Rightarrow\,  \dot \rho_\mathrm{
      prod} \to \frac{\gamma \rho_s}{\ell}  \mbox{~as~}  \rho
    \to \rho_s .
  \end{equation}
  In this case, the Universe reaches its singular state with the
  energy density due to particle production exactly balancing the
  density decrease caused by expansion, as given by
  (\ref{brane-prodrate}).

2.\,\, It is not clear whether the above regime will be realized or
  whether, if realized, it will be stable, since it requires the exact
  balancing of rates (\ref{brane-prodrate}) at the singularity. A
  second distinct possibility is that, due to the presence of particle
  production, the value of $\Delta \rho = \rho - \rho_s$ always
  \mbox{remains} bounded from below by a non-zero density. In this scenario,
  $\dot H$ initially decreases ($\vert \dot H\vert$ increases) under
  the influence of the increasing factor $1/\sqrt{\Delta \rho}$ in
  (\ref{brane-rate}). However, a large value of $\vert \dot H\vert$
  induces active particle production from the vacuum which leads to an
  increase in the value of $\dot \rho_\mathrm{ prod}$ in
  (\ref{brane-rate}). As the value of $\Delta \rho$ reaches its
  (non-zero) minimum, we have $\dot H = 0$ at this point, accor\-ding to
  (\ref{brane-cosmo1}), after which the rate $\dot H$ becomes positive
  due to self-sustained particle production that continues because of
  the large value of the second time derivative $\ddot H$. After a
  period of extensive particle production, the Universe reaches
  another turning point $\dot H = 0$ after which is continues to
  expand according to (\ref{brane-cosmo1}) with decreasing energy
  density. Thus, we arrive at a model of cyclic evolution with periods
  of extensive particle production alternating with periods of
  classical expansion during which quantum particle production is
  negligible. This scenario bears a formal resemblance to
  quasi-steady-state cosmology proposed in a very different context by
  Hoyle, Burbidge, and Narlikar [595---597]. The
  particle production rate in our case is estimated by the quantity
  $\dot \rho_\mathrm{ prod}$ given in (\ref{brane-prodrate}), which is
  approximately the value it takes at the turning points where $\dot H
  = 0$. The Hubble parameter in this scenario pe\-rio\-dically varies
  being of the order of $H \sim \ell^{-1}$, and the energy density is
  of the order $\rho_s$, so that particle production rate is
  \begin{equation}
    \dot \rho_\mathrm{ prod} \sim \frac{\gamma \rho_s}{\ell} .
  \end{equation}

Our discussion thus far was limited to quantum processes within a
single underdense region (void) which was assumed for simplicity to
be perfectly homogeneous. Let us now (qualitatively) discuss whether
this scenario can be generalized to the real (inhomogeneous)
Universe. Clearly, the particle production rate $\dot \rho_\mathrm{
  prod}$ in this case should be regarded as being averaged over the
entire Universe, to which several significantly underdense voids are
contributing.  Equation (\ref{brane-rate}) should therefore be
treated as an ensemble average, where the {\em mean\/} particle
production rate depends upon the distribution as well as dynamics of
{\em
  local\/} underdensity regions. As a result, equation
(\ref{brane-rate}) is not expected to explicitly depend upon the
behavior of the Hubble parameter and, in principle, particle
production can proceed even in a De~Sitter-like Universe, in which
the Hubble parameter $H$ remains constant in time. The rate of
particle production in this case is given by equality
(\ref{brane-rate}) with zero left-hand side:\vspace*{-1mm}
\begin{equation}
  \dot \rho_\mathrm{ prod} = \gamma H \rho .
\end{equation}\vspace*{-5mm}

The value of the Hubble parameter in such a steady-state Universe
can be related to the $\Omega$-parameter in matter\vspace*{-1mm}
\begin{equation}
  \Omega_\mathrm{ m} = \frac{\rho}{\rho_s} = \frac{\rho}{3 m^2 H^2} ,
\end{equation}
where we have used the basic Eq.~(\ref{brane-cosmo1}).  For the
average energy density, we obviously have $\rho - \rho_s \approx
\rho$.  Hence,\vspace*{-1mm}
\begin{equation}
  H = \frac 1 \ell + \sqrt{\frac{\Delta \rho}{3 m^2}} \approx \frac 1 \ell +
  \sqrt{\frac{\rho}{3 m^2}} = \frac 1 \ell + H \sqrt{\Omega_\mathrm{ m}} ,
\end{equation}\vspace*{-7mm}

\noindent or, finally\,\footnote{\,For comparison, the late-time
value of the
  Hubble parameter in LCDM is \cite{Sahni2000} $H =$ $=
  H_0\sqrt{1-\Omega_\mathrm{ m}}$.},\vspace*{-3mm}
\begin{equation}
  H \approx \frac{1}{\ell \left(\! 1 - \displaystyle \sqrt{\Omega_\mathrm{ m}} \right)
  } .
\end{equation}

\index{steady-state model} %
\index{singularity|)} %
In principle, one might use these preliminary results to construct a
brane\-world version of steady-state cosmology, in which matter is
being created at a steady rate in voids rather than in overdense
regions (as hypothesized in the original version
\cite{HBN-93,HBN-97}).  This would then add one more model to the
steadily growing list of dark-energy cosmologies
\cite{Sahni2000,sahni04}. These conclusions must, however, be
substantiated by a more detailed treatment which takes into account
the joint effect of vacuum polarization and particle production near
the quiescent singularity.

\section{\!Asymmetric branes} \label{brane-sec:asym}

\hspace*{3cm}\index{asymmetric branes}In this section, we consider
the properties of a more generic braneworld model with $N = 2$,
i.e., the physical case where the brane is a boundary of two bulk
spaces (has two sides). Two possibilities of principle exist in this
case: either the bulk space is constrained to be symmetric with
respect to the $Z_2$ group of reflections relative to the brane, or
such a symmetry is not imposed. The case where the bulk is symmetric
is equivalent to the geometrical setting with $N = 1$, which we were
considering up to now.  An embedded brane without the $Z_2$ symmetry
is, however, a more general case with rich possibilities for
cosmology [505, 598---614]).

We consider a braneworld model described by action
(\ref{brane-action}) with $N = 2$.  We shall focus on the asymmetric
case with $\Lambda_1 \neq \Lambda_2$ and $M_1 \neq M_2$, which
appears to be preferable from a string-theory perspective. For
instance, the dilaton stabilized in different vacuum states on
adjacent sides of the brane would lead to an effective
five-dimensional theory with $M_1 \neq M_2$. The string landscape is
likely to favor $\Lambda_1 \neq \Lambda_2$, which also occurs in
domain wall scenarios.

In the absence of $Z_2$ symmetry, cosmological evolution of the
brane is described by the general equation (\ref{brane-cosmol}), in
which we set $N = 2$. In that equation, $\rho$ is the total energy
density of matter on the brane, and $\zeta_i = \pm 1$, $i =$ $=
1,2$, correspond to the two possible ways of bounding each of the
bulk spaces ${\mathcal
  B}_i$, $i = 1,2$, by the brane. We classify the resulting four
branches according to the signs of $\zeta_1$ and $\zeta_2$ as
$(++)$, $(+-)$, $(-+)$, or $(--)$\,\footnote{\,Note that, in the
case of $Z_2$
  symmetry, there are only two ways of bounding the bulk by the brane,
  and these were called BRANE1 and BRANE2 in \cite{Sahni2003} and in
  Sec.~\ref{brane-sec:phantom}.  Of these, BRANE2 contains the
  self-accelerating DGP brane as a subclass, while BRANE1 can lead to
  phantom acceleration.}.  In the limit of $Z_2$ symmetry, the branches
$(--)$ and $(++)$ become the normal branch (BRANE1) and
self-accelerating branch (BRANE2), respectively (see
Sec.~\ref{brane-sec:phantom}).  The other two so-called {\em
mixed\/} branches are characterized by $\zeta_1 \zeta_2 = -1$.

In this section, we study the implications of (\ref{brane-cosmol})
for a spatially flat Universe ($\kappa = 0$) without dark radiation
($C_i = 0$, $i = 1,2$). Equation (\ref{brane-cosmol}) then
simplifies to\vspace*{-3mm}
\begin{equation} \label{brane-main} H^2 = \frac{\rho + \sigma}{3 m^2}
  + \frac{1}{m^2} \sum_{i = 1,2} \zeta^{}_i M_i^3 \sqrt{H^2 -
    \frac{\Lambda_i}{6} } = \frac{\rho + \sigma}{3 m^2} + \sum_{i =
    1,2} \frac{\zeta_i}{\ell_i} \sqrt{H^2 + \lambda_i^{-2} } ,
\end{equation}
where we have introduced the fundamental lengths
\begin{equation} \label{brane-lengths} \ell_i = \frac{m^2}{M_i^3} ,~ \lambda_i = \sqrt{ - \frac{6}{\Lambda_i}} ,~ i = 1,2
  ,
\end{equation}
assuming negative values of the bulk cosmological constants.

Note that (\ref{brane-main}) can be rewritten in terms of an {\em
  effective\/} cosmological constant, $\Lambda_\mathrm{ eff}$, %
\index{cosmological constant|(} %
as\vspace*{-3mm}
\begin{equation} \label{brane-LCDM} H^2 = \frac{\rho}{3m^2} +
  \frac{\Lambda_\mathrm{ eff}}{3} ,
\end{equation}\vspace*{-7mm}

\noindent where
\begin{equation} \frac{\Lambda_\mathrm{ eff}}{3} = \frac{\sigma}{3
    m^2} + \sum_{i = 1,2} \frac{\zeta_i}{\ell_i} \sqrt{H^2 +
    \lambda_i^{-2} }, \label{brane-lam}
\end{equation}
which is useful for the study of the cosmological properties of this
braneworld. A pictorial representation of the branches described by
(\ref{brane-main}) is given in Appendix~\ref{brane-app:asym}.

We consider the late-time evolution of the Universe, in which the
energy density $\rho$ is dominated by matter with the equation of
state $p = 0$.  Then, introducing the cosmological parameters as in
(\ref{brane-omegas}),
\begin{equation}
  \Omega_\mathrm{m}= \frac{\rho_0}{3 m^2 H_0^2} ,~
  \Omega_\sigma = \frac{\sigma}{3 m^2 H_0^2} ,~
  \Omega_{\ell_i} = \ell_i^{-2} H_0^{-2} ,~
  \Omega_{\lambda_i} = \lambda_i^{-2} H_0^{-2} ,
\end{equation}
where $\rho_0$ and $H_0$ are the current values of the matter
density
and Hubble parameter, %
\index{Hubble parameter|)} %
respectively, we rewrite (\ref{brane-main}) in terms of the
cosmological\linebreak \mbox{redshift~$z$:}\vspace*{-3mm}
\begin{equation} \label{brane-redshift} h^2 (z) \equiv \frac{H^2 (z)}{H_0^2}
= \Omega_\mathrm{m}(1 + z)^3 + \Omega_\sigma + \sum_{i =
    1,2} \zeta_i \sqrt{\Omega_{\ell_i}} \sqrt{h^2 (z) +
    \Omega_{\lambda_i}} .
\end{equation}
This equation implicitly determines the function $h(z)$, and
explicitly the inverse function $z (h)$. Note that the dimensionless
cosmological parameters are related through the constraint equation
\begin{equation}
  \Omega_\mathrm{m}+\Omega_\sigma + \sum_{i =
    1,2}\zeta_i\sqrt{\Omega_{\ell_i}}\sqrt{1+\Omega_{\lambda_i}} = 1.
\end{equation}

We now proceed to describe the specific features of braneworld
cosmology without $Z_2$ symmetry, some of which reproduce those
discussed in the pre\-ce\-ding sections for the $Z_2$-symmetric
case.


\subsection{\!Induced cosmological constant on the brane}

\hspace*{3cm}One way of accounting for cosmic acceleration within
the framework of brane\-world theory with mirror symmetry was
suggested in \cite{DDG,DLRZA} and described by
Eq.~(\ref{brane-DGP}). An extension of this model to the case when
mirror symmetry is absent is obtained by setting to zero the
cosmological constants on the brane and in the bulk, so that $\sigma
= 0$, $\Lambda_i = 0$, $i = 1,2$. The expansion law
(\ref{brane-main}) then simplifies to\vspace*{-2.5mm}
\begin{equation}
  \label{brane-dgp0}
  H^2 - H\sum_{i =
    1,2}\frac{\zeta_i}{\ell_i} = \frac{\rho}{3m^2} ,
\end{equation}\vspace*{-4mm}

\noindent which evolves to a De~Sitter limit at late
times\vspace*{-2.5mm}
\begin{equation} \label{brane-dgp} \lim_{z \to -1} H (z) = H_\mathrm{
    DS} = \sum_{i = 1,2}\frac{\zeta_i}{\ell_i} ,
\end{equation}\vspace*{-4mm}

\noindent provided $\sum\limits_{i=1,2}\zeta_i/\ell_i$ is positive,
which is true for branches $(++)$ and $(+-)$, provided $M_1 > M_2$
in the latter case.  If $m$ is of the order of the Planck mass
\mbox{$M_\mathrm{ P} \simeq 10^{19}~\mbox{GeV}$}, then the values of
\mbox{$M_i \sim 100~\mbox{MeV}$} can explain the observed cosmic
acceleration. [The self-accelerating solution (\ref{brane-DGP})
corresponds to the $(++)$ branch with $\ell_1 = \ell_2$.]

The absence of mirror symmetry provides a new avenue for this
mechanism. Specifically, the observed cosmic acceleration can be
produced on one of the mixed branches with {\em arbitrarily high\/}
values of the bulk Planck masses $M_1$ and $M_2$, provided these
values are sufficiently close to each other. If $0 < \Delta M \equiv
$ $\equiv M_1 - M_2 \ll M_1$, then we have\vspace*{-2mm}
\begin{equation} \label{brane-hds0} H_\mathrm{ DS} = \frac{M_1^3 -
    M_2^3}{m^2} \approx \frac{3 M_1^2}{m^2} \Delta M
\end{equation}\vspace*{-4mm}

\noindent on the $(+-)$ branch, and, by adjusting the value of
$\Delta M$, one can always achieve an observationally suitable value
of $H_\mathrm{
  DS}$.  For example, if $M_1, M_2 \sim m$, then one needs $\Delta M
\sim H_0$.

The previous model gave one example of late-time acceleration in the
absence of the (brane) cosmological constant. We now derive another
model with the same property but with a more flexible assumption
$\Lambda_i \ne 0$. Setting $\sigma = 0$ and $\Lambda_i \ne 0$ in
(\ref{brane-main}) leads to\vspace*{-2mm}
\begin{equation}
  H^2 - \sum_{i = 1,2}\frac{\zeta_i}{\ell_i}\sqrt{H^2 - \frac{\Lambda_i}{6}}
  = \frac{\rho}{3m^2} ,
\label{brane-newcosmos}
\end{equation}\vspace*{-4mm}

\noindent which evolves to a different De~Sitter limit, expressed by
the equation\vspace*{-2mm}
\begin{equation} \label{brane-hdesitter} \lim_{z \to -1} H^2 (z) =
  H_\mathrm{ DS}^2 = \sum_{i = 1,2}
  \frac{\zeta_i}{\ell_i}\sqrt{H_\mathrm{ DS}^2 + \lambda_i^{-2} } ,
\end{equation}\vspace*{-4mm}

\noindent where the length scales $\ell_i$ and $\lambda_i$ are
defined in (\ref{brane-lengths}).

It is interesting that a tiny asymmetry between the two bulk spaces
can lead to a small cosmological constant being induced on the brane
\cite{Shtanov:2009ss}.  Provided the bulk parameters $M_1$ and $M_2$
as well as $\Lambda_1$ and $\Lambda_2$ are close to each other, a
neat cancellation on the right-hand side of (\ref{brane-hdesitter}),
which occurs for $\zeta_1 \zeta_2 = -1$, leads to a small value of
$H_\mathrm{ DS}$.  Remarkably, this can happen even for very large
values of the bulk constants. In particular, assuming that
$\lambda_i \ll H_\mathrm{ DS}^{-1}$, we have\vspace*{-3mm}
\begin{equation}
  \label{brane-hds}
  H_\mathrm{ DS}^2  \approx  \left| \frac{1}{\ell_1 \lambda_1} - \frac{1}{\ell_2 \lambda_2} \right|
\end{equation}\vspace*{-3mm}

\noindent for one of the mixed branches.  Thus, for bulk parameters
of the order of a TeV, $M_i \sim 1\,\mbox{TeV}$, $\lambda_i \sim
1\,\mbox{TeV}^{-1}$, we recover the current value of the Hubble
parameter ($H_\mathrm{ DS} \sim H_0$) provided\vspace*{-1mm}
\begin{equation}
  \label{brane-lambda0}
  \left| \ell_1 \lambda_1 - \ell_2 \lambda_2 \right|^{1/2} \sim 10^{-13}\,\mbox{TeV}^{-1}
  \sim 10^{-30}\,\mbox{cm} .
\end{equation}

Equations such as (\ref{brane-hds0}) or (\ref{brane-hds}),
(\ref{brane-lambda0}) certainly represent fine tu\-ning, with a tiny
difference between bulk parameters only slightly breaking the
smoothness of the metric across the brane\,\footnote{\,Perhaps, the
small
  asymmetry in the fundamental constants characterizing the bulk can
  be explained by the presence of the brane itself. For instance, the
  presence of the brane could lead to a small difference in the
  quantum contribution to the effective action of the bulk on its two
  sides, inducing slightly different bulk constants.}. In the limit of
exact equality of the bulk constants on the two sides of the brane,
the branches with $\zeta_1 \zeta_2 = -1$ describe a smooth bulk
space, and the brane approaches the limit of a stealth brane
\cite{stealth}, evolving according to the usual Einstein equations
without affec\-ting the bulk space.\vspace*{-2mm}

\subsection{\!Cosmic mimicry}

\hspace*{3cm}For large values of the bulk parameters, we encounter
the phenomenon of {\em cosmic mimicry\/} which, in the context of
$Z_2$ symmetry, was described in Sec.~\ref{brane-sec:mimicry} and in
\cite{mimicry}. Note that, during the radiation and matter-dominated
epochs, the expansion of the Universe follows the
general-relativistic prescription\vspace*{-3mm}
\begin{equation}
  H^2 \approx \frac{\rho + \sigma}{3 m^2} ,
\end{equation}
where $\sigma/m^2$ plays the role of the cosmological constant on
the brane. However, at very late times, cosmic expansion gets
modified due to extra-dimensional effects. Indeed, if $\lambda_i \ll
H_0^{-1}$, then the square root in the last term of
(\ref{brane-main}) can be expanded in the small parameter
$\lambda_i^2 H^2$ at late times, and the braneworld expands
according to LCDM, namely\vspace*{-3mm}
\begin{equation}
  \label{brane-LCDM1}
  H^2 = \frac{8\pi G\rho}{3} + \frac{\Lambda}{3}
\end{equation}\vspace*{-7mm}

\noindent with\vspace*{-3mm}
\begin{equation}
  \label{brane-8piG} 8 \pi G = m^{-2} \left(\!\!1 - \sum_{i
      = 1,2} \frac{\zeta_i \lambda_i}{2 \ell_i} \!\!\right)^{\!\!-1}\!\!,
\end{equation}\vspace*{-3mm}
\begin{equation}
  \label{brane-mimic2} \Lambda = \left(\! \frac{\sigma}{ m^2} + \sum_{i
      = 1,2} \frac{3 \zeta_i}{\ell_i \lambda_i} \!\right)\! \left(\!\!1 -
    \sum_{i = 1,2} \frac{\zeta_i \lambda_i}{2 \ell_i} \!\right)^{\!-1}\!\!.
\end{equation}\vspace*{-3mm}

Note that both $G$ and $\Lambda$ are {\em independent of
  time\/}. Equations (\ref{brane-LCDM1})---(\ref{brane-mimic2}) have
important ramifications. They inform us that the `bare' value of the
cosmological constant on the brane, $\sigma$, is `screened' at late
times by extra-dimensional effects resulting in its effective value
$\Lambda$.  Thus, the early-time and late-time values of the
cosmological constant are likely to be different, and this makes our
model open to verification.

Also note that one can have $\Lambda \neq 0$ even if $\sigma=0$.
Then, a small $\Lambda$-term can be induced during late-time
evolution on the brane {\em solely by extra-dimensional effects\/},
as pointed out in the previous section.  The mechanism by which the
induced $\Lambda$-term becomes relatively small consists in a
compensation of two potentially large terms with opposite signs in
equation (\ref{brane-mimic2}).  Specifically, for small values of
$\lambda_i$ (which correspond to large values of $\Lambda_i$) such
that $\lambda_i / \ell_i \ll 1$, in the case $\sigma = 0$, we have,
approximately,
\begin{equation}
  \Lambda \approx \sum_{i = 1,2} \frac{3 \zeta_i}{\ell_i \lambda_i}  ,
\end{equation}
which is another form of the result (\ref{brane-hds}) for one of the
mixed branches.  What is remarkable here is that a {\em positive\/}
cosmological constant on the brane can be sourced by bulk
cosmological constants which are {\em negative\/}.

From (\ref{brane-LCDM1}), (\ref{brane-8piG}) we also find that the
effective gravitational constants during the early and late epochs
are related by a multiplicative factor
\begin{equation}
  1 - \sum_{i = 1,2} \frac{\zeta_i \lambda_i}{2 \ell_i}  ,
\end{equation}
which can be larger as well as smaller than unity, depending on the
braneworld branch.  This factor will be closer to unity for the
mixed branches ($\zeta_1 \zeta_2 =$ $= -1$) than it is for the usual
branches ($\zeta_1 \zeta_2 = 1$) which survive in the case of
\mbox{$Z_2$ symmetry.}

Focusing on the important case where $\sigma=0$ and the effective
four-di\-men\-sional cosmological constant is induced entirely by
five-dimensional effects, we find that, at redshifts significantly
below the {\em mimicry redshift\/}
\begin{equation}
  z_\mathrm{m}\simeq \left (\!\frac{\Omega_{\lambda_i}}{\Omega_{\rm m}}\!\right )^{\!\!1/3} - 1
  ,~
  \Omega_{\lambda_1} \simeq \Omega_{\lambda_2} ,
\end{equation}
the brane expansion mimics LCDM
\begin{equation}
  h^2(z) = \widetilde\Omega_{\rm m}(1+z)^3 + \Omega_{\Lambda} ,~  z \ll z_\mathrm{m},
\end{equation}
with `screened' values of the cosmological parameters:
\begin{equation}
  \widetilde\Omega_\mathrm{m}  = \Omega_{\rm m}\left (\!\!1-\sum_{i =
      1,2}\frac{\zeta_i}{2}
    \sqrt{\frac{\Omega_{\ell_i}}{\Omega_{\lambda_i}}}\right )^{\!\!-1} \!\!,
    \end{equation}\vspace*{-3mm}
\begin{equation}
  \Omega_{\Lambda}  = \sum_{i = 1,2}\zeta_i\sqrt{\Omega_{\ell_i}}
  \sqrt{\Omega_{\lambda_i}} \left (\!\!{1 - \sum_{i =
        1,2}\frac{\zeta_i}{2}\sqrt{\frac{\Omega_{\ell_i}}{\Omega_{\lambda_i}}}}\right
  )^{\!\!-1} \!\!. \label{brane-mimicry3}
\end{equation}

On the other hand, from (\ref{brane-LCDM1}) it follows that, at high
redshifts, the Universe expands as SCDM
\begin{equation}
  h^2(z) = \Omega_{\rm m}(1+z)^3 ,~ z\gg z_\mathrm{m}.
\end{equation}
An important distinguishing feature of this model is that the
(screened) matter density, $\widetilde\Omega_\mathrm{m}$, inferred
via geometrical tests based on standard candles and rulers, may not
match its (bare) dynamical value $\Omega_{\rm m}$. This allows
cosmic mimicry to be distinguished from other cosmological scenarios
by means of the {\em Om diagnostic\/} suggested in \cite{om}. The
fact that brane expansion also follows different laws at low and
high redshift provides another important observational test of this
model.

\subsection{\!Phantom branes}

\hspace*{3cm}In the presence of $Z_2$ symmetry, the BRANE1 branch of
the generic model (\ref{brane-cosmol}) exhibits phantom-like
behavior \cite{Sahni2003} (see Sec.~\ref{brane-sec:phantom}) which
is in excellent agreement with observations
\cite{brane_obs1,brane_obs2,Giannantonio:2008qr} (see also
\cite{Lue2004}). Let us see whether this behavior persists when
mirror symmetry is absent. Note first that the condition for phantom
acceleration, $w (z) < -1$, where $w (z)$ is given by
(\ref{brane-state}), has two equivalent formulations:
\begin{equation}
  \Omega_\mathrm{m}(z) > \frac{2}{3} \frac{d\log{H (z)}}{d \log (1+z)}~ \mbox{and}~ {\dot
    \Lambda_\mathrm{ eff}} > 0 ,
\end{equation}
where $\Lambda_\mathrm{ eff}$ is the effective cosmological constant %
\index{cosmological constant|)} %
in (\ref{brane-lam}), and differentiation is carried out with
respect to the physical time variable. In the case of the $(--)$
brane ($\zeta_1 = \zeta_2 = -1$), one has\vspace*{-4mm}
\begin{equation}
  \Lambda_\mathrm{ eff} = \frac{\sigma}{3 m^2} - \sum_{i = 1,2} \frac{\sqrt{H^2 +
      \lambda_i^{-2}}}{\ell_i} ,
\end{equation}\vspace*{-3mm}

\noindent and we find immediately that $\Lambda_\mathrm{ eff}$ {\em
increases
  with time\/} when the expansion rate, $H$, decreases.  It is also
quite clear that one (and only one) of the mixed branches will
necessarily have a negative value of the sum term in
(\ref{brane-lam}), again exhibiting phantom behavior.

It is straightforward to verify that ${\dot \Lambda_\mathrm{ eff}} >
0$ and ${\dot H} < 0$ on the two branches exhibiting phantom
behavior. Differentiating (\ref{brane-main}) and (\ref{brane-lam}),
we find\vspace*{-2mm}
\begin{equation}
  \dot{H} = - \frac{\rho}{m^2}\left \lbrack 2 -
    \sum_{i = 1,2}\frac{\zeta_i}{\ell_i\sqrt{H^2+\lambda_i^{-2}}}\right \rbrack^{-1}  <  0 ,
\end{equation}\vspace*{-3mm}
\begin{equation}
  {\dot \Lambda_\mathrm{ eff}} = 3H{\dot H}\, \sum_{i = 1,2}
  \frac{\zeta_i}{\ell_i\sqrt{H^2+\lambda_i^{-2}}} > 0
\end{equation}\vspace*{-5mm}

\noindent for\vspace*{-2mm}
\[
\sum_{i = 1,2} \frac{\zeta_i}{\ell_i\sqrt{H^2+\lambda_i^{-2}}} < 0 .
\]
Note that phantom models \cite{Caldwell2002} with constant equation
of state, $w<-1$, are marked by ${\dot \Lambda_\mathrm{ eff}} > 0$
and {\em super-acceleration}: ${\dot H} > 0$ at late
times\,\footnote{\,It is
  easy to show that, in phantom models, the turning point ${\dot H}=0$
  occurs at
\begin{equation}
  1+z_* \equiv \frac{a_0}{a(t_*)}=\left (\!\frac{1-\Omega_\mathrm{m}}{\Omega_\mathrm{m}}|1+w|\!\right
  )^{1/3|w|}\!,~ w < - 1 .
\end{equation}}.
This is related to the fact that the dark-energy (phantom) density
in such models {\em
  increases\/}, as the Universe expands, according to\vspace*{-2mm}
\begin{equation}
\rho_\mathrm{ phantom} \propto a^{3|1+w|} ,~  w < - 1 ,
\end{equation}\vspace*{-5mm}

\noindent which causes the Hubble parameter to grow at late times,
eventually
leading to a Big-Rip singularity %
\index{Hubble parameter} %
\index{singularity} %
at which $H$ diverges. By contrast, although the behavior of our
braneworld is phantom-like ($w_\mathrm{
  eff} < -1$), the Universe never super-accelerates since ${\dot H} <
0$ always holds. Furthermore, since $H$ decreases during expansion,
a Big-Rip-type future singularity which plagues phantom cosmology is
absent in the braneworld. From the definition of $q$ and property
${\dot H} < 0$, we find $q>-1$. In fact, the deceleration parameter
in our model always remains larger than the de~Sitter value of $q =
- 1$, approaching it only in the limit \mbox{of $t\to
\infty$.}

\subsection{\!Disappearing dark energy}

\hspace*{3cm}\index{disappearing dark energy}An important property
of this class of braneworld models is that the current acceleration
of the Universe need not be eternal. In other words, for a specific
relationship between the fundamental parameters in
(\ref{brane-main}), the acceleration of the Universe is a {\em
  transient\/} phenomenon, and the Universe reverts back to
matter-dominated expansion in the future. Within the context of
mirror symmetry, this scenario was called {\em disappearing dark
energy\/} and discussed in \cite{Sahni2003} and in
Sec.~\ref{brane-sec:dde}.  In the absence of mirror symmetry, it was
studied in \cite{CGP} under the name {\em stealth-acceleration\/}
(which should not be confused with the `stealth brane' of
\cite{stealth}).

Transient acceleration implies the property $H \to 0$ in the
asymptotic future, which requires the following condition to be
satisfied:
\begin{equation}
  \label{brane-dde}
  \frac{\sigma}{3 m^2} + \sum_{i = 1,2} \frac{\zeta_i}{\ell_i \lambda_i} =
  0
  \Rightarrow \Omega_\sigma + \sum_{i = 1,2}\zeta_i
  \sqrt{\Omega_{\ell_i}\Omega_{\lambda_i}} = 0 .
\end{equation}

On the $(--)$ and $(++)$ branches, this condition is realized with
the follo\-wing respective values of the brane tension:
\begin{equation} \frac{\sigma}{3 m^2} = \pm \left(\! \frac{1}{\ell_1
      \lambda_1} + \frac{1}{\ell_2 \lambda_2} \!\right)
  \Rightarrow \Omega_\sigma = \pm \left(\!
    \sqrt{\Omega_{\ell_1}\Omega_{\lambda_1}} +
    \sqrt{\Omega_{\ell_2}\Omega_{\lambda_2}}\right )
  . \label{brane-transient1}
\end{equation}

On the new mixed branches $(+-)$ and $(-+)$, the required brane
tension is smaller by absolute value:
\begin{equation}\label{brane-transient2}
\frac{\sigma}{3 m^2} = \pm \left| \frac{1}{\ell_1 \lambda_1} -
\frac{1}{\ell_2 \lambda_2} \right| \Rightarrow \Omega_\sigma = \pm
\left| \sqrt{\Omega_{\ell_1}\Omega_{\lambda_1}} -
  \sqrt{\Omega_{\ell_2}\Omega_{\lambda_2}}\right|\!.
\end{equation}

Under constraint (\ref{brane-dde}), the cosmological evolution
equation (\ref{brane-redshift}) becomes
\begin{equation} \label{brane-dde-evolution} h^2 (z) =
  \Omega_\mathrm{m}(1 + z)^3 + \sum_{i = 1,2} \zeta_i
  \sqrt{\Omega_{\ell_i}} \left(\! \sqrt{h^2 (z) + \Omega_{\lambda_i}} -
    \sqrt{\Omega_{\lambda_i}} \right)\!.
\end{equation}

Condition (\ref{brane-dde}) is necessary but not sufficient to speak
about transient acceleration on a particular branch.  A
distinguishing property of a transiently accelerating brane is that
$q(z) \to 0.5$ in the remote past ($10^5 \gg z \gg 1$) as well as in
the remote future ($z \to -1$), reflecting the fact that the
Universe is matter dominated in the past and in the future, while,
during the current phase, the deceleration parameter is negative,
$q_0 < 0$.  This last condition is realized only if the cosmological
expansion law $H^2 ( \rho)$ is convex upwards to a sufficiently high
degree. Specifically, in view of (\ref{brane-decel}), the condition
$q_0 < 0$ can be presented in the form
\begin{equation}
\frac{d H^2 (\rho_0)}{d \rho} < \frac{ 2 H_0^2}{3\rho_0} .
\end{equation}

Looking at figures B.1 and B.2 in Appendix~B.2, one can see that
this property can be realized only on two of the four branches: on
the $(++)$ branch and on one of the mixed branches.

\begin{figure}
\vskip1mm
    \includegraphics[width=13cm]{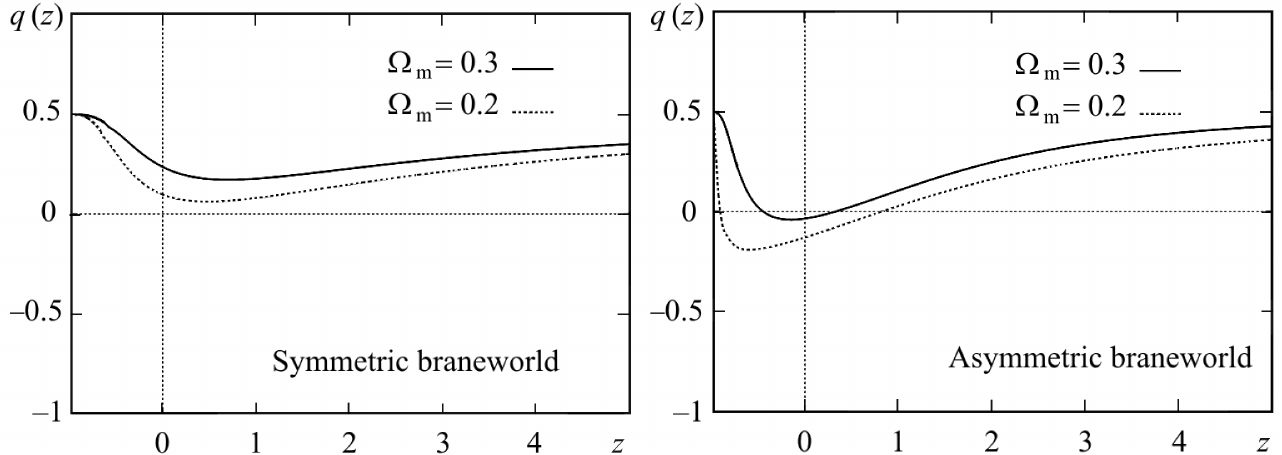}
\vskip-2mm
  \caption{The deceleration parameter versus redshift is plotted for
    the $(++)$ branch in the case of $Z_2$ symmetry. The model has the
    parameters $\Omega_{\lambda_1} = \Omega_{\lambda_2} = 2$.  We
    present plots for two different values of the matter density
    parameter: $\Omega_\mathrm{m}= 0.3$ and $\Omega_\mathrm{m}=
    0.2$. The sets of other parameters are calculated to be
    $\Omega_{\ell_1} = \Omega_{\ell_2} = 1.21$, $\Omega_\sigma = -
    3.11$ and $\Omega_{\ell_1} = \Omega_{\ell_2} = 1.58$,
    $\Omega_\sigma = - 3.56$, respectively. In this case, an
    accelerated regime is not realized, although deceleration is
    significantly slowed down at the present cosmological
    epoch. Figure taken from \cite{Shtanov:2009ss}
    \label{brane-fig:symmetric}}
    \vspace*{-2mm}
\caption{The deceleration parameter versus redshift is plotted for
    the $(+-)$ branch in the case of absence of $Z_2$ symmetry. The
    model has the parameters $\Omega_{\lambda_1} = 2$ and
    $\Omega_{\lambda_2} = 2.1$.  We present plots for two different
    sets of values of the remaining two independent parameters:
    $\left( \Omega_\mathrm{m}, \Omega_{\ell_1} \right) = (0.3,\,
    10000)$, which results in $\left(\Omega_{\ell_2}, \Omega_\sigma
    \right) = (9954.68,\, 3.16)$, and $\left( \Omega_\mathrm{m},
      \Omega_{\ell_1} \right) =$ $= (0.2,\, 5000)$, which results in
    $\left(\Omega_{\ell_2}, \Omega_\sigma \right) = (4840.15,\,
    0.82)$. Both plots show acceleration at the present cosmological
    epoch, which generically becomes more prominent for lower values
    of $\Omega_\mathrm{m}$. Figure taken from \cite{Shtanov:2009ss}
  \label{brane-fig:asymmetric}}
\end{figure}

The expression for the current value of the deceleration parameter
can be calculated by using the formula\vspace*{-3mm}
\begin{equation}
  q_0 =  \frac{ 3 \Omega_\mathrm{m}}{2 - \sum_i \zeta_i \sqrt{\frac{\displaystyle \Omega_{\ell_i}}
      {\displaystyle  1 + \Omega_{\lambda_i}}} } - 1 .
\end{equation}
One should note that only four out of five $\Omega$ parameters are
independent in this expression because of the normalization
condition $h^2(0) = 1$ applied to the evolution equation
(\ref{brane-dde-evolution}).  In the $Z_2$-symmetric case, there
remain only two independent $\Omega$ parameters.  It is clear then
that transient acceleration can be realized more easily in the
$Z_2$-asymmetric case.  This is illustrated in Figs.
4.\ref{brane-fig:symmetric} and 4.\ref{brane-fig:asymmetric}, which
show the corresponding behavior of the deceleration \mbox{parameter
$q(z)$.}

In a transiently accelerating Universe, cosmic acceleration is
sandwiched between two matter-dominated regimes. A transiently
accelerating braneworld clearly does not possess the Big Rip of
phantom cosmology, nor even the event horizon of de~Sitter space. %
\index{Big Rip} %
An in-depth study of this class of models \cite{CGP} has revealed
the existence of regions in parameter space which are stable
\mbox{(ghost-free).}

We have demonstrated that it is possible to construct braneworld
models with transient acceleration. What is less clear is whether
such transiently ac\-ce\-le\-ra\-ting branches will pass key
cosmological tests based on observations of high-red\-shift type Ia
supernovae, baryon acoustic oscillations, etc. This important issue
is open for further study.

\subsection{\!Quiescent singularities}

\hspace*{3cm}\index{singularity|(}As discussed in
Sec.~\ref{brane-sec:singular}, a new feature of brane cosmology is a
possible presence of {\em quiescent\/} singularities at which the
density, pressure and expansion rate remain finite, while the
deceleration parameter and the Kretchman invariant,
$R_{iklm}R^{iklm}$, diverge \cite{SS1,quantum}.  The Universe
encounters such a singularity in the future if a point is reached
during expansion where the derivative of $H^2$ with respect to
$\rho$ goes to infinity or, equivalently, where the derivative of
$\rho$ with respect to $H^2$ vanishes. Using (\ref{brane-main}), we
can express this condition as the existence of a positive root
$H_s^2$ of the equation
\begin{equation} \label{brane-sing} \sum_{i = 1,2} \frac{\zeta_i}{
    \ell_i \sqrt{H_s^2 + \lambda_i^{-2} } } = 2 ,
\end{equation}
and a quiescent singularity is approached as $H \to H_s$. At this
moment, expansion formally ceases, and one cannot extend the
classical evolution of the brane beyond this point.  Such a singular
point obviously exists on the $(++)$ branch if and only
if\vspace*{-3mm}
\begin{equation} \frac{\lambda_1}{\ell_1} + \frac{\lambda_2}{\ell_2} >
  2~ \Rightarrow~
  \sqrt{\frac{\Omega_{\ell_1}}{\Omega_{\lambda_1}}} +
  \sqrt{\frac{\Omega_{\ell_2}}{\Omega_{\lambda_2}}} > 2 ,
\end{equation}
and it is reachable on this branch if the brane tension $\sigma$ is
sufficiently negative:
\begin{equation} \frac{\sigma}{3 m^2} < H_s^2 - \sum_{i = 1,2}
\frac{1}{\ell_i} \sqrt{ H_s^2 + \lambda_i^{-2}} < 0 ,
\end{equation}\vspace*{-7mm}

\noindent or, equivalently,\vspace*{-3mm}
\begin{equation}
  \Omega_\sigma < \frac{H_s^2}{H_0^2} - \sum_{i = 1,2}
  \sqrt{\Omega_{\ell_i}}\sqrt{\frac{H_s^2}{H_0^2} + \Omega_{\lambda_i}} < 0 .
\end{equation}

Condition (\ref{brane-sing}) may or may not be realized on the mixed
branches. For example, in the simplifying case $\ell_1 = \ell_2 =
\ell$, condition (\ref{brane-sing}) is realized on the mixed branch
$(+-)$ provided $\lambda_1 > \lambda_2$.  One can show that the
values of the parameters $\ell_i$, $\lambda_i$, $i = 1,2$, in
principle can be chosen so that equation (\ref{brane-sing}) has
positive roots for three branches $(++)$, $(+-)$ and $(-+)$. To
achieve this, one only needs to satisfy the conditions $\ell_1 >
\ell_2$ and $\lambda_1 \ell_2 > \lambda_2 \ell_1$ and choose
sufficiently small values of $\ell_1$, $\ell_2$.

For a graphical presentation of the reasons for the existence of
quiescent singularities, the reader can look into Appendix~B2. As in
the case of mirror symmetry, quantum effects may play an important
role in the vicinity of a quiescent singularity \cite{quantum}; see
also [616---618].

We also note that, in the case of mirror symmetry, realization of
quiescent singularity requires either negative brane tension or
positive bulk cosmological constant (both conditions are suspicious
from the viewpoint of possible instabilities). However the quiescent
singularity can easily be realized without these assumptions in the
asymmetric case on a mixed branch.

The presence of a quiescent singularity in the future of the
cosmological evolution does not threaten the past cosmological
scenario.  Therefore, this issue, just like the issue of Big Rip of
phantom cosmology, is mainly of academic interest.  Here, we only
wish to point out that the possibility of quiescent
sin\-gu\-la\-ri\-ty can be realized rather easily in braneworld
theory in certain domain of its parameters without any additional
ingredients (such as phantom fields, which lead to Big Rip
singularities).
\index{singularity|)} %
\index{Big Rip} %

\subsection{\!Stability issues}

\hspace*{3cm}The stability issues of the class of braneworld models
without $Z_2$ mirror symmetry were studied in \cite{CGP,KPS}. It is
notable that ghost-free settings of the model with transient
acceleration (and phantom acceleration) appear to exist \cite{CGP}.
On the other hand, the analysis in paper \cite{KPS} reveals the
presence of ghosts on a background with a De~Sitter vacuum brane on
the three branches $(++)$, $(+-)$, $(-+)$ (i.e., which have at least
one `$+$', so that the bulk at least on one side of the brane has
`infinite volume' in terminology of \cite{KPS}). Whether this
situation is critical for the cosmology under investigation remains
to be seen.  In this connection, it should be noted that the $(++)$
branch, surviving in the $Z_2$ symmetric case, contains a ghost and
is, therefore, linearly unstable [517---521]. On the other hand, the
$(--)$ branch, responsible for `phantom acceleration' ($w_0 < -1$),
is ghost-free in the $Z_2$ symmetric case and, apparently, also in
the general case without $Z_2$ symmetry.


\section[\!Gravitational instability on the
  brane]{\!Gravitational\\ \hspace*{-1.2cm}instability on the
  brane} \label{brane-sec:stability}

\hspace*{3cm}In this section, we proceed to the study of
cosmological perturbations in a braneworld theory.  This subject is
not so well developed as the homogeneous cosmology because of
evident difficulties connected \mbox{with} the necessity of solving
the perturbations equations in the five-dimensional space and
imposing certain boundary conditions.  We will concentrate here on
the most general features of scalar perturbations, which are of most
relevance to the structure formation.
\index{scalar perturbations}%

\subsection{\!Scalar cosmological perturbations on the
  brane} \label{brane-sec:perturb}

\hspace*{3cm}\index{metric perturbations}The unperturbed metric on
the brane is described by the Ro\-bert\-son---Wal\-ker line element
and brane expansion is described by Eq.~(\ref{brane-cosmol2}).  The
two signs in (\ref{brane-cosmol2}) describe two different branches
corresponding to the two different ways in which a brane can be
embedded in the Schwarz\-schild---an\-ti-de~Sit\-ter bulk. In
Sec.~\ref{brane-sec:phantom}, we classified models with lower
(upper) \mbox{sign} as BRANE1 (BRANE2). Models with the upper sign
can also be called {\em
  self-accelerating\/} because they lead to late-time cosmic
acceleration even in the case of zero brane tension and bulk
cosmological constant \cite{DDG,DLRZA}. Throughout this section, we
consider the spatially flat case ($\kappa = 0$) for
\mbox{simplicity.}

Scalar metric perturbations of this cosmological solution are most
conveniently described by the relativistic potentials $\Phi$ and
$\Psi$ in the so-called longitudinal gauge:\vspace*{-3mm}
\begin{equation}
  \label{brane-longitudinal}
  ds^2 = - ( 1 + 2 \Phi) d t^2 + a^2 ( 1 - 2 \Psi) \gamma_{ij} d x^i dx^j ,
\end{equation}
where $\gamma_{ij} (x)$ is the spatial part of the metric, which is
used to raise and lower the spatial indices.  The components of the
linearly perturbed stress-energy tensor of matter in the coordinate
basis are defined by\vspace*{-1mm}
\begin{equation}
T^\alpha{}_\beta = \left(\!\!\!
\begin{array}{cc}
  \displaystyle - (\rho + \delta \rho) , &\!\!\!\! \displaystyle -
  \nabla_i v
  \medskip \\
  \displaystyle  \frac{\nabla^i v}{a^2} , &\!\!\!\! \displaystyle (p + \delta p)
  \delta^i{}_j + \frac{\zeta^i{}_j}{a^2}
\end{array}\!\!\!
\right)\!\!,
\end{equation}
where $\delta \rho$, $\delta p$, $v$, and $\displaystyle\zeta_{ij} =
\left(\!
  \nabla_i \nabla_j - \frac13 \gamma_{ij} \nabla^2 \!\right) \zeta$ are
small quantities.  The symbol $\nabla_i$ in this section
denotes\rule{0pt}{10pt} the spatial covariant derivative compatible
with the spatial metric $\gamma_{ij}$ present in
(\ref{brane-longitudinal}), $\nabla^2 \equiv \nabla^i \nabla_i$ is
the scalar Laplacian, and the spatial indices are raised and lowered
using $\gamma_{ij}$. Similarly, we introduce the scalar
perturbations $\delta \rho_{\mathcal{C}}$, $v_{\mathcal{C}}$, and
$\delta \pi_{\mathcal{C}}$ of the tensor $C_{ab}$ in the
coordinate\linebreak basis:\vspace*{-3mm}
\begin{equation}
  \label{brane-weyl projection definition}
  m^2 C^\alpha{}_\beta = \left(\!\!\!
  \begin{array}{cc}
    \displaystyle \frac{3 m^2 C}{a^4} - \delta \rho_{\mathcal{C}}
    , &\!\!\!\! \displaystyle - \nabla_i v_{\mathcal{C}} \medskip \\
    \displaystyle \frac{\nabla^i v_{\mathcal{C}}}{a^2} , &\!\!\!\! \displaystyle \left(\!\frac{\delta
        \rho_{\mathcal{C}}}{3 } - \frac{m^2 C}{a^4} \!\right) \delta^i{}_j + \frac{\delta \pi^i{}_j}{a^2}
  \end{array}\!\!\!\!
\right)\!\!,
\end{equation}
where $\displaystyle\delta \pi_{ij} = \nabla_i \nabla_j \delta
\pi_{\mathcal{C}} - \frac13 \delta_{ij} \nabla^2 \delta
\pi_{\mathcal{C}} $.  We call $v$ and $v_{\mathcal{C}}$ the momentum
potentials for matter\rule{0pt}{10pt} and dark radiation,
respectively.

Equation (\ref{brane-effective}) together with the stress-energy
conservation equation for matter and conservation equation %
\index{conservation law} %
(\ref{brane-conserv}) for dark radiation result in the follo\-wing
complete system of equations describing the evolution of scalar
perturba\-tions on the brane:\vspace*{-3mm}
\[
  \ddot \Psi + 3(1 + \gamma) H \dot \Psi + H \dot \Phi + \left[ 2
    \dot H + 3 H^2 (1 + \gamma) \right] \Phi - \frac{\gamma}{a^2} \nabla^2 \Psi -
    \frac{\kappa (1 + 3 \gamma)}{a^2} \Psi \,+ \]\vspace*{-3mm}
\begin{equation}
  +\, \frac{1}{3 a^2} \nabla^2 (\Phi - \Psi) = \left[ c_s^2 - \gamma +
    \frac{2}{\lambda} \left(\!c_s^2 - \frac13 \!\right)\!\right] \frac{\delta
    \rho}{2 m^2 } + \left(\! 1 + \frac{2}{\lambda} \!\right) \frac{\tau
    \delta S}{2 m^2} ,  \label{brane-psi}
\end{equation}\vspace*{-1mm}
\begin{equation}
\delta \dot \rho + 3 H ( \delta \rho + \delta p ) = \frac{1}{a^2}
\nabla^2 v   + 3 (\rho + p) \dot \Psi ,  \label{brane-rho}
\end{equation}\vspace*{-3mm}
\begin{equation}
\dot v + 3 H v = \delta p + (\rho + p) \Phi + \frac{ 2}{3 a^2}
\left( \nabla^2 + 3 \kappa \right) \zeta ,   \label{brane-v}
\end{equation}\vspace*{-3mm}
\begin{equation}
\delta \dot \rho_{\mathcal{C}} + 4 H \delta \rho_{\mathcal{C}} =
\frac{1}{a^2} \nabla^2 v_{\mathcal{C}} -  \frac{12 m^2 C}{a^4} \dot
\Psi ,  \label{brane-rho-c}
\end{equation}\vspace*{-3mm}
\[
  \dot v_{\mathcal{C}} + 3 H v_{\mathcal{C}} = \frac13 \delta \rho_{\mathcal{C}} - \frac{4 m^2 C}{a^4}
  \Phi + \frac16 \lambda (1 - 3 \gamma) \Delta_\mathrm{m}- \frac{2 +
    \lambda}{3 a^2}
  \left(\nabla^2 + 3 \kappa \right) \zeta \,- \]\vspace*{-3mm}
\begin{equation}
  -\, \frac{m^2 \lambda}{3 a^2} \left(\nabla^2 + 3 \kappa \right) \left[
    \Phi - 3 \gamma \Psi \right] \!, \label{brane-v-c}
\end{equation}\vspace*{-3mm}
\begin{equation}
\frac{1}{a^2} \left(\nabla^2 + 3 \kappa \right) \nabla^2 \Psi =
\left(\!1 + \frac{2}{ \lambda} \!\right) \frac{\Delta_\mathrm{m}}{2
m^2} + \frac{\Delta_{\mathcal{C}}}{m^2 \lambda } ,
\label{brane-con-delta}
\end{equation}\vspace*{-1mm}
\begin{equation}
m^2 \lambda \left( \dot \Psi + H \Phi \right) = \left(\!1 +
\frac{\lambda}{2} \!\right) v + v_{\mathcal{C}} ,
\label{brane-con-v}
\end{equation}\vspace*{-1mm}
\begin{equation}
\delta \pi_{\mathcal{C}} = - \frac{m^2}{4} \lambda (3 \gamma + 1)
\left(\! \Phi -
  \Psi + \frac{\zeta}{m^2} \!\right) - \zeta .  \label{brane-pi}
\end{equation}

Here, we use the following notation: $S$ is the entropy density of
the matter content of the Universe, $\tau = \left(
\partial p / \partial S \right)_\rho\,$, $c_s^2 = \left(
\partial p / \partial \rho \right)_S$ is the adiabatic sound velocity, the
time-dependent dimensionless functions $\lambda$ and $\gamma$ are
given by\vspace*{-1mm}
\begin{equation}
  \label{brane-lambda}
  \lambda \equiv \ell^2 \left( \!H^2 - \frac{\rho + \sigma}{3 m^2} \!\right) - 2 =
  \pm 2 \sqrt{1 + \ell^2 \left(\!\frac{\rho + \sigma}{3 m^2} - \frac{\Lambda_\mathrm{b}}{6} -
  \frac{C}{a^4} \!\right) } ,
\end{equation}\vspace*{-3mm}
\begin{equation}
  \label{brane-gamma}
  \gamma \equiv \displaystyle \frac13 \left(\!1 + \frac{\dot \lambda}{H \lambda}
  \!\right) = \frac13 \left[ 1 - \frac{\displaystyle \frac{\rho + p}{m^2} -
  \frac{4 C}{a^4} }{\displaystyle 2 \left(\! \frac{\rho + \sigma}{3m^2} + \frac{1}{\ell^2}
        - \frac{\Lambda_\mathrm{b}}{6} - \frac{C}{a^4}\! \right) } \right]\!\!,
\end{equation}
and the perturbations $\Delta_{\rm m}$ and $\Delta_{\mathcal{C}}$
are defined as\vspace*{-1mm}
\begin{equation}
  \Delta_\mathrm{m}= \delta \rho + 3 H v , \quad \Delta_{\mathcal{C}} = \delta \rho_{\mathcal{C}} + 3 H   v_{\mathcal{C}} .
\end{equation}
The overdot, as usual, denotes the partial derivative with respect
to \mbox{the time $t$.}

The system of equations (\ref{brane-psi})---(\ref{brane-pi})
generalizes the result obtained in \cite{Deffayet} (for the DGP
brane) to the case of a generic braneworld scenario described by
(\ref{brane-action}), which allows non-zero values for the brane
tension and bulk cosmological constant. %
\index{cosmological constant} %
It describes two dynamically coupled fluids: matter and dark
radiation. It is important to emphasize that the evolution equations
(\ref{brane-rho-c}), (\ref{brane-v-c}) for the dark-radiation
component are {\em not quite\/} the same as those for ordinary
radiation. Of special importance are the source terms on the
right-hand side of (\ref{brane-v-c}) which lead to non-conservation
of the dark-radiation density. Thus, the behavior of this component
is rather non-trivial, as will be demonstrated in next sections.

It is also interesting to note that the perturbations in dark
radiation for\-mal\-ly decouple from those in ordinary matter in the
important li\-mi\-ting case \mbox{$M \to 0$} (equivalently,
\mbox{$\ell \to \infty$}), for which the system
(\ref{brane-psi})---(\ref{brane-pi}) re\-pro\-duces the
corresponding equations of General Relativity (after setting $\gamma
= c_s^2$).

From equations (\ref{brane-rho})---(\ref{brane-con-v}), one can
derive the following useful system for perturbations in {\em
pressureless\/} matter and dark radiation in the important case $C =
0$:\vspace*{-1mm}
\begin{equation}
\ddot \Delta + 2 H \dot \Delta = \left(\!1 + \frac{6
\gamma}{\lambda} \!\right) \frac{\rho \Delta}{2 m^2} + (1 + 3
\gamma) \frac{\delta\rho_{\mathcal{C}}}{m^2 \lambda}  ,
\label{brane-one}
\end{equation}\vspace*{-1mm}
\begin{equation}
\dot v_{\mathcal{C}} + 4 H v_{\mathcal{C}} = \gamma
\Delta_{\mathcal{C}} + \left(\!\gamma - \frac13 \!\right)
\Delta_\mathrm{m} + \frac{4}{3 (1 + 3 \gamma) a^2 } \left(\nabla^2 +
3 \kappa \right) \delta \pi_{\mathcal{C}} , \label{brane-two}
\end{equation}\vspace*{-1mm}
\begin{equation}
\delta \dot \rho_{\mathcal{C}} + 4 H \delta \rho_{\mathcal{C}} =
\frac{1}{a^2} \nabla^2 v_{\mathcal{C}} \label{brane-three} ,
\end{equation}\vspace*{-5mm}

\noindent where\vspace*{-1mm}
\begin{equation}
\Delta \equiv \frac{\Delta_\mathrm{m}}{\rho}
\end{equation}
\index{matter perturbations}%
is the conventional dimensionless variable describing matter
perturbations.

\subsection{\!Simplified boundary conditions\\ \hspace*{-1.4cm}for scalar
  perturbations} \label{brane-subsec:bc}

\hspace*{3cm}The system of equations
(\ref{brane-psi})---(\ref{brane-pi}), or
(\ref{brane-one})---(\ref{brane-three}), de\-scri\-bing scalar
cosmological perturbations,\index{scalar perturbations} is not
closed on the brane sin\-ce the quantity $\delta \pi_{\mathcal{C}}$
in (\ref{brane-pi}) or (\ref{brane-two}) (hence, the difference
$\Phi \,-\, \Psi$) is un\-de\-ter\-mi\-ned and, in principle, can be
set arbitrarily from the brane view\-point.  For this reason, one
should also consider equations for perturbations in the bulk and
impose certain boundary conditions. This will be under
con\-si\-de\-ra\-ti\-on further in Sec.~\ref{brane-sec:bulk}.

From a broad perspective of solving equations in the brane-bulk
system, boundary conditions can be regarded as any conditions which
restrict the space of solutions. In view of the difficult problems
of solving the perturbation equations in the bulk, what some
researches usually do in practice is to specify such conditions
directly on the brane by making various reasonable assumptions (see,
e.g., \cite{Maartens,BM} for the issue of cosmological
perturbations; and \cite{bh} for the case of spherically symmetric
solutions on the brane). The behavior of the metric in the bulk is
of no further concern in this approach, since this metric is, for
all practical purposes, unobservable directly. The described
approach to the problem of boundary conditions effectively
``freezes'' certain degrees of freedom in the bulk; but its merit is
that it apparently leads to a well-defined closed, local, causal,
and, in principle, verifiable theory of gravity in four dimensions.
On the other hand, it corresponds to a certain class of
approximations to the perturbation equations on the brane
[621---623].

As first noted in \cite{SMS}, the intrinsic non-locality and
non-closure of the bra\-ne\-world equations is connected with the
dynamical properties of the bulk Weyl tensor projected on to the
brane. It, therefore, seems logical to impose certain restrictions
on this tensor in order to obtain a closed system of equations on
the brane.  A general family of boundary conditions on the brane is
obtained by relating the quantities $\pi_{\mathcal{C}}$ and
$\rho_{\mathcal{C}}$. As a simplest class of such relation which
does not involve dimensional parameters, one can set
\begin{equation} \label{brane-bc-linear} \frac{1}{a^2} \nabla^2 \delta
  \pi_{\mathcal{C}} = {{A}}\, \delta \rho_{\mathcal{C}} .
\end{equation}
In most of this section, ${{A}}$ shall be assumed to be a
(dimensionless) constant.  We also consider the simplifying case
where the spatial curvature is equal to zero ($\kappa = 0$) and the
matter anisotropic stresses are absent ($\zeta = 0$). Then, by
virtue of (\ref{brane-pi}), this relates the diffe\-rence $\Phi -
\Psi$ between the gravitational potentials to the perturbation of
the dark-radiation density $\delta \rho_{\mathcal{C}}$:
\begin{equation} \label{brane-phi-psi}
\frac{1}{a^2} \nabla^2 (\Phi - \Psi) = - \frac{4 {{A}}}{m^2 \lambda
(1 + 3 \gamma)} \delta \rho_{\mathcal{C}} .
\end{equation}

For the boundary condition (\ref{brane-bc-linear}), one can derive a
second-order dif\-fe\-ren\-tial equation for $\delta
\rho_{\mathcal{C}}$ by substituting for $v_{\mathcal{C}}$ from
(\ref{brane-three}) into (\ref{brane-two}):
\[
 \delta \ddot \rho_{\mathcal{C}} + (10 - \gamma)
H \delta \dot \rho_{\mathcal{C}} + 4 \left[
  \dot H + 3 (2 - \gamma)
  H^2 \right] \delta \rho_{\mathcal{C}}=
  \]\vspace*{-3mm}
\begin{equation}\label{brane-closed}
= \frac{1}{a^2} \nabla^2 \left[ \gamma \delta \rho_{\mathcal{C}} +
\frac{4 {{A}}}{3
    (1 + 3 \gamma)} \delta \rho_{\mathcal{C}} + \left(\! \gamma - \frac13 \!\right)
  \Delta_\mathrm{m}\right]\!.
\end{equation}
Equations (\ref{brane-one}) and (\ref{brane-closed}) then form a
closed system of two coupled second-order differential equations for
$\Delta$ and $\delta \rho_{\mathcal{C}}$. From the form of the
right-hand side of (\ref{brane-closed}) one expects this system to
have regions of stability as well as instability. Specifically, a
necessary condition for stability on small spatial scales is that
the sign of the coefficient of $\nabla^2 \delta \rho_{\mathcal{C}}$
on the right-hand side of (\ref{brane-closed}) be positive. This
leads to the condition\,\footnote{\,In the absence of matter on the
brane,
  $\Delta_\mathrm{m}= 0$, equation (\ref{brane-closed}) becomes a
  closed wave-like equation for the scalar mode of gravity, and
  condition (\ref{brane-stable}) becomes the boundary of its stability
  domain.  The existence of such a scalar gravitational mode is due to
  the presence of an extra dimension.}\vspace*{-1mm}
\begin{equation} \label{brane-stable}
{{A}} \ge - \frac34 \gamma (1 + 3 \gamma) .
\end{equation}
From (\ref{brane-gamma}), we find that $\gamma \approx - 1/6$ in a
matter-dominated Universe, and condition (\ref{brane-stable})
simplifies to\vspace*{-3mm}
\begin{equation} \label{brane-stable0}
{{A}} \ge \frac{1}{16} .
\end{equation}

We consider two important subclasses of (\ref{brane-bc-linear})
which we call the minimal boundary condition and the
Koyama---Maartens boundary condition, \mbox{respectively.}

{\bfseries\itshape Minimal boundary condition.} Our simplest
condition corresponds to setting ${{A}} = 0$.  In this case, from
(\ref{brane-pi}) we obtain the relation $\Phi = \Psi$, the same as
in General Relativity.  Under this condition, equations
(\ref{brane-psi})---(\ref{brane-v}) constitute a complete system of
equations for scalar cosmological perturbations on the brane in
which initial conditions for the relativistic potential $\Phi$,
$\dot\Phi$
and matter perturbations $\delta \rho$, $\delta p$, $v$ can be %
\index{matter perturbations}%
specified quite independently. Once a solution of this system is
given, one can calculate all components of dark-radiation
perturbations using (\ref{brane-con-delta}) and (\ref{brane-con-v}).
Thus, with this boundary condition, equations
(\ref{brane-rho-c})---(\ref{brane-con-v}) can be regarded as
auxiliary and can be used to felicitate and elucidate the dynamics
described by the main system (\ref{brane-psi})---(\ref{brane-v}). We
should stress that only the quantities pertaining to the induced
metric on the brane ($\Phi$, $\Psi$) and those pertaining to matter
($\delta\rho$, $\delta p$, $v$) can be regarded as directly
observable, while those describing dark radiation
($\delta\rho_{\mathcal{C}}$, $v_{\mathcal{C}}$) are not directly
observable.

{\bfseries\itshape Koyama---Maartens boundary condition.} In an
important paper \cite{KM}, Koyama and Maartens arrived at condition
(\ref{brane-bc-linear}) with ${{A}} = - 1/2$:
\begin{equation} \label{brane-KM}
\frac{1}{a^2} \nabla^2 \delta \pi_{\mathcal{C}}  = - \frac12 \delta
\rho_{\mathcal{C}} .
\end{equation}
This boundary condition was derived in \cite{KM,Koyama} as an
approximate relation in the DGP model valid only on small
(subhorizon) spatial scales under the assumption of quasi-static
behavior. It was later re-derived in \cite{SSH} under a similar
approximation. We call it, therefore, the Koyama---Maartens boundary
condition, although one should be aware of the different status of
this relation in the present section, where it is regarded as an
exact additional relation, and in [621---623], where it is derived
as an approximation.

In discussing the small-scale approximation in quasi-static regime,
it was argued in \cite{KM,Koyama} that equation (\ref{brane-three})
permits one to neglect the perturbation $v_{\mathcal{C}}$ in
(\ref{brane-two}), which, together with (\ref{brane-bc-linear}),
will then transform (\ref{brane-one}) into a closed equation for
matter perturbations\,\footnote{\,Equation (\ref{brane-KM-one}) was
derived in
  \cite{KM,Koyama} only for the DGP model and for the case ${{A}} = -
  1/2$; however, the argument can be extended to a general braneworld
  model and a general value of ${{A}}$ in (\ref{brane-bc-linear}).}:
\begin{equation}
\label{brane-KM-one} \ddot \Delta + 2 H \dot \Delta =
\Theta_\mathrm{ KM}\, \frac{\rho \Delta}{2 m^2} ,~ \Theta_\mathrm{
KM} = 1 + \frac{ 12 {{A}} \gamma + 3 \gamma + 1}{
  \lambda \left[2 {{A}} + \frac32 \gamma (1 + 3 \gamma) \right] } .
\end{equation}
Some cosmological consequences of this approach are discussed in
\cite{spergel}. As in General Relativity, this equation does not
contain spatial derivatives; hence, the evolution of $\Delta$ is
independent of the spatial scale.  We would obtain equation of the
type (\ref{brane-KM-one}) for perturbations had we followed the
route of [621---623] in finding approximate solutions of
perturbation equations in the bulk and using the quasi-static
approximation. However, in the approach adopted in this section, as
will be shown in the following two sections, numerical integration
of the exact linearized system
(\ref{brane-one})---(\ref{brane-three}) does not support
approximation (\ref{brane-KM-one}).  No matter what initial
conditions for dark radiation are set initially, one observes
a strong dependence of the evolution of matter perturbations on the %
\index{matter perturbations}%
wave number.  In particular, it is incorrect to neglect the quantity
$v_{\mathcal{C}}$ on small spatial scales, since it is precisely
this quantity which is responsible for the dramatic growth of
perturbations both in $\Delta_{\rm m}$ and in $\delta
\rho_{\mathcal{C}}$ on such scales.

From equations (\ref{brane-stable}) and (\ref{brane-stable0}), we
can see that the minimal and Koyama---Maartens boundary conditions
generally lead to unstable evolution.  This will be confirmed by
numerical simulations in the next section.

{\bfseries\itshape Scale-free boundary conditions.} Evolution of
perturbations in the sta\-bi\-li\-ty region (\ref{brane-stable}) \&
(\ref{brane-stable0}) shows little dependence on spatial scale. It
is interesting that there also exists an important class of boundary
conditions leading to {\em exact\/} scale-independence. We call
these, for simplicity, {\em scale-free
  boundary conditions.\/} To remove the dependence on wave number
altogether and thereby obtain a theory in which perturbations in
matter qualitatively evolve as in standard (post-recombination)
cosmology, it suffices to set the right-hand side of
(\ref{brane-two}) identically zero:
\begin{equation} \label{brane-bc-new}
\frac{1}{a^2 } \nabla^2 \delta \pi_{\mathcal{C}} = \frac{1 + 3
\gamma}{4} \Bigl[ (1 - 3 \gamma) \Delta_\mathrm{m}- 3 \gamma
\Delta_{\mathcal{C}} \Bigr]\!,
\end{equation}
which, in view of equation (\ref{brane-con-delta}), can also be
expressed in a form containing only the geometrical quantities
$\delta \pi_{\mathcal{C}}$, $\Delta_{\mathcal{C}}$, and $\Psi$.  In
this case, the perturbations $\delta v_{\mathcal{C}}$ and $\delta
\rho_{\mathcal{C}}$ in dark radiation decay very rapidly, according
to equations (\ref{brane-two}) and (\ref{brane-three}), and
(\ref{brane-one}) reduces to the simple equation
\begin{equation} \label{brane-matter} \ddot \Delta + 2 H \dot \Delta =
  \Theta\, \frac{\rho \Delta}{2 m^2} ,~ \Theta = \left(\!1 + \frac{6
      \gamma }{\lambda} \!\right)\!,
\end{equation}
valid on all spatial scales.  Equations (\ref{brane-con-delta}) and
(\ref{brane-pi}) then lead to simple relations between the
gravitational potentials $\Phi$ and $\Psi$ and matter
pertur\-bations:
\begin{equation} \label{brane-potentials} \frac{1}{a^2} \nabla^2 \Phi
  = \Theta \frac{\Delta_\mathrm{m}}{2 m^2} ,\quad \frac{1}{a^2}
  \nabla^2 \Psi = \left(\!1 + \frac{2}{\lambda} \!\right)
  \frac{\Delta_\mathrm{m}}{2 m^2} .
\end{equation}
The difference $\Phi - \Psi$ can be conveniently determined from
\begin{equation} \label{brane-difference} \frac{1}{a^2} \nabla^2 (\Phi
  - \Psi) = \frac{3 \gamma - 1}{m^2 \lambda} \Delta_\mathrm{m}.
\end{equation}

As can easily be seen from (\ref{brane-pi}) or
(\ref{brane-difference}), the general-relativistic relation $\Phi =
\Psi$ is not usually valid in braneworld models. An important
exception to this rule is provided by the mimicry models discussed
in Sec.~\ref{brane-sec:bc-mimic}.

One can propose other conditions of type (\ref{brane-bc-new}) that
lead to scale-in\-de\-pen\-dent behavior. For instance, one can
equate to zero the right-hand side of (\ref{brane-closed}). It
remains unclear how these conditions involving perturbations may be
generalized to the fully non-linear case. Nevertheless, in view of
the interesting properties of scale-independence and the fact that
perturbations in the stability region (\ref{brane-stable}) and
(\ref{brane-stable0}) behave in this manner, the consequences of
(\ref{brane-matter}) need to be further explored, and we shall
return to this important issue later on in this section.

Having described the system of linearized equations governing the
evolution of scalar perturbations in pressureless matter and dark
radiation, we now proceed to apply them to two important braneworld
models: the popular DGP model \cite{DGP,DDG,DLRZA} and the `mimicry'
model suggested in \cite{mimicry} and described in
Sec.~\ref{brane-sec:mimicry}. It should be noted that these two
models are complementary in the sense that the mimicry model arises
for {\em
  large values\/} of the bulk cosmological constant %
\index{cosmological constant} %
$\Lambda_\mathrm{b}$ and brane tension $\sigma$, whereas the DGP
cosmology corresponds to the opposite situation $\Lambda_\mathrm{b}
= 0$ and $\sigma = 0$.

\subsection{\!Scalar perturbations\\ \hspace*{-1.4cm}in the DGP model} \label{brane-sec:DGP}

\hspace*{3cm}\index{scalar perturbations}Amongst alternatives to LCDM, the %
\index{DGP model} %
Dvali---Gaba\-da\-dze---Por\-ra\-ti (DGP) model \cite{DGP} stands
out because of its stark simplicity. Like the cosmological constant
which features in LCDM, the DGP model too has an extra parameter
$\ell = {2 m^2 / M^3}$, the length scale beyond which gravity
effectively becomes five-dimensional. However, unlike the
cosmological constant whose value must be extremely small in order
to satisfy observations, the value \mbox{$\ell \sim cH_0^{-1}$},
required to explain cosmic acceleration, can be obtained by a
`reasonable' value of the five-dimensional Planck mass \mbox{$M \sim
10$}~MeV. As poin\-ted out earlier, DGP cosmology belongs to the
class of induced gravity models which we examine and is obtained
from (\ref{brane-action}) after setting to zero the brane tension
and
the cosmological constant in the bulk %
\index{cosmological constant} %
(i.e., \mbox{$\sigma = 0$} and \mbox{$\Lambda_\mathrm{b}=0$}). Under
the additional assumption of spatial flatness (\mbox{$\kappa = 0$}),
the modified Fried\-mann equation (\ref{brane-DGP}) for the upper
sign becomes \cite{DDG,DLRZA}
\begin{equation} \label{brane-friedman-DGP} H^2 - \frac{2H}{\ell} =
  \frac{\rho}{3m^2} .
\end{equation}
In a spatially flat Universe, given the current value of the matter
density and Hubble constant, %
\index{Hubble constant}%
$\ell$ ceases to be a free parameter and becomes related to the
matter density by the following relation
\begin{equation}
  \Omega_\ell \equiv \frac{1}{\ell^2 H_0^2} = \left (\!\frac{1-\Omega_\mathrm{m}}{2}\!\right
  )^{\!\!2}\!\!,
\end{equation}
which may be contrasted with $\Omega_{\Lambda_\mathrm{b}} = 1 -
\Omega_\mathrm{m}$ in LCDM.

Linear perturbation equations for this model were discussed in
\cite{LS,LSS,KM,Koyama,Mukohyama,Mukohyama2,Deffayet}. An
approximate
boundary condition for scalar perturbations was obtained by Koyama and %
\index{scalar perturbations}%
Maartens \cite{KM,Koyama} on subhorizon scales; it is described by
(\ref{brane-KM}).  For convenience, we present system
(\ref{brane-one})---(\ref{brane-three}) for this case:\vspace*{-3mm}
\begin{equation}
\ddot \Delta + 2 H \dot \Delta = \left(\!1 + \frac{6
\gamma}{\lambda} \!\right) \frac{\rho \Delta}{2 m^2} + (1 + 3
\gamma) \frac{\delta\rho_{\mathcal{C}}}{m^2 \lambda}  ,
\label{brane-DGP-one}
\end{equation}\vspace*{-1mm}
\begin{equation}
\dot v_{\mathcal{C}} + 4 H v_{\mathcal{C}} = \gamma
\Delta_{\mathcal{C}} + \left(\!\gamma - \frac13 \!\right)
\Delta_{\rm m} - \frac{2}{3 (1 + 3 \gamma) } \delta
\rho_{\mathcal{C}} , \label{brane-DGP-two}
\end{equation}\vspace*{-1mm}
\begin{equation}
\delta \dot \rho_{\mathcal{C}} + 4 H \delta \rho_{\mathcal{C}} =
\frac{1}{a^2} \nabla^2 v_{\mathcal{C}} \label{brane-DGP-three} .
\end{equation}

In the DGP model, the general expressions (\ref{brane-lambda}) and
(\ref{brane-gamma}) for $\lambda$ and $\gamma$ in the case of
pressureless matter reduce to\vspace*{-1mm}
\begin{equation} \label{brane-DGP-lagam}
\lambda = 2 \left( \ell H - 1 \right)\!, \quad \gamma = \frac{1}{2
(\ell H - 1)^2} - \frac16 .
\end{equation}

\begin{figure}
\vskip1mm
    \includegraphics[width=13cm]{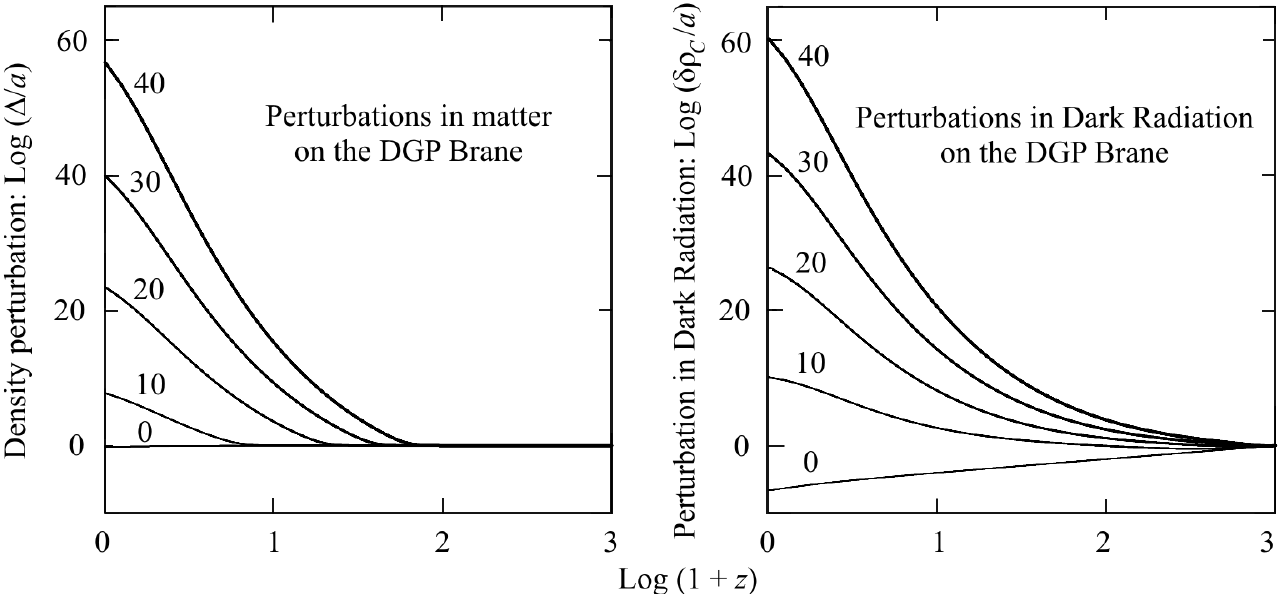}
    \vskip-2mm
  \caption{{The DGP brane with the Koyama---Maartens boundary
      condition ${{A}} = - 1/2$.} Growth of perturbations in matter
    (left) and dark radiation (right) on the DGP brane are shown for
    different values of the comoving wave number $k/a_0 H_0$
    (indicated by numbers above the corresponding curves). The current
    matter density is chosen to be $\Omega_\mathrm{m}= 0.22$, and the
    initial value of $v_{\mathcal{C}}$ is set to zero. (Our results remain
    qualitatively the same for other values of the density parameter.)
    Note the dramatic $k$-dependence in the growth of perturbations
    both in matter and in dark radiation. For comparison, $\Delta / a
    = 1$ at these redshifts in the standard CDM model with
    $\Omega_\mathrm{m}= 1$.  Figure taken from
    \cite{Shtanov:2007dh} \label{brane-fig:dgp}}\vspace*{-3mm}
\end{figure}

The results of a typical integration of the exact system of
equations (\ref{brane-DGP-one})---(\ref{brane-DGP-three}) for
different values of the wave number $k$ are shown in
Fig.~4.\ref{brane-fig:dgp}. We observe a dramatic escalation in the
growth of perturbations at moderate redshifts and a strong
$k$-dependence for perturbations in matter as well as in
dark-radiation (the y-axis is plotted in logarithmic units). These
results do not support the approximation made in \cite{KM,Koyama},
which assumes the left-hand side of equation (\ref{brane-DGP-two})
to be much smaller than individual terms on its right-hand side for
sufficiently large values of $k$, and which leads, subsequently, to
the scale-independent equation (\ref{brane-KM-one}).

We would like to stress that our conclusions themselves are not
based on the small-scale or quasi-static approximation.  Indeed, we
integrate the {\em exact\/} system of equations
(\ref{brane-one})---(\ref{brane-three}) on the brane, and the only
ansatz that we set in this system is the boundary condition
(\ref{brane-KM}).

The strong $k$-dependence of the evolution of perturbations can be
explained by the presence of the term $\nabla^2 v_{\mathcal{C}}$ on
the right-hand side of (\ref{brane-DGP-three}), \mbox{which} leads
to the generation of large perturbations of dark radiation $\delta
\rho_{\mathcal{C}}$. The quantity $v_{\mathcal{C}}$ is being
generated by the right-hand side of equation (\ref{brane-DGP-two}).
The instability in the growth of perturbations for the
Koyama---Maartens boundary condition is in agreement with the fact
that the value of ${{A}} = - 1/2$ lies well beyond the stability
domain (\ref{brane-stable0}).

\begin{figure}
\vskip1mm
    \includegraphics[width=13cm]{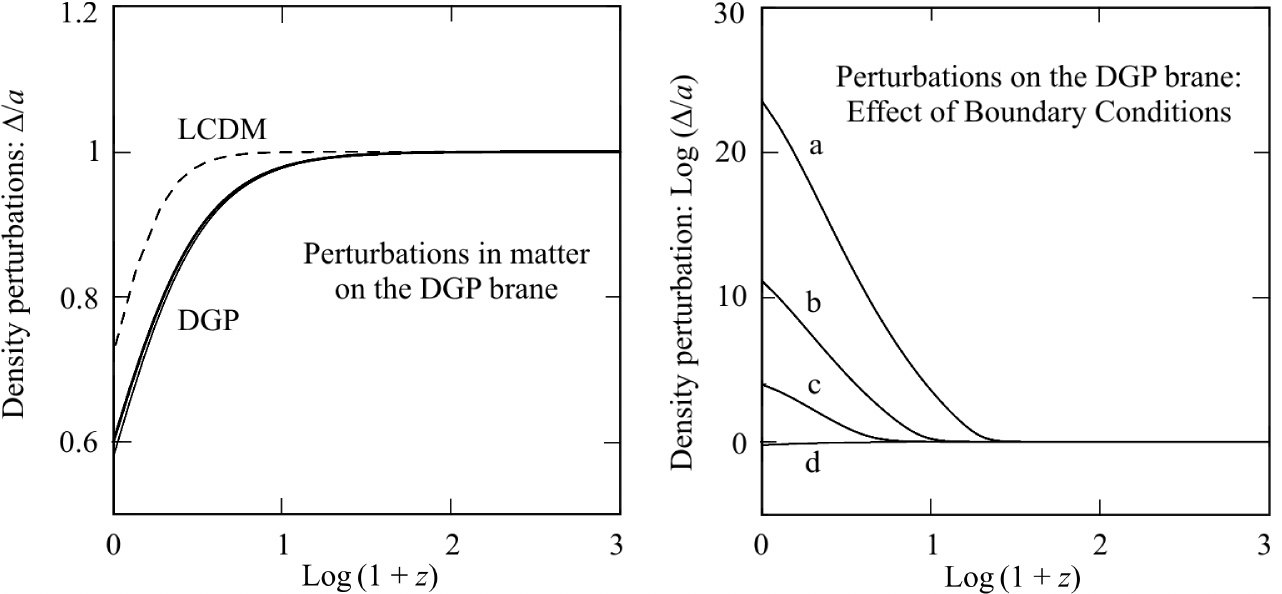}
    \vskip-2mm
        \caption{{The DGP brane with the boundary condition ${{A}}
        = 1/2$.} The parameters of the model are the same as in
      figure~4.\ref{brane-fig:dgp} but the y-axis is no longer plotted
      in logarithmic units. The two solid curves show the evolution of
      scalar perturbations corresponding to the como\-ving wave numbers
      $k / a_0 H_0 = 0$ (thin curve) and $40$ (thick curve). Note that
      these two curves are almost indistinguishable which illustrates
      that the growth of perturbations is virtually scale-independent
      in this case. The dashed line shows the behavior of scalar
      perturbations in the LCDM model. (In all cases
      $\Omega_\mathrm{m}= 0.22$ is assumed.) Figure taken \mbox{from
      \cite{Shtanov:2007dh}} \label{brane-fig:dgp-stable}}
\vspace*{-2mm} \caption{Growth of scalar perturbations in matter on
the DGP brane is
  shown for different boundary conditions in the brane-bulk
  system. (The comoving wave number $k/a_0 H_0 = 20$ in all cases.)
  Boundary conditions are specified by (\ref{brane-bc-linear}) and
  differ in the expression for ${{A}}$; namely: (a)~the Koyama---Maartens
  condition ${{A}} = - 1/2$, (b)~${{A}} = - (1 + 3 \gamma) / 4$, (c)~the
  minimal condition ${{A}} = 0$, and (d)~${{A}} = 1/2$.  Condition (b)
  with time-dependent value of ${{A}}$ was chosen because it simplifies
  equation (\ref{brane-v-c}), and condition (d) because it lies well
  inside the stability domain (\ref{brane-stable}). For comparison,
  note that $\Delta / a = 1$ for all values of $k$ in the standard CDM
  model with $\Omega_\mathrm{m}= 1$. Figure taken from
  \cite{Shtanov:2007dh}
  \label{brane-fig:bc}}\index{scalar perturbations|(}\vspace*{-3mm}
\end{figure}

As demonstrated earlier, depending upon the value of ${{A}}$,
perturbations on the brane can be either unstable or quasi-stable.
By unstable is meant $\Delta/a \gg 1$ while quasi-stability implies
$\Delta/a \sim O(1)$. The quasi-stable region (\ref{brane-stable})
is illustrated in Fig.~4.\ref{brane-fig:dgp-stable}, in which we
show the results of a numerical integration of equations
(\ref{brane-one})---(\ref{brane-three}) for ${{A}} = 1/2$.  It is
instructive to compare this figure with the left panel of
Fig.~4.\ref{brane-fig:dgp}. One clearly sees the much weaker growth
of perturbations as well as their scale-independence in this case.
We therefore conclude that boundary conditions can strongly
influence the evolution of perturbations on the brane. Our results
are summarized in Fig.~4.\ref{brane-fig:bc}, which shows the
evolution $\Delta/a$ obtained by integrating the system
(\ref{brane-one})---(\ref{brane-three}) for different boundary
conditions. (Results for the wave number $k/a_0 H_0 = 20$ are
shown.) We see that the growth of perturbations becomes weaker as
the value of ${{A}}$ approaches the stability domain
(\ref{brane-stable}), and quasi-stability is observed for ${{A}} =
1/2$.

The behavior of scalar perturbations on the DGP brane in the case of
scale-free boundary conditions (\ref{brane-bc-new}) is very similar
to that shown in Fig.~4.\ref{brane-fig:dgp-stable}. As in
\cite{LS,LSS,KM,Koyama}, we also find that perturbation growth on
the DGP brane is slower than that in LCDM.

\newpage

\subsection{\!Scalar perturbations in the mimicry
  model} \label{brane-sec:bc-mimic}
\vspace*{-0.5mm}

\hspace*{3cm}{\bfseries\itshape Scale-dependent boundary
conditions.} As expected, per\-tur\-ba\-ti\-ons in the mimicry model
crucially depend upon the type of boun\-da\-ry condition which has
been imposed. Generally speaking, brane per\-tur\-ba\-ti\-ons grow
moderately for BC's which lie in the stability domain
(\ref{brane-stable}) or (\ref{brane-stable0}) and more rapidly in
the instability region. This remains true for mi\-mic\-ry models. In
this section, we explore the behavior of perturbations in this
mo\-del for the boundary condition ${{A}} = 0$, which belongs to the
instability class. In the next section, we shall explore BC's which
give rise to more mo\-de\-ra\-te and sca\-le-in\-de\-pen\-dent
behavior.

\begin{figure}[b!]
\vspace*{-3.5mm}
    \includegraphics[width=13cm]{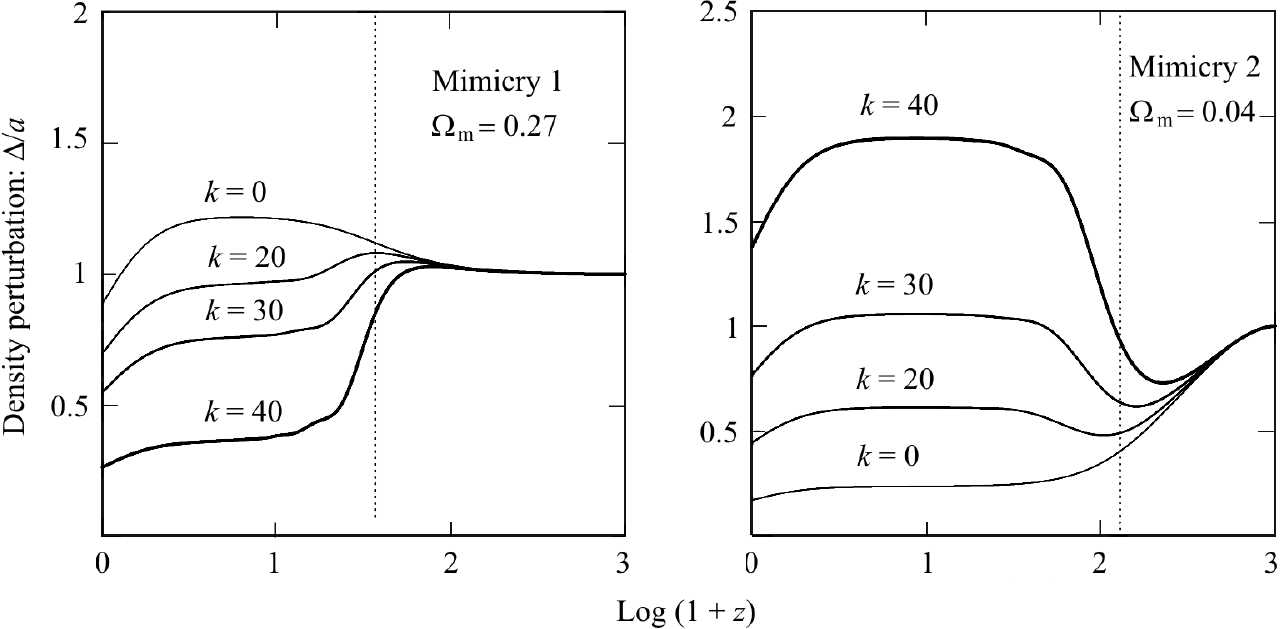}
    \vskip-2mm
  \caption{Growth of matter perturbations in the Mimicry\,1 and
    Mimicry\,2 models are shown for different values of the comoving
    wave number $k/a_0 H_0$ (indicated by numbers above the
    corresponding curves) and for the minimal boundary condition ${{A}}
    = 0$ in (\ref{brane-bc-linear}). Both models have the same
    effective parameter $\Omega^\mathrm{ LCDM}_\mathrm{m}= 0.22$,
    hence, quite different matter content, indicated by the parameter
    $\Omega_\mathrm{m}$. The value of $v_{\mathcal{C}}$ is set to zero
    initially. The position of the mimicry redshift $z_{\rm m}$ is
    indicated by the vertical dotted lines. Figure taken from
    \cite{Shtanov:2007dh}
    \label{brane-fig:mimicry12}}
  \index{matter perturbations}
  \index{scalar perturbations|)}\vspace*{-1.5mm}
\end{figure}

The growth of perturbations if ${{A}} = 0$ is substituted in
(\ref{brane-bc-linear}) is illustrated in
Fig.~4.\ref{brane-fig:mimicry12}. The $k$-dependence, clearly seen
in this figure, can be understood by inspecting the system of
equations (\ref{brane-one})---(\ref{brane-three}). Even if we start
with zero initial conditions for the dark-radiation components
$\delta \rho_{\mathcal{C}}$ and $v_{\mathcal{C}}$, the non-trivial
right-hand side of Eq.~(\ref{brane-two}) leads to the generation of
$v_{\mathcal{C}}$; then, via the $k$-dependent right-hand side of
(\ref{brane-three}), the density $\delta \rho_{\mathcal{C}}$ is
generated, which later influences the growth of perturbations of
matter in (\ref{brane-one}).  The instability in the growth of
perturbations is explained by the fact that the value of ${{A}} = 0$
lies outside the stability domain (\ref{brane-stable}) or
(\ref{brane-stable0}). However, the growth of perturbations is not
as dramatic in this case as in the DGP model with the
Koyama---Maartens BC's, mainly because the value ${{A}} = 0$ lies
much closer to the boundary (\ref{brane-stable0}) than the
Koyama---Maartens value ${{A}} = - 1/2$.

Qualitatively, the evolution of matter perturbations in mimicry
models can be understood as follows: during the early stages of
matter-domination the last term on the right-hand side of equation
(\ref{brane-one}) is not very important, which transforms
(\ref{brane-one}) into a closed equation for the matter
perturbation. Indeed, in the pre-mimicry regime, for $z \gg z_{\rm
m}$, we have $\left| \lambda \right| \gg 1$ for the quantity in the
denominator of the last term on the right-hand side of
(\ref{brane-one}), which makes this term relatively small for
moderate values of $\delta \rho_{\mathcal{C}}$. Thus, perturbations
in matter evolve according to (\ref{brane-matter}) on all spatial
scales, for redshifts greater than the mimicry redshift $z_{\rm m}$.
For $z \leq z_{\rm m}$, the quantity $\left| \lambda \right|$ is of
order unity. By this time, the perturbations $\delta
\rho_{\mathcal{C}}$ have grown large, and their amplitude strongly
depends on the wave number.  Through the last term in equation
(\ref{brane-one}), they begin to influence the growth of
matter perturbations for $z \sim z_{\rm m}$, %
\index{matter perturbations}%
resulting in the $k$-dependent growth of the latter. The reason for
the opposite $k$-dependence of matter perturbations in Mimicry\,1
and Mimicry\,2 shown in Fig.~4.\ref{brane-fig:mimicry12} is
connected with the difference in the sign of $\lambda$~--- defined
in (\ref{brane-gamma})~--- for the two models.  Thus, the last term
in (\ref{brane-one}) comes with opposite signs in Mimicry\,1 and
Mimicry\,2, and therefore works in opposite directions in these two
models.

Well inside the mimicry regime, for $z \ll z_{\rm m}$, we have
$\gamma \approx 1/3$, so that the second term on the right-hand side
of (\ref{brane-two}) can be ignored if matter perturbations are not
too large. Then equations (\ref{brane-two}), (\ref{brane-three}),
and (\ref{brane-bc-linear}) lead to a closed system of equations for
the evolution of dark-radiation per\-tur\-ba\-ti\-ons. Substituting
$\gamma = 1/3$ into this system, we obtain:
\begin{equation} \label{brane-vc} v_{\mathcal{C}} = a^{- 7/2} \xi ,~
  \delta \rho_{\mathcal{C}} = \frac{3}{(1 + 2 {{A}}) a^3} \frac{\partial}{
\partial t} \left(a^3 v_{\mathcal{C}} \right) \!,
\end{equation}
where we assumed ${{A}} \not \approx - 1/2$ to be constant. The
function $\xi$ obeys an oscillator-type equation
\begin{equation} \label{brane-xi} \ddot \xi - \left(\! \frac12 \dot H +
    \frac14 H^2 + \frac{1 + 2 {{A}}}{3 a^2} \nabla^2 \!\right) \xi = 0
  .
\end{equation}
This means that both $\delta\rho_{\mathcal{C}}$ and
$v_{\mathcal{C}}$ rapidly decay during the mimicry regime
(oscillating approximately in opposite phase) and the last term on
the right-hand side of (\ref{brane-one}) again becomes unimportant.
In particular, this will describe the behavior of the mimicry model
with the minimal boundary condition ${{A}} = 0$. The transient
oscillatory character of $\delta \rho_{\mathcal{C}}$ induces
transient oscillations with small amplitude in $\Delta$ through the
last term in (\ref{brane-three}). These small os\-cil\-la\-tions can
be noticed in Fig.~4.\ref{brane-fig:mimicry12} for $\log\, (1 + z)
\gtrsim 1$, particularly for values $40$ and $30$ of the comoving
wave number\,\footnote{\,For the Koyama---Maartens boundary
condition ${{A}} = -
  1/2$, the approximation described above is not valid during the
  mimicry stage.  Instead, during mimicry, the value of $v_{\mathcal{C}}$ decays
  without oscillating approximately as $v_{\mathcal{C}} \propto 1/a^3$, as can be
  seen from equations (\ref{brane-two}), (\ref{brane-bc-linear}), and
  the value of $\delta \rho_{\mathcal{C}}$ also decays, which follows from
  (\ref{brane-three}).}.

Two important features of mimicry models deserve to be highlighted:

1.\,\,As demonstrated in Fig.~4.\ref{brane-fig:mimicry12}, there is
a
  strong suppression of long-wavelength modes in Mimicry\,2.

2.\,\,From this figure, we also find that the growth of
short-wavelength modes in
    Mimicry\,2 can be substantial, even in a low-density Universe.

Both properties could lead to interesting cosmological consequences.
For instance, the relative suppression of low-$k$ modes may lead to
a correspon\-ding suppression of low-multipole fluctuations in the
CMB, while the increased amplitude of high-$k$ modes could lead to
an earlier epoch of structure formation. (Since the mimicry models
behave as LCDM at low redshifts, they satisfy the supernova
constraints quite well.) A detailed investigation of both effects,
however, requires that we know the form of the transfer function of
fluctuations in matter (and dark radiation) at the end of the
radiative epoch. This open issue lies outside the scope of the
present book\,\footnote{\,For simplicity, the
  amplitudes of all $k$-modes were assumed to be equal at high
  redshifts in figures 4.\ref{brane-fig:dgp}, 4.\ref{brane-fig:bc} and
  4.\ref{brane-fig:mimicry12}. A more realistic portrayal of $\Delta(k)$
  should take into consideration the initial spectrum and the
  properties of the transfer function for matter and dark
  radiation.}.

For the minimal boundary condition (${{A}} = 0$), assumed in this
section, equations (\ref{brane-con-delta}) and (\ref{brane-phi-psi})
imply $\Phi = \Psi$ and
\begin{equation}
  \frac{1}{a^2} \nabla^2 \Phi = \left(\!1 + \frac{2}{\lambda} \!\right)
  \frac{\Delta_\mathrm{m}}{2 m^2}  + \frac{\Delta_{\mathcal{C}}}{m^2 \lambda } ,
\end{equation}
which is a generalization of the Poisson equation for the mimicry
brane.

Mimicry models with the Koyama---Maartens boundary condition exhibit
much stronger instability in the growth of $\delta
\rho_{\mathcal{C}}$ for high values of $k$, enhancing the growth of
matter perturbations (not shown).  This can be explained by the fact
that ${{A}} = 0$ is much closer to the boundary of the stability
domain (\ref{brane-stable0}) than the Koyama---Maartens value ${{A}}
= - 1/2$. For the latter, the density perturbation $\Delta$ in the
Mimicry\,1 model grows to be large and negative, while the
perturbation $\delta \rho_{\mathcal{C}}$ becomes large and positive;
for instance, both $\Delta$ and the dimensionless quantity $\delta
\rho_{\mathcal{C}} / m^2 H^2$ grow by a factor of $10^{11}$ for $k /
a_0 H_0 = 40$. This provides another example of the very strong
dependence of perturbation evolution on boundary conditions.

In our calculations, we have not found any significant dependence of
the eventual growth of perturbations on initial conditions for dark
radiation specified in a reasonable range (at $z = 10^3$).

{\bfseries\itshape Scale-free boundary conditions.} As mentioned
earlier, BC's lying in the stability region (\ref{brane-stable})
lead to an almost scale-free growth of density perturbations. A
similar result is obtained if we assume the scale-free boundary
condition (\ref{brane-bc-new}) of Sec.~\ref{brane-subsec:bc}. In
this case, the momentum potential $v_{\mathcal{C}}$ decays as
$v_{\mathcal{C}} \propto a^{-4}$, and its spatial gradients in
(\ref{brane-three}) can therefore be neglected. The same is true, of
course, if one considers super-horizon modes with $k \ll aH$. In
both cases, we have approximately $\delta\rho_{\mathcal{C}} \propto
1/a^4$, suggesting that the dynamical role of perturbations in dark
radiation is unimportant. This results in a radical simplification:
as in the DGP model, for BC's lying in the stability region, the
growth of perturbations in matter can be effectively described by a
single second-order differential equation (\ref{brane-matter}),
namely,
\begin{equation} \label{brane-matter1} \ddot \Delta + 2 H \dot \Delta
  = \Theta\, \frac{\rho \Delta}{2 m^2} ,~ \Theta = \left(\!1 +
    \frac{6 \gamma}{\lambda} \!\right)\!,
\end{equation}
where $\lambda$ and $\gamma$ are defined in (\ref{brane-lambda}) and
(\ref{brane-gamma}), respectively. We shall call $\Theta(z)$ in
(\ref{brane-matter1}) the {\em `gravity term'\/} since it
incorporates the effects of modified gravity on the growth of
perturbations. The value of this term on the brane can depart from
the canonical $\Theta = 1$ in General Relativity.

\begin{figure}
\vskip1mm
    \includegraphics[width=13cm]{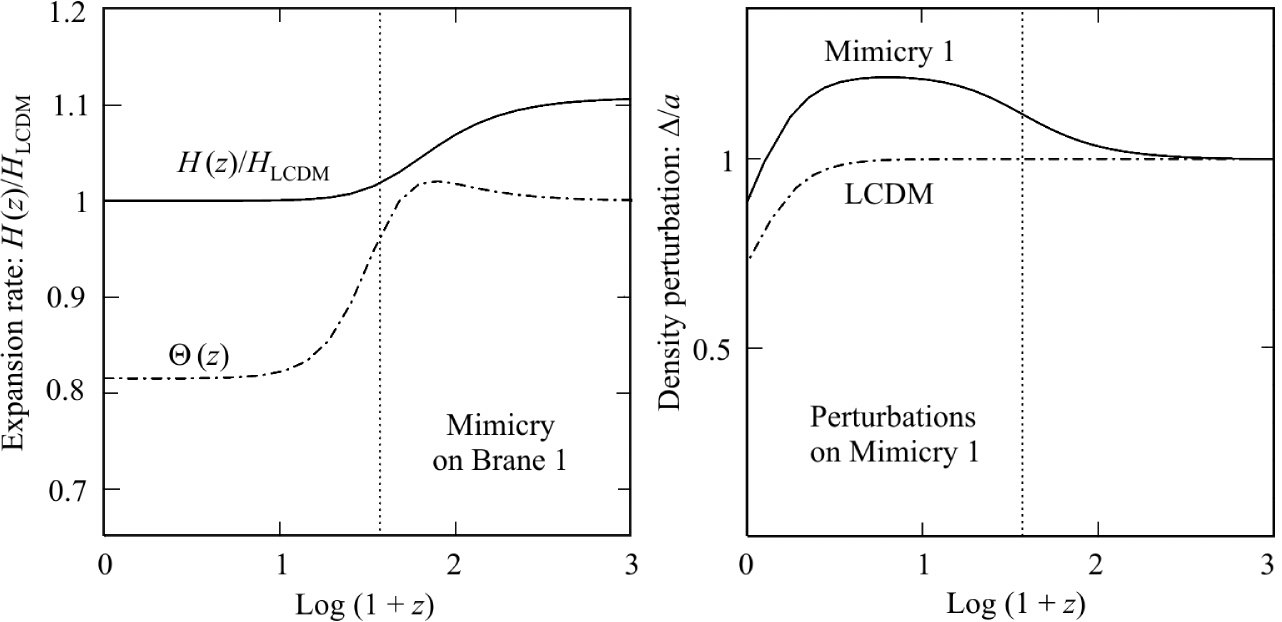}
\vskip-2mm
  \caption{The Hubble parameter in the braneworld `Mimicry~1' is shown
    relative to LCDM (solid). The LCDM model has $\Omega^\mathrm{
      LCDM}_\mathrm{ m} = 0.22$ while $\Omega_\mathrm{ m} = 0.27$ on
    the brane. Also shown is the `gravity term' $\Theta(z)$ defined in
    (\ref{brane-matter}) whose value diminishes from unity at high
    redshifts to the asymptotic form (\ref{brane-theta1}) at low
    redshifts. The dotted vertical line shows the mimicry redshift
    $z_\mathrm{m}\approx 37$. Figure taken from
    \cite{Shtanov:2007dh}\label{brane-fig:theta}}
    \vspace*{-2mm}
\caption{Density perturbations on the Mimicry\,1 brane (dot-dash)
    and in LCDM (solid). The evolution of perturbations in Mimicry\,1
    in this case is effectively described by (\ref{brane-matter}). In
    both cases, the perturbation amplitude is scaled by the expansion
    factor $a(t)$. (It may be noted that $\Delta/a = 1$ in standard
    CDM with $\Omega_\mathrm{m}= 1$.) The dotted vertical line shows
    the mimicry redshift $z_\mathrm{m}\approx 37$. The braneworld has
    $\Omega_\mathrm{m}= 0.27$ while $\Omega^\mathrm{ LCDM}_\mathrm{m}=
    0.22$. This leads to a moderate enhancement in the amplitude of
    brane perturbations over LCDM.  Figure taken from
    \cite{Shtanov:2007dh}
\index{matter perturbations} \label{brane-fig:growthB1}}
\end{figure}

Figure~4.\ref{brane-fig:theta} shows the behavior of $\Theta(z)$ for
a typical mimicry model. At redshifts significantly larger than the
mimicry redshift, $z \gg z_{\rm m}$, we have $\Theta(z) \simeq 1$,
whereas at low redshifts, $z \ll z_{\rm m}$, the value of
$\Theta(z)$ changes to
\begin{equation} \label{brane-theta1} \Theta(z) \simeq
  \frac{\Omega^\mathrm{ LCDM}_{\rm m}}{\Omega_\mathrm{m}} = \frac{\rho^\mathrm{
      LCDM}}{{\rho}}~ \mbox{for}~ z \ll z_\mathrm{m},
\end{equation}
where $\rho^\mathrm{ LCDM}$ is defined in (\ref{brane-lcdm}).  The
solid line in the same figure shows the ratio of the Hubble
parameter %
\index{Hubble parameter} %
on the brane to that in LCDM. The consequences of this behavior for
the growth equation (\ref{brane-matter}) are very interesting.
Substituting (\ref{brane-theta1}) into (\ref{brane-matter}) and
noting that $H(z) \simeq H^\mathrm{ LCDM}$ during mimicry, we
recover the standard equation describing perturbation growth in the
LCDM model
\begin{equation} {\ddot \Delta} + 2 H^\mathrm{ LCDM} \dot \Delta =
  \frac{\rho^\mathrm{ LCDM} \Delta}{2 m^2}  .
\end{equation}
Thus, ordinary matter in mimicry models gravitates in agreement with
the effective value of the gravitational constant which appears in
the cosmological relation (\ref{brane-omega-m}).

We therefore conclude that, deep in the mimicry regime ($z \ll
z_{\rm m}$), per\-tur\-ba\-ti\-ons grow {\em at the same rate\/} on
the brane and in LCDM. This is borne out by
Fig.~4.\ref{brane-fig:growthB1}, which shows the results of a
numerical integration of (\ref{brane-matter}) for Mimicry\,1
[integrating the exact system
(\ref{brane-one})---(\ref{brane-three}) gives indistinguishable
results]. Notice that the {\em total\/} amplitude of fluctuations
during mimicry in this model is {\em greater on the brane\/} than in
LCDM. Indeed, for mimicry models, we have
\begin{equation} \label{brane-ratio_omega} \frac{\Delta_\mathrm{
      brane}}{\Delta_\mathrm{ LCDM}} \simeq
  \frac{\Omega_\mathrm{m}}{\Omega^\mathrm{ LCDM}_{\rm m}}~ \mbox{for}~ z
  \ll z_\mathrm{m},
\end{equation}
and this ratio is greater than unity for Mimicry\,1.

Since the contribution from perturbations in dark radiation can be
neglected, the growth of matter perturbations in Mimicry\,2 is again
described by (\ref{brane-matter}) and by (\ref{brane-ratio_omega}).
However, since $\Omega_\mathrm{ m} < \Omega^\mathrm{ LCDM}_{\rm m}$
in this case, the final amplitude of perturbations will be {\em
smaller\/} in Mimicry\,2 than the corresponding quantity in LCDM,
which is the opposite of what we have for Mimicry\,1.

It is interesting that during mimicry, when $\gamma \approx 1/3$,
the relation between the gravitational potentials $\Phi$ and $\Psi$
reduces to the general-relativistic form $\Phi = \Psi$, as can be
seen from (\ref{brane-difference}), where $\Phi$ satisfies the
generalized Poisson equation (\ref{brane-potentials}), namely,
\begin{equation} \frac{1}{a^2} \nabla^2 \Phi = \Theta
  \frac{\Delta_\mathrm{m}}{2 m^2} .
\end{equation}

An interesting feature of Mimicry\,1 is that, at early times, the
expansion rate in this model {\em exceeds\/} that in LCDM, i.e.,
$H(z)\vert_\mathrm{ Mimicry 1} > H(z)\vert_\mathrm{ LCDM}$ for $z >
z_{\rm m}$ (see Figs.~4.\ref{brane-fig:mimic} and
4.\ref{brane-fig:theta}). [The opposite is the case for Mimicry\,2:
the expansion rate in this model is {\em lower\/} than that in LCDM
at early times, i.e., $H(z)\vert_\mathrm{ Mimicry\,2} <
H(z)\vert_\mathrm{ LCDM}$ for $z > z_{\rm m}$.]  As we can see, this
has important consequences for the growth of structure in this
model. The increase in the growth of perturbations in Mimicry\,1
relative to LCDM occurs during the period before and slightly after
the mimicry redshift has been reached, when the relative expansion
rate $H(z)/H^\mathrm{ LCDM}$ is declining while the `gravity term'
$\Theta(z)$ has still not reached its asymptotic form
(\ref{brane-theta1}). A lower value of $H(z)$ in
(\ref{brane-matter}) diminishes the damping of perturbations due to
cosmological expansion while a slower drop in $\Theta(z)$ signifies
a much more gradual decrease in the force of gravity. Consequently,
there is a net increase in the growth of perturbations on the
Mimicry\,1 brane relative to
LCDM\,\footnote{\,Figure~4.\ref{brane-fig:theta} clearly shows
  that $H(z)$ reaches its asymptotic form much sooner than
  $\Theta(z)$. Notice that, at redshifts slightly larger than $z_{\rm m}$,
  the value of $\Theta(z)$ exceeds unity.  The dependence of
  perturbation growth on the mimicry redshift $z_{\rm m}$ is very weak, and
  (\ref{brane-ratio_omega}) is a robust result which holds to an
  accuracy of better than $2\%$ for a wide range of parameter
  values.}.
For the models in Fig.~4.\ref{brane-fig:growthB1}, which have
$\Omega^\mathrm{ LCDM}_\mathrm{m}= 0.22$ and $\Omega_\mathrm{m}=
0.27$, the increase is about $20\%$. The increased amplitude of
perturbations in Mimicry\,1 stands in contrast to the DGP model as
well as Quintessence model, in both of which linearized
perturbations grow at a {\em slower\/} rate than in the LCDM
cosmology \cite{LS,LSS,KM,Koyama,WS,BB}.

It is important to note that observations of galaxy clustering by
the 2dFGRS survey provide the following estimate
\cite{Verde,Hawkins} for perturbation growth at a redshift $z =
0.15$:\vspace*{-3mm}
\begin{equation} \frac{d \log \delta}{d \log a} \equiv -
  \left. (1+z)\frac{d \log \delta}{dz} \right\vert_{z = 0.15} = 0.51
  \pm 0.11 , \label{brane-eq:2dF}
\end{equation}\vspace*{-4mm}

\noindent where $\delta \equiv \delta \rho / \rho$. Since the growth
of perturbations during the mimicry regime stays proportional to
that in the LCDM model ($\delta_\mathrm{ mimic} \propto
\delta_\mathrm{
  LCDM}$, $z \ll z_{\rm m}$), it follows that if perturbations in the LCDM
model satisfy (\ref{brane-eq:2dF}) (which they do), then so will
those in the mimicry scenario.  Nevertheless, as we have seen, the
{\em net
  increase\/} in the amplitude of perturbations on the brane is {\em
  larger than\/} that in the LCDM model. This clearly has important
cosmological consequences since it can enhance structure formation
at high redshifts as well as lead to higher values of $\sigma_8$.
Thus, while preserving the many virtues of the LCDM model, the
mimicry models add important new features which could be tested by
current and future observations.


\section[\!Perturbations of the bulk$_{ }$]{\!Perturbations of the bulk\vspace*{-1mm}} \label{brane-sec:bulk}

\subsection{\!General system of equations}

\hspace*{3cm}In this section, we consider the full physical problem
with perturbations described dynamically also in the bulk
\cite{ShVi}. The background bulk metric can be written in the form
(\ref{brane-AdS}), where $\gamma_{ij}$ is the metric of a maximally
symmetric space with coordinates $x^i$, and the function $f(r)$ in
our case of $C = 0$ is given by\vspace*{-3mm}
\begin{equation} \label{brane-bmetric_f} f(r) = \kappa -
  \frac{\Lambda_\mathrm{b}}{6} r^2 .
\end{equation}
The brane is moving along the trajectory (\ref{brane-trajec}), and
the relevant part of the bulk is given by $r \leq a(\tau)$. In what
follows, we will be interested in the case of a closed Universe
$\kappa = 1$ and $\Lambda_\mathrm{b} \leq 0$.  This makes the brane
a spherical boundary of a ball, and the boundary conditions in the
bulk are then specified simply as the requirement of regularity in
this ball (this can be called ``no-boundary conditions'').

It is convenient to present the first part of metric
(\ref{brane-AdS}) in the form $\gamma_{ab} dx^a dx^b$, where $x^a$,
$a = 1,2$ are arbitrary coordinates in place of $(\tau, r)$.  Thus,
for the background metric, we have\vspace*{-1mm}
\begin{equation} \label{brane-bmetric3}
ds^2_\mathrm{ bulk} = \gamma_{ab} dx^a dx^b + r^2 \gamma_{ij} dx^i
dx^j ,
\end{equation}\vspace*{-5mm}

\noindent where $r = r(x^a)$.  Some auxiliary expressions for the
curvature of this metric are presented in Appendix~B.3.

The scalar (with respect to the isometries of $\gamma_{ij}$)
perturbations of this metric can be described as in
\cite{Mukohyama}\,\footnote{\,In this subsection, we denote the
  five-dimensional space-time indices by capital Latin letters.}:\vspace*{-1mm}
  \[
\delta g_{AB} dx^A dx^B =  \sum_k \Big( h_{ab} Y dx^a dx^b + 2 h_a
V_i dx^a dx^i +
  \]\vspace*{-5mm}
\begin{equation}
\label{brane-bmetric_perturb}
  + \left[ h_L T_{(L)ij} + h_Y T_{(Y)ij} \right] dx^i dx^j
\Big),
\end{equation}\vspace*{-3mm}

\noindent where $Y$, $V_i \equiv \nabla_i Y$, $T_{(L)ij} \equiv 2
\nabla_i \nabla_j Y - \dfrac23 \gamma_{ij} \nabla^2 Y$, and
$T_{(Y)ij} \equiv \gamma_{ij} Y$ are the harmonics defined in
\cite{Mukohyama}, depending on the coordinates $x^i$ and all
expressible through the scalar harmonic $Y$, and $h_{ab}$, $h_a$,
$h_L$, and $h_Y$ are the perturbation coefficients depending on
$x^a$. Here, as before, $\nabla_i$ is the covariant derivative with
respect to the metric $\gamma_{ij}$. The number $k$ characterizes
the Laplacian eigenvalue of the scalar harmonics $Y$ as defined in
\cite{Mukohyama}.

Infinitesimal coordinate transformations of scalar type are defined
by the vector field $\xi^A$ which has the form\vspace*{-1mm}
\begin{equation}
  \xi_A dx^A = \sum_k \left( \xi_a Y dx^a + \xi V_i dx^i \right)\! .
\end{equation}\vspace*{-4mm}

\noindent Under diffeomorphisms, perturbations transform as
follows:\vspace*{-2mm}
\begin{equation}
  h_{ab} \to  h_{ab} - \nabla_a \xi_b - \nabla_b \xi_a , \label{brane-gauge-hab}
  \end{equation}\vspace*{-7mm}
  \begin{equation}
  h_a \to h_a - \xi_a - r^2 \nabla_a \left( r^{-2} \xi \right)\! , \label{brane-gauge-ha}
  \end{equation}\vspace*{-6mm}
  \begin{equation}
  h_L \to h_L - \xi , \label{brane-gauge-hL}
  \end{equation}\vspace*{-7mm}
  \begin{equation}
  h_Y \to h_Y - \xi^a \nabla_a r^2 + \frac23 k^2 \xi
  . \label{brane-gauge-hY}
\end{equation}
Here, $\nabla_a$ is the covariant derivative in the two-dimensional
space spanned by $(\tau, r)$, compatible with the metric
$\gamma_{ab}$. From these quantities, one can construct
gauge-invariant variables\vspace*{-3mm}
\begin{equation}
F_{ab}  =  h_{ab} - \nabla_a X_b - \nabla_b X_a ,
\end{equation}\vspace*{-5mm}
\begin{equation}
F  =  h_Y - X^a \nabla_a r^2 + \frac23 k^2 h_L ,
\end{equation}\vspace*{-6mm}

\noindent where\vspace*{-1mm}
\begin{equation}
  X_a = h_a - r^2 \nabla_a \left( r^{-2} h_L \right)
\end{equation}
is a gauge-dependent combination that transforms as $X_a \to X_a -
\xi_a$.

Note that perturbations of the tensors which are equal to zero for
the background solution are gauge-invariant: if
$T^{\cdots}_{\cdots}$ is any such tensor, then, under the
infinitesimal coordinate transformations $x^A \to x^A - \xi^A$, its
components transform as\vspace*{-3mm}
\begin{equation}
  \delta T^{\cdots}_{\cdots} = \mathcal{L}_\xi T^{\cdots}_{\cdots} = 0 ,
\end{equation}
i.e., are invariant.  In particular, perturbations of the Weyl
tensor $C^A{}_{BCD}$ (and all its contractions and derivatives) as
well as perturbations of the Einstein---De~Sitter
tensor\vspace*{-2mm}
\begin{equation}
  E_{AB} = G_{AB} + \Lambda_\mathrm{b} g_{AB}
\end{equation}
are gauge-invariant because these tensors are identically equal to
zero for the background solution (\ref{brane-AdS}).

Using the gauge transformations (\ref{brane-gauge-ha}),
(\ref{brane-gauge-hL}), one can fix the gauge in such a way as to
turn the coefficients $h_L$ and $h_a$ to zero (at least, this is
possible to do locally).  In this gauge, the coefficients $h_{ab}$
and $h_Y$ coincide with the gauge invariants $F_{ab}$ and $F$,
respectively, and the metric perturbation then simplifies
to\vspace*{-2mm}
\begin{equation}
\label{brane-gauge-inv} \delta g_{AB} dx^A dx^B =
  \sum_k Y \left( F_{ab} dx^a dx^b + F \gamma_{ij} dx^i dx^j
  \right)\!
  .
\end{equation}
Expression in this gauge can be used whenever one is to calculate
gauge-invariant perturbations, such as perturbations of the Weyl
tensor $C^A{}_{BCD}$, which is zero for the background solution.

Another set of simplifications can be obtained by taking into
account the equations of motions in the bulk (\ref{brane-bulk}),
which can be presented as
\begin{equation}
  \mathcal{R} = \frac{10}{3} \Lambda_\mathrm{b} , \quad \mathcal{R}_{AB} = \frac23 \Lambda_\mathrm{b} g_{AB} .
\end{equation}
Using these relations, one can express the Weyl tensor in the bulk
as follows:
\[
\mathcal{C}_{ABCD}  \equiv \mathcal{R}_{ABCD} -\frac23 \left(\!
g_{A[C}
    \mathcal{R}_{D]B} - g_{B[C} \mathcal{R}_{D]A}
  \!\right) + \frac16 \mathcal{R} g_{A[C} g_{D]B}=
\]\vspace*{-5mm}
\begin{equation}
    = \mathcal{R}_{ABCD} - \frac13 \Lambda_\mathrm{b} g_{A[C} g_{D]B}
   . \label{brane-weyl1}
\end{equation}

However, when calculating the curvature tensor $\mathcal{R}_{ABCD}$
to get the per\-tur\-bed equations of motion in the bulk, one needs
to deal with the complete met\-ric perturbation
(\ref{brane-gauge-inv}).

\index{Weyl tensor} %
Using \eqref{brane-weyl1} and equations (B.20), (B.21) and (B.22)
from Appendix~B.4, one easily computes the components of the
perturbed bulk Weyl tensor $\mathcal{C}_{ABCD}$ in the gauge
$h_L=0$, $h_a=0$. After that, the coefficients $h_{ab}$ and $h_Y$
can be replaced with gauge invariants $F_{ab}$ and $F$,
respectively. Thus, using (\ref{brane-AdS}) and
(\ref{brane-bmetric_f}), one can express the perturbed bulk Weyl
tensor as
\begin{equation}
\delta\mathcal{C}_{abcd} =
\sum_k\bigg[\frac{\Lambda_\mathrm{b}}{6}(\gamma_{d[a}F_{b]c}-
\gamma_{c[a}F_{b]d})+\nabla_{d}\nabla_{[a}F_{b]c}-
\nabla_{c}\nabla_{[a}F_{b]d}\bigg]Y , \label{brane-weyl_bulk1}
\end{equation}\vspace*{-3mm}
\begin{equation}
\delta\mathcal{C}_{iabc} = \sum_k
\left(\!\nabla_{[c}F_{b]a}-\frac{1}{r}F_{a[b}\nabla_{c]} r
\!\right)\nabla_i Y , \label{brane-weyl_bulk2}
\end{equation}\vspace*{-3mm}
\begin{equation}
\delta\mathcal{C}_{abij} =  0, \label{brane-weyl_bulk3}
\end{equation}
\[
\delta\mathcal{C}_{aibj} = -\frac{1}{2}\sum_k F_{ab} (\nabla_i
\nabla_j Y) +\]\vspace*{-5mm}
\[
 + \frac{1}{2}\gamma_{ij} \sum_{k} \bigg[ r (\nabla_e
r)\left(\nabla_a F^e{}_b + \nabla_b F^e{}_a - \nabla^e
F_{ab}\right)-
 \]\vspace*{-3mm}
\begin{equation}  -\, \frac{\Lambda_\mathrm{b}
r^2}{3}\,F_{ab}-\frac{\Lambda_\mathrm{b}}{6}\, \gamma_{ab} F-
r\,\nabla_a \nabla_b \left(\!\frac{F}{r}\!\right)\!\bigg]Y ,
\label{brane-weyl_bulk4}
\end{equation}
\begin{equation}
\delta\mathcal{C}_{aijk}  = \sum_k
\left[r^2\nabla_a\left(\frac{F}{r^2}\right)
-r(\nabla^{b}r)F_{ab}\right]\gamma_{i[j}\nabla_{k]}Y ,
\label{brane-weyl_bulk5}
\end{equation}
\[
\delta\mathcal{C}_{ijkl} = \sum_k F
\left(\gamma_{i[l}\nabla_{k]}\nabla_{j}Y-\gamma_{j[l}\nabla_{k]}\nabla_{i}Y
\right) +\]\vspace*{-3mm}
\begin{equation}
  +\, 2\gamma_{i[k}\gamma_{l]j}\sum_k
\left[r^2(\nabla_a r) (\nabla_b r) F^{ab}
+\left(\!\kappa-\frac{\Lambda_\mathrm{b} r^2}{3}\!\right)
F-r(\nabla_a r) (\nabla^a F) \right]Y .  \label{brane-weyl_bulk6}
\end{equation}

Expressed in this way, the perturbed bulk Weyl tensor is not
obviously traceless. In fact, as follows from \eqref{brane-weyl1},
$\delta\mathcal{C}=\delta\mathcal{R}$, and $\delta\mathcal{C}=0$
only if the perturbed equations of motion in the bulk are taken in
to account.

It was shown by Mukohyama (see \cite{Mukohyama}) that the gauge
invariants $F_{ab}$ and $F$, satisfying the perturbed bulk equations
of motion, can be expressed through a scalar master variable
$\Omega$ as
\begin{equation}
  rF_{ab} = \nabla_a \nabla_b \,\Omega-\frac{2}{3}\,\nabla^2\, \Omega\, \gamma_{ab}- \frac{\Lambda_\mathrm{b}}{18}\,\Omega\, \gamma_{ab} , \label{brane-master_fab}
\end{equation}\vspace*{-3mm}
\begin{equation}
  F = \frac{r}{3}\left(\!\nabla^2\,
    \Omega+\frac{\Lambda_\mathrm{b}}{3}\,\Omega\!\right)\!
  , \label{brane-master_f}
\end{equation}
where $\Omega$ is a solution of the master equation
\begin{equation}
  \nabla^2 \,\Omega-\frac{3}{r}\,\nabla_a r \nabla^a
  \Omega-\left(\!\frac{k^2-3\kappa}{r^2}+\frac{\Lambda_\mathrm{b}}{6}\!\right)\Omega+\frac{U}{r^2}=0
\label{brane-master equation}
\end{equation}
with some function $U$, which, in general case, is a solution of
\begin{equation} \label{U}
\nabla_a \nabla_b\, U+\frac{\Lambda_\mathrm{b}}{6}\gamma_{ab}\,U=0 .
\end{equation}

One can verify that the trace $\delta\mathcal{C}$ of the perturbed
bulk Weyl ten\-sor $\delta\mathcal{C}_{ABCD}$, defined by
\eqref{brane-weyl_bulk1}---\eqref{brane-weyl_bulk6}, can be
expressed through the Mu\-ko\-hya\-ma master variable $\Omega$ as
\begin{equation} \label{brane-weyl_trace_perturb} \delta \mathcal{C} = -
  \frac{1}{3r^3} \sum_k \left[ \nabla^2 \left( r^2 \Sigma \right) +
    \frac{\Lambda_\mathrm{b} r^2}{3} \, \Sigma \right] Y
  , \end{equation}\vspace*{-4mm}

\noindent where
\begin{equation} \label{sigma_def}
  \Sigma\equiv \nabla^2 \,\Omega-\frac{3}{r}\,\nabla_a r \nabla^a
  \Omega-\left(\!\frac{k^2-3\kappa}{r^2}+\frac{\Lambda_\mathrm{b}}{6}\!\right)\Omega.
\end{equation}
Obviously, the Mukohyama master equation \eqref{brane-master
equation} implies the condition\linebreak
\mbox{$\delta\mathcal{C}=0$.}

\subsection{\!Perturbations on the flat\\ \hspace*{-1.4cm}background bulk geometry}

\hspace*{3cm}The general problem of solving the Mukohyama master
equa\-tion and subsequent projection of the bulk Weyl tensor to the
brane is greatly simplified if the background bulk geometry is
simply a Minkowski space-time: $\gamma_{ab}=\eta_{ab}$ and
$\Lambda_\mathrm{b}=0$. In this case, the Mukohyama master equation
\eqref{brane-master equation} takes the form:\vspace*{-3mm}
\begin{equation}
  \label{brane-master equation flat} -\partial^{2}_{\tau} \Omega+\partial^{2}_{r}
  \Omega-\frac{3}{r}
  \partial_{r} \Omega-\frac{(n^2+2n-3)}{r^2}\, \Omega=0 ,
\end{equation}
where we have used the fact that, for a compact three-dimensional
manifold, the Laplacian eigenvalues $k$ of the scalar harmonics $Y$
are discrete: $k^{2}_{n}=n (n\,+$ $+\,2)$, $n=0,1,2 \mbox{...}$\,.
We also have restricted ourselves to the case $n\geq 2$, for which
the function $U$ from \eqref{brane-master equation} can be set to
zero \cite{Mukohyama}.

Equation \eqref{brane-master equation flat} is a partial
differential equation of hyperbolic type. Its simple form allows one
to separate variables\,\footnote{\,Such a simplification of the
problem becomes
  possible due to the choice of the coordinate system in the form
  \eqref{brane-AdS}. In the Gaussian normal coordinates, separation of
  variables can be performed for the special case of a de~Sitter brane
  (with constant Hubble parameter $H$), %
  \index{Hubble parameter} %
  which is of great interest because the bulk solution for the master
  variable in this case can be obtained analytically
  \cite{KLMW,KMiz}.}: $\Omega(\tau,\,r)=\xi(\tau)\,\chi(r)$ with the
functions $\xi(\tau)$ and $\chi(r)$ satisfying the ordinary
differential equations
\begin{equation} \label{brane-separation variables 1}
  \frac{d^2\xi(\tau)}{d\tau^2} + B \xi(\tau)=0,
\end{equation}
\begin{equation} \label{brane-separation variables 2}
  \frac{d^2\chi(r)}{dr^2} - \frac{3}{r} \frac{d\chi(r)}{dr} + \left[B
    - \frac{(n^2 + 2n - 3)}{r^2} \right] \chi(r)=0 ,
\end{equation}
where $B$ is some constant, which can be chosen arbitrary until some
boundary or regulatory conditions are specified.

Using expressions \eqref{brane-separation variables 1},
\eqref{brane-separation variables 2} and definitions
\eqref{brane-master_fab} and \eqref{brane-master_f}, one can easily
compute the components of the perturbed bulk Weyl tensor
\eqref{brane-weyl_bulk1}---\eqref{brane-weyl_bulk6}. Once this
operation is done, the projection $\delta C_{\mu\nu} = \delta
\mathcal{C}_{\mu
  A \nu B}\, n^A n^B$ of the bulk Weyl tensor to the brane is
trivial\,\footnote{\,The perturbation $\delta n^A$ of the unit
vector
  $n^A$ normal to the brane does not contribute to $\delta C_{\mu\nu}$
  because the Weyl tensor $\mathcal{C}_{MANB} = 0$ for the background
  solution.}. Setting the brane trajectory to be $r=a(\tau)$, we get
the components of $\delta C_{\mu\nu}$ [see definition
\eqref{brane-weyl projection definition}] to be
\begin{equation}
  \frac{1}{m^2}\delta \rho_{\mathcal{C}}  = -\,\frac{n (n+2)(n^2+2n-3)}{3 a^5}\,\Omega_\mathrm{b}   ,
  \label{brane-weyl projection rho master}
  \end{equation}\vspace*{1mm}
\begin{equation}
\frac{1}{m^2}{v_{\mathcal{C}}}(t)  = \,\frac{(n^2+2n-3)}{3 a^3}
  \left[ a H \left( \partial_{r} \Omega \right)_\mathrm{b} + \sqrt{1
      + a^2 H^2}\left(\partial_{\tau} \Omega\right)_\mathrm{b} - H\Omega_\mathrm{b}\right]\!\!,
      \label{brane-weyl projection upsilon master}
\end{equation}\vspace*{1mm}
\[
  \frac{1}{m^2}\delta \pi_{\mathcal{C}}  = - \,\frac{1}{2a} \bigg[\left(1 + 2 a^2
      H^2\right)\left(\partial^2_{\tau}\Omega\right)_\mathrm{b} + 2 a H \sqrt{1 + a^2
      H^2}\left(\partial^2_{\tau r} \Omega\right)_\mathrm{b}+
      \]\vspace*{1mm}
\begin{equation}  +  \frac{\left(1 + 3 a^2 H^2\right)}{a}\left(\partial_r
      \Omega\right)_\mathrm{b} + \frac{(n^2 + 2n - 3) \left(1 + 3 a^2 H^2\right)}{3a^2} \, \Omega_\mathrm{b}
  \bigg]\!. \label{brane-weyl projection pi master}
\end{equation}

\index{FRW metric} %
Here, $a=a(t)$ is a scale factor of the background
Friedmann---Robertson---Walker metric on the brane, $H=\dot{a}/a$ is
the
Hubble parameter on the brane, %
\index{Hubble parameter} %
and the function $\tau=\tau(t)$ is defined by the differential
equation $d \tau / d t =$ =~$ \sqrt{1+a^2 H^2}$.  The subscript
$(\mbox{...})_\mathrm{b}$ means that the corresponding function
should be evaluated at the brane. For example, $\Omega_\mathrm{b}
(t) \equiv \Omega(\tau(t), a(t))$.

Using the differentiation rule
\begin{equation}
\label{brane-rule of differentiation}
\dot{\Omega}_\mathrm{b}=\sqrt{1+a^2 H^2}\left(\partial_{\tau}
  \Omega\right)_\mathrm{b} + a H \left(\partial_{r} \Omega\right)_\mathrm{b} ,
\end{equation}
one can write \eqref{brane-weyl projection upsilon master} and
\eqref{brane-weyl projection pi master}, respectively, in the
following form:
\begin{equation}
{v_{\mathcal{C}}} = \,\frac{(n^2 + 2n-3)m^2}{3 a^3} \left[
    \dot \Omega_\mathrm{b} - H \Omega_\mathrm{b} \right]\! , \label{brane-weyl
    projection upsilon master t}
    \end{equation}
\[
  \delta \pi_{\mathcal{C}} = - \,\frac{m^2}{2a} \bigg[\ddot \Omega_\mathrm{b} -
    \frac{a^2 H (\dot H + H^2)}{(1 + a^2 H^2)}\, \dot
    \Omega_\mathrm{b}\,+\]
    \begin{equation}
    + \frac{(1 - a^2 \dot H)}{a(1 + a^2 H^2)}
    \left(\partial_r \Omega\right)_\mathrm{b}+ \frac{(n^2 + 2n - 3)}{3a^2}\, \Omega_\mathrm{b}
    \bigg]\!.  \label{brane-weyl projection pi master t}
\end{equation}

We observe that the function $v_{\mathcal{C}}(t)$ can be related to
the function $\delta\rho_{\mathcal{C}} (t)$, defined in
\eqref{brane-weyl projection rho
  master}, as\vspace*{-3mm}
\begin{equation} \label{brane-upsilon C} v_{\mathcal{C}} = -\, \frac{a^2}{n(n+2)}
  \left(\delta \dot \rho_{\mathcal{C}} + 4 H \delta \rho_{\mathcal{C}} \right)\!,
\end{equation}
which is in accordance with equation \eqref{brane-rho-c}, which we
obtained as one of the conservation equations on the brane. The
relation between functions $\delta\pi_{\mathcal{C}} (t)$ and
$\delta\rho_{\mathcal{C}} (t)$ is not so trivial due to the presence
of the third term in the square bracket on the right hand side of
\eqref{brane-weyl projection pi
  master t}.  To investigate the relation between $\delta\pi_{\mathcal{C}} (t)$
and $\delta\rho_{\mathcal{C}} (t)$, one should find the general
solution of the master equation \eqref{brane-master equation flat}
and specify $\Omega_\mathrm{b} (t)$.

This can easily be done. As we can see from \eqref{brane-separation
  variables 1}, the master variable $\Omega$ demonstrates an
oscillatory or exponential behavior depending on the sign of the
arbitrary constant $B$. In what follows, we consider the constant
$B$ to be positive to avoid problems with stability of our solution.
Setting $B \equiv \omega^2$, we get the solution of
\eqref{brane-separation variables 2} for a given $\omega$ in the
form
\begin{equation}
\label{brane-xi_solution}
  \chi(r)=r^2\left[A_\omega J_{n+1}(\omega r)+B_\omega Y_{n+1}(\omega
    r)\right]\! ,
    \end{equation}
where $A_\omega$ and $B_\omega$ are some constants that can be
chosen arbitrary until the boundary conditions are specified, and
$J_{n+1}(\omega r)$ and $Y_{n+1}(\omega r)$ are the Bessel and
Neumann functions, respectively.

The asymptotic behavior of the function $\chi(r)$ in the
neighborhood of the point $r=0$ is determined in the leading order
by the asymptotic of the Neumann functions:
\begin{equation} \label{brane-asymptN} \chi(r)\rightarrow -
  \frac{2^{n+1} n!\,B_\omega}{\pi \omega^{n+1}} \frac{1}{r^{n-1}} ,
  \quad r \rightarrow 0 .
\end{equation}
Our boundary condition at $r = 0$ is the absence of any
singularities; hence, $\chi(r)$ is regular at $r=0$, which implies
the condition $B_\omega = 0$ for all modes \mbox{with $n \geq 2$.}

Finally, the {\em general solution\/} of the master equation
\eqref{brane-master equation flat} can be written in the form of an
integral over all possible values of the parameter $\omega$:
\begin{equation}
\label{brane-master flat solution}
  \Omega(\tau, r)=\int\limits^{\infty}_{-\infty} d\omega\,\Omega(\omega)\,r^2
  J_{n+1}(\omega r)\,e^{i \omega \tau} ,
  \end{equation}
where $\Omega(\omega)$ is some complex function which we expect to
be specified from the boun\-da\-ry equations on the brane. We would
like to note that the same re\-sult may be obtained by applying the
method of Fourier transform to \mbox{equation \eqref{brane-master
equation flat}.}

Substituting \eqref{brane-master flat solution} in to
\eqref{brane-weyl projection rho master}, \eqref{brane-weyl
projection
  pi master}, we obtain
\begin{equation}
\frac{1}{m^2}\delta \rho_{\mathcal{C}} =  -\,\frac{n
    (n+2)(n^2+2n-3)}{3 a^3}  \int\limits^{\infty}_{-\infty}d\omega\,
  \Omega(\omega) J_{n+1}(\omega\,a) e^{i \omega \tau(t)} , \label{brane-weyl projection rho integral}
\end{equation}
\[
\frac{1}{m^2}\delta \pi_{\mathcal{C}} =
-\int\limits^{\infty}_{-\infty}d\omega\,
  \Omega(\omega) e^{i \omega \tau(t)} \times\]
  \[
  \times \bigg[\omega a\, J_{n}(\omega
    a)\left(\!\frac{1+3a^2 H^2}{2a} + \mathrm{ i} \,\omega a
      H\sqrt{1+a^2 H^2}\right) +\]
      \[
 +\, J_{n+1}(\omega a)\bigg(\!\frac{n(n-1)(1+3a^2 H^2)}{6a} -
    \frac{\omega^2 a (1+2a^2
      H^2)}{2} \,-\]
\begin{equation}-\, \mathrm{
        i} (n-1) \omega a H \sqrt{1+a^2 H^2}\!\bigg)\!\bigg]\!
  . \label{brane-weyl projection pi integral}
\end{equation}


\subsection{\!Quasi-static approximation}

\hspace*{3cm}The quasi-static approximation was proposed by Koyama
and Maartens in \cite{KM,Koyama} as a simplification of the general
equations describing the structure formation problem in the
braneworld molel. The main assumption of this approximation is that
the terms with time derivatives can be neglected relative to those
with spatial gradients. Specifically, this assumption implies $H
\dot{\Omega}_\mathrm{b},\,\ddot{\Omega}_\mathrm{b} \ll
(n^2/a^2)\,\Omega_\mathrm{b}$, where the values of $n$ should be
taken sufficiently lar\-ge ($n \gg 1$). In this case, our general
expressions \eqref{brane-weyl projection rho master} and
\eqref{brane-weyl
  projection pi master t} turn in\-to the approximate ones:
\begin{equation}
  \frac{1}{m^2}\delta \rho^{(qs)}_{\mathcal{C}} \approx  -\,\frac{n^4}{3 a^5}\,\Omega_\mathrm{b}
  , \label{brane-weyl projection rho qs}
\end{equation}
\begin{equation}
  \frac{1}{m^2} \delta \pi^{(qs)}_{\mathcal{C}}  \approx  - \, \frac{1}{6a} \left[\frac{n^2}{a^2}\,
    \Omega_\mathrm{b} + \frac{3(1 - a^2 \dot{H})}{a(1 + a^2 H^2)}\left(\partial_r \Omega\right)_\mathrm{b}
  \right]\! . \label{brane-weyl projection pi qs}
\end{equation}

This reproduces the result presented in \cite{KM,Koyama} in the
limit $H^2,\dot{H} \gg 1/a^2$ corresponding to the spatially flat
brane geometry. The regularity conditions imposed in
\cite{KM,Koyama} for the master variable in the bulk enabled the
authors of that work to neglect the term with $\left(\partial_r
\Omega\right)_b$ on the right hand side of \eqref{brane-weyl
projection pi qs} in the quasi-static approximation, arriving at a
relation between the functions $\delta \rho_{\mathcal{C}}(t)$ and
$\delta \pi_{\mathcal{C}}(t)$ in the form
\begin{equation} \label{brane-bc qs} \delta \rho^{(qs)}_{\mathcal{C}} \approx
  \frac{2n^2}{a^2}\delta \pi^{(qs)}_{\mathcal{C}} ,
\end{equation}
which is just the mode transform of (\ref{brane-KM}). Whether this
approximation can work for the general solution \eqref{brane-master
  flat solution} is a matter of debates and future
  \mbox{investigation.}


\section{\!Summary}

\hspace*{3cm}In this chapter, we have considered cosmological
implications of a popular braneworld model with a single extra
dimension, described by action (\ref{brane-action}), which contains
both bulk and brane curvature terms and cosmolo-\linebreak gical
constants.

We first considered an important case where the brane forms the
boundary of the five-dimensional bulk space. This is equivalent to
endowing the bulk with the $Z_2$ reflection isometry with respect to
the brane. The presence of the curvature term in the action leads to
two families of braneworld models, which we called BRANE1 \&
BRANE2\@. They differ in the manner in which the brane forms the
boundary of the five-dimensional bulk. Alternatively, the two
different families of braneworld models can be regarded as
corresponding to the two possible signs of the five-dimensional
Planck mass $M$.

Braneworld models of dark energy have an interesting and unusual
pro\-per\-ty that their luminosity distance $d_L$ can {\em exceed\/}
that in the LCDM model with the same matter content. This is unusual
since, within the general-re\-la\-ti\-vis\-tic framework, the
luminosity distance has this property {\em only if\/} the equation
of state of dark energy is strongly negative ($w < -1$). Phantom
dark energy, which realizes this feature \cite{Caldwell2002}, is
beset with a host of undesirable properties which makes this model
of dark energy unattractive. We have shown that braneworld models
have all the advantages and none of the disadvantages of phantom
models and therefore endow dark energy with exciting new
possibilities. A recent analysis of braneworld models in [546---548]
has demonstrated that BRANE1 models (which generically have $w \leq
-1$) are consistent \mbox{with} observations of supernovae combined
with baryon acoustic oscillations and integrated Sachs---Wolfe
effect.

Another feature of the braneworld scenario discussed in this chapter
is that it allows for a Universe which is {\em transiently
  accelerating\/}. Recent investigations indicate that an eternally
accelerating Universe, which possesses a cosmological event horizon,
prevents the construction of a conventional $S$-matrix descri\-bing
particle interactions within the framework of string or $M$-theory
[549---551]. We have demonstrated that braneworld models can enter
into a regime of accelerated expansion at late times {\em even if\/}
the brane tension and the bulk cosmological constant are tuned to
satisfy the Randall---Sundrum constraint (\ref{brane-RS}) on the
brane. In this case, braneworld dark energy and the acceleration of
the Universe are {\em transient\/} phenomena. In this class of
models, the Universe, after the current period of acceleration,
re-enters the matter-dominated regime.  We have shown that viable
models realizing this behavior are those of \mbox{BRANE2 type.}


We have shown that braneworld cosmology, in a certain broad region
of the values of fundamental constants, exhibits a property which we
called {\em cosmic mimicry\/}. During early cosmological epochs, the
braneworld behaves like a matter-dominated Friedmann Universe with
the value of the cosmological parameter $\Omega_{\rm m}$ that would
be inferred from observations of the local matter density.  At late
times, however, the Universe evolves almost exactly like in the LCDM
scenario but with a {\em renormalized\/} value of the cosmological
density parameter $\Omega^{\rm LCDM}_{\rm m}$. Specifically, a
positive-tension BRANE1 model, which at \mbox{high} redshifts
expands with density parameter $\Omega_{\rm m}$, at lower redshifts
mimics the LCDM cosmology with a {\em smaller value\/} of the
density parameter $\Omega^{\rm LCDM}_{\rm m} < \Omega_{\rm m}$. A
negative-tension BRANE2 model at low redshifts also mi\-mics LCDM
but with a {\em larger
  value\/} of the density parameter $\Omega^{\rm LCDM}_{\rm m} >
\Omega_{\rm m}$.

The cosmic-mimicry scenario has interesting cosmological properties.
For instance, in the case of BRANE1 (BRANE2), the Universe expands
faster (slower) than in the LCDM scenario at redshifts greater than
the mimicry redshift $z_{\rm m}$, whereas, for $z < z_{\rm m}$,
$H_{\rm
  brane}(z) \equiv H_{\rm LCDM}(z)$ in both models. The smaller value
of the Hubble parameter at intermediate redshifts ($z > \mbox{few}$)
in the case of BRANE2 leads to an older Universe and also to a
redshift of reionization which can be significantly lower than that
inferred for the LCDM model from the WMAP data \cite{WMAP7b}.

The effect of cosmic mimicry and the existence of two asymptotic
density parameters $\Omega_{\rm m}$ and $\Omega^{\rm LCDM}_{\rm m}$
is a consequence of the time-dependence of the {\em effective\/}
gravitational coupling in braneworld theory \cite{MMT}, which can be
related to the well known property of the scale-dependence of the
effective gravitational coupling in braneworld models [533---535].
On large spatial scales, $kr \gg 1$, the braneworld model with
positive brane tension (BRANE1) exhibits gravity with the
renormalized effective gravitational constant (\ref{brane-g-eff}),
and we have shown that this renormalization corresponds to a
renormalization of the cosmological density parameter
(\ref{brane-omega-m}).

Only two effective gravitational constants appear in the cosmology
under consideration, given by (\ref{brane-g-eff}) and
(\ref{brane-g-obs}), respectively, for low and high energy
densities. However, in the local gravitational physics, there also
appears the spatial distance $r_*$ defined in (\ref{brane-scale})
and depending upon the mass of the central source, so that gravity
in the range of distances\vspace*{-3mm}
\begin{equation}
r_* \lesssim r \lesssim \ell
\end{equation}

\noindent from the source has a different value of the gravitational
constant, given by (\ref{brane-g-tilde}), and, moreover, has a
scalar--tensor character manifest, e.g., in (\ref{brane-discrep}).
This may be important for the estimates of masses from the dynamics
of clusters of galaxies and from gravitational lensing on these
scales in the brane\-world theory [533---535, 541, 542].

On small distances from the central source, $r \ll r_*$, both
positive-tension and negative-tension branes apparently behave
similarly reproducing the Ein\-stein gravity to a high precision
with the gravitational constant $1/m^2$, which is the bare
gravitational coupling in the braneworld action
(\ref{brane-action}). However, this expectation is to be verified by
refined calculations in braneworld models with arbitrary sign of
brane tension and without the RS constraint (\ref{brane-RS}).  In
this respect, we should note that solution with a spherically
symmetric source (the analog of the Schwarzschild and interior
solution in General Relativity) largely remains to be an open
problem in braneworld theory (for recent progress in the DGP model,
see \cite{GI}).


\index{loitering Universe}%
We have demonstrated that the braneworld models of dark energy allow
one to construct a {\em loitering\/} Universe. An important aspect
of braneworld loitering is that, in contrast to the conventional
loitering scenarios that demand a closed Universe, loite\-ring on
the brane can easily occur in a spatially flat cosmological model. A
key role in making the brane loiter is the presence of (negative)
dark radiation --- a generic five-dimensional effect associated with
the projection of the bulk gravitational degrees of freedom onto the
brane. The Universe can loiter at large red\-shifts ($z \gtrsim 6$)
while accelerating at the present epoch. During loitering, the value
of the Hubble parameter %
\index{Hubble parameter|(} %
decreases steadily before increasing again. As a result, the age of
the loite\-ring braneworld is larger than that of a LCDM Universe at
a given red\-shift. This feature may help spur the formation of
$\sim $$10^9M_\odot$ black holes at red\-shifts $\gtrsim$$ 6$ whose
presence (within high red\-shift QSO's) could be problematic for
standard LCDM cosmology \cite{richards03,bh}. Loitering is also
expected to increase the growth rate of density inhomogeneities and
could, in principle, be used to reconcile structure formation models
which predict a lower amplitude of initial `seed' fluctuations with
the observed anisotropies in the cosmic microwave background (see
\cite{dodelson}).


\index{singularity} %
Braneworld models give rise to cosmological singularities which are
not commonly encountered in General Relativity. This is largely due
to the possibility of different kinds of embedding of the brane in
the higher dimensional (bulk) space-time. Singular embedding implies
that the expansion (in time) of the brane cannot be continued
indefinitely. The singularities which we have examined in this
chapter have the property that, while the density, pressure, and
Hubble parameter on the brane remain finite, higher derivatives of
the Hubble parameter blow up as the singularity is approached. For
this reason, we called them ``quiescent singularities.'' Despite its
deceptively mild nature, the quiescent singularity is a real
curvature singularity at which the Kretschmann invariant diverges
($R_{abcd}R^{abcd} \to \infty$). The importance of quantum effects
in regions of large space-time curvature has been demonstrated in a
number of papers \cite{zhBirrDav}, and it should therefore come as
no surprise that these effects can significantly alter the classical
behavior near the quiescent singularity, as demonstrated by us in
this chapter.

Unlike the classical Big-Bang singularity, the quiescent singularity
in brane\-world models is reached in regions of {\em low density\/}
and is therefore en\-coun\-te\-red during the course of the Universe
expansion rather than its col\-lap\-se. Densities lower than the
mean value are known to occupy a large fil\-ling fraction within the
cosmic web \cite{sheth,shandarin}. Therefore, if the braneworld
model is a reasonable representation of reality, one might expect
that it is likely to encounter the quiescent singularity (or its
quantum-corrected counterpart, the ``soft singularity'') within
large underdense regions, or voids. The rapidly va\-rying space-time
geometry near the quiescent singularity can, in addition to vacuum
polarization, also give rise to quantum creation of fields which do
not couple conformally to gravity. This allows one to suggest a
cosmological sce\-na\-rio which, at later times, is reminiscent of
quasi-steady-state cosmology, with the Hubb\-le parameter showing
oscillations about a constant value.\index{Hubble parameter|)}


In this chapter, we have also considered the braneworld model
without\linebreak the mirror ($Z_2$) symmetry of the bulk space with
respect to the brane. We find that, depending upon the choice of the
brane embedding, cosmological expansion on the brane can proceed
along four independent branches,
 two of which survive in the case of
$Z_2$ symmetry. An important property of this class of models is
that the four-dimensional gravitational and cos\-mological constants
are\index{cosmological constant} effective quantities derivable from
five-dimensional \mbox{physics.} In this case, brane expansion
mimics $\Lambda$CDM at low redshifts, but the `screened' matter
density parameter $\widetilde\Omega_{\rm m}$ does not equal to its
bare (dynamical) value $\Omega_{\rm m}$. This opens a new avenue for
testing such models against ob\-servations (see \cite{mimicry,om}).
Another important property of these models
 would be the growth of density perturbations which
is likely to differ from $\Lambda$CDM. Braneworld models can be
phantom-like and also exhibit transient acceleration. Thus, brane
phenomenology, with its basis in geometry, provides
 an
interesting alternative to `physical' dark energy scenarios such as
\mbox{quintessence.}


Braneworld theories with large extra dimensions have one common
pro\-per\-ty: while the dynamics of the higher-dimensional bulk
space needs to be taken into account in order to understand brane
dynamics, all observables are restricted to the four-dimensional
brane.  In field-theoretic language, the situation can be described
in terms of an {\em infinite (quasi)-continuum\/} of Kaluza---Klein
gra\-vi\-ta\-tio\-nal modes existing on the brane from the brane
viewpoint. This pro\-per\-ty makes braneworld theory complicated,
solutions on the brane non-uni\-que and evolution non-local.

Fortunately, in situations possessing a high degree of symmetry, the
above properties of braneworld theory do not affect its cosmological
solutions (at least, in the simplest case of one extra dimension).
Thus, homogeneous and isotropic cosmology on the brane is almost
uniquely specified since it involves only one additional integration
constant which is associated with the mass
 of a black hole in the
five-dimensional bulk space.  However, in order to turn a braneworld
model into a {\em complete\/} theory of gravity viable in all
physical circumstances, it is necessary to address the issue of
boundary conditions.

In this chapter, we adopted a different approach to the issue of
boundary conditions in the brane-bulk system.  From a broader
perspective, boundary conditions can be regarded as any conditions
restricting the space of solutions. Our approach is to specify such
conditions directly on the brane \mbox{which} represents the
observable world, in order to arrive at a local and closed system of
equations on the brane. The behavior of the metric in the bulk is of
no further concern in this approach, since this metric is for all
practical purposes unobservable. Since the non-locality of the
braneworld equations is known to be connected with the dynamical
properties of the bulk Weyl tensor projected onto the brane
\cite{SMS}, it is natural to consider the possibility of imposing
certain restrictions on this tensor. We have assumed the
one-pa\-ra\-me\-ter family of boundary conditions
(\ref{brane-bc-linear}) for perturbations. This family generalizes
the boundary condition derived by Koyama and Maartens \cite{KM}
(with ${{A}} = -1/2$) for the DGP model in the small-scale and
quasi-static \mbox{approximation.}\looseness=1

An important conclusion is that the growth of perturbations in
braneworld models strongly depends upon our choice of boundary
conditions.  This was illustrated in figure~\ref{brane-fig:bc} for
the DGP model. Specifying boundary conditions in the form
(\ref{brane-bc-linear}) allows us to determine regions of stability
and instability in terms of the single parameter ${{A}}$; they are
described by Eq.~(\ref{brane-stable}). In the DGP model,
perturbations are explicitly demonstrated to be quasi-stable for
${{A}} = 1/2$ (figure~4.\ref{brane-fig:dgp-stable}) and unstable for
${{A}} = -1/2$ (figure~4.\ref{brane-fig:dgp}). In the instability
domain, gradients in the momentum potential $v_{\mathcal{C}}$ of
dark radiation, lead to the creation of perturbations in this
quantity via equation (\ref{brane-two}). This effect can
significantly boost the growth of perturbations in matter. An
important implication of this effect is that perturbations in the
baryonic component might overcome the `growth problem,' which
plagues them in standard General Relativity, and grow to acceptable
values without requiring the presence of (deep potential wells in)
dark matter. The Mimicry\,2 model looks promi\-sing from this
perspective. Note that, in this model, the expansion of a
low-density Universe is virtually indistinguishable from that of a
(higher-density) \mbox{LCDM model.}

The values of ${{A}}$ lying in the stability region
(\ref{brane-stable}) or a scale-free boundary condition such as
(\ref{brane-bc-new}) may also be important. In this case,
perturbations in the DGP model grow {\em slower\/} than in LCDM
whereas perturbations in Mimicry\,1 grow somewhat faster. This
suggests that structure formation may occur slightly earlier in
Mimicry\,1 than it does in LCDM.

It is well known that the expansion history, $H(z)$, does not
characterize a given world model uniquely, and it is conceivable
that cosmological models having fundamentally different theoretical
underpinnings (such as dif\-ferent forms of the matter Lagrangian or
dif\-ferent field equations for gravity)\linebreak could have
identical expansion histories (some examples may be found
in\linebreak \cite{Sahni2000,Sahni:2006pa}).

Whether boundary conditions such as described by
Eq.~(\ref{brane-bc-new}) will remain in place for a more fundamental
extra-dimensional theory is presently a moot point. Perhaps, by
comparing the consequences of different boundary conditions with
observations we will gain a better understanding of the type of
braneworld theory most consistent with reality.


\setcounter{chapter}{4}
\chapter{ENERGY IN GENERAL\\[-0.5mm] RELATIVITY
  IN VIEW OF SPINOR\\[1mm] AND TENSOR METHODS}\markboth{CHAPTER 5.\,\,Energy in General Relativity}{CHAPTER 5.\,\,Energy in General Relativity}
\thispagestyle{empty}\vspace*{-12mm}

\begin{wrapfigure}{l}{2.6cm}
\vspace*{-6.45cm}{\includegraphics[width=3.0cm]{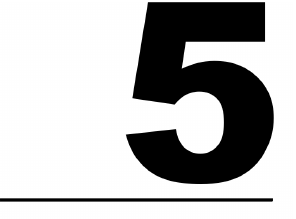}}\vskip17.2cm
\end{wrapfigure}
\vspace*{10mm}

 \setcounter{section}{1} \vspace*{-5mm}
\hspace*{3cm}\section*{\hspace*{-3cm}5.1.\,\,Introduction$_{}$}\label{ch5-51}\begin{picture}(10,10)
\put(0,-124){\bfseries\sffamily{302}}
\end{picture}

\vspace*{-1.1cm}\noindent \index{positive energy theorem (PET)}It is
known that positive energy theorem (PET) in General Relativity
\cite{yau, witten}, possessing independent principal significance,
also creates the basis for solving of other problems, for example,
stability of Minkowski space. Among assumptions at proving the
theorem there is hypothesis about sa\-tis\-fy\-ing of one of the
energy conditions~--- dominant energy condition
(DEC).\index{dominant energy condition (DEC)} But just in the
beginning of the proving PET searches the change of view of the
energy conditions has taken place.  Tipler \cite{tipler} had
analyzed the consequences of possible violation of strong energy
condition on existence of singularities in space-time.  Further
Vis\-ser \cite{viss, viss1}, Vis\-ser and Barcelo \cite{viss_barc}
pointed out the existence of quan-\linebreak tum and classical
effects which cause the violation of all energy conditions.
Possibility for proving the positive energy theorem at more weak
energy condition in comparison with standard one has been discussed
by Shiromizu and Sugai \cite{shiromizu}. Consequently the discovery
of accelerated expansi-\linebreak on\index{accelerated expansion} of
Universe, development of the dark ener\-gy con\-cept and radical
change of the energy conditions para\-digm re\-qui\-re detailed
analysis of influence of this change on all PET aspects, including
different methods of its proving.

From the moment of the first proof of this theorem \cite{yau} there
were presented many simplified and complement proofs [644---647]
within the limits of the Witten's spinor method, as well as
alternative to Wittenian one [648---650]. Among these alternative
tensor methods the most
developed one is Nester's method of special orthonormal frame (SOF), %
\index{special orthonormal frame (SOF)} %
on the basis of which there is a set of gauge conditions for the
choice of this orthonormal frame on a three-dimensional Riemannian
manifold \cite{nestjmp}. These conditions are purely geometrical,
because they are expressed in terms of teleparallel geometry.

Application of SOF is not limited by proof of PET and investigating
of quasilocalization.  In particular, by adopting four-dimensional
special or\-tho\-nor\-mal frames, the tetrad equations for vacuum
gravity are put into explicitly causal and symmetric hyperbolic
form, independent of any time slicing or other gauge or coordinate
specialization \cite{{Estabrook Robinson Wahlquist}}. Buchman and
Bardeen \cite{Buchman_Bardeen} within a first order symmetric
hyperbolic tetrad formulation of the Einstein equations developed by
Estabrook and Wahlquist obtained stable unconstrained evolution for
certain initial conditions in SOF, but not with some Lorentz gauge.

Dimakis and M${\rm {\ddot u}}$ller-Hoissen have shown
\cite{dim2,dim1} that Nester's gauge conditions are related to the
three-dimensional Dirac equation. Since the solutions of the latter
as elliptic equations can have zeros, as Ashtekar and Horowitz
\cite{ash} have noticed for the first time (see also \cite{dim2},
where the additional arguments for support of this statement are
presented together with cor\-res\-pon\-ding references to works, in
which zeros of elliptic equations are investigated), the special
orthonormal frame (triad) as well as Dimakis and M${\rm {\ddot
u}}$ller-Hoissen SOF (tetrad) are determined only almost everywhere
on space-like hy\-per\-sur\-fa\-ce in asymptotically Minkowskian
manifold. From uniqueness theorem for system of linear equations of
first order with smooth coefficients follows that nodal sets of
codimension 1 of Sen---Witten spinor field are absent, and B\"{a}r
\cite{bar} proved that nodal set of generalized Dirac equation is a
countably $(n-2)$-rec\-ti\-fi\-able set and thus has Hausdorff
dimension $(n-2)$  at most. But it is important to obtain conditions
for absence of all nodal points, because possible existence of zeros
of spinor field even on the set of zero measure is the barrier for
the correspondence of spinor and tensor methods as well as for
dis\-tin\-gui\-shing of SOF as constituent of frames of reference,
since the latter in some physically non-singular points does not
exist.

The connections between triads and Sen---Witten (SWE) equation %
\index{Sen---Witten equation (SWE)} %
were also investigated by Frauendiener \cite{Frau}. He obtained the
necessary and sufficient conditions that have to be satisfied by the
triad in order to correspond to the spinor that satisfies the
Sen---Witten equation. These conditions, as it was marked by
Frauendiener, are closely connected with Nester's conditions,
because they also include some cyclic conditions. But in the process
of obtaining of these conditions the possibility of the situation,
when the spinor equals zero in one or even on a set of points of
nonvanishing measure, is not taken into account, therefore
Frauendiener's theorem is correct only under the suitable additional
assumption. SOF reflects general features, domain of SOF application
is not limited by PET and asymptotically Minkowskian manifold. In
particular, SOF is used in problem of quasilocalization of
gravitational energy. Not long ago Frauendiener, Szabados and Nester
performed additional investigation of zeros absence conditions for
Sen---Witten (SWE) equation \cite{Frauendiener_Szabados_Nester}.
De\-ve\-lo\-ping our idea of elimination of zeros for Sen---Witten
equation by choice of appropriate boundary conditions [657---659]
firstly (2005 Annual Meeting of Chinese Physics Society, Taipei)
denied by Nester, they determined conditions for zeros absence in
class of spaces nearest to the flat one~--- Petrov class $O$, the
unique non-trivial representative of which is conformally flat
pseudorie\-mannian~\mbox{space.}

Our first purpose is to show that non-trivial solutions of SWE on
defined by some conditions hy\-per\-sur\-fa\-ce and, generally, in
algebraically more general space-time, will not have any node point,
and to prove on this basis the equivalence of SWE and Nester's gauge
and therefore the existence of correspondence of spinor and tensor
methods in investigations of the positive definition of the
gravitational energy and its quasilocalization.  Further we will
prove that in such spaces even at violation of WEC by dark energy,
the full energy can be positive definite.

\section[\!Connection
  between spinor and tensor methods] {\!\!Connection
  between spinor and tensor\\ \hspace*{-0.95cm}methods in the positive energy problem\label{p3ch5-2}}

\hspace*{3cm}In subsection 5.2.2 we review the ascertained by
Sko\-ro\-bo\-hat'ko properties of nodal sets of selfadjoint second
order elliptic equations and strong elliptic systems of equations of
second order.

In subsection 5.2.3 on the basis of these results we show at what
boundary values for the spinor field the solution for SWE has not
any nodal points in the bounded domain on the space-lake
spatial-constant mean curvature hypersurface in Petrov type  $N$
space-time. Then using the methods, introduced by Reula \cite{reu}
and Ashtekar and Horowitz \cite{ash}, we prove that in Petrov type
$N$ space-time non-trivial solutions of SWE with asymptotically flat
initial data set do not equal zero in any point of finite or
infinite domain on spatial-constant mean curvature (SCMC)
hypersurface  in asymptotically flat Petrov type $N$ space-time.

This allows us to prove in subsection 5.2.4 the equivalence of SWE
and Nester's gauge, and also to complement Nester's investigations,
showing that the local rotation to Nester's SOF exists not only for
geometries in a neighborhood of Euclidean space, but everywhere on
maximal hy\-per\-sur\-fa\-ce with good topological
properties\,\footnote{\,Maximal surface are space-like submanifold
which locally maximize the induced area functional.}. The latter
circumstance allows to take down fully the negation of Dimakis and
M${\rm {\ddot u}}$ller-Hoissen against the Nester's method
\cite{dim2}, which was taken down partly by Nester earlier
\cite{nes2}.

We use the Witten's method in the interpretation, given by Reula
\cite{reu}. In the basis of this interpretation lies the reduction
of $SL(2,C)$ spinors of space-time to $SU(2)$ spinors on space-like
hy\-per\-sur\-fa\-ce; this reduction was introduced by Sommers
\cite{som} and Sen \cite{sen}.

From the point of view of theory of differential equations with
partial derivatives the necessity for investigation of submanifolds,
on which the solutions of elliptic equations are equal to zero,
first of all is connected with the fact that the necessary and
sufficient conditions for absence of such closed submanifolds of
codimension one are simultaneously the necessary and sufficient
conditions for uniqueness of the Dirichlet problem for these
equations in the bounded
domain (the existence and stability of solutions at sufficiently
smoothed coefficients leads to its uniqueness). On the other hand,
the uniqueness theorems for Dirichlet problem define conditions for
absence of submanifolds of codimension one, on which the solutions
of elliptic equations are equal to zero.\vspace*{-1.5mm}


\subsection{\!Sen---Witten equation\\ \hspace*{-1.2cm}in Petrov type \boldmath$N$ space-time\label{p3ch5-sec21}}
\vspace*{-0.5mm}

\hspace*{3cm}In next sections we prove, that non-trivial solutions
of Sen---Witten equation with asympto\-tically flat data set in
Petrov type $N$ on SCMC hy\-per\-sur\-fa\-ce does do not equal zero
in any point of this hy\-per\-sur\-fa\-ce. On this basis we
ascertain the correspondence between Witten's spinor method and
Nester's tensor method.

Let $(M,g)$ be an asymptotically Minkowskian space-time  of Petrov
type $N$ with space-like foliation $\Sigma_t\times \{t\}$ and metric
$g$ of signature $(+,-,-,-)$. We assume initial data set $(\Sigma_t,
h_{\mu\nu}, \mathcal{K}_{\pi\rho})$ to be asymptotically flat in
introduced by Reula \cite{reu} sense and $h_{\mu\nu},\,
\mathcal{K}_{\pi\rho}$ to be of $C^\infty$ class on $C^\infty$
hy\-per\-sur\-fa\-ce $\Sigma_t$. Assumptions about asymptotical
properties (topology) are result of PET conditions; these are
important for existence of SOF, but, as it will be seen later, these
are not necessary for obtaining conditions of the nodal sets
absence.

The constraint equations of general relativity on the space-like
hy\-per\-sur\-fa\-ce $\Sigma_t$ are\vspace*{-5mm}
\begin{gather}
  -R-\mathcal{K}_{\mu\nu}\mathcal{K}^{\mu\nu} +{\cal K}^2=2\mu,
  \label{p1} \\
  D_\mu\left(\mathcal{K}^{\mu\nu}-\mathcal{K}h^{\mu\nu}\right)=
  \mathcal{J}^\nu\!, \label{ p2}
\end{gather}\vspace*{-5mm}

\noindent where $R$ is scalar curvature of $\Sigma_t$, $h=
g-n\otimes n$ is induced metric on $\Sigma_t$.  $D_\mu$ is induced
by connection $\nabla _\mu$ on $M$ connection on $\Sigma_t$, $
\mathcal{K}_{\mu\nu}$ is extrinsic curvature of $\Sigma_t$,
$\mathcal{K}= \mathcal{K}_\nu^\nu$.  $\mu$ and ${\cal J}^\nu$ are
the energy density and momentum density respectively of the matter
in the frame of reference of an observer, whose one-form of
4--velocity is $\xi=dt$. In the case of barionic matter $\mu$ and
$\mathcal{J}^\nu$ satisfies the dominant energy
condition\vspace*{-2mm}
\begin{equation}
\label{p3}
  \mu\geq\mid\mathcal{J}^\nu\mathcal{J}_\nu\mid^{1/2}\!\!.
\end{equation}\vspace*{-5mm}

As it was shown by Witten \cite{witten}, if on $\Sigma_t$ exists the
solution $\beta^{\,C}$ of Sen---Witten equation\vspace*{-1mm}
\begin{equation}
  \label{p3'}
  \mathcal{D}^B{}_C\beta^C=0
\end{equation}\vspace*{-5mm}

\noindent with $\beta^C$ going asymptotically to a constant spinor
$\beta_0{}^C$, then the total mass is non-negative.  The action of
Sen operator $\mathcal{D}_{AB}$ on spinor field is %
\index{action of spinor field}\vspace*{-1mm}
\begin{equation}
  \label{p4}
  \mathcal{D}_{AB}\lambda_C
  =D_{AB}\lambda_C+\frac{\sqrt2}{2}{\cal
    K}_{ABC}{}^D\lambda_D.
\end{equation}\vspace*{-4mm}

\noindent For solving the problem of zeros existence for $\beta^C$
solutions of the equation (\ref{p3'}), when $\beta^C$ is going
asymptotically to the constant spinor $ \lambda_0{}^C\neq0$, we
briefly review the results of Skorobohat'ko.

\subsection{\!Nodal surfaces of selfadjoint\\
  \hspace*{-1.2cm}elliptic  second order equations{\label{ ch5-subsec22}}}
\vspace*{-0.5mm}
%
%
%

\hspace*{3cm}The unique solvability of Dirichlet problem for
elliptic equa\-tions was studied in the works [663---666].  The
results of these works were ge\-ne\-ra\-li\-zed in \cite{aro,cor}
where the similar results were obtained independently.  In
particular, from the Aronszain \cite{aro} and Cordes \cite{cor}
theorem we see that non-tri\-vial solutions of the
equation\vspace*{-1mm}
\begin{equation*}
  a^{\mu\nu}\left(\!x, u, \frac{\partial u}{\partial x}\!\right)
  \frac{\partial^2 u}{\partial x^\mu\partial x^\nu }=f\left(\!x, u,
    \frac{\partial u}{\partial x}\!\right)\!,\quad u(x),\; a^{\mu\nu}(x)\in
  C^2
\end{equation*}\vspace*{-3mm}

\noindent in $R^n$, which is elliptic for all $x^1, \mbox{...}, x^n$
and $u$, in any point $\xi_0$ cannot have zero of infinite order.
Therefore this allows to exclude the existence of the solutions with
zeros of infinite order only.

In the works of Skorobohat'ko [672, 673] the known theorems about
the distribution of zeros for linearly independent solutions of the
ordinary differential second order equation $y''+c(x)y=0$ are
generalized for selfadjoint elliptic type equations\vspace*{-1mm}
\begin{equation}
\label{p5} \frac{\partial}{\partial
  x^\mu}\left[a^{\mu\nu}(x^\pi) \frac{\partial u}{\partial
    x^\nu}\right] +a(x^\pi)u=0,
\end{equation}
where $a^{\mu\nu}(x^\pi)\in C^2(\Omega),\: a(x^\pi)\in C^1(\Omega)$.
The point $x_0^\pi$, in which the solution $u$ equals zero, is said
to be a nodal point of this solution.  The important property of the
equation (\ref{p5}) is given in the following theorem.

{\itshape {\bfseries Theorem Skorobohat'ko~V.} The nodal
  points of any solution $u$ of the equation (\ref{p5}) aren't
  isolated in domain $\Omega$, but create the surfaces, which divide
  the domain $\Omega$}.

From the theorem we can see, that nodal surfaces are closed, or
their ends lie on the boundary of domain $\Omega$. The theorem is
simply generalized for the case of $n$-dimensional space, where it
takes the following form: nodal sets of codimension $(n-1)$ are not
isolated and create hypersurfaces, which divide the domain. The
theorem is simply generalized for the case of $n$-dimensional space,
where it takes the following form: nodal sets of codimension $(n-1)$
are not isolated and create hypersurfaces, which divide the domain.

\subsection{\!The solutions\\ \hspace*{-1.2cm}of Sen---Witten equation have no zeros{\label{ch5-subsec53}}}

%

 \hspace*{2.9cm}{\itshape{\bfseries Theorem 1.} Let $\lambda^C$ satisfies Reula's
   condition and is a solution of the equation (\ref{p3'}) with
   asymptotically flat initial data set  in
Petrov type $N$ space-time, that satisfies the dominant
   energy condition (\ref{p3}). Then the solution $\lambda^C$
   everywhere on the SCMC hy\-per\-sur\-fa\-ce $\Sigma_t$ has no zeros.}


\emph{Proof.} From Lemma \cite{ash} we obtain, that all solutions
$\lambda^C$ of equation
\begin{equation}
  \label{p6} \mathcal{D}_A{}^B\mathcal{D}_{BC}\lambda^C=0
\end{equation}
with the form $\lambda^C=\lambda_0{}^C+\beta^C $, where spinor field
$\lambda_0{}^C$ is asymptotically constant and $\beta^C$ is element
of Hilbert space $\mathcal{H}$, satisfy also the first order
equation (\ref{p3'}).  Here the space ${\cal H}$ is the Cauchy
completion of $C_0^\infty$ spinor fields under the norm
\begin{equation*}
  \mid\mid\!\beta^E\!\!\mid\mid^2_{\cal H}=\int\limits_{\Sigma_t}\left({\cal
      D}^A{}_B\beta^B\right)^+\left(\mathcal{D}_{AC}\beta^C\right)dV.
\end{equation*}

The equations (\ref{p6}) are elliptic system of equations. Indeed,
\begin{equation*}
  \mathcal{D}_A{}^B\mathcal{D}_{BC}\lambda^C=
  -D_{AB}D^B{}_C\lambda^C-\frac{\sqrt2}{2}
  D_{AB}\lambda^B-\frac{\sqrt2}{4}\lambda^BD_{AB}\mathcal{K}-\frac18{\cal
    K}^2\lambda_A;
\end{equation*}
taking into account that\vspace*{-3mm}
\begin{equation*}
- D_{AB}D^B{}_C\lambda^C=-\frac12
D^{BF}D_{BF}\lambda_A-\frac18R^{(3)}\lambda_A,
\end{equation*}
and using the equation (\ref{p1}), we obtain:\vspace*{-2mm}
\[
  \mathcal{D}_A{}^B\mathcal{D}_{BC}\lambda^C=\frac12D^{BC}D_{BC}\lambda_A-
  \frac{\sqrt2}{2}\mathcal{K}D_{AB}\lambda^B
  -\frac{\sqrt2}{2}\lambda^BD_{AB}\mathcal{K}\,-
  \]\vspace*{-3mm}
\begin{equation}
 - \frac14{\cal K}^2\lambda
  _A+\frac18\mathcal{K}_{\mu\nu}\mathcal{K}^{\mu\nu}\lambda_A
  +\frac14\mu\lambda_A.
\end{equation}

Let introduce on an open neighborhood of $\Sigma_t$ the Gau\ss{}
normal coordinates $(t,\,x^\alpha)$. Then\vspace*{-3mm}
\[
\mathcal{D}_A{}^B\mathcal{D}_{BC}\lambda^C=\frac1{2\sqrt{-h}}
\frac{\partial}{\partial
  x^\alpha}\left(\!\sqrt{-h}h^{\alpha\beta}\frac{\partial}{\partial
    x^\beta}\lambda_A\!\right)-
    \]\vspace*{-3mm}
\begin{equation}\label{p6'}
-\frac{\sqrt2}{2}\lambda^BD_{AB}\mathcal{K}+\frac18\mathcal{K}_{\alpha\beta}\mathcal{K}^{\alpha\beta}\lambda_A
+\frac14\mu\lambda_A=0,
\end{equation}
and ellipticity (\ref{p5}), as well as (\ref{p3'}), follows from the
negative definition of $\mid\mid\!h^{\alpha\beta}\!\mid\mid$ matrix.




On the hypersurface $\Sigma_t$ of spacial constant mean
curvature\,\footnote{\,It is known, that particular case of SCMC
hypersurfaces~--- hypersurfaces of constant mean curvature (CMC)~---
are favorized in cosmologies.} the system of equations (\ref{p8})
comes to two independent selfadjoint equations for the spinor in
abstract form
\begin{equation}
\label{p7}
  \frac1{\sqrt{-h}}\frac{\partial}{\partial
    x^\alpha}\left(\!\sqrt{-h}h^{\alpha\beta}\frac{\partial}{\partial
      x^\beta}\lambda_A\!\right)+\frac14 \mathcal{K}_{\alpha\beta}{\cal
    K}^{\alpha\beta}\lambda_A+ \frac12\mu\lambda_A=0.
\end{equation}

Taking into account that space-time belongs to Petrov type $N$, we
can make a conclusion that system of two equations for real and
imaginary parts for one of components $\lambda ^{\alpha}$ of spinor
in a spin-frame consists of two independent equations.
\index{Skorobohat'ko theorem}Then, applying Skorobohat'ko's theorem,
for every equation  we obtain, that nodal surfaces of every solution
are not isolated in arbitrary bounded domain $\Omega$ and divide
this domain. According to the dominant energy condition (\ref{p3})
\begin{equation}
  \label{p8}
  \mathcal{K}_{\alpha\beta}{\cal K}^{\alpha\beta}+ 2\mu\ge0,
\end{equation}
then for (\ref{p7}) the maximum principle is fulfilled, and the
solution of the Dirichlet problem for every from each of above
mentioned equation is unique; that is why the nodal surfaces set of
equation (\ref{p7}) does not include the closed surface. For the
case, when real or imaginary part of this component does not vanish
in any point on boundary, equation (5.12) has no node points. Since
in asymptotically flat space-time $\lambda^C$ asymptotically tends
to $\lambda_0{}^C\ne0$, then for the solution on $\Sigma_t$ the
nodal surfaces, which tend to infinity, are also absent.

Therefore the solution $\lambda^C$ of equation (\ref{p3'}) does not
have a zero on $\Sigma_t$. Note that maximality condition is not
necessary condition of splitting of equations. For splitting of
equations in  space-time of Petrov type $N$ it is sufficiently to
require for hypersurface to be spatial constant mean curvature
hypersurface, in other words, one of realizations of well-known
class of hypersurfaces of prescribed mean curvature
\cite{bar,malec}.

From Lemma 1 and Lemma \cite{sen} we can deduce, that the solution
of the equation (\ref{p3'}), which vanishes at a point on
$\Sigma_t$, vanishes everywhere on $\Sigma_t$. But during the
process of Lemma's 1 proof the Conjecture, which is based on the
observation of the properties of the solutions for the equation
(\ref{p3'}) in Minkowski space and at $\mathcal{K}_{\alpha\beta}=0$,
is used; these assumptions are too strong.

For the general case $D_{AB}\mathcal{K}\ne 0$ let us restrict by the
notice, that the solutions of Sen---Witten equation do not have zero
also everywhere on hy\-per\-sur\-fa\-ces in some neighborhood of the
SCMC hy\-per\-sur\-fa\-ce.  The proof can be obtained on the basis
of the Lopatynsky \cite{lop} theorem, according to which the
solutions of the Dirichlet problem for elliptic system of
second-order equations continuously depend on its right parts,
coefficients, $\Omega$-domain and values of the functions on
$\partial\Omega$. In subsection 5.3 we strengthen this result
significantly.

\subsection{\!Sen---Witten equations and SOF {\label{ ch5-subsec23}}}

\hspace*{3cm}{\itshape{\bfseries Definition 1.} A set of $N
  \,(0<N\le10)$ equations for the components of orthonormal vector
  basis $e_m{}^\mu$ (tetrad, vierbein)
  \begin{equation}
  \label{p9}
    \Phi_N(e_{m'}{}^{\mu'},\;\partial_{\nu'} e_{m'}{}^{\nu'},\;
    \partial^2_{\nu'\rho'}e_{p'}{}^{\pi'})=0,
  \end{equation}
  which are not covariant relatively to the local Lorentz
  transformations and (or) coordinate basis transformations, is said
  to be auxiliary conditions}.

 {\itshape{\bfseries Definition 2.} The auxiliary
conditions
  (\ref{p9}) are said to be gauge fixing conditions in some domain
  $\Omega$, if in this domain the solution
  $x^{\mu'}(x^\nu),\;L^{m'}_n(x)$ of the system of
  equations\vspace*{-3mm}
  \begin{equation}
  \label{p10}
    \Phi_N\left(\!e_{n}{}^\nu\frac{\partial x^{\mu'}}{\partial
        x^\nu}L^n{}_{m'},\,\mbox{...},\,\mbox{...}\!\right)=0
  \end{equation}
  with arbitrary coefficients $e_{n}{}^\nu$ exists.}

The sets of additional and gauge fixing conditions are not identical
[676---678]. This is caused firstly by the fact, that the
coefficients of the system (\ref{p10}) are considered in general as
the functions of $C^\infty$ class, but not $C^a$, and the solutions
of non-analytical equations may not exist; secondly, even the
conditions of integrability for equations (\ref{p10}) can be not
satisfied.

Nester \cite{nestjmp} introduced the additional conditions for the
choice of special orthonormal frame on three-dimensional Riemannian
manifold. Let $\theta^a$ denotes the corresponding orthonormal
coframe field. Nester's conditions, written in terms of the
differential forms\vspace*{-2mm}
 \begin{equation}
  \label{p11}
  \widetilde q=i_ad\theta^a,\;\;q=\theta_a\wedge d\theta^a,
  \end{equation}\vspace*{-5mm}

\noindent are given by\vspace*{-2mm}
  \begin{equation}
  \label{p12}
  d\widetilde q=0,\;\;d\ast q=0.
 \end{equation}

The system of equations (\ref{p10}), corresponding to these
additional conditions, is a non-linear second-order elliptic system
for the rotation $R^{a'}{}_b$. Nester proved the existence and
uniqueness of the solution of the linearization of this system for
geometries within a neighborhood of Euclidean space and therefore
the additional conditions (\ref{p12}) are gauge-fixing only
asymptotically.


Analogously to Nester under investigations and application of the
conditions (\ref{p12}) we would restrict ourselves to consideration
of the spaces with ``good'' topology, where the forms $\widetilde q$
and $\ast q$ are exact (vanishing of the first de Rahm cohomology
class of the three-manifolds is sufficient but not necessary).
Taking into account, that initial data set $(\Sigma_t, h_{\mu\nu},
{\cal
  K}_{\pi\rho})$ is asymptotically flat, the conditions (\ref{p12})
are replaced by their first integrals:\vspace*{-1mm}
\begin{equation}
  \label{ p13}
  \widetilde q=-4d\ln \rho, \; \ast q=0.
\end{equation}
Function $\rho$ everywhere on $\Sigma_t$ is positive.


{\itshape {\bfseries Theorem 2} \cite{peljmp}\,\footnote{\,Despite
reconfirm our priority in [659], the result of our Theorem 2 is
given by authors in [656, p.\,5]   as their own. In paper ``On the
zero set of solutions of the Wit\-ten equation on asymptotically
euclidean hypersurfaces'' (GR20/Amaldi10 Abstract Book, Warsaw July
6, 2013) Frauendiener repeatedly tried to declare the result of
Theorem 2 own, but the publication of the report on the conference
itself in our presence gave.}. Let an initial data set
  $(h_{\mu\nu}, {\cal K}_{\pi\rho})$ on maximal hy\-per\-sur\-fa\-ce
  $\Sigma_t$ in Petrov type $N$ space-time be asymptotically flat and satisfies the dominant energy
  condition.  Then everywhere on $\Sigma_t$ the Sen---Witten's
  equations (\ref{p3'}) with Reula conditions and Nester's condition
  (\ref{p12}) are equivalent.}

\emph{Proof.} Let assume firstly that on $\Sigma_t$ the equation
(\ref{p3'}) for $SU(2)$ spinor $\lambda_C$ is given. Then a spatial
one-form $L$ with components $L_\nu=-\lambda_A\lambda_B$ satisfies
``squared'' SWE\vspace*{-3mm}
\begin{equation}
  \label{p14}
  \langle{\widetilde L},\,D\otimes L\rangle-{\cal
    K}L+3!\,i*\left(n\land D\land L\right)=0,
\end{equation}
where $n$ is one-form of unit normal to $\Sigma_t$,
$\:\langle\widetilde L,\,D\otimes L\rangle$ is one-form \mbox{with}
com\-po\-nents $\widetilde L_\nu D_\mu L^\nu,\: \widetilde L=|
L|^{-1}*\left(L\land\overline L\right)$ is non-zero spatial
one-form. Because $\Sigma_t$ as three-dimensional orientable
manifold is parallelisable, it \mbox{admits} a globally defined
orthonormal 3-coframe $\theta^a$ and together with time-like unit
one-form $n$ of the normal to $\Sigma_t$ it forms 4-coframe
$\theta^m$.

Let us introduce $4$-coframe $\theta^m$ with the help of the
correlations
\begin{equation}
  L=\frac{\lambda}{\sqrt2}(\theta^1+i\theta^2),\;\theta^3=\widetilde
  L,\;\theta^0\equiv n=Ndt,
\label{p14'}
\end{equation}
where $\lambda=\lambda^{+A}\lambda_A,\,$ and let introduce metric
$\eta_{mn}=(1,-1,-1,-1)$ in which this frame is orthonormal. Let us
substitute the expression for $L$ from (\ref{p14'}) into (\ref{p14})
and take into account that in agreement with theorem 1 the spinor
$\lambda^A$ everywhere on $\Sigma_t$ does not equal zero. This
allows to write instead of (\ref{p14})
\begin{align}
  - & \left<\theta^1,\,D\otimes \theta^3\right>-{\cal
    K}\theta^1+3!\,i*\left[n\land(D+F)\land \theta^2\right]=0, \label{p15} \\
  { } & \left<\theta^2,\,D\otimes \theta^3\right>+{\cal
    K}\theta^3+3!\,i*\left[n\land(D+F)\land \theta^1\right]=0, \label{p16}
\end{align}
where we denoted $F=D\ln \lambda$. Contracting the left parts of
(\ref{p15}) and (\ref{p16})  with vector field $e_b$ and taking into
account that $\varepsilon_{(0)abc}=\varepsilon_{abc}$, we obtain
\begin{align}
  \label{p17}
  -\omega^{13}{}_b-\mathcal{K}\eta^1_b
  +\varepsilon_{bca}\omega^{a2c}+\varepsilon^{c2}{}_bF_c=0, \\
  \label{p18}
  \omega^{23}{}_b+\mathcal{K}\eta^2_b
  +\varepsilon_{bca}\omega^{a1c}+\varepsilon^{c1}{}_bF_c=0.
\end{align}

The connection one-forms coefficients $\omega^k{}_{mn}$ in
(\ref{p17})---(\ref{p18}) are defined as usually:
$\omega^k{}_{mn}=\langle\theta^k,\,\nabla_{e_m}e_n\rangle$. The
system of equations (\ref{p17}) and (\ref{p18}) includes only four
independent equations:
\begin{equation}
\begin{array}{c}
  \varepsilon^{abc}\omega_{abc}\equiv*q=0,~ \omega^a{}_{1a}\equiv-\widetilde
  q _1=F_1,\\[2mm] \omega^a{}_{2a}=-\widetilde q_2=F_2,~
\omega^a{}_{3a}=-\widetilde q   _3=\mathcal{K}+F_3.
\end{array}\label{ p19}
\end{equation}
Let us choose the hy\-per\-sur\-fa\-ce $\Sigma_t$ maximal and adjust
the parametrization in $V_3$ and on $\Sigma_t$ identifying $\rho$
with $\lambda\left(x^\mu\right)\mid_{\Sigma_t}$. After this the
direct conjecture of the theorem becomes evident.

The converse is obvious.\vspace*{-1mm}

\subsection{\!Conclusion\label{ ch5-subsec24}}
\vspace*{-0.5mm}

\hspace*{3cm}The possibility of the proof of the Theorem 1 is
provided by our splitting off two independent equations from SWE
four equations on SCMC hypersurface in Petrov type $N$ space-time.

\index{spinor fields} %
Evidently the Theorem 1, which plays auxiliary role in the proof of
the Theorem 2, possesses independent importance for spinor fields in
Riemannian space-time.  In particular, it is connected with the
fact, that SWE generalizes the equation for neutrino ``zero mode''
for the case of curved space. Jackiw and Rebbi \cite{jac1}
introduced this equation for investigations of vacuum state
structure in quantum gravity.

The main result of presented section~--- Theorem 2~--- solves (under
the defined conditions) the problem of relation between spinor and
tensor formalism, ascertaining that Witten's spinor formalisms and
Nester's tensor formalisms are isomorphic; this isomorphism is a
result of zeros absence for Witten's spinor (Theorem 1) and of
isomorphism between complexificated vector space $R^3$ and
three-dimensional complex vector space of symmetric second-rank
\mbox{$SU(2)$ spinors.}

We can also say, that on SCMC hypersurface in Petrov type $N$
space-time there exists globally defined (nowhere degenerate)
special orthonormal frame~--- Witten's orthonormal frame,~--- and
this SOF on maximal hypersurface is also Nester's
SOF.\vspace*{-1.0mm}

\section[Nodal points of elliptic equations] {\!Nodal
  points of elliptic\\ \hspace*{-0.95cm}equations and system of equations\label{ch5-sec3}}
\vspace*{-0.5mm}

\hspace*{3cm}From the mathematical point of view the necessity for
investigation of submanifolds, on which the solutions of elliptic
equations are equal to zero, is connected with the fact that the
necessary and sufficient conditions for absence of such closed
submanifolds of codimension one are simultaneously the necessary and
sufficient conditions for uniqueness of the Dirichlet problem for
these equations in the domain. The existence and stability of
solutions (at sufficiently high smoothness of known functions)
follows from its uniqueness. Since the elliptic equations refer to
the static solutions of the given hyperbolic field equations, the
non-uniqueness of solution for the boundary value problem from the
physical point of view defines the instability of ``zero modes'' of
given field equations. Additionally
there appears the necessity to study not only the closed
submanifolds and not only of codimension one, but all other ones, on
which zeros of solutions are located.

The purpose of this section is to develop a new approach for
establishing the conditions of solvability and zeros absence for
general from the physical point of view elliptic systems of
equations. This will give the possibility to prove the existence of
the wide class of hy\-per\-sur\-fa\-ces, in all points of which
there exists the two-to-one correspondence between Sen---Witten
spinor and a certain three-frame; we will name it Sen---Witten
orthonormal frame (SWOF). In all points on such
hy\-per\-sur\-fa\-ces there exist also the well defined lapses and
shifts, associated by Ashtekar and Horowitz \cite{ash} with
Sen---Witten spinor. On a subclass of this class, including also the
maximal hy\-per\-sur\-fa\-ces, we establish the existence of
two-to-one correspondence between Sen---Witten spinor and Nester
three-frame.



We introduce first three definitions.

{\itshape{\bfseries Definition 1.} The nodal point of the component
of the
  solution is a point, in which the component is equal to zero.}

{\itshape{\bfseries Definition 2.} The nodal point of the solution
for the
  elliptical system of equations is a point, in which the solution is
  equal to zero.}

From the general theory of elliptic differential equations it is
known that non-trivial solutions cannot vanish on an open subdomain,
but they can turn to zero on subsets of lower dimensions $k,
k=0,1,\mbox{...},n-1$, where $n$ is dimension of the domain.

{\itshape{\bfseries Definition 3.} The nodal submanifold of
dimension \mbox{$s,
  s=1,2,\mbox{...},n-1,$} is a maximal connected subset\,\footnote{\,Maximal
    connected subset $A$ is a non-empty connected subset such that the
    only connected subset containing $A$ is $A$.} of dimension $s$
  consisting of nodal points of \mbox{the solution.}}

Discrete set of nodal points is $0$-submanifold. We will show
further, that for the system of differential equations, interesting
for us, all nodal subsets are formed by intersection of nodal
surfaces of the components of the solution.

The connection between the unique solvability for the boundary value
problem in $\mathbf{R}^n$ and absence of $(n-1)$-dimensional closed
nodal submanifolds was established by Picone \cite{pic1, pic2}. The
existence of such connection follows from the next consideration: if
the boundary value problem in a certain domain $\Omega$ is uniquely
solvable, then the boundary value problem is also uniquely solvable
for any subdomain $\Omega_1\subseteq \Omega$. This excludes the
possibility of existence of non-trivial solutions which turn to zero
on the boundary of the arbitrary domain $\Omega_1$, i.e., excludes
the possibility of existence of closed nodal submanifolds of
codimension one, and vice versa.

The known investigation of elliptical equations of general form does
not allow to obtain the conditions for all nodal points absence.
For example, even in the case of the only single equation of general
form it is proved the absence of zeros only of infinite order is
proven.
\cite{aro, cor}. B\"{a}r \cite{bar}, continuing Aronszajn \cite{aro}
investigations, proved that the nodal et of Dirac equation is a
countable $(n-2)$ rectifiable set and thus has Hausdorff dimension
$(n-2)$ at most, but question about possibility of all nodal points
absence remains free-answer. That is why further we will examine
only such general equations, which possess also the necessary
physical properties, in particular, symmetry properties and reflects
embedding of the hypersurface into space-time.

Let $\Omega$ be a bounded closed spherical-type domain on space like
hy\-per\-sur\-fa\-ce in Petrov type $N$ space-time, otherwise, (i)
its boundary $\partial \Omega$ in every point has a tangent plane;
(ii) for every point $P$ on the boundary there exists a sphere,
which belongs to $\Omega$, and the boundary of sphere includes the
\mbox{point $P$.}

In the domain $\Omega$ let us consider the system of elliptic second
order \mbox{equations}\vspace*{-1mm}
\begin{equation}
  \label{p21}
      D^{AB}D_{AB}\,u_E+C_E{}^Bu_B=0,
\end{equation}
which generalized equation (\ref{p7}). The unknown functions $u_A$
of independent variables $x^\alpha$ are the elements of complex
vector space $\mathbf{C}^2$, in which the skew symmetric tensor
$\varepsilon^{AB}$ is defined, and the group $SU(2)$ acts. $C_A{}^B$
is Hermitian $(1,1)$ spinorial tensor.

The  equation (\ref{p21}) and every its summand  is invariant under
the arbit\-ra\-ry transformations of coordinates in $V^3$, and
covariant under the local $SU(2)$-trans\-formations of unknown
functions in a local space isomorphic to the complexified tangent
space in every point to $V^3$.

Long ago Picone had ascertained that at ar\-bit\-ra\-ry coefficients
of elliptic equations the boundary value problem is uniquely
solvable, and the closed node submanifolds of codimension one are
absent, respectively, only in the domains with enough small
intrinsic diameter.

The general conditions for the absence of closed nodal surfaces for
strong elliptic system (\ref{p21}) are ascertained by Theorem
\cite{sko1}.

  {\itshape{\bfseries Theorem.} If in domain $\Omega$
there
  exist symmetrical quadratic functional second-order matrices $B_1,
  B_2, B_3$ of $C^1$ class, such that matrix\vspace*{-2mm}
  \[
  \sqrt{-h}C-\sum_{\alpha=1}^3\frac{\partial B _\alpha}{\partial
    x^\alpha}+ B^TG^{-1}B
    \]\vspace*{-5mm}

\noindent is positive definite, where\vspace*{-2mm}
\[
B\!=\!(B_1,B_2,
  B_3), \quad G\!=\!\sqrt{-h}\, {\rm diag}(\|h_{\alpha\beta}\|,
  \|h_{\alpha\beta}\|, \|h_{\alpha\beta}\|),
  \]\vspace*{-5mm}

\noindent then the solutions of
  system of equations (\ref{p21}) with matrix $C=\|C_A{}^B\|$ of $C^1$
  class do not have the closed node surfaces in domain $\Omega$}.

The effective geometrical conditions of B-matrix existence and
corresponding unique solvability of Dirichlet problem in dependence
on the domain intrinsic diameter were obtained \cite{sko1} for the
Euclidean space.   Since the conditions of nodal manifolds absence
for quantum fields equations are the point of our interest, further
we will concentrate our attention on the conditions of nodal points
absence in the domains of arbitrary as well as infinite intrinsic
diameter.

Evidently, if matrix $C$ is positive definite, then the conditions
of Theorem 1 are fulfilled for $B\equiv0$, and closed nodal surfaces
are absent in the domain with arbitrary intrinsic diameter.
Simultaneously, the boundary value problem for the system of
equations (\ref{p21}) is uniquely solvable.

Above cite Theorem  does not indicate the conditions at which nodal
points, lines and node surfaces for solutions of equations
(\ref{p21}) are absent. We will obtain them in the next
subsection.\vspace*{-2mm}

\subsection{\!Conditions for the absence of nodal
points\label{ch5-subsec32}}

\hspace*{3cm}\index{elliptic equation}In the case of a single
selfadjoint elliptic equation in $V^3$ the nodal submanifolds can be
only the surfaces which divide the domain, but in the case of a
system of equations the topology of nodal submanifolds becomes more
various: they can be also the lines and the points.  We can take
this fact into account and ascertain the conditions for the nodal
manifolds absence exploiting the double covariance of the system of
equations (\ref{p21}) and using Zaremba---Giraud Lemma, generalized
at first by Keldysh and Lavrentiev \cite{kel} and later by Oleynik
\cite{ole}.

Let us introduce the matrix\vspace*{-3mm}
\begin{equation*}
  R:=||R_{A^\prime}{}^B||:=
  \left(\!\!\!\begin{array}{cc} \alpha & \beta
      \\ -\overline{\beta} & \overline{\alpha}
    \end{array}\!\!\right)\!,~ \alpha\overline\alpha+\beta\overline\beta=1,
\end{equation*}\vspace*{-5mm}

\noindent which is of the group $SU(2)$, and let its elements
additionally satisfy the \mbox{condition}\vspace*{-2mm}
\[
  C_0{}^1\beta^2+(C_0{}^0-C_1{}^1)\alpha\beta-C_0{}^1\alpha^2=0.\]\vspace*{-5mm}

\noindent Therefore,\vspace*{-2mm}
\[
C_{0^\prime}{}^{1^\prime}=R_{0^\prime}{}^AC_A{}^BR^{1^\prime}{}_B=0,
\]
and in accordance with Hermiticity of matrix $C$ also
\mbox{$\overline
C_{0^\prime}{}^{1^\prime}=C_{1^\prime}{}^{0^\prime}$}.
Then\linebreak \mbox{$C_0:=C_{0^\prime}{}^{0^\prime}$} and
\mbox{$C_1:=C_{1^\prime}{}^{1^\prime}$} are eigenvalues of matrix
\mbox{$C=||C_A{}^B||$}. This follows from a fact that for arbitrary
matrix $R\in SU(2)$ the identity\vspace*{-1mm}
\begin{equation*}
-\varepsilon R \varepsilon\equiv R^{T+}
\end{equation*}\vspace*{-6mm}

\noindent is valid, where $\varepsilon=\|\varepsilon^{AB}\|$.
Therefore\vspace*{-1mm}
\begin{equation}
  C^\prime=-\varepsilon
  R \varepsilon C R^{T+}=R^{T+} C R^T={\rm diag}(C_{0},\; C_{1}).
\end{equation}\vspace*{-7mm}

Let us denote\vspace*{-3mm}
\[
\Delta:=C_1{}^1-C_0{}^0-\left[\left(C_1{}^1-C_0{}^0\right)^2+
4|C_0{}^1|^2\right]^{1/2}\!\!,
\]\vspace*{-5mm}

\noindent and let us denote by $S$ a set of points in domain
$\Omega$, in all points of which $C_0{}^1$ does not equal to zero,
and let us denote by $T$ a set of points, in which $C_0{}^1$ is
equal to zero. Then the elements of the matrix $R$, which transforms
the matrix $C$ to diagonal form, satisfy on the set $S$ the
conditions\vspace*{-2mm}
\[
\alpha\overline\alpha(1+\Delta^2/4|C_0{}^1|^2)=1,~
\beta=\alpha\Delta/2C_0{}^1
\]
\vspace*{-7mm}

\noindent and on the set $T$ the conditions\vspace*{-4mm}
\[
\alpha\overline\alpha=1,~ \beta=0.
\]\vspace*{-7mm}

Functions $u_{0^\prime}$ and $u_{1^\prime}$ on the set $S$ will be
following:\vspace*{-2mm}
\begin{equation}
  u_{0^\prime}=\overline\alpha\left(\!u_0+\frac{\Delta}
    {2{\overline C}_0{}^1}\,u_1\!\right)\!,~
  u_{1^\prime}=\alpha\left(\!-\frac{\Delta }{2C_0{}^1}\,u_0
    +u_1\!\right)\!,\label{p22}
\end{equation}
\vspace*{-4mm}

\noindent  and on set $T$ they will be\vspace*{-5mm}
\begin{equation}
  u_{0^\prime}= \overline\alpha u_0,~
  u_{1^\prime}=\alpha u_1.\label{p23}
\end{equation}
 \vspace*{-7mm}

\noindent Respectively, eigenvalue $C_0$ on $S$ is:\vspace*{-1mm}
\[
C_{0}={ 4C_0{}^0|C_0{}^1|^4+\left(4
    \Delta\,|C_0{}^1|^2+C_1{}^1\Delta^2\right)
  \left(4|C_0{}^1|^2+\Delta^2 \right)\over
  4|C_0{}^1|^2\left(4|C_0{}^1|^2+\Delta^2 \right)}
\]
\vspace*{-4mm}

\noindent  and coincides with $C_0{}^0$ on the set $T$.

 {\itshape{\bfseries Lemma.} If real and imaginary parts of
functions
  $u_A$ and of elements of matrix $C_A{}^B$ are functions of class
  $C^2$ in domain $\Omega$, then the real and imaginary parts of
  functions $u_{A^\prime}$ defined by conditions
  (\ref{p22})---(\ref{p23}) are also the functions of class $C^2$ in
  this domain.}

\emph{Proof.} Taking into account that it is always possible to
choose $\mathfrak{Im}\,\alpha\!\in$ $\in C^2(\Omega)$, from direct
calculation we obtain that on the set $S$ there exists first and
second derivatives of real and imaginary parts of functions
$u_{A^\prime}$ and $\alpha$ with respect to arguments
$(\Delta^2/4|C_0{}^1|^2)$ and
$(\Delta/2|C_0{}^1|^2)$ and that\vspace*{-2mm} %
{
\begin{align}
  \lim_{P\ni S\rightarrow Q\in T}\mathfrak{Re}\,\alpha^{(m)}(P)=
  \mathfrak{Re}\,\alpha ^{(m)}(Q), &&\!\!\! \lim_{P\ni S\rightarrow Q\in T}
  \mathfrak{Im}\,\alpha^{(m)}(P)= \mathfrak{Im}\,\alpha^{(m)}(Q),
  \\[-1mm]
  \lim_{P\ni S \rightarrow Q\in T}\mathfrak{Re}\,
  u_{A^\prime}^{(m)}(P)= \mathfrak{Re}\, u_{A^\prime}^{(m)} (Q), &&\!\!\!
  \lim_{P\ni S\rightarrow Q\in T}\mathfrak{Im}\,
  u_{A^\prime}^{(m)}(P)=\mathfrak{Im}\, u_{A^\prime}^{(m)}(Q),
\end{align}}\vspace*{-5mm}

\noindent where symbol $f^{(m)}$ denotes arbitrary partial
derivatives of order $m=0,1,2$.

The following theorem is valid.

  {\itshape{\bfseries Theorem 3.} Let:

a) real and imaginary parts of the elements of matrix $C$ be
  of $C^2$ class in the domain $\Omega$ on space-lake hypersurface in Petrov $N$ space-time;

b) at least one eigenvalue of matrix $C$, for definiteness
  $C_0$, be non-negative everywhere in $\Omega$;

c) real or imaginary part of the function\vspace*{-3mm}
\[
v:= \begin{cases}
  \!\!\left. \left(u_0+\frac{\Delta} {2{\overline C}_0{}^1}\,u_1\right)\right|_{S\bigcap\partial\Omega} \!, &{}
  \\[-1mm]
  \!\!\left. u_0 \right|_{T\bigcap\partial\Omega} &{}
\end{cases}
\]\vspace*{-5mm}

\noindent be equal to zero in any point.}

Then solution $u_A$ of class $C^2$ for the system of equations
(\ref{p21}) does not have any nodal points in the domain $\Omega$ of
spherical type.

\emph{Proof.} The system of equations $(\ref{p21})$ is covariant
under the arbitrary transformations of coordinates and under the
local transformations from the group $SU(2)$ that allows to use them
independently. Let us apply at the \mbox{first} step the $SU(2)$
spinor transformation $u_{A}\rightarrow R_{A^\prime}{}^Bu_B$, which
transforms the matrix $C$ to the diagonal form, and under which the
equation (\ref{p21}) \mbox{is covariant.}

The eigenvalues of matrix $C$ are real, therefore, the resulting
system of equations (\ref{p21}) splits into two independent
subsystem for $u_{A^\prime}$. Taking into account that space-time is
of Petrov type $N$, we can make a conclusion that in a spin-frame
subsystem with $C_0$ splits into two independent equations for real
and imaginary parts of $u_{0^{\prime}}$ component. Taking into
account that $u_{A^\prime}$, $C_{0}$ and $C_{1}$ are scalars under
transformations of coordinates, and $C_0{}\geq0$, we can apply the
Zaremba---Giraud principle in the general form grounded by Oleynik
\cite{ole} to every equation containing $C_{0}$. According to this
principle, if in a certain point $P_0$ on the sphere the
non-constant function in the ball turns to zero, and everywhere in
the ball $\mathfrak{Re}\,u_{0^\prime}<0$, then $\left<\mathrm{ d}\,
\mathfrak{Re}\,u_{0^\prime},\,
  l\right>|{_{P_0}}<0$. Here $l$~--- arbitrary vector field, for which
$\left<n,\,\,l\right>|{_{P_0}}>0$, and $n$ is one-form of intrinsic
normal to the sphere in the point $P_0$.

Let us show further that a set of the nodal points for function
$\mathfrak{Re}\,u_{0^\prime}$ does not contain the isolated points.
Let us assume that such point exists, i.e.
$\mathfrak{Re}\,u_{0^\prime}=0$, and in a certain neighborhood of
the point $P_0$ the function has a constant sign. For definiteness
let in this neighborhood be $u_{0^\prime}<0$. Let us consider a
sphere, on which the point $P$ lies and which is so small that
completely belongs to the mentioned neighborhood of the point $P$.
Then, \mbox{using} Za\-rem\-ba---Gi\-raud principle, we obtain
$\left<\mathrm d\,
  \mathfrak{Re}\,u_{0^\prime},\,\,n\right>{|_{P_0}}>0$, and therefore
in any neighborhood of the point $P_0$, located outside the ball,
the function $\mathfrak{Re}\,u_{0^\prime}$ changes its sign, and
that is why its zeros are not isolated. Therefore, they form the
surfaces which divide $\Omega$. Since $C_0\geq0$ , then it follows
from the maximum principle that the closed nodal surfaces for the
components of solution $\mathfrak{Re}\,u_{0^\prime}$ are absent.
Analogous conclusion is true also for the component of solution
$\mathfrak{Im}\,u_{0^\prime}$. This means that the only surfaces
having common points with the boundary of domain $\Omega$ can be the
nodal surfaces of real or imaginary part of function $u_{0^\prime}$.
According to condition c), if, for definiteness,\vspace*{-2mm}
\begin{equation*}
  \mathfrak{Re}\!\left.\left(\!u_0+\frac{\Delta}{2{\overline C}_0{}^1}\,u_1 \!\right)
  \right|_{S\bigcap\partial\Omega} \neq0,\quad
  \mathfrak{Re}\,u_0 \Big|_{T\bigcap\partial\Omega} \neq 0,
\end{equation*}\vspace*{-5mm}

\noindent then we  can choose\vspace*{-3mm} %
{
\[
  \mathfrak{Re}\,\overline\alpha\,\Big|_{S\bigcap\partial\Omega} \neq
  \bigg\{\! \left[ \mathfrak{Re}
        \left(\!u_0+\frac{\Delta}{2{\overline
              C}_0{}^1}\,u_1\!\right)\!\right]^{-1}
      \mathfrak{Im}\overline\alpha\,\,\mathfrak{Im}\!\left(\!u_0+\frac{\Delta}
        {2{\overline
        C}_0{}^1}\,u_1\!\right)\!\bigg\}\bigg|_{S\bigcap\partial\Omega}\!\!,
        \]\vspace*{-4mm}
\[
  \mathfrak{Re}\,\overline\alpha \Big|_{T\bigcap\partial\Omega} \neq
  \left[\left(\mathfrak{Re}\,u_0\right)^{-1}
      \mathfrak{Im}\,\overline\alpha\,\,\mathfrak{Im}\,u_0\right]\Big|_{T\bigcap\partial\Omega}\]\vspace*{-5mm}

\noindent and obtain\vspace*{-3mm}
\[
  \left.\left[\mathfrak{Re}\,\overline\alpha\,\,
      \mathfrak{Re}\left(\!u_0+\frac{\Delta} {2{\overline
            C}_0{}^1}\,u_1\!\right)- \mathfrak{Im}\,\overline\alpha\,\,
      \mathfrak{Im}\left(\!u_0+\frac{\Delta} {2{\overline
      C}_0{}^1}\,u_1\!\right)\!\right]\right|_{S\bigcap\partial\Omega}\equiv\]\vspace*{-3mm}
\[  \equiv
  \mathfrak{Re}\,u_{0^\prime}\Big|_{S\bigcap\partial\Omega}\neq 0,
\]\vspace*{-2mm}
\[
\left(\!\mathfrak{Re}\,\overline\alpha\,\,\mathfrak{Re}\,u_0 -
    \mathfrak{Im}\,\overline\alpha\,\,\mathfrak{Im}\,u_0\!\right)
  \Big|_{T\bigcap\partial\Omega}\equiv \mathfrak{Re}\,
  u_{0^\prime}\Big|_{T\bigcap\partial\Omega} \neq 0.
\]

Therefore nodal surfaces as well as lines and points of the real (or
imaginary) part are absent, and that is why any nodal points of
complete solution $u_A$ are also absent. The statement of the
theorem is proven.

\emph{Note.} If the conditions a) and b) of the Theorem are
fulfilled, and the matrix $C$ is non-negative definite in domain
$\Omega$, then both eigenvalues are non-negative, and, therefore,
the boundary value problem for the system of equations (\ref{p21})
is uniquely solvable in arbitrary bounded domain, as it follows from
the classical maximum principle. Otherwise the solution in finite
domain exists only in the case when its intrinsic diameter does not
overcome a certain value.


\subsection{\!The conditions of nodal points\\ \hspace*{-1.2cm}absence for the solutions of
  Sen---Witten equation\label{ch5-subsec33}}

\hspace*{3cm}After Witten's positive energy proof the attempts of
de\-ve\-lop\-ment of tensor method for proof were performed along
two lines. The attempts of the tensor interpretation for
Sen---Witten spinor field belong to the first line. In particular,
Ashtekar and Horowitz \cite{ash} used Sen---Witten spinor field for
determination of a class of preferred lapses $T:=\lambda$ and shifts
$T^a:=-\sqrt2\,i\lambda^{+(A}\lambda^{B)}$.  Dimakis and
M\"uller-Hoissen \cite{dim1, dim2} have defined a preferred class of
orthonormal frame fields in which spinor field take a certain
standard form. Frauendiener \cite{Frau} has noticed a correspondence
between Sen---Witten spinor field and a triad. But, as it was shown
by Dimakis and M${\rm {\ddot u}}$ller-Hoissen, frame fields cannot
exist in the nodal points of the spinor field.

Let us weaken the conditions of zeros absence for SWE (\ref{p3'}) on
$\Sigma_t$ using the results of subsection 5.3.1. From equation
(\ref{p3'}), taking into account the equation of Hamiltonian
constraint on $\Sigma_t$, in Gauss normal coordinates we obtained
(see also \cite{peljmp}):\vspace*{-3mm}
\[
  \mathcal{D}_A{}^B\mathcal{D}_{BC}\lambda^C=\frac1{2\sqrt{-h}}
  \frac{\partial}{\partial
    x^\alpha}\left(\!\sqrt{-h}h^{\alpha\beta}\frac{\partial}{\partial
      x^\beta}\lambda_A\!\right)-\]\vspace*{-3mm}
\begin{equation}
  -\,\frac{\sqrt2}{2}\lambda^BD_{AB}\mathcal{K}+\frac14\mathcal{K}^2\lambda
  _A+\frac18\mathcal{K}_{\alpha\beta}\mathcal{K}^{\alpha\beta}\lambda_A
  +\frac14\mu\lambda_A=0. \label{p26}
 \end{equation}
Therefore, the system of equations (\ref{p26}) is a system of the
form (\ref{p21}); if it does not have the nodal points, the SWE also
does not have them.

Spinorial tensor\vspace*{-5mm}
\begin{equation}
\label{p27}
  C_A{}^B:=\frac{\sqrt2}{4}D_{A}{}^{B}\mathcal{K}+\frac14 \varepsilon
  _A{}^B\left(\!2\mathcal{K}^2+\frac12\mathcal{K}_{\pi\rho}{\cal
      K}^{\pi\rho}+\mu\!\right)
\end{equation}\vspace*{-6mm}

\noindent is Hermitian because\vspace*{-1mm}
\begin{equation*}
  (\mathcal{D}_A{}^B{ \cal
    K})^+=\left(\varepsilon^{BC}\mathcal{D}_{AC}\mathcal{K}\right)^+=
  \left(\varepsilon^{BC}\right)^+\left(\mathcal{D}_{AC}{\cal
      K}\right)^+=-\left(\mathcal{D}_{AC}{\cal
      K}\right)\varepsilon^{CB}=(\mathcal{D}_A{}^B\mathcal{K}).
\end{equation*}\vspace*{-6mm}

So, the SWE solutions of class $C^2$ do not have the nodal points in
a boun\-ded closed domain $\Omega$ of spherical type on $\Sigma_t$
in Petrov type $N$ space-time, if for spinorial tensor $C_A{}^B$ in
this domain and for the boundary values of the solution the
conditions of Theorem 3 are fulfilled.

Let us consider further a sequence $\Omega_n$ of increasing domains
of spherical type covering $\Sigma_t$. If in every domain the
conditions of Theorem 3 are fulfilled, then all solutions of class
$C^2$ do not have the nodal points in $\Omega_n$. According to
Reula, on $\Sigma_t$ there exists the SWE solution of
$\lambda^C=\lambda^C_{\infty}+\beta^C$ form, where
$\lambda^C_{\infty}$ is asymptotically covariant constant spinor
field on $\Sigma_t$, $\beta^C$ is an element of Hilbert space
$\mathcal{H}$, which is the Cauchy completion of $C_0^\infty$ spinor
fields under the norm\vspace*{-1mm}
\[\mid\mid\beta^E\mid\mid^2_{\cal
  H}=\int\limits_{\Sigma_t}\left({\cal
    D}^A{}_B\beta^B\right)^+\left(\mathcal{D}_{AC}\beta^C\right)dV.\]

Solution $\lambda^C$ belongs properly to $C^\infty$ class. From the
asymptotical flatness condition it follows that
$(\Delta^2/(4|C_0{}^1|^2)$, as well as real and imaginary parts of
functions $(\Delta/2{\overline C}_0{}^1)$ and $(\Delta/2 C_0{}^1)$
vanish asymptotically. Therefore, condition c) of Theorem 2
asymptotically takes the form:\; $\mathfrak{Re}\,
\lambda^0_\infty\neq0$ or $\mathfrak{Im}\, \lambda^0_\infty\neq0$.
In such a way we obtain the following theorem:

 {\itshape{\bfseries Theorem 4.} Let:

   a)~initial data set be asymptotically flat;

   b)~everywhere on space-like hypersurface $\Sigma_t$ in Petrov type $N$ space-time the matrix of spinorial tensor
    (\ref{p27}) have at least one non-negative eigenvalue, for
    definiteness $C_0$;

   c)~$\mathfrak{Re}\,\lambda^0_{\infty}$ or
    $\mathfrak{Im}\,\lambda^0_{\infty}$ equal zero asymptotically
    nowhere.

\noindent Then the asymptotically constant nontrivial solution
$\lambda^C$ to SWE does not have the nodal points on $\Sigma_t$.  }%

The conditions of Theorem 4 are fully admissible from the physical
point of view.\vspace*{-3mm}

\subsection{\!Towards Sen---Witten equation, special\\ \hspace*{-1.2cm}orthonormal frame and
  preferred time variables\label{ch5-subsec34}}

\hspace*{3cm}\index{Sen---Witten equation (SWE)}Usually the question
about existence of system of coordinates or orthonormal basis, which
satisfy certain gauge conditions, is reduced to the question about
existence of solution for non-linear system of differential
equations and often can be solved only at some additional
limitations and assumptions \cite{nes1}.

The existence theorem for Sen---Witten (linear) equation and the
Theorem 4 about their zeros (subsection 4) on surfaces, which can be
not maximal, allow us to prove the existence of a certain class of
orthonormal three-frames in all points of these
hy\-per\-sur\-fa\-ces which satisfies gauge conditions.
\begin{equation}
\begin{array}{c}
  \varepsilon^{abc}\omega_{abc}\equiv*q=0,\quad\omega^a{}_{1a}\equiv-\widetilde
  q _1=F_1,\\ \omega^a{}_{2a}=-\widetilde q_2=F_2, \quad
   \omega^a{}_{3a} =-\widetilde q_3=\mathcal{K}+F_3,
  \end{array}\label{p28}
\end{equation}
and generalizes Nester's SOF. Such three-frame we will name as
Sen---Witten orthonormal frame (SWOF)

  {\itshape{\bfseries Theorem 5.} Let the conditions of Theorem
  4 be fulfilled.  Then everywhere on $\Sigma_t$ there exists a
  two-to-one correspondence between Sen---Witten spinor and Sen---Witten
  orthonormal frame.}

\emph{Proof.} Really, let all conditions of Theorem 3 be fulfilled
on $\Sigma_t$. Then SWE solution $\lambda_A$ does not have the nodal
points anywhere on $\Sigma_t$.  This allows to prove on such
$\Sigma_t$ the Sommers \cite{som} assumption that spatial null
one-form $L=-\lambda_A\lambda_B$ on $\Sigma_t$ is non-zero, and
allows to turn everywhere on $\Sigma_t$ to the ``squared'' SWE
represented in the form:\vspace*{-2mm}
\begin{equation}
  \label{p29}
  \langle{\widetilde L},\,D\otimes L\rangle-{\cal
    K}L+3!\,i*\left(n\land D\land L\right)=0,
\end{equation}\vspace*{-5mm}

\noindent where $\langle\widetilde L,\,D\otimes L\rangle$ is
one-form with components $\widetilde L_\nu D_\mu L^\nu,\: \widetilde
L=\mid\! L\!\mid^{-1}*\left(L\land\overline L\right)$ is non-zero
spatial one-form, and $n$ is one form of unit normal to $\Sigma_t$.

The bilinear form\vspace*{-3mm}
\[
\frac1{\sqrt2}n^{A{\dot
    A}}\lambda_A\overline{\lambda}_{\dot
  A}=\lambda_A\lambda^{A+}\equiv \lambda,
\]\vspace*{-5mm}

\noindent where $n$ is one-form of a unit normal to $\Sigma_t$, is
Hermitian positive definite one, and the solution $\lambda_A$ does
not have the nodal points on $\Sigma_t$.  Consequently, we can
further introduce real nowhere degenerate orthonormal 4-coframe
$\theta^m$ as\vspace*{-1mm}
\begin{equation}
\label{p210} \theta^0\equiv n=N \mathrm{d}t,\quad
\theta^1=\frac{\sqrt2}{2\lambda}(L+\overline L),\quad
\theta^2=\frac{\sqrt2}{2\lambda i}(L-\overline L),
\quad\theta^3=\widetilde L
\end{equation}
and represent immediately (\ref{p29}) in the form\vspace*{-1mm}
\begin{align}
\label{p211} - & \left<\theta^1,\,D\otimes \theta^3\right>-{\cal
  K}\theta^1+3!\,*\left[n\land(D+F)\land\wedge \theta^2\right]=0, \\
\label{p212}
  & \left<\theta^2,\,D\otimes \theta^3\right>+{\cal
  K}\theta^3+3!\,*\left[n\land(D+F)\land \theta^1\right]=0,
\end{align}
where $F=D\ln \lambda$. The system of equations
(\ref{p211})---(\ref{p212}) includes only four independent
equations, and they are equations (\ref{p28}) for the connection
one-forms coefficients. From this it follows that if on $\Sigma_t$
the conditions of Theorem 3 and SWE are fulfilled, then on
$\Sigma_t$ there exists three-frame $\theta^a$ defined by
(\ref{p210}) in which conditions (\ref{p28}) are fulfilled.

Inversely, if on $\Sigma_t$ in some three-frame $\theta^a$ the
conditions of Theorem 4 and conditions (\ref{p28}) are fulfilled,
then it follows from condition of Theorem 3 that these one-forms
have a form $\theta^a=\theta^a_\infty+\phi^a$, where
$\theta^a_\infty$ tend asymptotically to the covariant constant
forms and $\phi^a$ belongs to $\mathcal{H}$.  We can turn from
four-frame $\theta^m\equiv\{n,\, \theta^a\}$ to one-forms
$\theta^0,\,L,\,\widetilde L$, assuming $\lambda_A |_
{{\scriptstyle{\Sigma}_{\scriptscriptstyle
      t}}}\neq0$. After this we obtain equation (\ref{p29}) and
further (\ref{p3'})\,\footnote{\,The equivalence of the SWE
(\ref{p3'})
  and of the equation (\ref{p26}) is proven by Reula \cite{reu}.} for
spinor field $\lambda^A$, which, as we have demonstrated previously,
indeed does not have the nodal points on selected
hy\-per\-sur\-fa\-ce $\Sigma_t$ and which together with asymptotical
conditions defines up to sign the spinor field $\lambda^A$.
Mentioned in conditions of the Theorem correspondence between
Sen---Witten spinor field and Nester's SOF is defined by
relationship (\ref{p210}).

We have proven in Section~5.2.4 that if initial set $(\Sigma_t,
h_{\mu\nu}, \mathcal{K}_{\pi\rho})$ on maximal hy\-per\-sur\-fa\-ce
$\Sigma_t$ is asymptotically flat and satisfies the dominant energy
condition, then everywhere on $\Sigma_t$ from existence of
Sen---Witten spinor field follows existence Nester's three-frame and
conversely.  Theorem 4 allows to strengthen significantly this
result taking away the assumption that $\Sigma_t$ is maximal.
Indeed, if all conditions of Theorem 3 are fulfilled on $\Sigma_t$,
and additionally the one-form $\mathcal{K}\widetilde L$ is globally
exact, we can perform in conditions the identification $F \equiv
{\rm d}\ln\lambda\,+\,\mathcal{K}\theta^3$ and obtain the Nester's
gauge (\ref{p11}), or to perform the inverse transition~--- from
Nester's gauge to SWE. Therefore, if on $\Sigma_t$ the conditions of
Theorem 3 are fulfilled, then SWE and Nester's gauge are equivalent
if and only if the one-form $\mathcal{K}\lambda^{+(A}\lambda^{B)}$
is exact.  In this case the correspondence between Sen---Witten
spinor and Nester's SOF is also ascertained by relationship
(\ref{p210}).

Ashtekar and Horowitz \cite{ash} have accented on the necessity of
zeros investigations for SWE solutions introducing the vector
interpretation of Sen---Wit\-ten's spinor which defines a preferred
lapse and shift. Evidently, the fulfilling of the Theorem 4
conditions ensures the existence of corresponding lapses and
\mbox{shifts} well defined everywhere on $\Sigma_t$.  And also the
preferred class of or\-tho\-nor\-mal four-frame fields introduced by
Dimakis and M\"{u}ller-Hoissen exists in all points of $\Sigma_t$
under fulfilling of the Theorem 3 conditions.

The presence of zeros in the solutions of elliptic equations is
rather ordinary than exceptional case, therefore, it is necessary to
prove the absence of zeros for concrete cases.

The represented investigation demonstrates the possibility for
obtaining the condition of the nodal manifolds absence for enough
general system of elliptic second order equations owing to its
double covariance.

The application of this result to SWE allows to prove the
equivalence of SWE and gauge conditions (\ref{p12}), and,
respectively, the existence of an everywhere well defined two-to-one
correspondence between Sen---Witten spinor field and the SWOF, which
is the Nester SOF in the particular case, when one of the one-forms
$\mathcal{K}\theta^a$ is exact. Therefore, the indicated
correspondence exists not only on the unique~--- maximal~---
hy\-per\-sur\-fa\-ce, but on the whole set of asymptotically flat
hy\-per\-sur\-fa\-ces.

Ashtekar and Horowitz \cite{ash} have shown that the Reula results
hold even if the energy condition is mildly violated. Also the
conclusion about existence of special three- and four-frames, as
well as preferred lapses and shifts, is stable under the violation
of the energy condition, because, as it is seen from (\ref{p27}),
there exist the hy\-per\-sur\-fa\-ces, on which this condition of
nodal points absence is fulfilled at violation of the energy
condition.


\section[\!Sen---Witten orthonormal three-frame] {\!Sen---Witten orthonormal
  three-frame\\ \hspace*{-0.95cm}and gravitational energy quasilocalization\label{ch5-54}}

\hspace*{3cm}The equivalence principle excludes a possibility for
existence of the gravitational energy density, however, in the
Penrose conception \cite{pen1} there is possible its
quasilocalization. This conception is realized in several proposals
for the quasilocal energy-momentum [684---689].

According to the Nester and coauthors approach, for each
gravitational energy-momentum pseudotensor there is Hamiltonian
boundary term, and the energy-momentum in a domain, bounded by close
2-surface, depends on the field values and the frame of reference on
the 2-surface. Various criteria are insufficient and, most probably
\cite{nes40}, will be insufficient for selecting a unique
Hamiltonian boundary term. Variety of these terms is characterized
by different choices of dynamic variables (metric, orthonormal
frame, spinors), boundary conditions and reference configurations.
According to this there a problem of the different Hamiltonians
comparing \cite{nestu,
  nestu30} appears. Among criteria that must be satisfied by the
quasilocal energy-momentum density, at least in asymptotically
Minkowskian space \cite{chr}, must be positivity. It can be ensured
by finding the locally non-negative Hamiltonian density dependent on
the Sen---Witten spinor according to Witten \cite{witten}, or by
applying the ADM Hamiltonian and the Nester special orthonormal
frame \cite{nes1}.

In the asymptotically flat space the Hamiltonian is of the general
\mbox{form \cite{isnes}}\vspace*{-2mm}
\begin{equation}
  \label{p31}
  H\left( {N} \right) = {\int\limits_{\Sigma} {\left( {N\mathcal{H} +
          N^{a}\mathcal{H}_{a}} \right)}} dV + {\oint\limits_{\partial \Sigma} {B}}
\end{equation}\vspace*{-3mm}

\noindent and includes the Regge---Teitelboim boundary term
\cite{regge} at spatial infinity.

Grounding and developing the Wittenian proof of the positive energy
theorem, Nester \cite{nesLec} proposed an expression for the
Hamiltonian density as the 4-covariant quadratic spinor
3-form:\vspace*{-1mm}
\begin{equation}
\label{p32} \mathcal{H}\left( {\psi} \right): = 2{\left[ {D\left(
{\overline
          {\psi}\land \gamma _{5} \gamma} \right)\land D\psi -
      D\overline {\psi} \land D\left( {\gamma _{5} \gamma \psi}
      \right)} \right]},
\end{equation}\vspace*{-7mm}

\noindent where\vspace*{-2mm}
\[
D\psi = \psi + {\frac{{1}}{{2}}}\omega ^{\mu \nu} \sigma _{\mu \nu}
\psi  ,~ \sigma _{\mu \nu} = {\frac{{1}}{{2}}}{\left[ {\gamma _{\mu}
      ,\gamma _{\nu }} \right]},~ \gamma = \gamma
_{\mu} \theta ^{\mu},
\]\vspace*{-5mm}
\[
 \gamma _{\mu} \gamma _{\nu} + \gamma
_{\nu} \gamma _{\nu} = 2g_{\mu
  \nu } ,~ \gamma _{5}^{2} = - E,~ \gamma _{5} = \gamma
^{0}\gamma ^{1}\gamma ^{2}\gamma^3  .
\]\vspace*{-5mm}

From expression (\ref{p32}) one can obtain the following expression
for $\mathcal{H} \left({\psi} \right)$:\vspace*{-1mm}
\begin{equation}
  \label{p33} \mathcal{H}\left( {\psi}\right) = 4D\overline {\psi}
  \land\gamma _{5} \gamma \land D\psi = 4\nabla _{\pi} \overline
  {\psi} \left( {\gamma ^{\mu} \sigma ^{\pi \rho} + \sigma ^{\pi
        \rho} \gamma ^{\mu} } \right)\nabla _{\rho} \psi d\Sigma
        _{\mu},
\end{equation}\vspace*{-5mm}

\noindent where $d\Sigma _{\mu} = {\dfrac{{1}}{{3!}}}\sqrt {{\left|
{g} \right|}} \varepsilon _{\mu \nu \pi \sigma} dx^{\nu} \land
dx^{\pi}\land dx^{\sigma} $.

In the Gaussian normal system of coordinates in the neighborhood of
arbitrary space-like hy\-per\-sur\-fa\-ce $\Sigma $, the Hamiltonian
density (\ref{p33}) can be written as a sum of positive and negative
definite components \cite{nesLec}\vspace*{-2mm}
\begin{equation}
  \label{p34}
  \mathcal{H}\left( {\psi}  \right) = - 4g^{ab} D_{a } \psi
  ^{ +} D_{b} \psi + D_{a} \overline {\psi}  \left(\! {\gamma ^{d}
      \gamma ^{a} \gamma ^{b} + \gamma ^{a} \gamma ^{b} \gamma ^{d} }
  \right)D_{b}  \psi\, d\Sigma _{0}
\end{equation}\vspace*{-5mm}

\noindent and be locally non-negative if $SL(4, \mathbb C)$-spinor
$\psi $ on the space-like hy\-per\-sur\-fa\-ce $\Sigma $ satisfies
the Sen---Witten equation\vspace*{-2mm}
\begin{equation}
  \label{p35}
  \gamma ^{a} D_{a}  \psi = 0.
\end{equation}\vspace*{-6mm}

Expression (\ref{p34}) cannot give the true positive energy density
for the gravitational field because $\psi $, as solution of the SWE,
is a non-local functional on the initial data $(h,K,\Sigma)$ set;
therefore, a concept of the locally non-negative density of the
gravitation energy is treated as the locally non-negative functional
on the set of initial data $(h,K,\Sigma)$ and the boundary values of
function $\psi $. The gravitational Hamiltonian density (\ref{p34})
has significant advantages in comparison with the other ones: except
a fact that it is explicitly 4-covariant, the gravitational
Hamiltonian, which includes it, allows to prove that the total
4-momentum and the Bondi 4-momentum are time-like. To its
liabilities Nester and Tung have referred the physical
mysteriousness of the Sen---Witten spinor field, and absence of the
direct relation to the Hamiltonian density in the SOF method
\cite{nestjmp, nes1, nesCQG}. For establishing such relation, Nester
and Tung \cite{nestu} had developed a new method of proving the PET
and the gravitational energy localization, which employs the
3-dimensional spinors and a new identity connecting the
3-dimensional scalar curvature to the spinor expression in the
Hamiltonian. The Einstein 3-spinor Hamiltonian with a zero shift the
authors obtained in the form\,\footnote{\,In this formula and in
some
  next formulas we change the signs, comparing with the original
  papers, according to the chosen here convention that a signature is
  $(+,-,-,-)$.}\vspace*{-1mm}
\[
  H = \int\limits_{\Sigma}  \bigg[ \varphi^{+} \varphi g^{-{{1}/{2}}}\left(\! { \pi^{ab} \pi_{ab}
  -\frac{1}{2}\pi^{2}}\!\right)-
  \]\vspace*{-3mm}
  \begin{equation}
  -\,4\left( g^{ab} \nabla_{a} \varphi^{+} \nabla_{b} \varphi +
      \nabla_{a}\varphi^{+}\sigma^{a} \sigma^{b} \nabla_{b}
      \varphi\right)\! \bigg] d^{3}x,
\label{p36}
\end{equation}\vspace*{-3mm}

\noindent from which follows a conclusion that the density is
non-negative definite, if on the maximal hy\-per\-sur\-fa\-ce the
asymptotically constant spinor $\varphi $ satisfies the Dirac
equation in the 3-dimensional space\vspace*{-1mm}
\begin{equation}
  \label{p37} \sigma ^{a}\nabla _{a} \varphi = 0.
\end{equation}

The main result of the Nester, Tung \cite{nestu} and the Nester,
Tung, Zhytnikov \cite{nestuzhy} works is formulated in the form of a
statement that between the localization method, based on the
4-covariant spinor Hamil\-tonian, and the SOF-ba\-sed method there
exists a close connection owing to the 3-spinor Hamil\-tonian
(\ref{p36}).

Such statement is grounded on the two circumstances: 1) among terms
of which the 4-covariant spinor density consists, the 3-spinorial
density is present; 2) between the 3-spinor field variables there
exists, as Nester and Tung declared, a close relation, since from
the 3-dimensional Dirac equation (\ref{p37}) it follows
that\vspace*{-4mm}
\begin{equation}
  \label{p38} \sigma ^{a} \nabla _{a}  \varphi = \sigma ^{a} \varphi
  _{,a} - {\frac{{1}}{{2}}}\tilde {q}_{b} \sigma ^{b} \varphi +
  {\frac{{1}}{{4}}}{\rm i}*q\varphi = 0,
\end{equation}\vspace*{-5mm}

\noindent where forms $q$ and $\tilde {q}$ are defined in the
following way:\vspace*{-1mm}
\begin{equation}
  \label{p39} q = \theta _{\hat{a}}  \land d\theta ^{\hat{a}}
    ,~ \tilde {q} = i_{\hat{a}} d\theta ^{\hat{a}}
\end{equation}\vspace*{-5mm}

\noindent and fix SOF on the asymptotically flat surfaces by means
of the Nester gauge\vspace*{-1mm}
\begin{equation}\label{p310}
  *q=0,~ \tilde{q} =\Phi,
\end{equation}\vspace*{-5mm}

\noindent where $\Phi$ is arbitrary exact one-form.  Nester, Tung
and Zhytnikov results do not connect the Dirac equation itself and
the Nester gauge by the equivalence relationship of a certain type,
and do not establish the explicit and unique connection in all
points of $\Sigma$ (see, for example, \cite{dim1,dim2,nesCQG})
between the variables of the 3-spinor field and the SOF variables.
That is why search for the valuable grounding of a statement about
existence of a close correlation between both approaches remains
topical.

We propose further a new insight into the problems of this
correlation that is fully correct for the case of maximal
hy\-per\-sur\-fa\-ce and is grounded asymptotically in the case of
quite arbitrary hy\-per\-sur\-fa\-ces. The reason is known: the
linear equations for the spinor fields become non-linear after
transition to the respective tensor functions.


\subsection{\!Direct link between the 4-covariant\\
    \hspace*{-1.2cm}spinor 3-form and the Einstein Hamiltonian{\label{ ch5-subsec41}}}

\hspace*{3cm}Taking into account that the Hamiltonian density
(\ref{p32}) and the SWE were obtained by the spinor parameterization
for the Hamiltonian displacement, we write in terms of the
Sommers---Sen spinors
\begin{equation}
  \label{p314+1}
  N^\mu=\lambda^A\lambda^{\dot{A}}=\lambda^{(A}\lambda^{B)+}+N^\mu
  n_\mu n^{AB}=
  \lambda^{(A}\lambda^{B)+}+\frac1{\sqrt2}\lambda_D\lambda^{D+}
  \varepsilon^{AB}.
\end{equation}
That is why $N\equiv N^0=\lambda_A\lambda^{A+}=\Phi$. Note, that the
Nester SOF approach does not limit a choice of the dependence
$\Phi=\Phi(N)$, but the only case of $\Phi\equiv N$ seems to be
meaningful and admissible \cite{nesCQG}.

We will further give the 4-covariant Hamiltonian density in terms of
the Sen---Witten spinor using the SWE in the form
\begin{equation}
  \label{p315}
  \mathcal{D}^B{}_C\lambda^C=0.
\end{equation}
An action of operator $\mathcal{D}_{AB}$ on the spinor fields is
\[
\mathcal{D}_{AB}\lambda_C =D_{AB}\lambda_C+\frac{\sqrt2}{2}{\cal
  K}_{ABC}{}^D\lambda_D,
\]
where $D_{AB}$ --- the spinorial form of the derivative operator
$D_\alpha$ compatible with metric $h_{\mu\nu}$ on the $C^\infty$
hy\-per\-sur\-fa\-ce $\Sigma_t$, $\mathcal{K}_{ABCD}$~--- the
spinorial tensor of the extrinsic curvature of hy\-per\-sur\-fa\-ce
$\Sigma$.

The standard substitution transforms (\ref{p33}) to the form
\[
 \mathcal{H}(\varphi,\chi) = \Big[- 2\sqrt {2} \left(
{n_{A\dot {A}}
      D_{\mu} \varphi ^{A}D^{\mu} \overline {\varphi} ^{\dot {A}} +
      n_{A\dot {A}} D_{\mu} \chi ^{A}D^{\mu} \overline
      {\chi} ^{\dot {A}}}\right)+\]\vspace*{-3mm}
\begin{equation}
+\, 2\left( {n^{B\dot {C}}D_{B\dot {A}} \overline {\varphi} ^{\dot
        {A}}D^{\mu} \varphi ^{A} + n^{B\dot {C}}D_{B\dot {A}}
      \overline {\chi} ^{\dot {A}}D^{\mu} \chi ^{A}}
  \right)\!\Big]d^{3}\Sigma.\label{p316}
\end{equation}

\noindent Let us take into account that
\[
  h^{\mu \nu} n_{A\dot {A}} D_{\mu} \varphi ^{A}
  D_{\nu} \overline {\varphi}^{\dot {A}} =
  \varepsilon ^{\dot {B}\dot {D}} \varepsilon^{BD}n_{A\dot {A}}
  \left( {D_{B\dot {B}} \varphi ^{A}} \right)
  \left({D_{D\dot {D}} \overline {\varphi} ^{\dot {A}}}
  \right)=\]\vspace*{-3mm}
  \begin{equation}\label{p317}
  = 2n^{R\dot {B}}n_{R}{}^{\dot {D}}\varepsilon^{BD}
   \left( {D_{B\dot {B}} \varphi ^{A}} \right)
    \left( {D_{D\dot {D}} \overline {\varphi} ^{\dot {A}}} \right) =
  \left( {\mathcal{D}_{B}{}^{R}\varphi ^{A}} \right)\left( {\mathcal{D}
      ^{B}{}_{R}\overline {\varphi} ^{\dot {A}}} \right)n_{A\dot {A}},
\end{equation}
and\vspace*{-1mm}
\begin{equation}\label{p318} \left( {\mathcal{D}_{R}{} ^{B}\overline {\varphi}
      ^{\dot {A}}} \right)n_{A\dot {A}} =
     \frac{{1}}{\sqrt {2}} \left[ - \mathcal{D}^{BR}\varphi ^{+}_{A} - \sqrt {2} \left( {\mathcal{D}^{BR}n_{A\dot {A}}} \right)\overline {\varphi}  ^{\dot
          {A}} \right]\!,
\end{equation}\vspace*{-3mm}
\begin{equation}
  \label{p319} \mathcal{D}^{BR}n_{A\dot {A}} =
\mathcal{K}^{BR}{}_{A\dot
    {A}} + {\frac{{\sqrt {2}} }{{2}}}F_{A\dot {A}} \varepsilon
  ^{(BR)}= \mathcal{K}^{BR}{}_{A\dot {A}}.%
\end{equation}\vspace*{-5mm}

\noindent Then\vspace*{-1mm}
\[
n_{A\dot {A}} h^{\mu \nu} \left( {D_{\mu} \varphi^{A} D_{\nu}
      \overline {\varphi} ^{\dot {A}} + D_{\mu} \chi ^{A}D_{\nu}
      \overline {\chi} ^{\dot {A}}} \right)  = \]\vspace*{-2mm}
\[
= \left( {\mathcal{D}_{BR}\varphi _{A}} \right)\left(\!
{{\frac{{\sqrt
            {2} }}{{2}}}\mathcal{D}^{BR}\varphi ^{
        +A} + \mathcal{K}^{BRA}_{S} \varphi ^{ + S}} \!\right)+\]\vspace*{-2mm}
\begin{equation}
  + \left( {\mathcal{D}^{BR}\chi ^{A}} \right) \left(\! {{\frac{{\sqrt
            {2} }}{{2}}}\mathcal{D}_{BR}\chi^+_{ A} +
      \mathcal{K}^{BRA}_{S} \chi ^{+S}} \!\right)\!\!.  \label{p320}
\end{equation}\vspace*{-3mm}

\noindent For transformation of the other terms we will use the
identity\vspace*{-3mm}
\[
D_{B\dot {A}} \overline {\varphi} ^{\dot {A}} =
  D_{B\dot {A}} \left( {2n^{\dot A {C}}n_{C\dot {C}} \overline
      {\varphi} ^{\dot {C}}} \right) = - {\frac{{2}}{{\sqrt {2}}
    }}D_{B\dot {A}} \left( {n^{\dot A {C}}\varphi ^{ + }_{C}} \right)
  =   \]\vspace*{-5mm}
\begin{equation}
  \label{p321}
  =- {\frac{{2}}{{\sqrt {2}} }}\left(\! {\mathcal{K}_{B\dot {A}}{} ^{\dot {A}C}\varphi ^{ +
      }_{C} + {\frac{{1}}{{\sqrt {2}} }}\mathcal{D}_{B}{} ^{C}\varphi ^{ +}
      _{C}} \!\right)
  =
  - {\frac{{2}}{{\sqrt {2}} }}\left(\! {\mathcal{K}_{B}{} ^{C}\varphi ^{ +} _{C} +
      {\frac{{1}}{{\sqrt {2}} }}\mathcal{D}_{B}{} ^{C}\varphi ^{ +} _{C}}
  \!\right)
\end{equation}

\noindent and, therefore,\vspace*{-3mm}
\begin{equation}
 \label{p322} n^{B\dot {C}}D_{B\dot {A}} \overline
{\varphi} ^{\dot
    {A}} D_{A\dot C} \varphi ^{A} = {\frac{{\sqrt {2}} }{{2}}}\left(
    {\mathcal{D}_{A}{} ^{B}\varphi ^{A}} \right)\left( {\mathcal{D}_{BC}
      \varphi ^{ +C}} \right) -{\frac{{1}}{{2}}}\mathcal{K}_{BC}\varphi ^{
    +C} \mathcal{D}_{A}{} ^{B}\varphi ^{A}.
\end{equation}\vspace*{-5mm}

\noindent The final expression for $\mathcal{H}(\varphi,\chi)$ we
will give in the form
\[
\label{p323} \mathcal{H}\left( {\varphi ,\chi} \right) = \sqrt
{2}\Big\{\!\left({\mathcal{D}_{BR}^{} \varphi_{A}} \right)\!\left(\!{\sqrt {2}\,\mathcal{D}^{BR}\varphi^{ + A} + %
    \mathcal{K}^{BRA}_{S} \varphi ^{ + S}} \right) +
    \]\vspace*{-3mm}
\[
+\left( {\mathcal{D}_{A}{}^{B}\varphi^A} \right)\!\left(\!-
\mathcal{D}_{BC} \varphi ^{ + C} +
  \mathcal{K}_{BC} \varphi^{ + C} \right) %
+ \left( {\mathcal{D}_{BR}^{} \chi_{A}} \right)\left( {\sqrt {2}
    \mathcal{D}^{BR}\chi ^{ + A} + \mathcal{K}^{BRA}_{S} \chi ^{ + S}} \right)
    +
\]\vspace*{-3mm}
\begin{equation}
+ \left( {\mathcal{D}_{A}{} ^{B}\chi^{A}} \right)\! \left(\!{
-\mathcal{D}_{BC}^{} \chi ^{ + C} +
    \mathcal{K}_{BC}^{} \chi ^{ + C}} \right)\!\Big\}dV.
\end{equation}

The Hamiltonian 3-form $\mathcal{H}\left( {\varphi ,\chi} \right)$
(\ref{p323}) in comparison with the Hamil\-to\-nian 3-form, obtained
by Ashtekar and Horowitz \cite{ash}, contains the terms with the
external curvature of hy\-per\-sur\-fa\-ce $\Sigma $.

The first and the second terms are positive definite, and the next
ones turn to zero if on hy\-per\-sur\-fa\-ce $\Sigma $ the spinor
fields $\varphi ^{A}$ and $\chi ^{A}$ satisfy \mbox{the SWE
(\ref{p315}).}

On the other hand, the ADM Hamiltonian density, parameterized with
orthonormal 3-frames $\theta^{\hat{a}}$, is of the form
\cite{nestu}\vspace*{-1mm}
\[
\mathcal{H} \left( {N} \right) = %
- 2\,{|h|}^{{1}/{2}}_{}\, {\widetilde q}^a
\partial_a N + N\,{|h|}^{{1}/{2}} %
\left( {\mathcal{K}^{ab} \mathcal{K}_{ab}  - \mathcal{K}^{2}}
\right) - \]\vspace*{-5mm}
\begin{equation}\label{p324}
-\,2\,{|h|}^{{1}/{2}}\left(\mathcal{K}^a{}_b-\delta^a{}_B\mathcal{K}\right)%
 D_aN^b + N\,{|h|}^{{1}/{2}}\left[ {q^{ab} q_{ab} +
    {\frac{1}{2}} {\widetilde q}^{a} {\widetilde q}_{a} -
    {\frac{1}{6}} (*q)^{2}} \right]\!,
\end{equation}\vspace*{-2mm}

\noindent where the symmetric tensor $q_{ab}$, vector $\widetilde
q_a$, and scalar $*q$ are defined by irreducible
decomposition\vspace*{-3mm}
\[
C ^{a}{} _{bc} = q^{ad} \varepsilon _{dcb} + {\frac{{1}}{{2}}}\left(
  {\delta ^{a} _{c} \widetilde q_{b} - \delta ^{a} _{b}\widetilde
    q_{c} } \right) + {\frac{{1}}{{3}}}*q\varepsilon ^{a}{} _{cb}.
\]\vspace*{-5mm}

\noindent Varying the lapse in (\ref{p324}), we obtain the
super-Hamiltonian constraint in the form\vspace*{-1mm}
\[
    2\partial_k\left({|h|}^{{1}/{2}} q_k\right)+ \frac12 \, {|h|}^{{1}/{2}}\,q^kq_k
    +\]\vspace*{-3mm}
\begin{equation} \label{p325}
    +\,\frac{|h|}{2} \left[
      \mathcal{K}_{}^{mn}\mathcal{K}^{}_{mn}-\mathcal{K}^2
      +q^{mn}_{} q^{}_{mn}+\frac12\tilde{q}^a\tilde{q}_a-\frac16(*q)^2\right]\!.
\end{equation}\vspace*{-3mm}

If the spinor fields $\varphi^A$ and $\chi^A$ satisfy the SWE and
conditions of Theorem 3, then condition (\ref{p12}) is fulfilled,
and vice versa. Then, on the one hand, $\mathcal{H}\left( {\varphi
,\chi} \right)$ will be positive, and, on the other hand, this will
permit us to write ${\mathcal H}(N)$ in the SWOF, under the
necessary in this context limitation for $\Phi$ and at $N^a=0$, in
the form\vspace*{-3mm}
\[
  \mathcal{H}^{SWOF}_{}\left( {N} \right) = N {|h|}^{{1}/{2}} %
  \bigg(\!{-\frac{3}{2}}\,h^{mn}\partial_m \ln N \partial^{}_n \ln
    N - \mathcal{K}\partial_{\hat{3}}\ln N\, - \]\vspace*{-5mm}
\begin{equation}\label{p326}
  -\frac32\,\mathcal{K}^2+\mathcal{K}_{}^{mn}\mathcal{K}^{}_{mn} +
    q^{mn}_{}q^{}_{mn} \!\bigg)\!.
\end{equation}

\noindent Here the lapse is determined by the super-Hamiltonian
constraint %
\[
  2\partial_m\!\left(
  {|h|}^{{1}/{2}} h^{mn}\partial_n \ln N\right) +
  2\partial_m\!\left(
  {|h|}^{{1}/{2}}\theta^{\hat{3}m}\mathcal{K}\right) +
  \]\vspace*{-5mm}
  \[
  +\,{|h|}^{{1}/{2}} \left(\!\frac12 {h}^{mn}\partial_m\ln N\partial_n \ln N\!+\!2 \mathcal{K}\theta^{\hat{3}m} \partial_m\ln N
  \!-\!\frac32\mathcal{K}^2\!+\!\mathcal{K}^{mn}\mathcal{K}_{mn}\!+\!
  q^{mn}q_{mn}\!\right)\!=\]\vspace*{-3mm}
  \[
  = 2\partial_m\left({|h|}^{{1}/{2}} h^{mn}\partial_m\ln N\right)
  +{|h|}^{{1}/{2}} \bigg(\!\frac12h^{mn}\partial_m\ln N\partial_n \ln N\,+ \]\vspace*{-3mm}
\begin{equation}
 +\,2\, \mathcal{K}\,
    \partial_{\hat{3}} \ln N-2 \partial_{\hat{3}}\mathcal{K}+\frac12{\cal
      K}^2 \label{p327}+\mathcal{K}^{mn}{\cal
      K}_{mn}+q^{mn}q_{mn}\!\bigg)=0.
\end{equation}

Let us consider first of all the especially simple case of a maximal
spatial Cauchy hy\-per\-sur\-fa\-ce.  Then the Hamiltonian density
(\ref{p326}) takes the form\vspace*{-2mm}
\begin{equation}
  \mathcal{H}^{SWOF}\left( {N} \right) = N|h|^{{1}/{2}}\left(\!{ {-\frac{{3}}{{2}}}}
    h^{mn}\partial_m\ln N\partial_n \ln N+\mathcal{K}^{mn}\mathcal{K}_{mn}
    +q^{mn}q_{mn} \!\right) \label{p328}
\end{equation}\vspace*{-3mm}

\noindent and will be everywhere positive definite if on $\Sigma$
exists an appropriate solution of the super-Hamiltonian constraint
\[
 2\partial_m\left({|h|}^{{1}/{2}} h^{mn}\partial_m\ln
  N\right) +\]\vspace*{-5mm}
\begin{equation}\label{p329}
+\,|h|^{{1}/{2}} \left(\!\frac12 {h}^{mn}\partial_m\ln N\partial_n
\ln
  N+\mathcal{K }^{mn}\mathcal{K}_{mn} +q^{mn}q_{mn}\!\right)=0.
\end{equation}
Unique positive solution $N$ of this equation exists because
Nester's gauge (\ref{p11}) has the property of conformal invariance
and thus fits into the Lichnerowicz---Choquet-Bruhat---York
initial-value problem analysis \cite{nesCQG}. Therefore, we
con\-clu\-de, that owing to the correspondence between the SWE and
the Nester gauge on the maximal hy\-per\-sur\-fa\-ce there exists
the direct relationship between the Hamiltonian based positivity
localization in the 4-covariant spinor method and in the ADM method
based on the SOF.

Now, let us consider the hy\-per\-sur\-fa\-ce $\Sigma$ which is not
maximal, and let it be asymptotically $N=a+O(r^{-1}),\, \partial_m
N=O(r^{-2})$. Then the super-Hamiltonian constraint (\ref{p327}) for
enough large $r$ can be written as
\begin{equation}
  \label{p330}
  2\partial_m\left(|h|^{{1}/{2}} h^{mn}\partial_m N\right)
  +N|h|^{{1}/{2}}
  \left(\!-2\partial_{\hat{3}}\mathcal{K}+\frac12\mathcal{K}^2
    +\mathcal{K}^{mn}\mathcal{K}_{mn}+q^{mn}q_{mn}\!\right)=0.
\end{equation}
The Dirichlet problem for equation (\ref{p330}) has the unique
solution, if
\begin{eqnarray}
  \label{p331}
  C(x)=|h|^{{1}/{2}}\left(\!-2\partial_{\hat{3}}\mathcal{K}+\frac12\mathcal{K}^2
    +\mathcal{K}^{mn}\mathcal{K}_{mn}+q^{mn}q_{mn}\!\right)\geq0.
\end{eqnarray}

The same condition and the condition that $N$ is positive on the
boundary or asymptotically ensure the non-occurrence of the nodal
points of equation (\ref{p330}), since the nodal submanifolds of
elliptic equation of second order are closed or have common points
with boundary.  That is why everywhere $N>0$, and we can choose
$a=1$.

Further, a general theorem for the elliptic second-order system
\mbox{claims} \cite{lop} that its solutions continuously depend on
coefficients, domain and va\-lues of the functions on the boundary,
therefore the Hamiltonian density $
\mathcal{H}^{SWOF}\!(N(\mathcal{K}), N )$ (\ref{p326}) continuously
depends on $\mathcal{K}$ and thus is non-negative on the
hy\-per\-sur\-fa\-ces which satisfy the condition\vspace*{-1mm}
\begin{equation}
  \label{p332}
  -2\partial_{\hat{3}}\mathcal{K}+\frac12\mathcal{K}^2\geq0
\end{equation}\vspace*{-5mm}

\noindent and lie in some neighborhood of the maximal one. The
presence of the terms $-2\partial_{\hat{3}}\mathcal{K}$ and
$\frac12\mathcal{K}^2$ in the right-hand side of relationship
(\ref{p331}) is caused just by the fact that we used the SWOF; the
application of Nester's gauge does not give a possibility to prove
the existence of this class of hy\-per\-sur\-fa\-ces, on which the
Hamiltonian density in the SOF is non-negative.

In order to establish a correspondence between condition b) of
Theorem 4 and (\ref{p332}), we write the following space spinors
definition
\begin{equation}
\label{p333}
  D_{A}{}^{B}\mathcal{K}=-\sqrt2
  n_{\hat{\alpha}}\sigma^{\hat{\alpha}}{}_{A\dot{A}}
  \sigma^{\hat{\beta}B\dot{A}}\partial_{\hat{\beta}}\mathcal{K}=
  -\sqrt2
  \sigma^{\hat{0}}{}_{A\dot{A}}
  \sigma^{\hat{\beta}B\dot{A}}\partial_{\hat{\beta}}\mathcal{K},
\end{equation}
and obtain that the diagonal elements of matrix\vspace*{-2mm}
\[
  \frac{\sqrt2}{4}D_{A}{}^{B}\mathcal{K}+\frac12 \varepsilon
  _A{}^B\mathcal{K}^2
\]
are\vspace*{-2mm}
\[
\frac14\partial_{\hat{3}}\mathcal{K}+\frac12\mathcal{K}^2~
  \mbox{and}~ -\frac14\partial_{\hat{3}}\mathcal{K}+\frac12\mathcal{K}^2.
\]

Therefore, the second of them is non-negative on the
hy\-per\-sur\-fa\-ces \mbox{which} satisfy condition (\ref{p332}).
This means that under fulfilling condition (\ref{p332}) condition b)
of Theorem 3 is also fulfilled.

So, if the SWE and conditions a) and c) of Theorem 3 are fulfilled,
then on hy\-per\-sur\-fa\-ces, which satisfy condition (\ref{p332})
and lie in some neigh\-bor\-hood of the maximal one, the
Hamil\-tonian density $\mathcal{H}\left( {\varphi ,\chi} \right)$
(\ref{p323}) and the ADM Hamil\-tonian density
$\mathcal{H}^{SWOF}\left( {N} \right)$ (\ref{p326}) are locally
non-negative simultaneously.

Let us note that just an absence of the result on connection between
the SWE equation and the Nester gauge (a theorem like Theorem 3 and
Theo\-rem~4) did not permit Nester and Tung to obtain a direct
relationship between the 4-spinor 3-form of the Hamiltonian density
under fulfilling the SWE and the Hamiltonian density in the SOF
formalism, both on the enough general hy\-per\-sur\-fa\-ces and even
on the maximal ones. The 3-spinor formalism, developed by these
authors and Zhytnikov, provides the partial solving of this problem;
in particular, the energy is guaranteed to be locally non-negative
only on the maximal hy\-per\-sur\-fa\-ces.

Generalization of the SOF by the SWOF allows us to remove two
liabilities of the SOF method: necessity of the restriction to the
maximal hy\-per\-sur\-fa\-ces, and impossibility of extension to the
future null infinity and, hence, description of the Bondi
4-momentum. Therefore, for the quasilocal Hamiltonian density
(\ref{p32}) investigation not the Dirac equation and the 3-spinors
are suitable, but the SWE and the space spinors introduced by
Sommers \cite{som}. Although the \mbox{3-di}\-men\-sional Dirac
equation and the SWE are very similar, we see that fi\-xing of the
spinor field by the Dirac equation or by the SWE leads to different
physical consequences. The mathematical consequences for application
of these gauge conditions for the spinor field are also different;
in particular, the conditions for existence of solutions differ in
domains of finite measure \cite{peljpa}.

The equivalence of the Sen---Witten spinor field and the SWOF, under
the reasonable from the physical point of view fulfilling of
conditions of Theorem 3, permits to establish that the method of the
4-covariant quadratic spinor Hamiltonian and the SOF method are very
close.  The spinor parameterization of the Hamiltonian displacement
and correlations (\ref{p210}) are the key for the orthonormal frame
interpretation of the Hamiltonian 4-covariant spinor form
(\ref{p32}) and the spinor interpretation of the ADM Hamiltonian
density even in the case when the spinor field or the orthonormal
frame are not fixed.

Note at the end that conditions of (\ref{p331}) and (\ref{p332})
type are the only sufficient ones, and we expect to weaken them
significantly or to exclude com-\linebreak pletely.\vspace*{-2mm}


\subsection{\!In which cases\\ \hspace*{-1.2cm}the conditions of
    Theorem 4 are fulfilled?\label{ ch5-subsec42}}

\hspace*{3cm}Theorem 4 implies that the absence of nodal
submanifolds can be ensured in accordance with condition 2) at least
for some hy\-per\-sur\-fa\-ces $\Sigma_t$ in Petrov type $N$
space-time with tensor of extrinsic curvature
$\mathcal{K}_{\pi\rho}$ and for some physical fields with density of
energy in comoving frame $\mu$. Further we will work out in detail
these constituents of condition 2).

Apparently, nodal sets are absent on SCMC hy\-per\-sur\-fa\-ces in
empty space and hence by continuity~--- in a neighborhood. However,
significantly more strong result takes place.

 {\itshape{\bfseries Theorem 6.} Let:

  a) everywhere in the  bounded domain on the    space-like   hypersurface on $\Sigma_t$ in Petrov type $N$
space-time the dominant energy conditions be
  fulfilled;

b) the functions ${\rm Re}\,\lambda^0$ or ${\rm Im}\,\lambda^0$,
which correspond to non-negative
  eigen\-value $C_0$,  nowhere on domain boundary equal zero.

  Then the  non-trivial solution $\lambda^C$ to
  SWE does not have the nodal points in domain $\Omega$.
}%

This theorem is a corollary of the following lemma:

 {\itshape{\bfseries Lemma.} If in a point of the domain $\Omega$ the
dominant
  energy condition is fulfilled,  then in this point the
  matrix of spinorial tensor}\vspace*{-2mm}
\begin{equation}\label{7}
  C_A{}^B:=\frac{\sqrt2}{2}D_{A}{}^{B}\mathcal{K}+\frac14 \varepsilon
  _A{}^B\left(\!2\mathcal{K}^2+\frac12\mathcal{K}_{\pi\rho}{\cal
      K}^{\pi\rho}+\mu\!\right)
\end{equation}\vspace*{-5mm}

\noindent {\itshape has at least one non-negative eigenvalue.}

\emph{Proof.} Applying strengthening of Descartes theorem we see
that this is possible in two cases:\vspace*{-2mm}
\[
\begin{array}{l}
\displaystyle\hspace*{5mm}1.\,\,\frac12\mathcal{K}^2+\frac18\mathcal{K}_{\pi\rho}{\cal
  K}^{\pi\rho}+\mu\geq0.\\[3mm]
\displaystyle\hspace*{5mm}2.\,\,\frac12\mathcal{K}^2+\frac18\mathcal{K}_{\pi\rho}{\cal
    K}^{\pi\rho}+\mu<0, \\[2mm]
\displaystyle\,\,\quad\frac12\left(\!\mathcal{K}^2+\frac18{\cal
    K}_{\pi\rho}\mathcal{K}^{\pi\rho}+\mu\!\right)^{\!\!2}-\frac{1}{4}\left[(\partial_1{\cal
    K})^2+
    (\partial_2\mathcal{K})^2+(\partial_3\mathcal{K})^2\right]\geq0.\hspace*{2.5cm}
\end{array}\]\vspace*{-3mm}

Condition 1 is fulfilled, if the dominant energy condition is
fulfilled.

Taking into account this lemma, let us substitute condition b) of
Theorem 5 by the dominant energy condition, let us denote
non-negative eigenvalue by $C_0$ and refine condition c). Then we
obtain Theorem 6.

Now we can strengthen significantly Theorem 6.

 {\itshape{\bfseries Theorem 7.} Let  on $\Sigma_t$ in Petrov type $N$
space-time Einstein constraints and the
  dominant energy condition be fulfilled. Then everywhere on
  $\Sigma_t$ there exists a two-to-one correspondence between
  Sen---Witten spinor and Sen---Witten orthonormal frame.}

Since for all physical fields, whose existence hitherto is confirmed
experimentally, the dominant energy condition is fulfilled, then
Theorem 7 permits to state that Sen---Witten spinor and Sen---Witten
orthonormal frame are equivalent in majority of physical models.
Nevertheless, in recent years the physical situations in which the
dominant energy condition will be violated (primarily of dark
energy, but also in wormhole space-time, in sudden future
singularity or in gravastar)\index{singularity} appear.  That is why
we note that Theorem 7 does not exhaust all cases, when Sen---Witten
spinor and Sen---Witten orthonormal frame are equivalent. In
particular, from condition 1) it follows, that even if the dominant
energy condition is violated ($\mu<0$), Sen---Witten spinor and
Sen---Witten orthonormal frame are equivalent on
hy\-per\-sur\-fa\-ces, which satisfy condition\vspace*{-2mm}
\begin{equation*}
  \frac12\mathcal{K}^2+\frac18\mathcal{K}_{\pi\rho}{\cal
    K}^{\pi\rho}\geq-\mu
\end{equation*}\vspace*{-5mm}

\noindent or simultaneously two conditions\vspace*{-2mm}
\[
  \frac12\mathcal{K}^2+\frac18\mathcal{K}_{\pi\rho}{\cal
    K}^{\pi\rho}\leq-\mu, \]\vspace*{-3mm}
\[
 \frac12\left(\!\mathcal{K}^2+\frac18{\cal
    K}_{\pi\rho}{\cal
    K}^{\pi\rho}+\mu\!\right)^{\!\!2}-\frac{1}{4}\left[(\partial_1{\cal
    K})^2+(\partial_2\mathcal{K})^2+(\partial_3\mathcal{K})^2\right]\geq0.
\]

The result of our investigations shows that mystery, above mentioned
by Nester and Bartnik, is eliminated for all spaces with typical
geometrical properties and fields with typical physical
properties.\vspace*{-1.5mm}

\section{\!Summary\label{ch5}}
\vspace*{-0.5mm}

\hspace*{3cm}Spinor methods in general relativity became not only a
tool for description of interaction between
  gravity field and particles with half-integer spin, but one of main methods for description of algebraic
   properties of curvature tensor and gravitational and electromagnetic radiation in the curved space-time.
   This confirmed their efficiency for proving the positive gravity energy theorem that became not only
   proof for correspondence of properties for gravity and other physical fields, but opened a way for
    establishing of conditions of stability or spontaneous compactification of Minkowski space in
     multidimensional theories of Kaluza---Klein type.  Discovery of Witten spinor method for positive
      energy theorem (PET) proof increased significantly an interest to study all aspects of spinor
       fields behavior in Riemannian spaces.

Schoen and Yau  proof of Einstein hypothesis about positive energy
of asymp\-totically flat space solved one of longstanding problems
in General Re\-la\-tivity, nevertheless its complexity stimulated an
appearance of alternative me\-thods of proving, and first of all the
most prospective one from the physical \mbox{point} of view~---
Wittenian spinor method (with significant contribution of Reula,
Ashtekar, Horovitz, Nester). Despite the fact that Wittenian proof
after its refinement obtained a perfect mathematical form, the
physical in\-ter\-pre\-ta\-ti\-on of proof is treated up to date
ambiguously because of de\-ci\-si\-ve part of auxiliary spinor field
in proof. According to common view (Gold\-berg, Nester, Bartnik)
this spinor field remains physically mysterious. \mbox{Establishing}
of correspondence between Witten spinor method and tensor methods
became a subject of many investigations. Taking into account that
local orthorepers may not exist on subsets of asymptotically flat
manifolds, Dimakis and M${\rm {\ddot u}}$ller-Hoissen made final
conclusion about impossibility of tensor method of \mbox{proving
PET.}

  In Chapter 5 we
ground the Nester tensor method for proving PET, ta\-king into
account its criticism by Dimakis and M${\rm {\ddot u}}$ller-Hoissen,
and generalize it for systems with radiation. Such grounding
establishes the connection between Witten spinor method and Nester
tensor method.

The main ideas of realized approach are the following. Firstly, we
take into account that existence of nodal points of Sen---Witten
equation is not a barrier for existence of correspondence between
the local orthoreper field and Witten spinor field, because for
Sommers transformation we can use such SWE solutions that do not
have the nodal points. Secondly, we show that the reason for
existence of the nodal points comes from the properties of SWE and
boundary values first of all, but not topological properties of
space. On this basis we develop the theory of nodal manifolds for
physically meaningful elliptic systems of equations, the most wide
class among which consists of double-covariant systems, and
establish conditions for coefficients and boundary values, for
\mbox{which} the nodal points are absent. Absence of nodal sets of
dimension $n-1$ for generalized Dirac (and SWE) equation is the
known fact, but for the case of nodal sets of lower dimensions the
consideration of the system of first order equations gives the
possibility to obtain the only conclusion that the nodal set of
solutions of generalized Dirac equation on an n-dimensional manifold
has the dimension at most $n-2$ and does not ascertain when the
nodal sets of such or lower dimension are absent. That is why we
consider the differential consequence of SWE in form of the
second-order equation and formulate the conditions, under which
zeros of solution are absent independently on dimension of nodal
sets. It is clear that for nodal sets of dimension 2 these
conditions are not necessary, but they are not too rigid from the
physical point of view.

After ensuring the absence of zeros of SWE we obtain the possibility
to use the Sommers transformation and prove the existence of
correspondence between Witten spinor field and some orthonormal
frame, which on maximal hy\-per\-sur\-fa\-ce coincides with Nester
special orthonormal frame.

Nester method did not cover the systems with radiation.
Generalization of tensor method on radiating systems, i.e.
nonmaximal hy\-per\-sur\-fa\-ces, is presented in Chapter~5, where
we prove the existence of tensor interpretation of Wittenian spinor
field also on nonmaximal hy\-per\-sur\-fa\-ces.

Equivalence principle excludes the possibility of existence of the
gravitational energy density, but one can describe the distribution
of energy as integrals over finite regions, as it was proposed by
Penrose and known as quasilocalization. As it was mentioned above,
/in theory of quazilocal values there appears a problem of
comparison of different Hamiltonian three-forms, in particular,
Hamiltonian density with spinor variables and
Arnowitt---Deser---Misner density, parametrized by special Nester
orthonormal frame. We establish existence and correspondence between
these forms.

One of necessary conditions for possibility of existence of tensor
interpreta\-tion of Wittenian spinor field and for existence of
correspondence between Hamiltonian density with spinor variables and
Arnowitt---Deser---Misner density in SOF is  fulfilling of dominant
energy condition. Existence of dark energy, which violates the
dominant energy condition, puts the question about possible change
of status for Wittenian spinor field in its presence. We show that
at violation of DEC the correspondence between Witten spinor field
and SOF continues to exist, i.e. Witten spinor field keeps its
geometrical (gauge) nature.

\newpage

\appendix
 {\parindent=0pt\hspace*{2.9cm}
\vbox{\vspace*{-1.6cm}\hbox{\large\raisebox{0pt}{{\par}}}
\vspace*{0.6cm}\hbox{$\blacksquare$}
 \vspace{-3.0mm}
\hbox{\rule{10cm}{1pt}}
\vspace{1.5mm}%
\raggedright{\bfseries\sffamily\Large  APPENDIXS}%
\vspace*{-0.5mm} \hbox{\rule{10cm}{1pt}}\par \vspace{39mm}}}

\begin{wrapfigure}{l}{2.6cm}
\vspace*{-5.9cm}{\includegraphics[width=3.0cm]{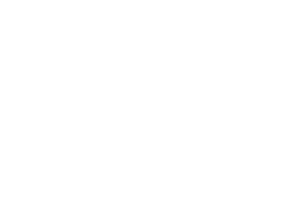}}\vskip17.2cm
\end{wrapfigure}
\vspace*{17mm}

\markboth{APPENDIXS$_{ }$}{APPENDIX A.\,\,Friedmann equations for
the scalar-field model$_{ }$} \vspace*{-12mm}

\begin{flushright}
   {\sffamily APPENDIX A}
\end{flushright}

 \setcounter{section}{1} \vspace*{-12mm}
\hspace*{3cm}\section*{\hspace*{-3cm}FRIEDMANN
  EQUATIONS\\ FOR THE MULTICOMPONENT\\ SCALAR-FIELD MODEL$_{}$}\label{sec:zhA}\begin{picture}(10,10)
\put(0,-113){\bfseries\sffamily{332}}
\end{picture}

\thispagestyle{empty}

\vspace*{-0.95cm} \noindent We consider $n$ scalar fields minimally
coupled to gravity in four dimensions.  The effective action of this
model reads as (see, e.g., (\ref{zh2.14}))
\[
S= \frac{1}{16\pi G_N}\int d^4x \sqrt{|\tilde {g}^{(0)}|}
  \Big(\!
    R\left[\tilde g^{(0)}\right] -
\]\vspace*{-3mm}
\begin{equation}
  \label{zha1}
  ~~~~~~~-\, G_{ij}\tilde g^{(0)\mu\nu}
    \partial_{\mu}\varphi^i\partial_{\nu}\varphi^j -2
    U(\varphi^1,\varphi^2,\mbox{...})\!\Big)\!,\tag{A.1}
\end{equation}
where the kinetic term is usually taken in the canonical form
$G_{ij}=\mbox{diag}(1,1,\mbox{...} )$ (flat $\sigma$ model). We use
the usual conventions \mbox{$c=\hbar =1$}, i.e. \mbox{$L_{\rm
Pl}=t_{\rm Pl}=1/M_{\rm Pl(4)}$} and $8\pi G_N = 8\pi /M_{\rm
Pl(4)}^2$. Here, scalar fields are dimensionless
and potential $U$ has dimension $[U] = \mbox{length}^{-2}$.

Because we want to investigate the dynamical behavior of our
Universe in the presence of scalar fields, we suppose that scalar
fields are homogeneous: $\varphi^i=\varphi^i(t)$ and
four-dimensional metric is spatially flat
Friedmann---Robertson---Walker one: $\tilde g^{(0)}= -dt\otimes dt +
a^2(t)d\vec{x}\otimes d\vec{x}$.

For the energy density and pressure we easily get (see also
Eqs.~(\ref{rho_de_L}) and (\ref{p_de_L})):
\begin{equation}
\begin{array}{c}
  \displaystyle\rho = \frac{1}{8\pi G_N}\left(\!\frac12 G_{ij}
    \dot\varphi^i\dot\varphi^j +U\!\right)\!, \\[3mm]
  \displaystyle P= \frac{1}{8\pi G_N}\left(\!\frac12 G_{ij} \dot\varphi^i\dot\varphi^j - U\!\right)

  \end{array}\!\!\Longrightarrow\label{zha2}\tag{A.2}
\end{equation}
\begin{equation}\hspace*{-2.5cm}\Longrightarrow\!\!
\begin{array}{c}
   \displaystyle\frac12 G_{ij} \dot\varphi^i\dot\varphi^j= 4\pi
  G_N(\rho+P), \\[3mm]
  \displaystyle U= 4\pi G_N(\rho-P),
\end{array}\label{zha3}\tag{A.3}
\end{equation}
where overdots denote the derivatives with respect to the
synchronous time $t$. In chapter 1, overdots denote the derivatives
with respect to the conformal time $\eta$.

The Friedmann equations for considered model are %
\index{Friedmann equations} %
\begin{equation}
  \label{zha4}
  3 \left(\!\frac{\dot a}{a}\!\right)^{\!\!2} \equiv 3 H^2 = 8\pi G_N \rho =
  \frac12 G_{ij} \dot\varphi^i\dot\varphi^j +U,\tag{A.4}
\end{equation}\vspace*{-5mm}

\noindent and\vspace*{-1mm}
\begin{equation}
  \label{zha5}
  \dot H = -4\pi G_N (\rho +P) = - \frac12 G_{ij}
  \dot\varphi^i\dot\varphi^j.\tag{A.5}
\end{equation}
From these two equations, we obtain the following expression for the
acceleration parameter:\vspace*{-3mm}
\[
 q_a \equiv \frac{\ddot a}{H^2 a} = 1 -\frac{4\pi G_N}{H^2} (\rho + P)
  = -\frac{8\pi G_N}{6H^2} (\rho +3P)=
\]\vspace*{-3mm}
\begin{equation}
  \label{zha6}
   = \frac{1}{6H^2} \left(\!-4
    \times \frac12 G_{ij} \dot\varphi^i\dot\varphi^j +
    2U\!\right)\!.\tag{A.6}
\end{equation}
It can be easily seen that the equation of state (EoS) parameter
$\omega = P/\rho$ and parameter $q_a$ are linearly connected:
\begin{equation}
\label{zha7} q_a=-\frac12(1+3\omega).\tag{A.7}
\end{equation}

From the definition of the acceleration parameter, it follows that
$q_a$ is constant in the case of the power-law and De~Sitter-like
behavior:
\begin{equation}
\label{zha8} q_a =
\begin{cases}
\!(s-1)/s ;~ a\propto t^s,\\
\!1 ;~ a\propto e^{H t}. \\
\end{cases}\tag{A.8}
\end{equation}
For example, $q_a=-0.5$ during the matter dominated (MD) stage where
\mbox{$s=2/3$.}

Because the minisuperspace metric $G_{ij}$ is flat, the scalar field
equa-\linebreak tions are:\vspace*{-1mm}
\begin{equation}
  \label{zha9}
  \ddot\varphi^i +3H\dot \varphi^i + G^{ij}\frac{\partial
    U}{\partial \varphi^j} = 0.\tag{A.9}
\end{equation}

\noindent For the action (\ref{zha1}), the corresponding Hamiltonian
is
\begin{equation}
  \label{zha10}
  \mathcal{H} = \frac{8\pi G_N}{2a^3}G^{ij}P_iP_j + \frac{a^3}{8\pi
    G_N} U,\tag{A.10}
    \end{equation}\vspace*{-5mm}

\noindent where\vspace*{-3mm}
\begin{equation}
  \label{zha11}
  P_{i} = \frac{a^3}{8\pi G_N} G_{ij} \dot \varphi^j\tag{A.11}
\end{equation}
are the canonical momenta and equations of motion have also the
canonical form
\begin{equation}
  \label{zha12}
  \dot \varphi^i = \frac{\partial \mathcal{H}}{\partial P_i}\,
  ,~ \dot P_i = - \frac{\partial \mathcal{H}}{\partial
    \varphi^i}.\tag{A.12}
\end{equation}
It can be easily seen that the latter equation (for $\dot P_i$) is
equivalent to the Eq.~(\ref{zha9}).

Thus, the Friedmann equations together with the scalar field
equations
can be replaced by the system of the first order ODEs: %
\index{Friedmann equations} %
\begin{equation}
  \dot \varphi^i  = \frac{8\pi G_N}{a^3} G^{ij}P_j, \label{zha13}\tag{A.13}
\end{equation}\vspace*{-3mm}
\begin{equation}
  \dot P_i = - \frac{a^3}{8\pi G_N}\frac{\partial U}{\partial
    \varphi^i}, \label{zha14}\tag{A.14}
\end{equation}
\begin{equation}
  \dot a = a H,\label{zha15}\tag{A.15}
\end{equation}\vspace*{-3mm}
\begin{equation}
  \dot H = \frac{\ddot a}{a} - H^2 = \frac16 \left(\!-4 \times \frac12
    G_{ij} \dot\varphi^i\dot\varphi^j + 2U\!\right) -
  H^2 \label{zha16}\tag{A.16}
\end{equation}
with Eq.~(\ref{zha4}) considered in the form of the  initial
conditions:
\begin{equation}
  \label{zha17}
  H(t=0) = \left.\sqrt{\frac13\left(\!\frac12 G_{ij}
        \dot\varphi^i\dot\varphi^j +U\!\right)}\right|_{t=0}\!.\tag{A.17}
\end{equation}
We can make these equations dimensionless:
\begin{equation}
  \label{zha18}
  \frac{d \varphi^i}{M_{\rm Pl(4)} dt} = \frac{8\pi }{M_{\rm Pl(4)}qb a^3} G^{ij}P_j~ \Rightarrow~ \frac{d \varphi^i}{dt} = \frac{8\pi }{a^3} G^{ij}P_j,\tag{A.18}
\end{equation}\vspace*{-3mm}
\begin{equation}
  \frac{d P_i}{M_{\rm Pl(4)} dt} = - \frac{a^3 M_{\rm Pl(4)}qb }{8\pi}\frac{\partial
    (U/M_{\rm Pl(4)}sq )}{\partial \varphi^i}~ \Rightarrow~ \frac{d
    P_i}{dt} = - \frac{a^3}{8\pi}\frac{\partial U}{\partial
    \varphi^i}\label{18}.\tag{A.19}
\end{equation}
That is to say the time $t$ is measured in the Planck times $t_{\rm
Pl}$, the scale factor $a$ is measured in the Planck lengths $L_{\rm
Pl}$ and the potential $U$ is measured in the $M_{\rm Pl(4)}^2$
units.

This system of dimensionless first order ODEs together with the
initial condition (\ref{zha17}) can be used for numerical
calculation of the dynamics of considered models with the help of a
Mathematica package applied to these equations in \cite{zhKP}.

\newpage


\begin{wrapfigure}{l}{2.6cm}
{\includegraphics[width=3.0cm]{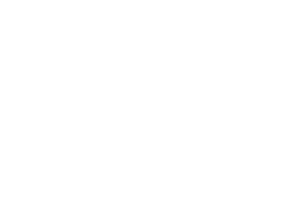}}\vskip16.2cm
\end{wrapfigure}


%
\begin{flushright}
   {\sffamily APPENDIX B}
\end{flushright}

 \setcounter{section}{1} \vspace*{-11mm}
\hspace*{3cm}\section*{\hspace*{-3cm}MATHEMATICAL DETAILS\\ OF THE
BRANEWORLD MODEL$_{}$}
\markboth{APPENDIXS$_{}$}{APPENDIX B.\,\,Mathematical details of the
braneworld model$_{}$}

\vspace*{3mm}

\section*{\hspace*{-3cm}B.1.\,\,Variational problem\\ in the presence of the
  brane} \label{brane-app:varia}

Here, we derive the expression for the first variation of the action
for gravity\vspace*{-3mm}
\begin{equation}
  \label{brane-a-action}
  S_g = \int\limits_\mathcal{B}
  \mathcal{R} - 2 \int\limits_\Sigma K ,\tag{B.1}
\end{equation}
the notation of which is the same as in (\ref{brane-action}).
Equations of this appendix are valid for arbitrary dimension of
space-time. In this derivation, we do not assume that the variation
of $g_{ab}$ vanishes at the boundary $\Sigma$, which is taken to be
\mbox{time-like}.

We start from the standard expression (see, e.g., Appendix~E of
\cite{Wald})
\begin{equation}
  \label{brane-var1} \delta \left( \int_\mathcal{B}
    \mathcal{R} \right) = \int\limits_\mathcal{B} \mathcal{G}_{ab}\, \delta
  g^{ab} + \int\limits_\mathcal{B} \nabla^a v_a ,\tag{B.2}
\end{equation}\vspace*{-5mm}

\noindent where
\begin{equation}
  v_a = \nabla^b \left(\delta g_{ab} \right) - g^{cd} \nabla_a \left(\delta g_{cd}
  \right)\!
  .\tag{B.3}
\end{equation}
The second integral in (\ref{brane-var1}) can be transformed with
the use of the Stokes theorem as
\begin{equation} \label{brane-a-vn} \int\limits_\mathcal{B} \nabla^a v_a = -
  \int\limits_\Sigma v_a n^a ,\tag{B.4}
\end{equation}
where we remember that we are using the {\em inner\/} unit normal
$n^a$ to $\Sigma$, and
\[
 v_a n^a = n^a g^{bc} \left[ \nabla_c \left( \delta g_{ab} \right) - \nabla_a \left(
      \delta g_{bc} \right) \right] =
\]\vspace*{-3mm}
\begin{equation}
 = n^a h^{bc} \left[ \nabla_c \left( \delta g_{ab} \right) -
    \nabla_a \left( \delta g_{bc} \right) \right]\! .\tag{B.5}
\end{equation}

\noindent Then we have
\begin{equation}
  \label{brane-a-deltaK}
  \delta K = \delta \left( h^a{}_b \nabla_a n^b \right) = \delta
  h^a{}_b \nabla_a n^b + h^a{}_b (\delta C)^b{}_{ac} n^c + h^a{}_b
  \nabla_a \delta n^b ,\tag{B.6}
\end{equation}
 where
\begin{equation}
  (\delta C)^b{}_{ac} =
  \frac{1}{2}\, g^{bd} \left[ \nabla_a \left(\delta g_{cd}\right) +
    \nabla_c \left(\delta g_{ad}\right) - \nabla_d \left(\delta
      g_{ac}\right) \right]\!.\tag{B.7}
\end{equation}
The first term on the right-hand side of (\ref{brane-a-deltaK}) is
identically zero.  Indeed, we have $\delta n_a = {}- n_a n_b \delta
n^b$, so that
\[
    \delta h^a{}_b \nabla_a n^b  = - \left( \delta n^a n_b + n^a
      \delta n_b \right) \nabla_a n^b = - \left( \delta n^a - n^a
      n_c \delta n^c
    \right) n_b \nabla_a n^b =\]\vspace*{-5mm}
 \begin{equation}
  = - \delta n^c h^a{}_c n_b \nabla_a n^b = - \delta n^c n_b
    K_c{}^b = 0 .\tag{B.8}
\end{equation}

Thus, variation of the second term of (\ref{brane-a-action}) is
\begin{equation}
\label{brane-a-varK} \delta \left(\! 2 \int\limits_\Sigma K
  \!\right) = \int\limits_\Sigma \left[ n^c h^{ab} \nabla_c \left( \delta
      g_{ab} \right) + 2 h^a{}_b \nabla_a \delta n^b - K h_{ab} \delta
    h^{ab} \right]\!,\tag{B.9}
\end{equation}
where the last term in the square brackets stems from the variation
of the volume element $\sqrt{- h}\, d^4 x$ in the integral over
$\Sigma$.

The total boundary term in the variation of action
(\ref{brane-a-action}) is given by the sum of (\ref{brane-a-vn}) and
(\ref{brane-a-varK}) with the result
\begin{equation}
  (\mbox{Boundary term}) = - \int\limits_\Sigma \left[ n^a h^{bc} \nabla_c \left( \delta g_{ab}
    \right) + 2 h^a{}_b \nabla_a \delta n^b - K h_{ab} \delta h^{ab} \right]\!.\tag{B.10}
\end{equation}

We transform the first term in the integrand of the last expression:
\[
 n^a h^{bc} \nabla_c \left( \delta g_{ab} \right) =  h^{bc} \nabla_c
  \left( n^a \delta
    g_{ab} \right) - h^{bc} \nabla_c n^a \delta g_{ab}=
\]\vspace*{-5mm}
\begin{equation}
=  h^{bc} \nabla_c \left( n^a \delta g_{ab} \right) + K_{ab} \delta
  h^{ab} .\tag{B.11}
\end{equation}
Then
\[
(\mbox{Boundary term})  = - \int\limits_\Sigma \left[ h^{bc}
\nabla_c \left(
    n^a \delta g_{ab} \right) + 2 h^a{}_b \nabla_a \delta n^b
    \right]-
\]
\begin{equation}
\label{brane-a-boundary}   - \int\limits_\Sigma \left(K_{ab} - K
h_{ab}\right) \delta h^{ab} .\tag{B.12}
\end{equation}

Now we show that the integrand of the first integral in
(\ref{brane-a-boundary}) is a divergence, so that this integral
vanishes for variations of $g_{ab}$ with compact support in
$\Sigma$. Indeed,
\[
h^{bc} \nabla_c \left( n^a \delta g_{ab} \right) + 2 h^a{}_b
  \nabla_a \delta n^b = h^{bc} \nabla_c \left( \delta n_b - g_{ab}
    \delta n^a \right)
  + 2 h^a{}_b \nabla_a \delta n^b=
\]\vspace*{-3mm}
\begin{equation}
  \label{brane-a-divergence}
    = h^a{}_b \nabla_a \left( g^{bc} \delta n_c + \delta n^b \right) =
  h^a{}_b \nabla_a \left(h^b{}_c \delta n^c \right) = D_b
  \left(h^b{}_c \delta n^c \right)\! ,
\tag{B.13}
\end{equation}
where $D_a$ is the (unique) derivative on $\Sigma$ associated with
the induced metric $h_{ab}$, and the last equality in
(\ref{brane-a-divergence}) is valid by virtue of Lemma~10.2.1 of
\cite{Wald}.

As a final result, we obtain\vspace*{-2mm}
\begin{equation}
  \delta S_g = \int\limits_\mathcal{B} G_{ab}\, \delta g^{ab} - \int\limits_\Sigma \left(K_{ab} - K h_{ab}\right) \delta h^{ab} .
\tag{B.14}
\end{equation}


\section*{B.2.\,\,Graphical representation\\ of the brane
  evolution} \markboth{APPENDIXS$_{}$}{B.2.\,\,Graphical representation of the brane
  evolution}\label{brane-app:asym}

\hspace*{3cm}The variables $\lbrace X,Y \rbrace$:\vspace*{-2mm}
\begin{equation}
  \label{brane-def-xy}
  X \equiv \frac{\rho_\mathrm{ tot}}{3 m^2} - H^2 ,~ Y \equiv H^2 ,~
  \rho_\mathrm{ tot} = \rho + \sigma ,\tag{B.15}
\end{equation}\vspace*{-4mm}

\noindent allow us to rewrite equation (\ref{brane-main}) in the
form\vspace*{-2mm}
\begin{equation}
  \label{brane-main1}
  X = - \sum_{i = 1,2} \frac{\zeta_i}{\ell_i} \sqrt{Y + \lambda_i^{-2}} ,\tag{B.16}
\end{equation}\vspace*{-4mm}

\noindent which has a convenient visual interpretation. Equation
(\ref{brane-main1}) describes four branches in the physically
restricted range $Y \ge 0$, with the symmetry of reflection with
respect to the $Y$ axis. If there exists a positive root $Y_c$ of
the right-hand side of (\ref{brane-main1}) for $\zeta_1 \zeta_2 =
-1$, then the two mixed branches intersect each other at the
point\vspace*{-4mm}
\begin{equation}
  \label{brane-intersect}
  Y_c = H_c^2 = \frac{\ell_1^2
    \lambda_2^{-2} - \ell_2^2 \lambda_1^{-2}}{\ell_2^2 -
    \ell_1^2} .\tag{B.17}
\end{equation}\vspace*{-4mm}

\noindent The condition for the existence of this intersection point
is that the constant on the right-hand side of
(\ref{brane-intersect}) be positive.

The four branches (with and without intersection) are shown in
Figs.~B.1 and B.2, respectively.

\begin{figure}[h!]
    \includegraphics[width=13cm]{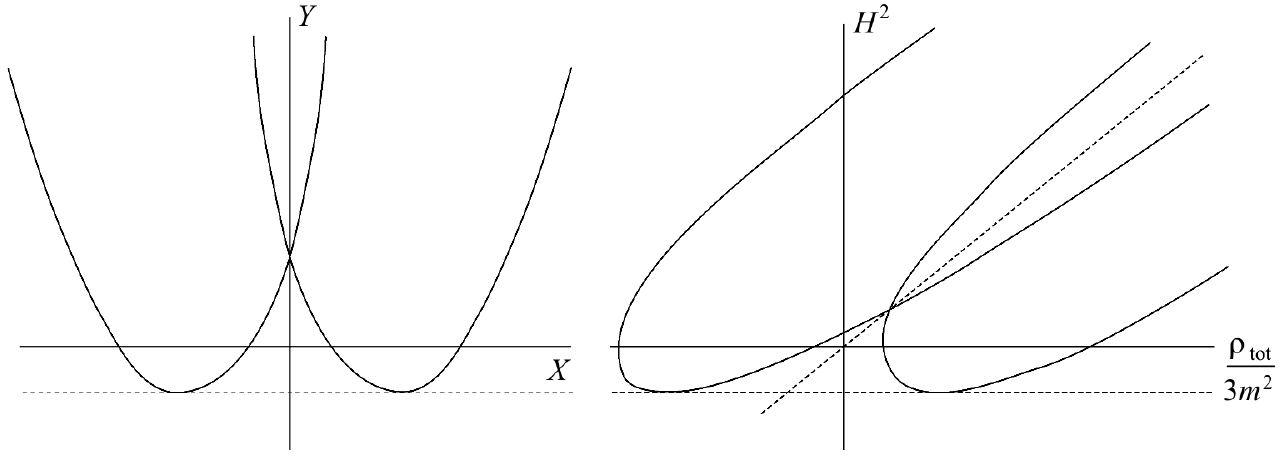}
    {\footnotesize {\bfseries\itshape Fig. B.1.} Four branches described\vspace*{-1.0mm} by Eq.~(\ref{brane-main1}) in the
    ($X, Y$) plane and in the ($\rho_\mathrm{ tot}, H^2$) plane in
    the\vspace*{-1.0mm}
    case where two of them intersect.  The horizontal dotted line
    indicates the position of the $H^2 = 0$ axis in the case
    $\Lambda_1 \Lambda_2 = 0$. The region below this axis is
    nonphysical \label{brane-fig:intersect}}\vspace*{-1mm}
\end{figure}

The brane evolves along one of the four branches towards decreasing
values of $\rho_\mathrm{ tot}$ (during expansion). Depending (in
particular) on the value of the brane tension $\sigma$, three
distinct possibilities can arise:

(i) The trajectory may reach the value of $H = 0$, after which
  the universe recollapses and evolves along the same branch in the
  opposite direction.  This happens when the value of $\rho_\mathrm{
    tot}$ at this point is greater than its minimum \mbox{value
    $\sigma$.}

\begin{figure}
  \vskip1mm
    \includegraphics[width=13cm]{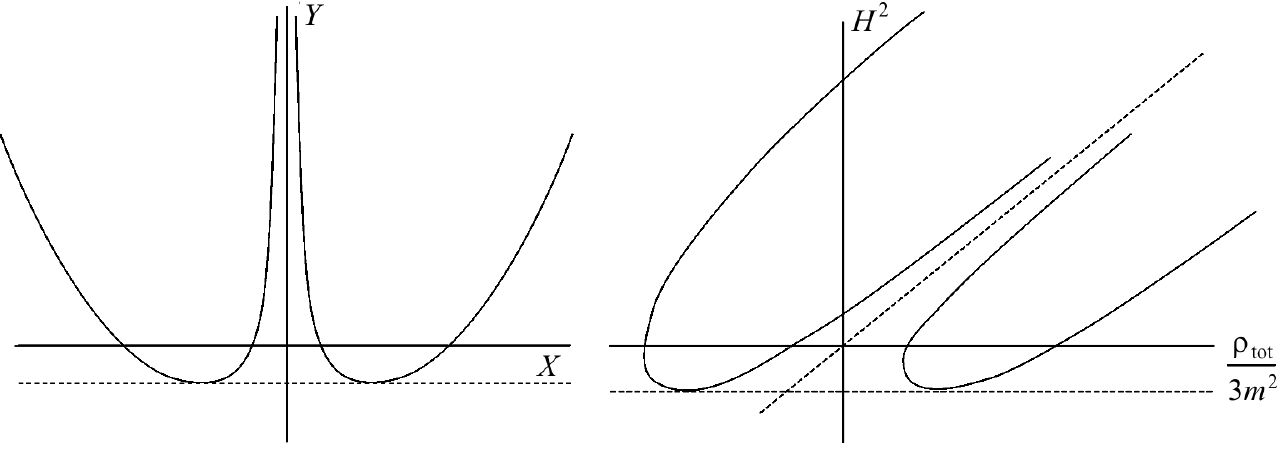}
     {\footnotesize {\bfseries\itshape Fig. B.2.} Four branches\vspace*{-1.0mm} described by Eq.~(\ref{brane-main1}) in the
    ($X, Y$) plane and in the ($\rho_\mathrm{ tot}, H^2$) plane in the\vspace*{-1.0mm}
    case of absence of intersection. The horizontal dotted line
    indicates the position of the $H^2 = 0$ axis in the case
    $\Lambda_1 \Lambda_2 = 0$. The region below this axis is
    nonphysical \label{brane-fig:nonintersect}}
\end{figure}

(ii) The trajectory may asymptotically tend to either De~Sitter
  space or the Minkowski universe with the minimum value
  $\rho_\mathrm{ tot} = \sigma$.  The second possibility occurs when
  the minimum value of $\rho_\mathrm{ tot} = \sigma$ is exactly the
  point where the corresponding graph crosses the axis $H^2 = 0$,
  which, therefore, requires some amount of fine tuning.  This
  possibility can be realized as {\em transient acceleration\/}.

(iii) The trajectory may end in a quiescent singularity at a
  finite value of $H$. This happens when the {\em critical minimum
    point\/} of $\rho_\mathrm{ tot}$ on the evolution curve is
  reached, and if this value of $\rho_\mathrm{ tot}$ is greater than
  its minimum value $\sigma$.  The reasons for the existence of
  quiescent singularities can be seen from the right panels in
  Figs.~B.1 and B.2.  They
  occur at the points of infinite derivative $d H^2 / d
  \rho_\mathrm{tot}$.\vspace*{-2mm}


\section*{B.3.\,\,Background\\ cosmological solution in the bulk}
 \markboth{APPENDIXS$_{}$}{B.3.\,\,Background cosmological solution in the
bulk}\label{app:background}

\hspace*{3cm}For the background bulk metric \eqref{brane-AdS} the
following relations\\ \hspace*{3cm}are satisfied:
\begin{equation}
\label{brane-riemann}
  \begin{array}{c}
    \mathcal{R}_{abcd}  = R_{abcd} , \\[3mm]
    \mathcal{R}_{aibj}  = -\gamma_{ij}\,r\nabla_{a}\nabla_{b}\,r  ,
    \\[3mm]
    \mathcal{R}_{ijkl}  = r^2 \left[\kappa - (\nabla_{a}r)^2 \right] \left(\gamma_{ik}\gamma_{jl} -
      \gamma_{il}\gamma_{jk}\right)\!,
  \end{array}\tag{B.18}
\end{equation}
where the curvature tensor $R_{abcd}$ and covariant derivative
$\nabla_{a}$ correspond to the two-dimensional metric tensor %
\index{metric tensor}%
$\gamma_{ab}$, defined by\vspace*{-1mm}
\begin{equation}
  ds_{(2)}^2=\gamma_{ab} dx^a dx^b=-f(r)d\tau^2+\frac{dr^2}{f(r)} .\tag{B.19}
\end{equation}\vspace*{-2mm}

For the case under consideration in this book,
$f(r)=\kappa-\Lambda_\mathrm{b} r^2/6$, the following expressions
can be verified:\vspace*{-1mm}
\begin{equation}
  \label{brane-bulk_background}
  \begin{array}{c}
    \displaystyle R_{abcd}  = \frac{\Lambda_\mathrm{b}}{6}
    \left(\gamma_{ac}\gamma_{bd}
      - \gamma_{ad} \gamma_{bc}\right)\! , \\[4mm]
    \displaystyle \nabla_{a}\nabla_{b}\,r  = -\frac{\Lambda_\mathrm{b} r}{6} \gamma_{ab}
    ,\quad
       \displaystyle (\nabla_{a}r)^2  =
    \kappa-\frac{\Lambda_\mathrm{b}}{6} r^2 .
  \end{array}\tag{B.20}
\end{equation}

\section*{B.4.\,\,Scalar perturbation of the bulk metric}
\markboth{APPENDIXS$_{}$}{B.4.\,\,Scalar perturbation of the bulk
metric} \label{app:perturb}

\hspace*{3cm}In the gauge $h_L=0$ and $h_a=0$, the general
expression for the perturbed bulk metric \eqref{brane-bmetric3},
\eqref{brane-bmetric_perturb} can be written in the
form\vspace*{-1mm}
\begin{equation}
  \label{brane-bmetric_perturb_gauge}
  ds^{2}_{bulk} =
  \left[ \gamma_{ab} + \sum_k h_{ab} Y \right] dx^a dx^b + \left[ r^2
    + \sum_k h_Y Y \right] \gamma_{ij} dx^i dx^j ,\tag{B.21}
\end{equation}\vspace*{-2mm}

\noindent where $r$, $h_{ab}$ and $h_Y$ depend on $x^a$, while $Y$
depends on $x^i$.

The perturbation of the five-dimensional Riemann curvature tensor
$\mathcal{R}_{ABCD}$ for the metric perturbation
\eqref{brane-bmetric_perturb_gauge} is found to be\vspace*{-1mm}
\begin{equation}
\label{brane-riemann_perturb}
  \begin{array}{c}
    \delta\mathcal{R}_{abcd} = \displaystyle  \sum_k \left( h^e{}_{[b}R_{a]ecd}+\nabla_{c} \nabla_{[b}  h_{a]d}-\nabla_{d} \nabla_{[b} h_{a]c} \right)Y
    ,\\[2mm]
    \delta\mathcal{R}_{iabc} = \displaystyle  \sum_k \left(\!\nabla_{[c}h_{b]a}-\frac{1}{r}h_{a[b}\nabla_{c]} r \!\right)\nabla_i Y ,  \\[3mm]
    \delta\mathcal{R}_{abij}= \displaystyle 0,  \\[2mm]
    \delta\mathcal{R}_{aibj}= \displaystyle {} -\frac{1}{2}\sum_k h_{ab} \nabla_i \nabla_j Y \,+\\[2mm]
     \displaystyle   +\, \frac{1}{2}\gamma_{ij}\sum_k \left[r(\nabla_e r)
      \left(\nabla_a h^e{}_b + \nabla_b h^e{}_a - \nabla^e h_{ab} \right)  \right. - \\
      \displaystyle  \left. -\,\frac{\nabla_a \nabla_b r}{r}\, h_Y-r\,\nabla_a \nabla_b
      \left(\!\frac{h_Y}{r} \!\right)\right]Y , \\[3mm]
    \delta\mathcal{R}_{aijk}= \displaystyle \sum_k \left[r^2\nabla_a\left(\!\frac{h_Y}{r^2}\!\right)-r(\nabla^{b}r)h_{ab}\right]\gamma_{i[j}\nabla_{k]}Y ,  \\[5mm]
    \delta\mathcal{R}_{ijkl} = \displaystyle \sum_k h_Y
    \left(\gamma_{i[l}\nabla_{k]}\nabla_{j}Y-\gamma_{j[l}\nabla_{k]}\nabla_{i}Y
    \right)+ \\[4mm]
    \displaystyle {} +  2\gamma_{i[k}\gamma_{l]j}\sum_k \left[r^2(\nabla_a r) (\nabla_b  r) h^{ab}+\kappa h_Y-r(\nabla_a r) (\nabla^a h_Y) \right]Y,
  \end{array}\tag{B.22}
\end{equation}\vspace*{-2mm}

\noindent where $R_{abcd}$ is the background curvature tensor,
defined in \eqref{brane-bulk_background}. \label{endpage}

\newpage

 {\parindent=0pt\hspace*{2.9cm}
\vbox{\vspace*{-1.6cm}\hbox{\large\raisebox{0pt}{{\par}}}
\vspace*{0.6cm}\hbox{$\blacksquare$}
 \vspace{-3.0mm}
\hbox{\rule{10cm}{1pt}}
\vspace{1.5mm}%
\raggedright{\bfseries\sffamily\Large  BIBLIOGRAPHY}%
\vspace*{-0.5mm} \hbox{\rule{10cm}{1pt}}\par \vspace{39mm}}}

\vspace*{-7mm}

\markboth{Bibliography}{Bibliography} 
\begin{picture}(10,10)
\put(-7.0,-151.5){\bfseries\sffamily{340}}
\end{picture}


\newpage


 {\parindent=0pt\hspace*{2.9cm}
\vbox{\vspace*{-1.6cm}\hbox{\large\raisebox{0pt}{{\par}}}
\vspace*{0.6cm}\hbox{$\blacksquare$}
 \vspace{-3.0mm}
\hbox{\rule{10cm}{1pt}}
\vspace{2mm}%
\raggedright{\bfseries\sffamily\Large  INDEX}\\%
\vspace*{-1mm} \hbox{\rule{10cm}{1pt}}\par \vspace{39mm}}}

\vspace*{-7mm}

\markboth{Index$_{ }$}{Index$_{ }$} 
\begin{picture}(10,10)
\put(-7.0,-151.5){\bfseries\sffamily{374}}
\end{picture}

\thispagestyle{empty}

\begin{multicols}{2}
{\small

\begin{list}{}{\itemindent=-3mm\parsep=-1mm\topsep=2mm\leftmargin=3mm}
\item \textbf{A}belian gauge fields, 152
  \item Abell-ACO, 51
  \item accelerated expansion, 13, 21, 74, 150, 178, 183, 192, 205, 303
  \item acoustic peaks, 15, 30, 35---37, 55, 59, 63---65, 70, 112, 129
  \item action of spinor field, 307
  \item ADD model, 207
  \item age of Universe, 14, 62---64, 157, 159, 164, 170, 230, 239---241,
        243, 247, 248, 251
  \item angular diameter distance, 21, 30, 38, 43---44, 55, 70, 132, 230,
        239
  \item asymmetric branes, 263
  \item Atacama Cosmology Telescope (ACT), 60
\end{list}

\begin{list}{}{\itemindent=-3mm\parsep=-1mm\topsep=2mm\leftmargin=3mm}
  \item \textbf{B}ardeen potentials, 32, 55
  \item baryon acoustic oscillations (BAO), 30, 38, 69
  \item best-fit parameters, 40, 66, 92, 107---110, 123---129, 133
  \item Big Rip, 118---119, 123---124, 132, 272, 273
  \item black branes, 199
  \item black strings, 199
  \item BOOMERanG, 30
  \item brane, 184, 210
  \item braneworld model, 207
  \item bulk, 210
  \item bulk flow, 14, 53
\end{list}

\begin{list}{}{\itemindent=-3mm\parsep=-1mm\topsep=2mm\leftmargin=3mm}
\item \textbf{C}AMB, 33, 67, 80, 107
  \item canonical Lagrangian, 83, 100, 113
  \item Chandra, 42
  \item CLASS, 33, 80
  \item classical Lagrangian, 80, 83, 86
  \item classical scalar field, 100, 102, 111
  \item CMB anisotropy, 15, 31, 33, 37, 42---45, 51, 54---57, 64, 66, 80,
        82, 105, 112, 128, 255
  \item CMB power spectrum, 33, 35, 52, 55, 69, 112
  \item CMB temperature fluctuations, 30---38, 51, 55---60, 65, 97, 107,
        112, 128
  \item CMBEasy, 33
  \item CMBFAST, 33, 80
  \item cold dark matter (CDM), 13, 79
  \item comoving distance, 22
  \item conformal Newtonian gauge, 33, 77, 81, 120
  \item conformal time, 16, 32, 167, 190
  \item conformal time gauge, 143
  \item conservation law, 17, 76, 82, 119, 261, 275
  \item correlation function, 30, 38---41, 61
  \item cosmic coincidence, 72
  \item cosmic microwave background (CMB), 15, 30---40, 42---44, 51---70,
        80, 82, 89, 97, 105, 107, 112---113, 125, 128---130, 184,
        236, 239, 243, 244, 254
  \item cosmic mimicry, 236
  \item cosmological constant, 13, 15, 16, 19, 33, 38, 72---74, 90, 117,
        124, 132, 136, 137, 140---142, 147---152, 160, 177,
        183, 184, 209---211, 214, 216, 220, 222, 223, 226, 228,
        233, 242, 243, 249, 260, 264---269, 276, 280, 281, 300
  \item cosmological distance, 43
  \item cosmological gravexcitons, 170
  \item cosmological parameters, 33, 37, 44, 80, 110, 225
  \item cosmological principle, 16
  \item cosmological recombination, 31, 40, 49
  \item CosmoMC, 33, 107
  \item CosmoRec, 31
  \item cross-correlation function (CCF), 57---59
  \item cur\-va\-tu\-re-non-linear models, 171
\end{list}

\begin{list}{}{\itemindent=-3mm\parsep=-1mm\topsep=2mm\leftmargin=3mm}
  \item \textbf{d}ark energy dominated (DED) epoch, 20, 21, 48, 49, 52, 55---56,
        120, 126
  \item dark energy dominated (DED) model, 26
  \item dark energy in KK models, 141, 146
  \item dark energy in multidimensional models, 171
  \item dark energy parameters, 64, 66
  \item dark matter in KK models, 141, 156
  \item dark radiation, 214
  \item deceleration parameter, 15, 19, 22, 227, 231
  \item deflection of light, 198
  \item density parameter, 18
  \item density perturbations, 14, 32, 39, 46, 77, 99
  \item DGP model, 279
  \item dimensional reduction, 137
  \item Dirac---Born---Infeld Lagrangian, 103
  \item Dirac-Born-Infeld Lagrangian, 83
  \item disappearing dark energy, 233, 270
  \item dominant energy condition (DEC), 303
\end{list}

\begin{list}{}{\itemindent=-3mm\parsep=-1mm\topsep=2mm\leftmargin=3mm}
  \item \textbf{e}ffective potential, 141
  \item Einstein equations, 56, 73, 76, 79, 91, 144, 194, 197, 211,
        215---216
  \item Einstein---Boltzmann equations, 33, 80
  \item Einstein-de Sitter model, 49
  \item elliptic equation, 315
  \item energy-momentum tensor, 16, 75, 174
  \item Enistein equations, 18
  \item EoS parameter, 17, 21, 22, 26, 37, 42, 50, 64, 69, 71, 75, 76,
        81, 85---104, 107---114, 123---126, 131
  \item equation of motion, 75, 83, 157, 172, 179
  \item equation of state (EoS), 17, 113, 145, 196, 235
  \item Equation of State: SupErNovae trace Cosmic Expansion   (ESSENCE),
        25, 27
  \item Euler equation, 54
  \item Euler---Lagrange equation, 75
  \item expansion dynamics, 101
  \item extra dimensions, 184
  \item extrinsic curvature, 257
\end{list}

\begin{list}{}{\itemindent=-3mm\parsep=-1mm\topsep=2mm\leftmargin=3mm}
  \item \textbf{f}ields of forms, 147, 178
  \item fine tuning, 72, 90, 142, 146, 149
  \item fine-structure constant, 164
  \item flat matter dominated (FMD) model, 26, 27, 32---37, 41---42,
        49---52, 56
  \item freezing model, 85
  \item Friedmann equations, 19, 75---77, 236, 334, 335
  \item FRW metric, 16, 30, 75, 134, 144, 157, 166, 212, 226, 294
\end{list}

\begin{list}{}{\itemindent=-3mm\parsep=-1mm\topsep=2mm\leftmargin=3mm}
  \item \textbf{g}alaxy clusters, 14, 42
  \item gamma-ray bursts (GRBs), 29, 70, 169---172
  \item Gaussian fluctuation, 54
  \item gravitational excitons, 137, 141, 142, 148, 153, 156---157, 159,
        164, 204
  \item growth factor, 14, 49---51, 90, 109, 132
  \item growth function, 53, 58
\end{list}

\begin{list}{}{\itemindent=-3mm\parsep=-1mm\topsep=2mm\leftmargin=3mm}
  \item \textbf{H}ALOFIT, 52
  \item harmonic time, 189, 191
  \item harmonic time gauge, 186, 188
  \item High Energy Astrophysical Observatory (HEAO), 58
  \item High-Z Supernova Search, 13, 15
  \item Hubble constant, 13, 33, 44, 45, 63, 64, 68, 108, 157, 214, 281
  \item Hubble parameter, 19, 92, 158, 161, 163, 189---192, 224, 226,
        229, 230, 233, 236---239, 243, 247, 248, 251, 253, 254,
        258, 260---264, 269, 287, 293, 294, 299---300
  \item Hubble Space Telescope (HST), 15, 25, 27, 66, 70, 107
  \item HyRec, 31
\end{list}

\begin{list}{}{\itemindent=-3mm\parsep=-1mm\topsep=2mm\leftmargin=3mm}
  \item \textbf{I}ntegrated Sachs---Wolfe effect, 55---61, 70
\end{list}

\begin{list}{}{\itemindent=-3mm\parsep=-1mm\topsep=2mm\leftmargin=3mm}
  \item \textbf{j}erk, 23, 126
\end{list}

\begin{list}{}{\itemindent=-3mm\parsep=-1mm\topsep=2mm\leftmargin=3mm}
 \item \textsc{K}aluza---Klein models, 134, 145, 191---205
  \item Kaluza---Klein particles, 135
  \item Klein---Gordon equation, 84
\end{list}

\begin{list}{}{\itemindent=-3mm\parsep=-1mm\topsep=2mm\leftmargin=3mm}
  \item \textbf{L}agrangian density, 75
  \item LambdaCDM model ($\Lambda $CDM), 118, 125, 127
  \item LambdaCDM model ($\Lambda$CDM), 14, 34, 44, 63, 70, 107, 134,
        159
  \item large scale structure, 14, 16, 30, 45, 47, 48, 51---53, 55, 57,
        60, 64---66, 70, 72, 75, 80, 82, 93, 97, 110, 119, 132
  \item latent solitons, 198---205
  \item Levi---Civita symbol, 182
  \item light-curve fitter, 28
  \item light-curve fitting, 25, 27---29, 66, 70, 71, 108
  \item light-curve width, 25
  \item likelihood function, 28, 37, 65, 112, 124
  \item loitering Universe, 240, 243, 246, 248---250, 299
  \item Lorentz invariance violation, 166
  \item luminosity distance, 22, 24
  \item Ly$\alpha$-clouds, 52, 54
\end{list}

\begin{list}{}{\itemindent=-3mm\parsep=-1mm\topsep=2mm\leftmargin=3mm}
  \item \textbf{M}-theory, 208
  \item Markov chain Monte Carlo (MCMC), 33, 44, 65, 68, 107, 108, 123---125,
        129, 132
   \item matter dominated (MD) epoch, 20, 158
  \item matter perturbations, 58, 276, 278, 279, 284, 285, 288
  \item MAXIMA, 15, 30
  \item metric perturbations, 77, 80---274
  \item metric tensor, 16, 340
  \item midisuperspace metric, 139
  \item Multicolor Light Curve Shape (MLCS), 25, 27---29, 66---71,
        108---133
  \item multicomponent Universe, 19, 33, 48
  \item multidimensional cosmological models, 134
\end{list}

\begin{list}{}{\itemindent=-3mm\parsep=-1mm\topsep=2mm\leftmargin=3mm}
  \item \textbf{N}ariai metric, 216
  \item no-go theorem, 143
  \item NRAO VLA Sky Survey (NVSS), 59
\end{list}

\begin{list}{}{\itemindent=-3mm\parsep=-1mm\topsep=2mm\leftmargin=3mm}
  \item \textbf{o}pen CDM (OCDM) model, 14
  \item open matter dominated (OMD) model, 26---31
\end{list}

\begin{list}{}{\itemindent=-3mm\parsep=-1mm\topsep=2mm\leftmargin=3mm}
  \item \textbf{p}arameter space, 251
  \item parametrized post-Newtonian (PPN) parameters, 194, 205
  \item perihelion shift, 193, 194, 198, 205
  \item perturbation theory, 33, 78
  \item phantom dark energy, 18, 68, 113, 225
  \item phantom scalar field (PSF), 94, 113---132
  \item Planck mission, 62
  \item positive energy theorem (PET), 303
  \item posterior function, 65---69
  \item potential $U(\phi )$, 76, 83---100, 181
  \item power spectrum, 30, 31, 33, 36---61, 65, 68, 77, 80, 92, 100,
        111, 112, 122, 125, 128---129
  \item primordial perturbation, 15, 45
\end{list}

\begin{list}{}{\itemindent=-3mm\parsep=-1mm\topsep=2mm\leftmargin=3mm}
  \item \textbf{q}uantum instability of the vacuum, 119
  \item quintessence, 13, 18, 68, 84, 107
  \item quintessential scalar field (QSF), 96, 100---113
  \item quintessential scalar fields (QSF), 94
\end{list}

\begin{list}{}{\itemindent=-3mm\parsep=-1mm\topsep=2mm\leftmargin=3mm}
  \item \textbf{r}adiation dominated (RD) epoch, 20, 158, 170
  \item Randall---Sundrum model, 213, 246
  \item RECFAST, 31
\end{list}

\begin{list}{}{\itemindent=-3mm\parsep=-1mm\topsep=2.5mm\leftmargin=3mm}
  \item \textbf{S}$p$-brane, 186
  \item S-brane, 185
  \item Sachs---Wolfe effect, 32, 55
  \item scalar field, 73---133, 136, 140---166, 174, 207
  \item scalar perturbations, 56, 81, 109, 174, 255, 274, 276, 280, 281,
        283---284
  \item scalaron, 157, 176---178, 180---184
  \item Sen---Witten equation (SWE), 304, 319
  \item Shapiro time-delay effect, 194, 199
  \item singularity, 108, 117, 119, 124, 131, 177, 230, 232, 235, 245,
        250, 255---263, 269, 272---273, 299, 330
  \item Skorobohat'ko theorem, 309
  \item Sloan Digital Sky Survey (SDSS), 39, 41, 58
  \item snap, 23
  \item soliton solutions, 195
  \item sound horizon, 40
  \item sound speed, 36, 76, 90, 97
  \item space-time manifold, 137
  \item space-time metric, 17
  \item special orthonormal frame (SOF), 303
  \item Spectral Adaptive Light curve Template (SALT), 25
  \item spinor fields, 311
  \item stable compactification, 140
  \item standard candles, 24
  \item standard CDM model (SCDM), 14
  \item standard rulers, 30
  \item steady-state model, 263
  \item Supernova Cosmology Project (SCP), 13, 15, 230
  \item Supernova Ia, 15, 24
  \item SuperNova Legacy Survey (SNLS), 25, 28
  \item SuperNova Sloan Digital Sky Survey (SN SDSS), 25
  \item synchronous gauge, 33, 46, 80, 109, 120
  \item synchronous time gauge, 143, 150, 187, 189
\end{list}

\begin{list}{}{\itemindent=-3mm\parsep=-1mm\topsep=2mm\leftmargin=3mm}
  \item \textbf{t}achyonic scalar field, 103---105
  \item tensor of electromagnetic field, 166
  \item tensor of extrinsic curvature, 215
  \item tensor Ricci, 18, 79, 196, 201, 202, 215
  \item thawing model, 85
  \item transfer function, 50
  \item two-component Universe, 80
  \item Two-degree Field Galaxy Redshift survey (2dFGRS), 41
  \item Two-Micron All-Sky Survey (2MASS), 58
\end{list}

\begin{list}{}{\itemindent=-3mm\parsep=-1mm\topsep=2mm\leftmargin=3mm}
  \item \textbf{U}nion sample, 25, 26, 69
  \item Universal extra dimension model, 151
\end{list}

\begin{list}{}{\itemindent=-3mm\parsep=-1mm\topsep=2mm\leftmargin=3mm}
  \item \textbf{v}acuum brane, 215
  \item vacuum energy, 73
  \item variable field, 82
  \item visibility function, 40
\end{list}

\begin{list}{}{\itemindent=-3mm\parsep=-1mm\topsep=2mm\leftmargin=3mm}
  \item \textbf{w}eak gravitational lensing, 60
  \item Weyl tensor, 292
  \item WMAP, 30, 34, 37, 57, 88
\end{list}

\begin{list}{}{\itemindent=-3mm\parsep=-1mm\topsep=2mm\leftmargin=3mm}
  \item \textbf{X}-ray gas mass function, 42
  \item X-ray luminosity, 43
\end{list}

}
\end{multicols}


\newpage


 {\parindent=0pt\hspace*{2.9cm}
\vbox{\vspace*{-1.6cm}\hbox{\large\raisebox{0pt}{{\par}}}
\vspace*{0.6cm}\hbox{$\blacksquare$}
 \vspace{-3.0mm}
\hbox{\rule{10cm}{1pt}}
\vspace{2mm}%
\raggedright{\bfseries\sffamily\Large  ABOUT THE AUTHORS}\\%
\vspace*{-1mm} \hbox{\rule{10cm}{1pt}}\par \vspace{39mm}}}

\vspace*{-7mm}

\markboth{About the authors$_{ }$}{About the authors$_{ }$} 
\begin{picture}(10,10)
\put(-7.0,-151.5){\bfseries\sffamily{378}}
\end{picture}

\thispagestyle{empty}

\vspace{-0.3cm} \mbox{ }
\begin{wrapfigure}{l}{3.5cm}
\vskip-4mm \includegraphics[width=3.5cm,height=4cm]{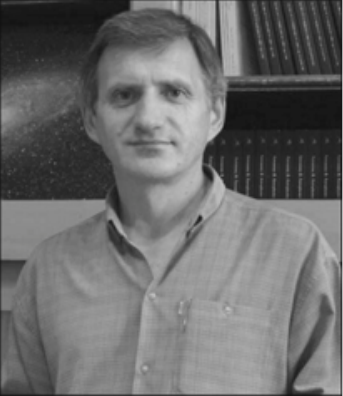}\\
\vskip4mm
\end{wrapfigure} \\\small \textbf{Bohdan NOVOSYADLYJ} is a
  leading researcher of the Astronomical Observatory (AO) and
  professor of the Astrophysics Department of the Ivan Franko National
  University of Lviv (LNU). After graduating from the Lviv State
  University in 1979 he started research activities in the AO LNU.  In
  1987---1991 he was a post-graduate student at the Astro Space Center
  of the Lebedev Physical Institute of the Russian Academy of Sciences
  (Moscow), where he received his PhD degree in 1991. Dr.~Novosyadlyj
  received his Habilitation Doctoral degree in 2007 from the Main
  Astronomical Observatory of the National Academy of Sciences of
  Ukraine. He was a visiting scientist at the Abdus Salam
  International Center for Theoretical Physics (Trieste, Italy), SISSA
  (Trieste, Italy), Potsdam Astrophysical Institute (Germany),
  Department of Theoretical Physics of Geneva University
  (Switzerland).\baselineskip=11pt

 His research interests include relativistic
  astrophysics, cosmology, formation of galaxies and large scale
  structure of the Universe, the nature of dark matter and dark
  energy.

\vspace*{-1mm}

\noindent\hrulefill

\vspace{0.2cm} \mbox{ }
\begin{wrapfigure}{l}{3.5cm}
\vskip-4mm \includegraphics[width=3.5cm,height=4cm]{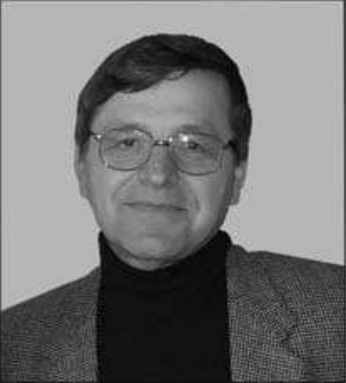}\\
\vskip-4mm
\end{wrapfigure} \\\small \textbf{Volodymyr PELYKH} is the head
  of the Department of Differential Equations at the Pidstryhach
  Institute for Applied Problems of Mechanics and Mathematics of the
  National Academy of Sciences of Ukraine in Lviv. After graduating
  from the Lviv State University in 1971 he started research
  activities at the Institute for Applied Problems of Mechanics and
  Mathematics and worked on his PhD thesis under the supervision of
  V.~Skorobohatko. He received his PhD degree in 1980 from the
  Institute of Phy\-sics of the Academy of Sciences of Belorussia.  In
  2006, he defended his Habilitation Doctoral thesis at the Bogolyubov
  Institute for Theoretical Physics of the National Academy of
  Sciences of Ukraine.\baselineskip=11.0pt

 His research interests include  mathematical physics,
differential equations and  general relativity, in particular, the
Cauchy problem, dark energy and gravitational energy.

\vspace{-0.1cm}

\mbox{ }\begin{wrapfigure}{l}{3.5cm} \vskip-4mm
\includegraphics[width=3.5cm]{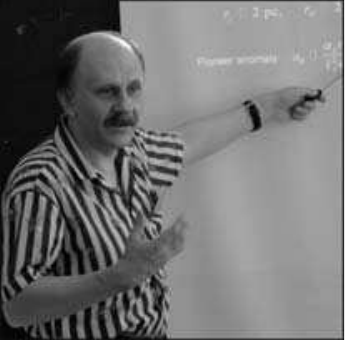}
\end{wrapfigure}  \\\small \textbf{Yuri SHTANOV} is the head of
  the Laboratory of Ast\-ro\-phy\-sics and Cosmology of the Department of
  Ast\-ro\-phy\-sics and Elementary Particles at the Bogolyubov Institute
  for Theoretical Physics of the National Academy of Sciences of
  Ukraine.  After graduating from the Moscow Institute of Physics and
  Technology in 1987, he worked on his PhD thesis at the Lebedev
  Physical Institute in Moscow under the supervision of G.V.~Chibisov
  and V.L.~Ginzburg.  He received his PhD degree in 1991 from the
  Moscow Institute of Physics and Technology.  Since 1991, he works at
  the Bogolyubov Institute for Theoretical Physics of the National
  Academy of Sciences of Ukraine. In 2012, he defended his
  Habilitation Doctoral thesis \mbox{at the same Institute.}\baselineskip=11pt

 Dr.~Shtanov was a visiting scientist at Brown
  University (USA), Inter-University Centre for Astronomy and
  Astrophysics (Pune, India), Nottingham University (UK), Perimeter
  Institute (Canada) and other international scientific institutions.

 Dr.~Shtanov's research interests include the theory
of
  inflationary and post-inflationary universe, the problems of dark
  matter and dark energy, multidimensional and modified gravity and
  cosmology.  His best known works are devoted to the theory of
  post-inflationary preheating and to the effects of dark energy in
  braneworld cosmology.

\vspace*{-1.5mm}

\noindent\hrulefill

\vspace{0.1cm} \mbox{ }
\begin{wrapfigure}{l}{3.5cm}
\vskip-4mm \includegraphics[width=3.5cm]{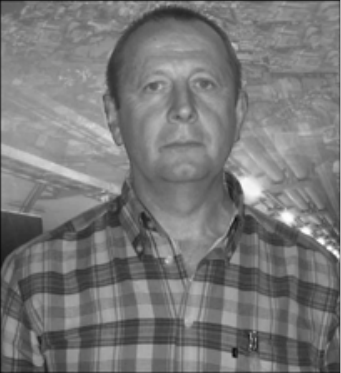}
\end{wrapfigure} \\\small \textbf{Alexander ZHUK} is a Principal
  Scientist at the Research Institute ``Astronomical Observatory'' of
  the Mechnikov National University of Odessa.  After graduating from
  the Moscow Engineering Physical Institute in 1977, he received his
  PhD degree in 1983 from the Lebedev Physical Institute of the
  Academy of Sciences of USSR in Moscow, and a Habilitation Doctoral
  degree in 1999 from the Bogolyubov Institute for Theoretical Physics
  of the National Academy of Sciences of Ukraine.  He worked at the
  Department of Theoretical Physics of the Mechnikov National
  University of Odessa during 1982---2007. During 1987, he was a
  visiting scientist at the Cambridge University where he worked under
  the supervision of Prof.~Steven Hawking. He was also a visiting
  scientist at several leading international scientific institutions
  (CERN (Switzerland), Free University (Germany),
  Albert-Einstein-Institute (Germany), University of Minnesota (USA),
  Princeton University (USA), Columbia University (USA), Tufts
  University (USA), Fermi National Accelerator Laboratory (USA), IMFF,
  CSIC (Spain), etc.). During 2003---2008 and during 2009---2014, he was
  elected a Senior Associate at the Abdus Salam International Center
  for Theoretical Physics in Trieste (Italy). His research interests
  include high energy physics, theories of gravity, multidimensional
  cosmological and gravitational models, structure of the Universe at
  large and small scales, dark matter and dark energy.  He has
  published more than 90 peer-reviewed papers in leading international
  physical journals. He has (co)organized over 10 international
  conferences and was a (co)editor of several conference Proceedings,
  including those published by the Cambridge Scientific Publishers and
  by the American Institute of Physics.\baselineskip=11pt

\newpage
\mbox{ }\thispagestyle{empty}

 \selectlanguage{ukrainian}

\vbox{\vspace*{-1.2cm}\hbox{\raisebox{0.6cm}{\parbox[b]{13cm}{\footnotesize
╩эшур тшёт│Єы■║ ёєўрёэшщ ёЄрэ яЁюсыхьш Єхьэю┐ хэхЁу│┐ Єр яЁхфёЄрты ║
Ёхчєы№ЄрЄш фюёы│фцхэ№ ртЄюЁ│т, ∙ю яЁрЎ■■Є№ т Ў│щ урыєч│. ┬ э│щ
яЁштхфхэ│ ёяюёЄхЁхцєтрэ│ рЁуєьхэЄш,  ъ│ търчє■Є№ эр │ёэєтрээ  Єхьэю┐
хэхЁу│┐, ьхЄюфш Єр юЎ│эъш чэрўхэ№ │ фют│Ёўшї │эЄхЁтры│т ┐┐ юёэютэшї
ярЁрьхЄЁ│т, ьюфхы■трээ  Єхьэю┐ хэхЁу│┐ ёъры Ёэшьш яюы ьш, ┐┐ яю тр т
ьюфхы ї яЁюёЄюЁє-ўрёє ч фюфрЄъютшьш яЁюёЄюЁютшьш тшь│Ёрьш, чюъЁхьр т
ьюфхы ї Єшяє ╩рыєЎш---╩ы щэр,   т ьюфхы ї сЁрэ ч ║фшэшь фюфрЄъютшь
тшь│Ёюь, р Єръюц рэры│чє■Є№ё  яЁюсыхьш, яют' чрэ│ ч яючшЄштэшь
тшчэрўхээ ь  уЁрт│ЄрЎ│щэю┐ хэхЁу│┐ т чруры№э│щ ЄхюЁ│┐ т│фэюёэюёЄ│,
хэхЁухЄшўэ│ єьютш Єр эрёы│фъш ┐ї яюЁє°хээ  т яЁшёєЄэюёЄ│ Єхьэю┐
хэхЁу│┐.
\\
\hspace*{7mm}╠юэюуЁрЇ│  яЁшчэрўхэр фы  эрєъютЎ│т, тшъырфрў│т,
рёя│ЁрэЄ│т Єр ёЄєфхэЄ│т,  ъ│ ёяхЎ│ры│чє■Є№ё  т чруры№э│щ ЄхюЁ│┐
т│фэюёэюёЄ│, ъюёьюыюу│┐ Єр Ї│чшЎ│ хыхьхэЄрЁэшї ўрёЄшэюъ │ яюы│т.

    }}}}

\vspace*{0.5cm} \noindent {\small\it ═рєъютх тшфрээ }

\vspace*{0.2cm} \noindent {\footnotesize ═└╓▓╬═└╦▄═└ └╩└─┼╠▓▀ ═└╙╩
╙╩╨└п═╚\\[0.5mm] \noindent\scriptsize ▓═╤╥╚╥╙╥ ╥┼╬╨┼╥╚╫═╬п ╘▓╟╚╩╚
│ь.\,╠.╠.\,┴╬├╬╦▐┴╬┬└\\[0.5mm]
\noindent ▓═╤╥╚╥╙╥ ╧╨╚╩╦└─═╚╒ ╧╨╬┴╦┼╠ ╠┼╒└═▓╩╚\\ ▓ ╠└╥┼╠└╥╚╩╚
│ь.\,▀.╤.\,╧▓─╤╥╨╚├└╫└\\[0.5mm]
╨└─▓╬└╤╥╨╬═╬╠▓╫═╚╔ ▓═╤╥╚╥╙╥\\[2mm]
╦▄┬▓┬╤▄╩╚╔ ═└╓▓╬═└╦▄═╚╔ ╙═▓┬┼╨╤╚╥┼╥ │ьхэ│\,▓┬└═└\,╘╨└═╩└\\[0.5mm]
╬─┼╤▄╩╚╔ ═└╓▓╬═└╦▄═╚╔ ╙═▓┬┼╨╤╚╥┼╥ │ьхэ│ ▓.▓.\,╠┼╫═╚╩╬┬└}\\[5mm]
\noindent {\normalsize\bfseries\sffamily ╥┼╠═└ ┼═┼╨├▓▀ ▓ ╥┼╠═└ ╠└╥┼╨▓▀ ╙ ┬╤┼╤┬▓╥▓}\\[1mm]
\noindent {\small ╟р ЁхфръЎ│║■ рърфхь│ър ═└═ ╙ъЁр┐эш ┬.\,╪╙╦▄├╚}\\[2mm]
\noindent {\footnotesize ╙ ╥╨▄╬╒ ╥╬╠└╒}\\[5mm]
\noindent {\small  ═╬┬╬╤▀─╦╚╔ ┴юуфрэ, ╧┼╦╚╒ ┬юыюфшьшЁ,\\  ╪╥└═╬┬ ▐Ё│щ, ╞╙╩ ╬ыхъёрэфЁ}\\[2mm]
\noindent{\normalsize\bfseries\sffamily ╥┼╠═└ ┼═┼╨├▓▀:
╤╧╬╤╥┼╨┼╞╙┬└═▓ ╧▓─╤╥└┬╚\\[0.5mm] ▓ ╥┼╬╨┼╥╚╫═▓ ╠╬─┼╦▓}\\[2mm]
\noindent {\footnotesize ╥╬╠ 1}\\[1.5mm]
\noindent {\small  └эуы│щё№ъю■ ьютю■\\[8mm]
\noindent\parbox[t]{6.2cm}{╦│ЄхЁрЄєЁэшщ ЁхфръЄюЁ
\textit{╬.~╤хЁу│║эъю}\\\noindent ╒єфюцэ║ юЇюЁьыхээ  \textit{к. ▓ы№эшЎ№ъюую}}\hspace*{0.6cm}\parbox[t]{6.2cm}{╥хїэ│ўэшщ ЁхфръЄюЁ  \textit{╥.~╪хэфхЁютшў}\\
\noindent ╩юья'■ЄхЁэр тхЁёЄър \textit{╦. ╪ьрурщыю}}\\[1.5mm] \noindent ╙
юЇюЁьыхээ│ юсъырфшэъш тшъюЁшёЄрэю Ёшёєэъш \textit{▓. ╞єър}
\\
\noindent \\[3mm]
\noindent ╧│фя. фю фЁєъє 06.12.2013. ╘юЁьрЄ 70$\times$100/16. ╧ря│Ё
юЇё.\\ ├рЁэ. Computer Modern. ─Ёєъ юЇё. ╙ь. фЁєъ. рЁъ. 31,04.
╬сы.-тшф. рЁъ. 31,29.\\ ╥шЁрц 300 яЁшь. ╟рь. 3764.\vspace*{-1mm}

\noindent\hrulefill

\noindent ┬шфртхЎ№ │ тшуюЄюты■трў ┬шфртэшўшщ ф│ь ``└ърфхьяхЁ│юфшър''
═└═
╙ъЁр┐эш\\
01004, ╩ш┐т-4, тєы. ╥хЁх∙хэъ│тё№ър, 4\\[1mm]
╤т│фюЎЄтю ёєс'║ъЄр тшфртэшўю┐ ёяЁртш ─╩ ╣ 544 т│ф 27.07.2001 Ё.}

\end{document}